\documentclass[12pt,a4paper, twoside]{report}
\usepackage{latexsym}
\usepackage{fancyhdr}
\usepackage{amsfonts}
\usepackage{amssymb}
\usepackage{amsmath}
\usepackage{amsthm}
\usepackage{xcolor}
\usepackage{graphicx} 
\usepackage{indentfirst}
\usepackage{bbm}
\usepackage{enumerate}
\usepackage{appendix}
\usepackage{chngcntr}
\usepackage{etoolbox}
\usepackage[colorlinks = true,
linkcolor = olive,
urlcolor  = blue,
citecolor = blue,
anchorcolor = blue]{hyperref}

\usepackage{verbatim}
\usepackage{mathrsfs}

\usepackage{dsfont}
\usepackage{cite}
\usepackage{xcolor}
\usepackage{enumitem}
\usepackage{ytableau}
\usepackage{tikz}

\usepackage{titlesec}

\allowdisplaybreaks

\AtBeginEnvironment{subappendices}{%
\chapter*{Appendix}
\addcontentsline{toc}{chapter}{Appendices}
\counterwithin{figure}{section}
\counterwithin{table}{section}
}

\newcommand\Tstrut{\rule{0pt}{2.6ex}}         
\newcommand\Bstrut{\rule[-0.9ex]{0pt}{0pt}}   

\graphicspath{{images/}}

\parskip=\medskipamount

\def\a{\alpha}
\def\b{\beta}
\def\c{\chi}
\def\d{\delta}
\def\e{\epsilon}
\def\f{\phi}
\def\g{\gamma}
\def\G{\Gamma}

\def\j{\psi}
\def\k{\kappa}
\def\l{\lambda}
\def\m{\mu}
\def\mub{\bar{\mu}}

\def\o{\omega}
\def\p{\pi}
\def\q{\theta}
\def\r{\rho}
\def\s{\sigma}
\def\t{\tau}

\def\x{\xi}
\def\z{\zeta}
\def\D{\Delta}
\def\F{\Phi}
\def\J{\Psi}
\def\L{\Lambda}
\def\O{\Omega}
\def\P{\Pi}

\def\U{\Upsilon}
\def\X{\Xi}
\newcommand{\ve}{\varepsilon}    
\def\mub{\bar{\mu}}

\newcommand {\cA}{{\cal A}}

\newcommand {\cC}{{\cal C}}
\newcommand {\cD}{{\cal D}}
\newcommand {\cE}{{\cal E}}
\newcommand {\cF}{{\cal F}}

\newcommand {\cH}{{\cal H}}
\newcommand {\cI}{{\cal I}}

\newcommand {\cL}{{\cal L}}
\newcommand {\cM}{{\cal M}}
\newcommand {\cN}{{\cal N}}

\newcommand {\cP}{{\cal P}}
\newcommand {\cQ}{{\cal Q}}
\newcommand {\cR}{{\cal R}}
\newcommand {\cS}{{\cal S}}

\newcommand {\cU}{{\cal U}}
\newcommand {\cV}{{\cal V}}
\newcommand {\cW}{{\cal W}}

\newcommand{\sSL}{\mathsf{SL}}

\newcommand{\sSO}{\mathsf{SO}}

\newcommand{\sOSp}{\mathsf{OSp}}

\newcommand{\ad}{{\dot{\alpha}}}                           
\newcommand{\bd}{{\dot{\beta}}}   
\newcommand{\mud}{{\dot{\mu}}} 
 
\newcommand{\gd}{{\dot\g}}
                              
\newcommand{\cDB}{{\bar\cD}}

\newcommand{\ab}{{\a\b}}

\newcommand{\pa}{\partial}                      
\newcommand{\hf}{\frac12}
\def\rd{{\rm d}}
\def\ri{{\rm i}}
\def\re{{\rm e}}

\newcommand{\od}{\dot{1}}
\newcommand{\td}{\dot{2}}
\newcommand{\FM}{$\mb{M}^4$}

\newcommand{\Po}{Poincar\'e~}
\newcommand{\PaF}{\mathfrak{iso}(3,1)}
\newcommand{\PaT}{\mathfrak{iso}(2,1)}

\newcommand{\HC}{{\mathrm{c.c.}}}
\newcommand{\Qb}{\bar{Q}}
\newcommand{\Psa}{$S\ms{P}$}

\newcommand{\tb}{{\bar{\theta}}}
\newcommand{\1}{\underline{1}}
\newcommand{\2}{\underline{2}}
\newcommand{\lb}{\lbrace}
\newcommand{\rb}{\rbrace}
\newcommand{\na}{\nabla}
\newcommand{\RomanNumeralCaps}[1]
{\MakeUppercase{\romannumeral #1}}

\newcommand{\AMT}{\ensuremath{\int \rd^3x~}}

\newcommand{\AMST}{\ensuremath{\int \rd^{3|2}z~}}

\newcommand{\vf}{\varphi}
\newcommand{\fb}{\bar{\f}}

\newcommand{\be}{\begin{equation}}
\newcommand{\ee}{\end{equation}}
\newcommand{\bsubeq}{\begin{subequations}}
\newcommand{\esubeq}{\end{subequations}}
\newcommand{\ba}{\begin{align}}
\newcommand{\ea}{\end{align}}
\newcommand{\bea}{\begin{eqnarray}}
\newcommand{\eea}{\end{eqnarray}}
\newcommand{\non}{\nonumber}

\newcommand{\mc}{\mathcal}
\newcommand{\mf}{\mathfrak}
\newcommand{\ms}{\mathscr}
\newcommand{\mb}{\mathbb}
\newcommand{\mds}{\mathds}

\newcommand{\ts}{{\tilde{\s}}}

\newcommand{\bm}[1]{\mbox{\boldmath$#1$}}

\titleformat{\paragraph}
{\normalfont\normalsize\bfseries}{\theparagraph}{1em}{}
\titlespacing*{\paragraph}
{0pt}{3.25ex plus 1ex minus .2ex}{1.5ex plus .2ex}

\begin{document}

\pagestyle{fancy}
\pagenumbering{Roman}

\begin{titlepage}

\begin{center}

	{ \LARGE \bf Superspin projection operators and off-shell higher-spin supermultiplets on Minkowski and anti-de Sitter superspace}

	\vspace{1cm}
	
	{\Large{\textbf{Daniel Hutchings}}}\\
	\vspace{0.2cm} 
	\large  
	\vspace{1cm}
	Supervisor: ~~~~~~~~~~~~~~~~Prof. Sergei M. Kuzenko\\
	Co-supervisor:   ~~~~~~~~~~~~A/Prof. Evgeny I. Buchbinder

	\vspace{1.65cm}

	\includegraphics[width=0.38\textwidth]{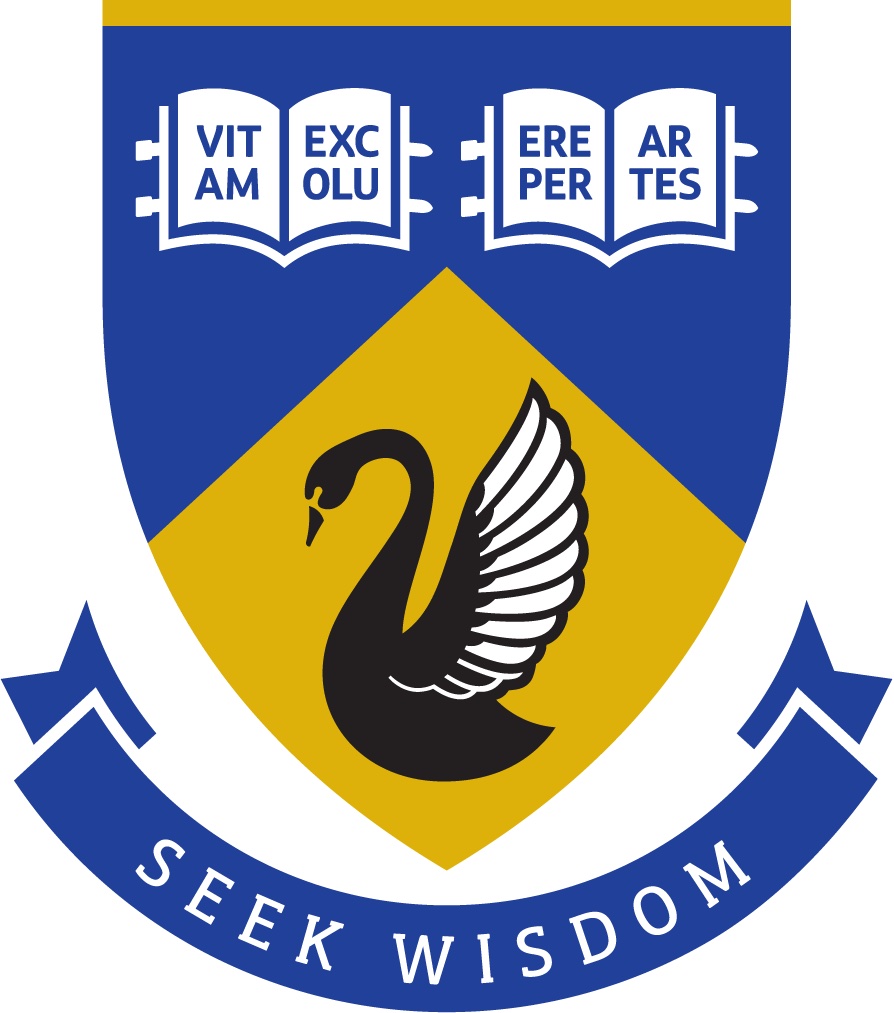}

	\vspace{1.65cm}

	This thesis is presented for the degree of Doctor of Philosophy\\
	The University of Western Australia\\
	Department of Physics\\
	April 2023
	
	\vspace{0.7cm}
	
	\flushleft{
		Examiners:\\
		
		Prof. Ruben Manvelyan ~ \hfill (Yerevan Physics Institute, Armenia) \\
		Prof. Rikard von Unge ~\hfill ~(Masaryk University, Czech Republic)}
	
\end{center}

\end{titlepage}

\newenvironment{changemargin}[3]{%
	\begin{list}{}{%
			\setlength{\topsep}{0pt}%
			\setlength{\leftmargin}{#1}%
			\setlength{\rightmargin}{#2}%
			\setlength{\listparindent}{\parindent}%
			\setlength{\itemindent}{\parindent}%
			\setlength{\parsep}{\parskip}%
		}%
		\item[]}{\end{list}}
\vspace{-1.4cm}
\chapter*{Abstract}\vspace{-1.15cm}
\begin{changemargin}{-0.1cm}{-0.1cm}
	
	This thesis is dedicated to the construction of (super)spin projection operators on maximally symmetric (super)space backgrounds in three and four dimensions. On such a background, the irreducible representations of the associated isometry (super)algebra may be realised on the space of tensor (super)fields satisfying certain differential constraints. The (super)spin projectors isolate the component of an unconstrained (super)field which furnishes the irreducible representation with maximal (super)spin. The explicit form of these (super)projectors are found and an array of novel applications are investigated.
	
	We begin the first half of this thesis by reviewing and elaborating upon the tale of spin projection operators in three-(3$d$) and four-(4$d$) dimensional Minkowski space and the supersymmetric generalisation of the latter. Special emphasis is placed on the interplay between (super)projectors and (super)conformal higher-spin theory as this is a recurring theme of the thesis. We derive the superprojectors in 3$d$ $\cN$-extended Minkowski superspace, for $1 \leq \cN \leq 6$, which are utilised to construct linearised off-shell actions for conformal higher-spin supergravity in terms of unconstrained prepotentials. These theories are then deformed to generate models for massive $\cN$-extended higher-spin supermultiplets.
	
	In the second half of this thesis we study extensions of the (super)spin projectors, and their applications detailed above, to anti-de Sitter (AdS) (super)space. In particular, the superspin projection operators in 4$d$ $\cN=1$ AdS superspace are derived. It is demonstrated that the poles of such superprojectors are naturally associated with partially massless supermultiplets. This allows us to provide a systematic discussion of how to realise all massive and (partially-)massless irreducible representations of the 4$d$ $\cN=1$ AdS superalgebra in terms of on-shell superfields. Similar results are derived in the case of 3$d$ AdS (super)space. We also compute the superprojectors in 4$d$ $\cN=2$ AdS superspace.
	
	Another major component of this thesis consists of a detailed study of massless higher-spin gauge models with $\cN=2$ AdS supersymmetry in three dimensions. Here, there exist two types of $\cN=2$ AdS supersymmetry which are referred to as $(1,1)$ and $(2,0)$ supersymmetry. To better understand the differences between $(1,1)$ and $(2,0)$ higher-spin supermultiplets, we complete the reduction of all known $(1,1)$ supersymmetric massless higher-spin models to $\cN=1$ AdS superspace. We find that every known higher-spin theory with $(1,1)$ AdS supersymmetry decomposes into a sum of two off-shell $(1,0)$ supermultiplets which belong to three series of inequivalent higher-spin gauge models. A similar analysis is performed for all massive higher-spin gauge models with $(1,1)$ AdS supersymmetry.
	
\end{changemargin}

\chapter*{Authorship Declaration}

This thesis is based on four published papers \cite{BHHK, HutchingsHutomoKuzenko, BHKP, HutchingsKuzenkoPonds2021}. Their details are as follows:
\begin{enumerate}
	
	\item E.~I.~Buchbinder, D.~Hutchings, J.~Hutomo and S.~M.~Kuzenko,  \\
	{\it Linearised actions for $ \mathcal{N} $-extended (higher-spin) superconformal gravity,}\\
	JHEP {\bf 1908}, 077 (2019) 
	\href{https://arxiv.org/abs/1905.12476}{[arXiv:1905.12476 [hep-th]]}. \\
 {\bf Location in thesis:} Chapter \ref{ChapThreeDimensionalExtendedMinkowskiSuperspace}.

	\item D.~Hutchings, J.~Hutomo and S.~M.~Kuzenko,  \\
	{\it Higher-spin gauge models with $(1, 1)$ supersymmetry in AdS$_3$: Reduction to $(1, 0)$ superspace,}\\
	Phys. Rev. D \textbf{103}, no.12, 126023 (2021)
	\href{https://arxiv.org/abs/2011.14294}{[arXiv:2011.14294 [hep-th]]}.\\
 {\bf Location in thesis:} Chapters \ref{Chapter2}, \ref{TAChapter3d(super)space} and \ref{TAS211AdS}.

	\item E.~I.~Buchbinder, D.~Hutchings, S.~M.~Kuzenko and M.~Ponds, \\
	{ \it AdS superprojectors,}\\
	JHEP {\bf 2104}, 074 (2021)
	\href{https://arxiv.org/abs/2101.05524}{[arXiv:2101.05524 [hep-th]]}.
{\bf Location in thesis:} Chapter \ref{Chapter4DAdS}.

	\item D.~Hutchings, S.~M.~Kuzenko and M.~Ponds,\\
	{ \it AdS (super)projectors in three dimensions and partial masslessness,}\\
	JHEP \textbf{2110}, 090 (2021), \href{https://arxiv.org/abs/2107.12201}{[arXiv:2107.12201 [hep-th]]}.\\
	{\bf Location in thesis:}  Chapters \ref{Chapter2} and \ref{TAChapter3d(super)space}.

	\end{enumerate}

Permission has been granted to use the above work.

\vspace{-0.2cm} \hspace{0.5cm}

Evgeny Buchbinder

\vspace{-0.2cm}  \hspace{0.5cm}

Jessica Hutomo

\vspace{-0.2cm} \hspace{0.5cm}

Sergei Kuzenko

\vspace{-0.2cm}  \hspace{0.5cm}

Michael Ponds

\vspace{-0.2cm}  \hspace{0.5cm}

\chapter*{Acknowledgements}
The completion of this thesis has been one of the most enjoyable, yet challenging, journeys that I have ever embarked on.
There are many people who have aided my academic pursuits and I would like to take a moment to thank them all for their support.

\subsubsection*{Field theory and quantum gravity group at UWA}
First and foremost, I would like to thank my supervisor Prof. Sergei Kuzenko for his continuous support and guidance throughout my candidature. Your passion and commitment to the pursuit of advancing theoretical physics has been inspiring and instrumental in shaping my academic journey.  Thank you for giving me the opportunity to explore the world of theoretical physics. I am also grateful to my co-supervisor Assoc Prof. Evgeny Buchbinder for the enjoyable collaborations over the whiteboard and his guidance on this thesis.

Notably, I would like to express my deepest gratitude to my dear friend, collaborator and mentor Dr. Michael Ponds. Your unwavering support, guidance and patience were invaluable in completing this thesis. I appreciate the time that you took to not only discuss the grandeurs of physics but also life with me.  A special thanks to my collaborator and friend Dr. Jessica Hutomo. Thank you for your mentorship and collaboration in the early stages of my candidature.

To my close friend and office companion Emmanouil Raptakis, I am grateful that we were able to share this academic journey together, which began in our first semester of university. I will fondly reminisce our office antics and discussions on physics and life. I would also like to thank Benjamin Stone for his friendship and support throughout my candidature. Finally, I would like to thank all current and past members of the field theory and quantum gravity group at The University of Western Australia. In particular Dr. Darren Grasso, James La Fontaine, Nowar Koning, Joshua Pinelli, Jake Stirling and Kai Turner.

\subsubsection*{Family and friends}

Firstly, thank you to my wonderful parents, Vicki and Richard.  I am forever indebted to you both for everything you have provided me in life. The sacrifices you have made to afford me every opportunity will never be forgotten. A special thanks to my brother Alex for always providing me with a well welcomed escape from the stresses of this thesis.

The completion of this thesis would not have possible without the endless support and encouragement from my beautiful partner Wahida. Thank you for always being by my side and supporting me through the challenging periods of my candidature. I am also thankful to my close friends, Conway, Lance, Luke, Mcleod, Samuel and Tyron for always supporting my academic pursuits and providing me with refreshing comedic relief.

It would be remiss of me to not thank my awesome housemates, Eliza, Michael and Sam for their constant encouragement and understanding. I am also grateful to Tahira and Sam for always providing me with good company and delicious food throughout my candidature. Finally, I would like to express my deepest gratitude to anyone whose support I have not mentioned.

\subsubsection*{Financial support}
This research was supported by a Jean Rogerson Postgraduate Scholarship and an Australian Government Research Training Program (RTP) Scholarship at The University of Western Australia.

\subsubsection*{Examiners}
I would like to thank Prof. Ruben Manvelyan and Prof. Rikard von Unge for examining my thesis and for their insightful and stimulating comments which have motivated future research directions. 

{\hypersetup{hidelinks}
	\tableofcontents
}
\clearpage\pagenumbering{arabic}

\chapter{Introduction} \label{Introduction}
Einstein's theory of special relativity revolutionised our theoretical understanding of the structure of the Universe, unifying the previously independent entities of space and time into the framework of spacetime. This profound result was founded on the two fundamental ideas that in any inertial system: (i) the laws of physics have the same form; and (ii)  the speed of light is constant. These principles are equivalent to the mathematical statement that the \Po group is a symmetry group of any closed system in Minkowski space. The successful use of symmetry in the formulation of special relativity ushered in a paradigm shift in theoretical physics, which saw this Poincar\'e, or relativistic, symmetry become the new guiding principle in the construction of theories attempting to describe the physical processes of nature.

Symmetry is ubiquitous in modern physics as all physical systems should be compatible with special relativity, i.e. possess \Po symmetry. This allows for the universal classification of elementary particles as unitary irreducible representations of the \Po group.  The fundamental correspondence between elementary particles propagating in four-dimensional Minkowski space $\mb{M}^4$ and unitary irreducible representations (UIRs) of the \Po algebra $\PaF$ was established by Wigner in his seminal work \cite{Wigner1939} in 1939. Here, Wigner completed the classification of the UIRs of $\PaF$ and showed that they are characterised by two parameters, mass ${m}$ and spin $s$. This groundbreaking identification resides at the heart of all modern relativistic field theories describing elementary particles.

Our best theoretical descriptions of the physical phenomenon that we observe in Nature are the Standard Model (SM) of particle physics and Einstein's General relativity (GR). These are examples of particle theories that realise UIRs of the \Po algebra at the linearised level. The Standard Model is a quantum field theory which describes the fundamental interactions of Nature dominant at the subatomic scale (the electromagnetic, the strong and the weak forces). 
The particle content encapsulated by the SM includes spin-$\hf$ fermions (matter), spin-$1$ gauge bosons which mediate interactions and the Higgs field (scalar boson).
On the other hand, GR is a classical gauge theory of the \Po group which concerns the physics governing Nature at large scales. Einstein's gravity is conjectured to be mediated by a massless spin-$2$ particle known as the graviton. Both of these theories have proven to be very successful, predicting many physical processes which have been experimentally verified to great accuracy.  It appears that the current theories which best describe the physical processes observed in Nature are realised in terms of particles carrying only spin $s \leq 2$. However, from a mathematical viewpoint, there is nothing  preventing the existence of particles with spin $s>2$ in Wigner's framework. A field (or particle) is said to be of \textit{higher-spin} if it possesses spin $s >2$.

The study of higher-spin fields has attracted much attention recently due to its inextricable connection to various topics in fundamental physics.\footnote{There have been extensive contributions to the development of higher-spin theory. For more information on these works, and also the current state of affairs, we recommend the review articles \cite{Vasiliev2003cph,Sorokin2004ie,Bouatta2004kk,Bekaert2004qos,Vasiliev2011zza,Bekaert2010hw,Vasiliev2014lt,Didenko2014dwa,Rahman2015pzl,Bekaert:2022poo,Ponomarev2022,Bengtsson} and references therein.}  
Its origin in four-dimensional Minkowski space can be attributed to the pioneering contributions of Majorana \cite{Majorana1932chs}, Dirac \cite{Dirac1936}, and Fierz \& Pauli \cite{Fierz1939, FierzPauli1939} in the 1930's, which culminated in the development of the relativistic field equations for massive fields of arbitrary spin \cite{Fierz1939, FierzPauli1939}.
The field-theoretic formulation of massive higher-spin fields \cite{Fierz1939, FierzPauli1939}
appeared before the seminal group-theoretic work of Wigner \cite{Wigner1939}. The field- and group-theoretic pictures were eventually unified in the later work of Bargmann and Wigner \cite{BargmannWigner1948} in 1948. Here, it was shown that the
wave equation that a field must satisfy in order to realise a massive UIR of $\PaF$ is equivalent to that appearing in the Fierz-Pauli equations \cite{Fierz1939, FierzPauli1939}. In this thesis, we will refer to the appropriate conditions which ensures that a tensor field furnishes a UIR of the isometry algebra as \textit{on-shell conditions}.\footnote{Some people pay homage by referring to these conditions as the Bargmann-Wigner equations, while others refer to them as the Fierz-Pauli equations. They are also known as irreducible field representations.}

The on-shell conditions are an essential ingredient in the formulation of any relativistic field theory in particle physics. Naturally, they should result from the Euler-Lagrange equations (the equations of motion) derived from an appropriate Lagrangian. 
The quest of seeking Lagrangian formulations which generate the Bargmann-Wigner equations for arbitrary spin was initiated by Fierz and Pauli in $\mb{M}^4$, who constructed Lagrangian theories for fields of spins $s=\frac{3}{2},2$ \cite{Fierz1939, FierzPauli1939}.\footnote{Fierz and Pauli \cite{Fierz1939, FierzPauli1939} did not comment on the gauge invariance of the massless spin $s=\frac{3}{2},2$ actions. The gauge symmetry for the massless spin $s=\frac{3}{2}$ action was first discussed by Rarita and Schwinger in \cite{Rarita1941mf}.} This program, which is aptly known as the Fierz-Pauli program, was completed for massive fields of arbitrary spin by Singh and Hagen \cite{SinghHagen1974Bos,SinghHagen1974Ferm} 35 years later in 1974. Building on this work, free field theories describing massless particles of arbitrary spin were later derived in four-dimensional Minkowski space \cite{Fronsdal1978Massless,FangFronsdal} and anti-de Sitter (AdS$_4$) space\cite{Fronsdal1979Sing,Fronsdal1979} by Fang and Fronsdal. Lagrangian descriptions of free field theories are essential as they are the starting point for:  constructing interacting (non-linear) extensions; and analysing the quantum properties of these theories. 


In this thesis, we will be interested in studying a certain type of non-local differential operator, the so-called spin projection operator, which extracts the physical part of an unconstrained field, i.e. the component satisfying the on-shell conditions with fixed maximal spin in $\mb{M}^4$. Our main objective is to construct novel generalisations of these operators in maximally symmetric backgrounds in three and four dimensions, and their supersymmetric analogues. Before describing the nature of these spin projection operators, it is necessary to first elaborate on the constraint space which these operators project onto.

\subsubsection{Irreducible (super)field representations}
The program of determining the on-shell conditions for (spin-)tensor fields of arbitrary rank in $\mb{M}^4$ was initiated by Bargmann and Wigner \cite{BargmannWigner1948}.\footnote{This program is aptly referred to as the Bargmann-Wigner program in the higher-spin community.}
Let us denote by $G(m,s)$ the massive UIR of $\PaF$ carrying mass $m>0$, integer spin $s \geq 0$ and positive definite energy. A totally symmetric real rank-$s$ tensor field $\f_{a_1 \ldots a_s}(x) = \f_{(a_1 \ldots a_s)}(x) \equiv \f_{a(s)}$ in $\mb{M}^4$ is said to furnish $G(m,s)$ if it satisfies the on-shell conditions \cite{BargmannWigner1948}
\bsubeq \label{FMBosonicOnShellFierz}
\bea 
\eta^{bc}\f_{bca(s-2)} &=& 0~, \label{FMBosonicOnShellFierzTraceless}\\
\pa^b \f_{b a(s-1)} &=& 0~, \label{FMBosonicOnShellFierzTransverse}\\
(\Box -  {m}^2) \f_{a(s)} &=& 0~. \label{FMBosonicOnShellFierzKG}
\eea
\esubeq
In other words, a field $\f_{a(s)}$ that is traceless \eqref{FMBosonicOnShellFierzTraceless}, transverse (divergenceless) \eqref{FMBosonicOnShellFierzTransverse} and satisfies the Klein-Gordon equation \eqref{FMBosonicOnShellFierzKG}  describes a bosonic particle carrying mass ${m}$ and spin $s$.
Note that in the case $s=0$, a massive field only satisfies  \eqref{FMBosonicOnShellFierzKG}, while for the case $s=1$ , it obeys \eqref{FMBosonicOnShellFierzTransverse} and \eqref{FMBosonicOnShellFierzKG}.

For integer $s>0$, a symmetric rank-$s$ Majorana tensor-spinor field $\bm{\j}_{a_1 \ldots a_s} (x)= \bm{\j}_{(a_1 \ldots a_s)}(x) \equiv \bm{\j}_{a(s)}$\footnote{We have omitted the spinor index of the Majorana spinor field $\bm{\j}_{a(s)} $.} realises $G({m}, s+\hf)$ if it obeys the on-shell conditions \cite{BargmannWigner1948}\bsubeq \label{FMFermionicOnShellFierz}
\bea
\pa^b \bm{\j}_{b a(s-1)} &=& 0~, \label{FMFermionicOnShellFierzTransverse}\\
\g^b \bm{\j}_{b a(s-1)}&=&0~, \label{FMFermionicOnShellFierzGammaTraceless} \\
(\ri \g^b \pa_b + {m}) \bm{\j}_{a(s)}&=& 0~. \label{FMFermionicOnShellFierzDE} 
\eea
\esubeq
\
Thus, a Majorana spinor-tensor field $\bm{\j}_{a(s)}$ which is transverse \eqref{FMFermionicOnShellFierzTransverse}, gamma traceless \eqref{FMFermionicOnShellFierzGammaTraceless} and obeys the Dirac equation \eqref{FMFermionicOnShellFierzDE}  describes a fermionic particle carrying mass ${m}$ and spin-$(s+\hf)$. A massive spin-$\frac{1}{2}$ field only satisfies the Dirac equation \eqref{FMFermionicOnShellFierzDE}.

In the massless case $m = 0$, the system of equations \eqref{FMBosonicOnShellFierz} and \eqref{FMFermionicOnShellFierz} admit a gauge symmetry, which accounts for the additional non-physical degrees of freedom. The  on-shell conditions for massless gauge fields of arbitrary spin in $\mb{M}^4$ were first studied in detail by Fronsdal and Fang in \cite{Fronsdal1978Massless,FangFronsdal}.\footnote{Bargmann and Wigner \cite{BargmannWigner1948} found that a field which furnishes the massless UIR of $\PaF$ with discrete spin must satisfy the massless Klein-Gordon equation. They did not detail the supplementary conditions, nor the gauge symmetry, associated with these fields.}

The on-shell conditions for a superfield with arbitrary superspin in four-dimensional $\cN=1$ Minkowski superspace $\mb{M}^{4|4}$ were first derived by Sokatchev \cite{Sokatchev1975} in 1975 (see also \cite{Sokatchev1981, SiegelGates1981}). Here, it was shown that chiral superfields realise irreducible representations of the super \Po algebra. In the same year, the on-shell conditions which select out the maximal superspin component of a higher-rank superfield were given by Howe, Stelle and Townsend \cite{HoweStelleTownsend1981} in the context of irreducible supercurrents. The on-shell conditions of \cite{Sokatchev1981,SiegelGates1981,HoweStelleTownsend1981} will be treated in more detail in section \ref{SecFourDimensionalMinkowskiSuperSpace}.

\subsubsection{(Super)spin projection operators}
A totally symmetric traceful field in $\mb{M}^4$ (which is otherwise unconstrained) satisfying the Klein-Gordon equation \eqref{FMBosonicOnShellFierzKG}  realises the  reducible representation of $\PaF$ 
\be \label{reducible}
G(m,s)\oplus G(m,s-1)\oplus 2 G(m,s-2) \oplus \cdots = \bigoplus^s_{j=0} \big ( \lfloor \frac{j}{2} \rfloor + 1 \big ) G(m, s-j) ~,
\ee
which contains a collection of UIRs with varying multiplicities.
The purpose of the spin projection operator is to select the maximal spin UIR $G(m,s)$ from this decomposition.

The spin projection operators, or transverse and traceless (TT) projectors, were
first derived in $\mb{M}^4$ by Fronsdal
in the late 1950's \cite{ Fronsdal1958} (see also \cite{BehrendsFronsdal1957}).\footnote{Isaev and Podoinitsyn \cite{IsaevPodoinitsyn2017,IsaevPodoinitsyn2018} refer to the spin projection operators, which appeared in the literature for the first time in the work of Behrends and Fronsdal \cite{BehrendsFronsdal1957}, as Behrends-Fronsdal spin projection operators. We will refrain from using this terminology as the authors of \cite{BehrendsFronsdal1957} credit the derivation of the spin projectors to Fronsdal, citing the work \cite{ Fronsdal1958} which was published in the following year.}
For any integer $s \geq 2$, let us denote ${V}^{\text{T}}_{(s)}$ the space of totally symmetric traceful rank-$s$ tensor fields ${\f}^{\text{T}}_{a(s)}$, and ${V}_{(s)}$ the space of symmetric traceless rank-$s$ tensor fields $\f_{a(s)}$. Fronsdal \cite{ Fronsdal1958} constructed the spin projection operator $\P^{\perp}_{(s)}$ which is defined by its action on ${V}^{\text{T}}_{(s)}$ via the rule 
\bea \label{FMBFProjectorMap}
\P^{\perp}_{(s)}: {V}^{\text{T}}_{(s)} &\longrightarrow& V_{(s)}~, \\
{\f}^{\text{T}}_{a(s)} &\longmapsto& \P^{\perp}_{a(s)}({\f}^{\text{T}}) ~, \qquad \eta^{bc}{\f}^{\text{T}}_{bca(s-2)} \neq 0~, \non
\eea
where the image $\P^{\perp}_{a(s)}({\f}^{\text{T}})  \equiv \P^{\perp}_{(s)} {\f}^{\text{T}}_{a(s)}$  satisfies the properties:
\bsubeq \label{FMBFProjectorPropeperties}
\begin{enumerate} 
	\item \textbf{Idempotent}: the operator $\P^{\perp}_{(s)}$ is a projector,
	\be
	\P^{\perp}_{(s)}\P^{\perp}_{(s)} {\f}^{\text{T}}_{a(s)} = \P^{\perp}_{(s)}{\f}^{\text{T}}_{a(s)}  ~.
	\ee
	\item \textbf{Symmetric}: the operator $\P^{\perp}_{(s)}$ maps ${\f}^{\text{T}}_{a(s)}$ to a totally symmetric field,
	\be \label{FMBFSymmetricProp}
	\P^{\perp}_{a_1 \cdots a_s}({\f}^{\text{T}}) =\P^{\perp}_{a(s)}({\f}^{\text{T}})~.
	\ee
	\item \textbf{Traceless}: the operator $\P^{\perp}_{(s)}$ maps ${\f}^{\text{T}}_{a(s)}$ to a traceless field,
	\be \label{FMBosonicProjectorTraceless}
	\eta^{b c}\P^{\perp}_{bc a(s-2)}({\f}^{\text{T}})  = 0~.
	\ee
	\item \textbf{Transverse}: the operator $\P^{\perp}_{(s)}$ maps ${\f}^{\text{T}}_{a(s)}$ to a transverse field,
	\be \label{FMBosonicTransverse}
	\partial^b\P^{\perp}_{b a(s-1)}({\f}^{\text{T}})  = 0~.
	\ee
\end{enumerate}
\esubeq
The constraints \eqref{FMBFSymmetricProp}, \eqref{FMBosonicProjectorTraceless} and \eqref{FMBosonicTransverse} eliminate all lower spin modes contained in the projected field $\f^{\text{T}}_{a(s)}$, thus ensuring  that the highest-spin component of the bosonic field is selected. In other words, the spin projection operator $\P^{\perp}_{(s)}$ maps an unconstrained tensor field $\f^{\text{T}}_{a(s)}$ to a pure spin-$s$ field. Moreover, 
if the field  $\f^{\text{T}}_{a(s)}$ is on the mass-shell \eqref{FMBosonicOnShellFierzKG}, it follows from \eqref{FMBosonicOnShellFierz}
that the operator $\P^{\perp}_{(s)}$ singles out the component of $\f^{\text{T}}_{a(s)}$ that furnishes the UIR $G({m},s)$ with maximal spin $s$.

The fermionic counterpart is computed in a similar manner. Let us denote
${\bm{V}}^{\text{T}}_{\hspace{-0.15cm}(s+\hf)}$ the space of four-component (gamma) traceful spin-tensor fields ${\bm{\j}}^{\text{T}}_{a(s)}$, and $\bm{{V}}_{\hspace{-0.15cm}(s+\hf)}$ the space of totally symmetric (gamma) traceless rank-$s$ tensor fields $\bm{{\j}}_{a(s)}$. For integer $s \geq 2$,  the spin projection operator  $\P^{\perp}_{(s+\hf)}$ acts on ${\bm{V}}^{\text{T}}_{\hspace{-0.15cm}(s+\hf)}$ according to the rule \cite{ Fronsdal1958}
\bsubeq \label{FMBFProjectorMapFermionic}
\bea 
\P^{\perp}_{(s+\hf)}: {\bm{V}}^{\text{T}}_{\hspace{-0.15cm}(s+\hf)} &\longrightarrow& {\bm{V}}_{\hspace{-0.15cm}(s+\hf)}~, \\
{\bm{\j}^{\text{T}}}_{\hspace{-0.25cm} a(s)} &\longmapsto& \P^{\perp}_{a(s)}({\bm{\j}}^{\text{T}}) ~, \qquad \g^{b}{\bm{\j}}^{\text{T}}_{ba(s-1)} \neq 0~.\non
\eea
By definition, the image $\P^{\perp}_{a(s)}(\bm{\j}^{\text{T}})  \equiv \P^{\perp}_{(s+\hf)} \bm{\j}^{\text{T}}_{a(s)}$ satisfies the properties \eqref{FMBFProjectorPropeperties}, but the traceless property \eqref{FMBosonicProjectorTraceless} is replaced by:

\begin{enumerate}
	\item \textbf{Gamma traceless}: the operator $\P^{\perp}_{(s+\hf)}$ maps $\bm{\j}^{\text{T}}_{a(s)}$ to a gamma traceless field
	\be \label{FMFermionicProjGammaTraceless}
	\g^{b}\P^{\perp}_{b a(s-1)}(\bm{\j}^{\text{T}})  = 0~.
	\ee
\end{enumerate}
\esubeq
Analogous to the bosonic case, the spin projector $\P^{\perp}_{(s+\hf)}$ maps an arbitrary spinor-tensor field ${\bm{\j}}^{\text{T}}_{a(s)}$ to a pure spin-$(s+\hf)$ state.
Let us consider a traceful spin-tensor field which satisfies the Dirac equation \eqref{FMFermionicOnShellFierzDE}. It follows from the defining properties of $\P^{\perp}_{(s+\hf)}$ that the spin projection operator $\P^{\perp}_{(s+\hf)}$ selects the physical component from ${\bm{\j}}^{\text{T}}_{a(s)}$ which furnishes $G({m}, s+\hf)$ with maximal spin $(s+\hf)$.

It was shown by Aurilia and Umezawa \cite{AuriliaUmezawa1967,AuriliaUmezawa1969} that the bosonic \eqref{FMBFProjectorPropeperties} and fermionic \eqref{FMBFProjectorMapFermionic} spin projection operators can be expressed  in terms of the Casimir operators of $\PaF$. In two-component spinor notation, the spin projection operators were generalised to spinor fields of arbitrary rank (including mixed symmetry fields) by Siegel and Gates \cite{SiegelGates1981} (see also \cite{GatesGrisaruRocekSiegel1983}). The TT projectors  were first generalised to $d$-dimensional Minkowski space for $d>2$ by Segal \cite{Segal2003} (see also \cite{FranciaMouradSagnotti2007, PonomarevTseytlin2016, Bonezzi2017,IsaevPodoinitsyn2017, IsaevPodoinitsyn2018}) in the bosonic case, and later by Isaev and Podoinitsyn \cite{IsaevPodoinitsyn2018} for fermionic fields.

Many applications for the TT projectors have been found within the landscape of high-energy physics. They are known to determine the structural form of propagators for massive fields of arbitrary spin \cite{Singh1981} (see also \cite{IsaevPodoinitsyn2018}).\footnote{The propagators for CHS fields were formulated in terms of the TT projectors in $\mb{M}^4$ in \cite{JoungNakachTseytlin}.} The spin projection operators were also essential in Fronsdal's Lagrangian formulation for a massive spin-$\frac{3}{2}$ particle \cite{Fronsdal1958}, which was later extended to spins $ s \leq 4$ by Chang \cite{Chang1967zzc}. Specifically, given a traceful bosonic field $\f^{\text{T}}_{a(s)}$, Fronsdal \cite{Fronsdal1958} introduced the equation
\be \label{IntroMassiveEquation}
m^2 \f^{\text{T}}_{a(s)} = \Box \P^{\perp}_{(s)} \f^{\text{T}}_{a(s)}~.
\ee
Making use of the projector properties \eqref{FMBFProjectorPropeperties}, it can be easily shown that equation \eqref{IntroMassiveEquation} encodes the massive on-shell conditions \eqref{FMBosonicOnShellFierz}. An analogous equation was also found for half-integer spin fields in \cite{Fronsdal1958} .

The spin projectors are also important tools which can be used to decompose any symmetric tensor field of arbitrary rank into irreducible components. For example, the irreducible decomposition of a rank-$2$ field was crucial in the canonical formulation of general relativity, see e.g. \cite{Arnowitt1960es, Deser67, York73,York74,GibbonsPerry}. Specifically, the projectors $\P^{\perp}_{(s)}$ with $s \leq 2$ can be used to decompose a rank-$2$ symmetric tensor field $g^{\text{T}}_{ab}$ in the following manner
\be
g^{\text{T}}_{a b} =   g^{\text{TT}}_{a b }  + \pa_{(a} g^{\perp}_{b)} + \pa_a \pa_b g + \eta_{ab}\tilde{g}~,
\ee
where $g$ and $\tilde{g}$ are scalar, $g^{\perp}_a$ is transverse and $g^{\text{TT}}_{a b }$ is traceless and transverse. The defining properties of these irreducible components are inherited immediately from  $\P^{\perp}_{(s)}$, thus illustrating their importance in the decomposition scheme. Furthermore, these decompositions prove invaluable in the framework of path-integral quantisation (see e.g. \cite{GGS,BKS}).

The superspin projection operators (or superprojectors) in four-dimensional $\cN=1$ Minkowski superspace were introduced by Salam and Strathdee \cite{SalamStrathdee1975} in the case of a scalar superfield. Shortly after, these operators were extended to certain constrained superfields of arbitrary rank by Sokatchev  \cite{Sokatchev1975} and were formulated in terms of Casimir operators.
A few years later Rittenberg and Sokatchev \cite{RittenbergSokatchev1981} made use of a similar method to construct the  superprojectors in $\cN$-extended Minkowski superspace $\mb{M}^{4|4\cN}$ (see also \cite{Sokatchev1981}).
An alternative powerful construction of the superprojectors
in $\mb{M}^{4|4\cN}$ was given in \cite{SiegelGates1981,GatesGrisaruRocekSiegel1983}, which requires no knowledge of the Casimir operators of the $\cN$-extended Minkowski superalgebra.  In the early days of supersymmetry, the superspin projection operators of \cite{RittenbergSokatchev1981,SiegelGates1981} were used in the construction of superfield equations of motion \cite{OS1,OS2} and  gauge-invariant  actions \cite{GS,GKP} in $\mb{M}^{4|4}$. One of the main goals of this thesis is to compute the (super)spin projection operators in three-dimensional $\cN$-extended superspace.

\subsubsection{(Super)conformal higher-spin theory}
One of the alluring applications of the spin projection operators is that they can be used as a tool to construct linearised conformal higher-spin theories. Conformal higher-spin (CHS) theories were first studied in the pioneering work of Fradkin and Tseytlin \cite{FradkinTseytlin1985} as generalisations of Maxwell's electrodynamics $(s=1)$, the conformal gravitino model $(s=\frac{3}{2})$ and conformal gravity $(s=2)$ in $\mb{M}^4$. By construction, the linearised CHS actions of \cite{FradkinTseytlin1985} describe  pure spin fields of arbitrary spin off the mass-shell (see also \cite{Fradkin1989md,Tseytlin2002gz,Segal2003}). 
For $s \geq 1$, the free bosonic CHS model of \cite{FradkinTseytlin1985} is described in terms of a totally symmetric real traceful rank-$s$ tensor field
$ {h}^{\text{T}}_{a(s)}$ which is defined modulo gauge transformations of the form\footnote{The gauge freedom of  $\l_{a(s-2)}$ can be used to make ${h}^{\text{T}}_{a(s)}$ traceless i.e. ${h}^{\text{T}}_{a(s)}=h_{a(s)}$, where $ \eta^{bc}h_{bca(s-2)}=0$. In this thesis, we will only work with totally symmetric and traceless fields.}
\be \label{FMBosCHSGaugeTrans}
\d_{\x,\l} {h}^{\text{T}}_{a(s)} = \pa_{(a_1} \x_{a_2 . . . a_s)} + \eta_{(a_1 a_2}\l_{a_3 . . .a_s)}~, \qquad \eta^{bc} \x_{b c a(s-3)} = 0~. 
\ee
The gauge parameters $\x_{a(s-1)}$ and $\l_{a(s-2)}$ are symmetric while $\x_{a(s-1)}$ is also traceless. The differential and algebraic gauge transformations in \eqref{FMBosCHSGaugeTrans} can be identified as higher-spin generalisations of linearised diffeomorphisms and Weyl transformations of conformal gravity respectively.

The spin-$s$ linearised conformal model of \cite{FradkinTseytlin1985} is given by 
\be \label{FMCHSBos}
S^{(s)}_{\text{CHS}} [ {h}] \propto \int \rd^4x~ {h}^{\text{T} a(s)} \Box^s \P_{(s)}^{\perp} {h}^{\text{T}}_{a(s)}~.
\ee
It must be noted that the spin projection operators $\P_{(s)}^{\perp}$ were not given for spins $s>2$ in \cite{FradkinTseytlin1985}.
The properties \eqref{FMBFProjectorPropeperties} of the projector ensure that the action \eqref{FMCHSBos} is invariant under the gauge transformations \eqref{FMBosCHSGaugeTrans}. 
The fermionic spin projection operator  $\P^{\perp}_{(s+\hf)}$ \eqref{FMBFProjectorMapFermionic} also appears explicitly in the fermionic CHS action \cite{FradkinTseytlin1985}.

It is apparent from the structural form of the action \eqref{FMCHSBos} that there is an intimate connection between CHS theories and spin projection operators.\footnote{The free CHS actions of \cite{FradkinTseytlin1985} were purely schematic for spins $s>2$, as only the explicit form of the spin projection operators were given for the cases $s \leq 2$.} This correspondence follows naturally from the definition of the spin projectors of \cite{Fronsdal1958}, which we recall are the unique differential operators which select the maximal pure spin component from an unconstrained (spin-)tensor. Thus, the presence of the spin projectors in the CHS actions ensure that the theory describes a single pure spin state off the mass-shell.\footnote{The 3$d$ CHS actions of \cite{PopeTownsend1989,Kuzenko2016} were recently constructed in terms of spin projectors in \cite{BuchbinderKuzenkoLaFontainePonds2018}. }

\sloppy{
	Many efforts have been invested into the development of CHS theories in the past decade due to their desirable properties. For example, in contrast to the massless case where no local non-linear Lagrangian description exists,\footnote{At the level of the equations of motion, a fully interacting theory of massless fields was derived in (A)dS space by Vasiliev \cite{Vasiliev1990,Vasiliev2003ev}.} Segal \cite{Segal2003} (see also \cite{Tseytlin2002gz,Bekaert2010ky,Bekaert2009ud,Bonezzi2017}) constructed the Lagrangian for a complete interacting bosonic CHS theory in even $d-$dimensional Minkowski space, for $d \geq 4$.
	The existence of a consistent non-linear action functional for CHS fields also allows for the study of its quantum properties \cite{Tseytlin2002gz,Bekaert2010ky,Tseytlin13,Beccaria2014jxa,Beccaria2014xda,Beccaria2015vaa,Beccaria2016syk,JoungNakachTseytlin,Hahnel2016ihf,Adamo2018srx}. In addition, CHS theories play a natural role in the context of the AdS/CFT correspondence \cite{Liu1998bu,Metsaev1999ui, Tseytlin2002gz,Segal2003,Metsaev2009ym,Giombi2013yva,Giombi2014iua,Beccaria2016tqy}.}

Supersymmetric generalisations of CHS fields were first discussed by Howe, Stelle and Townsend in the context of supercurrent multiplets via the Noether procedure in 1981 \cite{HoweStelleTownsend1981}. The SCHS superfields provided here were only for half-integer superspin. The first on-shell\footnote{A supermultiplet is said to be off-shell if the algebra of supersymmetry transformations closes off the equations of motion. Otherwise, the supermultiplet is called on-shell.} Lagrangian description of the half-integer supermultiplets of \cite{HoweStelleTownsend1981} was given by Fradkin and Tseytlin \cite{FradkinTseytlin1985} in 1985. Off-shell superconformal higher-spin multiplets and their corresponding actions were constructed over 30 years later in 2017 by Kuzenko, Manveylan and Theisen \cite{KuzenkoManvelyanTheisen2017}  in four-dimensional Minkowski  and anti-de Sitter superspace. 
The off-shell $\cN=1$ SCHS actions of \cite{KuzenkoManvelyanTheisen2017} were generalised two years later  to conformally-flat backgrounds by Kuzenko and Ponds \cite{KuzenkoPonds2019}. The $\cN$-extended superconformal free higher-spin models were recently formulated in conformally-flat backgrounds by Kuzenko and Raptakis in \cite{KR}. An interacting theory of SCHS fields was recently proposed in \cite{Kuzenko2022hdv} (see also \cite{Kuzenko2022qeq}).

All of the above SCHS theories were constructed independently of superspin projection operators. Recasting these theories in terms of superprojectors is beneficial for a variety of reasons. Firstly, it elucidates the defining feature of  SCHS theories that they describe maximal superspin states off the mass-shell. In addition, it also makes properties of the action such as gauge invariance manifest. The interplay of (S)CHS theories and (super)spin projection operators will be a prominent theme of this thesis.

\subsubsection{(Super)spin projection operators in (A)dS}
The (super)spin projection operators have proven to be very useful in Minkowski (super)space due to their many interesting applications.
For this reason, it is natural to consider generalisations of these operators to (super)space backgrounds other than Minkowski (super)space. The next logical step is to consider the formulation of the (super)spin projection operators in curved (super)space backgrounds, with the simplest starting point being  (anti-)de Sitter space. Upon initiating this program, one immediately encounters problems due to the presence of non-vanishing curvature, which makes constructing the (super)spin projectors technically challenging. Despite this, recent progress was made in this pursuit by Kuzenko and Ponds in 2020 \cite{KP20}, where the spin projection operators were computed in four-dimensional (anti-)de Sitter space (A)dS$_4$.

One of the important outcomes of \cite{KP20} was a new understanding of the
so-called partially massless fields in AdS$_4$. First discovered by Deser and Nepomechie \cite{DeserN1} in 1983, partially massless fields are an exotic species of on-shell fields in AdS$_4$ which have no direct analogue in Minkowski space. As the name suggests, the defining properties of these fields are intermediate of both massive and massless fields, in the sense that they possess a higher-depth gauge symmetry. Partially massless fields have been at the centre of extensive studies for the past 35 years \cite{DW2,DeserN1,DeserN2,Higuchi1,Higuchi2,Higuchi3,DeserW1,DeserW3,DeserW4,Zinoviev,Metsaev,BrinkMetsaevVasiliev,SV,BG}.
It was shown in \cite{KP20} that the poles of the spin projection operators correspond to partially massless fields. 
Thus, the spin projection operators provided a novel program to complete the dictionary of on-shell fields in AdS$_4$.

Another goal of this thesis is to study extensions of the spin projection operators to (anti-)de Sitter space, and their corresponding supersymmetric generalisations. In particular, we will compute the projection operators in three-dimensional $\cN=0$ and $\cN=1$ AdS (super)space and four-dimensional $\cN$-extended AdS superspace for $\cN=1$ and $\cN=2$. We will also initiate the construction of the superspin projectors in $3d$ $\cN=(1,1)$ AdS superspace.

\subsubsection{Free off-shell supersymmetric massless higher-spin theories}
The study of Lagrangian descriptions of free higher-spin massless fields,  particularly in three dimensions, will also be a point of interest in this thesis. There are two different approaches in which supersymmetric field theories can be formulated: the component (on-shell) formalism; and the superfield (off-shell) formalism.\footnote{For a pedagogical treatment of the superfield formalism, see e.g. \cite{BuchbinderKuzenko1998,GatesGrisaruRocekSiegel1983}.} The former deals with a multiplet of bosonic and fermionic fields on spacetime, known as component fields, which are related via supersymmetry transformations. In the component approach, supersymmetry is not manifest, and the supersymmetry algebra only closes on the equations of motion (EoM).\footnote{ In order for it to close off the EoM, one needs to introduce auxiliary (non-dynamical) fields.} The off-shell formalism deals with superfields which live on a superspace. The superfield is comprised of a multiplet of component fields, both physical and non-physical (auxiliary), which ensures that the superalgebra closes off the EoM. The main benefit of working in superspace is that supersymmetry is always manifest. In this thesis, we will always work in the superfield formulation, however, we will also make contact with the component formulation via a component reduction.

Four-dimensional supersymmetric field theories realising massless superspin-$\hat{s}$ multiplets, with $\hat{s}=\hf, 1, \frac{3}{2}, \cdots$, describe two massless fields which each carry spin $\hat{s}$ and $\hat{s} +\hf $, respectively.
These actions with on-shell \Po supersymmetry were first computed in the component formalism by Curtight \cite{Curtright1979uz} and Vasiliev \cite{Vasiliev1980as} in the metric- and frame- like formulations respectively.
A superfield realisation of the massless half-integer supermultiplet in $\mb{M}^{4|4}$ was first derived by Kuzenko, Postnikov and Sibiryakov \cite{KPS} in 1993. In the same year, the off-shell integer superspin massless multiplets were found by Kuzenko and Sibiryakov \cite{KS93}. The formulations of \cite{KPS,KS93} were extended to $\cN=1 $ anti-de Sitter superspace in \cite{KS94}. The Lagrangian formulation of free $\cN=2$ higher-spin supermultiplets in terms of $\cN=1$ superfields was achieved in \cite{Gates1996my,Gates1996xs}. Recently in 2021, Buchbinder, Ivanov and Zaigraev \cite{Buchbinder2021ite} computed an off-shell formulation for massless multiplets of integer superspin in $\cN=2$ harmonic superspace \cite{Galperin1984mln,Galperin1984av,Galperin2001seg}.

The story of off-shell higher-spin supersymmetric massless field theories in three dimensions is considerably different to its four-dimensional cousin. It is well known that these higher-superspin theories are purely kinematical, as a consequence of the fact that massless fields with spins $s \geq \frac{3}{2}$ do not to propagate any physical degrees of freedom. Despite this, the study of massless theories is attractive since they can be deformed into the so-called topologically massive higher-spin actions \cite{KuzenkoPonds2018} following the philosophy of the seminal works \cite{Siegel,JS,DJT1,DJT2}.
Off-shell  massless higher spin supermultiplets were first derived in $3d$ $\cN=2$ Minkowski superspace by Kuzenko and Ogburn \cite{KuzenkoOgburn2016} in 2016. The $\cN=1$ higher-spin supermultiplets were later computed in the same year by Kuzenko and Tsulaia \cite{KuzenkoTsulaia2017} by applying a $\cN=2 \rightarrow \cN=1$ superspace reduction to the massless supermultiplets of \cite{KuzenkoOgburn2016}. 

Three-dimensional supersymmetric field theories with $(p,q)$ AdS supersymmetry, where the integers $p \geq q \geq 0$ are such that $\cN = p + q$, can naturally be realised on the so-called $(p,q)$ AdS superspaces \cite{KLT-M12}. In the case of $\cN=2$ AdS supersymmetry, off-shell massless superspin-$\hat{s}$ theories can be formulated on either of the inequivalent $(2,0)$ or $(1,1)$ AdS superspaces respectively. All known (half-)integer off-shell massless supermultiplets in $(1,1)$ AdS superspace were computed by Hutomo, Kuzenko and Ogburn in \cite{HutomoKuzenkoOgburn2018}. Soon after, the massless multiplets of half-integer superspin were found by Hutomo and Kuzenko in $(2,0)$ AdS superspace \cite{HK18}.\footnote{The Lagrangian description of the massless integer superspin multiplet in $(2,0)$ AdS superspace is still unknown.} Direct comparison of the massless $(1,1)$ \cite{HutomoKuzenkoOgburn2018} and $(2,0)$ \cite{HK18}  theories describing the same superspin content is difficult as they are formulated in different superspaces. In order to make contact between these supermultiplets, the theories need to be reduced to $\cN=1$ AdS superspace. The massless multiplets with half-integer superspin on $(2,0)$ AdS superspace \cite{HK18} have already been reduced to $\cN=1$ AdS superspace by Hutomo and Kuzenko in \cite{HK19} in 2019.

In this thesis, we reduce all known massless supermultiplets on $(1,1)$ AdS superspace to $\cN=1$ AdS superspace. This will allow for the differences of three-dimensional $\cN=2$ massless theories living in $(1,1)$ and $(2,0)$ AdS superspace to be elucidated. To compliment this, we also reduce all known massless supermultiplets in $\cN=1$ AdS superspace to AdS space. Brief discussions concerning the massless on-shell conditions these theories realise will also be provided.

\subsubsection{Thesis structure}
This thesis consists of six chapters, excluding the introduction chapter \ref{Introduction}. The first chapter \ref{Chapter2} reviews essential background material, with emphasis placed on the notation and conventions which will be employed in this thesis. The original content of this thesis can be found in the research chapters \ref{ChapThreeDimensionalExtendedMinkowskiSuperspace}-\ref{TAS211AdS}. These chapters are based on the publications \cite{BHHK, HutchingsHutomoKuzenko,BHKP, HutchingsKuzenkoPonds2021}  and the current work in progress \cite{Hutchings2022}. A detailed description regarding the location of each of the above works can found in the authorship declaration section and at the beginning of each chapter. A summary of the general results of this thesis and discussion concerning possible future research directions is given in the final chapter \ref{TheEnd}. Each of the chapters \ref{Chapter2}-\ref{TAS211AdS} are accompanied by various technical appendices. 

In general, the structure of the chapters (and of particular sections) of this thesis are organised following a certain prescription, but each are clearly distinguished by the dimension and (super)space background being studied. These chapters (or sections) begin with a review of pertinent aspects concerning the representation and field theory of the (super)space backgrounds under consideration. In particular, the on-shell (super)fields which furnish irreducible representations of the corresponding isometry (super)algebra are presented. This provides all the ingredients necessary to introduce the (super)spin projection operators. With the (super)spin projectors at our disposal, we are able to study numerous applications. In particular, we establish their connection with theories describing the propagation of (S)CHS fields. The chapters (or sections) which are strictly based in three-dimensions also include a discussion concerning free theories that propagate massless and massive particles.

Let us specify the particular dimensions and backgrounds covered in each chapter (or section) of this thesis. In the mathematical background chapter \ref{Chapter2}, we will review and expand upon known results in four-dimensional Minkowski space, its $\cN=1$ supersymmetric extension and three-dimensional Minkowski space. These are studied in sections \ref{SecFourDimensionalMinkowskiSpace}, \ref{SecFourDimensionalMinkowskiSuperSpace} and \ref{SecThreeDimensionalMinkowskiSpace} respectively.
The first research chapter \ref{ChapThreeDimensionalExtendedMinkowskiSuperspace} is strictly based in three-dimensional $\cN$-extended Minkowski superspace. Chapter \ref{Chapter4DAdS} is set in four-dimensional anti-de Sitter backgrounds. The first section concerns AdS space \ref{FAec4dAdS}, with the subsequent sections \ref{FASsec4dAdS} and \ref{FAS2sec4dAdS} investigating $\cN=1$ and $\cN=2$ AdS superspace respectively. The first section of chapter \ref{TAChapter3d(super)space}  focuses on three-dimensional AdS space, while section \ref{TASsec3dAdS} deals with its $\cN=1$ extension. The final research chapter \ref{TAS211AdS} is concerned with three-dimensional $(1,1)$ AdS superspace. 

We have chosen to present the original results from the papers \cite{BHHK, HutchingsHutomoKuzenko,BHKP, HutchingsKuzenkoPonds2021}, which are essential to this thesis, as if it was the first time that they have appeared. In accordance with this, we will not provide citations for these results in the main body of this thesis. However, a summary of the main original results obtained, with relevant citations, is given in the last section of each chapter. As stated in the authorship declaration, some results from \cite{BHKP} are not strictly part of this thesis (e.g. the partially massless supermultiplets), however, they are included for completeness. These particular results will be credited accordingly.

\chapter{Background material} \label{Chapter2}
In this chapter we review essential background material pertaining to classical field theory and representation theory which is necessary for the study of (super)spin projection and their applications in three- and four-dimensional Minkowski (super)space. Particular emphasis is placed on the notations and conventions which will be employed throughout this thesis. Most of the content presented is well-established within the literature, thus important ideas are simply stated. The presentation of this chapter is not intended to be mathematically rigorous, nor complete, but written in such a way as to ensure that important themes are self-contained. 

\section{Four-dimensional Minkowski space} \label{SecFourDimensionalMinkowskiSpace}


In this section we will review facets of classical field theory in four-dimensional Minkowski space. We begin by introducing the unitary irreducible representations of the four-dimensional \Po algebra $\mf{iso}(3,1)$ and their corresponding field realisations on \FM. Once established, we study the spin projection operators \cite{FangFronsdal,Fronsdal1958,SiegelGates1981,GatesGrisaruRocekSiegel1983} and their corresponding applications. In particular, we will explore the intrinsic relationship between these operators and conformal higher-spin theory \cite{FradkinTseytlin1985}. The notation and conventions employed in this section coincide with those used in \cite{BuchbinderKuzenko1998}.

\subsection{Facets of four-dimensional Minkowski space} \label{Geometry4d}

Minkowski space in $d$-dimensions, which we denote by $\mb{M}^d$, is simply the manifold $\mb{R}^d$ endowed with the metric $\eta_{ab} = \text{diag} (-1, 1, \cdots, 1)$.
The spacetime $\mb{M}^d$ is parametrised by the local Cartesian coordinates $x^a = (x^0,x^1,\cdots, x^{d-1})$, where $x^0 =t$ is the time coordinate and $\vec{x} = (x^1, x^2, \cdots , x^{d-1})$ are the space coordinates.\footnote{From the onset and throughout, we will make use of natural units $c=\hbar = 1$.} In this thesis, we will use indices from the beginning of the Latin alphabet to denote Lorentz indices.

The isometry group of $\mb{M}^d$ is the \Po group $\mathsf{IO}(d-1,1)$, which is the group of coordinate transformations $(\L,a)$ 
\be \label{DMPoincareTransformation}
x'^a = \L^a{}_b x^b + a^a~,
\ee
that preserve the spacetime interval $\rd s^2 = \eta_{ab} \rd x^a \rd x^b$ on $\mb{M}^d$. Here, $a^a \in \mb{R}^d$ is a constant $d$-vector and $\L$ is an element of the Lorentz group $\mathsf{O}(d-1,1)$ which satisfies the defining master equation
\be \label{FMMasterequation}
\L^\text{T} \eta \L = \eta~,
\ee
where $\L^\text{T}$ is the matrix transpose of $\L$.

We are primarily interested in the subgroup of $\mathsf{IO}(d-1,1)$, known as the proper orthochronous \Po group $\mathsf{ISO}_0(d-1,1)$. The group $\mathsf{ISO}_0(d-1,1)$ describes the coordinate transformations $(\L,a)$,  where instead, the matrix $\L$ is an element of the proper orthochronous Lorentz group $\mathsf{SO}_0(d-1,1)$.\footnote{We will commonly refer to the proper orthochronous Lorentz group as the Lorentz group.} In addition to obeying \eqref{FMMasterequation}, every group element $\L \in \mathsf{SO}_0(d-1,1) $ also satisfies the properties
\be \label{DMLorentzconditions}
\text{det} (\L) =1~, \qquad \L^0{}_0 \geq 1~.
\ee
The conditions \eqref{DMLorentzconditions} ensure that the spatial orientation and the direction of time are preserved, respectively.

The Lorentz algebra $\mf{so}(d-1,1)$  is spanned by the antisymmetric generators $M_{ab} = - M_{ba}$ which satisfy  the commutation relation
\be \label{FMLorentzAlgebra}
[ M_{ab},M_{ce}] = \eta_{ac}M_{be}  - \eta_{bc}M_{ae}   + \eta_{be}M_{ac} - \eta_{ae}M_{bc} ~.
\ee
The quadratic Casimir operator of $\mf{so}(d-1,1)$ is $M^{ab}M_{ab}$. In four dimensions, there exists an additional quadratic Casimir operator of $\mf{so}(3,1)$, which takes the form $\ve_{abce}M^{ab}M^{ce}$.

The \Po algebra $\mf{iso}(d-1,1)$ is spanned by the generators of spacetime translations $P_a$ and Lorentz transformations $J_{ab}$. These generators satisfy the commutation relations
\bsubeq \label{FMPoincareAlgebra}
\bea
[P_a,P_b] &=& 0~, \\
\ [ J_{ab},P_c] &=& \ri \big (\eta_{ac}P_b  - \eta_{bc}P_a \big )~,\\
\ [ J_{ab},J_{ce}] &=& \ri \big (\eta_{ac} J_{b e} - \eta_{ae}J_{bc} +  \eta_{be}J_{ac} - \eta_{bc}J_{a e} \big )~.
\eea
\esubeq
These hold in an arbitrary representation of the \Po algebra.\footnote{In any unitary representation of $\mf{iso}(d-1,1)$, the energy-momentum $P_a$ and Lorentz  $J_{ab}$ generators are Hermitian.}

The quadratic Casimir operator $C_1$ of $\mf{iso}(d-1,1)$ takes the following form
\be \label{DMPoincareQuadCasimir}
C_1 = -P^aP_a~, \qquad  [C_1, P_a]=[C_1,J_{ab}] = 0~.
\ee
There exist higher-order Casimir operators of $\mf{iso}(d-1,1)$, however, they are dependent on the dimension $d$ of the spacetime under consideration (see e.g. \cite{KuzenkoPindur2020}). For the remainder of this section, we will restrict our attention to $4d$. In $4d$, there exists a Casimir operator $C_2$ which is quartic in the generators of $\mf{iso}(3,1)$,
\be \label{FMPoincareQuartCasimir}
C_2 = \mb{W}^a \mb{W}_a~, \qquad  [C_2 , P_a] = [C_2 , J_{ab}] = 0~,
\ee
where $\mb{W}_a$ is the Pauli-Lubanski pseudovector
\be \label{FMPauliLubanksiVector}
\mb{W}_a = \hf \ve_{abce}J^{bc}P^e~.
\ee
Here, $\ve_{abce}$ is the totally antisymmetric Levi-Civita tensor.\footnote{We choose the normalisation of the Levi-Civita tensor to be $\ve_{01\dots d-1}=-\ve^{01\dots d-1}=-1$.} Using the algebra \eqref{FMPoincareAlgebra}, it can be shown that the Pauli-Lubanski pseudovector \eqref{FMPauliLubanksiVector} satisfies the properties
\bsubeq \label{FMPauliLubanskiVectorProperties}
\bea
\mb{W}^a P_a &=& 0~, \\
\ [\mb{W}_a, P_b] &=& 0~, \\
\ [ J_{ab}, \mb{W}_c] &=& \ri \big ( \eta_{ac} \mb{W}_b -  \eta_{bc} \mb{W}_a \big )~, \\
\ [\mb{W}_a, \mb{W}_b] &=& \ri \ve_{abce}\mb{W}^c P^e~.
\eea
\esubeq
The properties \eqref{FMPauliLubanskiVectorProperties} are useful in showing that the quadratic Casimir operator $C_2$ commutes with the generators of $\mf{iso}(3,1)$.

\subsection{Irreducible representations of the \Po algebra} \label{Irreducible representations of the Poincare algebra}

In this section we review salient aspects concerning the classification of the massive and massless helicity UIRs of $\PaF$ which carry positive energy $p^0 \equiv E >0$.
The classification of these UIRs was first completed by Wigner \cite{Wigner1939} (see also \cite{Wigner1948}) via the method of induced representations.\footnote{The classification of the UIRs of the \Po algebra has been generalised to $d$ dimensions, for $d>2$. See the works \cite{BekaertBoulanger2003,BekaertBoulanger2007,BekaertBoulanger2021} for an extensive treatment.} 

The massive UIRs of $\PaF$ are labelled by the quantum numbers, mass $\bm{m}$ and spin $s$ \cite{Wigner1939}. It was shown by Bargmann and Wigner \cite{BargmannWigner1948}  that these quantum numbers are determined by the eigenvalues of the Casimir operators $C_1$ and $C_2$
\be \label{FMMassiveOnshellConditions}
C_1 = P^a P_a = -\bm{m}^2 \mds{1}~, \qquad C_2 = \mb{W}^a\mb{W}_a = s(s+1)\bm{m}^2\mds{1}~,
\ee
where the mass $\bm{m}>0$ is strictly positive and the spin $s$ takes positive (half-)integer values $s= 0, \hf, 1, \frac{3}{2}, \cdots$ in different representations. The massive irreducible representations of $\PaF$ are characterised by the conditions \eqref{FMMassiveOnshellConditions}. We will denote by $G(\bm{m},s)$ the massive UIR of $\PaF$ which carries mass $\bm{m}$ and spin $s$.\footnote{The notation $G(\bm{m},s)$ for a UIR of $\PaF$ is unconventional. We choose it to help distinguish between the UIRs of different isometry (super)algebras, which will appear in later sections.   } For a given $s$, the UIR $G(\bm{m},s)$ describes $(2s+1)$ physical degrees of freedom.

It was also shown in \cite{BargmannWigner1948} that the massless helicity UIRs cannot be characterised by the eigenvalues of the Casimir operators of $\PaF$, as they are both vanishing\footnote{There exists another class of massless UIRs, known as continuous-spin representations (see \cite{BekaertSkvortsov2017} for a modern review). These UIRs are characterised by $C_2= \m^2$, where $\m$ is a real parameter with unit mass dimension. These UIRs are not covered within the scope of this thesis.} 
\be \label{FMMasslessOnShellConditionsHelicity}
C_1=P^a P_a =0~, \qquad C_2=\mb{W}^a \mb{W}_a = 0~.
\ee 
The massless UIRs\footnote{In the hope that no confusion arises, we will refer to massless helicity representations as massless representations from this point onwards.} are instead characterised by Wigner's equation,
\be \label{FMMasslessWignersEquation}
\mb{W}_a = \l P _a~.
\ee
In every massless UIR, the Pauli-Lubanski pseudovector is proportional to the energy-momentum generator $P_a$ \eqref{FMMasslessWignersEquation} where the quantum number $\l$ is known as helicity. The helicity can take (half-)integer values $\l = 0, \pm \hf, \pm 1 , \pm \frac{3}{2}, \cdots$ in different massless UIRs. It is common to say that a massless UIR has spin $|\l|$. Thus for $s>0$, a massless representation carries a single physical degree of freedom corresponding to the helicity state it describes.

\subsection{Irreducible field representations}\label{FMIrrepField}

Let us denote by $V_{(m,n)}$ the space of totally symmetric complex spinor fields $\f_{\a(m)\ad(n)}(x)$ of Lorentz type $(\frac{m}{2}, \frac{n}{2})$ on $\mb{M}^4$.\footnote{ It is convenient to work in two-component spinor notation since fermionic and bosonic fields can both be treated in the same framework. See appendix \ref{FMAppendixA} for a review on the two-component spinor notation.
} For integers $m,n \geq 1$, a spinor field $\f_{\a(m)\ad(n)}(x)$ on $V_{(m,n)}$ is said to be an on-shell field if it satisfies the conditions
\bsubeq \label{FMOnshellConditions}
\bea \label{MinkOnshellTrans}
\pa^{\b\bd}\f_{\b\a(m-1)\bd \ad(n-1)}(x)&=&0~, \\
\label{MinkOnshellMass}
(\Box - \bm{m}^2)\f_{\a(m)\ad(n)}(x)&=&0~,
\eea
\esubeq
where mass $\bm{m} \geq 0$ is non-negative. The d'Alembertian operator $\Box = - \hf  \pa^{\b\bd} \pa_{\b\bd}  $ is defined in terms of the Lorentz covariant derivatives  $\pa_{\b\bd}$ of $\mb{M}^4$. In the case of a scalar field and spinor fields of Lorentz type $(\frac{m}{2},0)$ and $(0, \frac{n}{2})$, an on-shell field is defined to only satisfy the Klein-Gordon equation \eqref{MinkOnshellMass}. Note that from this point onwards, we will suppress the spacetime dependence of tensor fields $\f_{\a(m)\ad(n)}(x) \equiv \f_{\a(m)\ad(n)}$. In the following sections, we will demonstrate how on-shell fields realise massive and massless UIRs of $\PaF$, following the discussion of \cite{BuchbinderKuzenko1998}.

\subsubsection{Massive field representations} \label{MassiveFieldrepresentations}

An on-shell field \eqref{FMOnshellConditions} carrying strictly positive mass $\bm{m}>0$ furnishes the massive UIR $G(\bm{m}, s)$ of $\PaF$. We say that such an on-shell field is massive,  carrying spin $s = \hf (m+n)$ and mass $\bm{m}$.
We wish to demonstrate that a massive field \eqref{FMOnshellConditions} does indeed realise the massive UIR $G(\bm{m}, s)$. To do this, it suffices to show that the eigenvalues of the Casimir operators $C_1$ and $C_2$ of $\PaF$  take the fixed values \eqref{FMMassiveOnshellConditions}.

In the field representation, the \Po generators take the following form
\bsubeq \label{FMGeneratorsField}
\bea 
P_a &=& - \ri \pa_a~, \\
J_{ab} &=& \ri \big (x_b \pa_a -x_a \pa_b \big ) - \ri M_{ab}~. \label{FMLorentzGenerator}
\eea
\esubeq
One can check that the generators \eqref{FMGeneratorsField} do indeed satisfy the \Po algebra \eqref{FMPoincareAlgebra}. Note that in this thesis, we use the notation $M_{ab}$ to denote the Lorentz generators which, when acting on any tensor field, return the spin contribution only (and not the orbital part $L_{ab} = \ri (x_b \pa_a -x_a \pa_b)$ in \eqref{FMLorentzGenerator}). The action of the Lorentz generators $M_{ab}$ on a tensor field is given in appendix \ref{FMAppendixA}.

Using the generators \eqref{FMGeneratorsField}, it can be shown that the Pauli-Lubanski pseudovector \eqref{FMPauliLubanksiVector} takes the following form in the field representation
\be \label{FMPauliLubanskiField}
\mb{W}_{\a\ad} = - \ri \pa^\b{}_\ad M_{\b\a} + \ri \pa_\a{}^\bd \bar{M}_{\bd \ad}~.
\ee
It follows from \eqref{FMPauliLubanskiField} that the action of the Casimir operator $C_2$ \eqref{FMPoincareQuartCasimir} on $V_{(m,n)}$ is
\be \label{FMQuarticCasimirSpinorFields}
\mb{W}^2\f_{\a(m)\ad(n)} = s(s+1) \Box \f_{\a(m)\ad(n)} +mn \pa_{\a \ad}\pa^{\b\bd} \f_{\b \a(m-1)  \bd \ad(n-1) }~,
\ee
where we have made use of the identities
\be
\pa_{\a}{}^\bd \pa^\b{}_\bd =  \d_\a{}^\b \Box~, \qquad  \pa^\b{}_{\ad} \pa_\b{}^\bd =  \d_\ad{}^\bd \Box~.
\ee
One immediately sees that the action of the Casimir operator $C_2$ \eqref{FMQuarticCasimirSpinorFields} on a massive on-shell field \eqref{FMOnshellConditions} reduces to
\be \label{FMQuarticCasimirSpinorFieldsOnshell}
\mb{W}^2 \f_{\a (m) \ad (n)}= s(s+1)\bm{m}^2 \f_{\a (m) \ad (n)}~,
\ee
where $s = \hf (m+n)$. It follows from \eqref{MinkOnshellMass} and \eqref{FMQuarticCasimirSpinorFieldsOnshell} that the eigenvalues of the Casimir operators $C_1$ and $C_2$ on the space of massive fields are equivalent to those which characterise the massive UIR $G (\bm{m},s)$ (cf. \eqref{FMMassiveOnshellConditions}), thus demonstrating that massive on-shell fields do indeed realise the UIR $G (\bm{m},s)$. Note that in \eqref{FMQuarticCasimirSpinorFields} we have adopted the condensed notation \eqref{FMSym}.

Next we wish to show that a massive field describes $(2s+1)$ physical degrees of freedom, as required by a field which furnishes the massive UIR $G (\bm{m},s)$ with fixed spin $s$. Counting the physical degrees of freedom, a totally symmetric spinor field $\f_{\a(m)\ad(n)}$ has $(m+1)(n+1)$ independent components. The component dependence which arises from the transverse constraint \eqref{MinkOnshellTrans}  is equivalent to the number of independent components described by the tensor field $\f_{\a(m-1)\ad(n-1)}$, which is $mn$. Thus, there are $(m+1)(n+1) -mn = 2s+1$ independent components remaining, as required.

Let us denote by $V_{(m,n)}^{[\boldsymbol{m}]}$ the space of massive on-shell fields.
It turns out that the UIR $G(\bm{m},s)$ can be equivalently realised on spaces other than $V_{(m,n)}^{[\boldsymbol{m}]}$. In particular, it can be realised on the space of massive fields 
$V^{[\boldsymbol{m}]}_{(m+t,n-t)}$ for $1 \leq t \leq n$, or $V^{[\boldsymbol{m}]}_{(m-t,n+t)}$ for $1 \leq t \leq m$.  
To observe this, let us introduce the  operator $\D_{\a\ad}$ \cite{GatesGrisaruRocekSiegel1983}
\be \label{FMDeltaOperator}
\D_{\a\ad} = \frac{1}{\sqrt{\Box}} \pa_{\a\ad}= \frac{1}{m} \pa_{\a\ad} ~,
\ee
which is invertible
\be \label{FMDeltaOperatorProperties}
\D_\a{}^\bd \D^\b{}_\bd = \d_\a{}^\b~, \quad \D^\b{}_\ad \D_\b{}^\bd = \d_\ad{}^\bd~.
\ee
Since we are strictly working with massive fields \eqref{FMOnshellConditions}, the operators \eqref{FMDeltaOperator} will always be well defined. The operator $\D_{\a\ad}$ is important as it can be used to map the space $V^{[\boldsymbol{m}]}_{(m,n)}$ to either $V^{[m]}_{(m+t,n-t)}$ 
\bsubeq \label{FMDeltaMap1} 
\bea
\D^t : V^{[\boldsymbol{m}]}_{(m,n)} &\longrightarrow& V^{[\boldsymbol{m}]}_{(m+t,n-t)} ~, \qquad 1 \leq t \leq n~, \\
\f_{\a(m)\ad(n)} &\longmapsto& \f_{\a(m+t)\ad(n-t)} = \D_{\a_{1}}{}^{\bd_1} \cdots \D_{\a_{t}}{}^{\bd_t}\f_{\a(m)\ad(n-t)\bd(t)}~,
\eea
\esubeq
or $V^{[\boldsymbol{m}]}_{(m-t,n+t)}$
\bsubeq \label{FMDeltaMap2}
\bea
\widehat{\D}^t : V^{[\boldsymbol{m}]}_{(m,n)} &\longrightarrow& V^{[\boldsymbol{m}]}_{(m-t,n+t)} ~, \qquad 1 \leq t \leq m~,\\
\f_{\a(m)\ad(n)} &\longmapsto& \f_{\a(m-t)\ad(n+t)} =  \D^{\b_1}{}_ {\ad_{1}} \cdots \D^{\b_t}{}_ {\ad_t}  \f_{\a(m-t)\b(t)\ad(n)}~.
\eea
\esubeq
\sloppy{ The mappings \eqref{FMDeltaMap1} and \eqref{FMDeltaMap2} are invertible since $\D_{\a\ad}$ is invertible \eqref{FMDeltaOperatorProperties}. It is an easy exercise to show that if the field $\f_{\a(m)\ad(n)}$ furnishes the UIR $G(\bm{m},s)$, then so do the fields $\f_{\a(m+t)\ad(n-t)}$ and $\f_{\a(m-t)\ad(n+t)}$. Thus, the spaces of massive fields $V^{[\boldsymbol{m}]}_{(2s,0)}, V^{[\boldsymbol{m}]}_{(2s-1,1)}, \cdots , V^{[\boldsymbol{m}]}_{(1,2s-1)}, V^{[\boldsymbol{m}]}_{(0,2s)}$ can all furnish the massive UIR $G(\bm{m},s)$.}

\paragraph{Real massive field representations}
We have only treated massive UIRs of $\PaF$ on the space of complex fields in our discussions so far. 
A complex massive field $\f_{\a(m)\ad(n)}$ on $V^{[\boldsymbol{m}]}_{(m,n)}$ possesses a complex conjugate $\bar{\f}_{\a(n)\ad(m)}$, which is obtained from $\f_{\a(m)\ad(n)}$  via the operation of complex conjugation   
\vspace{-0.5cm}
\bsubeq\label{FMComplexConjugation}
\bea
{}^*:V_{(m,n)} &\longrightarrow& V_{(n,m)}  ~, \\
\f_{\a(m)\ad(n)} &\longmapsto& \big (\f_{\a(m)\ad(n)} \big )^* =  \fb_{\a(n)\ad(m)} ~.   
\eea
\esubeq
In general, a massive field \eqref{FMOnshellConditions} and its complex conjugate are independent. We will be interested in studying real massive fields, which are a class of complex massive fields $\f_{\a(m)\ad(n)}$ supplemented with some reality condition which ensures that $\f_{\a(m)\ad(n)}$ is related to its complex conjugate $\bar{\f}_{\a(n)\ad(m)}$. Note that this review of real massive fields follows the work of Siegel and Gates in \cite{SiegelGates1981} (see also \cite{GatesGrisaruRocekSiegel1983}).

We already have the tools at our disposal to find such suitable reality conditions on $V^{[\boldsymbol{m}]}_{(m,n)}$. Let us make use of the operator $\D_{\a\ad}$  \eqref{FMDeltaOperator} to construct the mapping 
\bsubeq\label{FMDeltaMapping}
\begin{gather}
\D: V^{[\boldsymbol{m}]}_{(m,n)} \longrightarrow V^{[\boldsymbol{m}]}_{(n,m)}~,  \\
\f_{\a(m)\ad(n)} \longmapsto \big (\D\f \big )_{\a(n)\ad(m)} = e^{\ri \l} \D_{\a_1}{}^{\bd_1} \cdots \D_{\a_n}{}^{\bd_n} \D^{\b_1}{}_{\ad_1} \cdots \D^{\b_m}{}_{\ad_m} \f_{\b(m)\bd(n)}~,
\end{gather}
\esubeq
for some $\l \in \mb{R}$. Note that the domain and co-domain of both ${}^*$ \eqref{FMComplexConjugation} and $\D$ \eqref{FMDeltaMapping} are identical, thus, it is natural to consider the subspace of $V^{[\boldsymbol{m}]}_{(m,n)}$ on which ${}^*$ and $\D$ coincide
\be \label{FMMassiveRealityConditions}
\fb_{\a(n)\ad(m)} =  e^{\ri \l}  \D_{\a_1}{}^{\bd_1} \cdots \D_{\a_n}{}^{\bd_n} \D^{\b_1}{}_{\ad_1} \cdots \D^{\b_m}{}_{\ad_m} \f_{\b(m)\bd(n)}~.
\ee
We will refer to \eqref{FMMassiveRealityConditions} as a reality condition, as it ensures that $\f_{\a(m)\ad(n)}$ and its complex conjugate  $\bar{\f}_{\a(n)\ad(m)}$ are dependent.

In accordance with the discussion above, a field $\f_{\a(m)\ad(n)}$ is said to be a real massive field if it satisfies the properties
\bsubeq  \label{TMRealMassiveFields}
\bea 
0&=&\pa^{\b\bd}\f_{\b\a(m-1)\bd \ad(n-1)}~, \label{TMRealtrans}\\
0&=&(\Box - \bm{m}^2)\f_{\a(m)\ad(n)}~, \label{TMRealMS}\\
\fb_{\a(n)\ad(m)} &=& e^{i \l }\D_{\a_1}{}^{\bd_1} \cdots \D_{\a_n}{}^{\bd_n} \D^{\b_1}{}_{\ad_1} \cdots \D^{\b_m}{}_{\ad_m} \f_{\b(m)\bd(n)}~. \label{FMRealityIrrep}
\eea
\esubeq
For different values of $\l$, the corresponding linear space of fields satisfying \eqref{TMRealMassiveFields} are all isomorphic. 
Let us look at the reality conditions \eqref{FMRealityIrrep} for the archetypal examples $m=n$ and $m=n+1$. 

In the case $m=n$, the reality condition \eqref{FMRealityIrrep} coincides with the expected result
\be
\f_{\a(n)\ad(n)} = \fb_{\a(n)\ad(n)} ~,
\ee
where we have made use of the properties \eqref{TMRealtrans}, \eqref{TMRealMS} and fixed $\l=0$.  Without loss of generality, the reality condition \eqref{FMMassiveRealityConditions} in the case $m=n+1$ coincides with the Dirac-type equation
\be
\bar{\f}_{\a(n)\ad(n)\bd} = e^{i \l } \D^\b{}_\bd \f_{\b \a(n)\ad(n)}  \quad \Longrightarrow \quad \bm{m} \bar{\f}_{\a(n)\ad(n)\bd} = e^{i \l } \pa^\b{}_\bd \f_{\b \a(n)\ad(n)}~.
\ee
For the case $\l=\frac{\p}{2}$, this simply reduces to the Dirac equation.

Let us show that a complex massive field $\f_{\a(m)\ad(n)}$ decomposes into two real massive fields when a reality condition is imposed. To do this, it is necessary to introduce the conjugation operator $K$ which is defined by its action on $V^{[\boldsymbol{m}]}_{(m,n)}$ \cite{SiegelGates1981, GatesGrisaruRocekSiegel1983}
\bea \label{FMConjugationOperator}
K \f_{\a(m)\ad(n)} = e^{i \l }\D_{\a_1}{}^{\bd_1} \cdots \D_{\a_m}{}^{\bd_m} \D^{\b_1}{}_{\ad_1} \cdots \D^{\b_n}{}_{\ad_n} \bar{\f}_{\b(n)\bd(m)}~.
\eea
It is a simple exercise to show that the conjugation operator is involutive, $K^2 \f_{\a(m)\ad(n)} = \f_{\a(m)\ad(n)}$.\footnote{Since the operator $K$ is involutive, it follows that the operators $\hf (1 \pm K)$ are orthogonal projectors.}
With this, let us bisect a complex massive field in the following manner
\be 
\f_{\a(m)\ad(n)} = \f^{(+)}_{\a(m)\ad(n)} + \f^{(-)}_{\a(m)\ad(n)} ~,
\ee
where we have introduced the fields
\be
\f^{(\pm)}_{\a(m)\ad(n)} = \hf (1 \pm K) \f_{\a(m)\ad(n)}~.
\ee
Using the on-shell properties \eqref{FMOnshellConditions} and the fact that $K$ is involutive, it can be shown that the fields $\f^{(\pm)}_{\a(m)\ad(n)} $ satisfy the conditions
\bsubeq
\bea
\pa^{\b\bd} \f^{(\pm)}_{\b\a(m-1) \bd\ad(n-1)} &=&0~, \\
\big ( \Box - \bm{m}^2 \big ) \f^{(\pm)}_{\a(m)\ad(n)} &=&0~, \\
\big (1 \mp K \big ) \f^{(\pm)}_{\a(m)\ad(n)} &=&0~. \label{FMBisectionReality}
\eea
\esubeq
The last equation \eqref{FMBisectionReality} is nothing more than the reality condition \eqref{FMRealityIrrep} written in terms of the conjugation operator. In accordance with \eqref{TMRealMassiveFields}, the fields $\f^{(\pm)}_{\a(m)\ad(n)} $ can be identified as real massive fields.

\subsubsection{Massless field representations} \label{FMMasslessfieldrepresentations}
The formulation of massless fields can be approached in two ways: in terms of gauge fields $\f_{\a(m)\ad(n)}$ or in terms of gauge-invariant field strengths $C_{\a(p)\ad(q)}(\f)$. We begin by reviewing the construction of massless gauge fields.

For integers $m,n \geq 1$, let us consider a field $\f_{\a(m)\ad(n)}$ on $V_{(m,n)}$ which satisfies the on-shell conditions 
\bsubeq \label{FMMasslessfields}
\bea 
\pa^{\b \bd} \f_{\b\a(m-1)\bd \ad(n-1)} &=& 0~,  \label{FMMasslessTransverse}\\
\Box \f_{\a(m)\ad(n)} &=& 0~. \label{FMMasslessMassShell} 
\eea
The equations \eqref{FMMasslessfields} are nothing more than the on-shell conditions  \eqref{FMOnshellConditions} in the massless limit $m \rightarrow 0$. It can be shown that these conditions are compatible with the gauge symmetry
\be \label{FMMasslessGT}
\d_\z \f_{\a(m)\ad(n)} = \pa_{\a \ad}\z_{\a(m-1) \ad(n-1)}~,
\ee
given that the complex gauge parameter $\z_{\a(m-1)\ad(n-1)}$ is itself on-shell 
\bea
\Box  \z_{\a(m-1)\ad(n-1)} &=& 0~, \label{FMMasslessTransverseGaugeKG} \\
\pa^{\b\bd} \z_{\b\a(m-2)\bd \ad(n-2)} &=&0~. \label{FMMasslessTransverseGaugeConditon}
\eea
\esubeq 
We say that a field satisfying the on-shell conditions \eqref{FMMasslessfields} is a massless (gauge) field. For the case $m=n=0$, $m = 0 , n =1$ and $m =1 , n=0$, the corresponding massless fields only satisfy the massless Klein-Gordon equation \eqref{FMMasslessMassShell}. It follows from \eqref{FMMasslessTransverse} and \eqref{FMMasslessMassShell}  that the eigenvalues of the Casimir operators vanish on the space of massless fields \eqref{FMMasslessfields}, which is consistent for a tensor field which realises a massless UIR (cf. \eqref{FMMasslessOnShellConditionsHelicity}).

However, studying the helicity content of the massless field \eqref{FMMasslessfields} reveals
\bea \label{FMMasslessPL}
&&\mb{W}_{\b \bd} \f_{\a(m)\ad(n)} \non \\ 
&&= \hf (m-n)P_{\b\bd}\f_{\a(m)\ad(n)} - m \ve_{\b \a} P^{\g}{}_\bd \f_{ \a(m-1) \g \ad(n)} + n \ve_{\bd \ad}P_\b{}^\gd \f _{\a(m) \ad(n-1) \gd}~.
\eea
Recall that in a massless UIR of $\PaF$, the helicity $\l$ is determined from the Wigner equation, $\mb{W}_{\b \bd} \f_{\a(m)\ad(n)}  = \l P_{\b\bd}\f_{\a(m)\ad(n)}$ (cf. \eqref{FMMasslessWignersEquation}), with $\l$ taking (half-)integer values. It is apparent from \eqref{FMMasslessPL} that this condition does not hold in the case of massless gauge fields \eqref{FMMasslessfields}. In other words, massless gauge fields do not furnish massless UIRs of $\PaF$.
This is consistent with the fact that massless gauge fields \eqref{FMMasslessfields} are not truly irreducible due to the presence of the gauge symmetry \eqref{FMMasslessGT}. It follows that the solution space of massless fields \eqref{FMMasslessTransverse} and \eqref{FMMasslessMassShell} possesses non-trivial invariant subspaces which are mapped to one another by the gauge transformations \eqref{FMMasslessGT}. To make the solution space irreducible, we need to factor out these redundant gauge degrees of freedom by completely fixing the gauge symmetry \eqref{FMMasslessGT}.

Going about this, it will be useful to switch to momentum space, i.e. $\f_{\a (m) \ad (n)} (x) \rightarrow \f_{\a (m) \ad (n)} (p)$. In momentum space, the constraints \eqref{FMMasslessTransverse} and \eqref{FMMasslessTransverseGaugeConditon} assume the following form
\be \label{FMMasslessTransverseMom}
p^{\b \bd} \f_{\b\a(m-1)\b \ad(n-1)} = 0~, \qquad p^{\b\bd} \z_{\b\a(m-2)\bd \ad(n-2)} =0~.
\ee
Since the conditions \eqref{FMMasslessONshellMomentum} are invariant under \Po transformations, we are free to transition to the standard momentum frame $p^a = (E, 0, 0 , E)$, where energy $E$ is strictly positive. Converting to two-component spinor notation yields
\be \label{FMMasslessrestframe}
p^a = (E, 0, 0 , E) \quad \Longrightarrow \quad p^{\a\ad} = 2E 
\begin{pmatrix}
	~0~ &0~\\
	~0~&1~ 
\end{pmatrix}~, 
\quad p_{\a\ad} = 2E 
\begin{pmatrix}
	~1~ &0~\\
	~0~&0~
\end{pmatrix}~.
\ee
The equations \eqref{FMMasslessTransverseMom} assume the following form in the rest frame 
\be \label{FMFieldConditionMomentum1}
\f_{2 \a(m-1) \td\ad(n-1)}(p)=0~, \qquad \z_{2 \a(m-2) \td \ad(n-2)}(p) = 0~.
\ee
Additionally, the gauge transformation \eqref{FMMasslessGT} takes the form,
\be
\d \f_{1 \a(m-1) \od \ad(n-1)}(p) \propto p_{1 \od} \z_{\a(m-1)\ad(n-1)}(p)~,
\ee
in the frame \eqref{FMMasslessrestframe}, which allows us to impose the gauge 
\be \label{FMGaugeCondMomentum2}
\f_{1\a(m-1)\od \ad(n-1)}(p)=0~.
\ee
It follows from \eqref{FMFieldConditionMomentum1} and \eqref{FMGaugeCondMomentum2} that the only non-vanishing components of $\f_{\a(m)\ad(n)}(p)$ are $\f_{1 \dots 1 \td \dots \td }(p) $ and $\f_{2 \dots 2 \od \dots \od }(p)$. Note that we have completely fixed the gauge freedom.

Elucidating the helicity content of the two non-vanishing components of $\f_{\a(m)\ad(n)}$ \eqref{FMMasslessPL}, we find
\bsubeq \label{FMWignEqMassless}
\bea
\mb{W}_{\b\bd} \f_{1 \dots 1 \td \dots \td }(p) &=& \phantom{-} \hf (m+n) \f_{1 \dots 1 \td \dots \td }(p) ~, \\
\mb{W}_{\b\bd}  \f_{2 \dots 2 \od \dots \od} (p) &=& -\hf(m+n) \f_{2 \dots 2 \od \dots \od} (p) ~.
\eea
\esubeq
Upon completely fixing the gauge symmetry, it follows from \eqref{FMMasslessMassShell} and \eqref{FMWignEqMassless} that the massless field \eqref{FMMasslessfields} furnishes a massless UIR of $\PaF$ (cf. \eqref{FMMasslessWignersEquation}). In particular, the massless field $\f_{\a(m)\ad(n)}$ describes two physical components, $\f_{1 \dots 1 \td \dots \td }$ and $\f_{2 \dots 2 \od \dots \od }$, which describe the two helicity modes  $\hf (m+n) $ and  $-\hf(m+n)$, respectively.

It is worth taking a moment to comment on the subtleties concerning the irreducibility of the massless field representations \eqref{FMMasslessfields}. In accordance with \eqref{FMMasslessWignersEquation}, massless irreducible representations of  the connected component of the Poincar\'e group $\textsf{ISO}_{0}(3,1)$ are characterised by a single helicity value $\l = 0, \pm \hf, \pm 1 , \pm \frac{3}{2}, \cdots$. Thus the massless fields \eqref{FMMasslessfields} realise a reducible representation of $\textsf{ISO}_{0}(3,1)$, since they describe  the two helicity modes $\l = \hf (m+n) $ and  $\l = -\hf(m+n)$ which differ by a sign factor. However, the massless fields \eqref{FMMasslessfields} realise an irreducible representation with respect to the full Poincar\'e group. $\textsf{ISO}(3,1)$. The reason for this is that  parity transformations change the sign of the helicity $\l$. Thus under a parity transformation, the massless representations characterised by the helicity $\hf (m+n) $ will map to the representation with helicity  $-\hf(m+n)$, and vice versa. Hence, a massless field which furnishes a UIR of $\textsf{ISO}(3,1)$ describes two helicity modes with identical spin $\l$, but different sign. The field equations \eqref{FMMasslessfields} are the desired on-shell conditions for a parity invariant theory that describing the propagation of a massless spin-$s$ field.

As seen, it is sometimes cumbersome to study massless dynamics in terms of the massless gauge fields, due to difficulties concerning irreducibility. However, these difficulties are immediately circumvented if massless dynamics are studied within the framework of gauge-invariant field strengths. Let us consider a field strength $C_{\a(p)\ad(q)}(\f)$, which is invariant under the gauge transformations \eqref{FMMasslessGT}
\be
\d_\z C_{\a(p)\ad(q)}(\f) = 0~.
\ee
For integers $p, q \geq 1$, the field strength $C_{\a(p)\ad(q)}(\f)$ furnishes a massless UIR of $\PaF$ if it obeys the constraints
\bsubeq \label{FMMasslessOnshell}
\bea 
\pa^{\b}{}_\bd C_{\b\a(p-1)\ad(q)}(\f) = 0~,  \label{FMMasslessOnShellFS1}\\
\pa_\b{}^{\bd} C_{\a(p)\bd\ad(q-1)}(\f) = 0~. \label{FMMasslessOnShellFS2}
\eea
\esubeq
Conditions \eqref{FMMasslessOnshell} imply that the the eigenvalues of the Casimir operators $C_1$ and $C_2$ vanish on the space of field strengths $C_{\a(p)\ad(q)}(\f)$. Note that for the cases $p > 0, q=0$ and $p = 0, q > 0$, the conditions \eqref{FMMasslessOnShellFS1} and \eqref{FMMasslessOnShellFS2} describe the massless UIRs, respectively. 

The helicity content of a massless field is easily attainable when working in terms of field strengths. Acting on $C_{\a(p)\ad(q)}(\f)$ with the Pauli-Lubanski pseudovector \eqref{FMPauliLubanskiField} yields
\be \label{FMWignerEqnMassless}
\mb{W}_{\b\bd}C_{\a(p)\ad(q)}(\f)= \hf \big ( p-q \big )P_{\b\bd} C_{\a(p)\ad(q)}(\f)~,
\ee
thus demonstrating that $C_{\a(p)\ad(q)}(\f)$ carries helicity $\l = \hf (p-q)$ (cf. \eqref{FMMasslessWignersEquation}). 

Next, we wish to show that $C_{\a(p)\ad(q)}(\f)$ describes the correct physical degrees of freedom for a massless field. Transitioning  to momentum space, the constraints \eqref{FMMasslessOnshell} take the following form
\bsubeq \label{FMMasslessONshellMomentum}
\bea
p^{\b}{}_\bd C_{\b\a(p-1)\ad(q)}(\f)=0~, \\
p_\b{}^\bd C_{\a(p)\ad(q-1)\bd}(\f)=0~.
\eea
\esubeq
Again,  we are free to move into the standard momentum frame \eqref{FMMasslessrestframe}, where the conditions \eqref{FMMasslessONshellMomentum} demonstrate that the only non-vanishing component of $C_{\a(p)\ad(q)}(\f)$ is $C_{{1 \ldots 1}{{\od\ldots\od}}}(\f)$. In accordance with the classification of massless UIRs (cf.\eqref{FMMasslessWignersEquation}), it follows from \eqref{FMWignerEqnMassless} that the field strength  $C_{\a(p)\ad(q)}(\f)$ describes a massless field with helicity $\l = \hf (p-q)$.

In the case $p>q$,  $C_{\a(p)\ad(q)}(\f)$ can always be expressed in terms of the gauge-invariant field strength $C_{\a(p-q)}(\f)$ which only carries undotted spinor indices,
\be
C_{\a(p)\ad(q)}(\f) = (\pa_{\a\ad})^q C_{\a(p-q)}(\f)~,
\ee
given that $C_{\a(p-q)}(\f)$ satisfies the on-shell condition
\be
\pa^{\b}{}_\bd C_{\b\a(p-q-1)}(\f)=0~.
\ee
Similarly for the case $q>p$, the field $C_{\a(p)\ad(q)}(\f)$ can always be written in terms of the gauge-invariant field strength $C_{\ad(q-p)}(\f)$ which only has dotted spinor indices 
\be
C_{\a(p)\ad(q)}(\f) = (\pa_{\a\ad})^p C_{\ad(q-p)}(\f)~,
\ee
given that $C_{\ad(q-p)}(\f)$ satisfies the on-shell condition
\be
\pa_{\b}{}^\bd C_{\bd\ad(q-p-1)}(\f)=0~.
\ee
In other words, we need only to work with constrained field strengths of a single index type to describe massless dynamics.

It proves instructive to perform an on-shell analysis of a massless theory in order to demonstrate how their dynamics can be studied in the two frameworks detailed above. As an example, we will study the action for a massless spin-$s$ field, which was formulated by Fronsdal \cite{Fronsdal1978Massless} in 1978. In two-component notation, Fronsdal's massless action is described in terms of two real bosonic fields $\f_{\a(s)\ad(s)}$ and $\vf_{\a(s-2)\ad(s-2)} $. Note that the analysis presented in this section follows that given by Buchbinder and Kuzenko in \cite{BuchbinderKuzenko1998}.

The Fronsdal action \cite{Fronsdal1978Massless} assumes the following form in two-component spinor notation
\bea \label{FMFronsAct}
S^{(s)}_{\text{F}}[\f,\vf] &=& \hf \Big ( -\frac{1}{2} \Big )^s \int \rd^4x~\Big \lb \f^{\a(s)\ad(s)}\Box \f_{\a(s)\ad(s)} - \frac{s}{2}\pa_{\b\bd}\f^{\b \a(s-1) \bd \ad(s-1)} \pa^{\g \gd} \f_{\g  \a(s-1) \gd \ad(s-1)} \non \\
&&-s(s-1)\vf^{\a(s-2)\ad(s-2)} \pa^{\b\bd}\pa^{\g\gd}\f_{\b\g\a(s-2)\bd\gd\ad(s-2)} -s(2s-1)\vf^{\a(s-2)\ad(s-2)}\Box\vf_{\a(s-2)\ad(s-2)} \non \\
&&-\hf s(s-2)^2\pa_{\b\bd}\vf^{\b\a(s-3)\bd\ad(s-3)}\pa^{\g\gd}\vf_{\g\a(s-3)\gd\ad(s-3)} \Big \rb ~.
\eea
The action \eqref{FMFronsAct} is invariant under the gauge transformations
\bsubeq
\bea
\d_\z \f_{\a(s)\ad(s)} &=& \pa_{\a\ad}\z_{\a(s-1)\ad(s-1)}~, \label{FMFronsGT1} \\
\d_\z \vf_{\a(s-2)\ad(s-2)} &=& \frac{1}{s^2}(s-1)\pa^{\b\bd}\z_{\b\a(s-2)\bd\ad(s-2)}~, \label{FMFronsGT2}
\eea
\esubeq
where the gauge parameter $\z_{\a(s-1)\ad(s-1)}$ is real unconstrained. The equations of motion corresponding to the action \eqref{FMFronsAct} read
\bsubeq \label{FMFronEoM1}
\bea
0 &=& 2\Box \f_{\a(s)\ad(s)} +s \pa_{\a\ad}\pa^{\b\bd}\f_{\b\a(s-1)\bd\ad(s-1)} -s(s-1)\pa_{\a\ad}\pa_{\a\ad} \vf_{\a(s-2)\ad(s-2)}~,  \\
0&=&(s-1)\pa^{\b\bd}\pa^{\g\gd}\f_{\b\g\a(2s-2)\bd\gd\ad(s-2)} + 2(2s-1)\Box \vf_{\a(s-2)\ad(s-2)} \non \\ &&-(s-2)^2\pa_{\a\ad}\pa^{\b\bd}\vf_{\b\a(s-3)\bd\ad(s-3)}~.
\eea
\esubeq
We can immediately make use of the gauge freedom \eqref{FMFronsGT2} to impose the gauge
\be \label{FMFronsG1}
\vf_{\a(s-2)\ad(s-2)}  =0~.
\ee
The gauge parameter of the residual symmetry is then constrained by 
\be \label{FMFronsTransGa}
\pa^{\b\bd}\z_{\b\a(s-2)\bd\ad(s-2)} = 0~.
\ee
In the gauge \eqref{FMFronsG1}, the equations of motion \eqref{FMFronEoM1} reduce to 
\bsubeq
\bea
0 &=& 2\Box \f_{\a(s)\ad(s)} +s \pa_{\a\ad}\pa^{\b\bd}\f_{\b\a(s-1)\bd\ad(s-1)} ~, \label{FMFrons2ndEoM1}\\
0&=&\pa^{\b\bd}\pa^{\g\gd}\f_{\b\g\a(2s-2)\bd\gd\ad(s-2)}~. \label{FMFrons2ndEoM2}
\eea
\esubeq
Computing the variation of $\pa^{\b\bd}\vf_{\b\a(s-1)\bd\ad(s-1)}$ yields
\be \label{FMFronsVariTranG}
\d_\z \big (\pa^{\b\bd}\vf_{\b\a(s-1)\bd\ad(s-1)} \big ) = -\frac{2}{s}\Box \z_{\a(s-1)\ad(s-1)}~.
\ee
Taking into account \eqref{FMFrons2ndEoM2}, it follows from \eqref{FMFronsVariTranG} that $\pa^{\b\bd}\vf_{\b\a(s-1)\bd \ad(s-1)}$ can be gauged away
\be \label{FMFronsFieldTrans}
\pa^{\b\bd}\vf_{\b\a(s-1)\bd \ad(s-1)} = 0 ~.
\ee
The gauge parameter of the residual gauge freedom is then constrained by
\be \label{FMFronsMasslessGa}
\Box \z_{\a(s-1)\ad(s-1)} = 0~.
\ee
In the gauge \eqref{FMFronsFieldTrans}, the equation of motion \eqref{FMFrons2ndEoM1} reduces to 
\be \label{FMFronsMasslesseqn}
0 = \Box \f_{\a(s)\ad(s)}~.
\ee
It follows from the on-shell analysis of the Fronsdal action \eqref{FMFronsAct} that the only surviving field $\f_{\a(s)\ad(s)}$ satisfies the massless on-shell conditions \eqref{FMMasslessfields} for the case $m=n=s$, as a consequence of the conditions \eqref{FMFronsGT1}, \eqref{FMFronsTransGa}, \eqref{FMFronsFieldTrans}, \eqref{FMFronsMasslesseqn} and \eqref{FMFronsMasslessGa}. Thus from the perspective of massless gauge fields, the model  \eqref{FMFronsAct} describes two massless spin-$s$ modes, with helicities $s$ and $-s$ respectively.

Moreover, it can be shown that Fronsdal's theory \eqref{FMFronsAct}  possesses two gauge-invariant field strengths associated with the field $\f_{\a(s)\ad(s)}$
\bsubeq \label{FMFronsFieldStrengths}
\bea
C_{\a(2s)}(\f) &=& \pa_{(\a_1}{}^{\bd_1} \cdots \pa_{\a_s}{}^{\bd_s} \f_{\a_{s+1} \ldots \a_{2s})\bd(s)}~, \\
\bar{C}_{\ad(2s)}(\f) &=& \pa_{(\ad_1}{}^{\b_1} \cdots \pa_{\ad_s}{}^{\b_s} \f_{\b(s)\ad_{s+1} \ldots \ad_{2s})}~.
\eea
\esubeq
By the on-shell analysis given above, it can be seen that the field strengths \eqref{FMFronsFieldStrengths} are non-vanishing, and satisfy the differential constraints
\be \label{FMFronsdalFSMassless}
\pa^{\b\gd}C_{\b\a(2s-1)}(\f) = 0~, \quad \pa^{\g\bd}\bar{C}_{\bd\ad(2s-1)} (\f)= 0~.
\ee
The constraints \eqref{FMFronsdalFSMassless} coincide with the conditions for a massless field within the framework of field strengths \eqref{FMMasslessOnshell}. In particular, the field strengths $C_{\a(2s)}(\f)$ and $\bar{C}_{\ad(2s)}(\f)$ can be identified as massless fields carrying helicity $s$ and $-s$ respectively. Thus in both frameworks, we consistently find that Fronsdal's action describes two massless spin-$s$ modes with helicities $s$ and $-s$.

\subsection{Spin projection operators} \label{Spin-projection operators}
In this section we review the spin projection operators on $\mb{M}^4$. Special attention is given to the study of the properties of these operators, and their corresponding applications. For integers $m,n \geq 1$, the spin projection operator $\P^{\perp}_{(m,n)}$ is defined by its action on $V_{(m,n)}$ via the rule
\bsubeq \label{FMProjector}
\bea
\P^{\perp}_{(m,n)}: {V}_{(m,n)} &\longrightarrow& {V}_{(m,n)}~, \\
\f_{\a(m)\ad(n)} &\longmapsto& \P^{\perp}_{(m,n)} \f_{\a(m)\ad(n)}  = : \P^{\perp}_{\a(m)\ad(n)}(\f) ~.
\eea
\esubeq
For fixed $m$ and $n$, the operator $\P^{\perp}_{(m,n)}$ satisfies the defining properties:
\begin{enumerate}
	\item \textbf{Idempotence}:  $\P^{\perp}_{(m,n)}$ is a projector in the sense that it squares to itself
	\bsubeq \label{FMBFProjectorsProp}
	\be
	\P^{\perp}_{(m,n)}	\P^{\perp}_{(m,n)}\f_{\a(m)\ad(n)}  = \P^{\perp}_{(m,n)}\f_{\a(m)\ad(n)}~. \label{FMIdempotence} 
	\ee
	\item \textbf{Transversality}: $\P^{\perp}_{(m,n)}$ maps any field $\f_{\a(m)\ad(n)}$ to a transverse field
	\be
	\pa^{\b\bd} \P^{\perp}_{\b\a(m-1)\bd\ad(n-1)}(\f) = 0~. \label{FMTransverse}
	\ee
	\esubeq
\end{enumerate}
It is easy to see that projector $\P_{(m,n)}^{\perp}$ maps an unconstrained field $\f_{\a(m)\ad(n)} \in V_{(m,n)}$ on the mass-shell \eqref{MinkOnshellMass} to a massive field \eqref{FMOnshellConditions},
\bsubeq \label{FMBehrendsFronsdalSpinor}
\bea
\pa^{\b\bd} \P^{\perp}_{(m,n)}  \f_{\b\a(m-1)\bd \ad(n-1)} = 0~, \\ 
(\Box - \bm{m}^2) \P^{\perp}_{(m,n)} \f_{\a(m)\ad(n)} = 0~.
\eea
\esubeq
Hence, the projector $\P_{(m,n)}^{\perp}$ selects the physical component which furnishes the massive UIR $G(\bm{m},\frac{1}{2}(m+n))$ from an unconstrained field $\f_{\a(m)\ad(n)}$ satisfying the Klein-Gordon equation.

The spin projection operators \eqref{FMBFProjectorsProp} take the following explicit form on $V_{(m,n)}$ \cite{GatesGrisaruRocekSiegel1983}
\bsubeq \label{FMBFprojectors}
\bea 
\P^{\perp}_{(m,n)}\f_{\a(m)\ad(n)} &=& \frac{1}{\Box^n}\pa_{(\ad_1}{}^{\b_1} \cdots \pa_{\ad_n)}{}^{\b_n} \pa_{(\b_1}{}^{\bd_1} \cdots \pa_{\b_n}{}^{\bd_n} \f_{\a_1 \ldots \a_m)\bd(n)}~, \label{FMBFProjector1}\\
\widehat{\P}^{\perp}_{(m,n)}\f_{\a(m)\ad(n)}  &=& \frac{1}{\Box^m}\pa_{(\a_1}{}^{\bd_1} \cdots \pa_{\a_m)}{}^{\bd_m} \pa_{(\bd_1}{}^{\b_1} \cdots \pa_{\bd_m}{}^{\b_m} \f_{\b(m)\ad_1 \ldots \ad_n)}~. \label{FMBFProjector2}
\eea
\esubeq
It is immediately apparent that for $\f_{\a(n)}$ off the mass-shell, the projected fields $\P^{\perp}_{(m,n)}\f_{\a(m)\ad(n)}$ and $\widehat{\P}^{\perp}_{(m,n)}\f_{\a(m)\ad(n)}$ are non-local due to the presence of the inverse d'Alembertian operators. Locality can be restored if the field being projected satisfies the massive Klein-Gordon equation \eqref{MinkOnshellMass}. In addition, it appears that the spin projection operators diverge if the field being mapped satisfies the massless Klein-Gordon equation \eqref{FMMasslessMassShell}. 

Thus we are able to draw some interesting conclusions from the structural form of the spin projection operators. Specifically, the relativistic wave equations which make the projectors local are characteristic of a massive field. On the other hand, the wave equations for which the projectors diverge are attributed to massless gauge fields. Accompanying the fields satisfying these wave equations with the defining constraint of the spin projection operator on $V_{(m,n)}$, which is the transverse condition, allows one to obtain the complete dictionary of on-shell fields in $\mb{M}^4$. This observation will be very useful in providing some intuition concerning the poles of the projection operators in AdS backgrounds, as we will see in the later sections of this thesis.

It can be shown after some work that the spin projection operators \eqref{FMBFProjector1} and \eqref{FMBFProjector2} are indeed equivalent on the space $V_{(m,n)}$
\be
\P^{\perp}_{(m,n)}\f_{\a(m)\ad(n)}= \widehat{\P}^{\perp}_{(m,n)}\f_{\a(m)\ad(n)}~.
\ee
Due to this equivalence, we will only consider $\P^{\perp}_{(m,n)}$ on $V_{(m,n)}$ unless stated otherwise. 

Additionally, the projector $\P^{\perp}_{(m,n)}$ acts like the identity operator (is surjective) on the space of transverse fields $\f^{\perp}_{\a(m)\ad(n)}$
\be \label{FMBFSurjectivity1}
\pa^{\b\bd}\f^{\perp}_{\b\a(m-1)\bd\ad(n-1)} = 0 \qquad \Longrightarrow \qquad {\P}^{\perp}_{(m,n)}\f^{\perp}_{\a(m)\ad(n)} = \f^{\perp}_{\a(m)\ad(n)}~.
\ee

The spin projection operators \eqref{FMBFprojectors} for fields of arbitrary rank, including mixed symmetry $m \neq n$, were first derived by Siegel and Gates \cite{SiegelGates1981} (see also \cite{GatesGrisaruRocekSiegel1983}). For the cases $m=n$ and $m=n+1$, the spin projection operators are equivalent to those derived in the pioneering work of Fronsdal \cite{Fronsdal1958}. In vector notation, the explicit form of these operators is very complicated and consequently proves very difficult  work with. However, in the two-component spinor formalism, the form of the spin projection operators take the remarkably simple and elegant form \eqref{FMBFprojectors} which is more tractable to use. This is due to the fact that in two component spinor notation, we work with irreducible spinor fields which are automatically traceless and symmetric.\footnote{We can start with a traceless tensor field in vector notation. For this to hold in general, it is necessary derive an operator which maps an arbitrary traceful tensor field to a traceless field.}

It is possible to derive an alternative set of spin projection operators which can be written solely in terms of the Casimir operators of $\PaF$. To this end, let us introduce the operator $\tilde{\P}^{\perp}_{(m,n)}$ which acts on $V_{(m,n)}$ in the following manner
\bsubeq \label{FMBFProjectorCasimir}
\bea
\tilde{\P}^{\perp}_{(m,n)}\f_{\a(m)\ad(n)}&=&\frac{m!}{(m+n)!n!}\frac{1}{\Box^n}\prod_{j=0}^{n-1}\big (\mb{W}^2 - (s-j)(s-j-1)\Box \big )\f_{\a(m)\ad(n)}~, \hspace{1cm} \label{FMBFProjectorCasimir1} \\
&=&\frac{n!}{(m+n)!m!}\frac{1}{\Box^m}\prod_{j=0}^{m-1}\big (\mb{W}^2 - (s-j)(s-j-1)\Box \big )\f_{\a(m)\ad(n)}~. \label{FMBFProjectorCasimir2} 
\eea
\esubeq
It can be shown that $\tilde{\P}^{\perp}_{(m,n)}$ satisfies the defining properties of  spin projection operator \eqref{FMBFProjectorsProp}.
The formulation of projection operators in terms of Casimir operators \eqref{FMBFProjectorCasimir} in $\mb{M}^4$ was first studied by Aurilia and Umezawa in \cite{AuriliaUmezawa1967,AuriliaUmezawa1969} for the cases of spin-$s$ and spin-$(s+\hf)$ fields. For $s \geq 1$, the projection operators given in \cite{AuriliaUmezawa1967,AuriliaUmezawa1969} take the explicit form\footnote{Note that the Casimir operator $W$ given in \cite{AuriliaUmezawa1967,AuriliaUmezawa1969} is equivalent to $W^2$ \eqref{FMPoincareQuartCasimir}. Additionally, we make the redefinition $p^2 \mapsto -p^2$ to ensure consistency between the conventions of \cite{AuriliaUmezawa1967,AuriliaUmezawa1969} and those employed in this thesis.}
\bsubeq
\bea
\tilde{\P}^{\perp}_{(s)} &=& \prod_{j=0}^{s-1} \frac{1}{\big (s(s+1)-j(j+1) \big )\Box} \big ( \mb{W}^2 - j(j+1) \Box \big )~, \label{FMAUBosonicProjector}\\
\tilde{\P}^{\perp}_{(s+\hf)} &=& \prod_{j=0}^{s-1} \frac{1}{\big ( (s+\hf)(s+\tfrac{3}{2}) - (j+\hf)(j + \frac{3}{2})\big )\Box} \big (\mb{W}^2 - \frac{1}{4}(2j+1)(2j+3)\Box \big )~. \hspace{1cm}\label{FMAUFermionicProjector}
\eea
\esubeq
Upon reversing the order of the products, the expressions \eqref{FMAUBosonicProjector} and \eqref{FMAUFermionicProjector} can be shown to coincide with the spin projection operators \eqref{FMBFProjectorCasimir1} for the cases $m=n=s$ and  $m=n+1 = s+1$, respectively.
The operator $\tilde{\P}^{\perp}_{(m,n)}$  also acts like the identity operator on the space of transverse fields $\f^{\perp}_{\a(m)\ad(n)}$
\be \label{FMBFSurjectivity}
\tilde{\P}^{\perp}_{(m,n)}\f^{\perp}_{\a(m)\ad(n)} = \f^{\perp}_{\a(m)\ad(n)}~.
\ee

Although it is not immediately obvious, it turns out that the spin projection operators $\P^{\perp}_{(m,n)}$ and $\tilde{\P}^{\perp}_{(m,n)}$ are equivalent on $V_{(m,n)}$. Proving this equivalence follows quite simply from the properties of the spin projectors. Specifically,
let us consider the action of $\tilde{\P}^{\perp}_{(m,n)}$, followed by the action of $\P^{\perp}_{(m,n)}$ on an arbitrary field $\f_{\a(m)\ad(n)}$. Since $\tilde{\P}^{\perp}_{(m,n)}\f_{\a(m)\ad(n)}$ is transverse, and $\P^{\perp}_{(m,n)}$ acts like the identity operator on this space, we find
\bsubeq \label{FMBFEquivalence1}
\be
\P^{\perp}_{(m,n)}\tilde{\P}^{\perp}_{(m,n)}\f_{\a(m)\ad(n)} = \tilde{\P}^{\perp}_{(m,n)}\f_{\a(m)\ad(n)}~. \label{FMBFEquivalenceProof1}
\ee
Next, we perform the same procedure but in the opposite order,
\be
\tilde{\P}^{\perp}_{(m,n)}\P^{\perp}_{(m,n)}\f_{\a(m)\ad(n)} = {\P}^{\perp}_{(m,n)}\f_{\a(m)\ad(n)}~. \label{FMBFEquivalenceProof2}
\ee
\esubeq
Since $\tilde{\P}^{\perp}_{(m,n)}$ is composed strictly from Casimir operators, and hence commutes  with $\P^{\perp}_{(m,n)}$, it follows from \eqref{FMBFEquivalenceProof1} and \eqref{FMBFEquivalenceProof2} that the two projectors are equivalent on $V_{(m,n)}$
\be \label{FMBFEquivalence}
\P^{\perp}_{(m,n)}\f_{\a(m)\ad(n)} = \tilde{\P}^{\perp}_{(m,n)}\f_{\a(m)\ad(n)}~.
\ee
This argument can be used to prove that the spin projection operators $\P^{\perp}_{(m,n)}$ are unique on $\mb{M}^4$. 

The form of the projector which one chooses to work with is dependent on the problem under consideration. It is natural to work with the projector in the form \eqref{FMBFProjector1} when proving that the operator satisfies the defining properties of idempotence and transversality. One of the immediate benefits of writing the projectors in the form \eqref{FMBFProjectorCasimir} is that they automatically factorise into a product of second-order differential operators. Additionally, the operator in the form $\tilde{\P}^{\perp}_{(m,n)}$ allows one to easily show that the spin projection operators act like the unit operator on the space of transverse fields.

The projectors $\tilde{\P}^{\perp}_{(m,n)}$ display some interesting properties if they are no longer restricted to the space $V_{(m,n)}$. The first is that the spin projection operators $\P^{\perp}_{(m,n)}$ \eqref{FMBFProjector1} and $\tilde{\P}^{\perp}_{(m,n)}$ \eqref{FMBFProjectorCasimir} are no longer equivalent on $V_{(j,k)}$, where $1 \leq k \leq m-1$ and $1 \leq j \leq n-1$. As a matter of fact, the operators $\P^{\perp}_{(m,n)}$ \eqref{FMBFProjector1} and $\tilde{\P}^{\perp}_{(m,n)}$ \eqref{FMBFProjectorCasimir} are no longer idempotent and transverse on $V_{(k,j)}$. It can be shown that $\tilde{\P}^{\perp}_{(m,n)}$ \eqref{FMBFProjectorCasimir} annihilates all rank-$(m-1,n-1)$ fields
\be \label{FMGeneralTransprop}
\tilde{\P}^{\perp}_{(m,n)}\f_{\a(m-1)\ad(n-1)} = 0~.
\ee
We wish to study the properties of the operator $\tilde{\P}_{(m,n)}$, for the archetypal cases of $m=n=s$ and $m=n+1=s$, on certain spaces $V_{(k,j)}$.

In the case $m=n=s$, the operator $\tilde{\P}_{(s,s)}^{\perp}$ takes the explicit form
\be
\tilde{\P}_{(s,s)}^{\perp} = \frac{1}{\Box^s(2s)!}\prod_{j=0}^{s-1} \big ( \mb{W}^2 - j(j+1) \Box \big )~.
\ee
It can be shown on $V_{(k,k)}$, where $1 \leq k\leq s-1$,  the operator $\tilde{\P}_{(s,s)}^{\perp}$ annihilates any lower rank field $\f_{\a(k) \ad(k)} $
\be \label{FMBehrendsFronsdalAnnihilatesLowBosons}
\tilde{\P}_{(s,s)}^{\perp}\f_{\a(k) \ad(k)} = 0~.
\ee
In the case $m=n+1=s$, the operator $\tilde{\P}^{\perp}_{(s,s-1)}$ assumes the form
\bsubeq \label{FMFermionicProjectorCasimir}
\bea
\tilde{\P}^{\perp}_{(s,s-1)} &=& \frac{s}{(2s-1)!}\frac{1}{\Box^{s-1}} \prod_{j=0}^{s-2} \Big ( \mb{W}^2 - \frac{1}{4}(2j+1)(2j+3)\Box \Big ) \\
&=&\frac{1}{s(2s-1)!}\frac{1}{\Box^s}\prod_{j=0}^{s-1} \Big ( \mb{W}^2 - \frac{1}{4}(2j-1)(2j+1)\Box \Big )~.
\eea
\esubeq
It can be shown that on $V_{(k,k-1)}$, where $1 \leq k \leq s-1$, the operator  $\tilde{\P}^{\perp}_{(s,s-1)} $ annihilates all lower rank fields $\f_{\a(k)\ad(k-1)} $
\be \label{FMBehrendsFronsdalAnnihilatesLowFerm}
\tilde{\P}^{\perp}_{(s,s-1)}\f_{\a(k)\ad(k-1)} = 0 ~.
\ee

The properties \eqref{FMBehrendsFronsdalAnnihilatesLowBosons} and \eqref{FMBehrendsFronsdalAnnihilatesLowFerm} have interesting implications on fields in (vector)four-component spinor notation. 
For example, let us consider a rank-$s$ traceful tensor fields $\f^{\text{T}}_{a(s)}$. Converting to two-component spinor notation yields
\be
\f^{\text{T}}_{\a_1 \ad_1 , \a_2 \ad_2 , \ldots , \a_s \ad_s} = \f_{\a(s)\ad(s)} + \text{trace contributions}~,
\ee
where $\f_{\a(s)\ad(s)}$ is the traceless component of  $\f^{\text{T}}_{a(s)}$, while the other terms consist of lower rank bosonic fields associated with the trace. Due to property \eqref{FMBehrendsFronsdalAnnihilatesLowBosons}, one can conclude that the operator $\tilde{\P}_{(s,s)}^{\perp}$ selects out the TT component of $\f^{\text{T}}_{a(s)}$
\be
\pa^b\f^{\text{TT}}_{ba(s-1)} = 0~, \qquad \eta^{bc}\f^{\text{TT}}_{bca(s-2)} = 0~, \qquad \tilde{\P}_{(s,s)}^{\perp}\f^{\text{T}}_{a(s)}:=\f^{\text{TT}}_{a(s)}~.
\ee
Therefore, the spin-$s$ projection operator is a TT projector when acting on a symmetric traceful rank-$s$ field $\f^{\text{TT}}_{a(s)}$. 

\subsubsection{Longitudinal projectors and lower-spin extractors}

Let us introduce the orthogonal complement of $\P^{\perp}_{(m,n)}$ on $V_{(m,n)}$
\be \label{FMLongProj}
\P_{(m,n)}^{\parallel} =  \mds{1} - \P_{(m,n)}^{\perp} ~,  
\ee
which automatically satisfies the following properties by construction
\bsubeq \label{FMLongProjProp}
\be \label{FMLongProjIdem}
\P_{(m,n)}^{\parallel} \P_{(m,n)}^{\parallel} =\P_{(m,n)}^{\parallel} ~, 
\ee
\be
\qquad \P_{(m,n)}^{\parallel} \P_{(m,n)}^{\perp} = \P_{(m,n)}^{\perp} \P_{(m,n)}^{\parallel} =0~.
\ee
\esubeq
Explicitly evaluating $\P_{(m,n)}^{\parallel}$ on $V_{(m,n)}$, it can be shown that $\P_{(m,n)}^{\parallel}$  projects onto the longitudinal part of  $\f_{\a(m)\ad(n)}$
\be \label{FMLongProjection}
\P_{(m,n)}^{\parallel} \f_{\a(m)\ad(n)} = \partial_{\a\ad} \f_{\a(m-1)\ad(n-1)}~,
\ee
where $\f_{\a(m-1)\ad(n-1)}$ is unconstrained. The operator $\P_{(m,n)}^{\parallel}$ is called the longitudinal projector, as a result of \eqref{FMLongProjIdem} and \eqref{FMLongProjection}.

Let  $\f^{\parallel}_{\a(m)\ad(n)} = \pa_{\a\ad} \f_{\a(m-1)\ad(n-1)}$ be some longitudinal field. Since the spin projector $\tilde{\P}_{(m,n)}^{\perp}$ annihilates any rank-$(m-1,n-1)$ field \eqref{FMGeneralTransprop}, it follows that it also annihilates any rank-$(m,n)$ longitudinal field
\be \label{FMTransKillsLong}
\tilde{\P}_{(m,n)}^{\perp} \f^{\parallel}_{\a(m)\ad(n)}  = 0~. 
\ee
An immediate consequence of the identity \eqref{FMTransKillsLong} is that the longitudinal projector $\P_{(m,n)}^{\parallel}$ acts as the identity operator on the space of rank-$(m,n)$ longitudinal fields 
\begin{alignat}{3} \label{key}
\tilde{\P}_{(m,n)}^{\perp} \f^{\parallel}_{\a(m)\ad(n)}  &=& ~ 0 \quad  &\Longrightarrow& \quad  \P_{(m,n)}^{\parallel} \f^{\parallel}_{\a(m)\ad(n)}  &=  \f^{\parallel}_{\a(m)\ad(n)} ~.
\end{alignat}
These properties will prove useful in the subsequent section.

We can make use of the fact that the projectors $\P^{\perp}_{(m,n)}$ and $\P_{(m,n)}^{\parallel}$ resolve the identity \eqref{FMLongProj} to decompose any tensor field $\f_{\a(m)\ad(n)}$ in the following manner
\be \label{FMDecomp}
\f_{\a(m)\ad(n)} = \big (\P^{\perp}_{(m,n)} + \P^{\parallel}_{(m,n)} \big )\f_{\a(m)\ad(n)} = \f^{\perp}_{\a(m)\ad(n)} + \pa_{\a\ad} \f_{\a(m-1)\ad(n-1)}~,
\ee
where $\f^{\perp}_{\a(m)\ad(n)} $ is transverse and $\f_{\a(m-1)\ad(n-1)}$ is unconstrained and thus reducible. It follows from properties the \eqref{FMBFSurjectivity} and \eqref{FMLongProjProp} that the transverse projector $\tilde{\P}^{\perp}_{(m,n)}$ selects the component $\f^{\perp}_{\a(m)\ad(n)} $  from the decomposition \eqref{FMDecomp}. In other words, the spin projection operators extract the pure spin state with maximal spin $s$ from an arbitrary field off the mass-shell. 

We can continue this prescription iteratively in order to obtain the general decomposition of any field $\f_{\a(m)\ad(n)}$ into irreducible components
\be \label{FMDecomposition}
\f_{\a(m)\ad(n)}  = \f^{\perp}_{\a(m)\ad(n)}  + \sum_{j=1}^{n-1}(\partial_{\a\ad})^j \f^{\perp}_{\a(m-j)\ad(n-j)} +(\partial_{\a\ad})^n\f_{\a(m-n)}~,
\ee
where we have assumed, without loss of generality, that $m \geq n$. The collection of fields $\big \{\f^{\perp}_{\a(m)\ad(n)}, \f^{\perp}_{\a(m-1)\ad(n-1)}, \cdots ,  \f^{\perp}_{\a(m-n+1)\ad} \big \}$ appearing in the decomposition are transverse, while $\f_{\a(m-n)}$ does not satisfy any differential constraints.

We can introduce operators which extract the bosonic $\f^{\perp}_{\a(s-k)\ad(s-k)}$ and fermionic $\f^{\perp}_{\a(s-k)\ad(s-k-1)} $ fields  from the decomposition \eqref{FMDecomposition}. In particular, the spin-$(s-k)$ and spin-$(s-k-\hf)$ components can be extracted via
\bsubeq
\bea
\f_{\a(s)\ad(s)} &\mapsto& \f^{\perp}_{\a(s-k)\ad(s-k)} =  \big (S^{\perp}_{(s-k)}\f \big )_{\a(s-k)\ad(s-k)} \equiv S^{\perp}_{\a(s-k)\ad(s-k)}(\f) ~, \\
\f_{\a(s)\ad(s-1)} &\mapsto& \f^{\perp}_{\a(s-k)\ad(s-k-1)} =  \big (S^{\perp}_{(s-k-\hf)}\f \big )_{\a(s-k)\ad(s-k-1)} \equiv S^{\perp}_{\a(s-k)\ad(s-k-1)}(\f) ~, \hspace{1cm}
\eea
\esubeq
where we have defined the operators
\bsubeq \label{FMExtractors}
\bea
S^{\perp}_{\a(s-k)\ad(s-k)}(\f) &=& a_k \Box^{-k}   \tilde{\P}^{\perp}_{(s-k,s-k)} \big (\pa^{\b\bd} \big )^k \f_{\b(k)\a(s-k)\bd(k)\ad(s-k)}~, \\
S^{\perp}_{\a(s-k)\ad(s-k-1)}(\f) &=& b_k \Box^{-k}   \tilde{\P}^{\perp}_{(s-k,s-k-1)} \big (\pa^{\b\bd} \big )^k \f_{\b(k)\a(s-k)\bd(k)\ad(s-k-1)}~,
\eea
with the constants $a_k$ and $b_k$ taking the form
\be
a_k = (-1)^k \binom{s}{k}^2 \binom{2s-k+1}{k}^{-1}~, \quad b_k = (-1)^k \binom{s}{k} \binom{s-1}{k} \binom{2s-k}{k}^{-1}~.
\ee
\esubeq
It is clear that  $S^{\perp}_{\a(s-k)\ad(s-k)}(\f)$ and $S^{\perp}_{\a(s-k)\ad(s-k-1)}(\f)$ are transverse 
\be
\pa^{\b\bd}S^{\perp}_{\b\a(s-k-1)\bd\ad(s-k-1)}(\f)=0~, \qquad \pa^{\b\bd}S^{\perp}_{\b\a(s-k-1)\bd\ad(s-k-2)}(\f)=0~,
\ee
due to the presence of the spin projection operators.
The operators $S^{\perp}_{\a(s-k)\ad(s-k)}(\f)$ and $S^{\perp}_{\a(s-k)\ad(s-k-1)}(\f)$ are called the spin-$(s-k)$ and spin-$(s-k-\hf)$ transverse extractors respectively. Note that the extractors \eqref{FMExtractors} are not idempotent because they are dimensionful. 

\subsection{Conformal higher-spin theory}\label{FMCHSSec}

In this section we illustrate an important application of the spin projection operators in the context of conformal higher-spin theory in $\mb{M}^4$.  In the spirit of the pioneering work of Fradkin and Tseytlin \cite{FradkinTseytlin1985}, we show that the actions for  CHS fields with mixed symmetry can be recast in terms of the spin projectors \eqref{FMBFprojectors}. Before doing this, it is necessary to review the construction of CHS theory in two-component spinor notation \cite{KuzenkoManvelyanTheisen2017}. 

For fixed integers $m,n \geq 1$, a CHS gauge field in $\mb{M}^4$ is described by a complex spinor field $h_{\a(m)\ad(n)}$ of Lorentz type $(\frac{m}{2},\frac{n}{2})$ which is defined modulo gauge transformations of the form \cite{KuzenkoManvelyanTheisen2017}
\be \label{FMCHSGT}
\d_\x h_{\a(m)\ad(n)} = \pa_{\a\ad} \x_{\a(m-1)\ad(n-1)}~.
\ee
Here, the gauge parameter $\x_{\a(m-1)\ad(n-1)}$ is complex unconstrained. The bosonic ($m=n$)  and fermionic  ($m=n+1$ and $n=m+1$) CHS fields were first given in \cite{FradkinTseytlin1985}. They were later extended to the cases of bosonic and fermionic fields with mixed symmetry in the works \cite{Vasiliev2009ck} and \cite{KuzenkoManvelyanTheisen2017}, respectively. 

Associated with the CHS field $h_{\a(m)\ad(n)}$ and its complex conjugate $\bar{h}_{\a(n)\ad(m)}$ are the linearised higher-spin Weyl tensors \cite{KuzenkoManvelyanTheisen2017}
\bsubeq \label{FMCHSWeylTensors}
\bea
{{W}}_{\a(m+n)}(h) &=& \pa_{(\a_1}{}^{\bd_1} \cdots \pa_{\a_n}{}^{\bd_n} h_{\a_{n+1} \ldots \a_{m+n}) \bd(n)}~, \\
{{W}}_{\a(m+n)}(\bar{h}) &=& \pa_{(\a_1}{}^{\bd_1} \dots \pa_{\a_m}{}^{\bd_m} \bar{h}_{\a_{m+1} \dots \a_{m+n})\bd(m)}~.
\eea
\esubeq
The complex conjugates of the field strengths \eqref{FMCHSWeylTensors}, $\overline{W}_{\ad(m+n)}(h) = \big (W_{\a(m+n)}(\bar{h}) \big )^* $ and  $\overline{W}_{\ad(m+n)}(\bar{h}) = \big (W_{\a(m+n)}(h) \big )^* $, can be computed via complex conjugation. The higher-spin Weyl tensors \eqref{FMCHSWeylTensors} are invariant under the gauge transformations \eqref{FMCHSGT}
\be \label{FMWeylGI}
\d_\x {{W}}_{\a(m+n)}(h) = 0 ~, \qquad \d_\x {{W}}_{\a(m+n)}(\bar{h}) = 0 ~.
\ee
Due to the properties \eqref{FMWeylGI}, it follows that the linearised conformal higher-spin action \cite{KuzenkoManvelyanTheisen2017}
\be \label{FMCHSActionWeyl}
S_{\text{CHS}}^{(m,n)}[h,\bar{h}] = \ri^{m+n} \int  \rd^4 x~ {{W}}^{\a(m+n)}(h) {{W}}_{\a(m+n)}(\bar{h})~+~\text{c.c.}~,
\ee
is invariant under the gauge transformations \eqref{FMCHSGT}. Note that by $+ \text{c.c.}$ we mean to add the complex conjugate. The presence of this complex conjugate contribution ensures that the actions being studied are real. The CHS actions were first expressed in terms of the linearised higher-spin Weyl tensors in \cite{FradkinLinetsky1991} by Fradkin and Linetsky, for the cases $m=n$ and $m=n+1$. 

As the name suggests, the CHS action \eqref{FMCHSActionWeyl} is also invariant under conformal transformations. In this thesis, we do not elaborate on the conformal properties of CHS theories, as in general, these are well studied. Instead, we will direct the interested reader to the appropriate resources which detail the conformal properties of the CHS actions under consideration. For example, for more information concerning the conformal properties of the CHS theories \eqref{FMCHSActionWeyl} in $\mb{M}^4$, see \cite{KuzenkoManvelyanTheisen2017,KuzenkoPonds2019}.  We are primarily interested in studying CHS theories due to their connection to spin projection operators.

Integrating by parts, the CHS action \eqref{FMCHSActionWeyl} can be written in the alternative form
\bsubeq
\bea
S_{\text{CHS}}^{(m,n)}[h,\bar{h}] &=& \ri^{m+n} \int  \rd^4 x~ \bar{h}^{\a(n)\ad(m)}{{B}}_{\a(n)\ad(m)}(h) +\HC~\\
&=& \ri^{m+n} \int  \rd^4 x~ \bar{h}^{\a(n)\ad(m)}\widehat{B}_{\a(n)\ad(m)}({h}) +\HC~, 
\eea
\esubeq
in which the linearised higher-spin Bach tensors appear 
\bsubeq \label{FMbachtensors}
\bea 
{{B}}_{\a(n)\ad(m)}(h) &=& \pa_{(\ad_1}{}^{\b_1} \cdots \pa_{\ad_m)}{}^{\b_m} {{W}}_{\a(n)\b(m)}(h)~,  \label{FMbachT1}\\
{\widehat{B}}_{\a(n)\ad(m)}({h}) &=& \pa_{(\a_1}{}^{\bd_1} \cdots \pa_{\a_n)}{}^{\bd_n} {\overline{W}}_{\ad(m)\bd(n)}({h})~. \label{FMbachT2}
\eea
\esubeq
The higher-spin Bach tensors \eqref{FMbachtensors} are gauge-invariant 
\be \label{FMBachGI}
\d_\x {{B}}_{\a(n)\ad(m)}(h) = 0 ~, \qquad \d_\x {\widehat{B}}_{\a(n)\ad(m)}({h}) = 0 ~,
\ee
and transverse
\be \label{FMBachTransverse}
\pa^{\b\bd}{{B}}_{\b\a(n-1)\bd\ad(m-1)}(h)  = 0~, \qquad \pa^{\b\bd}{\widehat{B}}_{\b\a(n-1)\bd\ad(m-1)}({h})  = 0~.
\ee
After some work, the Bach tensors \eqref{FMbachT1} and \eqref{FMbachT2} can be shown to be equivalent
\be
{{B}}_{\a(n)\ad(m)}(h)  = {\widehat{B}}_{\a(n)\ad(m)}({h}) ~.
\ee
For the remainder of this section, we will only work with ${{B}}_{\a(n)\ad(m)}(h)$.

Due to the transverse nature \eqref{FMBachTransverse} of ${{B}}_{\a(n)\ad(m)}(h)$, it is natural to recast the  Bach tensors in terms of the unique transverse projection operators \eqref{FMBFprojectors} as follows
\bsubeq \label{FMBachtensorsProj}
\bea
{B}_{\a(n)\ad(m)}(h) &=& \Box^n \pa_{(\ad_1}{}^{\b_1} \cdots \pa_{\ad_{m-n}}{}^{\b_{m-n}}\P_{(m,n)}^{\perp}h_{\b(m-n)\a(n)\ad_{m-n+1} \ldots \ad_m)} ~, ~~ m \geq n~, \\
{B}_{\a(n)\ad(m)}(h) &=& \Box^m \pa_{(\a_1}{}^{\bd_1} \cdots \pa_{\a_{n-m}}{}^{\bd_{n-m}}\P_{(m,n)}^{\perp}h_{\a_{n-m+1} \ldots \a_n)\ad(m)\bd(n-m)} ~, ~~ n \geq m~. \hspace{1cm}
\eea
\esubeq
One of the immediate benefits of expressing the Bach tensors \eqref{FMBachtensorsProj} in terms of spin projection operators is that the defining properties of divergenceless and gauge-invariance are made manifest.

Since the Bach tensor \eqref{FMbachtensors} is a descendent of the Weyl tensors \eqref{FMCHSWeylTensors}, it follows that they can also be realised in terms of the spin projection operators \eqref{FMBFprojectors}
\bsubeq
\bea
{W}_{\a(m+n)}(h) &=& \pa _{(\a_1}{}^{\bd_1} \cdots  \pa _{\a_n}{}^{\bd_n} \P^{\perp}_{(m,n)}h_{\a_{n+1} \ldots \a_{m+n})\bd(n) }~, \\
{W}_{\a(m+n)}(\bar{h}) &=& \pa _{(\a_1}{}^{\bd_1} \cdots  \pa _{\a_m}{}^{\bd_m}\P^{\perp}_{(n,m)}\bar{h}_{\a_{m+1} \ldots \a_{m+n})\bd(m) }~. 
\eea
\esubeq
It follows that the gauge-invariant CHS action \eqref{FMCHSActionWeyl} can be written in the form
\bsubeq \label{FMCHSProj}
\bea 
S^{(m,n)}_{\text{CHS}}[h,\bar{h}] &=& \ri^{m+n}\int \rd^4x~ \bar{h}^{\a(n)\ad(m)}\Box^n \big ( \pa_\ad{}^\b \big )^{m-n} \P_{(m,n)}^{\perp}h_{\b(m-n)\a(n)\ad(n)} + \text{c.c.} ~,  \label{FMCHSProj1}\\
S^{(m,n)}_{\text{CHS}}[h,\bar{h}] &=& \ri^{m+n}\int \rd^4x~ \bar{h}^{\a(n)\ad(m)}\Box^m \big ( \pa_\a{}^\bd \big )^{n-m} \P_{(m,n)}^{\perp}h_{\a(m)\bd(n-m)\ad(m)} + \text{c.c.}~, \hspace{1cm}  \label{FMCHSProj2}
\eea
\esubeq
where we have assumed  $m > n$ in \eqref{FMCHSProj1} and $n > m$ in  \eqref{FMCHSProj2}.
There are many benefits to recasting the CHS actions \eqref{FMCHSProj} in terms of the spin projection operators. Besides manifest gauge invariance, which follows from \eqref{FMTransKillsLong}, the presence of the projectors makes the defining feature of CHS theories, that they describe pure spin $s=\hf(m+n)$ states off the mass-shell, manifest. Moreover, they are useful in showing that the CHS actions factorise into products of second order differential equations. 

Let us demonstrate this property for the bosonic ($m=n=s$) CHS action. We begin by imposing the reality condition $h_{\a(s)\ad(s)} = \bar{h}_{\a(s)\ad(s)}$. Under reality, the bosonic action \eqref{FMCHSProj} becomes
\be
S^{(s,s)}_{\text{CHS}}[h] = (-1)^s \int \rd^4 x~h^{\a(s)\ad(s)} \Box^s \Pi^{\perp}_{(s,s)} h_{\a(s)\ad(s)}~,
\ee
which is equivalent to the conformal spin-$s$ action \eqref{FMCHSBos} given in \cite{FradkinTseytlin1985}. Next, let us decompose the CHS field $h_{\a(s)\ad(s)}$ into transverse and longitudinal parts using \eqref{FMDecomposition}. Since the spin projection operator $\Pi^{\perp}_{(s,s)}$ acts as the unit operator on the space of transverse fields \eqref{FMBFSurjectivity}, and annihilates all bosonic fields of lower rank \eqref{FMBehrendsFronsdalAnnihilatesLowBosons}, we find that the CHS action can be written in terms of the transverse component of the CHS field
\be \label{FMCHSBOSSO}
S^{(s,s)}_{\text{CHS}}[h^{\perp}] = (-1)^s \int \rd^4 x~h^{\perp \a(s)\ad(s)} \Box^s h^{\perp}_{\a(s)\ad(s)}~.
\ee
Thus we see that the bosonic CHS action \eqref{FMCHSBOSSO} can be factorised into second-order differential operators. This procedure utilised above is equivalent to fixing the gauge symmetry \eqref{FMCHSGT} by imposing the transverse gauge condition $h_{a(s)} \equiv h^{\perp}_{a(s)}$, since the action \eqref{FMCHSBOSSO} no longer has gauge symmetry. 

In the fermionic case $m=n+1=s+1$, the CHS action \eqref{FMCHSProj} takes the form
\be \label{FMCHSFermPro}
S^{(s+1,s)}_{\text{CHS}}[h,\bar{h}] = \ri (-1)^{s} \int \rd^4x~ \bar{h}^{\a(s)\ad(s+1)}\Box^s \P^{\perp}_{(s+\hf)} \pa_{(\ad_1}{}^{\b} h_{\b \a(s) \ad_2 \ldots \ad_{s+1})}~,
\ee
where we have defined $\P^{\perp}_{(s+1,s)}:=\P^{\perp}_{(s+\hf)}$. The action \eqref{FMCHSFermPro} is equivalent to the conformal spin-$(s+\hf)$ action given in \cite{FradkinTseytlin1985}. Following the argument presented in the bosonic case, it follows that the CHS action \eqref{FMCHSFermPro} cannot be factorised entirely in terms of second-order operators
\be \label{FMCHSFermProDec}
S^{(s,s+1)}_{\text{CHS}}[h^{\perp},\bar{h}^{\perp}] = \ri (-1)^s  \int \rd^4 x~\bar{h}^{\perp \a(s)\ad(s+1)} \Box^s \pa_{(\ad_1}{}^{\b}h^{\perp}_{\b \a(s)\ad_2 \ldots \ad_{s+1})}~.
\ee
However, it can be shown that the equation of motion of \eqref{FMCHSFermProDec} can be factorised purely in terms of second-order differential operators
\be
0 = \Box^{s+1}h^{\perp}_{\a(s+1)\ad(s)}~.
\ee
Analogously, it can be shown that the equations of motion for the CHS action \eqref{FMCHSProj} for mixed CHS fields can also be factorised in terms of second-order differential operators.

\section{Four-dimensional $\mc{N}=1$ Minkowski superspace} \label{SecFourDimensionalMinkowskiSuperSpace}


In this section we extend the discussion and results of section \ref{SecFourDimensionalMinkowskiSpace} to $\cN=1$ Minkowski superspace. We start by reviewing the unitary irreducible representations of the \Po superalgebra  \cite{Likhtman1971,SalamStrathdee1974,Grosser1975} and their corresponding realisations on the space of tensor superfields \cite{Sokatchev1975,Sokatchev1981,SiegelGates1981,HoweStelleTownsend1981}. This will provide the necessary background material to introduce supersymmetric generalisations of the spin projection operators in $\mb{M}^{4|4}$ \cite{Sokatchev1975, SiegelGates1981}. A variety of applications for these operators are explored, including their use in the formulation of superconformal higher-spin theory \cite{KuzenkoManvelyanTheisen2017}. The presentation and material covered in this section is inspired by the work \cite{BuchbinderKuzenko1998}, to which we direct the reader for a pedagogical treatment.

\subsection{Irreducible representations of the \Po superalgebra} \label{IrreduciblerepresentationsofthePoincaresuperalgebra}
The simplest supersymmetric extension of the \Po algebra \eqref{FMPoincareAlgebra},\footnote{It was shown by Haag, Lopuszanski and Sohnius \cite{HaagLopuszanskiSohnius1974} that this is the most general extension of $\PaF$ by fermionic generators.} known as the $\cN=1$ \Po superalgebra $S \ms{P}$,\footnote{The $\cN=1$ \Po superalgebra is simply referred to as the supersymmetry algebra. } was introduced in the pioneering work of Golfand and Likhtman \cite{GolfandLikhtman1971} in 1971. The $\cN=1$ \Po superalgebra is spanned by the \Po generators $P_a$ and  $J_{ab}$ and the supersymmetry generators $ Q_\a$ and $\bar{Q}_\ad$.\footnote{The $4d$ $\cN$-extended \Po superalgebra was introduced by Akulov and Volkov \cite{AkulovVolkov1974}, in which $\PaF$ is extended by including the supercharges $Q^I_\a$ and $\bar{Q}^I_\ad$, where $I = 1, 2, \ldots ,\cN$. }
The generators of the supersymmetry algebra satisfy the (anti-)commutation relations 
\bsubeq \label{FMSSuperPoincareAlgebra}
\begin{align}
[ P_a,P_b]&=0~, \qquad  \hspace{2cm}[J_{ab},P_c]=\ri \eta_{ac} P_b - \ri \eta_{bc} P_a~, \label{FMSMom}\\
\ [P_a,Q_\a] &= 0~, \qquad  \hspace{1.95cm} [P_a,\Qb_{\ad}] = 0~, \label{FMSMomSusyGenComm} \\
\ [ J_{ab},Q_\a]&= \ri (\s_{ab})_\a{}^\b Q_\b~, \qquad \ [ J_{ab},\Qb^{\ad}]= \ri (\ts_{ab})^{\ad}{}_{\bd} \Qb^\bd~, \label{FMSLorSusyGenComm}\\
\ [ J_{ab},J_{ce}] &= \ri \eta_{ac}J_{be} - \ri \eta_{ae}J_{bc} + \ri \eta_{be}J_{ac} - \ri \eta_{bc}J_{ae}~,  \label{FMSLor}\\
\{ Q_\a, Q_\b \} &= 0~, \qquad  \hspace{1.7cm} \{\Qb_{\ad},\Qb_{\bd} \}= 0~, \\
\{ Q_\a, \Qb_{\ad} \} &= 2(\s^a)_{\a\ad}P_a~. \label{FMSusyGenComm} 
\end{align}	
\esubeq
The relations \eqref{FMSSuperPoincareAlgebra} define an arbitrary representation of $S\ms{P}$. In a UIR of $S\ms{P}$, the \Po generators are Hermitian, while the supersymmetry generators are Hermitian conjugates to each other, $\bar{Q}_\ad = (Q_\a)^{+}$.  It is clear from the commutation relations \eqref{FMSMom} and \eqref{FMSLor} that the \Po  algebra $\PaF$ is indeed a subalgebra of $S\ms{P}$.

Since the momentum generator $P_a$ commutes with the supersymmetry generators $ Q_\a$ and $\bar{Q}_\ad$ \eqref{FMSMomSusyGenComm}, it follows that the quadratic Casimir operator $\mb{C}_1$ of $S\ms{P}$ has the same form as the quadratic Casimir operator $C_1$ of $\PaF$ \eqref{DMPoincareQuadCasimir}
\bsubeq \label{FMSCasimirOperators}
\begin{gather}\label{FMSCasimirQuadratic}
\mb{C}_1 = - P^aP_a~, \\ 
[\mb{C}_1, P_a]=[\mb{C}_1, J_{ab}]=[\mb{C}_1, Q_\a]=[\mb{C}_1, \Qb_\ad]=0~. \non
\end{gather}

However, the form of the Casimir operator $C_2$ of $\PaF$ \eqref{FMPoincareQuartCasimir} does not directly yield a Casimir operator of  $S\ms{P}$. This is due to the fact that $C_2$ contains the generator $J_{ab}$,  which does not commute with the fermionic generators $ Q_\a$ and $\bar{Q}_\ad$ (cf. \eqref{FMSLorSusyGenComm}). The Casimir operator $\mb{C}_2$ of $S\ms{P}$, which can be considered the supersymmetric generalisation of $C_2$, takes the following form 
\begin{gather} \label{FMSCasimirQuartic}
\mb{C}_2 = Z_aP^aZ_bP^b - Z_aZ^aP_bP^b~, \qquad  \\
[\mb{C}_2, P_a]=[\mb{C}_2, J_{ab}]=[\mb{C}_2, Q_\a]=[\mb{C}_2, \Qb_\ad]=0~. \non
\end{gather}
\esubeq
Here, $Z_a$ is the supersymmetric generalisation of the Pauli-Lubanski pseudovector \eqref{FMPauliLubanksiVector}
\be \label{FMSSUSYPLOP}
Z_a = \mb{W}_a - \frac{1}{8}(\ts_a)^{\bd\b}[Q_\b,\Qb_\bd]~, 
\ee
which can be shown to satisfy the properties
\be \label{FMSPLProp}
[Z_a, P_b] = 0~, \qquad [Z_a, Q_\a] = \hf Q_\a P_a~, \qquad [Z_a, Z_b] = \ri \ve_{abce} Z^c P^e~.
\ee
The commutation relations \eqref{FMSPLProp} are useful in proving that the operator $\mb{C}_2$ commutes with the generators of $S\ms{P}$. The Casimir operator $\mb{C}_2$ is often called the superspin operator.

Let us review the classification of the massive and massless UIRs of $S\ms{P}$. These particular representations are supersymmetric generalisations of the UIRs of $\PaF$ which were presented in section \ref{Irreducible representations of the Poincare algebra}. An important feature of the UIRs of $S \ms{P}$ is that they all possess positive energy by construction, in comparison to their non-supersymmetric counterparts of section \ref{Irreducible representations of the Poincare algebra}, for which the negative energy representations were discarded by hand. 

\subsubsection{Massive UIRs of $S \ms{P}$}\label{FMSMassiveUIRSec}
The massive UIRs of  $S \ms{P}$ are classified by the quantum numbers mass $\bm{m}$ and superspin $s$. It can be shown that the eigenvalues of the Casimir operators \eqref{FMSCasimirOperators} label the massive UIRs of $S \ms{P}$ as follows \cite{Likhtman1971,SalamStrathdee1974,Grosser1975}
\be \label{FMSMassiveFieldsRep}
\mb{C}_1 = - \bm{m}^2 \mds{1}~, \qquad \mb{C}_2=\bm{m}^4 s(s+1)\mds{1} ~,
\ee
where the mass $\bm{m}$ is strictly positive $\bm{m} > 0$ and the superspin $s$ takes positive (half-)integer values $s= 0 , \hf ,1 , \frac{3}{2}, \cdots$. We denote by $\mb{G}(\bm{m},s)$ a massive UIR of \Psa~carrying mass $\bm{m}$ and superspin $s$. Since the \Po algebra is a subalgebra of the supersymmetry algebra, it can be shown that the UIRs of $S \ms{P}$ describe a reducible representation of $\PaF$. Specifically for $s>0$, the UIR $\mb{G}(\bm{m},s)$ of $S \ms{P}$ can be decomposed into the direct sum of four UIRs of $\PaF$,\footnote{In the case $s=0$, the UIR $ G(\bm{m},s-\hf)$ is not present in the decomposition \eqref{FMSMassiveIrrepDecomp4}.}  with each carrying mass $\bm{m}$ and the respective spins $s -\hf, s, s, s+\hf$
\be \label{FMSMassiveIrrepDecomp4}
\mb{G}(\bm{m},s) = G(\bm{m},s-\hf) \oplus G(\bm{m},s) \oplus G(\bm{m},s) \oplus G(\bm{m},s+\hf)~.
\ee
The massive UIRs of $\PaF$ encoded in $\mb{G}(\bm{m},s)$  all possess the same mass since the Casimir operator $C_1$ of $\PaF$, which classifies the mass of a UIR of $\PaF$, shares the same structural form as the Casimir operator $\mb{C}_1$ of $S \ms{P}$. On the other hand, the Casimir operator $C_2$ of $\PaF$, which is responsible for labelling the spin of a UIR of $\PaF$, does not share the same form as the Casimir operator $\mb{C}_2$ of $S \ms{P}$. Thus the UIRs of $\PaF$  encoded in $\mb{G}(\bm{m},s)$ will carry different spins.
It follows from the direct sum \eqref{FMSMassiveIrrepDecomp4} that the UIR $\mb{G}(\bm{m},s)$ carries $4(2s+1)$ physical degrees of freedom, which can be computed by counting the independent components of each of the four UIRs of $\PaF$ encoded in $\mb{G}(\bm{m},s)$  (cf. section \ref{MassiveFieldrepresentations}).

\subsubsection{Massless UIRs of $S \ms{P}$ }
The eigenvalues of the Casimir operators \eqref{FMSCasimirOperators} no longer characterise a massless UIR of  $S \ms{P}$, since they  are known to vanish \cite{BuchbinderKuzenko1998}
\be \label{FMSMasslessIrrep}
\mb{C}_1 = 0~, \qquad \mb{C}_2 = 0~.
\ee
In order to label the massless UIRs,  let us introduce an alternative operator to the supersymmetric Pauli-Lubanski pseudovector \eqref{FMSSUSYPLOP}, which is known as the superhelicity operator $L_a$ \cite{BuchbinderKuzenko1998}
\bsubeq \label{FMSMSPauliLubanski}
\be
L_a = \mb{W}_a - \frac{1}{16}(\ts_a)^{\bd\b}[Q_\b,\bar{Q}_\bd]~.
\ee
Under the vanishing of the Casimir operator $\mb{C}_1$ in a massless UIR \eqref{FMSMasslessIrrep}, it can be shown that the superhelicity operator satisfies the properties
\be
L^aP_a = 0~, \qquad [L_a, P_b] = 0~, \qquad  [L_a,Q_\a] = 0~, \qquad [L_a,L_b] = \ri \ve_{abce} L^c P^e~.
\ee
\esubeq
In other words, the operator $L_a$ is translation and supersymmetry invariant.

It can be shown that in every massless UIR of $S \ms{P}$, the superhelicity $L_a $ is proportional to the translation generator $P_a$ via the relation
\be \label{FMSMasslessEquation}
L_a = \Big (\k + \frac{1}{4} \Big )P_a~,
\ee
where $\k$ takes positive or negative (half-)integer values $\k = 0, \pm \hf , \pm 1, \cdots$. Since the equation \eqref{FMSMasslessEquation} holds in every massless UIR, it follows that they are labelled by the  quantum number $\k$, which is referred to as superhelicity. Similar to the massive case, massless UIRs also describe a reducible representation of $\PaF$. In particular, for a given superhelicity $\k$,  a massless UIR of $S \ms{P}$ describes two massless UIRs of $\PaF$, each carrying the respective helicities $\k$ and $(\k+\hf)$.

\subsection{Irreducible superfield representations} \label{Irreducible superfield representations}
In this section we review pertinent aspects of classical field theory in $\mb{M}^{4|4}$. In particular, we will introduce the space of tensor superfields which furnish the massive and massless UIRs of $S \ms{P}$ outlined in \ref{IrreduciblerepresentationsofthePoincaresuperalgebra}. A pedagogical treatment of these topics can be found in \cite{BuchbinderKuzenko1998,GatesGrisaruRocekSiegel1983}.

\subsubsection{Facets of four-dimensional $\cN=1$ Minkowski superspace} \label{FMSMinkSupSec}
Four-dimensional $\cN=1$ Minkowski superspace \cite{SalamStrathdee19742,AkulovVolkov1974} is parametrised by the local coordinates $z^A = \lb x^a, \q^\a, \tb_\ad \rbrace$, where $x^a$ are real bosonic variables and $\q^\a$ and $\tb_\ad = (\q_\a)^{*}$ are complex Grassmann coordinates
\be \label{FMSAnticommutingCoordinates}
\q_\a \q_\b = - \q_\b \q_\a~, \qquad  \tb_\ad \tb_\bd = - \tb_\bd \tb_\ad~, \qquad \q_a \tb_\ad = - \tb_\ad\q_\a~.
\ee
The superspace $\mb{M}^{4|4}$ can be identified as the coset space $S\P/\mathsf{SO}_0(3,1)$, where $S\P$ is the $\cN=1$ super-\Po group. 
A superfield $\F_{\a(m)\ad(n)}(x, \q , \tb) \equiv \F_{\a(m)\ad(n)}(z) $ on $\mb{M}^{4|4}$ is called a tensor superfield\footnote{A tensor superfield $\F_{\a(m)\ad(n)}(z)$ of Lorentz type $(\frac{m}{2}, \frac{n}{2})$ carries the Grassmann parity $\e(\F) = m +n ~ (\text{mod 2})$. The superfield $\F_{\a(m)\ad(n)}$  is said to be  bosonic if it carries Grassmann parity $\e (\F) = 0$ or fermionic if it carries Grassmann parity $\e (\F) =1$.}  of Lorentz type $(\frac{m}{2}, \frac{n}{2})$ if it transforms under an infinitesimal super \Po transformation according to
\bea 
\d \F_{\a(m)\ad(n)}(z) &=& \ri \big (-b^a P_a + \hf K^{ab} J_{ab} + \e^\b Q_\b + \bar{\e}_\bd \bar{Q}^\bd \big ) \F_{\a(m)\ad(n)}(z)~.
\eea
Here the variables $b^a$, $K^{ab}$ are real bosonic and $\e^\a $ and $\bar{\e}_\ad$ are complex fermionic, and the generators of $S \ms{P}$ take the explicit form
\bsubeq \label{FMSGeneratorsSuperfield}
\bea 
P_a &=& \ri \pa_a~, \\
J_{ab}&=& \ri \big (x_b \pa_a -x_a \pa_b +(\s_{ab})^{\a\b}\q_\a \pa_\b -(\tilde{\s}_{ab})^{\ad\bd}\tb_\ad \bar{\pa}_\bd -M_{ab} \big )~, \\
Q_\a &=&  \ri \pa_\a + (\s^a)_{\a\ad} \tb^\ad \pa_a~, \\
\Qb_\ad &=& - \ri \bar{\pa}_\ad - (\s^a)_{\a\ad}\q^\a \pa_a~.
\eea
\esubeq
The differential operators $\pa_\a = \frac{\pa}{\pa \q^\a}$ and $\bar{\pa}_\ad= \frac{\pa}{\pa \tb^\ad}$  act solely on the Grassmann coordinates $\q_\b$ and $\tb_\bd$ respectively. The generators \eqref{FMSGeneratorsSuperfield} satisfy the commutation relations \eqref{FMSSuperPoincareAlgebra} and thus  form a representation of $S \ms{P}$ on the space of tensor superfields. From this point onwards, when referring to a superfield, we will implicitly mean a tensor superfield.

A superfield $\F_{\a(m)\ad(n)}(z)$ can be expanded in a finite power series about the fermionic coordinates of $\mb{M}^{4|4}$, as a consequence of the identities $\q_\a \q_\b \q_\g = 0$ and $\tb_\ad \tb_\bd \tb_\gd =0$  which follow from \eqref{FMSAnticommutingCoordinates}. For example, a real scalar field $\F(z)=\bar{\F}(z)$ can be expressed as
\bea \label{FMSExpansionScalarField}
\F(x,\q,\tb) =&& A(x)+\q^\a \j_\a(x) +\tb_\ad \bar{\j}^\ad (x) + \q^2 B(x) +\tb^2 \bar{B}(x) ~ \\
&&+ \q \s^a \tb C_a(x) + \tb^2 \q^\a \l_\a (x)+\q^2 \tb_\ad \bar{\l}^\ad (x) + \q^2 \tb^2 G(x) ~. \non 
\eea
Note we have introduced the notation $\q^2 = \q^\a \q_\a$ and $\tb^2 = \tb_\ad \tb^\ad$. The coefficients appearing in \eqref{FMSExpansionScalarField} are known as component fields. Tensor superfields transform in a certain manner under $S\P$ which ensures that the component fields are themselves tensor fields on $\mb{M}^4$.

An unconstrained tensor superfield on the mass-shell realises a reducible representation of $S \ms{P}$. This follows from the fact that the number of component fields that these superfields describe is far more than the four needed to realise a UIR of $S \ms{P}$.
In order to obtain a superfield that furnishes a UIR, we need to introduce covariant constraints on our superfields which will reduce the number of component fields. These constraints can be imposed by the spinor covariant derivatives of $\mb{M}^{4|4}$, which are given by 
\bea \label{FMSSpinorDerivatives}
\cDB_\ad = - \bar{\pa}_\ad - \ri \q^\a \pa_{\a\ad}~, \qquad
\cD_\a = \pa_\a + \ri \tb^\ad \pa_{\a\ad}~.
\eea
The covariant derivatives $\cD_A = \lb \pa_{\a \ad}, \cD_\a, \cDB^\ad \rb$ satisfy the (anti-)commutation relations
\be \label{FMSDerivativeAlgebra}
\lb \cD_\a, \cD_\b \rb = \lb \cDB_\ad , \cDB_\bd \rb = [\cD_\a , \pa_{\b\bd}] = [\cDB_\ad, \pa_{\b\bd}] = 0~, \qquad 
\lb \cD_\a , \cDB_\ad \rbrace = -2\ri\pa_{\a\ad}~.
\ee
The following important identities can be derived from \eqref{FMSDerivativeAlgebra}
\bsubeq \label{FMSCovariantDerivativesProperties}
\begin{alignat}{2}
\cD_\a \cD_\b &= \hf \ve_{\a\b} \cD^2~, & \cDB_\ad \cDB_\bd &= -\hf \ve_{\ad \bd} \cDB^2~, \\
\cD_\a \cD_\b \cD_\g &= 0 ~, & \cDB_\ad \cDB_\bd \cDB_\gd &=0~, \\
[\cD^2 , \cDB_\ad] &= - 4 \ri  \pa_{\b\ad} \cD^\b~, \qquad \quad & [\cDB^2, \cD_\a ] &= 4 \ri \pa_{\a \bd}\cDB^\bd~,
\end{alignat}
\vspace{-1.2cm}
\begin{gather}
\cD^\b \cDB^2 \cD_\b = \cDB_\bd \cD^2 \cDB^\bd~, \\
\cD^2\cDB^2 + \cDB^2\cD^2 - 2 \cD^\b \cDB^2 \cD_\b = 16\Box~, \label{FMSSSImporIdentity}
\end{gather}
\esubeq
where we have defined $\cD^2 = \cD^\b \cD_\b$ and $\cDB^2 = \cDB_\bd \cDB^\bd$. The spinor covariant derivatives satisfy the following conjugation property\footnote{The complex conjugate of the product of two superfields $\F_{\a(m)\ad(n)} $ and $\J_{\a(p) \ad(q)}$ is given by $(\F_{\a(m)\ad(n)} \J_{\a(p) \ad(q)})^*=( \J_{\a(p) \ad(q)})^* (\F_{\a(m)\ad(n)} )^* = \bar{\J}_{\a(q)\ad(p)} \bar{\F}_{\a(n)\ad(m)}$ $= (-1)^{\e(\F)\e(\J)} \bar{\F}_{\a(n)\ad(m)} \bar{\J}_{\a(q)\ad(p)}$.}
\be
(\cD_\a \F)^* = (-1)^{\e(\F)}\cDB_\ad \bar{\F}~, \qquad (\cD^2 \F)^* =\cDB^2 \bar{\F}~,
\ee
where $\F$ is some complex superfield with indices suppressed and $\F^* \equiv \bar{\F}$ is the complex conjugate of $\F$.

To extract the component fields encoded in the superfield \eqref{FMSExpansionScalarField}, we need to introduce the operation of bar-projection, which is defined as follows
\be \label{FMSBarProj}
\f(x):=\F(z) \big |_{\q = \tb = 0}~.
\ee
Using this, in conjunction with the covariant derivatives, we find that the component fields of the real scalar superfield \eqref{FMSExpansionScalarField} are defined as follows
\begin{alignat}{4}
A(x)&=\F(z)|~, & \j_{\a}(x) &= \cD_\a \F(z)|~, & \bar{\j}_\ad (x) &= \cDB_{\ad} \F(z)|~, \non \\
B(x) &= - \frac{1}{4} \cD^2 \F(z)|~, & \bar{B}(x) &= - \frac{1}{4} \cDB^2 \F(z)|~, & C_{\a\ad}(x) &= \frac{1}{2} [\cD_\a , \cDB_\ad] \F(z) |~,  \label{FMSComponentStructure} \\
\l_\a (x)& = - \frac{1}{4} \cD_\a \cDB^2 \F(z)|~, &\quad \bar{\l}_\ad(x) &= -\frac{1}{4} \cDB_\ad \cD^2 \F(z)|~, & G(x) &= \frac{1}{32} \lb \cD^2, \cDB^2 \rb \F(z)|~. \non
\end{alignat}
It is apparent from the component structure \eqref{FMSComponentStructure} that the spinor derivatives can be used to eliminate some of the component fields encoded within a superfield. However, these constraints may not be sufficient in eliminating the right number of component fields. Moreover, they may not even constrain the component fields, which we recall, is needed for them to furnish a UIR of $S \ms{P}$.  In the subsequent section, we will be show that certain constrained superfields realise UIRs of $S \ms{P}$.
Before doing this, let us list the most common constrained superfields in $\mb{M}^{4|4}$:
\begin{enumerate}[leftmargin=*]
	\item[] For integers $m,n \geq 0$, the complex superfields $\F_{\a(m)\ad(n)}$ and $G_{\a(m)\ad(n)}$ are said to be:
	
	\begin{itemize}
		\item 
		Chiral if it satisfies the condition
		\bsubeq
		\be \label{FMSChiralSuperfield}
		\cDB_\bd \F_{\a(m)\ad(n)} = 0 \qquad \Longrightarrow \qquad \cDB^2 \F_{\a(m)\ad(n)} = 0~.
		\ee
		\item   Anti-chiral if it obeys the constraint
		\be \label{FMSAntichiralSuperfield}
		\cD_\b \F_{\a(m)\ad(n)} = 0 \qquad \Longrightarrow \qquad \cD^2 \F_{\a(m)\ad(n)} = 0~.
		\ee
		\esubeq
		\item Linear if it satisfies the constraint
		\bsubeq \label{FMSLALSuperfield}
		\be \label{FMSLinearSuperfield}
		\cDB^2 \F_{\a(m)\ad(n)} = 0~.
		\ee
		\item  Anti-linear if it obeys the condition
		\be \label{FMSAntiLinearSuperfield}
		\cD^2 \F_{\a(m)\ad(n)} = 0~.
		\ee
		\esubeq
		\item  Longitudinal linear if it satisfies the constraint
		\bsubeq \label{FMSLongitudinalAntiLinearLinear}
		\be \label{FMSLongitudinalLinear}
		\cDB_\ad G_{\a(m)\ad(n)} = 0 \qquad \Longrightarrow \qquad \cDB^2 G_{\a(m)\ad(n)} = 0~.
		\ee
		\item  Longitudinal anti-linear if it obeys the condition
		\be \label{FMSLongitudinalAntiLinear}
		\cD_\a G_{\a(m)\ad(n)} = 0 \qquad \Longrightarrow \qquad \cD^2 G_{\a(m)\ad(n)} = 0~.
		\ee
		\esubeq
	\end{itemize}
	\item[] For integers $m,n >1$, a complex  superfield $\G_{\a(m)\ad(n)}$ is said to be:
	\begin{itemize}
		\item  Transverse linear if it satisfies the condition
		\bsubeq \label{FMSTLAL}
		\be
		\cDB^\bd \G_{\a(m)\bd\ad(n-1)} = 0 \qquad \Longrightarrow \qquad \cDB^2 \G_{\a(m)\ad(n)} = 0~. \label{FMSTransLinCons}
		\ee
		\item  Transverse anti-linear if it obeys the constraint
		\be
		\cD^\b \G_{\b\a(m-1)\ad(n)} = 0 \qquad \Longrightarrow \qquad \cD^2 \G_{\a(m)\ad(n)} = 0~. \label{FMSTransAntiLinCons}
		\ee
		\esubeq
	\end{itemize}
\end{enumerate}
Note that for $m=n=0$, the chiral \eqref{FMSChiralSuperfield} and longitudinal linear \eqref{FMSLongitudinalLinear} constraints are equivalent. Similarly, the anti-chiral \eqref{FMSAntichiralSuperfield} and longitudinal anti-linear \eqref{FMSLongitudinalAntiLinear} conditions are identical for the cases $m=n=0$.


\subsubsection{Massive superfield representations} \label{FMSMassiveSuperfieldRepresentations} 
We wish to realise the massive UIRs $\mb{G}(\bm{m},s)$ of $S \ms{P}$ on the space $\mb{V}_{(m,n)}$, where we denote  by $\mb{V}_{(m,n)}$ the space of superfields $\F_{\a(m)\ad(n)}$ of Lorentz type $(\frac{m}{2},\frac{n}{2})$ in $\mb{M}^{4|4}$.\footnote{A superfield is called a supermultiplet if it furnishes an irreducible representation of the corresponding isometry superalgebra.}
To achieve this, an appropriate set of differential constraints on $\F_{\a(m)\ad(n)}$ need to be found such that on the corresponding constraint space, the Casimir operators $\mb{C}_1$ \eqref{FMSCasimirQuadratic} and $\mb{C}_2$ \eqref{FMSCasimirQuartic} have the fixed eigenvalues  (cf. eq. \eqref{FMSMassiveFieldsRep})
\be \label{FMSCasimirFieldMass}
\mb{C}_1 \F_{\a(m)\ad(n)} = - \bm{m}^2 \F_{\a(m)\ad(n)}~, \qquad \mb{C}_2 \F_{\a(m)\ad(n)}= s(s+1)\bm{m}^4\F_{\a(m)\ad(n)}~.
\ee
Furthermore, as discussed in section \ref{FMSMassiveUIRSec}  (cf. \eqref{FMSMassiveIrrepDecomp4}), a superfield which furnishes $\mb{G}(\bm{m},s)$, where $s>0$, must also describe four massive UIRs of $\PaF$, each with spins $s-\hf,s,s$ and $s+\hf$. From the field-theoretic perspective, such a superfield will describe four independent component fields, with each carrying the aforementioned spins.

For integers $m,n \geq 0$, let us consider the superfield $\F_{\a(m)\ad(n)}$ on $\mb{V}_{(m,n)}$ which satisfies the conditions
\bsubeq \label{FMSOnshellConditions}
\bea 
\pa^{\b\bd}\F_{\b\a(m-1)\bd\ad(n-1)}&=&0~, \label{FMSTransverse} \\
(\Box -\bm{m}^2)\F_{\a(m)\ad(n)} &=& 0 ~,  \qquad \bm{m} >0~. \label{FMSKleinGordon}
\eea
\esubeq
Note that in the cases $m=n=0$, and $m=0$ or $n =0$, the transverse condition \eqref{FMSTransverse} is absent. We will denote by $\mb{V}^{[\boldsymbol{m}]}_{(m,n)}$ the space of superfields $\F_{\a(m)\ad(n)}$ satisfying the conditions \eqref{FMSOnshellConditions}. The superfields on the subspace $\mb{V}^{[\boldsymbol{m}]}_{(m,n)}$ will be the starting point for our discussion on massive superfield representations.\footnote{The constraints \eqref{FMSOnshellConditions} ensure that many of the component fields described by a superfield on $\mb{V}^{[\boldsymbol{m}]}_{(m,n)}$ realise a massive UIR of $\PaF$.  }

The first natural step is to check if the constraint space $\mb{V}^{[\boldsymbol{m}]}_{(m,n)}$ suffices in furnishing a single massive UIR $\mb{G}(\bm{m},s)$, which we recall was the case for its non-supersymmetric counterpart (cf. section \ref{MassiveFieldrepresentations}). Upon analysing the component structure of $\F_{\a(m)\ad(n)} \in \mb{V}^{[\boldsymbol{m}]}_{(m,n)}$, it is immediately apparent that $\F_{\a(m)\ad(n)}$ describes far more physical degrees of freedom than that of a superfield which realises the UIR $\mb{G}(\bm{m},s)$. In order to address this issue of excess physical components,  it is necessary to introduce further supplementary conditions on the space $\mb{V}^{[\boldsymbol{m}]}_{(m,n)}$. 

The guiding principle which aids in finding these additional constraints comes from the observation that the space $\mb{V}^{[\boldsymbol{m}]}_{(m,n)}$ forms a completely reducible representation of $S \ms{P}$. Specifically, representations furnished on $\mb{V}^{[\boldsymbol{m}]}_{(m,n)}$ can be expressed as the direct sum of the following four UIRs of $S \ms{P}$
\be \label{FMSMassiveIrrepDecomp}
\mb{G} \Big ( \bm{m} , s-\hf \Big ) \oplus \mb{G} \Big ( \bm{m} , s \Big ) \oplus \mb{G} \Big ( \bm{m} , s \Big ) \oplus \mb{G} \Big ( \bm{m} , s + \hf \Big ) ~.
\ee
Here, each of the massive UIRs of $S \ms{P}$ carry the same mass $\bm{m}$ with the respective superspins $s- \hf, s, s, s + \hf$,  where $s = \hf ( m + n)$. From a field-theoretic standpoint, it follows that a superfield on $\mb{V}^{[\boldsymbol{m}]}_{(m,n)}$ can be decomposed into four irreducible components, each of which realising one of the massive UIRs in \eqref{FMSMassiveIrrepDecomp}. Upon computing this superfield decomposition explicitly, one can then study the corresponding properties of each of the irreducible subspaces in order to find the additional constraints needed for a superfield on $\mb{V}^{[\boldsymbol{m}]}_{(m,n)}$ to furnish a massive UIR of $S \ms{P}$. 

In the spirit of  \cite{BuchbinderKuzenko1998}, let us go about decomposing the superfield $\F_{\a(m)\ad(n)}$ into irreducible components.
To get some intuition on this problem, it suffices to study the decomposition of a superfield which carries a single type of spinor index, i.e. only undotted or dotted indices. This is due to the fact that for positive (half-)integer $s=\hf(m+n)$, the spaces $\mds{V}^{[\boldsymbol{m}]}_{(2s,0)},\mds{V}^{[\boldsymbol{m}]}_{(2s-1,1)}, \cdots , \mds{V}^{[\boldsymbol{m}]}_{(1,2s-1)},\mds{V}^{[\boldsymbol{m}]}_{(0,2s)}$ describe equivalent representations of $S \ms{P}$. This result follows from the non-supersymmetric case (cf. section \ref{MassiveFieldrepresentations}) since the momentum and supersymmetry generators commute \eqref{FMSMomSusyGenComm}.

Pertinent to the decomposition of a superfield into irreducible components are the differential operators $\bm{\P}_{(i)} = \lbrace \bm{\P}_{(+)}, \bm{\P}_{(-)} , \bm{\P}_{(0)} \rbrace$, which are defined as follows
\bsubeq\label{FMSProjectors}
\bea
\bm{\P}_{(+)} &=& \frac{1}{16 \Box}\cDB^2 \cD^2~, \label{FMSChiralProjector} \\
\bm{\P}_{(-)} &=& \frac{1}{16 \Box}\cD^2 \cDB^2~, \label{FMSAntiChiralProjector} \\
\bm{\P}_{(0)} &=& - \frac{1}{8 \Box} \cD^\b \cDB^2 \cD_\b = - \frac{1}{8 \Box} \cDB_\bd \cD^2 \cDB^\bd~. \label{FMSLinearProjector}
\eea
\esubeq
The operators \eqref{FMSProjectors} are orthogonal projectors which resolve the unit operator\bsubeq \label{FMSProjectorProperties}
\bea 
\bm{\P}_{(i)}\bm{\P}_{(j)} &=& \d_{ij} \bm{\P}_{(i)}~, \label{FMSProjectorOrthogonality}\\
\bm{\P}_{(+)} + \bm{\P}_{(-)} + \bm{\P}_{(0)}  &=& \mds{1} ~. \label{FMSProjectorSumtoUnity}
\eea
\esubeq
The projectors $\bm{\P}_{(+)}$ and $\bm{\P}_{(-)}$ map any superfield on $\mb{V}^{[\boldsymbol{m}]}_{(m,n)}$ to a   chiral \eqref{FMSChiralSuperfield} and an anti-chiral \eqref{FMSAntichiralSuperfield} superfield on $\mb{V}^{[\boldsymbol{m}]}_{(m,n)}$ respectively. The operator $\bm{\P}_{(0)}$ projects a superfield on $\mb{V}^{[\boldsymbol{m}]}_{(m,n)}$ to a simultaneously linear \eqref{FMSLinearSuperfield} and anti-linear \eqref{FMSAntiLinearSuperfield} (LAL) superfield on $\mb{V}^{[\boldsymbol{m}]}_{(m,n)}$. We denote by $\mds{V}^{[\boldsymbol{m}](+)}_{(m,n)}$, $\mds{V}^{[\boldsymbol{m}](-)}_{(m,n)}$ and $\mds{V}^{[\boldsymbol{m}](0)}_{(m,n)}$ the space of superfields which in addition to satisfying \eqref{FMSOnshellConditions}, are also chiral, anti-chiral and LAL, respectively. Moreover, we denote by $\F^{(+)}_{\a(m)\ad(n)}$, $\F^{(-)}_{\a(m)\ad(n)}$ and $\F^{(0)}_{\a(m)\ad(n)}$ the superfields living on the spaces $\mds{V}^{[\boldsymbol{m}](+)}_{(m,n)}$, $\mds{V}^{[\boldsymbol{m}](-)}_{(m,n)}$ and $\mds{V}^{[\boldsymbol{m}](0)}_{(m,n)}$, respectively.

\paragraph{Massive irreducible representations on $\mb{V}^{[\boldsymbol{m}]}_{(2s,0)}$}\label{FMSMassiveUIRsUndotted}
Without loss of generality, let us consider the space of superfields $\mb{V}^{[\boldsymbol{m}]}_{(2s,0)}$. It follows from \eqref{FMSProjectorSumtoUnity} that any superfield $\F_{\a(2s)}$ on $\mb{V}^{[\boldsymbol{m}]}_{(2s,0)}$ can be decomposed into a chiral, anti-chiral and LAL part as follows
\be  \label{FMSDecompositionProjectors}
\F_{\a(2s)} = \big ( \bm{\P}_{(+)} + \bm{\P}_{(-)} + \bm{\P}_{(0)}  \big ) \F_{\a(2s)} ~.
\ee 
It turns out that the representations of $S \ms{P}$ realised on chiral or anti-chiral superfields on $\mb{V}^{[\boldsymbol{m}]}_{(2s,0)}$, which we collectively denote $\mds{V}^{[\boldsymbol{m}](\pm)}_{(2s,0)}$, are irreducible. By construction, superfields on $\mds{V}^{[\boldsymbol{m}](\pm)}_{(2s,0)}$ satisfy the condition
\be \label{FMSQuadraticCasimirChiral}
\mb{C}_1\F^{(\pm)}_{\a(2s)} = - \bm{m}^2 \F^{(\pm)}_{\a(2s)} ~.
\ee 

The superspin operator $\mb{C}_2$ \eqref{FMSCasimirQuartic} can be used to elucidate the superspin content associated with the spaces $\mds{V}^{[\boldsymbol{m}](\pm)}_{(2s,0)}$. In the superfield representation \eqref{FMSGeneratorsSuperfield}, the Casimir operator $\mb{C}_2$ takes the explicit form on  $\mds{V}^{[\boldsymbol{m}]}_{(m,n)}$
\be  \label{FMSQuarticCasimirOnShell}
\mb{C}_2\F_{\a(m)\ad(n)}= \bm{m}^4 \Big ( s \big (s+1 \big )\mds{1} + \Big (\frac{3}{4} + B \Big )\bm{\P}_{(0)} \Big ) \F_{\a(m)\ad(n)}~,
\ee 
where the operator $B$ is given by
\be
B = \frac{1}{4\Box} \mb{W}^{\b\bd}[\cD_\b, \cDB_\bd]~.
\ee
Here, $\mb{W}^{\b\bd}$ is the Pauli-Lubanski pseudovector \eqref{FMPauliLubanskiField}.

The action of the LAL projector $\bm{\P}_{(0)}$ on a (anti-)chiral superfield $\F^{(\pm)}_{\a(m)\ad(n)}$ is 
\be \label{FMSLALProjectorChiral}
\bm{\P}_{(0)} \F^{(\pm)}_{\a(m)\ad(n)} = 0~.
\ee
It follows from \eqref{FMSLALProjectorChiral} that the action of $\mb{C}_2$ \eqref{FMSQuarticCasimirOnShell} on the space superfields $\mds{V}^{(\pm)[\boldsymbol{m}]}_{(2s,0)}$ yields
\be \label{FMSQuarticCasimirChiral}
\mb{C}_2\F^{(\pm)}_{\a(2s)}= \bm{m}^4 s(s+1)\F^{(\pm)}_{\a(2s)}~.
\ee
Due to the conditions \eqref{FMSQuadraticCasimirChiral} and \eqref{FMSQuarticCasimirChiral}, it follows that the space of (anti-)chiral superfields $\mds{V}^{[\boldsymbol{m}](\pm)}_{(2s,0)}$ furnishes the massive UIR $\mb{G}(\bm{m},s)$ (cf. eq \eqref{FMSCasimirFieldMass}). Hence, the projection operators $\bm{\P}_{(\pm)}$ extract the physical component of an arbitrary field $\F_{\a(2s)}$ on $\mds{V}^{[\boldsymbol{m}]}_{(2s,0)}$ which realises the UIR $\mb{G}(\bm{m},s)$ .

However, the space of LAL superfields $\mds{V}^{[\boldsymbol{m}](0)}_{(2s,0)}$ does not describe a massive UIR of $S \ms{P}$. This follows from the fact that the action of $\mb{C}_2$ \eqref{FMSQuarticCasimirOnShell} on the space of LAL superfields is not proportional to the unit operator, as a consequence of the identity
\be \label{FMSLALIdentityProp}
\F^{(0)}_{\a(m)\ad(n)} = \bm{\P}_{(0)}\F^{(0)}_{\a(m)\ad(n)} ~.
\ee

Explicitly evaluating the projector $\bm{\P}_{(0)}$ in \eqref{FMSLALIdentityProp}, and reducing under $\mathsf{SL}(2,\mb{C})$ (by symmetrising and anti-symmetrising the spinor indices), it follows that any superfield on $ \mds{V}^{[\boldsymbol{m}](0)}_{(2s,0)}$ can be decomposed in the following manner
\be \label{FMSLALDecomposition}
\F^{(0)}_{\a(2s)} = \cD^\b \tilde{\F}^{(+)}_{\b\a(2s)} + \cD_{(\a_1} \tilde{\F}^{(+)}_{\a_2 \ldots \a_{2s})}~. 
\ee
Here, the superfields $\tilde{\F}^{(+)}_{\a(2s+1)}  $ and $\tilde{\F}^{(+)}_{\a(2s-1)} $ 
\bsubeq \label{FMSLALChiralSuperfields}
\bea
\tilde{\F}^{(+)}_{\a(2s+1)} &=& -\frac{1}{8 \Box}\cDB^2 \cD_{(\a_1} \F^{(0)}_{\a_2 \ldots \a_{2s+1}) }~, \label{FMSLALChiralSuperfieldsHighest}\\
\tilde{\F}^{(+)}_{\a(2s-1)} &=& \frac{1}{8 \Box}\frac{2s}{2s+1} \cDB^2 \cD^\b \F^{(0)}_{\b \a(2s-1)} ~, \label{FMSLALChiralSuperfieldsLowest}
\eea
\esubeq
are chiral
\be \label{FMSLALSuperfieldsIrrepsConditions}
\cDB_\bd \tilde{\F}^{(+)}_{\a(2s+1)} = 0~,   \qquad \cDB_\bd \tilde{\F}^{(+)}_{\a(2s-1)} =0~.
\ee
We see from \eqref{FMSLALSuperfieldsIrrepsConditions} that the superfields $\tilde{\F}^{(+)}_{\a(2s+1)}  $ and $\tilde{\F}^{(+)}_{\a(2s-1)} $ belong to the spaces $\mb{V}^{[\boldsymbol{m}](+)}_{(2s+1,0)}$ and $\mb{V}^{[\boldsymbol{m}](+)}_{(2s-1,0)}$, respectively. Thus for every positive (half-)integer $s$, the space of superfields $\mb{V}^{[\boldsymbol{m}](0)}_{(2s,0)}$ realises the reducible representation $\mb{G}( \bm{m} , s-\hf  ) \oplus \mb{G}  ( \bm{m} , s + \hf  )$.

The supplementary conditions needed to extract the superfields carrying the highest $\tilde{\F}^{(+)}_{\a(2s+1)} $ and lowest $\tilde{\F}^{(+)}_{\a(2s-1)}$ superspin from the LAL superfield $\F^{(0)}_{\a(2s)}$ are given by
\bsubeq
\bea
\cD_{(\a_1} \F^{(0)}_{\a_2 \ldots \a_{2s+1})} &=& 0~, \label{FMSLALHSCons}\\
\cD^\b \F^{(0)}_{\b\a(2s-1)} &=& 0~. \label{FMSLALLSCons}
\eea
\esubeq
The conditions \eqref{FMSLALHSCons} and \eqref{FMSLALLSCons} can be immediately read from the explicit form of the superfields $\tilde{\F}^{(+)}_{\a(2s+1)} $ \eqref{FMSLALChiralSuperfieldsHighest} and $\tilde{\F}^{(+)}_{\a(2s-1)}$ \eqref{FMSLALChiralSuperfieldsLowest} respectively.

It follows from the above analysis that the LAL projector $\bm{\P}_{(0)} $ does not extract out the physical component of an arbitrary field $\F_{\a(2s)}$. However, in accordance with \eqref{FMSLALDecomposition}, the projection operator $\bm{\P}_{(0)} $ can be bisected on $\mds{V}^{[\boldsymbol{m}]}_{(2s,0)}$ as follows
\be \label{FMSDecompLAL}
\bm{\P}_{(0)} = \bm{\P}^{(s+\hf)}_{(0)} + \bm{\P}^{(s-\hf)}_{(0)} ~.
\ee
Here the operators
$\bm{\P}^{(s+\hf)}_{(0)} $ and $\bm{\P}^{(s-\hf)}_{(0)} $ are also LAL projectors which act on $\mb{V}^{[\boldsymbol{m}]}_{(2s,0)}$ via the following rule
\bsubeq \label{FMSLALProjectors}
\bea
\bm{\P}^{(s+\hf)}_{(0)} {\F}_{\a(2s)} &=& -\frac{1}{8 \Box}\cD^\b\cDB^2 \cD_{(\b} \F_{\a_1 \ldots \a_{2s}) } \equiv \cD^\b \widehat{\F}^{(+)}_{\b\a(2s)}~, \label{FMSLALProjectors1}\\
\bm{\P}^{(s-\hf)}_{(0)} {\F}_{\a(2s)} &=& \frac{1}{8 \Box}\frac{2s}{2s+1} \cD_{(\a_1}\cDB^2 \cD^\b \F_{\a_2 \ldots \a_{2s}) \b} \equiv \cD_{(\a_1} \widehat{\F}^{(+)}_{\a_2 \ldots \a_{2s})}~.
\eea
\esubeq
The projectors $\bm{\P}^{(s+\hf)}_{(0)}$ and $\bm{\P}^{(s-\hf)}_{(0)}$ extract out the component of a superfield on $\mds{V}^{[\boldsymbol{m}]}_{(2s,0)}$ which realises the UIRs $\mb{G} ( \bm{m} , s + \hf  )$ and $\mb{G} ( \bm{m} , s - \hf  )$, respectively. This follows from the fact that projected superfields $\bm{\P}^{(s+\hf)}_{(0)} {\F}_{\a(2s)} $ and $\bm{\P}^{(s-\hf)}_{(0)} {\F}_{\a(2s)}$ can be
expressed in terms of the chiral superfields $\widehat{\F}^{(+)}_{\a(2s+1)}$ and $\widehat{\F}^{(+)}_{\a(2s-1)}$, which were shown previously to realise massive UIRs of $S\ms{P}$.

To summarise, the operators $\bm{\P}_{(+)} , \bm{\P}_{(-)} , \bm{\P}^{(s+\hf)}_{(0)}$ and $\bm{\P}^{(s-\hf)}_{(0)}$ can be shown to be orthogonal projectors on $\mb{V}^{[\boldsymbol{m}]}_{(2s,0)}$.  Accordingly, it follows that any superfield $\F_{\a(2s)}$ on $\mb{V}^{[\boldsymbol{m}]}_{(2s,0)}$ can be decomposed into irreducible components as follows
\bea \label{FMSDecomposition}
\F_{\a(2s)} = \Big ( \bm{\P}_{(+)} + \bm{\P}_{(-)} + \bm{\P}^{(s+\hf)}_{(0)}  +  \bm{\P}^{(s-\hf)}_{(0)} \Big ) {\F}_{\a(2s)} ~.
\eea 
It was shown that the projectors $\bm{\P}_{(+)}$,$\bm{\P}_{(-)}$, $\bm{\P}^{(s+\hf)}_{(0)} $ and  $\bm{\P}^{(s-\hf)}_{(0)}$ select out the components of ${\F}_{\a(2s)}$ which realise the massive UIRs  $\mb{G} ( \bm{m} , s  ), \mb{G} ( \bm{m} , s  ), \mb{G} ( \bm{m} , s + \hf  )$ and $\mb{G} ( \bm{m} , s - \hf  )$, respectively. 

Note that the irreducible decomposition \eqref{FMSDecomposition} can also be written purely in terms of chiral superfields as follows  
\bea \label{FMSDecompositionChiral}
\F_{\a(2s)} =   \check{\F}^{(+)}_{\a(2s)} + \cD^2\widehat{\F}^{(+)}_{\a(2s)} + \cD^\b \widehat{\F}^{(+)}_{\b\a(2s)} + \cD_{(\a_1} \widehat{\F}^{(+)}_{\a_2 \ldots \a_{2s})}~,
\eea 
where $\widehat{\F}^{(+)}_{\a(2s)}: = \frac{1}{16 \Box}\cDB^2 \F_{\a(2s)}$ and $\check{\F}^{(+)}_{\a(2s)}:=\bm{\P}_{(+)}\F_{\a(2s)}$. This decomposition can easily be uplifted to unconstrained superfields on $\mb{V}_{(m,n)}$ by composing the projectors in the decomposition \eqref{FMSDecomposition} with the transverse projectors ${\P}^{\perp}_{(m,n)}$ \eqref{FMBFprojectors}. The decomposition \eqref{FMSDecompositionChiral} is important in Siegel and Gate's \cite{SiegelGates1981} formulation of superprojectors, which will be elaborated upon in section \ref{SuperprojectorsGatesSiegel}.

It follows from the discussion above that the following families of constrained superfields on $\mds{V}_{(2s,0)}$ furnish massive UIRs of $S \ms{P}$:
\begin{enumerate}
	\item A chiral superfield on the mass-shell
	\be
	\cDB_\bd \F_{\a(2s)} =0~,  \qquad 	(\Box-\bm{m}^2)\F_{\a(2s)} =0~, \label{FMSChiralIrrep}
	\ee
	furnishes the massive UIR $\mb{G}(\bm{m},s)$.
	\item An anti-chiral superfield on the mass-shell
	\be
	\cD_\b \F_{\a(2s)} =0~, \qquad (\Box-\bm{m}^2)\F_{\a(2s)} =0~, \label{FMSAntiChiralIrrep}
	\ee
	furnishes the massive UIR $\mb{G}(\bm{m},s)$.
	\item A transverse anti-linear and linear superfield on the mass-shell
	\be
	\cD^\b \F_{\b\a(2s-1)} =0~, \qquad \bar{\cD}^2 \F_{\a(2s)} =0~, \qquad 	(\Box-\bm{m}^2)\F_{\a(2s)} =0~, \label{FMSTLCons}
	\ee
	furnishes the massive UIR $\mb{G}(\bm{m},s+\hf)$.
	\item A longitudinal anti-linear and linear superfield on the mass-shell
	\be
	\cD_{(\a_1} \F_{\a_2 \dots \a_{2s})} =0~, \qquad \cDB^2 \F_{\a(2s)} =0~, \qquad (\Box-\bm{m}^2)\F_{\a(2s)} =0~,
	\ee
	furnishes the massive UIR $\mb{G}(\bm{m},s-\hf)$.
\end{enumerate}

The case for the scalar superfield $\F$ needs to be treated separately. Again, using the fact that the projectors resolve the identity \eqref{FMSProjectorSumtoUnity}, we find that $\F$ decomposes into 
\be  \label{FMScalarDecomp}
\F = \big ( \bm{\P}_{(+)} + \bm{\P}_{(-)} + \bm{\P}_{(0)}  \big ) \F ~.
\ee 
For the scalar superfield $\F$, the LAL contribution $\bm{\P}_{(0)}\F $ does not need to be bisected, as it can be immediately expressed in terms of a chiral superfield $\F^{(+)}_\a$ as follows 
\be
\bm{\P}_{(0)}\F  = - \frac{1}{8 \Box} \cD^\a \cDB^2 \cD_\a \F \equiv \cD^\a \F^{(+)}_\a ~.
\ee
Hence the space of LAL scalar fields $\mb{V}^{[\boldsymbol{m}](0)}_{(0,0)}$ realises the massive UIR $\mb{G}(\bm{m}, \hf)$. Thus, every scalar field $\F$ can be decomposed \eqref{FMScalarDecomp} into three irreducible components: chiral; anti-chiral; and LAL. Since the projectors are orthogonal,  the reducible representation realised on $\mb{V}^{[\boldsymbol{m}]}_{(0,0)}$ can be expressed as the direct sum of the massive UIRs
\be
\mb{G}(\bm{m}, 0) \oplus \mb{G}(\bm{m}, 0) \oplus\mb{G}(\bm{m}, \hf)~.
\ee

\paragraph{Massive irreducible representations on $\mb{V}^{[\boldsymbol{m}]}_{(m,n)}$}\label{FMSMassiveFieldsSecGen}
We wish to find the necessary conditions to realise a massive UIR on the space of superfields $\mb{V}_{(m,n)}$, using the intuition gained from studying the analogous problem on $\mb{V}_{(2s,0)}$.
Recall that the representations of $S \ms{P}$ realised on $\mb{V}^{[\boldsymbol{m}]}_{(2s,0)}$ are equivalent to those on $\mds{V}^{[\boldsymbol{m}]}_{(m,n)}$. In accordance with this, there should exist four constrained subspaces of $\mds{V}^{[\boldsymbol{m}]}_{(m,n)}$ which realise each of the massive UIRs appearing in the reducible representation  \eqref{FMSMassiveIrrepDecomp}.

Following the discussion presented in section \ref{FMSMassiveUIRsUndotted}, it can be easily shown that both the space of chiral \eqref{FMSChiralSuperfield} and anti-chiral \eqref{FMSAntichiralSuperfield} superfields on $\mds{V}^{[\boldsymbol{m}]}_{(m,n)}$
furnish the massive UIR $\mb{G}(\bm{m},s)$. Next, we wish to find the constraints which select the component of a superfield on $\mds{V}^{[\boldsymbol{m}]}_{(m,n)}$ that realises the UIR $\mb{G}(\bm{m},s+\hf)$ with maximal superspin.

For integers $m,n \geq 1$, we say that a superfield $\F_{\a(m)\ad(n)}$ on $\mb{V}_{(m,n)}$ is on-shell if it satisfies the constraints
\bsubeq\label{FMSOnshellMassiveSuperfields}
\begin{gather}
\cD^\b\F_{\b\a(m-1)\ad(n)} = \cDB^{\bd} \F_{\a(m)\bd\ad(n-1)} =  0~,  \label{FMSOnshellTLAL} \\
(\Box -\bm{m}^2)\F_{\a(m)\ad(n)} = 0 ~. \label{FMSOnshellKG}
\end{gather}
\esubeq
Note that these constraints are inspired by the conditions \eqref{FMSTLCons} which ensured that a superfield on $\mds{V}_{(2s,0)}$ furnished the massive UIR $\mb{G}(\bm{m},s+\hf)$.  The simultaneous transverse linear and anti-linear (TLAL) constraint \eqref{FMSOnshellTLAL} implies 
\be \label{FMSTransverseTLAL}
\pa^{\b\bd}\F_{\b \a(m-1)\bd\ad(n-1)} =0~.
\ee
It can be shown that the action of the Casimir operator $\mb{C}_2$ \eqref{FMSQuarticCasimirOnShell}  on an on-shell superfield  \eqref{FMSOnshellMassiveSuperfields} is
\begin{align} \label{FMSSuperspinCasimirTLALOnshell}
\mb{C}_2\F_{\a(m)\ad(n)} = \tilde{s}(\tilde{s}+1) \bm{m}^4\F_{\a(m)\ad(n)}~,
\end{align}
where $\tilde{s}=\hf(m+n+1)$. In accordance with the massive UIR conditions \eqref{FMSCasimirFieldMass}, it follows from \eqref{FMSOnshellKG} and \eqref{FMSSuperspinCasimirTLALOnshell} that the space of on-shell superfields \eqref{FMSOnshellMassiveSuperfields} furnishes the massive UIR $\mb{G}(\bm{m},s+\hf)$, with $s=\hf(m+n)$, which carries maximal superspin.

To our knowledge, there are no constraints that can be imposed on a superfield in $\mb{V}^{[\boldsymbol{m}]}_{(m,n)}$ such that it furnishes the UIR $\mb{G}(\bm{m},s-\hf)$ with the lowest superspin. We come to this conclusion via the following argument. Recall that the irreducible components carrying the lowest and highest superspin are contained in the space of LAL superfields $\bm{\P}_{(0)} \F_{\a(m)\ad(n)} \in \mb{V}^{(0)}_{(m,n)}$. The only constraints which are consistent with the LAL condition are certain combinations of the longitudinal linear \eqref{FMSLongitudinalLinear}, longitudinal anti-linear \eqref{FMSLongitudinalAntiLinear}, transverse linear \eqref{FMSTransLinCons} and transverse anti-linear \eqref{FMSTransAntiLinCons} constraints. 

Upon performing a component analysis, it is easy to see that many of these constrained superfields do not describe the four component fields needed to realise a massive UIR of $S \ms{P}$. The exception to this are: i) TLAL superfields \eqref{FMSOnshellTLAL}; or ii) simultaneously longitudinal linear \eqref{FMSLongitudinalLinear} and the longitudinal anti-linear \eqref{FMSLongitudinalAntiLinear} superfields (LLLA). We demonstrated above that the TLAL superfields \eqref{FMSOnshellMassiveSuperfields} select out the UIR $\mb{G}(\bm{m}, s+ \hf)$ with maximal superspin. 

For $m,n >1$, a LLLA superfield on $\mds{V}^{[\boldsymbol{m}]}_{(m,n)}$ describes the four independent component fields $\lb A_{\a(m)\ad(n)}, B_{\a(m-1)\ad(n)} , C_{\a(m)\ad(n-1)} , F_{\a(m-1)\ad(n-1)}\rb$. These tensors fields have the correct spin content (index structure) required to furnish the UIRs of $\PaF$ encoded within the massive UIR $\mb{G}(\bm{m}, s- \hf)$ (cf. eq \eqref{FMSMassiveIrrepDecomp4}). However, it can be shown that these component fields are so strongly constrained that they must vanish. For example, the component field $A_{\a(m)\ad(n)}$ obeys the conditions
\be \label{FMSLongLinAntiLinCompConst}
\pa_{\a\ad}A_{\a(m)\ad(n)} = \pa^{\b\bd}A_{\b \a(m-1) \bd \ad(n-1)} = (\Box - \bm{m}^2) A_{\a(m)\ad(n)}= 0~.
\ee
It is easy to see that the constraints \eqref{FMSLongLinAntiLinCompConst} are too restrictive, forcing $A_{\a(m)\ad(n)}$ to vanish 
\be
A_{\a(m)\ad(n)} = 0~.
\ee
Thus, there are no differential constraints which extract the component of a superfield on $\mb{V}^{[\boldsymbol{m}]}_{(m,n)}$ which furnishes the massive UIR $\mb{G}(\bm{m}, s- \hf)$ with lowest superspin.

In summary, a superfield $\F_{\a(m)\ad(n)}$ on $\mb{V}^{[\boldsymbol{m}]}_{(m,n)}$ realises the completely reducible representation
\begin{align}\label{FMSMassiveIrrepDecompGeneral}
\mb{G} \Big ( \bm{m} , s-\hf \Big ) \oplus \mb{G} \Big ( \bm{m} , s \Big ) \oplus \mb{G} \Big ( \bm{m} , s \Big ) \oplus \mb{G} \Big ( \bm{m} , s + \hf \Big ) ~.
\end{align}
It was shown above that there exists three constrained subspaces of $\mb{V}_{(m,n)}$ which realise the massive UIRs $\mb{G} ( \bm{m} , s  ) , \mb{G}  ( \bm{m} , s  )$ and $\mb{G}  ( \bm{m} , s + \hf  )$. It follows that the following families of superfields $\F_{\a(m)\ad(n)}$ on $\mb{V}_{(m,n)}$ realise massive UIRs of $S\ms{P}$: 
\begin{enumerate}
	\bsubeq
	\item {Chiral transverse superfields on the mass-shell}
	\be \label{FMSOnShellChiralConditions}
	(\Box-\bm{m}^2)\F_{\a(m)\ad(n)} =0~, \quad  \pa^{\b\bd}\F_{\b\a(m-1)\bd\ad(n-1)} =0~, \quad \cDB_\bd \F_{\a(m)\ad(n)} =0~,
	\ee
	furnish the massive UIR $\mb{G}(\bm{m},s)$.
	\item {Anti-chiral transverse superfields on the mass-shell} 
	\be \label{FMSOnShellAChiralConditions}
	(\Box-\bm{m}^2)\F_{\a(m)\ad(n)} =0~, \quad  \pa^{\b\bd}\F_{\b\a(m-1)\bd\ad(n-1)} =0~, \quad \cD_\b \F_{\a(m)\ad(n)} =0~,
	\ee
	realise the massive UIR $\mb{G}(\bm{m},s)$.
	\item {TLAL superfields on the mass-shell} 
	\be  \label{FMSTLALChiralConditions}
	(\Box -\bm{m}^2)\F_{\a(m)\ad(n)} = 0 ~, \quad \cD^\b\F_{\b\a(m-1)\ad(n)} = \cDB^{\bd} \F_{\a(m)\bd\ad(n-1)} =  0~,
	\ee
	furnish the massive UIR $\mb{G}(\bm{m},s+\hf)$.
	\esubeq
\end{enumerate}

\paragraph{Real massive superfields}
Reality conditions can be introduced in the same way for superfields on $\mb{V}^{[\boldsymbol{m}]}_{(m,n)}$ as was done for on-shell fields in the non-supersymmetric case in subsection \ref{MassiveFieldrepresentations}. This follows from the fact that the momentum generators commute with the supersymmetry generators, thus the discussion holds automatically when uplifted to $\mb{M}^{4|4}$.  The conjugation operator \eqref{FMConjugationOperator} will take the same structural form on $\mb{V}^{[\boldsymbol{m}]}_{(m,n)}$  \cite{SiegelGates1981,GatesGrisaruRocekSiegel1983}
\be \label{FMSConjugationOperatorSUSYOnshell}
K \F_{\a(m)\ad(n)} = e^{i \l }\D_{\a_1}{}^{\bd_1} \cdots \D_{\a_m}{}^{\bd_m} \D^{\b_1}{}_{\ad_1} \cdots \D^{\b_n}{}_{\ad_n} \bar{\F}_{\b(n)\bd(m)}~.
\ee
The reality condition \eqref{FMSConjugationOperatorSUSYOnshell} can be used to introduce the notion of a real TLAL on-shell superfield.
This fact will be discussed in AdS$^{4|4}$ in section \ref{section 3.3}, with the analysis holding in the $\mb{M}^{4|4}$  case by taking the flat-superspace limit.

However the conjugation operator \eqref{FMSConjugationOperatorSUSYOnshell} cannot be applied naively to all constrained superfields, as the spinor differential constraints may not be preserved. For example, let us consider the simple case of a complex spinor field $\F_\a$ which is chiral $\cDB_\bd \F_\a  =0$. 
Applying the conjugation operator \eqref{FMSConjugationOperatorSUSYOnshell} to $\F_\a$ yields \cite{SiegelGates1981,GatesGrisaruRocekSiegel1983}
\be \label{FMSChiralConjugation}
K \F_\a = e^{i \l } \D_{\a}{}^{\bd} \bar{\F}_\bd~.
\ee
Let us impose the supersymmetric analogue of the reality condition \eqref{FMMassiveRealityConditions} for the case $m=1 , n=0$
\be \label{FMSSPinorreality}
K \F_\a = \F_\a~.
\ee 
We see that the right hand side of \eqref{FMSSPinorreality} is chiral, thus \eqref{FMSChiralConjugation} must also be chiral. However, upon inspection, one sees that this is not true. Thus, the conjugation operator \eqref{FMSChiralConjugation} needs to be altered in such a way that the chirality of the superfield is preserved. 

In accordance with this, let us introduce the chiral conjugation operator $\tilde{K}$ \cite{SiegelGates1981,GatesGrisaruRocekSiegel1983}
\be \label{FMSConjugationOperatorSUSYchiral}
\tilde{K} \F_{\a(m)\ad(n)} = \frac{1}{4 \sqrt{\Box}}e^{i \l } \cDB^2\D_{\a_1}{}^{\bd_1} \cdots \D_{\a_m}{}^{\bd_m} \D^{\b_1}{}_{\ad_1} \cdots \D^{\b_n}{}_{\ad_n} \bar{\F}_{\b(n)\bd(m)}~.
\ee
It can be shown that the operator \eqref{FMSConjugationOperatorSUSYchiral} preserves chirality and is involutive. 
Following a similar discussion given in subsection \ref{MassiveFieldrepresentations}, one can introduce the notion of real chiral superfields in $\mb{M}^{4|4}$. For further details concerning real chiral superfields, see \cite{SiegelGates1981}.

\subsubsection{Massless superfield representations} 
As in the non-supersymmetric setting (see section \ref{FMMasslessfieldrepresentations}), massless dynamics can be formulated in terms of gauge fields or gauge-invariant field strengths.

For integers $m, n \geq 1$, a superfield $\F_{\a(m)\ad(n)}$ on $\mb{V}_{(m,n)}$ is said to be a massless if it obeys the conditions
\bsubeq \label{FMSOnShellMasslessGF}
\begin{gather}
\cD^\b \F_{\b\a(m-1)\ad(n)} = \cDB^\bd \F_{\a(m)\bd\ad(n-1)} =0~, \label{FMSOnShellTrans} \\
\Box \F_{\a(m)\ad(n)} = 0~. \label{FMSMasslessKG}
\end{gather}
The system of equations \eqref{FMSOnShellTrans} and \eqref{FMSMasslessKG} are compatible with the gauge symmetry
\be \label{FMSMasslessGT}
\d_{\z,\x} \F_{\a(m)\ad(n)} = \cDB_{\ad} \z_{\a(m)\ad(n-1)} + \cD_{\a} \x_{ \a(m-1) \ad(n)}~,
\ee
where the gauge parameters are TLAL and satisfy the constraints
\begin{alignat}{2}
\pa_{\a}{}^\bd  \x_{\a(m-1) \bd \ad(n-1)}  &=0 \qquad & \Longrightarrow \qquad \Box \x_{\a(m-1)\ad(n)} &= 0~, \\
\pa^\b{}_{\ad}\z_{\b\a(m-1)\ad(n-1)} &= 0 \qquad & \Longrightarrow \qquad \Box \z_{\a(m)\ad(n-1)} &=0~.
\end{alignat}
\esubeq
A superfield satisfying the conditions \eqref{FMSOnShellMasslessGF} is said to be a massless gauge superfield. The space of massless gauge superfields \eqref{FMSOnShellMasslessGF} does not furnish massless UIRs of $S \ms{P}$. This becomes apparent upon examination of the superhelicity content of a massless gauge superfield. 

In the superfield representation, the superhelicity operator $L_{\b\bd}$ \eqref{FMSMSPauliLubanski}  takes the form
\be \label{FMSPauliLubasnkiFieldRep}
L_{\b\bd} = \mb{W}_{\b\bd} - \frac{1}{8} [ \cD_\b ,\cDB_\bd]~.
\ee
It is not difficult to see that the action of $L_{\b\bd}$ \eqref{FMSPauliLubasnkiFieldRep} on a massless superfield \eqref{FMSOnShellMasslessGF} is not of the desired form, $L_{\b\bd}\F_{\a(m)\ad(n)} \neq  \l P_{\b\bd}\F_{\a(m)\ad(n)}$, for some constant $\l$.
Thus in accordance with \eqref{FMSMasslessEquation}, massless gauge superfields do not furnish massless UIRs of $S \ms{P}$. This is due to the presence of gauge symmetry \eqref{FMSMasslessGT}, which leads to the existence of non-trivial invariant subspaces that are related via the gauge transformations \eqref{FMSMasslessGT}. In order to address this, we need to completely fix the gauge symmetry. As observed in $\mb{M}^4$, fixing the gauge symmetry required a lot of work. However, we can avoid all of these problems by working in the field strength framework instead.

Before elaborating upon the field strength framework, we note that the scalar superfield $m=n=0$ is not treated by \eqref{FMSOnShellMasslessGF}, thus it needs to be studied separately. There are two classes of scalar superfields that furnish massless UIRs of $S \ms{P}$. A superfield $\F$ which is both chiral and anti-linear is said to be massless
\be \label{FMSScalarMassless1}
\cDB_\bd \F =0~, \qquad \cD^2 \F =0~.
\ee
In addition, a superfield $\F$ which is both anti-chiral and linear is said to be massless
\be \label{FMSScalarMassless2}
\cD_\b \F =0~, \qquad \cDB^2 \F =0~.
\ee
The scalar superfields \eqref{FMSScalarMassless1} and \eqref{FMSScalarMassless2} furnish massless UIRs of $S\ms{P}$ which carry superhelicity $\l =0$.

Let us consider the field strength $\mds{W}_{\a(p)\ad(q)}(\F)$, with $p >0$ and $q > 0$, which is invariant under the gauge transformations \eqref{FMSMasslessGT}
\be
\d_{\z,\x}\mds{W}_{\a(p)\ad(q)}(\F) = 0~.
\ee
There exists two families of  field strengths $\mds{W}_{\a(p)\ad(q)}(\F)$ that furnish massless UIRs of $S \ms{P}$: massless chiral field strengths; and massless anti-chiral field strengths. 

A  field strength $\mds{W}_{\a(p)\ad(q)}(\F)$ is said to be massless chiral if it satisfies \cite{BuchbinderKuzenko1998}\bsubeq \label{FMSMasslessChiral}
\bea
\cDB_\bd \mds{W}_{\a(p)\ad(q)}(\F)&=& 0~, \label{FMSMasslessCChiralConstraint}\\
\cD^\b \mds{W}_{\b \a(p-1)\ad(q)}(\F)&=&0~, \label{FMSMasslessCTransverseConstraint}\\
\pa^{\b\bd} \mds{W}_{ \a(p) \bd \ad(q-1)}(\F) &=& 0~. \label{FMSMasslessCPartialConstraint}
\eea
\esubeq
Using equations \eqref{FMSMasslessChiral}, it can be shown that the action of the Casimir operators of $S \ms{P}$ on $\mds{W}_{\a(p)\ad(q)}(\F)$ vanishes
\be
\mb{C}_1 \mds{W}_{\a(p)\ad(q)}(\F) = 0~, \qquad \mb{C}_2 \mds{W}_{\a(p)\ad(q)}(\F)=0~.
\ee
In addition, it follows from \eqref{FMSMasslessChiral} that the superhelicity operator \eqref{FMSPauliLubasnkiFieldRep} acts appropriately on $\mds{W}_{\a(p)\ad(q)}(\F)$
\be
L_{\b\bd} \mds{W}_{\a(p)\ad(q)}(\F)= \Big (\hf(p-q)+\frac{1}{4} \Big )P_{\b\bd} \mds{W}_{\a(p)\ad(q)}(\F)~.
\ee
In accordance with \eqref{FMSMasslessEquation}, the massless chiral field strength \eqref{FMSMasslessChiral} furnishes the massless UIR with superhelicity $\k = \hf (p-q)$.

A field strength $\mds{W}_{\a(p)\ad(q)}(\F)$ is said to be massless anti-chiral if it satisfies the constraints \cite{BuchbinderKuzenko1998}
\bsubeq \label{FMSMasslessFieldStrengths}
\bea
\cD_\b \mds{W}_{\a(p)\ad(q)}(\F) &=& 0~, \\
\cDB^\bd \mds{W}_{\a(p) \bd \ad(q-1)}(\F) &=&0~, \\
\pa^{\b\bd} \mds{W}_{\b \a(p-1)\ad(q)}(\F) &=& 0~.
\eea
\esubeq
It can be shown that the eigenvalues of the Casimir operators on \eqref{FMSMasslessFieldStrengths}  are vanishing
\be
\mb{C}_1 \mds{W}_{\a(p)\ad(q)}(\F) = 0~, \qquad \mb{C}_2 \mds{W}_{\a(p)\ad(q)}(\F)=0~.
\ee
In addition, it follows from \eqref{FMSMasslessFieldStrengths} that the superhelicity operator \eqref{FMSPauliLubasnkiFieldRep} acts appropriately on $\mds{W}_{\a(p)\ad(q)}(\F)$
\be
L_{\b\bd} \mds{W}_{\a(p)\ad(q)}(\F) =  \Big (\hf (p-q-1) +\frac{1}{4} \Big )P_{\b\bd}\mds{W}_{\a(p)\ad(q)}(\F)~.
\ee
In accordance with \eqref{FMSMasslessEquation}, the massless  anti-chiral field strength \eqref{FMSMasslessFieldStrengths} realises the  massless UIR carrying superhelicity $\k = \hf (p-q-1)$.

Let us now consider the case when $p \geq 1 , q =0$. It can be shown via similar arguments used above that the there exists two classes of field strengths that furnish massless UIRs of $S \ms{P}$: massless chiral field strengths; and massless anti-chiral field strengths. 

A field strength $\mds{W}_{\a(p)}(\F)$ is said to be massless chiral if it satisfies the conditions \cite{BuchbinderKuzenko1998}
\vspace{-\baselineskip}
\bsubeq \label{FMSMasslessChiralOneIndex}
\bea
\cDB_\gd \mds{W}_{\a(p)}(\F)&=& 0~, \\
\cD^\b \mds{W}_{\b \a(p-1)}(\F)&=& 0~.
\eea
\esubeq
It can be shown that the massless chiral field strength \eqref{FMSMasslessChiralOneIndex} furnishes the massless UIR carrying superhelicity $\k = \frac{p}{2}$.

A field strength $\mds{W}_{\a(p)}(\F)$ is said to be massless anti-chiral if it obeys  \cite{BuchbinderKuzenko1998}
\bsubeq \label{FMSMasslessAntiChiralOneIndex}
\bea
\cD_\b\mds{W}_{\a(p)}(\F)&=& 0~, \\
\pa^{\b\bd}\mds{W}_{\b\a(p-1)}(\F)&=&0~, \\
\cDB^2 \mds{W}_{\a(p)}(\F)&=&0~.
\eea
\esubeq
It can be shown that the massless anti-chiral field strength \eqref{FMSMasslessAntiChiralOneIndex} furnishes the massless UIR carrying superhelicity $\k = \hf (p-1)$. For the case $m = 0, n \geq 1$, the analogous two  field strengths which furnish massless UIRs can be found by taking the complex conjugate of \eqref{FMSMasslessChiralOneIndex} and \eqref{FMSMasslessAntiChiralOneIndex}.

\subsection{Superprojectors}
We wish to construct the projectors which select out the multiplet encoded within an unconstrained superfield on $\mb{V}_{(m,n)}$ that furnishes a massive UIR of $S \ms{P}$. In 1975, Sokatchev \cite{Sokatchev1975} explicitly constructed the operators which extract the irreducible components encoded within a transverse superfield of arbitrary rank in $\mb{M}^{4|4}$. The prescription adopted in \cite{Sokatchev1975} relies heavily on the Casimir operators of the supersymmetry algebra. This method was later extended to $\mb{M}^{4|4\cN}$ by Rittenberg and Sokatchev in 1981 \cite{RittenbergSokatchev1981}. In the same year as \cite{RittenbergSokatchev1981}, an alternative prescription for deriving  projectors in $\mb{M}^{4|4\cN}$ was proposed by Siegel and Gates in \cite{SiegelGates1981} which did not require the use of Casimir operators. In addition, the projectors of \cite{SiegelGates1981}, which were coined superprojectors, extract the irreducible component from an unconstrained field on $\mb{V}_{(m,n)}$. In this section we outline and expand upon the prescriptions of Sokatchev \cite{Sokatchev1975}  and Siegel and Gates \cite{SiegelGates1981} to construct superprojectors in $\mb{M}^{4|4}$.

\subsubsection{Superprojectors \`a la Sokatchev} 
If a linear operator $F$ on a $n$-dimensional  vector space $\mb{V}$ takes distinct eigenvalues $f_1, f_2, \cdots , f_n$, with $f_i \neq f_j$ for $i \neq j$, such that $\mb{V}  = \oplus_{i=1}^n \mb{V}_{i}$ with ${F}\big |_{\mb{V}_i} = f_i \mds{1} \big |_{\mb{V}_i}$, then the projection operators $\bm{\P}_i$  onto the eigenspaces $\mb{V}_i$ take the form
\be
\bm{\P}_i =  \prod_{1 \leq j \leq n}^{j \neq i} \frac{(F -f_j)}{(f_i-f_j)}~, \qquad \bm{\P}_i \bm{\P}_j = \d_{ij} \bm{\P}_i~, \quad \sum_{i=1}^{n} \bm{\P}_i = \mds{1}~.  \label{FMSSokatvhevmethod}
\ee
This prescription is very powerful, since in essence, one can derive the (super)projectors for any (super)space background given that one knows the Casimir operators of the isometry (super)algebra, and their corresponding eigenvalues on the irreducible subspaces. This general method was employed in the construction of projection operators in $\mb{M}^{4|4}$ by Sokatchev  \cite{Sokatchev1975}.\footnote{To our knowledge, this method was first used by Aurilia and Umezawa \cite{AuriliaUmezawa1967} in 1967 to construct the $\mb{M}^4$ spin projection operators in terms of the Casimir operators of $\PaF$.}

All the necessary ingredients required to compute the projection operators following Sokatchev's recipe were computed in section \ref{FMSMassiveSuperfieldRepresentations}. In particular, the space of superfields $\mb{V}^{[\boldsymbol{m}]}_{(m,n)}$ decomposes into four irreducible subspaces which each furnish one of the massive UIRs appearing in the completely reducible representation \eqref{FMSMassiveIrrepDecomp}. It follows that the irreducible subspaces of $\mb{V}^{[\boldsymbol{m}]}_{(m,n)}$ which realise each of these massive UIRs are characterised by the following eigenvalues of the Casimir operator $\mb{C}_2$ of $S\ms{P}$
\be \label{FMSEV}
\big (s -\hf \big )\big (s + \hf \big ) \Box^2~, \quad   s \big ( s + 1\big )  \Box^2~, \quad  s \big ( s + 1\big ) \Box^2~, \quad  \big (s + \hf \big )\big (s + \frac{3}{2} \big ) \Box^2~,
\ee
where $s = \hf(m+n)$.

It was shown in section \ref{FMSMassiveSuperfieldRepresentations} that the space of LAL transverse superfields on $\mb{V}^{(0)}_{(m,n)}$  decomposes into two irreducible subspaces which carry the lowest and highest superspins (cf. \eqref{FMSDecompLAL}). Since we know the Casimir operator $\mb{C}_2$ of $S \ms{P}$ and the corresponding eigenvalues \eqref{FMSEV} on the irreducible subspace, we find the projectors via  \eqref{FMSSokatvhevmethod} to be
\bsubeq \label{FMSHighestLowestSuperprojectors}
\bea 
\widehat{\bm{\P}}^{(s-\hf)} &=& - \frac{1}{(2s+1)\Box^2} \Big ( \mb{C}_2 - \big (s+\hf \big ) \big (s + \frac{3}{2} \big )\Box^2 \Big ) ~, \\
\widehat{\bm{\P}}^{(s+\hf)}&=&\phantom{-}\frac{1}{(2s+1)\Box^2} \Big ( \mb{C}_2 -  \big (s-\hf \big ) \big (s+\hf \big )\Box^2 \Big ) ~.
\eea 
\esubeq
Here, the superscripts of the operators \eqref{FMSHighestLowestSuperprojectors} denote the superspin associated with the subspace on which the projectors map onto. The operators \eqref{FMSHighestLowestSuperprojectors} satisfy the following properties on the space of LAL transverse superfields
\be \label{FMSProjPropSok}
\widehat{\bm{\P}}_{(i)} \widehat{\bm{\P}}_{(j)}  = \d_{ij} \widehat{\bm{\P}}_{(i)} ~,  \qquad 
\sum_{i=1}^{n} \widehat{\bm{\P}}_{(i)} = \mds{1} ~, 
\ee
where $\widehat{\bm{\P}}_{(i)} = \lb \widehat{\bm{\P}}^{(s-\hf)}, \widehat{\bm{\P}}^{(s+\hf)} \rb$.

The superprojectors \eqref{FMSHighestLowestSuperprojectors} can immediately be extended to extract the lowest and highest superspin components of a (non-LAL) transverse superfield by composing \eqref{FMSHighestLowestSuperprojectors} with the LAL projector $\bm{\P}_{(0)}$ \eqref{FMSLinearProjector}
\bsubeq \label{FMSHighLowSueprspinProjector}
\bea 
{\bm{\P}}^{(s+\hf)}_{(0)} &:=& \widehat{\bm{\P}}^{(s+\hf)}  \bm{\P}_{(0)} ~, \label{FMSHighSueprspinProjector} \\
{\bm{\P}}^{(s-\hf)}_{(0)} &:=& \widehat{\bm{\P}}^{(s-\hf)}  \bm{\P}_{(0)}~. \label{FMSLowSueprspinProjector}
\eea
\esubeq 
It is easy to see that the operators \eqref{FMSHighLowSueprspinProjector} are orthogonal projectors, as a consequence of the properties \eqref{FMSProjectorOrthogonality} and \eqref{FMSProjPropSok}. The projection operators \eqref{FMSHighLowSueprspinProjector} are equivalent to those derived by Sokatchev \cite{Sokatchev1975}. It can be shown that the projectors \eqref{FMSHighLowSueprspinProjector} bisect the longitudinal projector $\bm{\P}_{(0)}$ \eqref{FMSLinearProjector}
\be \label{FMSLALSuperprojectorBisection}
\bm{\P}_{(0)}  = \bm{\P}^{(s-\hf)}_{(0)}  + {\bm{\P}}^{(s+\hf)}_{(0)} ~,
\ee
as a consequence of the property \eqref{FMSProjPropSok}.

The projection operators which select the chiral transverse  and anti-chiral transverse subspaces, which are both associated with the massive UIRs with superspin $s$, were also computed in \cite{Sokatchev1975}. These operators can be shown to be equivalent to the chiral  \eqref{FMSChiralProjector} and anti-chiral \eqref{FMSAntiChiralProjector} projectors. Since the projection operators computed by Sokatchev resolve the identity \eqref{FMSSokatvhevmethod}, one immediately finds that an arbitrary superfield on $\mb{V}^{[\boldsymbol{m}]}_{(m,n)}$ can be decomposed into irreducible components as follows
\be \label{FMSDecompositionSokatchev}
\F_{\a(m)\ad(n)} = \big (\bm{\P}^{(s)}_{(+)} + \bm{\P}^{(s)}_{(-)} +  {\bm{\P}}^{(s-\hf)}_{(0)} + {\bm{\P}}^{(s+\hf)}_{(0)} \big )\F_{\a(m)\ad(n)}~.
\ee
Here, we have denoted the (anti-)chiral projectors \eqref{FMSProjectors} in the new notation $\bm{\P}^{(s)}_{(\pm)} : = \bm{\P}_{(\pm)} $. 

The approach developed in \cite{Sokatchev1975} only holds on superfields which are assumed to be transverse \eqref{FMSTransverse} from the onset. It must be noted that the projection operators of Sokatchev can be promoted to superprojectors, i.e. they isolate the irreducible components of an unconstrained superfield. To do this, we simply need to compose the projectors appearing in the decomposition \eqref{FMSDecompositionSokatchev} with the non-supersymmetric transverse projectors ${\P}^{\perp}_{(m,n)}$ \eqref{FMBFprojectors}.

\subsubsection{Superprojectors \`a la Siegel and Gates }\label{SuperprojectorsGatesSiegel}
An alternative method for constructing superprojectors in $\mb{M}^{4|4}$ was proposed by Siegel and Gates in \cite{SiegelGates1981}. Their prescription is based on the result that an unconstrained superfield can be expanded in terms of chiral superfields, which we know to furnish massive UIRs of $S \ms{P}$ (cf. eq. \eqref{FMSOnShellChiralConditions}). Completing this decomposition explicitly generates a collection of superprojectors, each of which extract the irreducible components encoded within a general superfield. Without considering reality conditions, let us outline the main steps of \cite{SiegelGates1981} for the chiral decomposition of an unconstrained superfield $\F_{\a(m)\ad(n)}$ on $\mb{V}_{(m,n)}$. Before doing this, it proves instructive to perform the chiral decomposition on a vector superfield $\F_{\a\ad}$.

The most crucial step in the Siegel-Gates prescription \cite{SiegelGates1981} is the chiral expansion of an unconstrained superfield. In section \ref{FMSMassiveUIRsUndotted}, the chiral decomposition utilised in \cite{SiegelGates1981} was presented. Specifically, it was shown that a unconstrained symmetric superfield with only undotted spinor indices can be expanded \eqref{FMSDecompositionChiral} in terms of chiral superfields. Thus in order to make use of the decomposition \eqref{FMSDecompositionChiral}, it is necessary to convert all the dotted indices of the superfield under consideration into undotted ones via the operator $\D_{\a\ad}$, see \eqref{FMDeltaOperator}. This is the first step in the Siegel-Gates prescription \cite{SiegelGates1981}. For $\F_{\a\ad}$, the dotted index is converted to undotted one as follows
\be
\F_{\a \ad} \rightarrow \F_{\a, \b } = \D_\b{}^\ad \F_{\a\ad}~.
\ee
Recall that the equation \eqref{FMSDecomposition} which yields the chiral decomposition \eqref{FMSDecompositionChiral} only holds in the case for totally symmetric superfields. Thus, the second step in the Siegel-Gates prescription involves reducing $\F_{\a, \b } $ under $\mathsf{SL}(2,\mb{C})$ (symmetrising and anti-symmetrising the undotted indices) in order to express $\F_{\a, \b } $ in terms of totally symmetric superfields. Doing this yields
\be \label{FMSVectorReduced}
\D_\b{}^\ad \F_{\a\ad}  = \F_{\a\b} + \ve_{\a \b}\F~, \qquad \F_{\a\b}:= \D_{(\a}{}^\bd \F_{\b ) \bd}~, \quad \F := - \hf \D^{\b\bd} \F_{\b\bd}~.
\ee

The third step involves inverting the operator $\D_{\a\ad}$  in \eqref{FMSVectorReduced}, thus providing an expression for $\F_{\a\ad}$ in terms of totally symmetric superfields with only undotted spinor indices
\be \label{FMSVectorField2}
\F_{\a\ad} = \D_{\ad}{}^\b\F_{\a\b} + \D_{\a\ad}\F~.
\ee
We are now in the position to use the expansion \eqref{FMSDecomposition} to decompose the superfields $\F_{\a\b}$ and $\F$ into irreducible chiral superfields as follows
\bea \label{FMSVectoFieldDecomp}
\F_{\a\ad} =&& \frac{1}{4 \Box} \D_{\ad}{}^\b \Big (\frac{1}{4 }  \lbrace \cDB^2 , \cD^2 \rbrace \F_{\a \b} - \frac{1}{2}  \cD^\g \cDB^2  \cD_{(\g} \F_{\a \b)} + \frac{1}{3 }   \cD_{(\a} \cDB^2 \cD^\g \F_{\b ) \g}  \Big )\non \\
&&-\frac{1}{32 \Box} \D_{\a\ad}  \Big ( \lbrace \cDB^2 , \cD^2 \rbrace  -2 \cD^\b \cDB^2 \cD_\b  \Big) \F~. 
\eea

The final step in the Siegel-Gates prescription is to convert the superfields $\F_{\a \b}$ and $\F$ back in terms original vector field $\F_{\a\ad}$ via the relations \eqref{FMSVectorReduced}. Upon completing this, one arrives at the decomposition of an arbitrary vector field $\F_{\a\ad}$ into irreducible components 
\bea
\F_{\a\ad} =&& \frac{1}{16 \Box^2} \pa_{\ad}{}^\b  \cDB^2  \cD^2  \pa_{(\a}{}^\bd \F_{\b ) \bd} + \frac{1}{16 \Box^2} \pa_{\ad}{}^\b   \cD^2 \cDB^2 \pa_{(\a}{}^\bd \F_{\b ) \bd}  \non \\
&&- \frac{1}{8 \Box^2} \pa_\ad{}^\b \cD^\g \cDB^2 \pa_{(\a}{}^\bd \cD_\g \F_{\b)\bd} + \frac{1}{24 \Box^2} \pa_\ad{}^\b \cD_{(\a} \cDB^2 \cD^\g \big (\pa_{\b)}{}^\bd \F_{\g\bd} +\pa_{|\g|}{}^\bd \F_{\b)\bd} \big ) \non  \\
&&-\frac{1}{32 \Box^2} \pa_{\a\ad}  \cDB^2 \cD^2  \pa^{\b\bd}\F_{\b\bd} -\frac{1}{32 \Box^2} \pa_{\a\ad} \cD^2 \cDB^2    \pa^{\b\bd}\F_{\b\bd} + \frac{1}{16 \Box^2} \pa_{\a\ad} \cD^\g \cDB^2 \cD_\g \pa^{\b\bd}\F_{\b\bd} \non \\
&&= \Big ( \bm{\P}^{\text{T}}_{1} + \bm{\P}^{\text{T}}_{1} + \bm{\P}^{\text{T}}_{\frac{3}{2}} + \bm{\P}^{\text{T}}_{\hf}  + \bm{\P}^{\text{L}}_{0} + \bm{\P}^{\text{L}}_{0}+ \bm{\P}^{\text{L}}_{\hf} \Big ) \F_{\a \ad}~. \label{FMSScalarSuperfieldDecomposition}
\eea
Here, the operators $\bm{\P}_{(i)} = \big \lb  \bm{\P}^{\text{T}}_{1} , \bm{\P}^{\text{T}}_{1}, \bm{\P}^{\text{T}}_{\frac{3}{2}}, \bm{\P}^{\text{T}}_{\hf}, \bm{\P}^{\text{L}}_{0}, \bm{\P}^{\text{L}}_{0},  \bm{\P}^{\text{L}}_{\hf} \big \rb $ are orthogonal projectors 
$\bm{\P}_{(i)} \bm{\P}_{(j)} = \d_{ij} \bm{\P}_{(i)}$. The superscripts ${\text{T}}$ and ${\text{L}}$ indicate whether the projected superfield is transverse \eqref{FMSTransverse} or longitudinal  \eqref{FMLongProjection}, respectively. The subscript which labels the superprojectors  $\bm{\P}_{(i)} $ denotes the superspin of the component extracted from $\F_{\a\ad}$ by the projector $\bm{\P}_{(i)}$, which corresponds to the superspin of the chiral field strength encoded in each $\bm{\P}_{(i)} \F_{\a\ad}$.

Let us now sketch the Siegel-Gates prescription to compute the superprojectors on $\mb{V}_{(m,n)}$. We begin by converting all the indices of $\F_{\a(m)\ad(n)}$ into a single spinor type using the operator $\D_{\a\ad}$  \eqref{FMDeltaOperator}
\bea \label{FMSGSConversion}
\D^n: \mb{V}_{(m,n)} &\longrightarrow& \mb{V}_{(m,0)} \otimes \mb{V}_{(n,0)}~, \\
\F_{\a(m)\ad(n)} &\longmapsto& V_{\a(m),\b(n)} = \D_{\b_1}{}^{\ad_1} \cdots \D_{\b_n}{}^{\ad_n}  \F_{\a(m)\ad(n)} ~, \non
\eea
where we have assumed, without loss of generality that $m \geq n$.
{\sloppy 
	Next, we need to decompose $\mb{V}_{(m,0)} \otimes \mb{V}_{(n,0)}$ into the direct sum of subspaces $\oplus_{j=0}^n \mb{V}_{(m+n-2j,0)}$. This task is computationally expensive to complete explicitly for superfields of higher-rank. However schematically, the iterative decomposition of $ V_{\a(m),\b(n)} \in \mb{V}_{(m,0)} \otimes \mb{V}_{(n,0)}$ into symmetric and anti-symmetric parts will be of the general form}
\be \label{FMSIndexConversion}
\D_{\b_1}{}^{\ad_1} \cdots \D_{\b_n}{}^{\ad_n}  \F_{\a(m)\ad(n)} = \sum_{j=0}^{n} (\ve_{\a\b})^j V_{\a(m-j)\b(n-j)}(\F)~,
\ee
where $V_{\a(m-j)\b(n-j)}(\F)$ is a totally symmetric superfield which is a function of the operators $\D_{\a\ad}$. 

Using the invertible nature of $\D_{\a\ad}$ allows one to obtain the expression for $\F_{\a(m)\ad(n)} $ in terms of $\mathsf{SL}(2,\mb{C})$ irreducible superfields
\be \label{FMSGSAppraochStep2}
\F_{\a(m)\ad(n)} = \sum_{j=0}^{n} (\D_\ad{}^\b)^{n-j} (\D_{\a\ad})^j V_{\a(m-j)\b(n-j)}~.
\ee
Next, we use the resolution of the identity \eqref{FMSProjectorSumtoUnity} on the superfield $V_{\a(m-j)\b(n-j)}$ in \eqref{FMSGSAppraochStep2} to decompose $V_{\a(m-j)\b(n-j)}$ in terms of chiral superfields
\bea \label{FMSDecompostionintochiralsuperfields}
V_{\a(m-j)\b(n-j)} = &&A_{\a(m-j)\b(n-j)} + \cD^2 B_{\a(m-j)\b(n-j)} + \cD^\b  C_{ \a(m-j)\b(n-j+1)} \\
&&+  \cD_{(\a_1} F_{\a_2 \ldots \a_{m-j}\b_1 \ldots \b_{n-j})}~, \non
\eea
where 
\bsubeq
\bea
A_{\a(m-j)\b(n-j)}&=& \frac{1}{16 \Box}\cDB^2\cD^2 V_{\a(m-j)\b(n-j)}~, \\
B_{\a(m-j)\b(n-j)}   &=&\frac{1}{16 \Box} \cDB^2 V_{\a(m-j)\b(n-j)}~, \\
C_{ \a(m-j)\b(n-j+1)}  &=& -\frac{1}{8 \Box} \cDB^2 \cD_{(\b_1} V_{\a(m-j) \b_2 \ldots \b_{n-j+1})} ~, \\
F_{\a(m-j-1)\b(n-j)}&=&\frac{1}{8 \Box}\frac{m+n-2j}{m+n-2j+1}  \cDB^2 \cD^\g V_{\g \a{(m-j-1)} \b{(n-j)}}~.
\eea
\esubeq
The chiral superfields  $A_{\a(m-j)\b(n-j)}$, $B_{\a(m-j)\b(n-j)} $, $C_{ \a(m-j)\b(n-j+1)}$ and $F_{\a(m-j-1)\b(n-j)}$ carry the superspins $s,s,s+\hf$ and $s-\hf$ respectively, where $s=\hf(m+n-2j)$. 

It follows that $\F_{\a(m)\ad(n)}$ can be expressed in terms of chiral superfields by substituting  \eqref{FMSDecompostionintochiralsuperfields}  into \eqref{FMSGSAppraochStep2}
\bea \label{FMSFinal}
\F_{\a(m)\ad(n)} &=& \frac{1}{16 \Box} \sum_{j=0}^{n}(\D_\ad{}^\b)^{n-j} (\D_{\a\ad})^j \Big ( \lb \cDB^2 , \cD^2 \rb  V_{\a(m-j)\b(n-j)} -2 \cD^\g \cDB^2 \cD_{(\g} V_{\a(m-j)\b(n-j))} \non \\
&& + \frac{2(m+n-2j)}{m+n-2j+1} \cD_{(\a_1} \cDB^2 \cD^\g V_{\a_2 \ldots \a_{m-j}\b(n-j))\g} \Big )~.
\eea
The final step is to recast $V_{\a(m-j)\b(n-j)}$ in terms of the original superfield $\F_{\a(m)\ad(n)}$. Once completed, this expression can be substituted into \eqref{FMSFinal} in order to find the decomposition of a general superfield $\F_{\a(m)\ad(n)}$ in terms of irreducible chiral components. 

As previously stated, expressing $V_{\a(m-j)\b(n-j)}$ in terms of the original superfield $\F_{\a(m)\ad(n)}$ is computationally expensive for higher-rank superfields. Instead, let us consider the $j=0$ contribution of \eqref{FMSFinal}, since this corresponds to the subspace $\mb{V}_{(m+n,0)}$ in the decomposition \eqref{FMSFinal} which contains the highest superspin component of $\F_{\a(m)\ad(n)}$.\footnote{Moreover, if one implements the Siegel-Gates prescription on a superfield $\F_{\a(m)\ad(n)}$ on $\mb{V}^{[\boldsymbol{m}]}_{(m,n)}$, it can be shown that the only surviving term in the $\mathsf{SL}(2,\mb{C})$ decomposition \eqref{FMSIndexConversion} is the $j=0$ contribution}
For $j=0$, it can be shown from \eqref{FMSIndexConversion} that  $V_{\a(m)\a(n)}$ is related to the original superfield $\F_{\a(m)\ad(n)}$ via
\be
V_{\a(m)\a(n)} = \D_{(\a_1}{}^{\bd_1} \dots \D_{\a_n}{}^{\bd_n} \F_{\a_1 \ldots \a_m) \bd(n)}~.
\ee
Thus the $j=0$ contribution of the chiral expansion of $\F_{\a(m)\ad(n)}$ \eqref{FMSFinal} yields
\bsubeq \label{FMSGSJzeroContribution}
\bea
\bm{\P}^{\text{T}}_{s} \F_{\a(m)\ad(n)} &=& \frac{1}{16 \Box^{n+1}} \pa_{(\ad_1}{}^{\b_1} \dots \pa_{\ad_n)}{}^{\b_n}   \cDB^2  \cD^2 \pa_{(\b_1}{}^{\bd_1} \dots \pa_{\b_n}{}^{\bd_n} \F_{\a_1 \ldots \a_m)\bd(n)}~, \label{FMSGSJzeroContributionC}\\
\bm{\P}^{\text{T}}_{s} \F_{\a(m)\ad(n)} &=& \frac{1}{16 \Box^{n+1}} \pa_{(\ad_1}{}^{\b_1} \dots \pa_{\ad_n)}{}^{\b_n}  \cD^2  \cDB^2  \pa_{(\b_1}{}^{\bd_1} \dots \pa_{\b_n}{}^{\bd_n} \F_{\a_1 \ldots \a_m)\bd(n)}~,  \label{FMSGSJzeroContributionAC}\\
\bm{\P}^{\text{T}}_{s+\hf} \F_{\a(m)\ad(n)} &=& -\frac{1}{8\Box^{n+1}} \pa_{(\ad_1}{}^{\b_1}\dots \pa_{\ad_n)}{}^{\b_n}\cD^{\g}\cDB^2 \pa_{(\b_1}{}^{\bd_1}\dots\pa_{\b_n}{}^{\bd_n}\cD_\g \F_{\a_1\dots\a_m)\bd(n)} ~, \hspace{1cm}  \label{FMSGSJzeroContributionTLAL}\\
\bm{\P}^{\text{T}}_{s- \hf} \F_{\a(m)\ad(n)} &=&\frac{1}{8 \Box^{n+1}}\pa_{(\ad_1}{}^{\b_1} \dots \pa_{\ad_n)}{}^{\b_n} \Big ( m \cD_{(\a_1}\cDB^2 \cD^\g \pa_{\b_1}{}^{\bd_1} \dots \pa_{\b_n}{}^{\bd_n} \F_{\a_2 \ldots \a_m) \g \bd(n)} ~ \non\\
&&+ n \cD_{(\b_n} \cDB^2 \cD^\g \pa_{\b_1}{}^{\bd_1} \dots \pa_{\b_{n-1}}{}^{\bd_{n-1}} \pa_{|\g|}{}^{\bd_n} \F_{\a_1 \ldots \a_m)\bd(n)} \Big )~, \label{FMSGSJzeroContributionLow}
\eea
\esubeq
where $s=\hf(m+n)$. By an analogous argument, it can be shown that one can arrive at a similar result for the case $n \geq m$.  The operators \eqref{FMSGSJzeroContribution} are orthogonal superprojectors which single out irreducible subspaces with fixed superspins. It must be noted that the closed form expressions for the superprojectors which extract out the irreducible components of a higher-rank superfield $\F_{\a(m)\ad(n)}$ were not given \cite{SiegelGates1981}. The chiral decomposition was only computed explicitly for the cases of  a spinor $\F_\a$ and a vector $\F_{\a\ad}$ superfield in \cite{SiegelGates1981}.

The Siegel-Gates prescription is powerful as it decomposes any unconstrained superfield $\F_{\a(m)\ad(n)}$ into irreducible components. Recall that this was not the case in the work of Sokatchev \cite{Sokatchev1975}, for which the superfields were assumed to be transverse a priori. In accordance with this, it can be shown explicitly that the superprojectors of Siegel and Gates and Sokatchev, for the appropriate superspins, are equivalent on the space of transverse superfields.

In particular,
the  highest  \eqref{FMSGSJzeroContributionTLAL} and lowest \eqref{FMSGSJzeroContributionLow} superprojectors of Siegel and Gates are equivalent to their superspin counterparts \eqref{FMSHighSueprspinProjector} and \eqref{FMSLowSueprspinProjector} within the framework of Sokatchev on the space of transverse superfields 
\bea \label{FMSEquivProjsTrans}
\bm{\P}_{(0)}^{(s \pm \hf)} &\equiv& \bm{\P}^{\text{T}}_{(s\pm \hf)} ~. 
\eea
Moreover, composing both sides of  \eqref{FMSEquivProjsTrans} with the transverse projector $\P^{\perp}_{(m,n)}$, it can be shown that these superprojectors are equivalent on the space of unconstrained superfields $\mb{V}_{(m,n)}$ via the following relation
\bea
\bm{\P}_{(0)}^{(s \pm \hf)} \P^{\perp}_{(m,n)}& \equiv & \bm{\P}^{\text{T}}_{(s\pm \hf)} ~. 
\eea

\subsubsection{Superspin projection operators} 
Of particular interest are the superprojectors which extract the irreducible component of a superfield $\F_{\a(m)\ad(n)}$ on the mass-shell, but otherwise unconstrained, which furnishes the massive UIR  $\mb{G}(\bm{m}, s+ \hf)$ with maximal superspin. These operators, which we will refer to as superspin projection operators, are supersymmetric analogues of the spin projection operators \eqref{FMBFprojectors} on $\mb{M}^{4}$. As observed in $\mb{M}^4$ in sections \ref{Spin-projection operators} and \ref{FMCHSSec}, the applications of these operators proved very fruitful, thus motivating their study in $\mb{M}^{4|4}$. 

As discussed in section \ref{FMSMassiveSuperfieldRepresentations}, the superspin projection operator should map onto the space of TLAL superfields, as this was shown to be the constrained subspace of $\mb{V}_{(m,n)}$ which realises the massive UIR  $\mb{G}(\bm{m}, s+ \hf)$.
In accordance with this, let us introduce the superspin projection operator $\bm{\P}^{\perp}_{(m,n)}$ which is defined by its action on the superfield $\F_{\a(m)\ad(n)} \in \mb{V}_{(m,n)}$ via the rule
\bsubeq
\bea
\bm{\P}^{\perp}_{(m,n)}: \mb{V}_{(m,n)} &\longrightarrow& \mb{V}_{(m,n)}~, \\
\F_{\a(m)\ad(n)} &\longmapsto& \bm{\P}^{\perp}_{(m,n)} \F_{\a(m)\ad(n)}=: \bm{\P}^{\perp}_{\a(m)\ad(n)}(\F) ~.
\eea
\esubeq
For fixed integers $m,n \geq 1$, the operator $\bm{\P}^{\perp}_{(m,n)} $ satisfies the following properties on $\mb{V}_{(m,n)}$:
\vspace{-\baselineskip}
\bsubeq 
\begin{enumerate} \label{FMSBFProjectorProperties}
	\item \textbf{Idempotence}: The operator $\bm{\P}^{\perp}_{(m,n)}$ is idempotent on $\mb{V}_{(m,n)}$
	\be
	\bm{\P}^{\perp}_{(m,n)}	\bm{\P}^{\perp}_{(m,n)}\F_{\a(m)\ad(n)}  =  \bm{\P}^{\perp}_{(m,n)}\F_{\a(m)\ad(n)}~.
	\ee
	\item \textbf{TLAL}: The operator  $\bm{\P}^{\perp}_{(m,n)}$ maps $\F_{\a(m)\ad(n)}$ to a TLAL superfield
	\be \label{FMSTLALProjector}
	\cD^\b \bm{\P}^{\perp}_{\b\a(m-1)\ad(n)}(\F) = 0~, \qquad \cDB^\bd \bm{\P}^{\perp}_{\a(m)\bd\ad(n-1)}(\F) = 0~.
	\ee
\end{enumerate}
\esubeq
It follows immediately from the properties of $\bm{\P}^{\perp}_{(m,n)}$ that the action of the superspin projection operator on $\F_{\a(m)\ad(n)} $, given that  $\F_{\a(m)\ad(n)} $ is also on the mass-shell \eqref{FMSOnshellKG}, yields
\bsubeq
\begin{gather}
\cD^\b \bm{\P}^{\perp}_{(m,n)} \F_{\b\a(m-1)\ad(n)} = 0~, \qquad \cDB^\bd \bm{\P}^{\perp}_{(m,n)}\F_{\a(m)\bd\ad(n-1)} = 0~, \\
(\Box - \bm{m}^2) \bm{\P}^{\perp}_{(m,n)} \F_{\a(m)\ad(n)} = 0~.
\end{gather}
\esubeq
In accordance with \eqref{FMSTLALChiralConditions}, the superspin projection operator $\bm{\P}^{\perp}_{(m,n)}$ selects the component of $\F_{\a(m)\ad(n)}$ which furnishes the massive UIR $\mb{G}(\bm{m},s+\hf)$, where $s=\hf(m+n)$. Note that a superfield $\F_{\a(m)\ad(n)}$ which is TLAL  is said to describe a pure superspin state with maximal superspin $s+\hf$.

For integers $m,n \geq 1$, the superspin projection operators on $\mb{V}_{(m,n)}$ are given by
\bsubeq \label{FMSHighestSuperspinProjector}
\bea 
\bm{\P}^{\perp}_{\a(m)\ad(n)}(\F) &=& -\frac{1}{8\Box^{n+1}} \pa_{(\ad_1}{}^{\b_1}\dots \pa_{\ad_n)}{}^{\b_n}\cD^{\g}\cDB^2 \pa_{(\b_1}{}^{\bd_1}\dots\pa_{\b_n}{}^{\bd_n}\cD_\g \F_{\a_1\dots\a_m)\bd(n)}~,  \label{FMSBFProjectorExplicit}\hspace{0.5cm}\\
\widehat{\bm{\P}}{}^{\perp}_{\a(m)\ad(n)}(\F)&=&\frac{1}{8\Box^{m+1}}\pa_{(\a_1}{}^{\bd_1}\dots \pa_{\a_m)}{}^{\bd_m}\cDB^{\gd} \cD^2  \pa_{(\bd_1}{}^{\b_1}\dots\pa_{\bd_m}{}^{\b_m}\cDB_\gd \F_{\b(m)\ad_1\dots\ad_n)}~.\label{FMSBFProjectorExplicit2}
\eea
\esubeq
On the space $\mb{V}_{(m,n)}$, the superprojector $\bm{\P}^{\perp}_{(m,n)} $ \eqref{FMSBFProjectorExplicit} is equivalent to the operator $\bm{\P}^{\text{T}}_{s+\hf} $ \eqref{FMSGSJzeroContributionTLAL}, which was computed via the Siegel-Gates  prescription in section \ref{SuperprojectorsGatesSiegel}.\footnote{The alternative superprojector $\widehat{\bm{\P}}{}^{\perp}_{(m,n)}$ \eqref{FMSBFProjectorExplicit2} can also be derived using the Siegel-Gates prescription \cite{SiegelGates1981}. To do this, one instead needs to convert all of the undotted indices of $\F_{\a(m)\ad(n)}$ into dotted ones at the point \eqref{FMSIndexConversion} in the procedure.} It must be noted that no discussion concerning the TLAL nature of the superspin projection operator $\bm{\P}^{\text{T}}_{s+\hf} $ \eqref{FMSGSJzeroContributionTLAL} was provided in the work \cite{SiegelGates1981}. Their analysis primarily focused on irreducibility from the perspective of chiral superfields, not TLAL superfields. 

It can be shown that the superspin projection operators \eqref{FMSHighestSuperspinProjector} are equivalent on $\mb{V}_{(m,n)}$ 
\begin{align}\label{FMSProjEquiv}
\bm{\P}^{\perp}_{\a(m)\ad(n)}(\F) = \widehat{\bm{\P}}{}^{\perp}_{\a(m)\ad(n)}(\F) ~. 
\end{align}
It follows that $\bm{\P}^{\perp}_{(m,n)}$ is the unique superspin projection operator on $\mb{V}_{(m,n)}$. In accordance with the equivalence \eqref{FMSProjEquiv}, we will only consider $\bm{\P}^{\perp}_{(m,n)}$ when discussing superspin projection operators on $\mb{V}_{(m,n)}$.

The superspin projection operators also possess several other important properties. Firstly,  $\bm{\P}^{\perp}_{(m,n)}$  acts like the unit operator on the space of TLAL superfields
\be \label{FMSSurj}
\cD^\b\F^{\perp}_{\b\a(m-1)\ad(n)} = \cDB^\bd \F^{\perp}_{\a(m)\bd \ad(n-1)} =  0  \quad \Longrightarrow \quad  \bm{\P}^{\perp}_{(m,n)}\F^{\perp}_{\a(m)\ad(n)} = \F^{\perp}_{\a(m)\ad(n)} ~.
\ee
Secondly, the superspin projector is symmetric,
\be
\int \rd^{4|4}z~ \O^{\a(m)\ad(n)}\bm{\P}^{\perp}_{(m,n)} \F_{\a(m)\ad(n)} = \int \rd^{4|4}z~  \F^{\a(m)\ad(n)}\bm{\P}^{\perp}_{(m,n)} \O_{\a(m)\ad(n)} ~, \label{FMSSymmetry}
\ee
for arbitrary well-behaved superfields $\O_{\a(m)\ad(n)}$ and $\F_{\a(m)\ad(n)} $. Here, we have denoted the full superspace measure by $\rd^{4|4}z = \rd^4 x \rd^2 \q \rd^2 \tb$.

In the case when $m>n=0$, the superprojectors analogous to \eqref{FMSHighestSuperspinProjector} take the form  
\bsubeq
\begin{align}
\bm{\P}^{\perp}_{\a(m)}(\F)&:= -\frac{1}{8\Box}\cD^{\b}\cDB^2\cD_{(\b} \F_{\a_1\dots\a_m)}~,  \\
\widehat{\bm{\P}}{}^{\perp}_{\a(m)}(\F)&:=\frac{1}{8\Box^{m+1}}  \pa_{(\a_1}{}^{\bd_1}\dots \pa_{\a_m)}{}^{\bd_m}\cDB^{\gd} \cD^2  \pa_{(\bd_1}{}^{\b_1}\dots\pa_{\bd_m}{}^{\b_m}\cDB_{\gd)} \F_{\b(m)}~,
\end{align}
\esubeq
where $\bm{\P}^{\perp}_{(m)}:=\bm{\P}^{\perp}_{(m,0)}$\footnote{Note that the operator $\bm{\P}^{\perp}_{(m)}$ coincides with the projector $\bm{\P}^{(s+\hf)}$ \eqref{FMSLALProjectors1} for $s=\frac{m}{2}$.} and $\widehat{\bm{\P}}{}^{\perp}_{(m)}:= \widehat{\bm{\P}}{}^{\perp}_{(m,0)}$. These operators are idempotent and map onto the space of simultaneously transverse anti-linear and linear superfields 
\bsubeq 
\begin{alignat}{2}
\cD^{\b}\bm{\P}^{\perp}_{\b\a(m-1)}(\F)&=0~, \qquad &\cD^{\b}\widehat{\bm{\P}}{}^{\perp}_{\b\a(m-1)}(\F)=0~, \\
\cDB^2\bm{\P}^{\perp}_{\a(m)}(\F)&=0~, \qquad &\cDB^2\widehat{\bm{\P}}{}^{\perp}_{\a(m)}(\F)=0~.
\end{alignat}
\esubeq
Thus, the operator maps onto the constraint space which furnishes the massive UIR $\mb{G}(\bm{m}, s+ \hf)$ (cf. eq. \eqref{FMSTLCons}).
Once again, it can be shown that these two types of projectors coincide, $ \bm{\P}^{\perp}_{\a(m)}(\F) = \widehat{\bm{\P}}{}^{\perp}_{\a(m)}(\F)$. In the case when $n > m = 0$, the corresponding projectors $\bar{\bm{\P}}{}^{\perp}_{\ad(n)}(\bar{\F})$ and $\widehat{\bar{\bm{\P}}}{}^{\perp}_{\ad(n)}(\bar{\F}) $ can be  obtained by complex conjugation and similar comments apply.

\paragraph{Orthogonal complement of $\bm{\P}^{\perp}_{(m,n)}$ }
Let us introduce the orthogonal complement of $\bm{\P}^{\perp}_{(m,n)}$ on $\mb{V}_{(m,n)}$
\be \label{FMSLongProjDefn}
\bm{\P}^{||}_{(m,n)} = \mds{1} - \bm{\P}^{\perp}_{(m,n)}~,
\ee
which satisfies the properties\bsubeq
\begin{gather}
\bm{\P}^{||}_{(m,n)} \bm{\P}^{||}_{(m,n)}  = \bm{\P}^{||}_{(m,n)} ~, \label{FMSLongIdemp}\\
\bm{\P}^{||}_{(m,n)} \bm{\P}^{\perp}_{(m,n)} = \bm{\P}^{\perp}_{(m,n)} \bm{\P}^{||}_{(m,n)} =0~.
\end{gather}
\esubeq
For integers $m,n \geq 1$, it can be shown that the orthogonal complement $\bm{\P}^{||}_{(m,n)} $ projects a superfield $\F_{\a(m)\ad(n)} \in \mb{V}_{(m,n)}$ onto the union of spaces of longitudinal linear and longitudinal anti-linear (LLLA) superfields
\be \label{FMSLongProj}
\bm{\P}^{||}_{(m,n)}\F_{\a(m)\ad(n)} = \cDB_{(\ad_1}\J_{\a(m)\ad_2\dots\ad_{n})} + \cD_{(\a_1}\O_{\a_2\dots\a_m)\ad(n)}~,
\ee
for some complex unconstrained superfields $\J_{\a(m)\ad(n-1)}$ and $\O_{\a(m-1)\ad(n)}$. 

Let us denote by $\mb{V}^{\parallel}_{(m,n)}$ the space of LLLA superfields $\F^{\parallel}_{\a(m)\ad(n)} = \cDB_{(\ad_1}\J_{\a(m)\ad_2\dots\ad_{n})} + \cD_{(\a_1}\O_{\a_2\dots\a_m)\ad(n)}$, where the superfields $\J_{\a(m)\ad(n-1)}$ and $\O_{\a(m-1)\ad(n)}$ are unconstrained. It can be shown that the superprojector $\bm{\P}^{\perp}_{(m,n)}$ annihilates any LLLA superfield
\be\label{FMSSPKillsLong}
\bm{\P}^{\perp}_{(m,n)}  \F^{\parallel}_{\a(m)\ad(n)}  = 0~.
\ee
It follows immediately that the superprojector $\bm{\P}^{||}_{(m,n)}$ acts as the unit operator on the space of LLLA superfields
\be
\bm{\P}^{\parallel}_{(m,n)}\F^{\parallel}_{\a(m)\ad(n)} = \F^{\parallel}_{\a(m)\ad(n)} ~,
\ee
as a consequence of the properties \eqref{FMSLongProjDefn} and \eqref{FMSSPKillsLong}.

For $m \geq 1, n=0$, the orthogonal complement may be expressed as
\begin{align}
\bm{\P}^{||}_{(m)}\F_{\a(m)}= \cD_{(\a_1}\O_{\a_2\dots\a_m)}+\Lambda_{\a(m)}~, \qquad 
\cDB_{\bd}\Lambda_{\a(m)}=0~,
\end{align}
for some unconstrained $\O_{\a(m-1)}$ and chiral $\L_{\a(m)}$ superfields. On the other hand, if $n \geq 1$ and $m=0$, then the orthogonal complement can be written in the form
\begin{align}
\overline{\bm{\P}}^{||}_{(n)}\bar{\F}_{\ad(n)} = \cDB_{(\ad_1}\bar{\J}_{\ad_2\dots\ad_n)}+\bar{\Lambda}_{\ad(n)}~, \qquad \cD_{\b}\bar{\Lambda}_{\ad(n)}=0~,
\end{align}
for some unconstrained $\bar{\J}_{\ad(n-1)}$ and anti-chiral $\bar{\Lambda}_{\ad(n)}$ superfields.

Using the fact that the superprojectors $\bm{\P}^{\perp}_{(m,n)} $  and $\bm{\P}^{\parallel}_{(m,n)}$ resolve the identity operator \eqref{FMSLongProjDefn}, it follows that any unconstrained superfield $\F_{\a(m)\ad(n)}$ can be decomposed in the following manner
\begin{align}
\F_{\a(m)\ad(n)}=\F^{\perp}_{\a(m)\ad(n)}+\cDB_{(\ad_1}\z_{\a(m)\ad_2\dots\ad_n)}+\cD_{(\a_1}\x_{\a_2\dots\a_m)\ad(n)}~. \label{FMSdecomp0}
\end{align}
Here, $\F^{\perp}_{\a(m)\ad(n)}$ is TLAL and thus irreducible, whilst $\z_{\a(m)\ad(n-1)}$ and $\x_{\a(m-1)\ad(n)}$ are unconstrained and thus reducible. It is easy to see that $\bm{\P}^{\perp}_{(m,n)}$ selects the component $\F^{\perp}_{\a(m)\ad(n)}$ from the decomposition \eqref{FMSdecomp0} which describes the pure superspin state with maximal superspin $s+\hf$, by virtue of the properties \eqref{FMSSurj} and \eqref{FMSSPKillsLong}.  

One can then repeat the decomposition on the lower-rank unconstrained superfields in \eqref{FMSdecomp0}. After repeating this procedure a finite number of times, one eventually arrives at the decomposition of $\F_{\a(m)\ad(n)}$ solely in terms of irreducible superfields. This decomposition is completed explicitly in $\cN=1$ AdS$_4$ superspace in section \ref{FASOrthoSect} (cf. eq \eqref{decomp1}). Taking the flat-superspace limit of this result will provide the general irreducible decomposition of any superfield $\F_{\a(m)\ad(n)}$ in $\mb{M}^{4|4}$.

\subsection{Superconformal higher-spin theory}
We wish to elucidate the interplay between the highest superspin projection operator  $\bm{\P}^{\perp}_{(m,n)}$ \eqref{FMSHighestSuperspinProjector} and superconformal higher-spin (SCHS) theory in $\mb{M}^{4|4}$. For positive integers $m,n \geq 1$, the SCHS theory is described in terms of the SCHS gauge prepotential $H_{\a(m)\ad(n)}$, which is defined modulo the gauge transformations \cite{KuzenkoManvelyanTheisen2017}
\begin{align}
\delta_{\zeta,\xi}H_{\a(m)\ad(n)}=\cDB_{(\ad_1}\zeta_{\a(m)\ad_2\dots\ad_n)}+\cD_{(\a_1}\xi_{\a_2\dots\a_m)\ad(n)}~, \label{FMGT}
\end{align}
where the complex gauge parameters $\z_{\a(m)\ad(n-1)}$ and $\x_{\a(m-1)\ad(n)}$ are unconstrained.

Associated with  $H_{\a(m)\ad(n)}$ and its complex conjugate  $\bar{H}_{\a(n)\ad(m)}$
are the linearised higher-spin super-Weyl tensors \cite{KuzenkoManvelyanTheisen2017}
\begin{subequations} \label{FMSWeylTensors}
	\begin{align}
	\mds{W}_{\a(m+n+1)}(H)&= -\frac{1}{4}\cDB^2\pa_{(\a_1}{}^{\bd_1}\cdots\pa_{\a_n}{}^{\bd_n}\cD_{\a_{n+1}}H_{\a_{n+2}\dots\a_{m+n+1})\bd(n)}~, \\
	\mds{W}_{\a(m+n+1)}(\bar{H})&= -\frac{1}{4}\cDB^2\pa_{(\a_1}{}^{\bd_1}\cdots\pa_{\a_m}{}^{\bd_m}\cD_{\a_{m+1}}\bar{H}_{\a_{m+2}\dots\a_{m+n+1})\bd(m)}~,
	\end{align}
\end{subequations}
which are chiral 
\bsubeq
\be
\cDB_\bd \mds{W}_{\a(m+n+1)}(H)=0~, \qquad \cDB_\bd \mds{W}_{\a(m+n+1)}(\bar{H})=0~,
\ee
and invariant under the gauge transformations \eqref{FMGT}
\be
\delta_{\zeta,\xi}	\mds{W}_{\a(m+n+1)}(H)=0~,\qquad \delta_{\zeta,\xi}	\mds{W}_{\a(m+n+1)}(\bar{H})=0~.
\ee
\esubeq
Alternatively, the linearised higher-spin super-Weyl tensors \cite{KuzenkoManvelyanTheisen2017} with purely dotted indices can be obtained from \eqref{FMSWeylTensors} via complex conjugation
\bsubeq \label{FMSWeylTensorCC}
\begin{align}
\overline{\mds{W}}_{\ad(m+n+1)}({H})&= \frac{1}{4}(-1)^{m+n+1}\cD^2\pa_{(\ad_1}{}^{\b_1}\cdots\pa_{\ad_m}{}^{\b_m}\cDB_{\ad_{m+1}}{H}_{\b(m)\ad_{m+2}\dots\ad_{m+n+1})}~, \\
\overline{	\mds{W}}_{\ad(m+n+1)}(\bar{H})&= \frac{1}{4}(-1)^{m+n+1}\cD^2\pa_{(\ad_1}{}^{\b_1}\cdots\pa_{\ad_n}{}^{\b_n}\cDB_{\ad_{n+1}}\bar{H}_{\b(n)\ad_{n+2}\dots\ad_{m+n+1})}~.
\end{align}
\esubeq
The Weyl tensors \eqref{FMSWeylTensorCC} are still gauge-invariant, however, they are instead anti-chiral
\be
\cD_\b \overline{\mds{W}}_{\ad(m+n+1)}({H})=0~, \qquad \cD_\b 	\overline{	\mds{W}}_{\ad(m+n+1)}(\bar{H}) =0~.
\ee
The linearised higher-spin Weyl tensors \eqref{FMSWeylTensors} and their complex conjugates \eqref{FMSWeylTensorCC} are the building blocks of the gauge-invariant SCHS action\footnote{The SCHS gauge prepotential and its corresponding gauge parameters are also defined to be superconformal primary superfields in $\mb{M}^{4|4}$. For more information on the superconformal properties, see \cite{KuzenkoManvelyanTheisen2017}.}
\bea
S_{\text{SCHS}}^{(m,n)}[H,\bar{H}]&=&\frac{1}{2}\text{i}^{m+n}\int 
\text{d}^4x ~ \rd^2 \q \,  
\mds{W}^{\a(m+n+1)}(H)\mds{W}_{\a(m+n+1)}(\bar{H}) \non \\
&&+\frac{1}{2}(-\text{i})^{m+n}\int 
\text{d}^4x ~ \rd^2 \tb \,  
\overline{\mds{W}}_{\ad(m+n+1)}(H)\overline{\mds{W}}^{\ad(m+n+1)}(\bar{H})~.\label{FMSSCHSchiral}
\eea
Here, the first action is an integral over the chiral superspace, while the second is over the anti-chiral superspace. The chiral and anti-chiral superspaces can be written as integrals over the full superspace or reduced to the component space via the rules
\bsubeq
\bea
\int \rd^{4|4}z~\F(z) &=& - \frac{1}{4} \int \rd^4x \rd^2 \q~\cDB^2 \F(z) = \frac{1}{16} \int \rd^4x~\cD^2 \cDB^2 \F(z)|~, \\
&=& - \frac{1}{4} \int \rd^4x \rd^2 \tb~\cD^2 \F(z) = \frac{1}{16} \int \rd^4x~\cDB^2 \cD^2 \F(z)|~.
\eea
\esubeq
Note, the rule for integration by parts on $\mb{M}^{4|4}$ is given by
\be
\int \rd^{4|4}z~ \big (\cD_A \F^A(z) \big ) = 0~.
\ee 

Upon integrating by parts, the action \eqref{FMSSCHSchiral} may be written in the alternative forms
\vspace{-\baselineskip}
\begin{subequations} \label{FMSCHSAction}
	\begin{align}
	S_{\text{SCHS}}^{(m,n)}[H,\bar{H}]&= \text{i}^{m+n}\int \text{d}^{4|4}z \,  \bar{H}^{\a(n)\ad(m)}\mds{B}_{\a(n)\ad(m)}(H) +\text{c.c. } \\
	&= \text{i}^{m+n} \int \text{d}^{4|4}z \,  \bar{H}^{\a(n)\ad(m)}\widehat{\mds{B}}_{\a(n)\ad(m)}(H) +\rm{c.c.}
	\end{align}
\end{subequations}
These alternative forms for the SCHS actions \eqref{FMSCHSAction} are interesting because they are now formulated in terms linearised higher-spin super-Bach tensors
\begin{subequations} \label{FMSBach}
	\begin{align}
	\mds{B}_{\a(n)\ad(m)}(H)&= \pa_{(\ad_1}{}^{\b_1}\cdots\pa_{\ad_{m})}{}^{\b_m}\cD^{\b_{m+1}}\mds{W}_{\a(n)\b(m+1)}(H)~,\\
	\widehat{\mds{B}}_{\a(n)\ad(m)}(H)&=(-1)^{m+n+1}\pa_{(\a_1}{}^{\bd_1}\cdots\pa_{\a_{n})}{}^{\bd_n}\cDB^{\bd_{n+1}}\overline{\mds{W}}_{\ad(m)\bd(n+1)}(H)~.
	\end{align}
\end{subequations}
The super-Bach tensors are gauge-invariant and TLAL
\begin{subequations} \label{FMSBachTensorsTransverse}
	\begin{align}
	\cD^\b\mds{B}_{\b\a(n-1)\ad(m)}(H)&= 0~, \qquad \cDB^{\bd}\mds{B}_{\a(n)\bd\ad(m-1)}(H)=0~,\\
	\cD^\b\widehat{\mds{B}}_{\b\a(n-1)\ad(m)}(H)&= 0~, \qquad \cDB^{\bd}\widehat{\mds{B}}_{\a(n)\bd\ad(m-1)}(H)=0~.
	\end{align}
\end{subequations}

Since the higher-spin super-Bach tensors are TLAL \eqref{FMSBachTensorsTransverse}, it proves natural to formulate them in terms of the superspin projection operators \eqref{FMSHighestSuperspinProjector}
\bsubeq \label{FMSBachTensorsProjectors}
\bea
\mds{B}_{\a(n)\ad(m)}(\F) &=& 2 \Box^{n+1}\pa_{(\ad_1}{}^{\b_1} \dots \pa_{\ad_{m-n}}{}^{\b_{m-n}} \bm{\P}^{\perp}_{\b(m-n)\a(n)\ad_{m-n+1} \ldots \ad_m )}(H)~,\\
\widehat{\mds{B}}_{\a(n)\ad(m)}(\F) &=& 2 \Box^{n+1}\pa_{(\ad_1}{}^{\b_1} \dots \pa_{\ad_{m-n}}{}^{\b_{m-n}} \bm{\P}^{\perp}_{\a(n)\b(m-n)\ad_{m-n+1} \ldots \ad_m}(H)~.
\eea
\esubeq
The benefits of expressing the super-Bach tensors \eqref{FMSBachTensorsProjectors} in terms of superprojectors is that they are now  manifestly TLAL and gauge-invariant, as a consequence of the superprojector properties \eqref{FMSTLALProjector} and \eqref{FMSSPKillsLong}.

Similarly, we can express the super-Weyl tensors in terms of the superprojectors 
\bsubeq
\bea
\mds{W}_{\a(m+n+1)}(H)&=& \ri \pa_{(\a_1}{}^{\bd_1} \dots \pa_{\a_{n+1}}{}^{\bd_{n+1}} \cDB_{\bd_{n+1}} \bm{\P}^{\perp}_{\a_{n+2} \ldots \a_{m+n+1})\bd(n)}(H)~. \\
\mds{W}_{\a(m+n+1)}(\bar{H})&=& \ri \pa_{(\a_1}{}^{\bd_1} \dots \pa_{\a_{m+1}}{}^{\bd_{m+1}} \cDB_{\bd_{m+1}} \bm{\P}^{\perp}_{\a_{m+2} \ldots \a_{m+n+1})\bd(m)}(\bar{H})~.
\eea
\esubeq

It follows from \eqref{FMSBachTensorsProjectors} that the SCHS action \eqref{FMSCHSAction} can be recast in terms of superspin projection operators 
\bsubeq \label{FMSSCHSProj}
\bea
S_{\text{SCHS}}^{(m,n)}[H,\bar{H}] &=&\text{i}^{m+n} \int\text{d}^{4|4}z  \, \bar{H}^{\a(n)\ad(m)} \Box^{n+1}
(\pa_{\ad}{}^{\b})^{m-n} \bm{\P}^{\perp}_{(m,n)} H_{\a(n)\b(m-n)\ad(n)}+\text{c.c.}~,    \label{FMSSCHSProj1} \\
S_{\text{SCHS}}^{(m,n)}[H,\bar{H}] &=& \text{i}^{m+n}\int\text{d}^{4|4}z  \, \bar{H}^{\a(n)\ad(m)}\Box^{m+1} (\pa_{\a}{}^{\bd})^{n-m} \bm{\P}^{\perp}_{(m,n)}H_{\a(m)\bd(n-m)\ad(m)}+\text{c.c.}~, \label{FMSSCHSProj2}\hspace{1.5cm}
\eea
\esubeq
where we have assumed $m > n$ in \eqref{FMSSCHSProj1} and $n > m ~$ in \eqref{FMSSCHSProj2}.
The SCHS actions \eqref{FMSSCHSProj} can be considered  supersymmetric generalisations of the CHS actions \eqref{FMCHSProj}. In the form \eqref{FMSSCHSProj}, it is apparent that the SCHS actions \eqref{FMSSCHSProj} describe a pure superspin state with maximal superspin $s=\hf(m+n+1)$.

\section{Three-dimensional Minkowski space} \label{SecThreeDimensionalMinkowskiSpace}


In this section we generalise the analysis of section \ref{SecFourDimensionalMinkowskiSpace} to three-dimensional Minkowski space $\mb{M}^3$. We begin by reviewing the irreducible representations of the three-dimensional \Po algebra $\mf{iso}(2,1)$ \cite{Binegar1981}
and their corresponding field realisations in $\mb{M}^3$ \cite{GorbunovKuzenkoLyakhovich1996}. Once established, we will study the related topics of spin projection operators \cite{BuchbinderKuzenkoLaFontainePonds2018,IsaevPodoinitsyn2017},  conformal higher-spin theory \cite{KuzenkoPonds2018,KuzenkoPonds2019}, (Fang-)Fronsdal models \cite{KuzenkoOgburn2016} and topologically massive theories \cite{KuzenkoPonds2018}.

\subsection{Irreducible representations of the \Po algebra} \label{Irrepsof3dPoincare}
The three-dimensional \Po algebra $\mf{iso}(2,1)$ (cf. \eqref{FMPoincareAlgebra} for $d=3$) possesses two  quadratic Casimir operators $C_i$, for $i = 1,2$. The operators $C_i$ take the form
\be \label{TMCasimirOperators}
C_1 = - P^a P_a~, \qquad C_2 = W~,  \qquad 
\ [C_i , P_a] = 0~,   \quad[C_i , J_{ab}] = 0~,
\ee
where $W$ is the Pauli-Lubanski pseudoscalar given by
\be \label{TMPauliLubanksiScalar}
W= \ve^{abc}P_a J_{bc}~.
\ee
These Casimir operators are important in the classification of the massive and massless irreducible representations of $\mf{iso}(2,1)$.

\paragraph{Massive UIRs of $\mf{iso}(2,1)$}
The UIRs of $\mf{iso}(2,1)$ were classified via the method of induced representations in \cite{Binegar1981} (see also \cite{Grigore1993}). 
Here, it was shown that the massive irreducible representations of $\PaT$ are classified by the quantum numbers of mass $m > 0$ and helicity $\l \in \mb{R}$. The eigenvalues of the Casimir operators \eqref{TMCasimirOperators} label the massive UIRs as follows\footnote{Note that the factor of two appearing in the eigenvalue of $C_2$ is not common in the literature. This factor arises due to the different normalisation of $W$ \eqref{TMPauliLubanksiScalar} used in this thesis, which also differs by a factor of two.}
\be \label{TMMassiveOnshellConditions}
C_1 = -{m}^2 \mds{1}~, \qquad C_2 =2 {m} \l \mds{1}~.
\ee
We are only interested in the massive UIRs which are characterised by (half-)integer helicity values $\l = 0, \pm \hf, \pm 1,  \cdots$.\footnote{If the parameter $\l$ is not restricted to (half-)integer values, then one would obtain continuous spin representations which are associated with continuous spin particles called anyons. See the early works \cite{Binegar1981, JackiwNair1990,CortesPlyushchay1995} for  further discussions.} We denote by $D(m,\l)$ a massive UIR which carries mass $m$, helicity $\l$ and spin $s:=|\l|$. A massive UIR $D(m,\l)$ propagates a single physical degree of freedom, which corresponds to the single fixed helicity state it describes.

\paragraph{Massless UIRs of $\mf{iso}(2,1)$}
The landscape of massless UIRs in $3d$ is somewhat foreign when compared to the familiar case in $4d$. 
The massless UIRs \cite{Binegar1981} are labelled by a continuous parameter $\l \in \mb{R}$, and a discrete parameter $\e = \lb 0 , 1 \rb$. The case $\l \neq 0$ leads to continuous-spin massless particles, known as panyons \cite{SchusterToro2014}, which are usually ignored. We will be interested in studying the UIRs where $\l = 0$, which are known as massless helicity representations. These UIRs are labelled by the two parameters $\e = 0$ or $\e =1$ \cite{Binegar1981}. Thus there only exists two massless helicity UIRs in $3d$, which are referred to as spin-$0$ for $\e = 0$, and spin-$\hf$ for $\e = 1$.

\subsection{Irreducible field representations}
In this section we study the field realisations of the massive and massless UIRs of $\mf{iso}(2,1)$ which were discussed in section \ref{Irrepsof3dPoincare}. 

\subsubsection{Massive field representations}
We wish to realise the massive UIRs of $\PaT$ on the space of tensor fields in $\mb{M}^3$. 
For integer $n>1$, a real totally symmetric rank-$n$ spinor field $\f_{\a_1 \ldots \a_n} = \bar{\f}_{\a_1 \ldots \a_n}= \f_{\a(n)}$ furnishes the massive UIR $D(m,\frac{\s n}{2})$ if it obeys the on-shell conditions  \cite{GorbunovKuzenkoLyakhovich1996,Gitman1996fr} (see also \cite{TyutinVasiliev1997,BergshoeffHohmTownsend2010,KuzenkoPonds2018})
\bsubeq \label{TMOnshellField}
\bea
\pa^{\b\g}\f_{\b\g\a(n-2)} &=& 0~,  \label{TMOnshellPropTransverse}\\
\big (W - \s n m \big )\f_{\a(n)}&=&0~,  \qquad m>0~,\label{TMOnshellPropMassShell}
\eea
\esubeq
where $\s = \pm 1$. The constraint \eqref{TMOnshellPropMassShell} can also be written in the equivalent form 
\be
\pa_{(\a_1}{}^\b \f_{\a_2 \ldots \a_n ) \b} = \s m \f_{\a(n)}~.
\ee
A field which realises the UIR $D(m,\frac{\s n}{2})$ is said to be a massive field with mass $m$, helicity $\l = \frac{\s n}{2}$ and spin $s=\frac{n}{2}$. For the case $n=1$, a massive field only obeys the Dirac equation \eqref{TMOnshellPropMassShell}. Let us show explicitly that the on-shell field \eqref{TMOnshellField} does indeed describe the massive UIR $D(m,\frac{\s n}{2})$ of $\PaT$.

The generators of $\mf{iso}(2,1)$ take the following form in the field representation 
\be  \label{TMFieldGenerators}
P_a = -\ri \pa_a~, \qquad J_{ab} = \ri (x_b \pa_a -x_a \pa_b) - \ri M_{ab}~.
\ee
It follows that the Pauli-Lubanski pseudoscalar \eqref{TMPauliLubanksiScalar}  takes the form 
\be \label{TMPauliLubanksiCasimir}
W = \pa^{\b\g}M_{\b\g}~,
\ee
in the field representation. Note that the definition of the Pauli-Lubanski scalar \eqref{TMPauliLubanksiCasimir} differs to that given in \cite{KuzenkoOgburn2016} by a factor of 2. This normalisation factor was chosen to ensure that \eqref{TMPauliLubanksiCasimir} coincides with its AdS$_3$ analogue \eqref{QCF}.

The Casimir operators $C_i$ \eqref{TMCasimirOperators} of $\mf{iso}(2,1)$  are related to each other via the relation
\bea \label{TMWSquared}
W^2 \f_{\a(n)} = n^2  \Box   \f_{\a(n)} + n(n-1)\pa_{\a(2)}\pa^{\b(2)} \f_{\b(2)\a(n-2)}~,
\eea
where $\Box = -\hf \pa^{\a \b} \pa_{\a \b}$. 
Acting on a transverse field \eqref{TMOnshellPropTransverse}, the equation \eqref{TMWSquared} becomes
\be \label{TMPauliLubanskiScalarTransverseField}
W^2 \f_{\a(n)} = n^2  \Box   \f_{\a(n)}~.
\ee
Using  \eqref{TMPauliLubanskiScalarTransverseField}, it can be shown that the on-shell equations \eqref{TMOnshellField} lead to the Klein-Gordon equation
\be \label{TMOnshellFieldKGE}
(\Box - m^2)\f_{\a(n)} = 0~.
\ee
It follows from the relation \eqref{TMOnshellPropMassShell} and \eqref{TMOnshellFieldKGE} that an on-shell field \eqref{TMOnshellField} furnishes the UIR $D(m, \frac{\s n}{2})$ of $\PaT$ (cf. eq \eqref{TMMassiveOnshellConditions}). It can be shown that a massive field \eqref{TMOnshellField} on $\mb{M}^3$ describes a single propagating degree of freedom. For an explicit derivation of this result, see \cite{KuzenkoPonds2018}.

In place of the on-shell conditions \eqref{TMOnshellField}, one may instead consider tensor fields $\f_{\a(n)}$ constrained by the equations \eqref{TMOnshellPropTransverse} and \eqref{TMOnshellFieldKGE}\footnote{The conditions \eqref{TMFierzPauli} are equivalent to the 3$d$ Fierz-Pauli equations  \cite{FierzPauli1939}}
\bsubeq \label{TMFierzPauli}
\bea
\pa^{\b\g}\f_{\b\g\a(n-2)} &=& 0~,  \\
(\Box - m^2)\f_{\a(n)} &=& 0~. \label{TMKG}
\eea
\esubeq
In this case, the equation \eqref{TMWSquared} becomes
\be
\big ( W -nm \big ) \big (W+nm \big ) \f_{\a(n)} = 0~.
\ee
This indicates that a field satisfying the conditions \eqref{TMFierzPauli} furnishes the reducible representation of $\PaT$
\be \label{TMMassiveRep} 
D \Big ( m , - \frac{n}{2} \Big ) \oplus D \Big (m , \frac{n}{2} \Big )~,
\ee
which describes two massive UIRs which carry the  helicities $-\frac{n}{2}$ and $\frac{n}{2}$, respectively. 

Note that the representation \eqref{TMMassiveRep} is reducible with respect to the connected part of the \Po group $\mathsf{ISO}_0(2,1)$. However, since a parity transformation flips the sign of the helicity, it follows that the representation \eqref{TMMassiveRep} is irreducible with respect to the full \Po group $\mathsf{IO}(2,1)$. Therefore,  \eqref{TMFierzPauli} are the correct on-shell conditions if one desires a parity invariant theory. Note that the conditions \eqref{TMOnshellField} break parity since they select a single helicity state.

\subsubsection{Massless field representations} \label{TMMasslessFieldsRepsSec}
Recall that higher-spin massless fields in $\mb{M}^4$ furnish massless UIRs of $\PaF$, and thus are characterised by helicity $\l$ (and spin $|\l|$) which takes (half-)integer values $\l = 0, \pm \hf , \pm 1, \cdots$. Thus the notion of spin comes from the label of the massless UIR it realises. However the story is considerably different in $\mb{M}^3$, as there only exists two massless UIRs of $\PaT$: the spin-$0$ UIR; and the spin-$\hf$ UIR. Thus, the notion of massless `higher-spin' fields is non-existent in $\mb{M}^3$ as a higher-rank field cannot furnish a higher-spin massless UIR of $\PaT$. 

In accordance with this, let us consider a higher-rank field $\f_{\a(n)}$ in $\mb{M}^3$. For integers $n \geq 3$, a field $\f_{\a(n)}$ is said to be a massless higher-spin field if it does not propagate any physical degrees freedom. Such a field $\f_{\a(n)}$ is said to carry spin $s=\frac{n}{2}$, in which we refer to spin as a property related to the rank of the massless field. Elaborating on the notion of spin, there are two ways in which massless dynamics can be described in three dimensions: in terms of gauge fields; or in terms of gauge-invariant field strengths.

Let us begin by introducing massless gauge fields in $\mb{M}^3$.
For an integer $n \geq 2$, we say that a real totally symmetric rank-$n$ field $\f_{\a(n)}$ is massless if it satisfies the conditions
\bsubeq \label{TMMasslessFieldW}
\bea 
\pa^{\b(2)}\f_{\b(2)\a(n-2)} &=& 0~,  \label{TMMasslessWTransverse} \\
W \f_{\a(n)}  &=& 0~. \label{TMMasslessWKG}
\eea
The equations \eqref{TMMasslessWTransverse} and \eqref{TMMasslessWKG} are compatible with the gauge transformations
\be \label{TMMasslessWGaugeTransformation}
\d_\z \f_{\a(n)} = \pa_{\a(2)}\z_{\a(n-2)}~,
\ee
given that the real gauge parameter $\z_{\a(n-2)}$ is constrained by
\bea
\pa^{\b\g}\z_{\b\g \a(n-4)} &=& 0~, \\
W \z_{\a(n-2)} &=& 0~.
\eea
\esubeq
The constraints \eqref{TMMasslessWTransverse} and \eqref{TMMasslessWKG} naturally arise from the massive on-shell equations \eqref{TMOnshellField} in the massless limit $m \rightarrow 0$. It follows from the conditions \eqref{TMMasslessWTransverse} and \eqref{TMMasslessWKG} that 
\be \label{TMmass1KG}
\Box \f_{\a(n)} = 0~.
\ee

Let us show that the massless gauge fields \eqref{TMMasslessFieldW} propagate no physical degrees of freedom. To do this, it proves useful to transition to momentum space, $\f_{\a(n)}(x) \rightarrow \f_{\a(n)}(p)$. Since the equations \eqref{TMMasslessFieldW} are Lorentz invariant, we are free to transition the standard momentum frame
\be \label{TMMasslessRestFrame}
p^a = \Big (\hf, 0, \hf \Big ) \quad \Longrightarrow \quad p^{\a\b} = 
\begin{pmatrix}
	~0~ &0~\\
	~0~&1~ 
\end{pmatrix}~, \quad p_{\a \b} =
\begin{pmatrix}
	~1~ &0~\\
	~0~&0~ 
\end{pmatrix}~.
\ee 
In this frame, the transverse constraint \eqref{TMMasslessWTransverse} implies 
\be \label{TMMasslessGC1}
p^{22}\f_{22 \a(n-2)}(p) = 0 \quad \Longrightarrow \quad \f_{22 \a(n-2)}(p) = 0~.
\ee
The gauge transformation \eqref{TMMasslessWGaugeTransformation} also takes the following form in \eqref{TMMasslessRestFrame}
\be
\d_\z \f_{11 \a(n-2)}(p) \propto p_{11} \z_{\a(n-2)}(p)~.
\ee
This allows us to impose the gauge 
\be \label{TMMasslessGC2}
\f_{11 \a(n-2)}(p) = 0~.
\ee
It follows from conditions \eqref{TMMasslessGC1} and \eqref{TMMasslessGC2} that for $n \geq 3$, the field $\f_{\a(n)}(p)$ describes no physical degrees of freedom. As we will see in subsection \ref{TMFangFronsdalModels}, the conditions \eqref{TMMasslessFieldW} appear in the on-shell analysis of the Fang-Fronsdal model \eqref{TMFangFronsdalAction}.

Special attention needs to be given to the $n=2$ case. It follows from the counting argument above that the massless field $\f_{\a(2)}$ carries a single propagating degree of freedom, $\f_{12}=\f_{21}$. However, this component can be shown to vanish in the standard momentum frame \eqref{TMMasslessRestFrame}, as a consequence of the constraint \eqref{TMMasslessWKG}. In the $n=2$ case, the massless conditions \eqref{TMMasslessFieldW} can be identified as the on-shell equations obtained from the Chern-Simons type action \eqref{TMCHSAction}.

We can also introduce an alternative definition for a massless gauge field in $\mb{M}^3$. That is, for integers $n \geq 2$, we say that a real  totally symmetric rank-$n$ field  $\f_{\a(n)}$ is massless if it satisfies the conditions
\bsubeq \label{TMMasslessField}
\bea 
\pa^{\b(2)}\f_{\b(2)\a(n-2)} &=& 0~,  \label{TMMasslessTransverse} \\
\Box \f_{\a(n)}  &=& 0~. \label{TMMasslessKG}
\eea
The equations \eqref{TMMasslessTransverse} and \eqref{TMMasslessKG} are compatible with the gauge symmetry
\be \label{TMMasslessGaugeTransformation}
\d_\z \f_{\a(n)} = \pa_{\a(2)}\z_{\a(n-2)}~,
\ee
given that the real gauge parameter $\z_{\a(n-2)}$ satisfies the conditions
\bea
\Box \z_{\a(n-2)} &=& 0~, \\
\pa^{\b\g}\z_{\b\g \a(n-4)} &=& 0~.
\eea
\esubeq
By a similar counting argument, it can be shown that the massless gauge field \eqref{TMMasslessField} propagates no physical degrees of freedom when $ n \geq 3$. Moreover, the constraints \eqref{TMMasslessField} coincide with the on-shell conditions obtained from the Fronsdal action \eqref{TMFronsdalAction}.

Again, special attention needs to be given to the case $n=2$. It can be shown that the massless vector field $\f_{\a(2)}$ \eqref{TMMasslessField}
describes a single independent component $\f_{12}=\f_{21}$, by virtue of the constraints \eqref{TMMasslessGC1} and \eqref{TMMasslessGC2}. It must be noted that this degree of freedom cannot be eliminated. For $n=2$, the constraints \eqref{TMMasslessField} coincide with the on-shell conditions arising from Maxwell's theory. 

Demonstrating this, the gauge-invariant Maxwell action is given by
\be \label{MAXWELL}
S_{\text{Maxwell}} = -\frac{1}{4}\int \rd^3 x~F^{\a \b} F_{\a \b}~.
\ee
Here $F_{\a \b} = \pa_{(\a}{}^\g A_{\b)\g}$ is the (dual of the) Maxwell field strength which is written in terms of the gauge field $A_{\a(2)}$. The field strength $F_{\a \b}$ is transverse off-shell and is invariant under gauge transformations of the form 
\be \label{TMSMAxwell}
\d_\z A_{\a \b} = \pa_{\a\b}\z~, 
\ee
where the gauge parameter $\z$ is real unconstrained. 

The equation of motion corresponding to the action \eqref{MAXWELL} is
\be \label{MAXWELLSEOM}
0 = \Box A_{\a\b}  + \hf \pa_{\a \b}\pa^{\g \d}A_{\g\d}~.
\ee
The gauge freedom \eqref{TMSMAxwell} allows us to impose the Lorenz gauge 
\be \label{TMLorenzGauge}
\pa^{\b\g}A_{\b \g} = 0~,
\ee 
where the residual gauge freedom is described by 
\be
\Box \z = 0~.
\ee 
In the gauge \eqref{TMLorenzGauge}, the equation of motion \eqref{MAXWELLSEOM} reduces to
\be
\Box A_{\a\b} = 0~.
\ee

In accordance with the massless on-shell conditions \eqref{TMMasslessField}, it follows from the above analysis that the Maxwell action describes a massless gauge field on-shell.
It is well known that Maxwell electrodynamics can be recast in terms of a scalar field via the relation $F_{\a\b} = \pa_{\a\b} \f$. In accordance with this, it follows that the vector field $A_{\a \b}$ furnishes the spin-$0$ massless UIR \cite{Binegar1981}.

Note that fields $\f$ and $\f_\a$ are not treated within the frameworks of \eqref{TMMasslessFieldW} and \eqref{TMMasslessField}, and thus need to be studied separately. We say that a scalar field $\f$ is massless if it satisfies the condition
\be \label{TMMasslessScalar}
\Box \f =0~.
\ee
The action of $W$ \eqref{TMPauliLubanksiCasimir} on a scalar field is automatically zero. A massless scalar field  \eqref{TMMasslessScalar} can be shown to carry a single physical degree of freedom, which is consistent with the fact that it furnishes the massless spin-$0$ UIR of $\PaT$. 

Moreover, a spinor field $\f_\a$ is said to be massless if it satisfies the condition
\be \label{TMMasslessSpinorConstraint}
W \f_\a = 0 \quad \Longrightarrow \quad \Box \f_\a = 0~.
\ee
It can be shown that the massless spinor field \eqref{TMMasslessSpinorConstraint} furnishes the massless spin-$\hf$ UIR of $\PaT$.
To see that the field \eqref{TMMasslessSpinorConstraint} carries a single physical degree of freedom, it proves useful to transition to the frame \eqref{TMMasslessRestFrame}. In this frame, the condition \eqref{TMMasslessSpinorConstraint} yields $\f_2 = 0$, thus leaving one physical degree of freedom in the form of $\f_1$.

Let us now formulate massless dynamics in terms of the gauge-invariant field strengths $C_{\a(n)}(\f)$. For $n \geq 2$, the field strengths $C_{\a(n)}(\f)$ associated with the field $\f_{\a(n)}$ are given by \cite{PopeTownsend1989, Kuzenko2016}
\bsubeq \label{TMFangFieldStrength1}
\bea
{C}_{\a(2s)}(\f)&=&\frac{1}{2^{2s-1}}\sum_{j=0}^{s-1}2^{2j+1}\binom{s+j}{2j+1}\Box^j
\pa_{\a(2)}^{s-j-1}\pa_{\a}{}^{\b}\big(\pa^{\b(2)}\big)^{s-j-1}\f_{\a(2j+1)\b(2s-2j-1)}~, \hspace{1.3cm}\\
{C}_{\a(2s+1)}(\f)&=&\frac{1}{2^{2s}}\sum_{j=0}^{s}2^{2j}\binom{s+j}{2j}\frac{(2s+1)}{(2j+1)}\Box^j  \pa_{\a(2)}^{s-j}\big(\pa^{\b(2)}\big)^{s-j}\f_{\a(2j+1)\b(2s-2j)}~.
\eea
\esubeq
These field strengths are invariant under the gauge transformations \eqref{TMMasslessGaugeTransformation}
\be \label{TMMasslessFieldStrengthsProp1}
\d_\z C_{\a(n)}(\f) = 0~.
\ee
The field strengths $C_{\a(n)}(\f)$ are known as the higher-spin Cotton tensors and will be discussed in more detail in section \ref{TMsecCT}.
For an integer $n>2$, the field strengths $C_{\a(n)}(\f)$ \eqref{TMFangFieldStrength1} vanish when the field $\f_{\a(n)}$ belongs to either one of the massless families \eqref{TMMasslessFieldW} or \eqref{TMMasslessField}. This follows directly from the properties \eqref{TMmass1KG} and \eqref{TMMasslessKG}. Due to gauge completeness \eqref{TMGaugeCompleteness},\footnote{The gauge completeness property is elaborated upon in section \ref{TMsecCT}.} it follows that $\f_{\a(n)} $ is pure gauge.

Special attention needs to be given to the field strength $C_{\a(2)}(\f)$. For $n=2$, it follows from the first set of massless conditions \eqref{TMMasslessFieldW} that $C_{\a(2)}(\f)$ vanishes. Thus, it does not describe any propagating degrees of freedom due to the gauge completeness property. On the other hand, it can be shown that the field strength $C_{\a(2)}(\f)$ is non-vanishing given that $\f_{\a(2)}$ satisfies the other massless field conditions \eqref{TMMasslessField}. Moreover, it can be shown to satisfy the properties
\bsubeq \label{FMN2FieldStrength}
\bea
\pa^{\b(2)}C_{\b(2)}(\f) &=& 0~, \label{TMMasslessFSN2Prop1}\\
W C_{\a(2)}(\f) &=& 0~. \label{TMMasslessFSN2Prop2}
\eea
\esubeq
The field strength $C_{\a (2)}(\f)$ coincides with the (dual of the) electromagnetic field strength $F_{\a \b}$ from Maxwell's theory \eqref{MAXWELL}. Hence, the  condition \eqref{TMMasslessFSN2Prop1} can be identified as the off-shell transverse constraint for $F_{\a \b}$, while \eqref{TMMasslessFSN2Prop2} coincides with the equation of motion
\eqref{MAXWELLSEOM} in the Lorenz gauge \eqref{TMLorenzGauge}.

\subsection{Spin projection operators} \label{TMSectionSPO}
Let us denote by $V_{(n)}$ the space of totally symmetric real rank-$n$ spinor fields $\f_{\a(n)}$ on $\mb{M}^3$. The spin projection operator $\P^{\perp}_{(n)}$ is defined by its action on $V_{(n)}$ according to the rule
\begin{align}
\P^{\perp}_{(n)}: V_{(n)} &\longrightarrow V_{(n)}~, \\
\f_{\a(n)} &\longmapsto \P^{\perp}_{(n)} \f_{\a(n)} =: \P^{\perp}_{\a(n)}(\f)~. \non
\end{align}
For fixed integer $n \geq 2$,  the operator  $\P^{\perp}_{(n)}$ satisfies the following properties on $V_{(n)}$:
\bsubeq
\begin{enumerate} \label{TMBFProjectorProperties}
	\item \textbf{Idempotence:} The operator $\P^{\perp}_{(n)}$ squares to itself
	\be \label{TMBFProjectorIdempotence}
	\P^{\perp}_{(n)}\P^{\perp}_{(n)} \f_{\a(n)} = \P^{\perp}_{(n)} \f_{\a(n)}~.
	\ee
	\item \textbf{Transversality:} The operator $\P^{\perp}_{(n)}$ maps $\f_{\a(n)}$ to a transverse field
	\be \label{TMBFProjectorTransverse}
	\pa^{\b(2)}\P^{\perp}_{\b(2)\a(n-2)}(\f) = 0~.
	\ee
	\item \textbf{Surjectivity:} Every transverse field $\f^{\perp}_{\a(n)}$ belongs to the image of $\P^{\perp}_{(n)}$
	\be \label{TMBFProjectorSurjective}
	\pa^{\b(2)}\f^{\perp}_{\b(2)\a(n-2)} =0 \quad \Longrightarrow \quad \P^{\perp}_{(n)} \f^{\perp}_{\a(n)} = \f^{\perp}_{\a(n)}~.
	\ee
	In other words, $\P^{\perp}_{(n)}$ acts as the identity operator on the space of transverse fields.
\end{enumerate}
\esubeq

The spin projection operator $\Pi^{\perp}_{(n)}$ maps any field on $V_{(n)}$ obeying the first-order equation \eqref{TMOnshellPropMassShell} to a massive on-shell field \eqref{TMFierzPauli}
\bsubeq \label{TMBFProjectorOnshell}
\bea
\pa^{\b(2)}\P^{\perp}_{(n)}\f_{\b(2)\a(n-2)} &=&0~, \\
\big (W - \s n m \big )\P^{\perp}_{(n)} \f_{\a(n)}&=&0~.
\eea
\esubeq
In accordance with \eqref{TMOnshellField}, the projected field $\P^{\perp}_{\a(n)}(\f)$ \eqref{TMBFProjectorOnshell} realises the massive UIR $D(m, \frac{n\s}{2})$.

For integers $n \geq 2$, the spin projection operator \eqref{TMBFProjectorProperties} takes the explicit form on $V_{(n)}$ 
\bea \label{TMTransverseProjector}
\widehat{\Pi}^{\perp}_{(n)}\f_{\a(n)}  = \frac{1}{\Box^{\lfloor \frac{n}{2} \rfloor}} \sum_{j=0}^{\lfloor \frac{n}{2} \rfloor} 2^{2j-2\lfloor \frac{n}{2} \rfloor} \frac{n(\lceil \frac{n}{2} \rceil + j -1)!}{(n+2j-2\lfloor \frac{n}{2} \rfloor)!(\lfloor \frac{n}{2} \rfloor-j)!} \non
\Box^j  \pa_{\a(2)}^{\lfloor \frac{n}{2} \rfloor-j}\big (\pa^{\b(2)}\big )^{\lfloor \frac{n}{2} \rfloor-j}\\
\times \f_{\a(n+2j -2\lfloor  \frac{n}{2} \rfloor)\b(2\lfloor \frac{n}{2} \rfloor-2j)}~. 
\eea
Specifically, for the bosonic $(n=2s)$ and fermionic $(n=2s+1)$ cases, the projectors \eqref{TMTransverseProjector} reduce to
\bsubeq 
\bea 
\widehat{\Pi}^{\perp}_{(2s)}\f_{\a(2s)}  &=& \frac{1}{\Box^s} \sum_{j=0}^{s} 2^{2j-2s}\frac{2s}{s+j}\binom{s+j}{2j}
\Box^j  \pa_{\a(2)}^{s-j}\big (\pa^{\b(2)}\big )^{s-j}\f_{\a(2j)\b(2s-2j)}~, \label{TMBosonicProjector} \\
\widehat{\P}^{\perp}_{(2s+1)} \f_{\a(2s+1)}&=&\frac{1}{\Box^s}  \sum_{j=0}^{s} 2^{2j-2s}\frac{2s+1}{2j+1}\binom{s+j}{2j} \Box^j   \pa_{\a(2)}^{s-j}\big (\pa_{\b(2)}\big )^{s-j}\f_{\a(2j+1)\b(2s-2j)}~. \label{TMFermionicProjector} \hspace{1cm}
\eea
\esubeq
The spin projection operators \eqref{TMBFProjectorOnshell} were recently derived for the first time by Isaev and Podoinitsyn in \cite{IsaevPodoinitsyn2017} (see also \cite{BuchbinderKuzenkoLaFontainePonds2018}). These projectors can be made local if the field being projected satisfies either the first-order constraint \eqref{TMOnshellPropMassShell}\footnote{Recall that the constraint \eqref{TMOnshellPropMassShell} implies \eqref{TMKG}.} or the second-order constraint \eqref{TMKG}.

It is possible to construct a spin projection operator strictly in terms of the Casimir operators \eqref{TMCasimirOperators} of $\PaT$  \cite{HutchingsKuzenkoPonds2021}. To this end, we introduce the following operator on $V_{(n)}$
\be \label{TMTransverseProjectorCasimir}
\P^{\perp}_{(n)} = \frac{1}{2^{n-1} (2\lfloor \frac{n}{2} \rfloor )! \Box^{\lfloor \frac{n}{2} \rfloor}} \prod_{j=0}^{\lfloor \frac{n}{2} \rfloor -1 } \Big ( W^2 - \big ( n -2 \Big \lfloor \frac{n}{2} \Big \rfloor + 2j \big )^2 \Box \Big )~.
\ee
Proving that $\P^{\perp}_{(n)}$ satisfies the defining properties \eqref{TMBFProjectorProperties} of a spin projection operator is difficult due to the presence of the Casimir operators \eqref{TMCasimirOperators}. However, this task is more tractable if one makes use of the generating function formalism, which was developed recently in \cite{KP21}. This formalism is introduced in section \ref{TAProjectors} to demonstrate that the AdS$_3$ analogues of $\P^{\perp}_{(n)}$ are indeed spin projection operators. Taking the flat-space limit of these results will yield the proofs which show that $\P^{\perp}_{(n)}$ obeys the constraints \eqref{TMBFProjectorProperties}.

It can be shown that the projectors $\widehat{\Pi}^{\perp}_{(n)}$ and $\P^{\perp}_{(n)}$ are equivalent on $V_{(n)}$  \cite{HutchingsKuzenkoPonds2021}
\be \label{EBP}
\widehat{\P}^{\perp}_{(n)}\f_{\a(n)}=\P^{\perp}_{(n)}\f_{\a(n)} ~.
\ee
The proof for this is identical to that used to show the equivalence of the spin projection operators in $\mb{M}^4$ (cf. \eqref{FMBFEquivalence1}). In accordance with \eqref{EBP}, we will only deal with the spin projection operator $\P^{\perp}_{(n)}$ on $V_{(n)}$.

An important property of the projectors \eqref{TMTransverseProjectorCasimir} is that they are symmetric operators, that is \cite{HutchingsKuzenkoPonds2021}
\bea
\int\text{d}^3x\, \vf^{\a(n)} \P^{\perp}_{(n)} \f_{\a(n)} 
= \int\text{d}^3x \, \f^{\a(n)} \P^{\perp}_{(n)} \vf_{\a(n)} ~,
\eea
for arbitrary well-behaved fields $\vf_{\a(n)} $ and $\f_{\a(n)} $. 

The spin projection operators $\P^{\perp}_{(n)}$ display some interesting features when they are no longer restricted to the space $V_{(n)}$. To explore these properties further, it is necessary to study the bosonic and fermionic projectors separately.

For $n=2s$, the bosonic spin projection operator \eqref{TMTransverseProjectorCasimir} takes the form
\be  \label{TMBosonicProjectorCasimir}
\P^{\perp}_{(2s)}
= \frac{1}{2^{2s-1}(2s)!  \Box^s }  \prod_{j=0}^{s-1} \Big ( W^2 -(2j)^2 \Box \Big )  
~.
\ee

For fixed $s$,
the operator $\Pi_{(2s)}^{\perp}$  is also defined 
to act on the linear spaces ${V}_{(2s')}$ with $s' < s$. 
In fact, it is possible to show that the following holds true  \cite{HutchingsKuzenkoPonds2021}
\be \label{TMBosProjProp2}
\P^{\perp}_{(2s)} \f_{(2s')}=0~, \qquad 0 \leq s' \leq s-1 ~.
\ee 
This important identity states that $\Pi^{\perp}_{(2s)}$ annihilates any lower-rank bosonic field $\f_{\a(2s')}\in V_{(2s')}$. It should be mentioned that $\P^{\perp}_{(2s)} $ does not annihilate lower-rank fermionic fields $\f_{\a(2s'+1)}$. When acting on $V_{(2s')}$, the two operators $\widehat{\Pi}_{(2s)}^{\perp}$ and $\Pi_{(2s)}^{\perp}$ are no longer equivalent, and in particular $\widehat{\Pi}_{(2s)}^{\perp}\phi_{(2s')}\neq 0 $. It is for this reason that we will continue to use different notation for the two operators on $V_{(2s')}$. 

For $n=2s+1$, the fermionic spin projection operator \eqref{TMTransverseProjectorCasimir} takes the form
\be
\P^{\perp}_{(2s+1)} 
=\frac{1}{2^{2s}(2s)!\Box^s}\prod_{j=0}^{s-1} \Big ( W^2 -(2j+1)^2 \Box \Big ) 
~.
\ee

Stepping away from $V_{(2s+1)}$, one can show that for fixed $s$, the projector $\Pi_{(2s+1)}^{\perp}$ annihilates any lower-rank fermionic field $\f_{(2s'+1)}\in {V}_{(2s'+1)}$ \cite{HutchingsKuzenkoPonds2021}
\be \label{TMFermProjProp2}
\P^{\perp}_{(2s+1)} \f_{(2s'+1)}=0~, \qquad 1 \leq s' \leq s-1 ~.
\ee 
The two operators $\widehat{\Pi}_{(2s+1)}^{\perp}$ and $\Pi_{(2s+1)}^{\perp}$ are no longer equivalent on ${V}_{(2s'+1)}$. We remark that $\P^{\perp}_{(2s+1)} $ does not annihilate lower-rank bosonic fields $\f_{\a(2s'+2)}$.

\subsubsection{Helicity projectors}
Suppose that $\phi_{\a(n)} \in V_{(n)}$ satisfies the mass-shell equation \eqref{TMOnshellFieldKGE} instead of the differential constraint \eqref{TMOnshellPropMassShell}. Then $\Pi^{\perp}_{(n)}$ maps $\phi_{\a(n)}$ to the field which obeys the conditions
\vspace{-\baselineskip}
\bsubeq \label{TMBFProjectorOnshellKG}
\bea
\pa^{\b(2)}\P^{\perp}_{\b(2)\a(n-2)}(\f) &=&0~, \\
\big (\Box- m^2\big )\P^{\perp}_{\a(n)}(\f)&=&0~.
\eea
\esubeq 
In accordance with \eqref{TMMassiveRep}, it follows that the projected field $\P^{\perp}_{\a(n)}(\f)$ satisfying \eqref{TMBFProjectorOnshellKG} furnishes the reducible representation $D(m,-\frac{n}{2})\oplus D(m,\frac{n}{2})$ of $\PaT$. In particular, representations with both signs of helicity $\pm\frac{n}{2}$ appear in this decomposition. In order to isolate the component of $\phi_{\a(n)}$ describing a UIR of $\PaT$, it is necessary to bisect the spin projection operators $\Pi_{(n)}^{\perp}$ according to
\begin{align}
\Pi_{(n)}^{\perp}=\mc{P}^{(+)}_{(n)}+\mc{P}^{(-)}_{(n)}~.\label{TMHelicityProjectors}
\end{align}    
The operators $\mc{P}^{(\pm)}_{(n)}$, which we refer to as helicity projectors, should satisfy the properties  \eqref{TMBFProjectorIdempotence} and \eqref{TMBFProjectorTransverse}. As the name suggests, these operators should also project out the component of $\phi_{\a(n)}$ carrying a single value of helicity. The last two requirements are equivalent to the equations
\begin{subequations}
	\begin{align}
	\pa^{\b(2)}\f^{(\pm)}_{\b(2)\a(n-2)}&=0~,\\
	\big (W - \s n m \big )\f^{(\pm)}_{\a(n)}&=0~,  \label{TMHelicityConstraint2}
	\end{align}
\end{subequations}
where we have denoted $\f^{(\pm)}_{\a(n)}:=\mc{P}^{(\pm)}_{(n)} \f_{\a(n)}$. By virtue of \eqref{TMOnshellField}, it follows that $\f^{(\pm)}_{\a(n)}$ furnishes the UIR $D(m, \pm \frac{n}{2})$.

It is not difficult to show that the following operators satisfy these requirements\footnote{The helicity projector $\mc{P}^{(\pm)}_{(n)}$ is equivalent to the spin projection operator $\Pi_{(n)}^{\perp}$ when acting on a field on $V_{(n)}$ which satisfies the constraint \eqref{TMOnshellPropMassShell}.}
\be \label{helicityproj}
\mc{P}^{(\pm)}_{(n)} :=\hf \bigg (\mathds{1} \pm \frac{W}{n\sqrt{\mathcal{\Box}}} \bigg ) {\Pi}^{\perp}_{(n)}~.
\ee 
Here ${\Pi}^{\perp}_{(n)}$ are the spin projectors written in terms of Casimir operators, and are given by 
\eqref{TMTransverseProjectorCasimir}. 
Of course, on $V_{(n)}$, one could instead choose to represent the latter in their alternate form \eqref{TMTransverseProjector}, in which case they coincide with those derived in \cite{BuchbinderKuzenkoLaFontainePonds2018}.
Using the defining features of ${\Pi}^{\perp}_{(n)}$, it can be shown that the operators $\mc{P}^{(+)}_{(n)} $ and $\mc{P}^{(-)}_{(n)} $ are orthogonal projectors when restricted to ${V}_{(n)}$
\be 
\mc{P}^{(\pm)}_{(n)} \mc{P}^{(\pm)}_{(n)}  = \mc{P}^{(\pm)}_{(n)} ~, \qquad \mc{P}^{(\pm)}_{(n)} \mc{P}^{(\mp)}_{(n)}  = 0~.
\ee
It follows from \eqref{helicityproj} that $\mc{P}^{(\pm)}_{(n)}$ inherits transversality from ${\Pi}_{(n)}$. Moreover, the field $\f^{(\pm)}_{\a(n)}$ satisfies the constraint
\be \label{TMhelcityprojprop}
\Big(W \mp n \sqrt{\Box } \Big)\f^{(\pm)}_{\a(n)} =0 ~,
\ee
off the mass-shell.
However, if $\f^{(\pm)}_{\a(n)}$ is placed on the mass-shell \eqref{TMOnshellFieldKGE}, then \eqref{TMhelcityprojprop} reduces to \eqref{TMHelicityConstraint2}. Thus given a field $\f_{\a(n)}$ satisfying \eqref{TMOnshellFieldKGE}, it follows that the helicity projectors $\mc{P}^{(\pm)}_{(n)}$ \eqref{helicityproj} isolate the component $\f^{(\pm)}_{\a(n)}$ which furnishes the massive UIR $D(m, \pm \frac{n}{2})$.\footnote{Although $\mc{P}^{(\pm)}_{(n)}$ singles out the UIR $D(m, \pm \frac{n}{2})$ from $\f_{\a(n)}$ on the mass-shell, they are not spin projection operators in our framework as they do not satisfy property \eqref{TMBFProjectorSurjective}. But this is just semantics. } 

\subsubsection{Longitudinal projectors and lower-spin extractors}
Let us introduce the operator $\Pi^{\parallel}_{(n)}$ which is the orthogonal complement of $\P^{\perp}_{(n)}$
\be \label{LongProj}
\Pi^{\parallel}_{(n)}  :=  \mathds{1} - \P^{\perp}_{(n)} ~.
\ee
By construction, the operators $\Pi^{\perp}_{(n)}$ and $\Pi^{\parallel}_{(n)}$ resolve the identity, $\mathds{1} = \Pi^{\parallel}_{(n)} + \Pi^{\perp}_{(n)}$, and form an orthogonal set of projectors
\begin{subequations} \label{OrthoProjProp}
	\begin{align}
	\Pi^{\perp}_{(n)}\Pi^{\perp}_{(n)} &= \Pi^{\perp}_{(n)}~,\qquad \Pi^{\parallel}_{(n)}\Pi^{\parallel}_{(n)} = \Pi^{\parallel}_{(n)}~, \\
	\Pi^{\parallel}_{(n)} \Pi^{\perp}_{(n)} &=0~, \qquad  ~~~\phantom{.} \Pi^{\perp}_{(n)} \Pi^{\parallel}_{(n)}=0 ~.
	\end{align}
\end{subequations}
Moreover, it can be shown that $\Pi^{\parallel}_{(n)}$ projects a field $\f_{\a(n)}$ onto its longitudinal component. A rank-$n$ field $\f^{\parallel}_{\a(n)}$ is said to be longitudinal if there exists a rank-$(n-2)$ field $\f_{\a(n-2)}$ such that $\f^{\parallel}_{\a(n)}$ may be expressed as $\f^{\parallel}_{\a(n)} = \pa_{\a(2)}\f_{\a(n-2)}$. Such fields are sometimes referred to as being pure gauge.
Therefore, we find that \cite{HutchingsKuzenkoPonds2021}
\begin{align}
\f^{\parallel}_{\a(n)}:=\Pi^{\parallel}_{(n)} \f_{\a(n)}=\pa_{\a(2)}\f_{\a(n-2)}~,
\end{align}
for some unconstrained field $\phi_{\a(n-2)}$. 

Let $\f^{\parallel}_{\a(n)}$  be some longitudinal field, 
for which we do not assume to be in the image of $\Pi^{\parallel}_{(n)}$. 
Since $\P_{(n)}^{\perp}$ commutes with $\pa_{\a(2)}$ and annihilates all lower-rank fields \eqref{BosProjProp}, it follows that it also annihilates any rank-$n$ longitudinal field \cite{HutchingsKuzenkoPonds2021}\footnote{This also implies that $\widehat{\P}^{\perp}_{(n)}\psi_{\a(n)}=0$, since both $\widehat{\P}^{\perp}_{(n)}$ and $\P^{\perp}_{(n)}$ are equal on ${V}_{(n)}$. }
\begin{align} \label{TMProjKillsLong}
\f^{\parallel}_{\a(n)} =\pa_{\a(2)}\f_{\a(n-2)}\qquad \implies \qquad \P^{\perp}_{(n)} \f^{\parallel}_{\a(n)} =0~.
\end{align}
As a consequence, given two integers $n, k$ satisfying $2\leq k \leq n$, it immediately follows that $\Pi^{\parallel}_{(n)}$ acts as the identity operator on the space of rank-$k$ longitudinal fields $\f^{\parallel}_{\a(k)}$ \cite{HutchingsKuzenkoPonds2021},
\begin{align} \label{Transkillslong}
\f^{\parallel}_{\a(k)}=\pa_{\a(2)}\f_{\a(k-2)} \qquad \implies \qquad \Pi^{\parallel}_{(k+2s)}\f^{\parallel}_{\a(k)}=\f^{\parallel}_{\a(k)}~,
\end{align}
with $s$ a non-negative integer.
These properties will be useful in the subsequent section.

Using the fact that $\Pi^{\perp}_{(n)}$ and $\Pi^{\parallel}_{(n)}$ resolve the identity, one can decompose an arbitrary field $\f_{\a(n)}$ as follows
\be
\f_{\a(n)} = \f^{\perp}_{\a(n)} + \pa_{\a(2)}\f_{\a(n-2)}~.
\label{2.51}
\ee
Here $\f^{\perp}_{\a(n)}$ is transverse and $\f_{\a(n-2)}$ is unconstrained. Repeating this process iteratively, we obtain the following decomposition \cite{HutchingsKuzenkoPonds2021}
\bea\label{TMDecomp2}
\f_{\a(n)} &=& \sum_{j=0}^{\lfloor n/2 \rfloor }  \big (\pa_{\a(2)} \big )^j \f^{\perp}_{\a(n-2j)}~. 
\eea
Here each of the fields $\f^{\perp}_{\a(n-2j)}$ are transverse, except of course $\phi^{\perp}$ and $\f^{\perp}_{\a}$. Note that the spin
projection operators \eqref{TMTransverseProjectorCasimir} will only select the component in the decomposition \eqref{TMDecomp2} with the highest spin, as a consequence of the properties \eqref{TMBosProjProp2} and \eqref{TMFermProjProp2}. We note that, using \eqref{TMHelicityProjectors}, one may take the decomposition \eqref{TMDecomp2} a step further and bisect each term into irreducible components which are transverse and have positive or negative helicity \cite{HutchingsKuzenkoPonds2021},
\begin{align}
\f_{\a(n)} = \sum_{j=0}^{\lfloor n/2 \rfloor }  \big (\pa_{\a(2)} \big )^j\Big( \f^{(+)}_{\a(n-2j)}+\f^{(-)}_{\a(n-2j)}\Big)~. 
\end{align}

Making use of the spin projectors \eqref{TMTransverseProjectorCasimir} and their corresponding properties, one can construct operators which extract the component $\phi_{\a(n-2j)}^{\perp}$ from the decomposition \eqref{TMDecomp2}, where $1\leq j \leq \lfloor n/2 \rfloor$. In particular, we find that the spin $\frac{1}{2}(n-2j)$ component may be extracted via \cite{HutchingsKuzenkoPonds2021}
\begin{align}
\phi_{\a(n)}\mapsto \phi^{\perp}_{\a(n-2j)}=\big({S}_{(n-2j)}^{\perp}\phi\big)_{\a(n-2j)}\equiv {S}_{\a(n-2j)}^{\perp}(\phi)~,
\end{align}
where we have defined
\begin{align}
{S}_{\a(n-2j)}^{\perp}(\phi)&=\frac{(-1)^j}{2^{2j}}\binom{n}{j} \frac{1}{\Box^j} \P^{\perp}_{(n-2j)}\big(\partial^{\b(2)}\big)^j\f_{\a(n-2j)\b(2j)}~.
\end{align}
From this expression, it is clear that ${S}_{\a(n-2j)}^{\perp}(\phi)$ is transverse,
\be
0=\pa^{\b(2)}{S}_{\b(2)\a(n-2j-2)}^{\perp}(\phi)~.
\ee
Therefore it is appropriate to call ${S}_{(n-2j)}^{\perp}$ the transverse spin $\frac{1}{2}(n-2j)$ extractor. It is not idempotent, since it is dimensionful and reduces the rank of the field on which it acts.

\subsection{Conformal higher-spin theory} \label{TMsecCT}
In this section we explore applications for the spin projection operators in modern conformal higher-spin theories in $\mb{M}^3$. For $n \ge 2$, a real rank-$n$ spinor field $h_{\a(n)}$ is said to be a CHS field if it is defined modulo gauge transformations of the form \cite{KuzenkoPonds2018}\footnote{For a detailed discussion concerning the conformal nature of CHS fields in $\mb{M}^3$, see \cite{KuzenkoPonds2018}. }
\be \label{TMCHSGS}
\d_\z h_{\a(n)} = \pa_{\a(2)} \z_{\a(n-2)}~,
\ee
where $\z_{\a(n-2)}$ is a real unconstrained gauge parameter.

Associated with the gauge prepotentials $h_{\a(n)}$ are the real primary descendents $C_{\a(n)}(h)$, which are known as the linearised higher-spin Cotton tensors. The Cotton tensors $C_{\a(n)}(h)$ are uniquely defined by the properties:
\begin{enumerate}
	\item $C_{\a(n)}(h)$ is transverse
	\bsubeq \label{TMCottonTensor}
	\be
	\pa^{\b\g}C_{\b\g\a(n-2)} (h)= 0~. \label{TMCottonTransverse}
	\ee
	\item $C_{\a(n)}(h)$ is gauge-invariant
	\be
	\d_\z C_{\a(n)}(h) = 0~. \label{TMCottonGaugeInvariance}
	\ee
	\esubeq
\end{enumerate}
Modulo an overall normalisation factor, the linearised higher-spin bosonic and fermionic Cotton tensors  take the respective closed forms
\begin{subequations} \label{TMCottonTensorsExplicit}
	\bea
	{C}_{\a(2s)}(h)&=&\frac{1}{2^{2s-1}}\sum_{j=0}^{s-1}2^{2j+1}\binom{s+j}{2j+1}\Box^j
	\pa_{\a(2)}^{s-j-1}\pa_{\a}{}^{\b}\big(\pa^{\b(2)}\big)^{s-j-1}h_{\a(2j+1)\b(2s-2j-1)}~,  \label{TMBosonicCottonTensor} \hspace{1.4cm}\\
	{C}_{\a(2s+1)}(h)&=&\frac{1}{2^{2s}}\sum_{j=0}^{s}2^{2j}\binom{s+j}{2j}\frac{(2s+1)}{(2j+1)}\Box^j  \pa_{\a(2)}^{s-j}\big(\pa^{\b(2)}\big)^{s-j}h_{\a(2j+1)\b(2s-2j)}~.  \label{TMFermionicCottonTensor}
	\eea
\end{subequations}
The linearised higher-spin bosonic Cotton tensor \eqref{TMBosonicCottonTensor} was first derived by Pope and Townsend  \cite{PopeTownsend1989} in 1989, while its fermionic counterpart \eqref{TMFermionicCottonTensor} was only recently derived by Kuzenko \cite{Kuzenko2016} in 2016.

Since the Cotton tensors \eqref{TMCottonTensorsExplicit} are transverse by definition, it proves natural to recast them in terms of the transverse spin projection operators \eqref{TMTransverseProjector}. Making use of the projectors $\widehat{\Pi}^{\perp}_{(n)}$ \eqref{TMTransverseProjector}, it can be shown that the higher-spin Cotton tensors \eqref{TMCottonTensorsExplicit} can be reformulated in the simple form \cite{BuchbinderKuzenkoLaFontainePonds2018}
\begin{subequations} \label{CT}
	\bea
	{C}_{\a(2s)} (h) &=& \frac{1}{2s} \Box^{s-1} W \widehat{\Pi}^{\perp}_{(2s)}h_{\a(2s)}~, \\
	{C}_{\a(2s+1)}(h) &=&\Box^s\widehat{\Pi}^{\perp}_{(2s+1)}h_{\a(2s+1)}~.
	\eea 
\end{subequations}
Moreover, we can make use of the equivalent family of projectors $\P^{\perp}_{(n)}$ to recast ${C}_{\a(n)}(h)$ \eqref{CT} purely in terms of the Casimir operators \eqref{TMCasimirOperators}. Explicitly, they read \cite{HutchingsKuzenkoPonds2021}\footnote{It can be shown that the Cotton tensors are equivalent to those derived in \cite{HHL, HHL2}.}
\begin{subequations} \label{TMCottonTensorsCasimir}
	\bea
	C_{\a(2s)} (h) &=& \frac{W}{2^{2s-1}(2s-1)!}  \prod_{j=1}^{s-1} \Big (W^2 -(2j)^2\Box \Big )   h_{\a(2s)}~, \\
	C_{\a(2s+1)}(h) &=&\frac{1}{2^{2s}(2s)!}\prod_{j=0}^{s-1} \Big ( W^2 -(2j+1)^2 \Box  \Big ) h_{\a(2s+1)}~. 
	\eea
\end{subequations}

There are many advantages to expressing the Cotton tensors in terms of spin projection operators. Firstly, in both \eqref{CT} and \eqref{TMCottonTensorsCasimir}, the properties of transversality \eqref{TMCottonTransverse} and gauge invariance \eqref{TMCottonGaugeInvariance} are made manifest, as a consequence of the projector properties \eqref{TMBFProjectorTransverse}  and \eqref{TMProjKillsLong}. 

Using the gauge freedom \eqref{TMCHSGS},  one may impose the transverse gauge condition 
\begin{align}
h_{\a(n)}\equiv h^{\perp}_{\a(n)}~, \qquad 0=\pa^{\b(2)}h^{\perp}_{\b(2)\a(n-2)}~.
\end{align}
In this gauge, the Cotton tensors \eqref{CT} become manifestly factorised into products of (at most) second-order differential operators
\bsubeq
\bea
{C}_{\a(2s)}(h^{\perp}) &=&  \frac{1}{2s} \Box^{s-1} W h^{\perp}_{\a(2s)} ~, \label{TMCottonTensorTransverseGauge} \\
{C}_{\a(2s+1)}(h^{\perp}) &=&\Box^s h^{\perp}_{\a(2s+1)}~, \label{TMCottonFermionicTensorTransverseGauge}
\eea
\esubeq
on account of the projector property \eqref{TMBFProjectorSurjective}.
An interesting feature of the new realisation \eqref{TMCottonTensorsCasimir} is that the Cotton tensors are manifestly factorised in terms of (at most) second-order differential operators without having to enter the transverse gauge.

By virtue of the above observations, it follows that the conformal higher-spin action 
\begin{align}  \label{TMCHSAction}
S_{\text{CHS}}^{(n)}[h]=\frac{\text{i}^n}{2^{\lceil n/2 \rceil+1}}\int\text{d}^3x\, h^{\a(n)}{C}_{\a(n)}(h)~,
\end{align}
is manifestly gauge invariant and factorised when ${C}_{\a(n)}(h)$ is expressed as  \eqref{TMCottonTensorsCasimir}.

It is well known that in $\mb{M}^3$, the Cotton tensor $C_{\a(n)}(h)$ vanishes if and only if the gauge field $h_{\a(n)}$ is pure gauge
\be \label{TMGaugeCompleteness}
C_{\a(n)}(h) = 0 \qquad \Longleftrightarrow \qquad h_{\a(n)} = \pa_{\a(2)}\z_{\a(n-2)}~.
\ee
This property, which is known as gauge completeness, was first proved in \cite{HHL,HHL2}.\footnote{Note that a sketch of this proof which utilises the spin projection operators \eqref{TMTransverseProjector} was given in \cite{KP21}. } Gauge completeness is useful in determining whether a higher-spin theory in $\mb{M}^3$ is massless. Specifically, if it can be shown that the Cotton tensor of the surviving gauge field $h_{\a(n)}$ vanishes on the equations of motion, then $h_{\a(n)}$ is massless since it propagates no physical degrees of freedom.

\subsection{Massless higher-spin theories} \label{TMFangFronsdalModels}
We wish to study free theories which, at the level of the equations of motion, generate the massless conditions \eqref{TMMasslessField}.  These actions were first computed by Fronsdal \cite{Fronsdal1978Massless} and Fang and Fronsdal \cite{FangFronsdal} in $\mb{M}^4$ for a massless spin-$s$ and spin-$(s+\hf)$ field, respectively. In this section we review generalisations of these massless models to $\mb{M}^3$, which were recently constructed by Kuzenko and Ogburn \cite{KuzenkoOgburn2016}.

\subsubsection{Fronsdal action}
Given an integer $s \geq 2$, the Fronsdal action is described by the real bosonic fields $h_{\a(2s)}$ and $y_{\a(2s-4)}$ on $\mb{M}^3$, which are defined modulo gauge transformations of the form
\bsubeq \label{TMFronsdalGT}
\bea 
\d_{\z} h_{\a(2s)} &=& \pa_{\a(2)} \z_{\a(2s-2)}~, \label{TMFronsdalGT1} \\
\d_{\z} y_{\a(2s-4)} &=& \frac{2s-2}{2s-1} \pa^{\b(2)} \z_{\b(2)\a(2s-4)}~. \label{TMFronsdalGT2}
\eea
\esubeq
Here the gauge parameter $\z_{\a(2s-2)}$ is real unconstrained. The Fronsdal action is given by 
\bea \label{TMFronsdalAction}
S^{\text{F}}_{(s)}[h,y] =&& \hf \Big (- \hf \Big )^{s}\AMT \Big \{ h^{\a(2s)}\Box h_{\a(2s)} - \frac{s}{2}\pa_{\b(2)}h^{\b(2)\a(2s-2)}\pa^{\g(2)}h_{\g(2)\a(2s-2)}  \non  \\
&&-\hf (2s-3)  \Big ( y^{\a(2s-4)}\pa^{\b(2)}\pa^{\g(2)}h_{\b(2)\g(2)\a(2s-4)} +\frac{2}{s} y^{\a(2s-4)}\Box y_{\a(2s-4)} \non \\
&&+ \frac{1}{4s(s-1)}(s-2)(2s-5)\pa_{\b(2)} y^{\b(2)\a(2s-6)}\pa^{\g(2)} y_{\g(2)\a(2s-6)} \Big ) \Big \rbrace~,
\eea
which is invariant under the gauge transformations \eqref{TMFronsdalGT}. 

It is an instructive exercise to show that the action \eqref{TMFronsdalAction} does indeed describe a massless spin-$s$ field. The equations of motion corresponding to \eqref{TMFronsdalAction} are
\bsubeq
\bea
0 = && 2\Box h_{\a(2s)} + s \pa_{\a(2)} \pa^{\b(2)} h_{\b(2)\a(2s-2)} - \hf (2s-3) \pa_{\a(2)} \pa_{\a(2)} y_{\a(2s-4)}~, \label{TMFronsdalEoM1} \\
0 =&& 8 (s-1)\Box y_{\a(2s-4)} -(s-2)(2s-5) \pa_{\a(2)}\pa^{\b(2)}y_{\b(2)\a(2s-6)} \non \\
&& + 2s(s-1)\pa^{\b(2)}\pa^{\g(2)} h_{\b(2)\g(2)\a(2s-4)}  ~. \label{TMFronsdalEoM2}
\eea
\esubeq

The gauge transformation \eqref{TMFronsdalGT2} allows us to impose the gauge condition
\be \label{TMFronsdalGC1}
y_{\a(2s-4)} = 0~.
\ee
In accordance with \eqref{TMFronsdalGT2}, the residual gauge freedom which preserves  \eqref{TMFronsdalGC1}  is described by the parameter  $\z_{\a(2s-2)} $ constrained by
\be \label{TMFronsdalResGF}
\pa^{\b(2)}\z_{\b(2)\a(2s-4)} = 0~.
\ee
In the gauge \eqref{TMFronsdalGC1}, the equation of motion \eqref{TMFronsdalEoM2} reduces to
\be \label{TMFronsdalEoM2G}
\pa^{\b(2)}\pa^{\g(2)}h_{\b(2)\g(2)\a(2s-4)} = 0~.
\ee
Using \eqref{TMFronsdalGT1}, the divergence of $h_{\a(2s)}$ can be shown to transform as
\be \label{TMFronsdalGaugeFreedom2}
\d_\z \big (\pa^{\b(2)} h_{\b(2)\a(2s-2)} \big ) = - \frac{2}{s}\Box \z_{\a(2s-2)}~,
\ee
where we have also used the fact that the gauge parameter is constrained by \eqref{TMFronsdalResGF}.
Taking into account \eqref{TMFronsdalEoM2G}, the gauge freedom \eqref{TMFronsdalGaugeFreedom2} can be used to gauge away $\pa^{\b(2)}h_{\b(2)\a(2s-2)}$
\be \label{TMFronsdalGaugeCond2}
\pa^{\b(2)}h_{\b(2)\a(2s-2)} =0 ~.
\ee
The remaining residual gauge symmetry is described by the gauge parameter satisfying \eqref{TMFronsdalResGF}, and the second-order equation
\be  \label{TMFronGaugePar}
\Box \z_{\a(2s-2)} =0~.
\ee
Upon imposing the gauges \eqref{TMFronsdalGC1} and \eqref{TMFronsdalGaugeCond2}, it follows that the equation of motion \eqref{TMFronsdalEoM1} reduces to
\be \label{TMFronsdalMassShell}
\Box h_{\a(2s)}=0~.
\ee
In accordance with \eqref{TMMasslessField}, it follows from the above analysis that the Fronsdal action \eqref{TMFronsdalAction} describes a massless spin-$s$ field on-shell.

\subsubsection{Fang-Fronsdal  action}
For $s \geq 2$, the Fang-Fronsdal model is described in terms of the real fields $h_{\a(2s+1)}$, $y_{\a(2s-1)}$ and $y_{\a(2s-3)}$, which are defined modulo gauge transformations of the form
\bsubeq \label{TMFangFronsdalGT}
\bea
\d_\z h_{\a(2s+1)} &=& \pa_{\a(2)}\z_{\a(2s-1)}~, \label{TMFangFronsdalGT1}\\
\d_\z y_{\a(2s-1)} &=& \frac{1}{2s+1}\pa^\b{}_{(\a_1} \z_{ \a_2 \ldots \a_{2s-1})\b}~ , \label{TMFangFronsdalGT2} \\
\d_\z z_{\a(2s-3)} &=& \pa^{\b(2)}\z_{\b(2) \a(2s-3)}~.\label{TMFangFronsdalGT3}
\eea
\esubeq
Here, the gauge parameter $\z_{\a(2s-1)}$ is real and unconstrained. The Fang-Fronsdal action is given by
\begin{align} \label{TMFangFronsdalAction}
S^{\text{FF}}_{(s+\hf)}[h,y,z] &= \frac{\ri}{2}\Big ( - \hf \Big )^s \AMT \Big \{ h^{\b\a(2s)}\pa_\b{}^\g h_{\g\a(2s)} + 2(2s-1) y^{\a(2s-1)}\pa^{\b(2)}h_{\b(2)\a(2s-1)} \non \\
&+  4(2s-1) y^{\b\a(2s-2)}\pa_\b{}^\g y_{\g\a(2s-2)} + \frac{2}{s}(s-1)(2s+1)z^{\a(2s-3)} \pa^{\b(2)} y_{\b(2)\a(2s-3)} \non \\
&-\frac{1}{s(2s-1)}(s-1)(2s-3)z^{\b\a(2s-4)}\pa_\b{}^\g z_{\g\a(2s-4)}  \Big \rbrace~,
\end{align}
which is invariant under the gauge transformations \eqref{TMFangFronsdalGT}. 

Let us show that the Fang-Fronsdal model \eqref{TMFangFronsdalAction} describes a massless spin-$(s+\hf)$ field on-shell. 
The equations of motion corresponding to the model \eqref{TMFangFronsdalAction} are
\bsubeq
\bea
0&=&\pa_\a{}^\g h_{\g\a(2s)} - (2s-1)\pa_{\a(2)} y_{\a(2s-1)}~, \label{TMFangFronsdalEoM1}\\
0&=&\pa^{\b(2)}h_{\b(2)\a(2s-1)} + 4\pa_\a{}^\b y_{\b\a(2s-2)} - \frac{1}{s(2s-1)}(s-1)(2s+1)\pa_{\a(2)}z_{\a(2s-3)}~, \hspace{0.65cm} \label{TMFangFronsdalEoM2}\\
0&=&(2s-1)(2s+1)\pa^{\b(2)}y_{\b(2)\a(2s-3)}-(2s-3)\pa_\a{}^\b z_{\b\a(2s-4)}~. \label{TMFangFronsdalEoM3}
\eea
\esubeq

We can use the gauge transformation \eqref{TMFangFronsdalGT3} to impose the gauge
\be \label{TMFangFronsdalGC1}
z_{\a(2s-3)} =0~.
\ee
In accordance with \eqref{TMFangFronsdalGT3}, the residual gauge symmetry that preserves the condition \eqref{TMFangFronsdalGC1} is
\be \label{TMFangFronsdalRG1}
\pa^{\b(2)}\z_{\b(2)\a(2s-3)} = 0~.
\ee
In the gauge \eqref{TMFangFronsdalGC1}, the equation of motion \eqref{TMFangFronsdalEoM3} reduces to 
\be \label{TMFangFronsdalEoMG1}
\pa^{\b(2)}y_{\b(2)\a(2s-3)} = 0~.
\ee
Hence, due to \eqref{TMFangFronsdalRG1} and the gauge transformation \eqref{TMFangFronsdalGT2}, we are allowed to impose the additional gauge
\be \label{TMFangFronsdalGC2}
y_{\a(2s-1)} = 0~,
\ee
which is consistent with \eqref{TMFangFronsdalEoMG1}. The residual gauge freedom which preserves \eqref{TMFangFronsdalGC2} is given by
\be \label{TMGaugeParCons2}
W \z_{\a(2s-1)}~  = 0 \quad \Longrightarrow \quad  \Box \z_{\a(2s-1)}=0~.
\ee
In the gauges \eqref{TMFangFronsdalGC1} and \eqref{TMFangFronsdalGC2}, the EoMs \eqref{TMFangFronsdalEoM1} and \eqref{TMFangFronsdalEoM2} reduce to 
\bsubeq \label{TMFieldConsMassless}
\bea 
W h_{\a(2s+1)}&=&0~, \\
\pa^{\b(2)} h_{\b(2)\a(2s-1)} &=& 0~.
\eea
\esubeq
The conditions  \eqref{TMFangFronsdalGT1}, \eqref{TMFangFronsdalRG1} \eqref{TMFangFronsdalEoMG1}, \eqref{TMGaugeParCons2}  and \eqref{TMFieldConsMassless} coincide with the massless on-shell conditions \eqref{TMMasslessFieldW} for the case $n=2s+1$, thus confirming that the action \eqref{TMFangFronsdalAction} describes a massless spin-$(s+\hf)$ field.

\subsubsection{Dual formulation of the Fronsdal-type action in $\mb{M}^3$} \label{apB2}
In this section we introduce a one-parameter family of dual formulations \cite{HutchingsHutomoKuzenko} for the Fronsdal-type action \cite{KuzenkoPonds2018}. This Fronsdal-type action was shown in \cite{KuzenkoPonds2018} to take the following form\footnote{In the $n=2s$, the action \eqref{flat-n} coincides with the Fronsdal action \eqref{TMFronsdalAction}. }
\begin{align}
S_{(\frac{n}{2})}^{\rm F} [h,y]
=\frac{\ri^n}{2^{\lfloor n/2 \rfloor +1}}&\int \rd^3x\,\bigg \{ 
\hf \pa^{\b\g} h^{\a(n)} \pa_{\b\g} h_{\a(n)}-\frac{n}{4}\pa_{\b \g}h^{\b\g\a(n-2)}\pa^{\d\l}h_{\d\l\a(n-2)}~  \non\\
&+\frac{n-3}{n} \Big( \frac{n}{2} \pa^{\b\g} y^{\a(n-4)}\pa^{\d\l}h_{\b\g\d\l\a(n-4)}
- \pa^{\b\g}  y^{\a(n-4)} \pa_{\b\g} y_{\a(n-4)}
~ \non \\
& -\frac{(n-4)(n-5)}{4(n-2)}\pa_{\b\g}y^{\b\g\a(n-6)}\pa^{\d\l}y_{\d\l\a(n-6)} \Big ) \bigg \}~.
\label{flat-n}
\end{align}
It is invariant under the gauge transformations
\begin{subequations} \label{fg00-flat}
	\bea
	\d_\z h_{\a(n)}&=&\pa_{\a(2)}\z_{\a(n-2)}~ , \label{fg00-flat-a} \\
	\d_\z y_{\a(n-4)}&=&\frac{n-2}{n-1}\pa^{\b(2)}\z_{\b(2) \a(n-4)}~. \label{fg00-flat-b}
	\eea
\end{subequations}

Both fields $h_{\a(n)}$ and $y_{\a(n-4)}$ appear in \eqref{flat-n} with derivatives, 
\bea
S^{\rm F}_{(\frac{n}{2})} =\frac{\ri^n}{2^{\lfloor n/2 \rfloor +1}}  \int \rd^{3} x \, \cL \big( \pa_{\b\g} h_{\a(n)}, \pa_{\b \g} y_{\a(n-4)} \big)~. 
\label{b00}
\eea 
Therefore a duality transformation may be performed upon each of them. Here we will only dualise $y_{\a(n-4)}$ and keep $h_{\a(n)}$ intact.\footnote{The gauge transformation law \eqref{fg00-flat-a} allows us to interpret $h_{\a(n)}$ as a conformal 
	spin-$\frac{n}{2}$ gauge field, while \eqref{fg00-flat-b} is compatible with the interpretation of 
	$y_{\a(n-4) }$ as a conformal compensator (following the modern supergravity terminology). Performing a duality transformation on 
	$y_{ \a(n-4) }$ is equivalent to the introduction of  an alternative conformal compensator.   }

Our dual formulation for \eqref{b00} is obtained by introducing the following first-order action 
\bea \label{b01}
S_{\text{first-order}} = \frac{\ri^n}{2^{\lfloor n/2 \rfloor +1}} \int \rd^{3}x \,  \Big\{ \cL \big( \pa_{\b\g} h_{\a(n)}, \cH_{\b \g; \, \a(n-4)} \big)
\non\\
-(n-3) \cH^{\b \g; \, \a(n-4)} F_{\b \g;\, \a(n-4)} \Big\}~, 
\eea
where the Lagrangian $\cL ( \pa_{\b\g} h_{\a(n)}, \cH_{\b \g; \, \a(n-4)} )$ explicitly reads
\bea
&& \cL \big( \pa_{\b\g} h_{\a(n)}, \cH_{\b \g; \, \a(n-4)} \big) = \hf \pa^{\b\g} h^{\a(n)} \pa_{\b\g} h_{\a(n)}-\frac{n}{4}\pa_{\b \g}h^{\b\g\a(n-2)}\pa^{\d\l}h_{\d\l\a(n-2)}~  \non\\
&&+\frac{n-3}{n} \Big( \frac{n}{2} \cH^{\b\g;\, \a(n-4)} \pa^{\d\l}h_{\b\g\d\l\a(n-4)}
- \cH^{\b\g;\, \a(n-4)}  \cH_{\b\g;\, \a(n-4)}
~ \non \\
&&+ A \, \cH^{\b\g;\, \d \l \a(n-6)} \cH_{\d\l;\, \b \g \a(n-6)} + B\, \cH_{\b \g;}\,^{\b \g \a(n-6)} \cH^{\d \l;}\,_{\d \l\a(n-6)}\Big )~.
\eea
Here the real coefficients $A$ and $B$ are constrained by
\bea
A+ B = -\frac{(n-4)(n-5)}{4(n-2)}~.
\label{B.17}
\eea
The field $\cH_{\b \g;\, \a(n-4)}$ in \eqref{b01}  is unconstrained, whilst the Lagrange multiplier $ F_{\b \g; \, \a(n-4)} $ is given by 
\bea
F_{\b \g; \, \a(n-4)} = \pa_{(\b}{}^{\d} \f_{\g) \d; \, \a(n-4)} \quad \implies \quad 
\pa^{\b \g}F_{\b \g; \, \a(n-4)}=0~,
\label{b02}
\eea
for some unconstrained field $\f_{\b \g; \, \a(n-4)}$. 

Varying the first-order action \eqref{b01} with respect to $\f_{\b \g;\, \a(n-4)}$ yields
\bea
\pa_{(\d}{}^{\b} \cH_{\g) \b; \, \a(n-4)} = 0
\quad \implies \quad \cH_{\b \g; \,\a(n-4)} =   \pa_{\b \g} y_{\a(n-4)}~,
\eea
thus $S_{\text{first-order}} $ reduces to the original  action \eqref{flat-n}.
On the other hand, we can integrate out
$\cH_{\b \g; \,\a(n-4)}$ via its equation motion, 
\bea
\frac{\pa}{\pa \cH^{\b \g; \,\a(n-4)}} \cL \big( \pa_{\d\l} h_{\a(n)}, \cH_{\d \l; \, \a(n-4)} \big) + F_{\b \g;\, \a(n-4)} = 0~,
\eea
which is equivalent to 
\bea
&&\cH_{\b \g; \, \a(n-4)}
-\frac{B}{2} \Big( \ve_{\b (\a_1} \ve_{|\g| \a_2} \cH^{\d \l;}\,_{\a_3 \dots \a_{n-4})\d \l}+ \ve_{\g (\a_1} \ve_{|\b| \a_2} \cH^{\d \l;}\,_{\a_3 \dots \a_{n-4})\d \l} \Big)\non\\
&&\qquad \quad - A \cH_{(\a_1 \a_2; \, \a_3 \dots \a_{n-4}) \b \g}
= \frac{n}{4} \pa^{\d \l} h_{\d \l \b \g \a(n-4)} - \frac{n}{2} F_{\b \g; \, \a(n-4)}~.
\eea
This equation allows us to express $\cH_{\b \g;\, \a(n-4)}$ in terms of $h_{\a(n)}$ and $F_{\b \g;\, \a(n-4)}$,
\bea
&&\cH_{\b \g; \a(n-4)} = \frac{n}{4(1-A)}\Big\{ \pa^{\d \l} h_{\d \l \b \g \a(n-4)} -2 F_{\b \g; \a(n-4)} \non\\
&&+ c_1 (A) \Big( \ve_{\b (\a_1} F^{\d}{}_{\a_2; \,\a_3 \dots \a_{n-4})\g\d} + \ve_{\g (\a_1} F^{\d}{}_{\a_2; \,\a_3 \dots \a_{n-4})\b\d} \Big) \non\\
&&+ c_2 (A) \Big( \ve_{\b (\a_1} \ve_{|\g| \a_2} F^{\d \l;}\,_{\a_3 \dots \a_{n-4})\d \l}+ \ve_{\g (\a_1} \ve_{|\b| \a_2} F^{\d \l;}\,_{\a_3 \dots \a_{n-4})\d \l} \Big) \Big\} ~,~~~~~~
\label{eomcH}
\eea
where we have defined
\begin{subequations}
	\bea
	c_1 (A) &=& \frac{2A(n-4)}{(n+2(A-2))}~,\\
	c_2 (A) &=& -\frac{(n-4) \Big( 2A \,\big(2A+B(n-3)\big)+ (n-4)(B-A)\big(n+2(A-2)\big) \Big)}{(n+2(A-2)) \Big((n-4)\big(n-5-B(n-3)\big)-2A \Big) }~. \hspace{1cm}
	\eea
\end{subequations}
Substituting \eqref{eomcH} into $S_{\text{first-order}}$ yields the following one-parameter family of dual actions
\bea
S_{\rm A}^{(\frac{n}{2})}[ h, \f]= \frac{\ri^n}{2^{\lfloor n/2 \rfloor +1}}  \int \rd^{3}x \, 
\cL_{\rm dual} \Big(  \pa_{\b \g} h_{\a(n)}, F_{\b \g;\, \a(n-4)}  \Big)~.
\label{b03}
\eea

The action \eqref{b01} is invariant under the gauge $\l$-transformations \eqref{fg00-flat-a}, which act on $\mathcal{H}_{\b \g; \, \a(n-4)}$ and $\f_{\b\g;\a(n-4)} $ as follows:
\begin{subequations}
	\bea
	\d_{\z} \mathcal{H}_{\b \g; \, \a(n-4)}&=&\frac{n-2}{n-1}\partial_{\b\g}\partial^{\d\rho}\z_{\d \rho \a(n-4)}~, \\
	\d_{\z} \f_{\b\g;\a(n-4)} &=& -\frac{2}{n(n-1)}  \Big( \big( 2n-5-2B(n-2)\big) \pa_{(\b}{}^{\d} \z_{\g) \d \a(n-4)} \non\\
	&& -\big( n-4-2B(n-2)\big) \partial_{(\a_1}{}^{\d} \z_{\a_2 \dots \a_{n-4}) \b \g \d } \Big) ~.
	\eea
	\label{b04}
\end{subequations}
The dual action \eqref{b03} is also invariant under  the following gauge $\rho$-transformations
\begin{subequations}
	\bea
	\d_{\rho} {\f}_{\g \d; \,\a(n-4)} &=& \pa_{\g \d} \rho_{\a(n-4)} 
	\quad \implies \quad \d_{\rho} {F}_{\b \g; \,\a(n-4)} = 0~,\\
	\d_{\rho} h_{\a(n-4)} &=& 0~.
	\eea
\end{subequations}

\subsection{New topologically massive theories}
The so-called `new topologically massive' (NTM) models were first formulated in $\mb{M}^3$ for fields with integer spin by Bergshoeff, Kovacevic, Rosseel, Townsend and Yin in \cite{BergshoeffKovacevicRosseelTownsendYin2011}. They were recently extended to the case of half-integer spin fields by Kuzenko and Ponds in \cite{KuzenkoPonds2018}.\footnote{Note that the spin-$3$ case was first treated in \cite{BergshoeffHohmTownsend2010}.} These NTM models are constructed purely in terms of the gauge prepotentials $h_{\a(n)}$  and the associated Cotton tensors ${C}_{\a(n)}(h)$.

Specifically, given an integer $n\geq 2$, the gauge-invariant NTM action for the field $h_{\a(n)}$ given in \cite{KuzenkoPonds2018}  is 
\begin{align}
S_{\text{NTM}}^{(n)}[h]=\frac{\text{i}^n}{2^{\lceil n/2 \rceil+1}}\frac{1}{m}\int\text{d}^3x\, h^{\a(n)}
\big(W-\s n m \big) {C}_{\a(n)}(h)
~, \label{TMHSNTMG}
\end{align}
where $m$ is some positive mass parameter and $\s:=\pm 1$. Making use of  the representation \eqref{TMCottonTensorsCasimir} leads to a manifestly gauge invariant and factorised form of the action \eqref{TMHSNTMG} \cite{HutchingsKuzenkoPonds2021}. 

The equation of motion obtained by varying \eqref{HSNTMG} with respect to the field $h_{\a(n)}$ is 
\begin{align}
0=\big(W - \s n m \big){C}_{\a(n)}(h)~. \label{TMEOM1}
\end{align}
By analysing \eqref{TMEOM1}, it can be shown that on-shell, the  action \eqref{TMHSNTMG} describes a propagating mode with pseudo-mass $m$, spin $\frac{n}{2}$ and helicity $\frac{\s n}{2}$ (cf. eq \eqref{TMOnshellPropTransverse}).

Recently, a new variant of the NTM model for bosonic fields in $\mb{M}^3$ was given in \cite{DalmaziSantos2021}. This model also does not require auxiliary fields, but is of order $2s-1$ in derivatives, as compared to those given in \cite{BergshoeffKovacevicRosseelTownsendYin2011} which are of order $2s$. 

Given an integer $s\geq 1$,  the actions of \cite{DalmaziSantos2021} are given by
\bea \label{TANNTM}
\widetilde{S}_{\text{NTM}}^{(2s)}[h] =\int \rd^3 x~h^{\a(2s)} \big ( W - \s n m \big ) F_{\a(2s)}(h)  ~,
\eea
where $m$ is a positive mass parameter, $\s:=\pm 1$, and  $F_{\a(2s)}(h)$ is the field strength,
\be
F_{\a(2s)}(h):=\Box^{s-1}  {\Pi}^{\perp}_{(2s)}h_{\a(2s)}~.
\ee
Due to the properties of ${\Pi}^{\perp}_{(2s)}$, the action \eqref{TANNTM} is manifestly gauge invariant and factorised.  The descendent $F_{\a(2s)}(h)$ may be obtained by stripping off a $W$ from ${C}_{\a(2s)}(h)$:
\begin{align}
C_{\a(2s)}(h)=\frac{1}{2s}WF_{\a(2s)}(h)~.
\end{align}
A similar construction does not appear to be possible in the fermionic case. 

The equation of motion obtained by varying \eqref{TANNTM} with respect to the field $h_{\a(2s)}$ is
\bea \label{TAEoM}
0 = (W - \s n m)F_{\a(2s)}(h)~.
\eea
By analysing \eqref{TAEoM}, it can be shown that on-shell, the model \eqref{TANNTM} has the same particle content as the NTM model \eqref{TANNTM}.

\begin{subappendices}
\section{Four-dimensional notation and identities} \label{FMAppendixA}
This appendix serves as a practical compendium on the $4d$ two-component spinor notation which is employed in this thesis. We follow the notation and conventions used in \cite{BuchbinderKuzenko1998}.\footnote{We direct the interested reader to \cite{BuchbinderKuzenko1998} for a pedagogical treatment.}

The double-covering group of $\mathsf{SO}_0 (3,1)$ in four dimensions is the spin group $\mathsf{SL}(2,\mb{C})$. The irreducible representations of $\mathsf{SL}(2,\mb{C})$ are labelled by $(\frac{m}{2},\frac{n}{2})$ for integers $m,n \geq 0$. These irreducible representations are naturally realised on the space of complex rank-$(m,n)$ two-component spinor fields $\F_{\a_1 \ldots \a_m \ad_1 \ldots \ad_n}$ 
\be
\F_{\a_1 \ldots \a_m \ad_1 \ldots \ad_n} = \F_{(\a_1 \ldots \a_m) (\ad_1 \ldots \ad_n)} \equiv \F_{\a(m)\ad(n)}~,
\ee
where the two-component spinor indices are independently symmetric in their dotted and undotted indices, respectively. The spinor indices are always denoted by Greek letters from the beginning of the alphabet, with $\a = 1,2$ and $\ad = \od , \td$.
The fields $\f_{\a(m)\ad(n)}$ will be called spinor fields of Lorentz type $(\frac{m}{2},\frac{n}{2})$ in this thesis.  

The $\mathsf{SL}(2,\mb{C})$ invariant tensors
\bsubeq
\begin{align}
\ve_{\a\b}=\left(\begin{array}{cc}0~&-1\\1~&0\end{array}\right)~,\qquad
\ve^{\a\b}=\left(\begin{array}{cc}0~&1\\-1~&0\end{array}\right)~,\qquad
\ve^{\a\g}\ve_{\g\b}=\d^\a_\b~,\\
\ve_{\ad\bd}=\left(\begin{array}{cc}0~&-1\\1~&0\end{array}\right)~,\qquad
\ve^{\ad\bd}=\left(\begin{array}{cc}0~&1\\-1~&0\end{array}\right)~,\qquad
\ve^{\ad\gd}\ve_{\gd\bd}=\d^{\ad}_{\bd}~,
\end{align}
\esubeq
are used to raise and lower the spinor indices:
\begin{align}
\F^{\a}=\ve^{\a\b}\F_{\b}~,\qquad \F_{\a}=\ve_{\a\b}\F^{\b}~,\qquad \F^{\ad}=\ve^{\ad\bd}\F_{\bd}~,\qquad \F_{\ad}=\ve_{\ad\bd}\F^{\bd}~.
\end{align}
Since the metric tensors $	\ve_{\a\b}$ and $\ve_{\ad\bd}$ are  antisymmetric, it follows that all spinor fields of Lorentz type $(\frac{m}{2},\frac{n}{2})$ are also traceless in their spinor indices.

Let us introduce the $4d$ sigma matrices which are defined as $\s_a \equiv (\s_a)_{\a\ad} = (\mathds{1},\vec{\s})$, where $\vec{\s}$ are the usual Pauli matrices. Explicitly, they read
\bsubeq
\begin{align}
\s_{0}=\left(\begin{array}{cc}1~&0\\0~&1\end{array}\right)~,\quad
\s_{1}=\left(\begin{array}{cc}0~&1\\1~&0\end{array}\right)~,\quad
\s_{2}=\left(\begin{array}{cc}0~&-\ri\\\ri~&0\end{array}\right)~,\quad
\s_{3}=\left(\begin{array}{cc}1~&0\\0~&-1\end{array}\right)~.\quad
\end{align}
\esubeq
The sigma matrices satisfy the following useful properties 
\bsubeq
\bea
(\s_{a} \tilde{\s}_{b} + \s_{b} \tilde{\s}_{a})_\a{}^\b = - 2 \eta_{ab}\d_\a{}^\b~, \qquad (\tilde{\s}_{a} {\s}_{b} + \tilde{\s}_{b} {\s}_{a})^\ad{}_\bd = - 2\eta_{ab}\d^\ad{}_\bd~, \\
\text{Tr}(\s_a \tilde{\s}_b) = (\s_a)_{\a \ad} (\tilde{\s}_b)^{\ad \a} = - 2 \eta_{ab}~, \qquad (\s^a)_{\a \ad} (\tilde{\s}_a)^{\bd \b} = -2 \d_\a{}^\b \d^\bd{}_\ad~,
\eea
\esubeq
where $\tilde{\s}_{a} \equiv (\tilde{\s}_{a})^{\ad\a}:=\ve^{\a\b}\ve^{\ad\bd}(\s_a)_{\b\bd}$.

A four-vector $V_{a}$ can be equivalently realised as a rank-$(1,1)$ spinor field $V_{\a \ad}$ by
\begin{align}
V_{\a\ad}=(\s^a)_{\a\ad}V_{a}~,\qquad V_{a}=-\frac{1}{2}(\tilde{\s}_a)^{\ad\a}V_{\a\ad}~.
\end{align}

Let us introduce the antisymmetric double sigma matrices
\begin{align}
\big(\s^{ab}\big)_{\a}{}^{\b}:=-\frac{1}{4}\big(\s^a\tilde{\s}^b-\s^b\tilde{\s}^a\big)_{\a}{}^{\b}~,\qquad \big(\tilde{\s}^{ab}\big)^{\ad}{}_{\bd}:=-\frac{1}{4}\big(\tilde{\s}^a\s^b-\tilde{\s}^b\s^a\big)^{\ad}{}_{\bd}~.
\end{align}
These matrices may be used to relate a real antisymmetric tensor $F_{ab}$ to a pair of symmetric two-spinors $\bar{F}_{\ad(2)}=\big(F_{\a(2)}\big)^*$ via
\begin{align}
F_{ab}=&~(\s_{ab})^{\a\b}F_{\a\b}-(\tilde{\s}_{ab})^{\ad\bd}\bar{F}_{\ad\bd}~,\non\\
F_{\a\b}=\frac{1}{2}(\s^{ab})_{\a\b}&F_{ab}~,\qquad \bar{F}_{\ad\bd}=-\frac{1}{2}(\tilde{\s}^{ab})_{\ad\bd}F_{ab}~.
\end{align}

The Lorentz generators $M_{\a\b} $ and ${\bar M}_{\ad \bd}$ 
act on two-component spinors as follows: 
\begin{subequations} 
	\bea
	M_{\a\b} \,\F_\g=
	\hf(\ve_{\g\a}\F_{\b}+\ve_{\g\b}\F_{\a})
	~,\quad&&\qquad M_{\a\b}\, {\bar \F}_{\gd}=0~,\\
	{\bar M}_{\ad\bd} \,{\bar \F}_{\gd}=
	\hf(\ve_{\gd\ad}{\bar \F}_{\bd}+\ve_{\gd\bd}{\bar \F}_{\ad})
	~,\quad&&\qquad {\bar M}_{\ad\bd}\, \F_{\g}=0~.
	\eea
\end{subequations} 
It can be shown that the following identities hold:
\begin{subequations}
	\bea
	M_{\a_1}{}^{\b}\F_{\b \a_2 ... \a_{m}\ad(n)} &=& - \hf (m+2)\F_{\a(m)\ad(n)}~,\\
	\bar{M}_{\ad_1}{}^{\bd}\F_{\a(m)\bd \ad_2 ... \ad_{n}} &=& - \hf (n+2)\F_{\a(m)\ad(n)}~, \\
	M^{\b\g}M_{\b\g}  \F_{\a(m)\ad(n)} &=& -\hf m(m+2) \F_{\a(m)\ad(n)}~, \\
	\bar{M}^{\bd\gd} \bar{M}_{\bd\gd} \F_{\a(m)\ad(n)} &=& - \hf n(n+2)\F_{\a(m)\ad(n)}~.
	\eea
\end{subequations}

In this thesis, there will always be an implicit symmetrisation over the indices of any tensor field which share the same Greek letter
\be \label{FMSym}
U_{\a(m)\ad(n)} V_{\a(p)\ad(q)} = U_{(\a_1 \ldots \a_m (\ad_1 \ldots \ad_n} V_{\a_{m+1} \ldots \a_{m+p})\ad_{n+1} \ldots \ad_{n+q})}~.
\ee

The (brackets) parentheses correspond to (anti-)symmetrisation of tensor or spinor indices, which encode a normalisation factor, for example
\bea
V_{[a_1 a_2 \dots a_n]} := \frac{1}{n!}\sum_{\pi \in S_n} \mbox{sgn}(\pi) V_{a_{\pi(1)} \dots a_{\pi(n)}}~, \quad V_{(\a_1 \dots \a_n )} := \frac{1}{n!} \sum_{\pi \in S_n} V_{\a_{\pi(1)} \dots \a_{\pi(n)}}~,~~~~~~
\eea
with $S_n$ being the symmetric group of $n$ elements.

\section{Three-dimensional notation and identities} \label{TMAppendixA}

We follow the notation and conventions adopted in
\cite{KLT-M11}. 
The  $\rm SL(2,{\mathbb R})$ invariant tensors 
\bea
\ve_{\a\b}=\left(\begin{array}{cc}0~&-1\\1~&0\end{array}\right)~,\qquad
\ve^{\a\b}=\left(\begin{array}{cc}0~&1\\-1~&0\end{array}\right)~,\qquad
\ve^{\a\g}\ve_{\g\b}=\d^\a_\b~,
\eea
are used to raise and lower spinor indices:
\bea
\psi^{\a}=\ve^{\a\b}\psi_\b~, \qquad \psi_{\a}=\ve_{\a\b}\psi^\b~.
\label{A2}
\eea

We make use of real Dirac gamma-matrices,  $\g_a := \big( (\g_a)_\a{}^\b \big)$, 
which are defined by
\be
(\g_a)_\a{}^\b := \ve^{\b\g}(\g_a)_{\a\g} = (\ri \s_2, \s_3,\s_1)~.
\ee

They have the following properties:
\bsubeq
\bea
\gamma_a \gamma_b &=& \eta_{ab}{\mathbbm{1}} + \varepsilon_{abc}
\gamma^c~,\\
(\gamma^a)_{\alpha\beta}(\gamma_a)^{\rho\sigma}
&=&-(\delta_\alpha^\rho\delta_\beta^\sigma
+\delta_\alpha^\sigma\delta_\beta^\rho)~, \\
\ve_{abc}(\g^b)_{\a\b}(\g^c)_{\g\d}&=&
\ve_{\g(\a}(\g_a)_{\b)\d}
+\ve_{\d(\a}(\g_a)_{\b)\g}
~,
\\
\text{tr}[\g_a\g_b\g_{c}\g_d]&=&
2\eta_{ab}\eta_{cd}
-2\eta_{ac}\eta_{db}
+2\eta_{ad}\eta_{bc}
~.
\eea
\esubeq

A three-vector $x_a$ can be equivalently realised as a symmetric second-rank spinor $x_{\a\b}$
defined as
\bea
x_{\a\b}:=(\g^a)_{\a\b}x_a=x_{\b\a}~,\qquad
x_a=-\hf(\g_a)^{\a\b}x_{\a\b}~.
\eea

The relationships between Lorentz generators with two vector indices ($M_{ab} =-M_{ba}$), one vector index ($M_a$)
and two spinor indices ($M_{\a\b} =M_{\b\a}$) are:
\bea
M_{ab} = -\ve_{abc}M^c~, \,\,\, M_a=\hf \ve_{abc}M^{bc}~, \,\,\, M_{\a\b}=(\g^a)_{\a\b}M_a~, \,\,\, M_{a}= -\hf (\g_a)^{\a \b} M_{\a \b}~.~~~~~
\eea
These generators 
act on a vector $V_c$ 
and a spinor $\J_\g$ by the rules
\bea
M_{ab}V_c=2\eta_{c[a}V_{b]}~, ~~~~~~
M_{\a\b}\J_{\g}
=\ve_{\g(\a}\J_{\b)}~.
\label{generators}
\eea

\end{subappendices}

\chapter{Three-dimensional $\cN$-extended Minkowski superspace} \label{ChapThreeDimensionalExtendedMinkowskiSuperspace}
In this chapter we derive the transverse superspin projection operators in three-dimensional
$\cN$-extended Minkowski superspace $\mb{M}^{3|2 \mc{N}}$, for $1 \leq \cN \leq 6$. These operators are supersymmetric generalisations of the spin projectors in $\mb{M}^3$, which were reviewed in section \ref{TMSectionSPO}. The defining characteristic of a superspin projection operator in $\mb{M}^{3|2 \mc{N}}$ is that it extracts out the irreducible component of a superfield with maximal superspin. These superprojectors have numerous applications. Most importantly, we will see that these operators are vital in the construction of linearised higher-spin super-Cotton tensors.

These linearised super-Cotton tensors are the main ingredients in the off-shell formulation of linearised $\cN$-extended (higher-spin) conformal supergravity. For $1 \leq \mc{N} \leq 6$, the complete non-linear
actions for $\mc{N}$-extended conformal supergravity were derived in \cite{BKNT-M2,KuzenkoNovakTartaglino-Mazzucchelli2014} using the off-shell
formulation for $\mc{N}$-extended conformal supergravity  developed in \cite{BKNT-M1}.\footnote{The
	$\cN=1$ and $\cN=2$ conformal supergravity theories were constructed for the first time by  van Nieuwenhuizen \cite{vN} and  
	Ro\v{c}ek and  van Nieuwenhuizen  \cite{RvN}, respectively. 
	The off-shell action for
	$\cN=6$ conformal supergravity was independently derived by 
	Nishimura and Tanii \cite{NT}.
	On-shell  formulations for $\cN$-extended conformal supergravity with  $\cN>2$
	were given in \cite{LR89,NG}.} Since the complete nonlinear actions are known, it is natural to ask the following question: What is the
point of constructing linearised conformal supergravity actions? The answer is that the
supergravity actions proposed  in \cite{BKNT-M2} are realised using certain closed super three-forms, which are themselves constructed in terms of the constrained geometry of $\mc{N}$-extended conformal superspace \cite{BKNT-M1}. However, it may be shown that these constraints can be solved in terms of unconstrained gauge prepotentials $H$ (with indices suppressed). Thus, the action for conformal supergravity, $S_{\text{CSG}}$, may be reformulated in terms of $H$, $S_{\text{CSG}} = S_{\text{CSG}}[H]$.  In order to determine the structure of $S_{\text{CSG}}$, it is necessary to construct a linearised conformal supergravity action. One of the aims of this chapter is to construct these actions, and their higher-spin extensions.

This chapter is based on the publication \cite{BHHK} and is organised as follows. In section \ref{TMSIrreducibleRepsPo} we review salient facts concerning massive UIRs of the three-dimensional super-\Po algebra. In section \ref{TMSIrreducibleFieldRepresentations} we discuss how these massive UIRs can be realised on the space of real superfields in $\mb{M}^{3|2 \mc{N}}$.  The construction of the superspin projection operators in $\mb{M}^{3|2 \mc{N}}$, for the cases $1 \leq \cN \leq 6$, is completed in section \ref{TMSSecSpinProjectors}.  The transverse superprojectors are then utilised in section \ref{TMSSecCHS} to construct the linearised higher-spin super-Cotton tensors, and subsequently linearised actions for $\cN$-extended superconformal gravity, in terms of unconstrained gauge prepotentials. In section \ref{TMSSecMassless} we review massless (half-)integer superspin theories in $\mb{M}^{3|2}$, which were first derived by Kuzenko and Tsulaia \cite{KuzenkoTsulaia2017}. The explicit computation of the linearised super-Cotton tensors enables the derivation of  new topologically massive actions. These models are investigated  in section \ref{TMSSecMassive}. A summary of the results obtained is given in section \ref{TMSConclusion}. The chapter is accompanied by a single technical appendix \ref{TASappendixSCHSPrimary} which contains technical details concerning the $\cN$-extended superconformal group.

\section{Irreducible representations of the super-Poincar\'e algebra} \label{TMSIrreducibleRepsPo}

The three-dimensional $\cN$-extended super-\Po algebra is spanned by the \Po generators $P_{\a \b}$ and $J_{\a \b}$ and the supersymmetry generators $Q^I_\a$. Here the $R$-symmetry index takes $\cN$ values, $I = \1, \2, \cdots , \underline{\cN}$.\footnote{Since the $R$-symmetry group is $\mathsf{SO}(\cN)$, the corresponding indices are raised and lowered using the Kronecker delta. Hence we do not distinguish between upper and lower $\mathsf{SO}(\cN)$ indices.} The generators of the $\cN$-extended super-\Po algebra satisfy the (anti-)commutation relations
\bsubeq \
\begin{align}
[ P_{\a\b} , P_{\g \d} ] &= 0~,  &  \ [ J_{\a\b} , P_{\g \d} ] &= \ri \ve_{\g(\a} P_{\b)\d} + \ri \ve_{\d(\a} P_{\b)\g}~, \\
\ [J_{\a\b},J_{\g\d}] &=\ri \ve_{\g(\a} J_{\b)\d} + \ri \ve_{\d(\a} J_{\b)\g}~, & \ [ P_{\a\b}, Q^I_\g ] &= 0~, \\
\ [ J_{\a\b} , Q^I_\g ] &= \ri \ve_{\g(\a} Q^I_{\b)} ~, & \{  Q^I_\a ,Q^J_\b \} &=2 \d^{IJ} P_{\a\b}~.
\end{align}
\esubeq

The $\cN$-extended super-\Po algebra has two quadratic Casimir operators, which we label $C_i$ for $i=1,2$. These operators take the form\cite{MT}
\bsubeq
\be
C_1 := - P^a P_a = \hf P^{\a \b}P_{\a \b}~,  \qquad  C_2 :=   2W -\frac{\ri}{2} Q^{\a I} Q_\a^I ~,  \label{TMCasimiroperators}
\ee
\be
[C_i,P_{\a\b}] = [C_i, J_{\a \b}] = [C_i, Q^I_\a] = 0~,
\ee
\esubeq
where $C_2$ is the supersymmetric extension of the Pauli-Lubanski scalar $W$ \eqref{TMPauliLubanksiScalar}. 
Note that the Casimir operator $C_2$ is commonly referred to as the superhelicity operator and is often denoted $Z \equiv C_2$.

The massive UIRs of the three-dimensional $\cN$-extended super-\Po algebra are classified by the quantum numbers mass $m$ and superhelicity $\k$. Given a massive UIR, these quantum numbers are defined by the eigenvalues of the Casimir operators \eqref{TMCasimiroperators} \cite{MT}
\bea \label{TMSMassiveIrrepCasimir}
C_1 = -m^2 {\mathbbm 1} ~, \qquad C_2= 4 m  \k {\mathbbm 1}~.
\eea
Note that the eigenvalue of $C_2$ in a massive UIR has been adjusted by a factor of four, which is not common practice in the literature. This factor was introduced in order for our definition of superhelicity to match that given in \cite{MT}.
We will denote  by $\mb{D}(m, \k)$ a massive UIR carrying strictly positive mass $m > 0$ and superhelicity $\k$. In principle, the superhelicity can be any real number \cite{MT}. However, we will only be interested in particular values for $\k$ which ensure that the UIRs of $\mf{iso}(2,1)$ encoded in $\mb{D}(m, \k)$ carry (half-)integer helicity. We recall that these were the types of massive UIRs of $\mf{iso}(2,1)$ we were interested in section \ref{Irrepsof3dPoincare}.  

\section{Irreducible superfield representations} \label{TMSIrreducibleFieldRepresentations}

Three-dimensional $\cN$-extended Minkowski superspace $\mb{M}^{3|2\cN}$ is parametrised by the real coordinates $z^A= (x^a, \theta^{\alpha}_I)$\cite{KPT-MvU2011}.\footnote{We make use of the notations and conventions adopted in \cite{KPT-MvU2011}.}
The covariant derivatives $D_A = ( \pa_{\a \b}, D^I_\a )$ obey the (anti-)commutation relations 
\bea
[\pa_{\a\b} , \pa_{\g\d}] = 0~, \qquad [\pa_{\a\b} , D^{I}_{\g}] = 0~, \qquad \{ D^{I}_{\alpha}, D^{J}_{\beta}\} =2 \ri\, \d^{IJ} \partial_{\alpha \beta}~.
\label{TMSNDerivativeAlgebra}
\eea
The relations \eqref{TMSNDerivativeAlgebra} lead to the following useful identities
\bsubeq\label{TMSDerivativeIdentities}
\bea
D^{I}_\a  D^{I}_\b  &=& \ri \pa_{\a \b} +\frac{1}{2}\ve_{\a \b}  (D^{I} )^2~, \label{TMSDerivativeIdentities1}\\
D^{\a I} D^{I}_\b D^{I}_\a &=& 0 \quad \implies \quad [D^{I}_{\a} D^{I}_{\b}, D^{I}_{\g} D^{I}_{\d}]=0~, \label{TMSDerivativeIdentities2} \\
D^{\b I} D^I_\b D^{I}_{\a} &=& -D^{I}_{\a} D^{\b I} D^I_\b = 2\ri \pa_{\a\b}D^{\b I}~, \label{TMSDerivativeIdentities3}\\
D^{\b I} D^I_\b D^{\g I} D^I_\g &=& -4 \Box ~. \label{TMSDerivativeIdentities4}
\eea
\esubeq
It must be emphasised that the Einstein summation convention was not employed over the $R$-symmetry indices in the above identities \eqref{TMSDerivativeIdentities}. This will be the only instance where this occurs.

For superfield representations of  the $\cN$-extended super-\Po group, the  infinitesimal super-Poincar\'e transformation of a tensor superfield $\F$ (with indices suppressed) is given by 
\bea
\d \F =
\ri \Big(\hf b^{\a\b} P_{\a\b}  +\hf \o^{\a \b} J_{\a \b} + \ri \e^{\a I} Q^I_\a \Big)\F~,
\eea
where the generators of the three-dimensional $\cN$-extended super-\Po group take the following form 
\bsubeq \label{TMSFieldGenerators}
\bea
P_{\a\b}&=& -\ri \pa_{\a\b} ~,\qquad \pa_{\a\b}= (\g^a)_{\a\b} \pa_a~, \\
J_{\a\b} &=& -\ri x^\g{}_{(\a} \pa_{\b)\g } 
+ \ri \q^I_{(\a} \pa^I_{\b)} 
-\ri M_{\a\b}~,\\
Q^I_\a &=& \pa^I_\a -\ri \q^{\b I} \pa_{\a\b} ~,\qquad \pa^I_\a =\frac{\pa}{\pa \q^{\a}_I}~.
\eea
\esubeq

For completeness, we also give the explicit form of the spinor covariant derivative  \cite{KPT-MvU2011}
\bea
D^I_\a = \pa^I_\a +\ri \q^{\b I} \pa_{\a\b} ~.
\eea
The spinor covariant derivative $D^I_\a$ possesses unusual complex conjugation properties which differ to those in four dimensions. In particular, given an arbitrary superfield $\F$ (with indices suppressed), its complex conjugate is defined via the rule 
\be
(D^I_\a \F)^* = - (-1)^{\e (\F)}D^I_\a \bar{\F}~,
\ee
where $(\F)^{*}:= \bar{\F}$ and $\e (\F)$ denotes the Grassmann parity of $\F$.

In the superfield representation, it follows from \eqref{TMSFieldGenerators} that the superhelicity operator can be written in the manifestly supersymmetric form
\bea \label{TMSSuperhelicityCasimir}
Z = 2\pa^{\a \b} M_{\a\b} + \cN \D~,
\qquad [  Z , D^I_\a ]=0~,
\eea
where the scalar operator $\D$ is defined by 
\bea \label{TMDeltaoperator}
\D := -\frac{\ri}{2\cN} D^{\b I} D^I_{\b } ~.
\eea
The operator \eqref{TMDeltaoperator} possesses the important property
\be
\big[\Delta \,, D^{\b I}D^{I}_{\a} \big] = 0~.
\label{TMSDeltaOperator}
\ee
The normalisation factor of $Z$, which differs to that given in \cite{KuzenkoTsulaia2017} by a factor of four, was chosen to ensure that the AdS$^{3|2}$ analogue of $C_2$, which is $\mb{C}_2$ \eqref{FQC}, reduces to $C_2$  in the flat superspace limit.

Let $\mb{V}^{[\cN]}_{(n)}$ denote the space of totally symmetric real rank-$n$ superfields $ \F_{\a(n)} = \bar{\F}_{\a(n)}$ on $\mb{M}^{3|2\cN}$. The two Casimir operators \eqref{TMCasimiroperators} are related to each other on $\mb{V}^{[\cN]}_{(n)}$ as follows
\bea \label{TMSCasimirRelation}
Z^2 \F_{\a(n)} &=& 4n^2\Box \F_{\a(n)} + \cN^2 \D^2 \F_{\a(n)} + 4n \cN \D \pa_\a{}^\b \F_{\b\a(n-1)} \non \\
&&+ 4n(n-1) \pa_{\a(2)}\pa^{\b(2)} \F_{\b(2)\a(n-2)} ~.
\eea

For integers $n>1$, a superfield $\F^{\perp}_{\a(n)} $ on $\mb{V}^{[\cN]}_{(n)}$  is called
transverse (or divergenceless) if it obeys the constraint
\bea
D^{\b I} \F^{\perp}_{\b \a(n-1)} =0 ~.
\label{TAMTransverseSuperfield}
\eea
The transverse condition \eqref{TAMTransverseSuperfield} implies 
that $\F^{\perp}_{\a(n)}$ is transverse in the usual sense, that is $\pa^{\b(2)} \F^{\perp}_{\b(2) \a(n-2)} =0$. 
On the space of transverse superfields $\F^{\perp}_{\a{(n)}}$, the scalar operator $\D$ can be shown to possess the properties
\bsubeq
\bea
D^{\b I} \D \F^{\perp}_{\b\a(n-1)} &=& 0~,  \label{TMSDeltaProperty2}\\
\D^2 \F^{\perp}_{\a(n)} &=& \Box \F^{\perp}_{\a(n)} ~. \label{TMSDeltaProperty3}
\eea
\esubeq
The relation \eqref{TMSDeltaProperty2} states that the $\D$ operator preserves the transverse nature of $\F^{\perp}_{\a(n)}$.

\subsection{Massive superfields} \label{TMSMassivefields}

For integer $n>0$, a superfield $\F_{\a(n)}$ on $\mb{V}^{[\cN]}_{(n)}$  is said to be an on-shell massive supermultiplet if it satisfies the conditions\footnote{This definition generalises those given earlier in the $\cN=1$ and $\cN=2$ cases \cite{KuzenkoTsulaia2017,KuzenkoOgburn2016,KuzenkoNovakTartaglino-Mazzucchelli2015}.}
\begin{subequations} \label{TMSMassiveOnshell}
	\bea
	D^{\b I} \F_{\b \a(n-1)} &=& 0 ~ , \label{TMSMassiveTransverse} \\
	\D  \F_{\a(n)} &=& m \s \F_{\a(n)}~, \qquad \s =\pm 1 ~,\label{TMSMassiveMassShell}
	\eea
\end{subequations}
where $m > 0$ is a real mass.
Making use of the condition \eqref{TMSMassiveTransverse}, it can be shown that the first-order constraint \eqref{TMSMassiveMassShell} can be expressed in the equivalent form 
\be \label{TMSSuperhelicity}
\big ( Z - 4 m \k \big ) \F_{\a(n)} = 0 ~, \qquad \k = \frac{\s}{2} \Big ( n + \frac{\cN}{2} \Big ) ~.
\ee
Any field $\F_{\a(n)}$ satisfying both constraints \eqref{TMSMassiveTransverse} and \eqref{TMSMassiveMassShell} is an eigenvector of the Casimir operator $\Box$
\be \label{TMSKG}
\big (\Box -m^2 \big )\F_{\a(n)} = 0~.
\ee
In accordance with \eqref{TMSMassiveIrrepCasimir}, it follows from \eqref{TMSSuperhelicity} and \eqref{TMSKG} that the on-shell superfield \eqref{TMSMassiveOnshell} furnishes the massive UIR $\mb{D}(m, \k)$. Such massive supermultiplets are said to carry mass $m$, superhelicity $\k = \frac{\s}{2}(n+\frac{\cN}{2})$ and superspin $|s| :=\frac{n}{2}$. Note that in the superfield representation we choose to label the massive UIR by $\mb{D}(m: \s , |s| , \cN)$.\footnote{The notation  {\sloppy $\mb{D}(m : \s , |s| , \cN)$} is not conventional, since the UIR is not labelled strictly in terms of mass $m$ and superhelicity $\k$. However, the parameters $\s$, $|s|$ and  $\cN$ prove sufficient  in determining $\k$ exactly (cf. \eqref{TMSSuperhelicity}).   }

Studying the $\cN=0$ component structure of a massive superfield $\F_{\a(n)}$, it follows from the constraints \eqref{TMSMassiveOnshell} that $\F_{\a(n)}$ describes the independent component fields
\bea \label{TMSMassiveComp}
\f^{I_1 \dots I_k}_{\a_1 \dots \a_{n+k} } (x) :&=& 
\ri^{ nk + \hf k(k+1) }D^{[I_1}_{\a_1} \dots D^{I_k]}_{\a_k} 
\F_{\a_{k+1} \dots \a_{n+k} } \Big|_{\q=0}~, \qquad 0 \leq k \leq \cN~.
\eea
Here we have made use of the $3d$ analogue of bar projection $ V|:= V(x,\q)|_{\q=0}$, in which all the Grassmann coordinates $\q_a^I$ are set to zero.
Each of the component fields \eqref{TMSMassiveComp} are completely symmetric in their spinor indices
$\f^{I_1 \dots I_k}_{\a_1 \dots \a_{n+k} }(x) =\f^{I_1 \dots I_k}_{(\a_1 \dots \a_{n+k} )}(x) $,
and are also transverse
\bea  \label{TMSMassiveCompTrans}
\pa^{\b\g} \f^{I_1 \dots I_k}_{\b\g \a(n+k-2) }(x) =0~, \qquad n+k>1~.
\eea

Recall that in order to elucidate the helicity content of $\f^{I_1 \dots I_k}_{\a(n+k) }$, it is necessary to compute the eigenvalues of the helicity operator $W$ \eqref{TMPauliLubanksiCasimir} on $\f^{I_1 \dots I_k}_{\a(n+k) }$.
Given $\F^{I_1 \ldots I_K}_{\a(n+k)} $ satisfies the massive on-shell conditions \eqref{TMSMassiveOnshell}, one can show that $\f^{I_1 \dots I_k}_{\a(n+k) }$ obeys 
\begin{align} \label{TMSMassiveCompMassCondition}
\big ( W - \s(n+k) m \big )\f^{I_1 \dots I_k}_{\a(n+k) }(x)  =0 ~.
\end{align}
In accordance with the on-shell conditions \eqref{TMOnshellField} on $\mb{M}^3$, it follows from \eqref{TMSMassiveCompTrans} and \eqref{TMSMassiveCompMassCondition} that the component field $\f^{I_1 \dots I_k}_{\a(n+k) }(x)$ is a massive field carrying  helicity $\frac{\s}{2} ( n + k )$. As expected, a massive supermultiplet describes a multiplet of massive fields.

In place of the on-shell conditions \eqref{TMSMassiveOnshell}, one may instead consider a superfield $\F_{\a(n)}$ on $\mb{V}^{[\cN]}_{(n)}$  which satisfies the equations \eqref{TMSMassiveTransverse} and \eqref{TMSKG}
\bsubeq \label{TMSMassiveOnshell2}
\bea
D^{\b I} \F_{\b \a(n-1)} &=& 0 ~ , \label{TMSMassiveTransverse2} \\
\big ( \Box - m^2 \big ) \F_{\a(n)} &=& 0~.\label{TMSMassiveMassShell2}
\eea
\esubeq
In this case, the relation \eqref{TMSCasimirRelation} can be expressed in the form
\begin{align}
\big ( Z + 4 m \k \big ) \big ( Z - 4 m \k \big ) \F_{\a(n)} = 0~, \qquad \k= \frac{1}{2}\Big (n+\frac{\cN}{2} \Big )~.
\end{align}
It follows that such a field $\F_{\a(n)}$ realises the reducible representation
\begin{align} \label{TMSMassiveReducibleRep}
\mb{D} \Big (m:-,\frac{n}{2}, \cN \Big ) \oplus \mb{D}\Big (m:+,\frac{n}{2}, \cN  \Big )~,
\end{align}
in which both signs of the superhelicity $\pm \frac{1}{2}\big (n+\frac{\cN}{2} \big )$ are present.


\subsection{Massless superfields} \label{TMSMasslessSupermultiplets}
As is well known, any massless superfield $\F_{\a(n)}$ in $\mb{M}^{3|2\cN}$, with rank $n>2$, carries no propagating degrees of freedom, and thus the notion of superspin is purely kinematical. Thus when speaking of a massless higher-superspin model in $\mb{M}^{3|2\cN}$, we will refer to the kinematic structure of the field variables, their gauge transformation laws and the gauge-invariant action which ensure that the theory describes no propagating degrees of freedom on the equations of motion.
Let us elaborate on this point in the context of $\mb{M}^{3|2}$. 

Given an integer $n \geq 1$, we say that a superfield $\F_{\a(n)}$ on $\mb{V}^{[1]}_{(n)}$ is massless if it satisfies the constraints\footnote{Similar to the story in $\mb{M}^3$, there exists an alternative set of conditions which define a massless higher-spin superfield in  $\mb{M}^{3|2}$.  These constraints were first given in \cite{KuzenkoTsulaia2017}, where they were shown to describe the superfield content of the massless half-integer superspin theory \eqref{TMSSecondOrderMassless}.}
\bsubeq \label{TMSMasslessOnshellConditions1}
\bea
D^{\b } \F_{\b \a(n-1)} &=& 0~, \label{TMSMasslessConstraint1}\\
\D \F_{\a(n)} &=& 0~. \label{TMSMasslessConstraint2}
\eea
It can be shown that system of equations \eqref{TMSMasslessConstraint1} and \eqref{TMSMasslessConstraint2} are compatible with the gauge symmetry
\be \label{TMSMasslessGaugeSymmetry}
\d_\z \F_{\a(n)} = \ri^{n} D_{\a} \z_{\a(n-1)}~,
\ee
given that the real gauge parameter $\z_{\a(n-1)}$ is also on-shell
\label{TMSMasslessGaugeParameterConstraints1}
\bea \label{TMSMasslessGaugeParameter1}
D^{\b } \z_{\b \a(n-2)} &=& 0~, \\
\D \z_{\a(n-1)} &=& 0~. \label{TMSMasslessGaugeParameter2}
\eea
\esubeq
Note that the massless constraint \eqref{TMSMasslessConstraint2} is obtained from the massive on-shell condition \eqref{TMSMassiveMassShell} by taking the massless limit $m \rightarrow 0$. The conditions \eqref{TMSMasslessConstraint1} and \eqref{TMSMasslessConstraint2} lead to the massless Klein-Gordon equation $\Box \F_{\a(n)} = 0$. 
We will see in section \ref{TMSSecMassless} that the massless superfields \eqref{TMSMasslessOnshellConditions1}, for the case $n \geq 4$, describe the superfield content of the massless integer superspin theory \eqref{TASFirstOrderAction}.  We say that a massless superfield $\F_{\a(n)}$ carries superspin $s:=\frac{n}{2}$.

In order to show that a superfield in $\mb{M}^{3|2}$ is massless, it suffices to show that at the component level, the fields surviving upon implementing an appropriate Wess-Zumino gauge do not carry any propagating degrees of freedom. Let us show that this is indeed true for the  massless superfield \eqref{TMSMasslessOnshellConditions1}. 

A superfield $\F_{\a(n)}$ satisfying the conditions \eqref{TMSMasslessOnshellConditions1} can be shown to describe two independent component fields, which we choose to define as
\be
A_{\a(n)} := \F_{\a(n)}|~, \qquad B_{\a(n+1)} :=\ri^{n+1} D_{\a}\F_{\a(n)}|~.
\ee
The gauge freedom \eqref{TMSMasslessGaugeSymmetry} allows us to implement the following Wess-Zumino gauge
\be \label{TMSWZGauge}
\F_{\a(n)}| =0~.
\ee
In this gauge, the component field $A_{\a(n)}$ vanishes, leaving $B_{\a(n+1)} $ as the only remaining field. The residual gauge freedom is characterised by the condition 
\be \label{TMSWZGaugeResidualgauge}
D_\a \z_{\a(n-1)}| = 0~.
\ee
It  follows from   \eqref{TMSMasslessGaugeParameterConstraints1} and \eqref{TMSWZGaugeResidualgauge}  that there is only one independent gauge parameter at the component level,  which we choose to define as
\be \label{TMSIndependetGaugeparameter}
\x_{\a(n-1)}: = -(-1)^n \z_{\a(n-1)}|~.
\ee
The sole surviving field $B_{\a(n+1)} $ can be shown to satisfy the conditions
\be \label{TMSComponentMassless}
\pa^{\b(2)}B_{\b(2) \a(n-1)} = 0~, \qquad W B_{\a(n+1)} = 0 \quad \Longrightarrow \quad  \Box B_{\a(n+1)} =0~,
\ee
as a consequence of the conditions \eqref{TMSMasslessConstraint1} and \eqref{TMSMasslessConstraint2}.
In accordance with \eqref{TMSMasslessGaugeSymmetry} and \eqref{TMSIndependetGaugeparameter}, it can be shown that the gauge transformation associated with $B_{\a(n+1)}$ is given by
\be \label{TMSComponentMasslessGS}
\d_\x B_{\a(n+1)} = \pa_{\a(2)}\x_{\a(n-1)}~,
\ee
where the gauge parameter $\x_{\a(n-1)}$ obeys the conditions 
\be \label{TMSComponentMasslessGP}
\pa^{\b(2)}\x_{\b (2) \a(n-3)} = 0~, \qquad W \x_{\a(n-1)}  = 0 \quad \Longrightarrow \quad  \Box \x_{\a(n-1)} =0~.
\ee
Here, we have made use of the results \eqref{TMSMasslessGaugeParameter1}, \eqref{TMSMasslessGaugeParameter2}, \eqref{TMSWZGaugeResidualgauge} and \eqref{TMSIndependetGaugeparameter}. It follows from  \eqref{TMSComponentMassless}, \eqref{TMSComponentMasslessGS} and \eqref{TMSComponentMasslessGP} that  $B_{\a(n+1)}$ describes a massless field (cf. \eqref{TMMasslessFieldW}), which was shown to be pure gauge in section \ref{TMMasslessFieldsRepsSec}. Hence a superfield satisfying the conditions \eqref{TMSMasslessOnshellConditions1} is indeed massless, since it describes a single massless field at the component level which propagates no physical degrees of freedom.

\section{Superspin projection operators} \label{TMSSecSpinProjectors}

In this section we compute the superspin projection operators on $\mb{M}^{3|2\cN}$ for $1\leq \cN \leq 6$.\footnote{Note that there is nothing preventing the computation of the superspin projection operators on $\mb{M}^{3|2\cN}$ for generic $\cN$. However, they must computed on a case by case basis for each $\cN$.} The rank-$n$ superspin projection operator $\bm{\P}^{\perp}_{(n)}$ is defined by its action on $\mb{V}^{[\cN]}_{(n)}$ according to the rule
\bsubeq \label{TMSProjectionOperator}
\bea
\bm{\bm{\P}}^{\perp}_{(n)}: \mb{V}^{[\cN]}_{(n)} &\longrightarrow& \mb{V}^{[\cN]}_{(n)}~, \\
\F_{\a(n)} &\longmapsto&  \bm{\P}^{\perp}_{(n)} \F_{\a(n)} =: \bm{\P}^{\perp}_{\a(n)}(\f)~.
\eea
\esubeq
For fixed integer $n \geq 1$,  the differential operator $\bm{\P}^{\perp}_{(n)}$ satisfies the properties:
\bsubeq \label{TMSProjectorProperties}
\begin{enumerate}
	\item \textbf{Idempotence:} The operator $\bm{\P}^{\perp}_{(n)}$ squares to itself
	\be
	\bm{\P}^{\perp}_{(n)}\bm{\P}^{\perp}_{(n)} = \bm{\P}^{\perp}_{(n)} ~. \label{TMSGeneralProjectorIdempotent}
	\ee
	\item \textbf{Transversality:} The operator $\bm{\P}^{\perp}_{(n)}$ maps $\F_{\a(n)}$ to a transverse superfield,
	\be
	D^{\b I}\bm{\P}^{\perp}_{\b \a(n-1)}(\F) = 0~. \label{TMSGeneralProjectorTransverse}
	\ee
	\item \textbf{Surjectivity:} Every transverse superfield $\F^{\perp}_{\a(n)}$ belongs to the image of $\bm{\P}^{\perp}_{(n)}$ 
	\be 
	D^{\b I} \F^{\perp}_{\b \a(n-1)} = 0 \quad \Longrightarrow \quad \bm{\P}^{\perp}_{(n)} \F^{\perp}_{\a(n)} = \F^{\perp}_{\a(n)}~. \label{TMSGeneralProjectorSurjective}
	\ee
\end{enumerate}
\esubeq
By construction, the superspin projection operator $\bm{\P}^{\perp}_{(n)}$ maps any superfield $\F_{\a(n)}$ on $\mb{V}^{[\cN]}_{(n)}$, which is supplemented with the first-order differential constraint \eqref{TMSMassiveMassShell}, to a massive field \eqref{TMSMassiveOnshell} which furnishes the UIR $\mb{D}(m:\s , \frac{n}{2}, \cN)$.

For arbitrary supersymmetry type $\cN$, the superspin projection operator $\bm{\P}^{\perp}_{(n)}$ acts on $\mb{V}^{[\cN]}_{(n)}$ according to the rule
\be \label{TMSGeneralProjector}
\bm{\P}^{\perp}_{(n)} \F_{\a(n)}:= \bm{\P}^{\b_1}{}_{(\a_1}{} \cdots \bm{\P}^{\b_n}{}_{\a_n)} \F_{\b(n)}~,
\ee
where $\bm{\P}^{\b}{}_{\a}$ is a differential operator which should possess the general properties 
\bsubeq\label{TMSProjectorPropertiesIndividual}
\bea
D^{\b I} \bm{\P}^{\a}{}_{\b} &=& 0~,  \label{5.2a}\\
\bm{\P}^{\b}{}_{\a} D^{I}_{\b}&=& 0~, \label{5.2b}\\
\bm{\P}^{\b}{}_{\a}\bm{\P}^{\g}{}_{\b} &=& \bm{\P}^{\g}{}_{\a}~, \label{5.2c} \\
\big[ \bm{\P}^{\b}{}_{\a}  , \bm{\P}^{\d}{}_{\g} \big]&=&0~. \label{5.2d} 
\eea
\esubeq

We propose the ansatz for the general structure of $\bm{\P}^\b{}_\a$ to be
\be \label{TMSSpinorProjector}
\bm{\P}^\b{}_\a = D^{\b I} D^I_\a F(\D,\Box)~,
\ee
where $F(\D,\Box)$ is some function which needs to be computed independently for each $\cN$. The ansatz \eqref{TMSSpinorProjector} was chosen on account that it automatically satisfies property \eqref{5.2d}, as a consequence of the identities 
\be
D^{\b I} D^I_\a = \ri \cN \big ( \pa_\a{}^\b + \D \d_\a{}^\b  \big )~,
\ee
and  \eqref{TMSDeltaOperator}.
We will see in the subsequent sections that the conditions \eqref{TMSProjectorPropertiesIndividual} suffice in determining $\bm{\P}^\b{}_\a$ uniquely for each $\cN$. Furthermore, each of these $\bm{\P}^\b{}_\a$ can be shown to satisfy the additional property
\bea \label{TMSTransverseProj}
D^{\b I} \F^{\perp}_\b=0 \quad \implies \quad \bm{\P}^{\b}{}_{\a} \F^{\perp}_{\b} =\F^{\perp}_\a~.
\eea

It follows from \eqref{TMSProjectorPropertiesIndividual} and \eqref{TMSTransverseProj} that the operator $\bm{\P}^{\b}{}_{\a}$ is a superspin projection operator \eqref{TMSProjectorProperties} on the space of spinor superfields $\mb{V}^{[\cN]}_{(1)}$. Given that $\bm{\P}^{\b}{}_{\a}$ is a superspin projection operator, it follows that its higher-rank generalisation $\bm{\P}^{\perp}_{(n)}$ also satisfies the defining conditions \eqref{TMSProjectorProperties} of a superspin projection operator. These properties are immediately inherited from $\bm{\P}^{\b}{}_{\a}$. Thus it suffices to only compute $\bm{\P}^{\b}{}_{\a}$ for each $\cN$, since these operators are the building blocks of the higher-rank superprojectors via \eqref{TMSGeneralProjector}. The explicit form of the spinor superprojector for $1 \leq \cN \leq 6$, and their higher-rank extensions, will be presented in sections \ref{TMSN1SUSY}-\ref{TMSN6SUSY} for each supersymmetry type, respectively.
Before giving the explicit form of these superprojectors, it is worth taking a moment to detail some of their properties which hold for generic $\cN$.

\subsubsection{Orthogonal complement of $\bm{\P}^{\perp}_{(n)}$ }
The orthogonal complement $\bm{\P}^{\parallel}_{(n)}$ of $\bm{\P}^{\perp}_{(n)}$  is defined to act on $\mb{V}^{[\cN]}_{(n)}$ via the rule
\be \label{TMSLongitudinalProjector}
\bm{\P}^{\parallel}_{(n)} \F_{\a(n)} = \big ( \mds{1} - \bm{\P}^{\perp}_{(n)} \big )\F_{\a(n)}~.
\ee
By construction, $\bm{\P}^{\perp}_{(n)}$ and $\bm{\P}^{\parallel}_{(n)}$ are orthogonal projectors
\be \label{TMSLongProjectorProp}
\bm{\P}^{\parallel}_{(n)} \bm{\P}^{\parallel}_{(n)}  = \bm{\P}^{\parallel}_{(n)} ~, \qquad \bm{\P}^{\parallel}_{(n)}\bm{\P}^{\perp}_{(n)}   = \bm{\P}^{\perp}_{(n)} \bm{\P}^{\parallel}_{(n)}  = 0~.
\ee
Using the explicit form of the superprojectors given in the subsequent sections \ref{TMSN1SUSY}-\ref{TMSN6SUSY}, it can be shown that the action of the orthogonal complement $\bm{\P}^{\parallel}_{(n)}$ on $\mb{V}^{[\cN]}_{(n)}$ takes the general form
\be \label{TMSLongProj}
\bm{\P}^{\parallel}_{(n)} \F_{\a(n)} = \ri^n D^I_\a \F^{I}_{\a(n-1)}~,
\ee
where $\F^{I}_{\a(n-1)}$ are  unconstrained superfields. Any symmetric rank-$n$ spinor superfield $\F^{\parallel I}_{\a(n)} = D^{I}_{\a} \F_{\a(n-1)}$ is said to be longitudinal. In accordance with \eqref{TMSLongProjectorProp} and \eqref{TMSLongProj}, the operator $\bm{\P}^{\parallel}_{(n)}$ projects onto the space of rank-$n$ longitudinal superfields, and thus is called the longitudinal superprojector.  

Using the fact that  $\bm{\P}^{\perp}_{(n)}$ and $\bm{\P}^{\parallel}_{(n)}$ resolve the identity operator \eqref{TMSLongitudinalProjector}, it follows from \eqref{TMSLongProj} that any unconstrained superfield $\F_{\a(n)}$ on $\mb{V}^{[\cN]}_{(n)}$ can be decomposed in the following manner
\be \label{TMSDecomp}
\F_{\a(n)} = \F^{\perp}_{\a(n)} +  \ri^n D^I_\a \F^{I}_{\a(n-1)}~.
\ee
Here $\F^{\perp}_{\a(n)} $ is irreducible, whilst $\F^{I}_{\a(n-1)}$ is unconstrained.

The superprojector $\bm{\P}^{\perp}_{(n)}$ annihilates any longitudinal superfield $\F^{\parallel I}_{\a(n)}$ in $\mb{M}^{3|2 \cN}$
\be \label{TMSTransKillsLong}
\bm{\P}^{\perp}_{(n)} \F^{\parallel I}_{\a(n)} = 0~,
\ee
as a direct consequence of \eqref{5.2b}. The property \eqref{TMSTransKillsLong} ensures that the $\bm{\P}^{\perp}_{(n)}$ selects  the pure superspin component $\F^{\perp}_{\a(n)}$ from the decomposition \eqref{TMSDecomp}.

It follows immediately from \eqref{TMSLongitudinalProjector} and \eqref{TMSTransKillsLong} that the longitudinal superprojector acts like the identity operator on the space of longitudinal superfields
\be
\bm{\P}^{\parallel}_{(n)} \F^{\parallel I}_{\a(n)} = 0~.
\ee
\subsubsection{Superhelicity projection operators}
Let us consider a superfield $\F_{\a(n)}$ satisfying the mass-shell equation \eqref{TMSMassiveMassShell2}. The superspin projection operator $\bm{\P}^{\perp}_{(n)}$ maps $\F_{\a(n)}$ to the superfield obeying the conditions 
\bsubeq
\bea
D^{\b I} \bm{\P}^{\perp}_{\b \a(n-1)}(\F) &=& 0~, \\
(\Box -m^2) \bm{\P}^{\perp}_{\a(n)}(\F) &=& 0~. 
\eea
\esubeq
In accordance with \eqref{TMSMassiveOnshell2}, it follows that the projected superfield  $\bm{\P}^{\perp}_{\a(n)}(\F)$ realises the reducible representation $\mb{D}  (m:-,\frac{n}{2}, \cN) \oplus \mb{D} (m:+,\frac{n}{2}, \cN )$.

In order to extract the on-shell massive supermultiplet of definite superhelicity from $\F_{\a(n)}$, we must split $\bm{\P}^{\perp}_{(n)}$ into the superhelicity projection operators ${\mathbb P}^{(\pm)}_{(n)}$ according to 
\be
\bm{\P}_{(n)}^{\perp} = \mb{P}^{(+)}_{(n)} + \mb{P}^{(-)}_{(n)}~.
\ee
Each of the superhelicity projectors ${\mathbb P}^{(\pm)}_{(n)}$ should satisfy \eqref{TMSGeneralProjectorIdempotent} and \eqref{TMSGeneralProjectorTransverse}. In order to project out the component carrying a single value of superhelicity, the projected superfield $\F^{(\pm)}_{\a(n)}:=\mb{P}^{(\pm)}_{(n)} \F_{\a(n)}$ must obey the conditions
\bsubeq \label{TMSConditionsHelciity}
\bea
D^{\b I} \F^{(\pm)}_{\b \a(n-1)} &=& 0~, \\
(\D \mp m)\F^{(\pm)}_{\a(n)}&=& 0 ~. \label{TMSDeFHelicityCondition2}
\eea
\esubeq
It follows that $\F^{(\pm)}_{\a(n)}$ furnishes the massive UIR $\mb{D}(m:\pm , \frac{n}{2} , \cN)$ (cf. \eqref{TMSMassiveOnshell}).

The operator projecting to the subspace \eqref{TMSConditionsHelciity} is given by
\bea \label{TMSSuperhelicityprojectors}
{\mathbb P}^{(\pm)}_{(n)}:= \hf \Big(\mathbbm{1} \pm \frac{\D}{\sqrt{\Box}}\Big) \bm{\P}^{\perp}_{(n)}~.
\eea
By construction, the operators ${\mathbb P}^{(+)}_{(n)}$ and ${\mathbb P}^{(-)}_{(n)}$ are orthogonal projectors on $\mb{V}^{[\cN]}_{(n)}$\footnote{The superhelicity projectors \eqref{TMSSuperhelicityprojectors} are strictly not superspin projection operators, since they do not satisfy the property \eqref{TMSGeneralProjectorSurjective}, but this is just semantics. }
\bea
{\mathbb P}^{(+)}_{(n)}{\mathbb P}^{(-)}_{(n)}=0~, \qquad 
{\mathbb P}^{(+)}_{(n)}{\mathbb P}^{(+)}_{(n)}={\mathbb P}^{(+)}_{(n)}~, \qquad 
{\mathbb P}^{(-)}_{(n)}{\mathbb P}^{(-)}_{(n)}={\mathbb P}^{(-)}_{(n)}~.
\eea
They also map a superfield $\F_{\a(n)}$ on $\mb{V}^{[\cN]}_{(n)}$ to a transverse superfield, since the operator $\D$ preserves the transverse nature of the superprojector $\bm{\P}^{\perp}_{(n)}$, by virtue of \eqref{TMSDeltaProperty2}. Moreover, they satisfy the additional constraint
\bea \label{TMSSuperHelicityProjectorProperty}
\D \F^{(\pm)}_{\a(n)}= \pm \sqrt{\Box} \F^{(\pm)}_{\a(n)}~.
\eea
If $\F^{(\pm)}_{\a(n)}$ is on the mass-shell  \eqref{TMSMassiveMassShell2}, then 
\eqref{TMSSuperHelicityProjectorProperty} reduces to \eqref{TMSDeFHelicityCondition2}. Thus, 
the superhelicity projectors ${\mathbb P}^{(\pm)}_{(n)}$ map a superfield $\F_{\a(n)}$ obeying \eqref{TMSMassiveMassShell2} to a massive superfield which furnishes the UIR $\mb{D}(m:\pm , \frac{n}{2} , \cN)$ with definite superhelicity.


\subsection{Superspin projection operators in $\mb{M}^{3|2}$} \label{TMSN1SUSY}
The superspin projection operator on the space of spinor superfields $\mb{V}_{(1)}^{[1]}$ is  given by
\bea
\bm{\P}^\b{}_\a= -\frac{D^2}{4 \Box} D^{\b}D_{\a} = -\frac{\ri}{2} \frac{\D}{ \Box} D^{\b}D_{\a} ~. \label{TMSN1Projector}
\eea
Proving that the operator $\bm{\P}^\b{}_\a$ satisfies the defining properties \eqref{TMSProjectorPropertiesIndividual} and \eqref{TMSTransverseProj} is simple if one makes use of  the identities \eqref{TMSDeltaOperator} and 
\be \label{TMSDeltaSquaredProperties}
\D^2 = \Box~.
\ee
Note that in $\mb{M}^{3|2}$, the identity \eqref{TMSDeltaSquaredProperties} is equivalent to \eqref{TMSDerivativeIdentities4}.
We can immediately build the higher-rank superprojector $\bm{\P}^{\perp}_{(n)}$  according to \eqref{TMSGeneralProjector}
\bea \label{TMSN1ProjectorE}
\bm{\P}_{(n)}^{\perp} \F_{\a(n)}=\Big ( -\frac{\ri \D}{2 \Box} \Big )^n  D^{\b_1}D_{\a_1} \cdots D^{\b_n}D_{\a_n} \F_{\b(n)}~,
\eea
where we have made use of the property \eqref{TMSDeltaOperator}. Note that the right hand side of \eqref{TMSN1ProjectorE} is totally symmetric in its $\a$ indices, as a consequence of the projector property \eqref{5.2d}. 

Studying $\bm{\P}^{\perp}_{(n)}$ for the particular cases $n=2s$ and $n=2s+1$, the superprojectors can be shown to take the simplified form
\bsubeq
\begin{align}
\bm{\P}^{\perp}_{(2s)} \F_{\a(2s)} &= \Big ( - \frac{1}{4 \Box} \Big )^s  D^{\b_1}D_{\a_1} \cdots D^{\b_{2s}}D_{\a_{2s}} \F_{\b(2s)}~, &  s &\geq 1~,\\
\bm{\P}^{\perp}_{(2s+1)} \F_{\a(2s+1)} &= - \frac{\ri \D}{2 \Box}\Big ( - \frac{1}{4 \Box} \Big )^s D^{\b_1}D_{\a_1} \cdots D^{\b_{2s+1}}D_{\a_{2s+1}} \F_{\b(2s+1)}~, & s &\geq 0~,
\end{align}
\esubeq
as a consequence of the identity \eqref{TMSDeltaSquaredProperties}.

We now turn our attention to computing the explicit form of the longitudinal superprojector $\bm{\P}^{\parallel}_{(n)}$, which is defined according to \eqref{TMSLongitudinalProjector}. Given an arbitrary superfield $\F_{\a(n)}$ on $\mb{V}^{[1]}_{(n)}$, it can be shown using \eqref{TMSN1ProjectorE} that the action of $\bm{\P}^{\parallel}_{(n)}$ on $\F_{\a(n)}$ is given by
\bsubeq
\bea
\bm{\P}^{\parallel}_{(n)}\F_{\a(n)}
&=& \ri^n D_{\a} \U_{\a(n-1)}~,
\eea
where the real superfield $\U_{\a(n-1)}$ takes the explicit form
\bea
\U_{\a(n-1)}&:=& -(-\ri)^n\sum_{j=1}^{n} \frac{1}{(4\Box)^j} \binom{n}{j} 
(D^2)^j
D^{\b_{n-1}}  D_{(\a_{n-1}} \dots D^{\b_{n-j+1}}  D_{\a_{n-j+1}}
\non\\
&&\times D^{\b_n}\F_{\a_{1} \dots \a_{n-j}) \b_{n-j+1} \dots \b_n}~.
\eea
\label{TMS1Long}
\esubeq
In order to compute \eqref{TMS1Long}, it proves useful to rewrite $\bm{\P}^{\b}{}_{\a}$ in  the form
\bea
\bm{\P}^{\b}{}_{\a} = \d_{\a}\,^{\b}- \frac{D^2}{4\Box}D_{\a}D^{\b}~.
\eea

Since the superprojectors $\bm{\P}_{(n)}^{\perp}$ and $\bm{\P}_{(n)}^{\parallel}$ resolve the unit operator \eqref{TMSLongitudinalProjector} , it follows that any superfield on $\mb{V}^{[1]}_{(n)}$ can be decomposed in the following manner
\be
\F_{\a(n)} = \big ( \bm{\P}_{(n)}^{\perp} + \bm{\P}_{(n)}^{\parallel} \big ) \F_{\a(n)} = \F^{\perp}_{\a(n)} + \ri^n D_{\a} \F_{\a(n-1)}~,
\ee
where $\F^{\perp}_{\a(n)}$ is transverse and $\F_{\a(n-1)}$ is unconstrained. Repeating this procedure iteratively yields the decomposition of $\F_{\a(n)}$ into irreducible components
\be \label{TMSIrredDecomp}
\F_{\a(n)} = \sum_{j=0}^{\lfloor n/2 \rfloor} \big ( D_{\a(2)} \big )^j \F^{\perp}_{\a(n-2j)} + \ri^n \sum_{j=0}^{\lceil n/2 \rceil -1}  ( D_{\a(2)} \big )^j D_\a\F^{\perp}_{\a(n-2j-1)} ~.
\ee
Here the superfields $\F^{\perp}_{\a(n-2j)}$ and $\F^{\perp}_{\a(n-2j-1)}$ are transverse, except for $\F^{\perp}$.

\subsection{Superspin projection operators in $\mb{M}^{3|4}$}\label{TMSN2SUSY}
In the case of $\cN=2$ supersymmetry, it is often useful to work in 
a complex basis for the spinor covariant derivatives. 
Such a basis was first introduced in \cite{KPT-MvU2011} by replacing 
the covariant derivatives  $D^{I}_{\a} = (D^{\1}_{\a}, D^{\2}_{\a})$ 
with $D_{\a}$ and $\bar D_{\a}$, which are defined by 
\bea
D_{\a} = \frac{1}{\sqrt{2}}(D^{\1}_{\a} - \ri D^{\2}_{\a})~, \qquad \bar D_{\a} = -\frac{1}{\sqrt{2}}(D^{\1}_{\a} + \ri D^{\2}_{\a})~.
\label{D2real}
\eea
It follows from  \eqref{TMSNDerivativeAlgebra} that
$D_{\a}$ and $\bar D_{\a}$ satisfy the anti-commutation relations\footnote{These anti-commutation relations are reminiscent to those in $\mb{M}^{4|4}$ \eqref{FMSDerivativeAlgebra}.  }
\bea
\{D_{\alpha}, D_{\beta}\}=0\,, \qquad
\{\bar D_{\alpha}, \bar D_{\beta}\}=0\,, \qquad
\{ D_{\alpha}, \bar D_{\beta}\} =-2 \ri \partial_{\alpha \beta}\,.
\label{3.2}
\eea

In this section we will construct the superspin projection operators on $\mb{M}^{3|4}$ in both the complex \eqref{3.2} and real \eqref{TMSNDerivativeAlgebra} bases for the spinor covariant derivatives.

\subsubsection{Superspin projection operators in the complex basis}

Before diving into the construction of the superprojectors, it is worth pausing to discuss the types of constrained superfields which will be pertinent to our subsequent analysis. According to the terminology of \cite{KuzenkoOgburn2016}, for integers $n \geq 1$,
a real symmetric rank-$n$ superfield $T_{\a(n)}= \bar T_{\a(n)}$ constrained by
\bea
\bar D^{\b} T_{ \b \a (n-1)} =0 \quad \Longleftrightarrow \quad
D^{\b}
T_{\b \a(n-1)} =0~,
\label{4.60}
\eea
is said to be  real transverse linear. It is linear since the above constraint implies
\bea \label{TMSRealLinearFields}
\bar D^{2}  T_{\a(n)}=0 \quad \Longleftrightarrow \quad D^{2} T_{\a(n)}=0~,
\eea
as a consequence of  \eqref{3.2}. 

Moreover, a complex superfield $\F_{\a(n)}$ is said to be simultaneously transverse linear and transverse anti-linear if it obeys the constraints
\bsubeq \label{TMSTLAL2}
\bea 
\bar D^{\b} \G_{\b \a(n-1)} =0 \quad &\implies& \quad
\bar D^2 \G_{\a(n)} =0~, \label{TMSTransverseLinear}\\
D^{\b} {\G}_{\b \a(n-1)} =0 \quad &\implies& \quad
D^2 {\G}_{\a(n)} =0~. \label{TMSTransverseAntiLinear}
\eea
\esubeq

A complex superfield $G_{\a(n)}$ is said to be chiral if it satisfies the condition 
\be
\bar{D}_\b G_{\a(n)} =0~, \label{TMSChiralSuperfields}
\ee
or anti-chiral if it obeys the constraint
\be
D_\b {G}_{\a(n)} = 0~.  \label{TMSAntiChiralSuperfields}
\ee

In the case of $\cN=1$ supersymmetry, $\D$ is an invertible operator (cf. \eqref{TMSDeltaSquaredProperties}). This is no longer true for $\cN=2$, due to the relation 
\bea
\Box = \tilde{\D}^2 + \frac{1}{16} \big\{ \bar D^2 , D^2 \big\}~,
\label{Salam}
\eea
where $\tilde{\D}$ denotes the $\cN=2$  version of  \eqref{TMDeltaoperator}, which written in the complex basis is
\bea \label{TM2DeltaOperator}
\tilde{\D} &=& \frac{\ri}{2} D^{\b} \bar{D}_{\b} = \frac{\ri}{2} \bar D^{\b} D_{\b}~.
\eea
The identity \eqref{Salam} can be expressed in the alternative form
\be \label{TMSProjResIde}
\mds{1}= \cP_{(0)} + \cP_{(+)} +\cP_{(-)}~, 
\ee
where the differential operators $\cP_i =\big(\cP_{(0)}, \cP_{(+)},\cP_{(-)}\big)$
\be
\cP_{(0)} = \frac{1}{\Box} \tilde{\D}^2~, 
\qquad \cP_{(+)}=\frac{1}{16\Box}  \bar D^2 D^2 ~, \qquad 
\cP_{(-)}=\frac{1}{16\Box}  D^2 \bar D^2 ~,
\label{4.100}
\ee
are orthogonal projectors, $\cP_i \cP_j = \d_{ij} \cP_i$.
The operator $\cP_{(0)}$ projects onto the space of real linear superfields \eqref{TMSRealLinearFields}, while $\cP_{(+)}$ and $\cP_{(-)}$ project onto the space of chiral \eqref{TMSChiralSuperfields} and anti-chiral superfields \eqref{TMSAntiChiralSuperfields}, respectively. Note that the projector $\cP_{(0)}$ maps a complex superfield to the space of simultaneously linear and anti-linear superfields.

Let us now introduce the $\cN=2$ spinor superprojector 
\bea \label{3.3a}
\bm{\P}^\b{}_\a= \frac{\ri}{4 \Box} {\tilde{\D}}\big( \bar D^{\b} D_{\a} + D^{\b} \bar D_{\a}\big)~,
\eea
which is written in terms of the covariant derivatives in the complex basis.
One may check that $\bm{\P}^\b{}_\a$ satisfies the conditions \eqref{TMSProjectorPropertiesIndividual} and \eqref{TMSTransverseProj}, keeping in mind that $D^I_\a$ now encodes $D_\a$ and $\bar D_\a$. Making use of the anti-commutation relations \eqref{3.2}, $\bm{\P}^{\b}{}_\a$ can be rewritten in the form
\bea
\bm{\P}^\b{}_\a= \frac{\tilde{\D}}{2\Box} \big( \pa_{\a}\,^{\b} + \d_{\a}\,^{\b} \tilde{{\Delta}} \big)~.
\label{TMSProjComp2}
\eea

The higher-rank generalisation $\bm{\P}^{\perp}_{(n)} $ of  \eqref{TMSProjComp2} is given by 
\be \label{TMSN2ProjectorComplex}
\bm{\P}^{\perp}_{(n)} \F_{\a(n)} = \Big ( \frac{\tilde{\D}}{2\Box} \Big )^n \big (  \pa_{(\a_1}{}^{\b_1} + \tilde{\D} \d_{(\a_1}{}^{\b_1} \big ) \big (  \pa_{\a_2}{}^{\b_2} + \tilde{\D} \d_{\a_2}{}^{\b_2} \big ) \cdots \big (  \pa_{\a_n)}{}^{\b_n} + \tilde{\D} \d_{\a_n)}{}^{\b_n} \big ) \F_{\b(n)}~.
\ee
The superspin projection operator $\bm{\P}^{\perp}_{(n)} $ maps any real symmetric rank-$n$ superfield $\F_{\a(n)} = \bar{\F}_{\a(n)}$ to the subspace of real transverse superfields \eqref{4.60}. If we instead consider the space of complex superfields $\F_{\a(n)}$, then $\bm{\P}^{\perp}_{(n)} $ projects onto the space of simultaneously transverse linear and transverse anti-linear superfields \eqref{TMSTLAL2}. 

We now have the tools at our disposal to construct the longitudinal superprojector \eqref{TMSLongitudinalProjector} explicitly. For an arbitrary real symmetric rank-$n$ superfield $\F_{\a(n)}$, we find\bsubeq
\bea\label{longH}
\bm{\P}^{\parallel}_{(n)} \F_{\a(n)}
&=& \bar D_{\a} \L_{\a(n-1)} 
- (-1)^n D_{\a} \bar \L_{\a(n-1)}~,
\eea
where the complex superfield $\L_{\a(n-1)}$ can be shown to take the form
\bea
\L_{\a(n-1)} &=& 
- \sum_{j=1}^{n} \bigg(\frac{\ri}{4\Box}\bigg)^j \binom{n}{j} 
D^{\b_n} A_{(\a_{n-1}}^{\quad \b_{n-1}} \dots A_{\a_{n-j+1}}^{\quad \b_{n-j+1}}
\tilde{\Delta}^{j}\F_{\a_{1} \dots \a_{n-j}) \b_{n-j+1} \dots \b_n} \non\\
&&+ \frac{1}{8\Box} D^{\b} \bar D^2 \F_{\b \a(n-1)}~.
\eea
\esubeq
Here the operator $A_{\a}{}^{\b}$ is defined by
\bea
A_{\a}{}^{\b} := D_{\a} \bar D^{\b} + \bar D_{\a} D^{\b}~,
\eea
and satisfies the  property
\bea
[\tilde{\D}, A_{\a}\,^{\b}]=0~,
\eea
which is crucial to the analysis above.

The relation \eqref{longH} naturally leads to the gauge transformation law
\bea
\d_\z H_{\a(n)} = g_{\a(n)} + \bar g_{\a(n)}~, \qquad 
g_{\a(n)} = \bar{D}_{\a}\z_{\a(n-1)} ~,
\label{4.1888}
\eea
where $\z_{\a(n-1)}$ is complex unconstrained. The transformation law \eqref{4.1888} was postulated in \cite{KuzenkoOgburn2016} to describe the gauge freedom of a superconformal gauge prepotential $H_{\a(n)}$.

\subsubsection{Superspin projection operators in the real basis}
The $\cN=2$ superspin projection operator \eqref{3.3a} takes the following form in the real basis for the spinor covariant derivatives \eqref{TMSNDerivativeAlgebra}
\bea
\bm{\P}^{\b}{}_{\a} = -\frac{\ri \Delta}{4\Box} D^{\b I}D^{I}_{\a}~.
\label{N2real}
\eea
It follows from \eqref{TMSGeneralProjector} that the higher-rank superprojector $\bm{\P}^{\perp}_{(n)}$ takes the following form on the space of real superfields $\F_{\a(n)}$
\be
\bm{\P}^{\perp}_{(n)}\F_{\a(n)} = \Big ( - \frac{\ri \D}{4 \Box} \Big )^n D^{\b_1}D_{\a_1} \cdots D^{\b_n}D_{\a_n} \F_{\b(n)}~. \label{TMSHRprojectorN1}
\ee
Note that this expression can be simplified in the bosonic $n=2s$ and fermionic cases $n=2s+1$, as a result of the identity \eqref{TMSDeltaProperty3}. In the subsequent sections,  the superspin projection operators, in the case of $ 3 \leq {\cal N} \leq 6$ supersymmetry, will be computed in terms of the spinor covariant derivatives in the real basis \eqref{TMSNDerivativeAlgebra}.


\subsection{Superspin projection operators in $\mb{M}^{3|6}$}\label{TMSN3SUSY}

The ${\cal N}=3$ superspin projection operator $\bm{\P}^{\b}{}_{\a}$ takes the form
\bea
\bm{\P}^\b{}_\a&=& -\frac{\ri \Delta}{48\Box^2} D^{\b I} D_{\a}^{I}\big(9 \Delta^2 - \Box \big)~.
\label{TMS3Proj}
\eea
To show that $\bm{\P}^\b{}_\a$ satisfies the properties \eqref{TMSProjectorPropertiesIndividual} and \eqref{TMSTransverseProj}, it is beneficial to use the identities \eqref{TMSDeltaOperator}, \eqref{TMSDeltaProperty3} and 
\bea
\Delta^4 &=& \frac{1}{9} \Box \big ( 10  \Delta^2 - \Box \big )~. \label{4.3d}
\eea
The higher-rank generalisation $\bm{\P}^{\perp}_{(n)}$ of the superprojector \eqref{TMS3Proj} is 
\be
\bm{\P}^{\perp}_{(n)}\F_{\a(n)} = \Big (-\frac{\ri \Delta}{48\Box^2} \Big )^n  \big (9 \Delta^2 - \Box \big)^n D^{\b_1 I_1} D_{\a_1}^{I_1} \cdots D^{\b_n I_n} D_{\a_n}^{I_n}  \F_{\b(n)}~.
\ee

Given an integer $n \geq 1$,  the following identities can be derived in $\mb{M}^{3|6}$ 
\bsubeq
\bea
(9 \Delta^2 - \Box)^n &=& (8\Box)^{n-1} (9 \Delta^2 - \Box)~,  \label{4.6a}\\
\Delta^{2n}( 9 \Delta^2 - \Box) &=& \Box^n (9 \Delta^2 - \Box)~,\label{4.6b}
\eea
\esubeq
as a result of relation \eqref{4.3d}. Utilising \eqref{4.6a}, the superspin projection operator $\bm{\P}^{\perp}_{(n)}$ can be reduced to the simpler form on $\mb{V}_{(n)}^{[3]}$
\be
\bm{\P}^{\perp}_{(n)}\F_{\a(n)} = \frac{1}{8 \Box }\Big (-\frac{\ri \Delta}{6\Box} \Big )^n  \big (9 \Delta^2 - \Box \big ) D^{\b_1 I_1} D_{\a_1}^{I_1} \cdots D^{\b_n I_n} D_{\a_n}^{I_n}  \F_{\b(n)}~. \label{TMSHNProjN3}
\ee


\subsection{Superspin projection operators in $\mb{M}^{3|8}$}\label{TMSN4SUSY}

The ${\cal N}=4$ spinor superprojector $\bm{\P}^{\b}{}_{\a}$  assumes the form
\bea
\bm{\P}^\b{}_\a&=& -\frac{\ri \Delta}{24 \Box^2} D^{\b I} D_{\a}^{I}\big( 4 \Delta^2 - \Box \big)~.
\label{5.1}
\eea
It is possible to show that the operator $\bm{\P}^\b{}_\a$  satisfies the defining properties \eqref{TMSProjectorPropertiesIndividual} and \eqref{TMSTransverseProj}, given that one utilises the identities \eqref{TMSDeltaOperator}, \eqref{TMSDeltaProperty3} and 
\bea
\Delta^5 &=& \frac{1}{4} \D \Box \big ( 5  \Delta^2 - \Box  \big )~. \label{5.3d}
\eea
The higher-rank extension of the superprojector $\bm{\P}^\b{}_\a$ can be shown to take the form
\be
\bm{\P}^{\perp}_{(n)}\F_{\a(n)} = \frac{1}{3 \Box }\Big (-\frac{\ri \Delta}{8\Box} \Big )^n  (4 \Delta^2 - \Box) D^{\b_1 I_1} D_{\a_1}^{I_1} \cdots D^{\b_n I_n} D_{\a_n}^{I_n}  \F_{\b(n)}~. \label{TMSHRProjN4}
\ee
In order to arrive at this result, it is important to make use of the identity
\be
\Delta (4 \Delta^2 - \Box)^n = (3\Box)^{n-1} \Delta (4 \Delta^2 - \Box)~, \qquad n \geq 1~,
\label{5.7}
\ee
which can be derived from \eqref{5.3d}.


\subsection{Superspin projection operators in $\mb{M}^{3|10}$}\label{TMSN5SUSY}
The ${\cal N}=5$ spinor superprojector $\bm{\P}^{\b}{}_{\a}$ takes the form
\bea
\bm{\P}^\b{}_\a&=& -\frac{\ri \Delta}{3840 \Box^3} D^{\b I} D_{\a}^{I}\big ( 25 \Delta^2 -  \Box \big ) \big (25 \D^2 - 9 \Box \big)~.
\label{6.1}
\eea
The projector properties \eqref{TMSProjectorPropertiesIndividual} and \eqref{TMSTransverseProj} can be proved using  \eqref{TMSDeltaOperator} and \eqref{TMSDeltaProperty3}, in conjunction with the  identity
\bea
\Delta^6 &=& \frac{1}{625} \Box \big ( 875 \Delta^4 -259 \Box \D^2 + 9 \Box^2 \big )~. \label{6.3d}
\eea
The higher-rank generalisation of \eqref{6.1} is given by
\be
\bm{\P}^{\perp}_{(n)}\F_{\a(n)} = \Big (-\frac{\ri \Delta}{3840 \Box^3} \Big )^n  \big ( 25 \Delta^2 -  \Box \big )^n \big (25 \D^2 - 9 \Box \big)^n D^{\b_1 I_1} D_{\a_1}^{I_1} \cdots D^{\b_n I_n} D_{\a_n}^{I_n}  \F_{\b(n)}~. \label{TMSHRPROHN5}
\ee
This expression can be simplified if one studies the particular cases when $n$ is either even $n=2s$ or odd $n=2s+1$.


\subsection{Superspin projection operators in $\mb{M}^{3|12}$}\label{TMSN6SUSY}

The ${\cal N}=6$ spinor superprojector $\bm{\P}^{\b}{}_{\a}$  is given by 
\be
\bm{\P}^\b{}_\a= -\frac{\ri \Delta}{480 \Box^3} D^{\b I} D_{\a}^{I}\big( 9 \Delta^2 -  \Box \big ) \big ( 9 \D^2 - 4 \Box \big)~.
\label{7.1}
\ee
The superprojector properties \eqref{TMSProjectorPropertiesIndividual} and \eqref{TMSTransverseProj} can be proved using the identities \eqref{TMSDeltaOperator},  \eqref{TMSDeltaProperty3} and
\bea
\Delta^7 &=& \frac{1}{81} \Box \D \big ( 126 \Delta^4 -49 \Box \D^2 + 4 \Box^2  \big )~. \label{7.3d}
\eea
In accordance with \eqref{TMSGeneralProjector}, the higher-rank generalisation of \eqref{7.1} is 
\be
\bm{\P}^{\perp}_{(n)}\F_{\a(n)} = \Big (-\frac{\ri \Delta}{480 \Box^3}\Big )^n  \big ( 9 \Delta^2 -  \Box \big )^n \big ( 9 \D^2 - 4 \Box \big)^n D^{\b_1 I_1} D_{\a_1}^{I_1} \cdots D^{\b_n I_n} D_{\a_n}^{I_n}  \F_{\b(n)}~. \label{TMSN6Superprojector}
\ee
It can be shown that the form of the superprojector \eqref{TMSN6Superprojector} simplifies if one analyses the cases $n=2s$ and $n=2s+1$ separately.

\section{Superconformal higher-spin theory} \label{TMSSecCHS}

The off-shell actions for $\cN$-extended conformal supergravity theories in $3d$ were formulated in \cite{BKNT-M2,KuzenkoNovakTartaglino-Mazzucchelli2014} for $1 \leq \cN \leq 6$ using the off-shell
formulation for $\cN$-extended conformal supergravity developed in \cite{BKNT-M1}. Each action is generated by a closed super three-form which is constructed in terms of the constrained geometry of $\cN$-extended conformal superspace. In this section we initiate a program to recast these actions, and their higher-spin extensions, in terms of unconstrained gauge prepotentials. In order to achieve this, we first need to compute the linearised higher-spin super-Cotton tensors in $\mb{M}^{3|2\cN}$, for $1 \leq \cN \leq 6$, in terms of unconstrained gauge prepotentials.
Before doing this, it is worth commenting on the super-Cotton tensor in the context of conformal supergravity.

The geometry of three-dimensional $\cN$-extended conformal superspace \cite{BKNT-M1} 
is formulated in terms of a single curvature superfield, which is known as the super-Cotton 
tensor $\cW$ (with suppressed indices). 
The functional structure of $\cW$ is dependent on the choice of $\cN$. 
The $\cN=1$ super-Cotton tensor \cite{KT-M12}
is a primary symmetric rank-3 spinor superfield
$\cW_{\a\b\g}$ of dimension $\frac{5}{2}$, which obeys the conformally invariant constraint
\cite{BKNT-M1}
\bea
\nabla^\a \cW_{\a \b\g} = 0 ~. 
\label{WW2.2}
\eea
In the $\cN=2$ case,  the super-Cotton tensor \cite{ZP,Kuzenko12} 
is a primary symmetric rank-2 spinor superfield 
$\cW_{\a\b}$ of dimension 2, which obeys the Bianchi identity  \cite{BKNT-M1}
\bea
\nabla^{\a I} \cW_{\a\b} = 0 ~. 
\label{WW2.3}
\eea
In the $\cN=3$ case,  the super-Cotton tensor is a  primary spinor superfield
$\cW_{\a}$ of dimension $\frac{3}{2}$ constrained by \cite{BKNT-M1}
\bea
\nabla^{\a I} \cW_\a = 0 ~.
\label{WW2.4}
\eea
In the $\cN=4$ case, the super-Cotton tensor is a primary  scalar
superfield  $\cW$ of dimension 1
constrained by \cite{BKNT-M1}
\bea
\nabla^{\a I}\nabla_{\a}^J \cW=\frac{1}{4}\d^{IJ}\nabla^{\a K}\nabla_{\a}^K \cW~.
\label{WW2.5}
\eea
For $\cN>4$,  the super-Cotton tensor \cite{HIPT,KLT-M11}
is a completely antisymmetric tensor $\cW^{IJKL}$
of dimension 1 constrained by \cite{BKNT-M1}
\bea \nabla_{\a}^I \cW^{JKLP} = \nabla_\a^{[I} \cW^{JKLP]} 
- \frac{4}{\cN - 3} \nabla^Q_{\a} \cW^{Q [JKL} \d^{P] I} \ .
\label{2.666}
\eea
In the above relations, $ \nabla^I_\a$ denotes the spinor covariant derivative
of $\cN$-extended conformal superspace \cite{BKNT-M1}.

The super-Cotton tensor is encoded in the action for 
conformal supergravity $S_{\rm CSG}$ by the rule \cite{BKNT-M1,KuzenkoNovakTartaglino-Mazzucchelli2014}
\bea \label{TMSCottonTensorsSupergravity}
\cW \propto \frac{\d S_{\rm CSG} [H]}{\d H}~,
\eea
where $H$ denotes an unconstrained gauge prepotential for conformal supergravity (with indices suppressed).
Modulo pure gauge degrees of freedom,
the structure of the unconstrained conformal gauge prepotentials are as follows:
$H_{\a\b\g} $ 
for $\cN=1$ \cite{GatesGrisaruRocekSiegel1983}, $H_{\a\b}$ for $\cN=2$ \cite{ZP,Kuzenko12}, 
$H_\a$ for $\cN=3$ \cite{BKNT-M1}, and $H$ for $\cN=4$ \cite{BKNT-M1}. 
The above consideration implies that we need expressions for linearised super-Cotton
tensors in terms of the gauge prepotentials, $W = W(H)$, in order to obtain linearised
actions for conformal supergravity actions for $1 \leq \cN \leq 4$.

As a generalisation of the earlier $\cN=1$ \cite{Kuzenko2016,KuzenkoTsulaia2017} and $\cN=2$ \cite{KuzenkoOgburn2016} results, 
we propose a $\cN$-extended superconformal gauge-invariant action of the form
\bea
{S}^{(n|\cN)}_{\text{SCHS}} [  H_{\a(n)}] = \frac{ \ri^n}{2} 
\int \rd^{3|2\cN}z \, H^{\a(n)} 
{W}_{\a(n)}(H) ~, \qquad n>0~,
\label{action}
\eea
where the real unconstrained superfield $H_{\a(n)}$ is defined modulo gauge transformations
\bea
\d_\z H_{\a(n )} &=& \ri^n D^I_{\a } \z^I_{\a(n-1)} ~.
\label{TMSSCHSGT}
\eea
Here the gauge parameter $\z_{\a(n-1)}$ is also real unconstrained.
The super-Cotton tensor $W_{\a(n)}(H)$ in \eqref{action} is a local descendant of $H_{\a(n)}$.
It is a  totally symmetric real
rank-$n$ superfield, which is defined to obey the following conditions
\begin{enumerate} \label{TMSCTProp}
	\item $W_{\a(n)}$ is gauge invariant 
	\bea
	\d_\z {W}_{\a(n) }\big( H \big) =0~. \label{1.7}
	\eea
	\item $W_{\a(n)}$ is transverse
	\bea
	D^{\b I} W_{\b \a(n-1)}(H) =0~. \label{TMSCottTrans}
	\eea
	\item The field strength $W_{\a(n)}(H)$ is a primary superconformal multiplet
	(all relevant technical details about the $\cN$-extended superconformal 
	group are collected in Appendix \ref{TASappendixSCHSPrimary}). 
\end{enumerate}

The latter fact, in conjunction with  \eqref{TMSCottTrans}, uniquely fixes the dimension $d_{W_{\a(n)}}$
of ${W}_{\a(n)}(H) $.
The dimension $d_{H_{\a(n)}}$
of ${H}_{\a(n) }$ is also fixed uniquely if we require $H_{\a(n)}$ and the associated 
gauge parameter $\z_{\a(n-1)}$ in \eqref{TMSSCHSGT} to be superconformal primary.
The mass dimensions are: 
\bea
d_{H_{\a(n)}} = 2-\cN -\frac n2~, \qquad
d_{W_{\a(n)}} = 1 +\frac n2~.
\label{1.9}
\eea

For integers $n \geq 1$, the above conditions suffice in determining
$W_{\a(n)}(H)$, modulo an overall numerical factor, in the form
\bea
W_{\a(n)} \big(H\big) =  \D^{n+\cN-1}  \bm{\P}^{\perp}_{[n]}H_{\a(n)}~.
\label{2.16}
\eea
Here $ \bm{\P}^{\perp}_{[n]}$ is the superspin projection operator, which was computed explicitly in section \ref{TMSSecSpinProjectors} for the supersymmetry types $1 \leq \cN \leq 6$. Note that the super-Cotton tensor in the form \eqref{2.16} is automatically invariant under the gauge transformations \eqref{TMSSCHSGT}, due to the superprojector property \eqref{TMSTransKillsLong}.  Below, we will utilise the superspin projection operators to compute closed form expressions for the linearised super-Cotton tensors in terms of unconstrained gauge prepotentials.


\subsection{Linearised higher-spin super-Cotton tensor in $\mb{M}^{3|2}$}

Given an integer $n \geq 1$, the rank-$n$ 
super-Cotton tensor \cite{Kuzenko2016} (see also \cite{KuzenkoTsulaia2017,KuzenkoPonds2018}) is
\bea
W_{\a(n)} (H)= \Big(-\frac{\ri}{2}\Big)^n D^{\b_1}D_{\a_1} \cdots D^{\b_n} D_{\a_n}
H_{\b(n)}~.
\label{N1CT}
\eea
The choice  $n=1$  in  \eqref{N1CT} 
corresponds to the gauge-invariant field strength
of an Abelian vector multiplet   \cite{Siegel}. 
The case $n=2$ corresponds to the super-Cottino tensor \cite{Kuzenko2016}, 
which is the gauge-invariant field strength of a superconformal gravitino 
multiplet.\footnote{Among the component fields of $W_{\a\b}$ is the so-called 
	Cottino tensor $C_{\a\b\g} =C_{(\a\b\g)}$, which is
	the gauge-invariant field strength of a conformal gravitino 
	\cite{DK,GPS,ABdeRST}.
}
Choosing $n=3$ in \eqref{N1CT} gives the linearised version 
of the $\cN=1$ super-Cotton tensor \cite{KT-M12}.

In accordance with \eqref{2.16}, the super-Cotton tensor \eqref{N1CT} can be recast in terms of the 
superspin projection operator $ \bm{\P}_{[n]}^{\perp}$ \eqref{TMSN1ProjectorE} as follows
\bea
W_{\a(n)}(H) =  \D^n  \bm{\P}^{\perp}_{[n]}H_{\a(n)}~.
\label{3.99}
\eea
In order to demonstrate  that \eqref{3.99} is equivalent to \eqref{N1CT}, one simply needs to recall the identity \eqref{TMSDeltaSquaredProperties}.
This identity also allows us to simplify  $W_{\a(n)}$ when considering the cases of $n$ is even or odd.
Specifically, $W_{\a(2s)}$ and $W_{\a(2s+1)}$ take the following form 
\begin{subequations}
	\begin{align}
	W_{\a(2s)}(H) &= \Box^{s}  \bm{\P}^{\perp}_{[2s]}H_{\a(2s)}~, & s&\geq 1 ~,\\
	W_{\a(2s+1)}(H) &= 
	\Box^{s}\D  \bm{\P}^{\perp}_{[2s+1]}H_{\a(2s+1)} ~, & s&\geq 0~ .
	\end{align}
\end{subequations}


\subsection{Linearised higher-spin super-Cotton tensor in $\mb{M}^{3|4}$}
In this section we derive a new representation for the 
linearised  ${\cal N}=2$ rank-$n$ super-Cotton tensor $W_{\a(n)}$, with $n > 1$.
The real tensor superfield  $W_{\a(n)}$ was recently shown in \cite{KuzenkoOgburn2016}  to take the following form  
\bea
W_{\a(n)} (H) &:=& \frac{1}{2^{n-2}}  \sum_{j=0}^{\lfloor\frac{n}{2}\rfloor} \Bigg\{\binom{n}{2j} {\tilde{\D}} \Box^j \pa_{(\a_1}^{\,\,\,\b_1} \cdots \pa_{\a_{n-2j}}^{\,\,\,\b_{n-2j}} H_{\a_{n-2j+1} \ldots \a_{n}) \b(n-2j)}
\non \\
&&+ \binom{n}{2j+1} {\tilde{\D}}^2 \Box^j \pa_{(\a_1}^{\,\,\,\b_1} \cdots \pa_{\a_{n-2j-1}}^{\,\,\,\b_{n-2j-1}} H_{\a_{n-2j} \ldots \a_{n})  \b(n-2j-1)} \Bigg\}~. 
\label{TM2CT}
\eea
Note that $W_{\a(n)} (H)$ \eqref{TM2CT} is constructed in terms of the spinor covariant derivatives in the complex basis. The super-Cotton tensor \eqref{TM2CT} satisfies the following defining properties:
\begin{enumerate}
	\item $W_{\a(n)}(H) $ is invariant under the gauge transformations \eqref{4.1888}
	\be
	\d_\z	W_{\a(n)}(H) = 0~.
	\ee
	\item $W_{\a(n)}(H) $  is real transverse linear 
	\bea
	D^{\b}W_{\b \a(n-1)}(H)= \bar D^{\b}W_{\b \a(n-1)}(H)= 0~.
	\eea
\end{enumerate}
The case $n=1$ corresponds to the super-Cottino tensor \cite{Kuzenko2016}, 
which is the gauge-invariant field strength of a superconformal gravitino multiplet.
The choice $n=2$ gives the linearised version 
of the $\cN=2$ super-Cotton tensor \cite{ZP,Kuzenko12}.

One may show that $W_{\a(n)} (H)$ \eqref{TM2CT} can be recast in terms of the superprojectors  \eqref{TMSN2ProjectorComplex} in the complex basis as follows
\bea
W_{\a(n)} (H) = {\tilde{\D}}^{n+1}  \bm{\P}_{[n]}^{\perp}H_{\a(n)}~.
\label{4.188}
\eea
Note that \eqref{4.188} is the analogue of \eqref{2.16} in the complex basis. 
The properties of the operator $\tilde{\D}$
\bea
\tilde{\D}^{2k}= \Box^{k-1} \tilde{\D}^2~, \qquad
\tilde{\D}^{2k+1}= \Box^{k} \tilde{\D}~, \qquad k = 1,2, \dots ~,
\label{delta2}
\eea
prove useful in showing that the super-Cotton tensors \eqref{TM2CT} and \eqref{4.188} are equivalent.

A new representation of $W_{\a(n)} (H)$ \eqref{TM2CT}
can be found using $\bm{\P}^{\perp}_{[n]}$ \eqref{N2real} in the real basis. It follows from \eqref{2.16} that $W_{\a(n)} (H)$ assumes the form
\bea
W_{\a(n)} \big(H\big) =  \D^{n+1} \, \bm{\P}^{\perp}_{[n]}H_{\a(n)}~.
\label{W2}
\eea
Substituting the explicit form of the superprojector  \eqref{TMSHRprojectorN1} 
into \eqref{W2} yields
\bea
W_{\a(n)}(H) = \Big(-\frac{\ri}{4 \Box}\Big)^n \Delta^{2n+1} D^{\b_1 I_1}D^{I_1}_{\a_1} \cdots 
D^{\b_n I_n} D^{I_n}_{\a_n} H_{\b(n)}~.
\label{W2.1}
\eea
It can be shown that the above is equivalent to
\bea
W_{\a(n)}(H) = \Big(-\frac{\ri}{4} \Big)^n \D  D^{\b_1 I_1}D^{I_1}_{\a_1} \cdots D^{\b_n I_n} 
D^{I_n}_{\a_n} H_{\b(n)}~,
\label{W2.2}
\eea
where we have made use of the property \eqref{delta2}.
The new representation \eqref{W2.2} for the rank-$n$ super-Cotton tensor 
is clearly much  simpler than the  expression \eqref{TM2CT} originally given in \cite{KuzenkoOgburn2016}.


\subsection{Linearised higher-spin super-Cotton tensor in $\mb{M}^{3|6}$}
A linearised version $W_\a(H)$ of the $\cN=3$ super-Cotton tensor \cite{BKNT-M1}
has never been computed.
Our goal in this section is to construct $W_\a(H)$
and its higher-rank extension $W_{\a(n)} (H)$
using the superspin projection operator \eqref{TMSHNProjN3}.

In accordance with \eqref{1.9}, the super-Cotton tensor $W_{\a(n)}(H)$ in $\mb{M}^{3|6}$ is given by
\bea
W_{\a(n)} \big(H\big) =  \D^{n+2} \, \bm{\P}^{\perp}_{[n]}H_{\a(n)}~.
\label{W3}
\eea
When $W_{\a(n)}(H)$ is represented in the form \eqref{W3}, both gauge invariance \eqref{1.7} and transversality \eqref{TMSCottTrans} are made manifest as a consequence of the properties  \eqref{TMSGeneralProjectorTransverse}, \eqref{TMSProjectorPropertiesIndividual} and  \eqref{TMSTransverseProj}.

Making use of explicit form for the superprojector \eqref{TMSHNProjN3}, the expression \eqref{W3} becomes
\bea
W_{\a(n)}(H) = \frac{1}{8 \Box }\Big(-\frac{\ri }{6 \Box}\Big)^n \Delta^{2(n+1)} \big(9 \Delta^2 -\Box \big) D^{\b_1 I_1}D^{I_1}_{\a_1} \cdots D^{\b_n I_n} D^{I_n}_{\a_n} H_{\b(n)}~.
\label{4.7}
\eea
It appears that the super-Cotton tensor is not local, due to the presence of inverse $\Box$. However, making use of the identity \eqref{4.6b}, it follows that the super-Cotton tensor \eqref{4.7} can be written in the manifestly local form
\bea
W_{\a(n)}(H) = \frac{1}{8}\Big(-\frac{\ri}{6} \Big)^n \big(9 \Delta^2 - \Box\big) D^{\b_1 I_1}D^{I_1}_{\a_1} \cdots D^{\b_n I_n} D^{I_n}_{\a_n} H_{\b(n)}~.
\label{TMS4Cott}
\eea

In the $n=1$ case, the field strength $W_{\a}$ corresponds to the linearised version of
the ${\cal N}=3$ super-Cotton tensor. Thus, the field strength \eqref{W3} (or equivalently \eqref{TMS4Cott}) can be referred to as the rank-$n$ super-Cotton tensor. 

\subsubsection{Superconformal gravitino multiplet}
In accordance with the analysis of $\cN=3$ supermultiplets of conserved currents  \cite{BKS2}, the superconformal gravitino multiplet should be described by a real scalar 
gauge prepotential $H$ of dimension $-1$, which is defined modulo 
the gauge transformations 
\bea \label{TSMCottinoGT}
\d_\z H = \ri D^{\a I} D^{J}_{\a} \z^{IJ} ~, 
\qquad \z^{IJ} = \z^{JI}~, \quad \z^{II}=0~.
\eea
Associated with $H$ is a primary descendant $W(H) $ of dimension $+1$, which 
has the following properties: 
\begin{enumerate}
	\item $W(H)$ is gauge invariant
	\bsubeq \label{TMSN3CottinoConstraints}
	\bea
	\d_\z W(H)=0~.
	\eea
	\item $W(H)$ obeys the constraint 
	\bea
	(D^{\a I} D^{J}_{\a} - 2 \ri \d^{IJ} \Delta)W(H)=0~.
	\label{511}
	\eea
	\esubeq
\end{enumerate}
We normalise this super-Cottino tensor as
\bea
W(H)= \frac{1}{8}(9\D^2 -\Box)H~.
\label{512}
\eea
It can be shown that the super-Cottino tensor \eqref{512} is consistent with the conditions \eqref{TMSN3CottinoConstraints}. 
Note that $W(H)$ naturally follows from the definition of the $\cN=3$ rank-$n$ super-Cotton tensor \eqref{TMS4Cott} in the case $n=0$, with $\bm{\P}^{\perp}_{(0)}$ being identified as the scalar operator $\mc{P}$ 
\bea \label{TMSGravitinoProjector}
\cP = \frac{1}{8\Box}(9\D^2 -\Box)~.
\eea
It follows from \eqref{4.6a} that $\cP$ is indeed a projector, $ \cP^2 = \cP$. 

A linearised gauge-invariant action for the $\cN = 3$ superconformal gravitino multiplet
is fixed up to an overall constant. We propose the following action
\bea
S [  H] = \hf
\int \rd^{3|6}z \, H W (H) ~,
\label{514}
\eea
to describe the dynamics of the $\cN=3$ superconformal gravitino multiplet.


\subsection{Linearised higher-spin super-Cotton tensor in $\mb{M}^{3|8}$}

In accordance with \eqref{2.16}, the super-Cotton tensor $W_{\a(n)}(H)$ in $\mb{M}^{3|8}$ is fixed, modulo an overall factor, in the form
\bea
W_{\a(n)} \big(H\big) =  \D^{n+3} \, \bm{\P}^{\perp}_{[n]}H_{\a(n)}~,
\label{W4}
\eea
where $\bm{\P}^{\perp}_{[n]}$ is the $\cN=4$ superprojector \eqref{TMSHRProjN4}.
Substituting $\bm{\P}^{\perp}_{[n]}$ into  \eqref{W4} gives
\bea
W_{\a(n)}(H) =  \frac{1}{3 \Box }\Big (-\frac{\ri}{8\Box} \Big )^n  \D^{2n+3} (4 \Delta^2 - \Box) D^{\b_1 I_1} D_{\a_1}^{I_1} \cdots D^{\b_n I_n} D_{\a_n}^{I_n}  \F_{\b(n)}~.
\label{TMSCT5}
\eea

The super-Cotton tensor \eqref{TMSCT5} can be simplified in such a way that the resulting expression is manifestly local. 
Using the identity \eqref{TMSDeltaProperty3}, it follows that the super-Cotton tensor \eqref{TMSCT5} can be written in the manifestly local form
\bea
W_{\a(n)}(H) = \frac{1}{3}\Big(-\frac{\ri}{8} \Big)^n \D \big(4 \Delta^2 - \Box\big) 
D^{\b_1 I_1}D^{I_1}_{\a_1} \cdots D^{\b_n I_n} D^{I_n}_{\a_n} H_{\b(n)}~.
\label{5.9}
\eea
The superconformal field strength \eqref{W4}, or equivalently \eqref{5.9},
can be called the higher-rank $\cN =4$ super-Cotton tensor.

\subsubsection{Linearised $\cN=4$ conformal supergravity}
In the ${\cal N}=4$ case, the  linearised super-Cotton tensor proves to be a real scalar superfield, which obeys the following condition \cite{BKNT-M1}
\bea
(D^{\a I} D^{J}_{\a} - 2 \ri \d^{IJ} \Delta)W(H) =0~.
\label{TMS4Cont}
\eea
In accordance with \eqref{1.9}, both $W$ and $H$ are primary superfields 
of dimension 1 and $-2$, respectively.
The super-Cotton tensor is given by 
\bea
W (H) = \frac{\D}{3}(4 \Delta^2 - \Box) H~.
\label{6.122}
\eea
It can be shown that \eqref{6.122} is a solution to \eqref{TMS4Cont}. 
The super-Cotton tensor $W (H)$ can be obtained from the rank-$n$ super-Cotton tensor \eqref{5.9} in the case $n=0$. Note that in \eqref{5.9}, the superprojector $\bm{\P}_{(0)}$ is given by the scalar operator $\cP$
\bea \label{TMSN4Projector}
\cP = \frac{\D^2}{3\Box^2} (4\D^2 - \Box)~.
\eea
As a direct consequence of \eqref{5.3d} and \eqref{5.7}, we deduce that the $\cP $ a projector, $\cP^2 = \cP$.  

We define the linearised action for $\cN=4$ conformal supergravity to be
\bea
S [  H] = \hf
\int \rd^{3|8}z \, H W (H) ~,
\label{6.133}
\eea
which is invariant under the gauge transformations 
\bea
\d_\z H = \ri D^{\a I} D^{J}_{\a} \z^{IJ} ~, 
\qquad \z^{IJ} = \z^{JI}~, \quad \z^{II}=0~.
\eea


\subsection{Linearised higher-spin super-Cotton tensor in $\mb{M}^{3|10}$}\label{TMSCottN5}
In accordance with \eqref{2.16}, the higher-spin super-Cotton tensor $W_{\a(n)}(H)$ assumes the following form in $\mb{M}^{3|10}$
\bea
W_{\a(n)} \big(H\big) =  \D^{n+4} \, \bm{\P}^{\perp}_{[n]}H_{\a(n)}~,
\label{W5}
\eea
where $\bm{\P}^{\perp}_{[n]}$ is the superprojector \eqref{TMSHRPROHN5}. The explicit form of \eqref{W5} reads
\bea
W_{\a(n)}(H) &=& \Big(-\frac{\ri}{3840\Box^3}\Big)^n  \Delta^{2(n+2)}\big ( 25 \Delta^2 -  \Box \big )^n \big (25 \D^2 - 9 \Box \big)^n
\non\\
\qquad &&\times D^{\b_1 I_1}D^{I_1}_{\a_1} \cdots D^{\b_n I_n} D^{I_n}_{\a_n} H_{\b(n)}~.
\label{6.6}
\eea
It is possible to express \eqref{6.6} in a manifestly local form. 
First, equation \eqref{6.3d} leads to the following relation 
\bea
\D^{2n} \big ( 25 \Delta^2 -  \Box \big )^n \big (25 \D^2 - 9 \Box \big)^n &=& 384^{n-1} \Box^{3(n-1)} \D^2 \big ( 25 \Delta^2 -  \Box \big ) \big (25 \D^2 - 9 \Box \big)~, \hspace{0.5cm}
\label{6.7}
\eea
for $n \geq  1$.
As a result, using \eqref{6.7} and \eqref{TMSDeltaProperty3}, one may show 
that \eqref{6.6} reduces to 
\bea
W_{\a(n)}(H) = \frac{1}{384} \Big( -\frac{\ri}{10} \Big)^{n} 
\big(25 \Delta^2 - \Box \big ) \big ( 25 \D^2 -9\Box \big)  D^{\b_1 I_1}D^{I_1}_{\a_1} \cdots D^{\b_n I_n} D^{I_n}_{\a_n} H_{\b(n)}~. \hspace{0.7cm}
\label{7.77}
\eea


\subsection{Linearised higher-spin super-Cotton tensor in $\mb{M}^{3|12}$} \label{TMSCottN6}
In accordance with \eqref{2.16}, the super-Cotton tensor $W_{\a(n)}(H)$ takes the following form
\bea
W_{\a(n)} \big(H\big) =  \D^{n+5} \, \bm{\P}^{\perp}_{[n]}H_{\a(n)}~,
\label{W6}
\eea
where $\bm{\P}^{\perp}_{[n]}$ is the higher-rank superprojector \eqref{TMSN6Superprojector}.
The field strength $W_{\a(n)}$  \eqref{W6} takes the explicit form
\bea
W_{\a(n)}(H) &=& \Big(-\frac{\ri}{480\Box^3}\Big)^n \Delta^{2n+5}\big(9 \Delta^2 - \Box \big )^n \big ( 9 \D^2 - 4\Box \big)^n
\non\\
\qquad &&\times D^{\b_1 I_1}D^{I_1}_{\a_1} \cdots D^{\b_n I_n} D^{I_n}_{\a_n} H_{\b(n)}~.
\label{7.6}
\eea
Again, we wish to express \eqref{7.6} in a manifestly local form. To do this, we need to make use of the important relation
\bea
\Delta^2 \big(9 \Delta^2 - \Box \big )^n \big ( 9 \D^2 - 4\Box \big)^n = 40^{n-1} \Box^{2(n-1)} \D^2 \big(9 \Delta^2 - \Box \big ) \big ( 9 \D^2 - 4\Box \big)~,
\label{7.7}
\eea
which follows from the identity \eqref{7.3d}. 
As a result, using \eqref{7.7} in conjunction with \eqref{TMSDeltaProperty3}, one may show that $W_{\a(n)}(H) $ can be expressed in the manifestly local form
\bea
W_{\a(n)}(H) = \frac{1}{40} \Big(-\frac{\ri}{12} \Big)^n \D \big(9 \Delta^2 - \Box \big ) \big ( 9 \D^2 - 4\Box \big) D^{\b_1 I_1}D^{I_1}_{\a_1} \cdots D^{\b_n I_n} D^{I_n}_{\a_n} H_{\b(n)}~.
~~~
\label{8.66}
\eea

Recall that in the context of conformal supergravity with $\cN \geq 5$, the super-Cotton tensors $\cW_{IJKL}$ \eqref{2.666} possess only isospinor indices. However, it is clearly evident that the proposed $\cN=5$ \eqref{7.77}  and $\cN=6$ \eqref{8.66} higher-spin Cotton tensors do not have the correct tensorial form. We still call them linearised Cotton-tensors, in the sense that they satisfy the defining conditions \eqref{TMSCTProp}. It appears that the super-Cotton tensors $\cW_{IJKL}$ cannot be derived via superprojectors, thus an alternative method is required for their computation.

\section{Massless higher-spin theories in $\mb{M}^{3|2}$} \label{TMSSecMassless}


In this section we review the (half-)integer massless theories in $\mb{M}^{3|2}$, which were first studied by Kuzenko and Tsulaia  in \cite{KuzenkoTsulaia2017}. As discussed in section \ref{TMSMasslessSupermultiplets}, an off-shell higher-spin gauge theory in $\mb{M}^{3|2}$ is said to be massless if it does not propagate any physical degrees of freedom on the equations of motion. These theories can be constructed solely in terms of two superfields:  a superconformal gauge prepotential $H_{\a(n)}$, which was introduced in section \ref{TMSSecCHS};\footnote{ Note that when we refer to a massless theory in $\mb{M}^{3|2}$ as integer or half-integer, we are referring to the index structure of $H_{\a(n)}$ possessing an even or odd number of indices respectively.} and some compensating multiplet. Let us begin by analysing the equations of motion for the action describing a massless integer superspin multiplet.


\subsection{Massless integer superspin action in $\mb{M}^{3|2}$}
For integers $s \geq 2$, the model for the massless integer superspin multiplet is realised in terms of the real unconstrained superfields \cite{KuzenkoTsulaia2017}
\be
\cV_{\text{FO}} = \big \lb H_{\a(2s)} , Y_{\a(2s-2)} \big \rb ~,
\ee
which are defined modulo gauge transformations
\bsubeq \label{TMSMasslessIntegerGT}
\bea
\d_\L H_{\a(2s)} &=& D_\a \L_{\a(2s-1)}~, \label{TMSMasslessIntegerActionGT1} \\
\d_\L Y_{\a(2s-2)} &=& \frac{1}{2s} D^\b \L_{\b\a(2s-2)}~. \label{TMSMasslessIntegerActionGT2}
\eea
\esubeq 
Here the gauge parameter $\L_{\a(2s-1)}$ is real unconstrained. The massless integer superspin multiplet is described by the  gauge-invariant action \cite{KuzenkoTsulaia2017}
\bea \label{TASFirstOrderAction}
S_{(s)}^{\text{FO}}[H,Y] &&= \bigg (-\frac{1}{2} \bigg )^s \frac{\ri}{2} \AMST \Big \lb H^{\a(2s)} D^2 H_{\a(2s)} -D_\b H^{\b\a(2s-1)}D^\g H_{\g\a(2s-1)} \non \\
&&+(2s-1) \Big ( Y^{\a(2s-2)} D^2 Y_{\a(2s-2)} +2(s-1)D_\b Y^{\b \a(2s-3)} D^\g Y_{\g \a(2s-3)} \Big )  \non \\
&&+ 2 \ri (2s-1) Y^{\a(2s-2)}\pa^{\b(2)}H_{\b(2)\a(2s-2)} \Big \rb ~. \hspace{0.5cm}
\eea

Let us show that the model \eqref{TASFirstOrderAction} describes no propagating degrees of freedom by performing an on-shell analysis at the level of superfields. The equations of motion corresponding to \eqref{TASFirstOrderAction} are 
\bsubeq \label{TMSMasslessIntegerEOM}
\bea
0 &=& 2\ri \pa_\a{}^\b H_{\b\a(2s-1)}+D^2H_{\a(2s)}-2\ri(2s-1)\pa_{\a(2)}Y_{\a(2s-2)}~, \\
0 &=& \ri \pa^{\b(2)} H_{\b(2)\a(2s-2)} + s D^2 Y_{\a(2s-2)} -2\ri(s-1) \pa_\a{}^\b Y_{\b\a(2s-3)}~.
\eea
\esubeq
The gauge freedom \eqref{TMSMasslessIntegerActionGT2} can be used to gauge away the superfield $Y_{\a(2s-2)}$
\be
Y_{\a(2s-2)} = 0~. \label{TMSMasslessIntegerGC1}
\ee
The residual gauge freedom is described by the parameter $\L_{\a(2s-1)}$ constrained by
\be
D^\b \L_{\b \a(2s-2)} = 0 \quad \Longrightarrow \quad \pa^{\b\g}\L_{\b\g\a(2s-3)} = 0~, \quad  \ri \pa_\a{}^\b \L_{\b\a(2s-2)} = \hf D^2 \L_{\a(2s-1)}~. \label{TMSMasslessIntegerRG}
\ee
In the gauge \eqref{TMSMasslessIntegerGC1}, the equations of motion \eqref{TMSMasslessIntegerEOM} reduce to 
\bsubeq \label{TMSMasslessIntegerEOMGC1}
\bea
0 &=& 2\ri \pa_\a{}^\b H_{\b\a(2s-1)}+D^2H_{\a(2s)}~, \label{TMSMasslessIntegerEOMGC1a}\\
0 &=&  \pa^{\b(2)} H_{\b(2)\a(2s-2)} ~.
\eea
\esubeq

According to \eqref{TMSMasslessIntegerActionGT1}, in conjunction with the constraint \eqref{TMSMasslessIntegerRG}, it can be shown that $D^\b H_{\b\a(2s-1)}$ transforms as
\be
\d_\L (D^\b H_{\b\a(2s-1)}) = D^2 \L_{\a(2s-1)}~. \label{TMSMasslessIntegerGTR}
\ee
It follows from \eqref{TMSMasslessIntegerGTR} that $D^\b H_{\b\a(2s-1)}$ can be completely gauged away
\be
D^\b H_{\b\a(2s-1)} = 0 \quad \Longrightarrow \quad \ri \pa_\a{}^\b H_{\b\a(2s-1)} = \hf D^2 H_{\a(2s)} ~. \label{TMSMasslessIntegerGC2}
\ee
The corresponding residual gauge freedom further constrains $\L_{\a(2s-1)}$  by 
\be
D^2 \L_{\a(2s-1)} = 0 \quad \Longrightarrow \quad \Box \L_{\a(2s-1)} = 0~.
\ee
In the gauges \eqref{TMSMasslessIntegerGC1} and \eqref{TMSMasslessIntegerGC2}, the equation of motion \eqref{TMSMasslessIntegerEOMGC1a} reduces to
\be
0 = D^2 H_{\a(2s)}~.
\ee

In summary, upon imposing the gauges \eqref{TMSMasslessIntegerGC1} and \eqref{TMSMasslessIntegerGC2}, we are left with the  field $H_{\a(2s)}$ which satisfies the conditions
\bsubeq \label{TMSOnshellanalysis}
\be 
D^\b H_{\b\a(2s-1)} = 0~, \qquad D^2 H_{\a(2s)} = 0~.
\ee
The superfield $ H_{\a(2s)}$ supports the gauge symmetry 
\be
\d_\L H_{\a(2s)} = D_\a \L_{\a(2s-1)}~,
\ee
where the gauge parameter $\L_{\a(2s-1)}$ is constrained via the residual gauge symmetry
\be
D^\b \L_{\b \a(2s-2)} = 0~, \qquad D^2 \L_{\a(2s-1)} = 0~.
\ee
\esubeq
The conditions \eqref{TMSOnshellanalysis} coincide with those describing a massless superfield \eqref{TMSMasslessOnshellConditions1} for $n=2s$. Thus, in accordance with the analysis of section \ref{TMSMasslessSupermultiplets}, the superfield $H_{\a(2s)}$ carries no physical propagating degrees of freedom on-shell and hence the action \eqref{TASFirstOrderAction} describes a massless superfield with superspin $s$.

Alternatively, one could analyse the component structure of the massless integer superspin multiplet in order to determine that it propagates no physical degrees of freedom.
The component analysis of the model describing the massless integer superspin multiplet \eqref{TASFirstOrderAction} was completed in \cite{KuzenkoPonds2018}. Here it was shown that the gauge freedom \eqref{TMSMasslessIntegerGT} can be used to impose the following Wess-Zumino gauge
\be  \label{TMSGCLI}
H_{\a(2s)}| = 0~, \qquad D^\b H_{\b\a(2s-1)}|=0~, \qquad Y_{\a(2s-2)}| =0~.
\ee
The residual gauge symmetry which preserves  \eqref{TMSGCLI} are
\begin{subequations} \label{TMSRGLI}
	\bea
	0 &=& D_{(\a_1}\z_{\a_2 ... \a_{2s})}|~,\\
	0 &=& D^\b \z_{\b\a(2s-2)}|~,\\
	D^2 \z_{\a(2s-1)}| &=&- \frac{2 \ri }{2s+1}(2s-1)\pa_{(\a_1}{}^\b \z_{\b\a_2 ... \a_{2s-1})} |~.
	\eea
\end{subequations}
The relations \eqref{TMSRGLI} indicate that there is only one independent gauge parameter, which we choose to define as
\be
\x_{\a(2s-1)} := \z_{\a(2s-1)}|~.
\ee
Thus in the gauge \eqref{TMSGCLI}, we are left with the remaining independent component fields:
\begin{subequations} 
	\bea
	h_{\a(2s+1)}&:=&- \ri D_{(\a_1}H_{\a_2 ... \a_{2s+1})}|~, \\
	h_{\a(2s)}&:=&\frac{\ri}{4}D^2 H_{\a(2s)}|~, \\
	y_{\a(2s-2)}&:=&\frac{\ri}{4}D^2 Y_{\a(2s-2)}|~, \\
	y_{\a(2s-1)} &:=& \frac{\ri}{2} D_{(\a_1}Y_{\a_2 ... \a_{2s-1})}|~, \\
	z_{\a(2s-3)}&:=&-2\ri s D^\b Y_{\b \a(2s-3)}|~. 
	\eea
\end{subequations}
It is easy to show that the gauge transformation laws for the bosonic fields are
\bea
\d h_{\a(2s)} = 0~, \qquad \d y_{\a(2s-2)} = 0~.
\eea

Applying the component reduction procedure to the action \eqref{TASFirstOrderAction}  yields the following decoupled theories in $\mb{M}^3$
\bea \label{TMSMasslessIntegerCompAction}
S^{\text{FO}}_{(s)}[H ,Y ] &=& \Big ( -\frac{1}{2} \Big )^s \int \rd^3 x \,  \Big \{ h^{\a(2s)} h_{\a(2s)} + 2s(2s-1)y^{\a(2s-2)}y_{\a(2s-2)} \Big \} \non \\
&&+ S^{\text{FF}}_{(s+\hf)}\,[h,y,z]~.
\eea
The bosonic sector is auxiliary, since the fields $h_{\a(2s)}$ and $y_{\a(2s-2)}$ appear with no derivatives in the action \eqref{TMSMasslessIntegerCompAction}. The action $S^{\text{FF}}_{(s+\hf)}\,[h,y,z]$ coincides with the Fang-Fronsdal action \eqref{TMFangFronsdalAction}, which was shown in section \ref{SecThreeDimensionalMinkowskiSpace} to be a massless theory. Thus, the massless integer superspin action \eqref{TMSMasslessIntegerCompAction} describes no propagating degrees of freedom at the component level, as required. From this point onwards, we will always determine whether a higher-spin gauge theory in $3d$ (super)space is massless by analysing its component structure.

\subsection{Massless half-integer superspin action in $\mb{M}^{3|2}$}
For $s \geq 2$, the model for the massless half-integer superspin multiplet is realised in terms of the real unconstrained superfields \cite{KuzenkoTsulaia2017}
\be
\cV_{\text{SO}} = \big \lb H_{\a(2s+1)} , Y_{\a(2s-2)} \big \rb ~.
\ee
These superfields are defined modulo gauge transformations
\bsubeq \label{TSMMasslessGT}
\bea
\d_\L H_{\a(2s+1)} &=& \ri D_\a \L_{\a(2s)}~, \label{TMSSOGTH}\\
\d_\L Y_{\a(2s-2)} &=& \frac{s}{2s+1} \pa^{\b(2)}\L_{\b(2)\a(2s-2)}~, \label{TMSSOGTY}
\eea
\esubeq
where the gauge parameter $\L_{\a(2s)}$ is real unconstrained. 

The massless half-integer superspin multiplet is described by the gauge-invariant action \cite{KuzenkoTsulaia2017}
\begin{align}\label{TMSSecondOrderMassless}
S_{(s+\hf)}^{\text{SO}}&[H,Y]= \bigg (-\frac{1}{2} \bigg )^s \AMST \Big \lb - \frac{\ri}{2} H^{\a(2s+1)} \Box H_{\a(2s+1)} - \frac{\ri}{8}D_\b H^{\b \a(2s)} D^2 D^\g H_{\g\a(2s)} \non \\
&+ \frac{\ri s}{4} \pa_{\b\g} H^{\b \g \a(2s-1)}\pa^{\d \l} H_{\d \l \a(2s-1)} + (2s-1) Y^{\a(2s-2)} \pa^{\b\g}D^\d H_{\b\g\d \a(2s-2)} \non \\
&+ 2\ri (2s-1) \Big ( Y^{\a(2s-2)}D^2 Y_{\a(2s-2)} - \frac{1}{s}(s-1) D_\b Y^{\b\a(2s-3)}D^\g Y_{\g \a(2s-3)} \Big) \Big \rb ~. 
\end{align}
As stated in the previous section, we will study the component structure of the action \eqref{TMSSecondOrderMassless} in order to determine that it does not describe any physical degrees of freedom on-shell.\footnote{The superfield analysis of the massless half-integer superspin action \eqref{TMSSecondOrderMassless} was completed in \cite{KuzenkoTsulaia2017}.} Note that the component analysis for the massless half-integer superspin multiplet in $\mb{M}^{3|2}$ was first completed in \cite{KuzenkoPonds2018}.

The gauge freedom \eqref{TSMMasslessGT} can be used to impose the following Wess-Zumino gauge
\be \label{TMSgc}
H_{\a(2s+1)}| = 0~, \qquad D^\b H_{\b\a(2s)}|=0~.
\ee
The residual gauge freedom which preserves the conditions \eqref{TMSgc} is described by 
\begin{subequations}
	\bea
	0 &=& D_{(\a_1}\z_{\a_2 ... \a_{2s+1})}|~,\\
	D^2 \z_{\a(2s)}| &=&- \frac{2 \ri s}{s+1} \pa^{\b}{}_{(\a_1} \z_{\a_2 \ldots \a_{2s})\b}|~.
	\eea
\end{subequations}
These conditions imply that there are only two independent gauge parameters, which we define as follows:
\be
\x_{\a(2s)} := \z_{\a(2s)}|~, \qquad \l_{\a(2s-1)}:=-\frac{\ri s}{2s+1}D^\b \z_{\b\a(2s-1)}|~.
\ee
Upon imposing the gauge \eqref{TMSgc}, we are left with the following  component fields:
\begin{subequations} \label{cf}
	\bea
	h_{\a(2s+1)}&:=&\frac{\ri}{4}D^2 H_{\a(2s+1)}|~, \\
	h_{\a(2s+2)}&:=&-D_{(\a_1}H_{\a_2 ... \a_{2s+2})}|~, \\
	y_{\a(2s-2)}&:=&-4Y_{\a(2s-2)}|~, \\
	y_{\a(2s-1)} &:=& (2s+1)\ri D_{(\a_1}Y_{\a_2 ... \a_{2s-1})}|~, \\
	z_{\a(2s-3)}&:=&2\ri D^\b Y_{\b \a(2s-3)}|~, \\
	F_{\a(2s-2)}&:=&\frac{\ri}{4}D^2 Y_{\a(2s-2)}|~.
	\eea
\end{subequations} 

Performing a component reduction, 
we find that the higher-spin model \eqref{TMSSecondOrderMassless} decouples into a bosonic and a fermionic sector. The bosonic action is given by
\bea 
&&S_{\text{bos}}[h,y,F] = \Big ( -\frac{1}{2} \Big )^s \int \rd^3 x\, \bigg \{ - \frac{1}{4}h^{\a(2s+2)} \Box h_{\a(2s+2)} +\frac{1}{8}(s+1)\pa_{\b\g}h^{\b\g\a(2s)} \pa^{\d\l}h_{\d\l\a(2s)} ~ \non \\
&&+\frac{1}{8}(2s-1)y^{\a(2s-2)}\pa^{\b\g}\pa^{\d\l}h_{\b\g\d\l\a(2s-2)}+\frac{1}{16s}(2s-1)(s+1)y^{\a(2s-2)}\Box y_{\a(2s-2)}~ \non \\
&&+ \frac{1}{s}(s-1)(2s-1)F^{\b\a(2s-3)}\pa_\b{}^\g y_{\g\a(2s-3)}+\frac{4}{s}(s+1)(2s-1)F^{\a(2s-2)}F_{\a(2s-2)}~  \bigg \}~. 
\eea
Here the field $F_{\a(2s-2)}$ is auxiliary, so upon elimination via its equation of motion
\be 
F_{\a(2s-2)}=-\frac{1}{8(s+1)}(s-1)\pa_{(\a_1}{}^\b y_{\a_2 ... \a_{2s-2})\b}  ~,
\ee
we find that the resulting action coincides with the massless spin-$(s+1)$ action
$S_{(s+1)}^{\text{F}}[h,y] $, which is given by \eqref{TMFronsdalAction} when setting  $s \rightarrow s+1$.

It can be shown that the fermionic action emerging from the reduction procedure coincides with the Fang-Fronsdal type action $S^{\text{FF}}_{(s+\hf)}[h, y, z]$ \eqref{TMFangFronsdalAction}. 
In summary, the massless half-integer superspin action \eqref{TMSSecondOrderMassless} decouples into the Fronsdal  $S_{(s+1)}^{\text{F}}[h, y ] $ and Fang-Fronsdal  $S^{\text{FF}}_{(s+\hf)}[h, y, z]$ actions  at the component level
\bea
S^{\text{SO}}_{(s+\hf)}[H ,Y] = S_{(s+1)}^{\text{F}}[h, y] + S^{\text{FF}}_{(s+\hf)}[h, y, z]~.
\label{b40}
\eea
It was shown in section \ref{SecThreeDimensionalMinkowskiSpace} that such theories describe no propagating degrees of freedom on-shell, hence the action \eqref{TMSSecondOrderMassless} is indeed massless.

\section{New topologically massive theories} \label{TMSSecMassive}
New topological massive (NTM) theories were first proposed in $\mb{M}^{3|2}$ by Kuzenko and Ponds in \cite{KuzenkoPonds2018} as supersymmetric generalisations of the NTM models \eqref{TMHSNTMG}  in $\mb{M}^3$. These theories are generated by taking an appropriate deformation of the superconformal action \eqref{action} such that the corresponding equations of motion coincide identically with the massive on-shell conditions \eqref{TMSMassiveOnshell}.

Given an integer $n \geq 1$, we propose that the NTM actions take the following form in $\mb{M}^{3|2\cN}$ 
\bea
{S}^{(n|\cN)}_{\rm NTM} [  H_{\a(n)}] \propto \frac{ \ri^n}{2} \frac{1}{m}
\int \rd^{3|2\cN}z \, H^{\a(n)} \big( \D - m \s \big)
{W}_{\a(n)}\big(H\big) ~,
\label{TMSNewTopoMassive}
\eea
where $\s := \pm 1$ and $m$ is an arbitrary mass parameter. Here, the superfield $H_{\a(n)}$ is the superconformal gauge prepotential and $W_{\a(n)}(H)$ is the higher-spin super-Cotton tensor \eqref{2.16}. The action \eqref{TMSNewTopoMassive} is invariant under the the gauge transformations \eqref{TMSSCHSGT}. 

The equation of motion obtained by varying \eqref{TMSNewTopoMassive} with respect to $H_{\a(n)}$ is
\be \label{TMSMassiveCottonTensor}
\big( \D - m \s \big){W}_{\a(n)}(H) = 0~.
\ee
In accordance with ${W}_{\a(n)}(H)$ being transverse \eqref{1.8} by definition, in conjunction with \eqref{TMSMassiveCottonTensor}, it follows from \eqref{TMSMassiveOnshell} that the field strength $W_{\a(n)}(H)$ describes a massive $\cN$-extended supermultiplet which carries mass $m$, spin $\frac{n}{2}$ and superhelicity $\k = \frac{\s}{2} \big (n + \frac{\cN}{2} \big )$.

It is worth taking a moment to comment on an alternative formulation for off-shell theories describing massive supermultiplets in $\mb{M}^{3|2}$, known as topologically massive theories. The corresponding actions of these theories are built by appropriately coupling the off-shell superconformal action \eqref{action} with the actions describing massless integer \eqref{TASFirstOrderAction}  and half-integer \eqref{TMSSecondOrderMassless} superspin multiplets. In topologically massive theories,  the massless actions are supplemented with a coupling constant which carries mass dimension.

Topologically massive higher-spin theories were recently formulated in $\mb{M}^3$ by Kuzenko and Ponds in \cite{KuzenkoPonds2018}\footnote{Topologically massive higher-spin theories in AdS$_3$ were also computed in \cite{KuzenkoPonds2018}.} as higher-spin extensions
of topologically massive gravity \cite{DJT1,DJT2}.
Similarly, off-shell\footnote{The  on-shell formulations for
	massive higher-spin $\cN=1$ supermultiplets in 
	${\mathbb M}^{3}$ and AdS${}_3$ were  developed in \cite{BSZ3,BSZ4}
	by combining the massive bosonic 
	and fermionic higher-spin actions described in \cite{BSZ1,BSZ2}.
	The formulations given in \cite{BSZ1,BSZ2,BSZ3,BSZ4}
	are based on the gauge-invariant approaches 
	to  the dynamics of massive higher-spin fields, which were advocated by Zinoviev \cite{Zinoviev} and Metsaev \cite{Metsaev}.}   topologically massive higher-spin gauge theories were first constructed in $\mb{M}^{3|2}$ by Kuzenko and Tsulaia \cite{KuzenkoTsulaia2017} as extensions of $\cN=1$ topologically massive supergravity \cite{DK,Deser84}. These actions were shown to take the form\footnote{These topologically massive  theories \eqref{TMSMassiveActions} were extended to $\mb{M}^{3|4}$ in \cite{KuzenkoOgburn2016,HutomoKuzenkoOgburn2018}.} 
\bsubeq \label{TMSMassiveActions}
\bea
S^{\parallel}_{(s)}[H_{\a(2s)} ,V_{\a(2s-2)} |m]
&=&  {S}^{(2s|1)}_{\text{SCHS}} [ H_{\a(2s)}] 
+m^{2s-1}S_{(s)}^{\text{FO}}[\cV_{\text{FO}} ]~, \label{TMSMassiveActionInteger}\\
S^{\parallel}_{(s+\hf)}[H_{\a(2s+1)} ,Y_{\a(2s-2)} |m]
&=& {S}^{(2s+1|1)}_{\text{SCHS}} [ H_{\a(2s+1)}] 
+m^{2s-1}S_{(s+\hf)}^{\text{SO}}[\cV_{\text{SO}}]~, \label{TMSMassiveActionHalfInteger}
\eea
\esubeq
where $m$ is a coupling constant with mass dimension one. The superfield analysis of the topologically massive theories \eqref{TMSMassiveActions} was conducted in \cite{KuzenkoTsulaia2017}, where it was shown that on the equations of motion, the integer   \eqref{TMSMassiveActionInteger} and half-integer  \eqref{TMSMassiveActionHalfInteger} superspin theories  each describe the massive supermultiplet \eqref{TMSMassiveOnshell}, with $n=2s$ and $n=2s+1$, respectively. The same conclusion was arrived at via a component analysis  in \cite{KuzenkoPonds2018}.

\section{Summary of results} \label{TMSConclusion}
This chapter was dedicated to the construction of superspin projection operators in $\mb{M}^{3|2 \cN}$, for $1 \leq \mc{N} \leq 6$. In section \ref{TMSIrreducibleRepsPo}, we reviewed massive UIRs of the three-dimensional super-\Po algebra. In section \ref{TMSIrreducibleFieldRepresentations}, we presented the $\cN$-extended generalisation of the massive on-shell supermultiplets \eqref{TMSMassiveOnshell} which realise the massive UIR $\mb{D}(m : \s , \frac{n}{2} , \cN)$. These $\cN$-extended supermultiplets appeared for the first time in \cite{BHHK}. We also studied higher-spin massless superfields \eqref{TMSMasslessOnshellConditions1} in $\mb{M}^{3|2}$, which we defined as gauge superfields which do not propagate any physical degrees on the equations of motion. This feature was shown explicitly for massless superfields \eqref{TMSMasslessOnshellConditions1} at the level of superfields and component fields.

In section \ref{TMSSecSpinProjectors}, we computed the superspin projection operators $\mb{M}^{3|2 \cN}$, for $1 \leq \mc{N} \leq 6$ \cite{BHHK}. These superprojectors map any unconstrained superfield to the space of transverse superfields (pure superspin states) \eqref{TMSMassiveTransverse}. If the superspin projection operator is applied to a superfield which satisfies the first-order differential equation \eqref{TMSMassiveMassShell}, then the resulting massive superfield \eqref{TMSMassiveOnshell} furnishes the massive UIR $\mb{D}(m :\s, \frac{n}{2}, \cN)$. If instead the superfield satisfies the Klein-Gordon equation, then the projected field realises the reducible representation $\mb{D}  (m:-,\frac{n}{2}, \cN) \oplus \mb{D} (m:+,\frac{n}{2}, \cN )$. In order to extract out the UIR with definite superhelicity, one needs to make use of the superhelicity projectors \eqref{TMSSuperhelicityprojectors}.

In section \ref{TMSSecCHS},  we made use of the superspin projection operators to construct the linearised rank-$n$ super-Cotton tensor, for $n \geq 1$, in terms of the unconstrained gauge prepotential $H_{\a(n)}$. The general form of these super-Cotton tensors, which were presented for generic $\cN$ in \cite{BHHK}, are given by eq. \eqref{2.16}. In particular, a new expression for the rank-$n$ super-Cotton tensor in the case of $\mc{N} = 2$ supersymmetry was computed. This expression is much simpler than the one originally given in \cite{KuzenkoOgburn2016}. The rank-$n$ super-Cotton tensors, for $3 \leq  \cN  \leq 6$, were derived for the first time in \cite{BHHK}. The corresponding results are given by eqs. \eqref{TMS4Cott}, \eqref{5.9},  \eqref{7.77} and \eqref{8.66}, respectively. 

The linearised super-Cotton tensor of $\cN = 4$ conformal supergravity required special attention since it is a scalar superfield $W$. It was also computed in \cite{BHHK} for the first time, and is given by eq. \eqref{6.122}. Making use of $W$ allowed us to construct the linearised action for $\cN = 4$ conformal supergravity, which is given by eq. \eqref{6.133}. In the $\mc{N} = 3$ case, we also constructed the gauge-invariant action \eqref{514} which describes the dynamics of the superconformal gravitino multiplet. In the case of conformal supergravity with $\mc{N} \geq 5$, the super-Cotton tensor has a
different tensorial form than that of $W_{\a(n)}$, see eq. \eqref{2.666}. Consequently, it cannot be directly obtained from our results.

In section \ref{TMSSecMassless} we reviewed the formulation of massless (half-)integer superspin actions in $\mb{M}^{3|2 \cN}$, which were first given in \cite{KuzenkoTsulaia2017}. In section \ref{TMSSecMassive} we proposed $\cN$-extended generalisations for new topologically massive theories \cite{BHHK}, see eq. \eqref{TMSNewTopoMassive}. The computation of the linearised higher-spin Cotton tensors lead to explicit realisations of these models for $3 \leq \cN \leq 6$ for the first time  \cite{BHHK}.

\begin{subappendices}
	\section{Superconformal primary multiplets} \label{TASappendixSCHSPrimary} 
	The $\cN$-extended superconformal symmetry 
	in three dimensions was studied in detail by Park \cite{Park}.
	In this appendix our presentation follows \cite{KPT-MvU2011,BKS1}.
	In $\cN$-extended Minkowski superspace ${\mathbb M}^{3|2\cN}$, 
	superconformal transformations, $z^A \to z^A +\d z^A = z^A +\x^A(z)$, 
	are generated by superconformal Killing vectors. By definition, a superconformal Killing vector 
	\bea
	\x  = \x^{a} (z) \, \pa_{ a} 
	+  \x^{\a}_I(z) \, D_{\a}^I 
	\eea
	is a real vector field obeying the condition
	$[\x, D_\a^I ] \propto D_\b^J$. This condition implies
	\bea
	[\x, D_\a^I ] = -(D^I_\a \x^\b_J) D^J_\b =\hf  \o_\a{}^\b (z)D_\b^I + \L^{IJ}(z) D_\a^J -\hf \s (z) D^I_\a~,
	\label{master1}
	\eea
	where we have defined 
	\begin{subequations}
		\bea
		\o_{\a\b} &:=& -\frac{2}{\cN} D^J_{(\a} \x_{\b)}^J =-\frac{1}{2} \pa^\g{}_{(\a} \x_{\b )\g} ~,
		\label{Lorentz}\\
		\L^{IJ} &:=& -2 D_\a^{ [I} \x^{J]\a}~, \label{rotation}\\
		\s&:=& \frac{1}{\cN} D^I_\a \x^{\a I} =\frac{1}{3} \pa_a \x^a~.
		\label{scale}
		\eea
	\end{subequations}
	Here the parameters $\o_{\a \b} =\o_{\b \a}$, $\L^{IJ}=- \L^{JI}$ and $\s$ correspond
	to  $z$-dependent Lorentz, ${\rm SO} (\cN )$ and scale transformations.
	These transformation parameters are related to each other as follows:
	\begin{subequations}
		\bea
		D^I_\a \o_{\b \g} &=& 2 \ve_{ \a ( \b} D^I_{ \g)} \s~,  \label{Lorentz-scale}\\
		D^I_\a \L^{JK} &=& -2 \d^{ I [ J } D_\a^{ K] } \s~. \label{rotation-scale}
		\eea
	\end{subequations}
	
	Let  $\Phi_{\cal A}^{\cal I}(z)$ be a superfield that transforms
	in a
	representation $T$ of the Lorentz group with respect to its index $\cA$
	and in a representation $D$ of the $R$-symmetry group ${\rm SO}(\cN)$
	with respect to the index  $\cI$.
	Such a superfield is called primary of dimension $d$ if its
	superconformal transformation law is
	\bea
	\delta\Phi_{\cal A}^{\cal I} =
	-\xi\Phi_{\cal A}^{\cal I}-d \,\sigma \Phi_{\cal A}^{\cal I}
	+\hf \o^{\alpha\beta} (M_{\alpha\beta})_{\cal A}{}^{\cal B}
	\Phi_{\cal B}^{\cal I}
	+\hf \Lambda^{IJ}(R^{IJ})^{\cal I}{}_{\cal J}\Phi_{\cal A}^{\cal J}~.~~
	\eea
	The ${\rm SO}(\cN)$  generator $R^{IJ}$ acts on an ${\rm SO}(\cN)$-vector $V^K$ as
	\bea
	R^{IJ} V^K = 2 \d^{K[I} V^{J]}~.
	\eea
	Making use of this transformation law allows one to determine the dimensions in
	\eqref{1.9}.

\end{subappendices}

\chapter{Four-dimensional anti-de Sitter (super)space} \label{Chapter4DAdS}
It was demonstrated in chapters \ref{Chapter2} and \ref{ChapThreeDimensionalExtendedMinkowskiSuperspace} that the (super)spin projection operators possess many interesting applications in three- and four-dimensional Minkowski (super)space.
For example, they can be used to decompose an arbitrary (super)field into its irreducible components. Moreover, the (super)spin projection operators were also shown to be fundamental in the formulation of (S)CHS theory. Thus it is natural to consider extensions of these operators to curved backgrounds, with the simplest case being four-dimensional (anti-)de Sitter space (A)dS$_4$. However, when deriving the spin projection operators in (A)dS$_4$, one immediately encounters computational difficulties due to the presence of non-vanishing curvature. 

Despite these challenges, the spin projection operators were recently found in (A)dS$_4$ by Kuzenko and Ponds \cite{KP20} in 2020. One of the important outcomes of \cite{KP20} was a new understanding of the so-called partially massless fields in AdS$_4$. Specifically, it was shown that all necessary information required to generate partially massless fields is encoded within the poles of the spin projection operators. This chapter is devoted to the study of spin projection operators in four-dimensional anti-de Sitter space, and their supersymmetric generalisations to $\cN=1$ and $\cN=2$ AdS superspace.

This chapter is based on the publication \cite{BHKP}, and the work in progress \cite{Hutchings2022}. It is organised as follows. Section \ref{FAec4dAdS} is devoted to 
reviewing and expanding upon the spin projection operators and their corresponding applications in AdS$_4$. In subsection \ref{FARepTheory}, we provide a brief overview of the UIRs of the AdS$_4$ algebra $\mathfrak{so}(3, 2)$. Next, we introduce salient facts concerning field theory in AdS$_4$. In particular, we discuss how the UIRs of $\mathfrak{so}(3, 2)$ are realised on the space of totally symmetric fields. This is given in subsection \ref{FAIrredFieldReps}. In section \ref{FASpinProjectors}, we study the spin projection operators and their applications. Here, we also introduce an alternative, yet equivalent, form of the spin projection operators of \cite{KP20}. The novelty of these operators is that they are expressed solely in terms of the Casimir operators of $\mathfrak{so}(3, 2)$.  These operators are then used to recast the known CHS theory in a manifestly factorised form. This is accomplished in subsection \ref{FACHS}. 

Section \ref{FASsec4dAdS} is devoted to extending the results of section \ref{FAec4dAdS} to four-dimensional $\cN=1$ anti-de Sitter superspace AdS$^{4|4}$. In section \ref{FASRepTheory} we review the UIRs of the  $\cN=1$ AdS$_4$ superalgebra $\mathfrak{osp} (1|4)$. Essential aspects of AdS$^{4|4}$ are detailed in section \ref{N1AdSSuperspaceSec}, which will allow for a discussion of on-shell supermultiplets in section \ref{FASOnShellSupermultiplets}. The component structure of these on-shell supermultiplets are analysed in section \ref{FASComponentAnalysis}. We compute the superspin projection operators in AdS$^{4|4}$ and their corresponding applications in section \ref{Superspin-projection operators}.
In section \ref{FMSSCHSTheory} we elucidate the relationship between the superspin projection operators and SCHS theory.  In section \ref{Off-shellmodels} we study several off-shell models in AdS$^{4|4}$. Most notably, we present an off-shell model for the massive gravitino (superspin-$1$) multiplet.

Section \ref{FAS2sec4dAdS} is devoted to extending various results from sections \ref{FAec4dAdS} and \ref{FASsec4dAdS} to $\cN=2$ anti-de Sitter superspace AdS$^{4|8}$. In section \ref{FAS2Superspace}, we introduce pertinent aspects concerning  AdS$^{4|8}$. In section \ref{FAS2SCHS}, we review the free SCHS action in AdS$^{4|8}$, which was recently derived by Kuzenko and Raptakis in \cite{KR}. The superspin projection operators are then extracted from this SCHS theory in section \ref{FAS2Superprojectors}. Finally, we propose a dictionary for on-shell superfields in section \ref{FAS2Onshell} using the poles of the superspin projection operators. 
A summary of the results obtained is given in section \ref{FASsecSummary}. This chapter is also accompanied by two technical appendices, \ref{TASappendixA} and \ref{TASappendixB}. 
Appendix \ref{TASappendixA} contains a list of identities that are indispensable for the derivation of many of the results in section \ref{FASsec4dAdS}. Appendix \ref{TASappendixB} is devoted to the derivation of partially massless gauge transformations in AdS$^{4|4}$.

\section{Four-dimensional anti-de Sitter space}\label{FAec4dAdS}
In this section we review and elaborate upon the spin projection operators in AdS$_4$ \cite{KP20} and their corresponding applications. These operators are AdS$_4$ generalisations of the spin projectors in $\mb{M}^4$, which were detailed in section \ref{SecFourDimensionalMinkowskiSpace}. Before introducing the spin projection operators, it is necessary to introduce the irreducible representations of $\mathfrak{so}(3, 2)$.


\subsection{Irreducible representations of the anti-de Sitter algebra}\label{FARepTheory}
Anti-de Sitter space in $d$ dimensions AdS$_d$ can be identified as a hyperboloid embedded in $(d+1)$-dimensional ambient space $\mb{R}^{d+1}$
\be
\eta^{AB}X_A X_B = - l^2~.
\ee
Here the index of the ambient space coordinate $X_A$ take the values $A=0,1,2, \cdots ,d$ and $l$ is the radius of AdS$_d$, with $l^2 > 0$. The indices are raised and lowered by the ambient metric tensor $\eta^{AB} = \eta_{AB} = \text{diag}(-1,1, \cdots ,1,-1)$. 

The Lie algebra $\mathfrak{so}(d-1, 2)$ of the AdS$_d$ isometry group is spanned by the antisymmetric generators $J_{AB} = -J_{BA}$, which obey the commutation relations
\be \label{FAAlgebra}
[J_{AB},J_{CD}] = \ri \big (  \eta_{AC} J_{BD} - \eta_{AD} J_{BC}  + \eta_{BD} J_{AC} - \eta_{BC} J_{AD} \big )~.
\ee
The normalisation factor of the commutation relation  \eqref{FAAlgebra} is chosen to ensure that in the limit where the AdS$_d$ radius tends to infinity $l \rightarrow \infty$, the algebra \eqref{FAAlgebra} contracts to the \Po algebra \eqref{FMPoincareAlgebra}. This can be checked directly by splitting the generators of $\mf{so}(d-1,2)$ as follows
\be \label{FASplitGenerators}
J_{AB} \longrightarrow \big ( J_{ab}, J_{a d} \big ) \equiv  \big ( J_{ab}, l P_a\big ) ~,
\ee
where $a,b = 0,1, \cdots ,d-1$.\footnote{Note that the index denoted $d$ refers to the dimension $d$ and is not to be confused with a free index.}
The generators $J_{0d}$ and $J_{ab}$ in \eqref{FASplitGenerators} can be identified as the AdS$_d$ analogues of the energy and angular momentum operators respectively.

Let us now restrict our attention to four dimensions $d=4$.
The quadratic $C_1$ and quartic $C_2$ Casimir operators  of $\mf{so}(3,2)$ are \cite{Evans,Fronsdal:1974ew}
\be \label{FACasimirOperatorsRep}
C_1 = \hf J^{AB}J_{AB}~, \qquad C_2 = - W^A W_A~, \qquad [C_1,J_{AB}] = [ C_2, J_{AB} ] = 0~, 
\ee
where $W_A = \frac{1}{8} \ve_{ABCDE}J^{BC}J^{DE}$.

The irreducible representations  $\mathfrak{so}(3, 2)$ have been extensively studied in the pioneering works \cite{Dirac:1935zz, Dirac:1963ta, Fronsdal:1965zzb, Fronsdal:1974ew, Fronsdal:1975eq, Fronsdal:1975ac, Evans,Angelopoulos,AFFS}. We are interested in the lowest-energy (or lowest-weight) irreducible representations\footnote{For a modern discussion on the lowest-energy UIRs of $\mathfrak{so}(3, 2)$, see the works \cite{Nicolai:1984hb,deWit:1999ui,BrinkMetsaevVasiliev,Ponomarev2022}.} of $\mathfrak{so}(3, 2)$ since the energy they carry is bounded from below. The irreducible representations of $\mathfrak{so}(3, 2)$ are specified by the lowest value $E_0$ of the energy $E$\footnote{The energy $E$ is chosen to be dimensionless; it can be restored by rescaling $E$ by $E \to l^{-1} \,E$.} and spin $s$, and are commonly denoted by $D (E_0, s)$.
The parameters $E_0$ and $s$ are determined by the eigenvalues of the Casimir operators \eqref{FACasimirOperatorsRep}  \cite{Fronsdal:1974ew,Evans} 
\bsubeq
\bea \label{FACasimirRep}
C_1  &=&  E_0(E_0 -3) +s(s+1) ~,\label{FAQuadCasimirRep}\\
C_2 &=& s(s+1)(E_0-1)(E_0-2) \label{FAQuarticCasimirEvans}~.
\eea
\esubeq

Unlike in Minkowski space, unitarity imposes a bound on the allowed values of energy in AdS$_4$. According to the theorems proved in \cite{Evans,Angelopoulos}, the irreducible representation $D(E_0, s)$ is unitary\footnote{In any UIR of $\mf{so}(d-1,2)$, the generators $J_{AB}$ are Hermitian and the norm for the lowest-energy states is positive definite, i.e. see \cite{Ponomarev2022}.} if and only if one of the following conditions hold:
\be \label{FASUnitaryBound}
\text{ (i)} \quad s=0~, ~ E_0 \geq \hf~, \qquad \text{(ii)} \quad s=\hf~, ~ E_0\geq 1~, \qquad \text{(iii)} \quad s\geq 1~,~ E_0 \geq s+1~. 
\ee
The representations $D\big(\hf , 0\big) = {\rm Rac} $ and $D\big( 1, \hf \big) = {\rm Di} $ 
are known as the Dirac singletons \cite{Dirac:1963ta}.\footnote{It was found by Flato and Fronsdal \cite{Flato:1980zk,FF78} that the singletons are the square roots of massless particles in the sense that all two-singleton states are massless.} 
The representations 
$D(s+1, s) $ for $s>0$ and $D( 1, 0)  \oplus D (2, 0)$ are called massless since they contract  to the massless discrete helicity representations of the Poincar\'e group 
\cite{AFFS}. These representations prove to be restrictions of certain unitary representations of the conformal algebra   
$\mathfrak{so}(4, 2)$  to  $\mathfrak{so}(3, 2)$ \cite{AFFS,Barut:1970kp}. The remaining representations $D(E_0, s)$  are usually referred to as the massive AdS$_4$ representations. 

\subsection{Irreducible field representations}\label{FAIrredFieldReps}

The geometry  of ${\rm AdS}_4$ can be described in terms of torsion-free Lorentz-covariant derivatives of the form 
\bea \label{FACovDeriv}
\nabla_a = e_a + \o_a~, \qquad e_a= e_a{}^m \partial_m  ~, 
\eea
where $e_a{}^m $ is the inverse vielbein and $\o_a$ is the Lorentz connection
\bea
\o_a = \frac{1}{2}\,\o_a{}^{bc} M_{bc}
= \o_a{}^{\b \g} M_{\b \g}
+\bar \o_a{}^{\bd \gd} \bar M_{\bd \gd} ~.
\eea	
The covariant derivative \eqref{FACovDeriv} satisfies the commutation relation\footnote{Note that we use the convention $[\na_a , \na_b] = \hf \cR_{ab}{}^{ce} M_{ce}$, where $\cR_{abce}$ is the Riemann tensor which we normalise as $\cR_{abce} = -2 \m \mub \eta_{a[c} \eta_{e]b}$. }
\begin{align}
\big[\nabla_a,\nabla_b\big]=-\mu\mub M_{ab} \quad \Longleftrightarrow \quad \big[\nabla_{\a\ad},\nabla_{\b\bd}\big]=-2\mu\mub\big(\ve_{\a\b}\bar{M}_{\ad\bd}+\ve_{\ad\bd}M_{\a\b}\big)
~,
\label{FASAlgebra}
\end{align} 
where the parameter $\mu\mub > 0$ is related to the AdS$_4$ scalar curvature $\mathcal{R}$ via $\mathcal{R}=-12\mu\mub$, and to the AdS$_4$ radius $l$ via $l^{-2} = \m \mub$.\footnote{The complex parameter $\m$ appears explicitly only in the algebra of AdS$^{4|4}$ covariant derivatives \eqref{FASDerivativeAlgebra}.} Note that changing the sign of the curvature $\mathcal{R}$ yields de-Sitter space, while Minkowski space is obtained in the limit $\mathcal{R}\rightarrow 0$.

We will make extensive use of the Casimir operators \eqref{FACasimirOperatorsRep} of $\mathfrak{so}(3, 2)$ in the subsequent sections, which take the following form in the field representation
\bsubeq \label{FACasimirOperators}
\begin{align}
\mathcal{Q}=&~\Box-\mu\mub \big(M^2
+\bar{M}^2\big)~,\ & \big[\mathcal{Q},\nabla_{\a\ad}\big]&=0~,  \label{FAQuadratic Casimir} \\
\mc{W}^2=&-\frac{1}{2}\big(\mc{Q}+2\m \mub \big)\big(M^2+\bar{M}^2\big)+\na^{\a\ad}\na^{\b\bd}M_{\a\b}\bar{M}_{\ad\bd} \non && \\
&-\frac{1}{4}\m \mub \big(M^2M^2+\bar{M}^2\bar{M}^2+6M^2\bar{M}^2\big)~, \label{FAQuarticCasimir} & \big[\mc{W}^2,\nabla_{\a\ad}\big]&=0~. 
\end{align}
\esubeq
Here $\Box := -\frac{1}{2}\nabla^{\a\ad}\nabla_{\a\ad} $ is the d’Alembert operator in AdS$_4$. Note that we have  introduced the notation $\mc{Q} \equiv  l^{-2} C_1$ and $\mc{W}^2 \equiv  l^{-2} C_2$ for the Casimir operators \eqref{FACasimirOperatorsRep} of $\mathfrak{so}(3, 2)$ in the field representation, with dimensions being restored.

Given two positive integers $m$ and $n$, a tensor field $\f_{\a(m)\ad(n)}$ of Lorentz type $(\frac{m}{2}, \frac{n}{2})$ in AdS$_4$ is said to be on-shell if it satisfies the conditions
\begin{subequations}\label{FAMassiveOnShellConditions}
	\begin{align}
	\nabla^{\b\bd}\f_{\b \a(m-1)\bd \ad(n-1)}&=0~, \label{FAOnshellTransverseCondition} \\
	\big(\mathcal{Q}-\rho^2\big)\f_{\a(m)\ad(n)}&=0~. \label{FAOnshellMasseqn}
	\end{align}
\end{subequations}
Here the real parameter $\r$, which carries mass dimension one, is called the pseudo-mass. 

It must be noted that there is an ambiguity in the way mass is defined in AdS$_4$. There exist additional terms proportional to the scalar curvature $\m\mub$ in AdS$_4$, which have no direct counterpart in the flat-space limit $\m \mub \rightarrow 0$. Since these contributions also have mass dimension two, it follows that the mass-shell equation \eqref{FAOnshellMasseqn} can be redefined modulo $\m\mub$ contributions without any effect on the corresponding mass-shell equation in the flat space limit. In other words, there exists an infinite family of on-shell fields in AdS$_4$ which reduce to the same on-shell field in $\mb{M}^4$.

To address this ambiguity, let us introduce the notion of physical mass $\r_{\text{phys}}$ whose meaning is derived from gauge symmetry. In particular, the definition of physical mass is related to pseudo-mass via the relation
\begin{align}
\rho^2_{\text{phys}}=\rho^2-\tau_{(1,m,n)}\mu\mub~, \qquad
\tau_{(1,m,n)} = \hf (m+n+2)(m+n-2)  ~, \label{FAPhysicalPseudoRel}
\end{align}
where $\t_{(1,m,n)}$ is a partially massless value \eqref{FAPMValue}. As we will see, this mass convention ensures that an on-shell field \eqref{FAMassiveOnShellConditions} with physical mass $\r^2_{\text{phys}}=0$, which we will call a massless field, is compatible with the maximal gauge symmetry \eqref{FAMasslessGaugeSymmetry}. Hence in the flat-space limit, a massless field in AdS$_4$ reduces to a massless gauge field  \eqref{FMMasslessfields} in $\mb{M}^4$.

Contact with the representation theory can be established by comparing the eigenvalue \eqref{FAOnshellMasseqn} of the quadratic Casimir operator $\mc{Q}$ associated with the on-shell field \eqref{FAMassiveOnShellConditions} with the eigenvalue \eqref{FAQuadCasimirRep} of the Casimir operator $l^{-2}C_1$ in the lowest-energy irreducible representation. Doing this yields the following relationship\footnote{See \cite{Nicolai:1984hb, deWit:1999ui} for pedagogical derivations of the result \eqref{FARelationbetweenFieldsandReps}. } between the pseudo-mass $\r$ and minimal energy $E_0$
\bea \label{FARelationbetweenFieldsandReps} 
\rho^2 =\big [ E_0(E_0-3) +s(s+1)\big ]\mu\mub~,
\eea
where $s=\hf(m+n)$. 
It follows from \eqref{FAPhysicalPseudoRel} that the minimal energy is related to the physical mass  through
\begin{align}
\rho_{\text{phys}}^2 
&=\big[E_0(E_0-3)-(s+1)(s-2)\big]\mu\mub~. \label{FASPhysicalMassEnergy}
\end{align}

The action of the Casimir operator $\mc{W}^2$ \eqref{FAQuarticCasimir} on an unconstrained field yields
\bea \label{FAQuarticCasimirOffShellFields}
\mc{W}^2 \f_{\a(m)\ad(n)} &=& s(s+1)\big ( \mc{Q} - (s-1)(s+2) \m \mub  \big ) \f_{\a(m)\ad(n)} \\ &&+ m n  \nabla_{\a\ad} \nabla^{\b\bd} \f_{\b \a(m-1) \bd \ad(n-1)}~. \non
\eea
If the field satisfies the on-shell conditions \eqref{FAMassiveOnShellConditions}, then \eqref{FAQuarticCasimirOffShellFields} reduces to
\be \label{FAQuarticCasimirOn-ShellFields}
\mc{W}^2 \f_{\a(m)\ad(n)} = s(s+1)\big ( \r^2 - (s-1)(s+2) \m \mub  \big ) \f_{\a(m)\ad(n)}~.
\ee
Expressing \eqref{FAQuarticCasimirOn-ShellFields} in terms of the minimal energy via the relation \eqref{FARelationbetweenFieldsandReps} gives
\be \label{FAQuadCasOnshell}
\mc{W}^2 \f_{\a(m)\ad(n)} = s(s+1) ( E_0-2)(E_0-1) \mu \mub \f_{\a(m)\ad(n)}~.
\ee
Note that the eigenvalue \eqref{FAQuadCasOnshell} of the Casimir operator $\mc{W}^2$ on the space of on-shell fields coincides identically with the eigenvalue \eqref{FAQuarticCasimirEvans} of the quartic Casimir operator $l^{-2}C_2$ in the lowest-energy irreducible representation. It follows from \eqref{FAOnshellMasseqn} and \eqref{FAQuadCasOnshell} that an on-shell field \eqref{FAMassiveOnShellConditions} furnishes the irreducible representation $D(E_0,s)$ of $\mf{so}(3,2)$. Thus, we say that an on-shell field carries spin $s=\hf(m+n)$ and pseudo-mass $\rho$. From a field-theoretic perspective, is natural to use the pseudo-mass $\r$ as a representation label in place of $E_0$. However, we choose to stay with the label $E_0$ in this chapter.

It follows from \eqref{FASPhysicalMassEnergy} that there are two distinct values of $E_0$ leading to the same value of $\rho_{\text{phys}}^2$, 
\begin{align}
\big(E_0\big)_{\pm}=\frac{3}{2}\pm \frac{1}{2}
\sqrt{
	4\frac{\rho^2_{\text{phys}}}{\mu\mub}+(m+n-1)^2
}
~.
\label{FAEnergyModesOnshell}
\end{align}
Hence the minimal energy carried by an on-shell field \eqref{FASOnShellConditions} is not unique. However, we have yet to take unitarity into account. For $m+n>1$, the unitarity bound $E_0\geq  \frac{1}{2}(m+n+2)$ (cf. \eqref{FASUnitaryBound}) can be expressed in terms of pseudo-mass and physical mass as follows 
\begin{align}
\rho^2 \geq \tau_{(1,m,n)}\mu\mub \quad \implies \quad \rho^2_{\text{phys}}\geq 0 ~.
\label{FAUnitarityBoundsMasses}
\end{align}
In accordance with \eqref{FAUnitarityBoundsMasses}, the solution $(E_0)_-$  always violates the unitarity bound for $m+n>1$. Thus when referring to a unitary representation with $s >\hf$ and pseudo-mass $\rho^2$, we are implicitly referring to the representation corresponding to $(E_0)_+$\footnote{Using the properties of gauge symmetry and unitarity to uniquely determine the minimal energy $E_0$ for massless fields in AdS$_4$ was first completed in \cite{Fronsdal:1975ac,Fronsdal1979Sing}. Such studies were later generalised to AdS$_d$ in the case of bosonic \cite{Metsaev:1995re} and fermionic \cite{Metsaev:1998xg} fields. }
\begin{align}
E_0 =\frac{3}{2} + \frac{1}{2}
\sqrt{
	4\frac{\rho^2_{\text{phys}}}{\mu\mub}+(m+n-1)^2}
~, \qquad m+n>1~.
\end{align}

For $s= 0, \hf $ (or, equivalently, $m+n=0,1$),  the unitary bound is $E_0\geq s +\hf$, 
and the solution  $(E_0)_-$ in \eqref{FAEnergyModesOnshell} does not violate the unitarity bound for certain values of $\rho^2_{\text{phys}}$. For $s=0$ the values of $\rho^2_{\text{phys}}$ leading to $(E_0)_-\geq \hf$ are  restricted by
the condition
$- \frac 14 \m \mub \leq \rho^2_{\text{phys}} \leq  \frac 34 \m \mub $, which is known as  
the Breitenlohner-Freedman bound \cite{BreitenF}. 

Below we review the different species of on-shell fields in AdS$_4$, which are characterised by the pseudo-mass they carry.

\subsubsection{Massless fields}
Given positive integers $m$ and $n$, an on-shell field \eqref{FASOnShellConditions} is said to be massless if it carries the pseudo-mass 
\begin{align} 
\r^2=\t_{(1,m,n)}\mu\mub ~, \qquad \t_{(1,m,n)} = \hf (m+n-2)(m+n+2)~.\label{FASMasslessGaugeField}
\end{align}
The system of equations \eqref{FAMassiveOnShellConditions} with $\r^2$ \eqref{FASMasslessGaugeField} is compatible with the gauge symmetry 
\begin{align}
\delta_{\zeta}\f_{\a(m)\ad(n)}=\nabla_{\a\ad}\zeta_{\a(m-1)\ad(n-1)}~, \label{FAMasslessGaugeSymmetry}
\end{align}
where the gauge parameter $\zeta_{\a(m-1)\ad(n-1)}$ is complex unconstrained.  The maximal gauge symmetry \eqref{FAMasslessGaugeSymmetry} is the AdS$_4$ analogue of the gauge symmetry \eqref{FMMasslessGT} associated with massless fields in $\mb{M}^4$.
Note that the relationship  \eqref{FAPhysicalPseudoRel} between physical mass and pseudo-mass was chosen to ensure that  massless fields in AdS$_4$  carry vanishing physical mass $\r^2_{\text{phys}} = 0$. It is easy to see that this physical mass saturates the bound \eqref{FAUnitarityBoundsMasses} 
and hence defines a unitary representation of $\mathfrak{so}(3, 2)$. It follows from \eqref{FARelationbetweenFieldsandReps} and \eqref{FASMasslessGaugeField} that a massless field has minimal energy $E_0 = s+1$. 

Analogous to the discussion on massless fields in $\mb{M}^4$ (see section \ref{FMMasslessfieldrepresentations}), it follows that massless gauge fields in AdS$_4$ do not realise irreducible representations of $\mathfrak{so}(3, 2)$ due to the presence of gauge symmetry. These redundant gauge degrees of freedom are automatically eliminated if one  works with the gauge-invariant field strengths \eqref{FAWeylTensor1},
\bsubeq \label{FASFieldStrengthMassless}
\bea
\mf{W}_{\a(m+n)}(\f) &=& \na_{(\a_1}{}^{\bd_1} \cdots \na_{\a_n}{}^{\bd_n}\f_{\a_{n+1} \ldots \a_{m+n}) \bd(n)}~, \\
\bar{\mf{W}}_{\ad(m+n)}({\f}) &=& \na_{(\ad_1}{}^{\b_1} \cdots \na_{\ad_m}{}^{\b_m}{\f}_{\b(m) \ad_{m+1} \ldots \ad_{m+n})}~,
\eea
\esubeq
instead of the gauge field $\f_{\a(m)\ad(n)}$.
Making use of the constraint \eqref{FAOnshellTransverseCondition} satisfied by $\f_{\a(m)\ad(n)}$, it may be shown that the field strengths \eqref{FASFieldStrengthMassless} are constrained by
\bsubeq  \label{FASFieldStrengthOnshell}
\bea
\na_{\bd}{}^\b \mf{W}_{\b\a(m+n-1)}(\f) =0 \qquad  &\Longrightarrow& \qquad \big ( \mc{Q}- \t_{(1,m,n)} \m \mub \big )\mf{W}_{\a(m+n)}(\f) = 0~,  \label{FASFS1}\\
\na_{\b}{}^\bd \bar{\mf{W}}_{\bd\ad(m+n-1)}(\f) =0 \qquad &\Longrightarrow& \qquad \big ( \mc{Q}- \t_{(1,m,n)} \m \mub \big )\bar{\mf{W}}_{\ad(m+n)}(\f) = 0~. \label{FASFS2}
\eea
\esubeq
The latter constraints are consistent with \eqref{FAOnshellMasseqn} only if the  pseudo-masses obey \eqref{FASMasslessGaugeField}.

The actions describing massless spin-$s$ and spin-$(s+\hf)$ particles in AdS$_4$ were computed in the seminal works \cite{Fronsdal1979Sing} and \cite{Fronsdal1979}, respectively.\footnote{See \cite{KS94} for the corresponding actions in two-component spinor notation.} On the equations of motion, it can be shown that the only surviving field strengths for the massless spin-$s$ theory are \eqref{FASFieldStrengthOnshell} for the cases $m=n=s$. While for the massless spin-$(s+\hf)$ model, the non-vanishing field strengths are \eqref{FASFS1} with $n = m-1 = s$  and \eqref{FASFS2} with $m = n-1 = s$. As was shown in \cite{KS94}, these field strengths satisfy the constraints on the equations of motion
\bsubeq
\begin{align}
\na_{\bd}{}^\b \mf{W}_{\b\a(2s-1)}(\f) &=0 ~, & \na_{\b}{}^\bd \bar{\mf{W}}_{\bd\ad(2s-1)}(\f) &=0~,  \\
\na_{\bd}{}^\b \mf{W}_{\b\a(2s)}(\f) &=0 ~, & \na_{\b}{}^\bd \bar{\mf{W}}_{\bd\a(2s)}(\f) &=0~,
\end{align}
\esubeq
which are exactly the on-shell conditions \eqref{FASFieldStrengthOnshell}.

\subsubsection{Partially massless fields}\label{FASPMSec}

Given two positive integers $m$ and $n$, the tensor field $\f^{(t)}_{\a(m)\ad(n)}$ is said to be partially massless with depth-$t$ if it satisfies the on-shell conditions \eqref{FAMassiveOnShellConditions} such that its pseudo-mass
takes the special value \cite{DeserW4, Zinoviev, Metsaev, KP20}
\begin{align}
\rho^2=\tau_{(t,m,n)}\mu\mub~,\qquad 1 \leq t \leq \text{min}(m,n)~, \label{FMPartiallyMasslessPseudoMass}
\end{align}
where the constants $\tau_{(t,m,n)}$ are the partially massless values defined by 
\begin{align}
\tau_{(t,m,n)}:=\frac{1}{2}\Big((m+n-t+3)(m+n-t-1)+(t-1)(t+1)\Big) ~.\label{FAPMValue}
\end{align} 
The attributing feature of partially massless fields is that,
for a fixed depth $t$, the system of equations \eqref{FAMassiveOnShellConditions} with $\r^2$ \eqref{FMPartiallyMasslessPseudoMass} admit a depth-$t$ gauge symmetry of the form
\begin{align}
\delta_{\zeta}\f^{(t)}_{\a(m)\ad(n)}=\nabla_{(\a_1(\ad_1}\cdots \nabla_{\a_t\ad_t}\zeta^{(t)}_{\a_{t+1}\dots\a_{m})\ad_{t+1}\dots\ad_{n})}~. \label{FADepthGaugeSymmetry}
\end{align}
This is true as long as the gauge parameter $\zeta^{(t)}_{\a(m-t)\ad(n-t)}$ is also on-shell with the same pseudo-mass,
\begin{subequations}\label{GPP}
	\begin{align}
	\big(\mathcal{Q}-\tau_{(t,m,n)}\mu\mub\big)\zeta^{(t)}_{\a(m-t)\ad(n-t)} &= 0~,\label{FAPMGaugeparameterMass}\\
	\nabla^{\b\bd}\zeta^{(t)}_{\b \a(m-t-1) \bd \ad(n-t-1)}&=0~.\label{FAPMPMGaugeparametertransverse}
	\end{align}
\end{subequations}
Note that strictly massless fields \eqref{FASMasslessGaugeField} carry depth $t=1$. 

On account of the inequality
\begin{align}
\tau_{(1,m,n)}> \tau_{(t,m,n)}~,\qquad 2 \leq t \leq \text{min}(m,n)~,
\end{align}
it follows from \eqref{FAUnitarityBoundsMasses} that true partially massless representations are non-unitary. In particular, this means that there are two minimal energy values, 
\begin{align}
(E_0)_{\pm}=\frac{3}{2}\pm\frac{1}{2}(m+n-2t+1)~,
\end{align}
which are equally valid since they both violate the unitarity bound. In this chapter, whenever this ambiguity arises, we always implicitly choose the positive branch, $E_0\equiv (E_0)_+$.   To distinguish the partially massless fields, we denote true partially massless fields as those with depth $2 \leq t \leq \text{min}(m,n)$ and are thus non-unitary.
To distinguish the true partially-massless representations with depth $t$ and Lorentz type $(\frac{m}{2},\frac{n}{2})$, we will employ the notation  
\begin{align}
P\big(t,m,n\big)~,  \qquad2\leq t \leq \text{min}(m,n)~. 
\label{NSPMrep}
\end{align}
Such a representation carries minimal energy $E_0=\frac{1}{2}(m+n)-t+2$.

It is common in the higher-spin literature to define the partially massless on-shell conditions in terms of the AdS$_4$ d'Alembertian $\Box$, rather than the Casimir operator $\mc{Q}$. Converting to these conventions, we find that the conditions \eqref{FAOnshellMasseqn} and \eqref{FAPMGaugeparameterMass} take the respective forms
\bsubeq
\begin{align}
\big(\Box - \big [ (m-t)(n-t)-(t+2) \big ] \m \mub  \big )\f^{(t)}_{\a(m)\ad(n)} &=0~,\\
\big(\Box  - (mn+t-2) \m \mub  \big )\zeta^{(t)}_{\a(m-t)\ad(n-t)}&=0~.
\end{align}
\esubeq

\subsubsection{Massive fields}
An on-shell field \eqref{FAMassiveOnShellConditions} is said to be massive if its pseudo-mass is constrained by 
\begin{align}
\rho^2>\tau_{(1,m,n)}\mu\mub \quad\implies\quad \rho_{\text{phys}}^2>0~,
\end{align}
but is otherwise arbitrary. This restriction ensures the unitarity of the representation. 

It must be noted that there also exists massive fields which carry negative physical mass $\rho_{\text{phys}}^2<0$, given that $\rho_{\text{phys}}^2$ does not coincide with the partially massless modes $\r^2 \neq \tau_{(t,m,n)}\mu\mub$, where $ 2 \leq t \leq \text{min}(m,n)$. In accordance with \eqref{FAUnitarityBoundsMasses}, such a field will realise a non-unitary irreducible representation of $\mf{so}(3,2)$.

\subsection{Spin projection operators} \label{FASpinProjectors}
In this section we review, and expand upon, the spin projection operators in AdS$_4$. Let us denote by $V_{(m,n)}$ the space of  tensor fields $\f_{\a(m)\ad(n)}$ of Lorentz type $(\frac{m}{2},\frac{n}{2})$ in AdS$_4$. For integers $m,n \geq 2$, we define the spin projection operator $\P^{\perp}_{(m,n)}$ by its action on $V_{(m,n)}$ via the rule
\bsubeq \label{FASpinProjectorStructure}
\bea
\P^{\perp}_{(m,n)}: V_{(m,n)} &\longrightarrow& V_{(m,n)}~, \\
\f_{\a(n)} &\longmapsto& \P^{\perp}_{(m,n)}\f_{\a (m) \ad (n)} =: \P^{\perp}_{\a(m)\ad(n)}(\f)~,
\eea
\esubeq
For fixed $m$ and $n$, the operator $\P^{\perp}_{(m,n)}$ is defined by the following properties:
\begin{enumerate}
	\item \textbf{Idempotence:} The operator $\P^{\perp}_{(m,n)}$ squares to itself on $V_{(m,n)}$, 
	\bsubeq  \label{FASpinProjectorProperties}
	\be
	\P^{\perp}_{(m,n)}\P^{\perp}_{(m,n)} = \P^{\perp}_{(m,n)} ~. 
	\ee
	\item \textbf{Transversality:} The operator $\P^{\perp}_{(m,n)}$ maps $\f_{\a(m)\ad(n)}$ to a transverse field,
	\be
	\na^{\b\bd} \P^{\perp}_{\b \a(m-1) \bd \ad(n-1)}(\f)= 0~. \label{FASProjectorTransverse}
	\ee
	\esubeq
\end{enumerate}
The spin projection operator $\P^{\perp}_{(m,n)}$ maps a field $\f_{\a(m)\ad(n)}$ on $V_{(m,n)}$, where $\f_{\a(m)\ad(n)}$ is  on the mass-shell \eqref{FAOnshellMasseqn}, to an on-shell field \eqref{FAMassiveOnShellConditions}
\bsubeq
\bea
( \mc{Q} - \r^2) \P^{\perp}_{\a(m)\ad(n)}(\f) &=& 0~, \\
\na^{\b\bd} \P^{\perp}_{\b \a(m-1) \bd \ad(n-1)}(\f) &=& 0~.
\eea
\esubeq
Thus the spin projection operator $\P^{\perp}_{(m,n)}$ singles out the component of such a $\f_{\a(m)\ad(n)}$ which furnishes the irreducible representation $D(E_0,s)$ of $\mf{so}(3,2)$.

The spin projection operators take the following form on $V_{(m,n)}$\footnote{Switching the curvature $\m \mub$ to $- \m \mub$ in \eqref{FAProjectionOperators} yields the projection operators on de-Sitter space \cite{KP20}.  }
\bsubeq \label{FAProjectionOperators}
\bea
\P^{\perp}_{(m,n)}\f_{\a(m)\ad(n)} &=& \Big [ \prod_{t=1}^{n} \big ( \mc{Q} - \t_{(t,m,n)} \m \mub \big ) \Big ]^{-1} \mb{P}_{\a(m)\ad(n)}(\f)~,  \label{FASSpinProjector1}\\
\widehat{\P}^{\perp} _{(m,n)}\f_{\a(m)\ad(n)} &=& \Big [ \prod_{t=1}^{m} \big ( \mc{Q} - \t_{(t,m,n)} \m \mub \big ) \Big ]^{-1} \widehat{\mb{P}}_{\a(m)\ad(n)}(\f)~, \label{FASSpinProjector2}
\eea
\esubeq
where the projected fields $\mb{P}_{\a(m)\ad(n)}(\f)$ and $\widehat{\mb{P}}_{\a(m)\ad(n)}(\f)$ are
\bsubeq \label{FASTransverseOperator}
\bea
\mb{P}_{\a(m)\ad(n)}(\f) &=& \na_{(\ad_1}{}^{\b_1} \cdots \na_{\ad_n)}{}^{\b_n} \na_{(\b_1}{}^{\bd_1} \cdots \na_{\b_n}{}^{\bd_n} \f_{\a_1 \ldots \a_m)\bd(n)}~, \\
\widehat{\mb{P}}_{\a(m)\ad(n)}(\f) &=& \na_{(\a_1}{}^{\bd_1} \cdots \na_{\a_m)}{}^{\bd_m} \na_{(\bd_1}{}^{\b_1} \cdots \na_{\bd_m}{}^{\b_m} \f_{\b(m)\ad_1 \ldots \ad_n)}~.
\eea
\esubeq
It was shown in \cite{KP20} that the projectors \eqref{FASSpinProjector1} and \eqref{FASSpinProjector2} are equivalent on $V_{(m,n)}$
\be \label{FAProjectorEquivalence}
\P^{\perp}_{\a(m)\ad(n)}(\f)  = \widehat{\P}^{\perp}_{\a(m)\ad(n)}(\f)~.
\ee
In accordance with this, we are free to work with either the spin projection operator $\P^{\perp}_{(m,n)}$ \eqref{FASSpinProjector1} or $\widehat{\P}^{\perp}_{(m,n)}$  \eqref{FASSpinProjector2} on ${V}_{(m,n)}$.  From this point onwards, we will only consider the spin projection operator $\P_{(m,n)}^{\perp}$ on $V_{(m,n)}$.
Furthermore, it can be shown that $\P_{(m,n)}^{\perp}$ acts like the unit operator on the space of transverse fields
\be
\na^{\b \bd }\f^{\perp}_{\b \a(m-1) \bd \ad(n-1)} =0 \qquad \Longrightarrow \qquad \P_{(m,n)}^{\perp}\f^{\perp}_{\a(m)\ad(n)} = \f^{\perp}_{\a(m)\ad(n)} ~.
\ee

The projectors \eqref{FAProjectionOperators} were recently derived by Kuzenko and Ponds in \cite{KP20}. 
They demonstrated that the parameter $\t_{(t,m,n)}$, which determines the poles of the spin projection operators, coincides with the partially massless value \eqref{FAPMValue}.\footnote{The partially massless values \eqref{FAPMValue} appearing in the denominator of the projectors \eqref{FAProjectionOperators} differ to those given in \cite{KP20} by a factor of a half. This is due to the fact that this thesis uses a different normalisation factor for the AdS$_4$ curvature.} This observation provided a new understanding of partially massless fields. Thus, the spin projection operators provide an alternative method for computing partially massless values. This result will be essential in the formulation of partially massless superfields in AdS$^{4|4}$, since the partially massless values are unknown.

We would like to point out that $\t_{(t,m,n)}$ appears in the poles of the projector \eqref{FAProjectionOperators} with $1 \leq t \leq \text{max}(m,n)$. In accordance with the pseudo-mass \eqref{FMPartiallyMasslessPseudoMass} of a partially massless field, it is apparent that the poles of the projector coincide with partially massless modes for $1 \leq t \leq \text{min}(m,n)$ and additional (non-unitary) massive modes for $\text{min}(m,n) < t \leq \text{max}(m,n)$. Depending on the integers $m$ and $n$,  one can always choose to work with a particular projector such that its poles only coincide with the partially massless values \eqref{FAPMValue}. Specifically, given $m > n$, it proves natural to work with the projector $\P^{\perp}_{(m,n)}$, as the bounds of the product in \eqref{FASSpinProjector1} includes all values of $t$ within the range $1 \leq t \leq \text{min}(m,n)$. Hence, the poles of the projector only contains partially massless modes. Using a similar argument, if $n > m$, it is natural to work with  $\widehat{\P}^{\perp}_{(m,n)}$  \eqref{FASSpinProjector2}.

The spin projection operators \eqref{FAProjectionOperators} can be recast solely in terms of the Casimir operators \eqref{FAQuadratic Casimir} and \eqref{FAQuarticCasimir} of $\mf{so}(3,2)$ as follows
\bsubeq \label{FAProjectorsCasimir}
\bea
\tilde{\P}^{\perp}_{(m,n)}&=& \frac{m!}{n!(m+n)!}  \prod_{j=0}^{n-1}  \frac{\big (\mc{W}^2 - (s-j)(s-j-1) [\cQ - (s-j-2)(s-j+1)\m \mub] \big ) }{\big (\cQ - \t_{(j+1,m,n)}\m \mub \big ) }~, \hspace{1.5cm} \label{FAProjectorsCasimir1}\\ 
&=& \frac{n!}{m!(m+n)!}  \prod_{j=0}^{m-1}  \frac{\big (\mc{W}^2 - (s-j)(s-j-1) [\cQ - (s-j-2)(s-j+1)\m \mub] \big ) }{\big (\cQ - \t_{(j+1,m,n)}\m \mub \big ) }~. \label{FAProjectorsCasimir2}
\eea
\esubeq
The spin projection operator $\tilde{\P}^{\perp}_{(m,n)}$ is a novel realisation of the projectors \eqref{FAProjectionOperators} given in \cite{KP20}.
Alike the projection operators $\P_{(m,n)}^{\perp}$, all information concerning partially massless fields is encoded within the poles of the projectors $\tilde{\P}^{\perp}_{(m,n)}$. Furthermore, it can be shown that the zeros of $\tilde{\P}^{\perp}_{(m,n)}$ actually coincide with the eigenvalues generated by $\cW^2$ when acting on the space of transverse fields $\f^{\perp}_{\a(m-j)\ad(n-j)}$, with $1\leq j \leq n$ and $m \geq n$. It can be shown that this also holds in the case $m \leq n$ with $1\leq j \leq m$.

One can show that the projection operators \eqref{FAProjectorsCasimir} act as the identity operator on the space of transverse fields $\f^{\perp}_{\a(m)\ad(n)}$
\be
\na^{\b\bd}\f^{\perp}_{\b\a(m-1)\bd\ad(n-1)}=0 \qquad \Longrightarrow \qquad \tilde{\P}^{\perp}_{(m,n)} \f^{\perp}_{\a(m)\ad(n)} =\f^{\perp}_{\a(m)\ad(n)}  ~. \label{FASurjectivity}
\ee

The projectors $\tilde{\P}^{\perp}_{(m,n)}$ exhibit some non-trivial properties when acting on fields that belong to spaces other than $V_{(m,n)}$. Specifically, the projectors $\P^{\perp}_{(m,n)}$ and $\tilde{\P}^{\perp}_{(m,n)}$ are no longer equivalent on the space $V_{(j,k)}$, where $2 \leq j \leq m-1 $ and $2 \leq k \leq n-1$. Additionally, these operators are no longer idempotent and transverse on $V_{(j,k)}$. It can be shown that the  $\tilde{\P}^{\perp}_{(m,n)}$ annihilates any field belonging to $V_{(j,k)}$
\be \label{FAProjAnnihilatesLowBosons}
\tilde{\P}_{(m,n)} \f_{\a(j) \ad(k)} = 0~, 
\ee
where $1 \leq j \leq m-1 $ and $1 \leq k \leq n-1$.

\subsubsection{Longitudinal projectors and lower-spin extractors}
The orthogonal complement of $\P^{\perp}_{(m,n)}$ is defined on $V_{(m,n)}$ by the rule
\be \label{FAOrthogonalComplement}
\P^{\parallel} _{(m,n)} = \mds{1} - \P^{\perp}_{(m,n)}~,
\ee
which, by construction, satisfies the properties
\bsubeq \label{FALongProjProp}
\be \label{FALongProjIdem}
\P_{(m,n)}^{\parallel} \P_{(m,n)}^{\parallel} =\P_{(m,n)}^{\parallel} ~, 
\ee
\be
\qquad \P_{(m,n)}^{\parallel} \P_{(m,n)}^{\perp} = \P_{(m,n)}^{\perp} \P_{(m,n)}^{\parallel} =0~.
\ee
\esubeq
It can be shown that $\P_{(m,n)}^{\parallel}$  projects onto the longitudinal part of  $\f_{\a(m)\ad(n)}$
\be \label{FALongProjection}
\P_{(m,n)}^{\parallel} \f_{\a(m)\ad(n)} = \nabla_{\a\ad} \f_{\a(m-1)\ad(n-1)}~,
\ee
where $\f_{\a(m-1)\ad(n-1)}$ is an unconstrained field.
Due to properties \eqref{FALongProjIdem} and \eqref{FALongProjection}, the operator $\P_{(m,n)}^{\parallel}$ is called the longitudinal projector. 

Let $\f^{\parallel}_{(m,n)} = \nabla_{\a\ad} \f_{\a(m-1)\ad(n-1)}$ be some longitudinal field. Since the spin projection operator $\tilde{\P}^{\perp}_{(m,n)}$ \eqref{FAProjectorsCasimir} annihilates all fields of rank-$(m-1,n-1)$, it follows that $\tilde{\P}^{\perp}_{(m,n)}$ annihilates any rank-$(m,n)$ longitudinal field 
\be \label{FATransKillsLong}
\tilde{\P}^{\perp}_{(m,n)} \f^{\parallel}_{\a(m)\ad(n)} = 0~.
\ee
Making use of this result, it can be shown that $\P_{(m,n)}^{\parallel}$ acts like the unit operator on the space of longitudinal fields 
\be
\tilde{\P}^{\parallel}_{(m,n)} \f^{\parallel}_{\a(m)\ad(n)} = 0~.
\ee

Since the projectors  $\P^{\perp}_{(m,n)}$ and $\P_{(m,n)}^{\parallel}$ resolve the identity operator \eqref{FAOrthogonalComplement}, it follows that  any unconstrained field  $\f_{\a(m)\ad(n)}$ on $V_{(m,n)}$ can be decomposed in the following manner
\be \label{FAFirstDecomp}
\f_{\a(m)\ad(n)} = \f^{\perp}_{\a(m)\ad(n)} + \na_{\a\ad} \f_{\a(m-1)\ad(n-1)}~,
\ee
where $\f^{\perp}_{\a(m)\ad(n)} $ is transverse and $\f_{\a(m-1)\ad(n-1)}$ is unconstrained and thus reducible. Due to properties \eqref{FASurjectivity} and \eqref{FATransKillsLong}, it follows that the spin projection operator $\P^{\perp}_{(m,n)}$ selects out the pure spin $s=\hf(m+n)$ component from the decomposition \eqref{FAFirstDecomp}. We can continue this prescription iteratively to obtain the following decomposition 
\be \label{FADecomposition}
\f_{\a(m)\ad(n)}  =  \sum_{j=0}^{n-1}(\na_{\a\ad})^j \f^{\perp}_{\a(m-j)\ad(n-j)} +(\na_{\a\ad})^n\f_{\a(m-n)}~,
\ee
where we have assumed, without loss of generality, that $m \geq n$. The set of fields $\big \{\f^{\perp}_{\a(m)\ad(n)}, \f^{\perp}_{\a(m-1)\ad(n-1)}, \cdots ,  \f^{\perp}_{\a(m-n+1)\ad} \big \}$ are transverse and $\f_{\a(m-n)}$ does not satisfy any differential constraints, thus irreducible.

We can introduce operators which extract the bosonic $\f^{\perp}_{\a(s-k)\ad(s-k)}$ and fermionic $\f^{\perp}_{\a(s-k)\ad(s-k-1)} $ fields  from the decomposition \eqref{FADecomposition}. In particular, the spin-$(s-k)$ and spin-$(s-k-\hf)$ components can be extracted via
\bsubeq
\bea
\f_{\a(s)\ad(s)} &\mapsto& \f^{\perp}_{\a(s-k)\ad(s-k)} =  \big (S^{\perp}_{(s-k)}\f \big )_{\a(s-k)\ad(s-k)} \equiv S^{\perp}_{\a(s-k)\ad(s-k)}(\f) ~, \\
\f_{\a(s)\ad(s-1)} &\mapsto& \f^{\perp}_{\a(s-k)\ad(s-k-1)} =  \big (S^{\perp}_{(s-k-\hf)}\f \big )_{\a(s-k)\ad(s-k-1)} \equiv S^{\perp}_{\a(s-k)\ad(s-k-1)}(\f) ~, \hspace{0.6cm}
\eea
\esubeq
where we have defined the operators
\bsubeq \label{FAExtractors}
\bea
S^{\perp}_{\a(s-k)\ad(s-k)}(\f) &=& a_k \Big [ \prod_{j=1}^{k} \big (\mc{Q} - \t_{(j,s-k+j,s-k+j)} \m \mub \big ) \Big ]^{-1} \non \\
&&\times \big (\na^{\b\bd} \big )^k \tilde{\P}^{\perp}_{(s-k,s-k)} \f_{\b(k)\a(s-k)\bd(k)\ad(s-k)}~, \\
S^{\perp}_{\a(s-k)\ad(s-k-1)}(\f) &=& b_k \Big [ \prod_{j=1}^{k} \big (\mc{Q} - \t_{(j,s-k+j,s-k+j-1)} \m \mub \big ) \Big ]^{-1}  \non \\
&& \times \big (\na^{\b\bd} \big )^k  \tilde{\P}^{\perp}_{(s-k,s-k-1)} \f_{\b(k)\a(s-k)\bd(k)\ad(s-k-1)}~. 
\eea
The constants $a_k$ and $b_k$ take the form
\be
a_k = (-1)^k \binom{s}{k}^2 \binom{2s-k+1}{k}^{-1}~, \quad b_k = (-1)^k \binom{s}{k} \binom{s-1}{k} \binom{2s-k}{k}^{-1}~.
\ee
\esubeq
It is clear that the $S^{\perp}_{\a(s-k)\ad(s-k)}(\f)$ and $S^{\perp}_{\a(s-k)\ad(s-k-1)}(\f)$ are transverse 
\bea
\na^{\b\bd}S^{\perp}_{\b\a(s-k-1)\bd\ad(s-k-1)}(\f)=0~, \qquad \na^{\b\bd}S^{\perp}_{\b\a(s-k-1)\bd\ad(s-k-2)}(\f)=0~.
\eea
The operators $S^{\perp}_{\a(s-k)\ad(s-k)}(\f)$ and $S^{\perp}_{\a(s-k)\ad(s-k-1)}(\f)$ are called the spin-$(s-k)$ and spin-$(s-k-\hf)$ transverse extractors,\footnote{Note that the extractors \eqref{FAExtractors} are not projectors because they are dimensionful.} respectively. These extractors were derived for the first time in this thesis. In the flat-space limit, the operators \eqref{FAExtractors} reduce to \eqref{FMExtractors}.


\subsection{Conformal higher-spin theory}\label{FACHS}
For integers $m,n \geq 1$, the CHS action in AdS$_4$ is described in terms of a totally symmetric complex spinor field $h_{\a(m)\ad(n)}$, which is defined modulo gauge transformations of the form\footnote{For a modern treatment on CHS theory in a conformally flat background, see \cite{KuzenkoPonds2019}.}
\be \label{FACHSGT}
\d_\z h_{\a(m)\ad(n)} = \na_{\a \ad} \z_{\a(m-1)\ad(n-1)}~,
\ee
where the gauge parameter $\z_{\a(m-1)\ad(n-1)}$ is complex unconstrained.

Associated with $h_{\a(m)\ad(n)}$ and its complex conjugate $\bar{h}_{\a(n)\ad(m)}$ are the linearised higher-spin Weyl tensors 
\bsubeq \label{FAWeylTensors}
\bea
\mf{W}_{\a(m+n)}(h) &=& \na_{(\a_1}{}^{\bd_1} \cdots \na_{\a_n}{}^{\bd_n}h_{\a_{n+1} \ldots \a_{m+n}) \bd(n)}~, \label{FAWeylTensor1}\\
\mf{W}_{\a(m+n)}(\bar{h}) &=& \na_{(\a_1}{}^{\bd_1} \cdots \na_{\a_m}{}^{\bd_m}\bar{h}_{\a_{m+1} \ldots \a_{m+n}) \bd(m)}~,
\eea
\esubeq
which are invariant under the gauge transformations \eqref{FACHSGT}
\be
\d_\z \mf{W}_{\a(m+n)}(h) = 0~, \qquad \d_\z \mf{W}_{\a(m+n)}(\bar{h}) = 0~.
\ee
The Weyl tensors \eqref{FAWeylTensors} are the fundamental building blocks of the gauge-invariant CHS action, which is given by
\be \label{FACHSActionWeyl}
S_{\text{CHS}}^{(m,n)}[h,\bar{h}] = \ri^{m+n} \int  \rd^4 x~e~  {\mf{W}}^{\a(m+n)}(h) {\mf{W}}_{\a(m+n)}(\bar{h})~+~\text{c.c.}
\ee

Integrating by parts, the action \eqref{FACHSActionWeyl} can be written in the alternative form 
\bsubeq \label{FACHSActionBach}
\bea
S_{\text{CHS}}^{(m,n)}[h,\bar{h}] &=& \ri^{m+n} \int  \rd^4 x~e~ \bar{h}^{\a(n)\ad(m)}{\mf{B}}_{\a(n)\ad(m)}(h) +\HC~, \\
&=& \ri^{m+n} \int  \rd^4 x~e~  \bar{h}^{\a(n)\ad(m)}\widehat{\mf{B}}_{\a(n)\ad(m)}({h}) +\HC~, 
\eea
\esubeq
in which the linearised higher-spin Bach tensors appear
\bsubeq \label{FAbachtensors}
\bea 
{\mf{B}}_{\a(n)\ad(m)}(h) &=& \na_{(\ad_1}{}^{\b_1} \cdots \na_{\ad_m)}{}^{\b_m} {\mf{W}}_{\a(n)\b(m)}(h)~, \\
\widehat{\mf{B}}_{\a(n)\ad(m)}({h}) &=& \na_{(\a_1}{}^{\bd_1} \cdots \na_{\a_n)}{}^{\bd_n} \overline{\mf{W}}_{\ad(m)\bd(n)}({h})~.
\eea
\esubeq
The linearised higher-spin Bach tensors \eqref{FAbachtensors} are descendants of the Weyl tensors \eqref{FAWeylTensors},  which in addition to being gauge-invariant 
\be \label{FABachGI}
\d_\z {\mf{B}}_{\a(n)\ad(m)}(h) = 0 ~, \qquad \d_\z \widehat{\mf{B}}_{\a(n)\ad(m)}(\bar{h}) = 0 ~,
\ee
they are also transverse
\be \label{FABachTransverse}
\na^{\b\bd}{\mf{B}}_{\b\a(n-1)\bd\ad(m-1)}(h)  = 0~, \qquad \na^{\b\bd}\widehat{\mf{B}}_{\b\a(n-1)\bd\ad(m-1)}(\bar{h})  = 0~.
\ee
One can show that the linearised higher-spin Bach-tensors \eqref{FAbachtensors} are equivalent \cite{KP20}
\be
{\mf{B}}_{\a(n)\ad(m)}(h)   = \widehat{\mf{B}}_{\a(n)\ad(m)}({h}) ~.
\ee

Since the higher-spin Bach tensors \eqref{FAbachtensors} are transverse by definition, it is natural to recast them in terms of the unique  transverse spin projection operators \eqref{FAProjectionOperators} \cite{KP20}
\bsubeq \label{FABachTensors}
\bea
\mf{B}_{\a(n)\ad(m)}(h) &=& \prod_{j=1}^{n} \big ( \mc{Q} - \t_{(j,m,n)} \m \mub \big  ) \big (\na_\ad{}^\b \big )^{m-n}\P_{(m,n)}^{\perp}h_{\a(n)\b(m-n)\ad(n)}~, ~~ m \geq n~.  \label{FABachTensorProjectorMGreaterN}\\
{\mf{B}}_{\a(n)\ad(m)}(h) &=&  \prod_{j=1}^{m} \big ( \mc{Q} - \t_{(j,m,n)} \m \mub \big  )  \big (\na_\a{}^\bd \big )^{n-m} \P_{(m,n)}^{\perp}h_{\a(m) \bd(n-m)\ad(m)} ~, ~~ n \geq m~. \hspace{1cm}
\eea
\esubeq
The Weyl tensors \eqref{FAWeylTensors} can also be realised in terms of the spin projectors \eqref{FAProjectionOperators},
\bsubeq \label{FASWeylTensors}
\bea
\mf{W}_{\a(m+n)}(h) &=& \na _{(\a_1}{}^{\bd_1} \cdots  \na _{\a_n}{}^{\bd_n} \P^{\perp}_{(m,n)}h_{\a_{n+1} \ldots \a_{m+n})\bd(n) }~, \\
\mf{W}_{\a(m+n)}(\bar{h}) &=& \na _{(\a_1}{}^{\bd_1} \cdots  \na _{\a_m}{}^{\bd_m}\P^{\perp}_{(n,m)}\bar{h}_{\a_{m+1} \ldots \a_{m+n})\bd(m) }~.
\eea
\esubeq
It follows from \eqref{FABachTensors} that the CHS action \eqref{FACHSActionBach} can be recast in terms of the spin projection operators \cite{KP20}
\bsubeq \label{FACHSProj}
\begin{align} 
S^{(m,n)}_{\text{CHS}}[h,\bar{h}] = \ri^{m+n} &\int \rd^4x~e~ \bar{h}^{\a(n)\ad(m)} \prod_{j=1}^{n}\big ( \mc{Q} - \t_{(j,m,n)} \m \mub \big ) \big ( \na_\ad{}^\b \big )^{m-n}   \non \\
&\times \P_{(m,n)}^{\perp}h_{\a(n)\b(m-n)\ad(n)}  + \HC  ~, \qquad m \geq n~, \label{FACHSProj1} \\
S^{(m,n)}_{\text{CHS}}[h,\bar{h}] = \ri^{m+n} &\int \rd^4x~e~ \bar{h}^{\a(n)\ad(m)} \prod_{j=1}^{m}\big ( \mc{Q} - \t_{(j,m,n)} \m \mub \big ) \big ( \na_\a{}^\bd \big )^{n-m}   \non \\
&\times \P_{(m,n)}^{\perp}h_{\a(m)\bd(n-m)\ad(m)}  + \HC  ~, \qquad n \geq m~. \label{FACHSProj2}
\end{align}
\esubeq
Note that all the results given above can be expressed in terms of the projector $\tilde{\P}^{\perp}_{(m,n)}$ \eqref{FAProjectorsCasimir}, due to the fact that $\tilde{\P}^{\perp}_{(m,n)}$ and ${\P}^{\perp}_{(m,n)}$ are equivalent on $V_{(m,n)}$. This will give rise to novel realisations of the results  given in \cite{KP20}.

In particular, one of the benefits of writing the CHS action \eqref{FACHSActionWeyl} in terms of $\tilde{\P}^{\perp}_{(m,n)}$ is that they can easily be factorised into second order differential equations. In the cases $m=n=s$, we find that the CHS action \eqref{FACHSProj} takes the following form,
\be \label{FACHSBOS}
S^{(s,s)}_{\text{CHS}}[h] = 2(-1)^s \int \rd^4 x~e~h^{\a(s)\ad(s)} \prod_{j=1}^{s}\big ( \mc{Q} - \t_{(j,s,s)} \m \mub \big ) \tilde{\Pi}^{\perp}_{(s,s)} h_{\a(s)\ad(s)}~,
\ee
where we have imposed the reality condition $h_{\a(s)\ad(s)} = \bar{h}_{\a(s)\ad(s)}$. Explicitly evaluating $\tilde{\Pi}^{\perp}_{(s,s)}$ in \eqref{FACHSBOS} yields
\begin{align}
S^{(s,s)}_{\text{CHS}}[h] = \frac{2(-1)^s}{(2s)!} &\int \rd^4 x~e~h^{\a(s)\ad(s)} \prod_{j=0}^{s-1}\big (\mc{W}^2 - (s-j)(s-j-1)  \non \\
&\times [\cQ - (s-j-2)(s-j+1)\m \mub] \big )   h_{\a(s)\ad(s)}~.
\end{align}
Thus the CHS action factorises wholly into second-order differential operators without having to enter the transverse gauge, which was necessary in the analysis of \cite{KP20}. However, for arbitrary $m$ and $n$, it is easy to see that the CHS actions \eqref{FACHSProj}, whether expressed in terms of  ${\P}^{\perp}_{(m,n)}$ or $\tilde{\P}^{\perp}_{(m,n)}$, do not completely factorise into products of second-order operators. 

The factorisation of the conformal operators \eqref{FAbachtensors} was observed long ago in \cite{DeserN1,Tseytlin5, Tseytlin6} for the lower-spin values $s= \frac{3}{2} $ and $s= 2$. The factorisation of the higher-spin conformal operators was conjectured in \cite{Tseytlin13,Karapet1} (see also \cite{Karapet2}), and later proved by several groups \cite{Metsaev2014, NutmaTaronna, GH, KP20}.

\section{Four-dimensional $\cN=1$ anti-de Sitter superspace} \label{FASsec4dAdS}
This section is devoted to the novel formulation of superspin projection operators in AdS$^{4|4}$. These operators can be considered supersymmetric generalisations of the AdS$_4$ spin projection operators of section \ref{FAec4dAdS}. Additionally, they can also be viewed as AdS$_4$ extensions of the superspin projection operators in $\mb{M}^{4|4}$, which were reviewed in section \ref{SecFourDimensionalMinkowskiSuperSpace}. Before studying these superprojectors and their applications, it is first necessary review the irreducible representations of the $\cN=1$ AdS$_4$ superalgebra  $\mathfrak{osp} (1|4)$.


\subsection{Irreducible representations of  $\cN=1$ AdS$_4$ superalgebra  } \label{FASRepTheory}
The unitary irreducible representations of  $\mathfrak{osp} (1|4)$ were first studied in  \cite{Heidenreich:1982rz} (see also \cite{Nicolai:1984hb,deWit:1999ui} for a detailed review). They are conveniently described by their decomposition into irreducible representations of $\mathfrak{so}(3, 2)$ in analogy with the case of the super-Poincar\'e algebra in $\mb{M}^{4|4}$ (cf. \eqref{FMSMassiveIrrepDecomp4}). Thus from a field theoretic standpoint, each term in the decomposition can be identified with a particle (or field)
carrying definite energy and spin. Unlike in  $\mb{M}^4$, particles in AdS${}_4$ belonging to the same supermultiplet do not carry the same energy. 

The irreducible representations of  $\mathfrak{osp} (1|4)$  are labelled by the quantum numbers $E_0$ and superspin $s$, which we encode in the notation $\mf{S}(E_0,s)$.
There exist four types of unitary supermultiplets in AdS${}_4$.\footnote{Our notation for the AdS$_4$ supermultiplets follows the one introduced in \cite{Sibiryakov}. Fronsdal 
	\cite{Fronsdal:1985pc} used a different notation for these representations, 
	$D^S(E_0, s)$.
} 
The first are known as massive representations and have the following decomposition\cite{Heidenreich:1982rz}
\begin{subequations} \label{E1}
	\begin{align}
	\mathfrak{S}\big({E}_0,s\big):=
	D\big(E_0+ \frac{1}{2}, s- \frac{1}{2}\big)  \oplus &D\big( E_0, s\big) \oplus  D \big(E_0+1, s\big) \oplus  D\big(E_0+ \frac{1}{2}, s+ \frac{1}{2}\big)~, \label{E1.a} \\
	&\quad s>0\,, \quad E_0 >s+1\,, 
	\end{align}
\end{subequations}
where the last inequality is the unitarity bound. This decomposition implies that a massive supermultiplet describes four particles in AdS${}_4$. 

When $s=0$ the first term in the right-hand side of \eqref{E1.a} is absent in the decomposition and the unitarity bound is also modified. 
The corresponding representations are referred to as the Wess-Zumino representations: 
\bea
\mathfrak{S}\big({E}_0,0\big):=D\big( E_0, 0 \big)  \oplus D \big(E_0+1, 0\big)
\oplus D\big(E_0+ \frac{1}{2},  \frac{1}{2}\big)  \,, \qquad E_0 >\frac{1}{2}\,. 
\label{E2}
\eea
When the unitarity bound is saturated, $E_0= s+1$ in eq.~\eqref{E1} or $E_0=\frac{1}{2}$ in eq.~\eqref{E2}, the supermultiplets get shortened. 
This yields  the other two types of representations. 

For $s>0$ the decomposition~\eqref{E1} contains fields of spin $s \geq 1$ and the shortening 
can be attributed to the appearance of  gauge invariance at the field theoretic level. The corresponding representations are called massless and are of the form 
\bea
\mathfrak{S}\big(s+1,s\big):=
D\big(s+1, s \big) \oplus  D\big(s+ \frac{3}{2}, s+ \frac{1}{2}\big) \,, \qquad s>0\,.
\label{E3}
\eea
This implies that a massless supermultiplet in AdS$_4$ consists of two physical component fields. 

Finally, the fourth type of representation occurs 
for $s=0, \ E_0= \frac{1}{2}$ and is called the Dirac supermultiplet 
\cite{Flato:1980zk,Fronsdal:1981gq}
(or super-singleton) 
\bea 
\mathfrak{S}\big( \hf, 0\big) = D\big(\hf , 0 \big) \oplus  D\big(1, \frac{1}{2}\big) \,.
\label{E4}
\eea
It unifies the bosonic and fermionic singletons.

\subsection{$\cN=1$ AdS superspace}\label{N1AdSSuperspaceSec}
Historically,  $\cN=1$ AdS superspace AdS$^{4|4}$ was originally introduced 
\cite{Keck,Zumino77,IS} as the coset space
${\rm AdS^{4|4} } := {{\sOSp}(1|4)}/{{\sSO}(3,1)}$. 
The same superspace is equivalently realised as a unique maximally symmetric solution of 
the two off-shell formulations for $\cN=1$ AdS supergravity:
(i) the well-known minimal theory (see, e.g., \cite{GatesGrisaruRocekSiegel1983,BuchbinderKuzenko1998} for reviews); and 
(ii)  the more recently discovered non-minimal theory \cite{ButterK}. Here we prefer to use the second definition, since it allows us to obtain all information about the geometry of 
${\rm AdS}^{4|4}$ from the well-known supergravity results. 

As usual, we denote by  $z^M =(x^m , \q^\m, \bar \q_{\dot \m} ) $ the local coordinates
of ${\rm AdS}^{4|4}$. 
The geometry  of ${\rm AdS}^{4|4}$ is described in terms of covariant derivatives
of the form
\bea
\cD_A = (\cD_a , \cD_\a ,\bar \cD^\ad ) = E_A + \O_A~, \qquad
E_A = E_A{}^M \partial_M  ~,
\label{33.1}
\eea
where $E_A{}^M $ is the inverse superspace vielbein, and 
\bea
\O_A = \frac{1}{2}\,\O_A{}^{bc} M_{bc}
= \O_A{}^{\b \g} M_{\b \g}
+\bar \O_A{}^{\bd \gd} \bar M_{\bd \gd} ~,
\eea	
is the Lorentz connection. 
The covariant derivatives $\cD_A$
obey the following graded commutation relations:
\begin{subequations}\label{FASDerivativeAlgebra}
	\bea
	\{\cD_\a,\cDB_{\ad} \} &=& -2 \ri \cD_{\a\ad}~, \\
	\{ \cD_\a, \cD_\b \} &=& -4\mub M_{\a\b}~, \qquad \{ \cDB_\ad, \cDB_\bd \} = 4 \m \bar{M}_{\ad \bd}~, \\
	\, [\cD_\a , \cD_{\b\bd} ] &=& \ri \mub \varepsilon_{\a\b}\cDB_{\bd},  \qquad  \, \hspace{0.22cm} [\cDB_\ad, \cD_{\b\bd} ] = - \ri \m \varepsilon_{\ad\bd}\cD_\b, \\
	\, [\cD_{\a\ad},\cD_{\b\bd}] &=& - 2\mub \m (\varepsilon_{\a\b}\bar{M}_{\ad\bd} + \varepsilon_{\ad\bd}M_{\a\b} )~.
	\eea
\end{subequations}
Here $\m\neq 0$ is a  complex parameter, which we recall is related to the scalar curvature $\cR$ of  AdS$_4$ by the rule $\cR = -12 \m \bar \m$.

In what follows, we make extensive use of the quadratic Casimir of $\mathfrak{osp} (1|4)$, whose realisation on superfields takes the form \cite{BKS}
\be \label{Cas1}
\mathbb{Q}:=\Box+\frac{1}{4}\big(\mu\mathcal{D}^2+\mub\bar{\mathcal{D}}^2\big)
-\mu\mub\big(M^{\a\b}M_{\a\b}+\bar{M}^{\ad\bd}\bar{M}_{\ad\bd}\big)~, 
\quad 
\big[\mathbb{Q},\mathcal{D}_{A}\big]=0~. 
\ee
Here $\Box = -\hf \cD^{\a\ad}\cD_{\a\ad}$ is the d'Alembert operator in AdS$^{4|4}$.

Let us introduce several types of constrained superfields on AdS$^{4|4}$ which will be crucial to our subsequent analysis. 
For integers $m,n \geq 1$, a totally symmetric complex tensor superfield $\G_{\a(m)\ad(n)}$ is said to be transverse linear if it satisfies the constraint 
\bea
\cDB^\bd {\G}_{\a(m)\bd\ad(n-1)} &=& 0 \quad \Longrightarrow \quad (\cDB^2-2(n+2)\mu){\G}_{\a(m)\ad(n)} = 0 ~. \label{FASTransverseLinearConstraint}
\eea
On the other hand, the superfield $\G_{\a(m)\ad(n)}$ is said to be transverse anti-linear if it obeys the constraint
\bea
\cD^\b {\G}_{\b\a(m-1)\ad(n)} &=& 0 \quad \Longrightarrow \quad (\cD^2-2(m+2)\mub){\G}_{\a(m)\ad(n)} = 0~. \label{FASTransverseAntiLinearConstraint}
\eea

Given integers $m,n \geq 0$, a complex superfield $G_{\a(m)\ad(n)}$ is said to be longitudinal linear if it satisfies the constraint 
\bea
\cDB_{(\ad_1}{G}_{\a(m)\ad_2 \ldots \ad_{n+1})}&=&0 \quad \Longrightarrow \quad (\cDB^2+2n\m){G}_{\a(m)\ad(n)}=0~. \label{FASLongitudinalLinearCon}
\eea
Similarly, the complex superfield $G_{\a(m)\ad(n)}$ is said to be longitudinal anti-linear if it obeys
\bea
\cD_{(\a_1}{G}_{\a_2\dots\a_{m+1})\ad(n)}&=&0 \quad \Longrightarrow \quad (\cD^2+2m\mub){G}_{\a(m)\ad(n)}=0~. \label{FASLongitduinalAntiLinear}
\eea
The transverse (anti-)linear and longitudinal (anti-)linear superfields were first introduced in AdS${}^{4|4}$ by Ivanov and Sorin in \cite{IS}.
For $n=0$, the first constraint in \eqref{FASTransverseLinearConstraint} is not defined, while the latter condition
\be
(\cDB^2- 4 \mu){\G}_{\a(m)\ad(n)} = 0~, \label{FASLinearSupCon}
\ee
defines a linear superfield. Note that the definition for an anti-linear superfield follows from \eqref{FASLinearSupCon} via complex conjugation.

For $n=0$, the constraint \eqref{FASLongitudinalLinearCon} defines a chiral superfield, 
\bsubeq
\be
\cDB_{\ad}G_{\a(m)}=0 \quad  \Longrightarrow \quad \cDB^2 G_{\a(m)}=0~, 
\ee
whilst for $m=0$, the condition \eqref{FASLongitduinalAntiLinear} defines an  anti-chiral superfield 
\be
\cD_{\a}{G}_{\ad(n)}=0 \quad \Longrightarrow \quad \cD^2{G}_{\ad(n)}=0~.
\ee
\esubeq
Unlike in $\mb{M}^{4|4}$, chiral and anti-chiral superfields only exist in AdS$^{4|4}$ for the cases $m \geq 0,~n=0$ and $n \geq 0,~m=0$, respectively. It is easy to see that the other cases are inconsistent, as a consequence of the algebra \eqref{FASDerivativeAlgebra}. 

Given an integer $n \geq 1$, a complex superfield ${\mathfrak V}_{\a(m)  \ad(n)} $ can be uniquely expressed
as the sum of a transverse linear and a longitudinal linear superfield
\bea
\mathfrak{V}_{\a(m) \ad(n)} = &-& 
\frac{1}{2 \mu (n+2)} \cDB^\bd \cDB_{(\bd} \mathfrak{V}_{\a(m) \ad_1 \dots  \ad_n)} 
- \frac{1}{2 \mu (n+1)} \cDB_{(\ad_1} \cDB^{|\bd|} \mathfrak{V}_{\a(m) \ad_2 \dots \ad_{n} ) \bd} 
~ . ~~~
\label{UniqueDecom}
\eea
Note that we have introduced the notation where an index placed between vertical bars, for example $|\bd|$, implies that it is not subject to symmetrisation. 
Choosing $\mathfrak{V}_{\a(m) \ad(n)} $ to be  transverse linear ($\G_{\a(m) \ad(n)} $)
or longitudinal linear ($G_{\a(m) \ad(n)} $), the above relation
gives
\vspace{-\baselineskip}
\begin{subequations}
	\bea	\label{SolveTLCon}
	\G_{\a(m)\ad(n)} &=& \cDB^\bd \x_{\a(m)\bd\ad(n)}~,\\
	G_{\a(m)\ad(n)} &=& \cDB_{(\ad_1}\z_{\a(m)\ad_2...\ad_n)}~, \label{SolveLLCon}
	\eea
\end{subequations}
for some complex prepotentials $ \x_{\a(m) \ad(n+1)}$ and $  \z_{\a(m) \ad(n-1)}$.
These relations provide  general solutions to the constraints \eqref{FASTransverseLinearConstraint}  and \eqref{FASLongitudinalLinearCon} respectively.

Ivanov and Sorin also introduced the projectors $\mathcal{P}^{\perp }_{(n)}$ and 
$\mathcal{P}^{\parallel}_{(n)}$  in \cite{IS}
which map the space of unconstrained superfields $\mathfrak{V}_{\a(m)\ad(n)}$ to the space of transverse linear \eqref{FASTransverseLinearConstraint} and longitudinal linear superfields \eqref{FASLongitudinalLinearCon}, 
respectively. These operators take the form
\begin{subequations}\label{TLLL}
	\bea
	\mathcal{P}^{\perp }_{(n)} &=& \frac{1}{4(n+1)\m} (\cDB^2+2n\m)~, \qquad \qquad \quad \cDB^\bd \mathcal{P}^{\perp }_{(n)} \mathfrak{V}_{\a(m)\bd\ad(n-1)} = 0~, \\
	\mathcal{P}^{\parallel}_{(n)} &=& -\frac{1}{4(n+1)\mu}(\cDB^2 - 2(n+2)\m)~, \qquad \cDB_{(\ad_1}\mathcal{P}^{\parallel}_{(n)} \mathfrak{V}_{\a(m)\ad_2 ... \ad_{n+1})} = 0 ~,
	\eea
\end{subequations}
and satisfy the projector properties
\be
\mathcal{P}^{\perp}_{(n)} \mathcal{P}^{\parallel}_{(n)} =0~,\qquad \mathcal{P}^{\parallel}_{(n)} \mathcal{P}^{\perp}_{(n)}  = 0~, \qquad \mathcal{P}^{\perp}_{(n)} + \mathcal{P}^{\parallel }_{(n)} =\mathds{1}~.
\ee
It follows that 
that any superfield $\mathfrak{V}_{\a(m)\ad(n)}$ can be uniquely represented as a sum of transverse linear and longitudinal linear superfields,
\bea
\mathfrak{V}_{\a(m)\ad(n)} = \G_{\a(m)\ad(n)} + G_{\a(m)\ad(n)}~,
\eea
which is consistent with \eqref{UniqueDecom}. Note that the projectors 
\eqref{TLLL} are non-analytic in $\m$.   

Finally, let us introduce the following operators:
\begin{subequations} \label{FASScalarProjectors}
	\begin{align}
	\mathcal{P}_{(0)}&:=-\frac{1}{8\mathbb{Q}}\mathcal{D}^{\b}\big(\bar{\mathcal{D}}^2-4\mu\big)\mathcal{D}_{\b}~,\\
	\mathcal{P}_{(+)}&:=\frac{1}{16\mathbb{Q}}\big(\bar{\mathcal{D}}^2-4\mu\big)\mathcal{D}^2~,\\ 
	\mathcal{P}_{(-)}&:=\frac{1}{16\mathbb{Q}}\big(\mathcal{D}^2-4\mub\big)\bar{\mathcal{D}}^2~.
	\end{align}
\end{subequations}
By making use of the identities in appendix \ref{TASappendixA}, one may show that when restricted to the space of scalar superfields, the operators \eqref{FASScalarProjectors}  satisfy the projector properties 
\begin{align}
\mathds{1}=\sum_{i}\mathcal{P}_{(i)}~,\qquad \mathcal{P}_{(i)}\mathcal{P}_{(j)}=\delta_{ij}\mathcal{P}_{(j)}~, \label{projprop}
\end{align}
for $i=0,+,-$. Here $ \mathcal{P}_{(+)}$ and $ \mathcal{P}_{(-)}$ are the chiral and anti-chiral projectors, respectively, while $ \mathcal{P}_{(0)}$ projects onto  the space of LAL superfields. 
Using these projectors, it is always possible to decompose an unconstrained complex scalar superfield $\mathfrak{V}$  as
\begin{align}
\mathfrak{V}=\mathfrak{L}+\sigma+\bar{\rho}~,
\end{align}
where the complex superfield $\mathfrak{L}$ is simultaneously linear and anti-linear, $\sigma$ is chiral and $\bar{\rho}$ is anti-chiral
\begin{align}
\big(\mathcal{D}^2-4\mub\big)\mathfrak{L}=\big(\bar{\mathcal{D}}^2-4\mu\big)\mathfrak{L} =0~,\qquad \bar{\mathcal{D}}_{\ad}\s=0~,\qquad \mathcal{D}_{\a}\bar{\rho}=0~.
\end{align}
If $\mathfrak{V}$ is real, then $\mathfrak{L}$ is also real and $\bar{\rho}=\bar{\s}$.
In the flat superspace limit  $\m \rightarrow 0$, the projectors reduce to those constructed by Salam and Strathdee \cite{SalamStrathdee1975}.

\subsection{Irreducible superfield representations} \label{FASOnShellSupermultiplets}
Given integers $m,n \geq 1$, a tensor superfield $\F_{\a(m)\ad(n)}(z) \equiv \F_{\a(m)\ad(n)}$ on AdS$^{4|4}$ is said to be on-shell if it satisfies the conditions 
\begin{subequations}\label{FASOnShellConditions}
	\begin{gather}
	\mathcal{D}^{\b}\F_{\b\a(m-1)\ad(n)}=0~,\qquad \bar{\mathcal{D}}^{\bd}\F_{\a(m) \bd \ad(n-1)}=0 ~, \label{FASOnShellTLAL} \\
	\big(\mathbb{Q}-M^2\big)\F_{\a(m)\ad(n)}=0~. \label{FASOnShellMass} 
	\end{gather}
\end{subequations}
Hence, an on-shell superfield is simultaneously transverse linear \eqref{FASTransverseLinearConstraint} and transverse anti-linear \eqref{FASTransverseAntiLinearConstraint} (TLAL). It follows from the TLAL condition \eqref{FASOnShellTLAL} that an on-shell superfield is also transverse
\be
\cD^{\b\bd}\F_{\b \a(m-1) \bd \ad(n-1)} =0~.
\ee
By going to components, we will see in section \ref{FASComponentAnalysis} that an on-shell superfield $\F_{\a(m)\ad(n)}$  \eqref{FASOnShellConditions} furnishes the irreducible representation $\mathfrak{S}\big(E_0(M),\frac{1}{2}(m+n+1)\big)$ of $\mathfrak{osp} (1|4)$, where
\begin{align}
E_0 (M) =1  + \frac{1}{2}
\sqrt{ 4\frac{M^2}{\mu\mub}-(m+n)(m+n+4) +1} ~. \label{Energy}
\end{align}
In accordance with this, an on-shell supermultiplet \eqref{FASOnShellConditions} is said to carry superspin $s=\frac{1}{2}(m+n+1)$ and pseudo-mass $M$. 

In the case when $m>n=0$ and $n>m=0$, the condition \eqref{FASOnShellTLAL} needs to be changed to
\begin{subequations} \label{FASOnShellConditionsOneIndexType}
	\begin{align}
	\mathcal{D}^{\b}\F_{\b \a(m-1)}&=0~, \qquad \big(\bar{\mathcal{D}}^2-4\mu\big)\F_{\a(m)}=0~,  \label{FASOnShellConditionsOneIndexTypeMGN}\\
	\bar{\mathcal{D}}^{\bd}\bar{\F}_{\bd \ad(n-1)}&=0~, \qquad \big(\mathcal{D}^2-4\mub\big)\bar{\F}_{\ad(n)}=0~,  \label{FASOnShellConditionsOneIndexTypeNGM}
	\end{align} 
\end{subequations}
respectively, whilst the mass-shell condition \eqref{FASOnShellMass} remains the same. In the scalar case $m=n=0$, the on-shell superfields are discussed in sections
\ref{WZ section} and \ref{section5.5}.

In analogy with the story in AdS$_4$ (cf. \eqref{FAPhysicalPseudoRel}), we define the physical mass $M_{\text{phys}}$ according to the rule
\begin{align}
M^2_{\text{phys}}= M^2-\lambda_{(1,m,n)}\mu\mub~, \qquad \l_{(1,m,n)} := \hf  \big (m+n \big ) \big (m+n+3 \big )~,\label{FASPhysicalMass}
\end{align}
where the constant $\lambda_{(1,m,n)}$ is a partially massless value.
As discussed in section \ref{FAIrredFieldReps}, the pseudo-mass of an on-shell field in AdS$_4$ is subject to a unitarity bound (cf. \eqref{FAUnitarityBoundsMasses}). Since the on-shell supermultiplet $\F_{\a(m)\ad(n)}$ contains a multitude of such fields, this in turn induces a unitarity bound on the pseudo-mass of $\F_{\a(m)\ad(n)}$:
\be
M^2_{\text{phys}}\geq 0  \quad \implies \quad M^2\geq \lambda_{(1,m,n)}\mu\mub~. \label{Sbound}
\ee

Recall in section \ref{FMSMassiveFieldsSecGen}, for $m,n \geq 1$, there also existed other constraint spaces on which tensor superfields realised irreducible representations of the super-\Po algebra. Namely, the space of chiral \eqref{FMSOnShellChiralConditions} and anti-chiral \eqref{FMSOnShellAChiralConditions} transverse superfields. However  for $m, n \geq 1$, there does not exist a direct analogue of such superfields in AdS$^{4|4}$, since (anti-)chiral superfields do not exist.


\subsubsection{Massless supermultiplets}\label{section 3.1}

An on-shell supermultiplet \eqref{FASOnShellConditions} is said to be massless if it  carries the pseudo-mass
\begin{align}
M^2=\l_{(1,m,n)}\mu\mub~, \qquad \l_{(1,m,n)} := \hf   (m+n  )  (m+n+3  )  ~. \label{FASMasslessField}
\end{align} 
It can be shown that the system of equations \eqref{FASOnShellConditions}, with pseudo-mass \eqref{FASMasslessField}, is compatible with the gauge symmetry 
\begin{subequations} \label{FASMasslessGaugeParameters}
	\begin{align}
	\delta_{\zeta,\xi}\F_{\a(m)\ad(n)}=\bar{\mathcal{D}}_{(\ad_1}
	\zeta_{\a(m)\ad_2\dots\ad_n)}+\mathcal{D}_{(\a_1}\xi_{\a_2\dots\a_m)\ad(n)}~,\label{FASMasslessGaugeSymmetry}
	\end{align} 
	where the complex gauge parameters $\z_{\a(m)\ad (n-1)}$ and 
	$\x_{\a(m-1) \ad (n)}$ are TLAL and satisfy the reality conditions
	\bea 
	\cD_{(\a_1}{}^\bd \x_{\a_2 \dots \a_m) \bd \ad (n-1)} &=&\phantom{-} \ri (n+1) \m \zeta_{\a(m)\ad (n-1)}~, \label{FASMasslessRealityCond1}\\
	\cD^\b{}_{(\ad_1} \z_{\b \a(m-1)  \ad_2 \dots  \ad_n)} 
	&=& -\ri (m+1) \bar \m \x_{\a(m-1)\ad (n)}~.\label{FASMasslessRealityCond2}
	\eea
\end{subequations}
These on-shell conditions imply that $\z_{\a(m)\ad (n-1)}$ and 
$\x_{\a(m-1) \ad (n)}$ satisfy the equations  
\bea
\big(\mathbb{Q}-\lambda_{(1,m,n)}\mu\mub\big)\z_{\a(m)\ad(n-1)}=0~, \qquad
\big(\mathbb{Q}-\lambda_{(1,m,n)}\mu\mub\big)\x_{\a(m-1)\ad(n)}=0~.\label{FASMasslessMassShellGauge}
\eea
It follows from \eqref{FASMasslessMassShellGauge} and the gauge variation of \eqref{FASOnShellMass} that the gauge parameters are only non-zero if $M$ obeys \eqref{FASMasslessField}. The mass convention \eqref{FASPhysicalMass} was chosen to ensure that a massless gauge superfield in AdS$^{4|4}$ carries zero physical mass $M^2_{\text{phys}} = 0$. In accordance with the unitarity bound, it follows that massless representations of  $\mathfrak{osp} (1|4)$ are unitary.

In the case $m=n=s$, one can consistently impose the reality condition $H_{\a(s)\ad(s)}:=\F_{\a(s)\ad(s)}=\bar{H}_{\a(s)\ad(s)}$, whereupon the gauge transformations \eqref{FASMasslessGaugeSymmetry} take the form
\begin{subequations} \label{FASMasslessReal}
	\begin{align}
	\delta_{\zeta}H_{\a(s)\ad(s)}=\bar{\mathcal{D}}_{(\ad_1}
	\zeta_{\a(s)\ad_2\dots\ad_s)}-\mathcal{D}_{(\a_1}\bar \z_{\a_2\dots\a_s)\ad(s)}~,\label{FASMasslessGaugeSymmetryReal}
	\end{align} 
	whilst the reality conditions \eqref{FASMasslessRealityCond1} and \eqref{FASMasslessRealityCond2} for the gauge parameters become
	\begin{align}
	\cD^\b{}_{(\ad_1} \z_{\b\a(s-1)\ad_2 \dots \ad_s)} &= \phantom{-}\ri (s+1) \mub \bar{\zeta}_{\a(s-1)\ad (s)}~,\label{Blah1}\\
	\cD_{(\a_1}{}^\bd \bar{\z}_{\a_2 \dots \a_s) \bd \ad (s-1)} &= -\ri (s+1) \m \zeta_{\a(s)\ad (s-1)}~. \label{Blah2}
	\end{align}
\end{subequations}

The presence of the gauge freedom \eqref{FASMasslessGaugeSymmetry} means that the 
$\sOSp (1|4)$ representation  realised on the on-shell superfield
\eqref{FASOnShellConditions} with $M^2$ given by \eqref{FASMasslessField} is not irreducible. To obtain an irreducible representation,  the space of TLAL superfields 
\eqref{FASOnShellConditions} and \eqref{FASMasslessField} has to be factorised
with respect to the gauge modes. More specifically, two superfields $\F_{\a(m)\ad(n)}$ and $\widetilde \F_{\a(m)\ad(n)}$ are said to be equivalent if they differ by a gauge transformation \eqref{FASMasslessGaugeSymmetry}. The genuine massless representation is realised on the quotient space of
the space of on-shell superfields 
\eqref{FASOnShellConditions} and \eqref{FASMasslessField} with respect to this equivalence relation. 

Alternatively, the gauge degrees of freedom are automatically eliminated if one works with 
the gauge-invariant chiral field strength \eqref{HSW1}, 
\bea
\mathfrak{W}_{\a(m+n+1)}(\F)&= -\frac{1}{4}\big(\bar{\mathcal{D}}^2-4\mu\big)\mathcal{D}_{(\a_1}{}^{\bd_1}\cdots\mathcal{D}_{\a_n}{}^{\bd_n}\mathcal{D}_{\a_{n+1}}\F_{\a_{n+2}\dots\a_{m+n+1})\bd(n)} ~,
\label{FASFieldStrength}
\eea
instead of the prepotential $\F_{\a(m)\ad(n)}$. Assuming the descendent field $\F_{\a(m)\ad(n)}$ is on-shell, it follows from the constraint \eqref{FASOnShellTLAL} that $\mathfrak{W}_{\a(m+n+1)}(\f)$ is constrained by
\begin{align}
0=\mathcal{D}^{\b}\mathfrak{W}_{\b \a(m+n)}(\F)\quad\implies\quad 0=\big(\mathbb{Q}-\lambda_{(1,m,n)}\mu\mub\big)\mathfrak{W}_{\a(m+n+1)}(\f)~,
\label{4.9}
\end{align}
where the second relation follows from  the chirality of the field strength. Once again, the latter is consistent with \eqref{FASOnShellMass} only if $M$ satisfies \eqref{FASMasslessField}.
The description in terms of $\mathfrak{W}_{\a(m+n+1)}(\F)$ provides the second way to formulate massless dynamics. 

Let us return to the gauge-invariant chiral field strength \eqref{FASFieldStrength} for $n=2s+1$ and $n=2s$, which 
originate in the massless models for the superspin-$(s+\hf)$ and superspin-$s$ multiplets in AdS$^{4|4}$ \cite{KS94}, respectively. As was demonstrated in \cite{KS94}, on the mass shell they satisfy the equations 
\begin{subequations}
	\bea
	\mathcal{D}^{\b}\mathfrak{W}_{\b \a(2s)}(\F) &=&0~, \\
	\qquad 
	\mathcal{D}^{\b}\mathfrak{W}_{\b \a(2s-1)}(\F) &=&0~,
	\eea
\end{subequations}
which are exactly the on-shell constraints \eqref{4.9}.


\subsubsection{Partially massless supermultiplets} \label{section 3.2}
The results presented in this section were obtained by Michael Ponds in \cite{BHKP}.

An on-shell supermultiplet $\F^{(t)}_{\a(m)\ad(n)}$ \eqref{FASOnShellConditions} is said to be partially massless with super-depth $t$ if it carries the pseudo-mass
\begin{align}
M^2=\lambda_{(t,m,n)}\mu\mub~,\qquad 1\leq t \leq \text{min}(m+1,n+1)~, \label{FASPMMass}
\end{align}
where the constants $\lambda_{(t,m,n)}$ are the partially massless values defined by
\begin{align}
\l_{(t,m,n)} &= \hf \Big ( (m+n-t+1)(m+n-t+4) + t(t-1) \Big )~.
\label{FASSuperPartiallyMasslessValues}
\end{align}
We will demonstrate in section \ref{FASComponentAnalysis} that the surviving component fields of an on-shell partially massless superfield with super-depth $t>1$ are themselves partially massless with depths $t$ or $t-1$.  The reason that we have chosen the bound on $t$ to be  $1\leq t \leq \text{min}(m+1,n+1)$ in the definition \eqref{FASPMMass} is because, as we will see, 
these are the only values for $t$ which have a gauge symmetry.

From \eqref{FASPMMass}, we see that strictly massless superfields carry super-depth $t=1$. Partially massless supermultiplets whose super-depth lies within the range $2\leq t \leq \text{min}(m+1,n+1)$ have negative physical mass and describe non-unitary representations of $\sOSp(1|4)$. This may be seen from the inequality
\begin{align}
\lambda_{(1,m,n)} > \lambda_{(t,m,n)}~ , \qquad 2\leq t \leq \text{min}(m+1,n+1)~.
\label{PMbound}
\end{align}
To distinguish the latter we may sometimes refer to them as true partially massless supermultiplets. In analogy with \eqref{NSPMrep}, we will denote 
the true partially massless representation with super-depth $t$ and Lorentz type $(\frac{m}{2},\frac{n}{2})$
by 
\begin{align}
\mathfrak{P}(t,m,n)~, \qquad 2\leq t \leq \text{min}(m+1,n+1)~.
\end{align}
According to \eqref{Energy}, such a representation carries minimal energy $E_0=\frac{1}{2}(m+n+1)-t+2$.

In the previous section it was observed that on-shell, \eqref{FASMasslessField} is the only pseudo-mass value compatible with the massless gauge symmetry \eqref{FASMasslessGaugeSymmetry}. For true partially massless supermultiplets the story is considerably more complicated. This is because the correct gauge symmetry at the superspace level is not yet known, much less the gauge invariant actions and field strengths.
However, by making use of the superprojectors which we will introduce in section \ref{Superspin-projection operators}, it is possible to systematically derive the most general gauge symmetry
compatible with the on-shell conditions \eqref{FASOnShellConditions}.\footnote{To emphasise, the partially massless values \eqref{FASPMMass}, which characterise partially massless supermultiplets in AdS$^{4|4}$, were extracted from the superspin projection operators \eqref{FASSuperprojectors}. Nothing was known about partially massless superfields prior to the creation of these superprojectors. } In appendix \ref{TASappendixB} this procedure is carried out in detail for the real supermultiplet $\F_{\a\ad}$. The results of this analysis, and that of the supermultiplet $\F_{\a(2)\ad(2)}$ obtained by analogy, are summarised below 
\begin{table}[h]
	\begin{center}
		\begin{tabular}{|c|c|c|c|}
			\hline
			~Super-depth~  & $t=1$     & $t=2$ & $t=3$  \Tstrut\Bstrut\\ \hline
			$\delta \F_{\a\ad}$        &  $\bar{\mathcal{D}}_{\ad}\z_{\a}-\mathcal{D}_{\a}\bar{\z}_{\ad}$   & $\mathcal{D}_{\a\ad}(\s+\bar{\s})$  & -- \Tstrut\\
			\hline
			$\delta \F_{\a(2)\ad(2)}$    & $\bar{\mathcal{D}}_{\ad}\z_{\a(2)\ad}-\mathcal{D}_{\a}\bar{\z}_{\a\ad(2)}$   & $\mathcal{D}_{\a\ad}\big(\bar{\mathcal{D}}_{\ad}\xi_{\a}-\mathcal{D}_{\a}\bar{\xi}_{\ad}\big)$     & $\mathcal{D}_{\a\ad}\mathcal{D}_{\a\ad}(\eta+\bar{\eta})$\\ \hline
		\end{tabular}
	\end{center} 
	\vspace{-15pt} 
	\caption{Gauge symmetry for lower-rank PM supermultiplets.}
	\label{table 3}
\end{table}

In table \ref{table 3}, the most general gauge symmetries for the aforementioned supermultiplets are given. In both cases it turns out that a gauge symmetry is present only when the pseudo-mass takes the values specified in \eqref{FASPMMass}. The gauge parameters are on-shell in the sense that they each possess the same pseudo-mass as its parent gauge field and that they satisfy certain irreducibility constraints. Specifically, in the super-depth $t=1$ case, the gauge parameter $\z_\a$ in table \ref{table 3} is simultaneously linear \eqref{FASLinearSupCon} and transverse anti-linear \eqref{FASTransverseAntiLinearConstraint} (LTAL), while $\z_{\a(2) \ad} $ is TLAL. They also obey the reality conditions \eqref{Blah1} and \eqref{Blah2}. The gauge parameter $\x_\a$, which corresponds to $\F_{\a(2)\ad(2)}$ with super-depth $t=2$, is LTAL and obeys the reality condition 
\bea
\mathcal{D}_{\ad}{}^{\b}\xi_{\b}=3\text{i}\mub\bar{\xi}_{\ad}~.
\eea
In the case of maximal super-depth, 
the gauge parameters $\s$ and $\eta$ are chiral and  obey 
the  reality conditions
\begin{subequations} \label{suppGb}
	\bea
	-\frac{1}{4}\big(\mathcal{D}^2-4\mub\big)\s+2\mub\bar{\s}&=&0~, \\
	-\frac{1}{4}\big(\mathcal{D}^2-4\mub\big)\eta+3\mub\bar{\eta}&=&0~.
	\eea
\end{subequations}
These conditions are the equations of motion which follow from various Wess-Zumino models (c.f. \eqref{EoMWZ}).

The method used to derive the above results (see appendix \ref{TASappendixB} for the details) is quite general in that it deduces all types of gauge symmetry an on-shell supermultiplet can possess, and at which mass values they appear. However, for higher-rank multiplets it quickly becomes infeasible due to the computational expense. Nevertheless, from the results of the above analysis, we are now in a position to make an ansatz for the gauge transformations of the  half-integer superspin-$(s+\frac{1}{2})$ multiplets  $\F^{(t)}_{\a(s)\ad(s)}$.\footnote{The gauge transformations for partially massless superfields when $m \neq n$ have yet to be derived.  } For true partially-massless multiplets with super-depth $2\leq t \leq s+1$, they are given by\footnote{Interestingly, the higher-depth gauge transformations \eqref{Blah7} are different to those proposed for the generalised superconformal multiplets in \cite{KPR2}. The latter transformations possess the same functional form as the second group of terms in \eqref{decomp2}.
}
\begin{subequations}
	\begin{align}
	2\leq t \leq s~&: ~\qquad \delta_{\xi}\F^{(t)}_{\a(s)\ad(s)}=\big(\mathcal{D}_{\a\ad}\big)^{t-1}\big(\bar{\mathcal{D}}_{\ad}\xi_{\a(s-t+1)\ad(s-t)}-\mathcal{D}_{\a}\bar{\xi}_{\a(s-t)\ad(s-t+1)}\big)~,\label{Blah7}\\[5pt]
	t=s+1~&:~\qquad  \delta_{\s}\F^{(s+1)}_{\a(s)\ad(s)}=\big(\mathcal{D}_{\a\ad}\big)^{s}\big(\s+\bar{\s}\big)~. \label{Blah8}
	\end{align}
\end{subequations}
The system of equations \eqref{FASOnShellConditions} and \eqref{FASPMMass} is invariant under the transformations \eqref{Blah7} only if the gauge parameters are TLAL and satisfy the reality conditions
\begin{subequations}\label{Blah5}
	\begin{align}
	\cD^\b{}_{\ad} \xi_{\b\a(s-t)\ad(s-t)} &= \phantom{-}\ri (s+1) \mub \bar{\xi}_{\a(s-t)\ad (s-t+1)}~, \\
	\cD_{\a}{}^{\bd} \bar{\xi}_{\a(s-t)\ad(s-t)\bd} &= -\ri (s+1) \mu \xi_{\a(s-t+1)\ad (s-t)}~. 
	\end{align}
\end{subequations}
The same is true for the gauge transformations \eqref{Blah8} given that $\s$ is chiral, $\bar{\mathcal{D}}_{\ad}\s=0$, and that it obeys the conditions
\begin{subequations}\label{BLAH7}
	\begin{align}
	-\frac{1}{4}\big(\mathcal{D}^2-4\mub\big)\s+(s+1)\mub\bar{\s}&=0~,\\
	-\frac{1}{4}\big(\bar{\mathcal{D}}^2-4\mu\big)\bar{\s}+(s+1)\mu\s &=0~.
	\end{align}
\end{subequations}
It may be shown that \eqref{Blah5} and \eqref{BLAH7} imply that each gauge parameter satisfies the same mass-shell equation as its parent gauge field, as required. We note that upon substituting $t=1$ into \eqref{Blah7} and \eqref{Blah5}, one recovers the massless gauge transformations \eqref{FASMasslessGaugeSymmetryReal} with $\xi_{\a(s)\ad(s-1)}\equiv \z_{\a(s)\ad(s-1)}$.


\subsubsection{Massive supermultiplets} \label{section 3.3}

An on-shell superfield \eqref{FASOnShellConditions} is said to be massive if it has a pseudo-mass that satisfies
\begin{align}
M^2 \geq  \lambda_{(1,m,n)}\mu\mub~,   \label{FASMassiveSuperfields}
\end{align}
but is otherwise arbitrary. By virtue of the unitarity bound \eqref{Sbound}, this definition ensures that the physical mass is positive and that the corresponding representation of $\sOSp(1|4)$ is unitary. Note that we can also introduce (non-unitary) massive fields in AdS$^{4|4}$, which carry pseudo-mass $M^2 \leq  \lambda_{(1,m,n)}\mu\mub$ such that , $M^2 \neq \lambda_{(t,m,n)}\mu\mub$, for $2\leq t \leq \text{max}(m+1,n+1)$. However, we will not be interested in these types of massive supermultiplets.

Given integers $m,n \geq 1$, with $m+n \equiv 2s$, we denote by  $\cL_{(m,n)}^{[M]} $ the space of on-shell superfields \eqref{FASOnShellConditions}.
The following proposition holds: 
Provided the pseudo-mass satisfies
\begin{align}
M^2 \neq \l_{(t, m,n)}~,\qquad 1\leq t \leq \text{max}(m+1,n+1)~, \label{equiv55}
\end{align}
the $\mf{osp}(1|4)$ representations on the functional spaces  $\cL_{(2s,0)}^{[M]},~ 
\cL_{(2s-1,1)}^{[M]} , \dots 
,\cL_{(1,2s-1)}^{[M]}, ~  \cL_{(0, 2s)}^{[M]} $ are equivalent to 
$\mathfrak{S} \big( E_0 (M) , s +\hf \big)$, where $E_0(M)$ is given by \eqref{Energy}.

To prove the above claim, we consider  an arbitrary superfield $\F_{\a(m)\ad(n)} \in \cL_{(m,n)}^{[M]}$ and  associate with it
the following descendants 
\begin{subequations}\label{FASMassiveReality}
	\bea
	\F_{\a(m)\ad(n)} ~& \to &~  \j_{\a(m+1)\ad(n-1)}
	:= \cD_{\a_{m+1} }{}^\bd  \F_{\a(m)\bd \ad(n-1)} 
	\label{FASMassiveReality1}
	~, \\
	\F_{\a(m)\ad(n)} ~& \to &~  \c_{\a(m-1)\ad(n+1)}
	:= \cD^{\b }{}_{\ad_{n+1}}  \F_{\b\a(m-1)\ad(n)} ~.
	\label{FASMassiveReality2}
	\eea
\end{subequations} 
It is obvious that $ \j_{\a(m+1)\ad(n-1)}$ is completely symmetric in its undotted indices, 
while  $\c_{\a(m-1)\ad(n+1)} $ is completely symmetric in its dotted indices. 
The descendant $ \j_{\a(m+1)\ad(n-1)}$ proves to obey the conditions \eqref{FASOnShellConditions}  
if $n\neq 1$. In the $n=1$ case,  $ \j_{\a(m+1)\ad(n-1)}$ is an on-shell superfield of the type \eqref{FASOnShellConditionsOneIndexTypeMGN}. The descendant $ \c_{\a(m-1)\ad(n+1)}$ has analogous properties. 
Therefore the relations \eqref{FASMassiveReality1} and \eqref{FASMassiveReality2} define linear mappings
from  $\cL_{(m,n)}^{[M]} $ to 
$\cL_{(m+1,n-1)}^{[M]} $ and  $\cL_{(m-1,n+1)}^{[M]} $, respectively. 

Making use of the identities \eqref{A.4}
leads to the relations
\begin{subequations}\label{FASVectIdenExpl}
	\bea
	\cD^\b{}_\gd \cD_{\b }{}^\bd  \F_{\a(m)\bd \ad(n-1)} &=& 
	\big(M^2 - \l_{(n+1, m, n)} \m \bar \m  \big)\F_{\a(m)\gd \ad(n-1)} ~,\\
	\cD_\g{}^\bd \cD^{\b }{}_\bd  \F_{\b\a(m-1) \ad(n)} &=& 
	\big(M^2 - \l_{(m+1, m, n)} \m \bar \m  \big)\F_{\g \a(m-1) \ad(n)}~.
	\eea
\end{subequations}
In conjunction with the identities
\bea
\l_{(n+1, m, n)} &=& \l_{(m+2,m+1,n-1)}~, \qquad
\l_{(m+1, m, n)} = \l_{(n+2,m-1,n+1)}~,
\eea
the relations \eqref{FASVectIdenExpl} tell us  
that the linear transformations \eqref{FASMassiveReality1} and \eqref{FASMassiveReality2} 
are one-to-one and onto, as long as 
$M^2 \neq \l_{(n+1, m, n)} \m \bar \m$ and $M^2 \neq  \l_{(m+1, m, n)} \m \bar \m$.

Let us introduce the linear maps
\bea
\D_{(m,n)} : \cL_{(m,n)}^{[M]}  \to 
\cL_{(m+1,n-1)}^{[M]} ~, \qquad 
\widetilde  \D_{(m,n)} : \cL_{(m,n)}^{[M]}  \to 
\cL_{(m-1,n+1)}^{[M]} ~,
\eea
defined by
\begin{subequations}
	\bea
	\F_{\a(m)\ad(n)} ~& \to &~  \F_{\a(m+1)\ad(n-1)}
	:=  \big(\D_{(m,n)} \big)_{\a_{m+1} }{}^\bd  \F_{\a(m)\bd \ad(n-1)} ~, \\
	\F_{\a(m)\ad(n)} ~& \to &~  \F_{\a(m-1)\ad(n+1)}
	:=  \big(\widetilde \D_{(m,n)} \big)^\b{}_{\ad_{n+1} }  \F_{\b\a(m-1)\bd \ad(n)} ~, 
	\eea
\end{subequations}
where we have introduced the dimensionless operators
\bea
\big(\D_{(m,n)} \big)_{\a\ad} = \frac{\mathcal{D}_{\a\ad}}{\sqrt{\mathbb{Q}-\lambda_{(n+1,m,n)}\mu\mub}}~, \qquad 
\big(\widetilde{\D}_{(m,n)} \big)_{\a\ad} = \frac{\mathcal{D}_{\a\ad}}{\sqrt{\mathbb{Q}-\lambda_{(m+1,m,n)}\mu\mub}}~.~~~ \label{indexconverters}
\eea
The above relations are equivalent to 
\bea
\widetilde \D_{(m+1,n-1)} \D_{(m,n)} \Big|_{\cL_{(m,n)}^{[M]}  } = {\mathbbm 1}~,\qquad 
\D_{(m-1,n+1)} \widetilde \D_{(m,n)} \Big|_{\cL_{(m,n)}^{[M]}  } = {\mathbbm 1}~.
\eea
Therefore, the superfields 
\bea
\F_{\a(2s)}, ~ \F_{\a(2s-1)\ad} ,~ \dots   , ~\F_{\a(m)\ad(n)} ,~ \dots , 
~  \F_{\a\ad(2s-1)}, ~ \F_{\ad(2s)} ~,
\eea
\sloppy{ are also on-shell  \eqref{FASOnShellConditions} and hence furnish the representation $\mathfrak{S} \big( E_0 (M) , s +\hf \big)$.}

In proving the above assertion it was assumed that the pseudo-mass of $\F_{\a(m)\ad(n)}$
satisfies \eqref{equiv55}. This was to ensure that when $\F_{\a(m)\ad(n)}$ is on the mass-shell, the operators \eqref{indexconverters} are well defined. When $\F_{\a(m)\ad(n)}$ is partially-massless with super-depth $t$, at some stage the maps between tensor types become ill defined. Nevertheless, in this case it may be shown that the corresponding representation may be equivalently realised on any of the functional spaces $\mathcal{L}^{[M]}_{(m+n-t+1,t-1)},\mathcal{L}^{[M]}_{(m+n-t,t)},\dots,\mathcal{L}^{[M]}_{(t,m+n-t)},\mathcal{L}^{[M]}_{(t-1,m+n-t+1)}$.

In general, $ \F_{\a(m)\ad(n)} $ and its conjugate $\bar  \F_{\a(n)\ad(m)} $ 
describe two equivalent representations of  $\mf{osp}(1|4)$. In order to obtain a single 
representation, we need to impose a reality condition on $ \F_{\a(m)\ad(n)} $. 
In the case when $m+n \equiv 2s $ is even,  $ \F_{\a(m)\ad(n)} $ 
can represented  by  $ \F_{\a(s)\ad(s)} $ which we require to be real,
\bea
\bar  \F_{\a(s)\ad(s)}  =\F_{\a(s)\ad(s)}~.
\eea
If $m+n \equiv 2s-1 $ is odd,  $ \F_{\a(m)\ad(n)} $ can be represented 
by $\F_{\a(s)\ad(s-1)}$. In this case a reality condition can be chosen in the form of a
Dirac-type pair of equations
\begin{subequations} \label{Dirac}
	\begin{align}
	\mathcal{D}_{(\ad_1}{}^{\b}\F_{\b\a(s-1)\ad_2\dots\ad_s)}&=\re^{\ri \g} 
	M^2_{(s+1)}\bar{\F}_{\a(s-1)\ad(s)}~,\\
	\mathcal{D}_{(\a_1}{}^{\bd}\bar{\F}_{\a_2\dots\a_s)\ad(s-1)\bd}&=
	\re^{-\ri \g} M^2_{(s+1)}\F_{\a(s)\ad(s-1)}~,
	\end{align}
\end{subequations}
where $M^2_{(s+1)}:=M^2-\lambda_{(s+1,s,s-1)}\mu\mub$ and $\g$ is a constant real phase. 
The pair of equations \eqref{Dirac} lead to the same mass-shell equation \eqref{FASOnShellConditions}.

\subsection{Component analysis of on-shell supermultiplets} \label{FASComponentAnalysis}
We now turn to the component analysis of the on-shell superfield $\F_{\a(m)\ad(n)}$ satisfying the conditions  \eqref{FASOnShellConditions} with arbitrary $M^2$. 
For this, we make use of the bar-projection of a tensor superfield ${\mathfrak V} ={\mathfrak V}(x,\q, \bar \q)$ (with suppressed indices)  in AdS$^{4|4}$, which is defined in the same way as its counterpart in $\mb{M}^{4|4}$ (cf. \eqref{FMSBarProj}).
The supermultiplet of fields associated with $\mathfrak V$ is defined to consist 
of all independent fields contained in 
$\big\{ {\mathfrak V}|, \cD_{ \a} {\mathfrak V}|, \bar \cD_{ \ad} {\mathfrak V}|, \dots \big\}$.

The covariant derivative of AdS$_4$, $\nabla_{a}$,  
is related to $\cD_{a}$ according to the rule
\begin{align}
\nabla_{a} \cV  :=(\mathcal{D}_{a} {\mathfrak V} )|~, \qquad \cV := {\mathfrak V}|~.
\end{align}
In practice, we assume a Wess-Zumino gauge condition to be imposed on the geometric objects in \eqref{33.1}
such that the background geometry 
is purely bosonic, 
\bea
\cD_a| :=  E_a{}^M |\partial_M  + \frac{1}{2}\,\O_a{}^{bc} | M_{bc}
=e_a{}^m \partial_m  + \frac{1}{2}\,\o_a{}^{bc} M_{bc} = \nabla_a~.
\eea
The supersymmetry transformation of the component fields of $\mathfrak V$ 
is computed according to the rule \eqref{A.14}

Below we make use of the non-supersymmetric AdS quadratic Casimir operator 
\eqref{FAQuadratic Casimir}
which is related to the Casimir \eqref{Cas1} via $\mathcal{Q} \cV =\Big[ \Big( \mathbb{Q}-\frac{1}{4}\big(\mu\mathcal{D}^2+\mub\bar{\mathcal{D}}^2\big) \Big) {\mathfrak V}\Big]\Big|$.


\subsubsection{Massive supermultiplets}\label{FASMassiveSupermultipletsComp}

In general there are four non-vanishing independent complex component fields,
\begin{subequations}
	\begin{align}
	A_{\a(m)\ad(n)}&:=\F_{\a(m)\ad(n)}|~,\\
	B_{\a(m+1)\ad(n)}&:=\mathcal{D}_{\a}\F_{\a(m)\ad(n)}|~,\\
	C_{\a(m)\ad(n+1)}&:=\bar{\mathcal{D}}_{\ad}\F_{\a(m)\ad(n)}|~,\\
	E_{\a(m+1)\ad(n+1)}&:=\bigg(\frac{1}{2}\big[\mathcal{D}_{\a},\bar{\mathcal{D}}_{\ad}\big]-\text{i}\frac{m-n}{m+n+2}\mathcal{D}_{\a\ad}\bigg)\F_{\a(m)\ad(n)}|~.
	\end{align}
\end{subequations}
They have each been defined in such a way that they are transverse 
\begin{subequations}
	\begin{align}
	0&=\nabla^{\b\bd}A_{\b\a(m-1)\bd\ad(n-1)}~,\label{PMA1}\\
	0&=\nabla^{\b\bd}B_{\b\a(m)\bd\ad(n-1)}~,\label{PMB1}\\
	0&=\nabla^{\b\bd}C_{\b\a(m-1)\bd\ad(n)}~,\label{PMC1}\\
	0&=\nabla^{\b\bd}E_{\b\a(m)\bd\ad(n)}~,\label{PMD1}
	\end{align}
\end{subequations}
which may be shown by making use of the on-shell conditions \eqref{FASOnShellConditions}. Furthermore, one may show that the bottom and top components satisfy the mass-shell equations
\begin{subequations} \label{MS1}
	\begin{align}
	0&=\Big(\mathcal{Q}-\big[M^2-\frac{1}{2}(m+n+4)\mu\bar{\mu}\big]\Big)A_{\a(m)\ad(n)}~,\\
	0&=\Big(\mathcal{Q}-\big[M^2+\frac{1}{2}(m+n)\mu\bar{\mu}\big]\Big)E_{\a(m+1)\ad(n+1)}~,
	\end{align}
\end{subequations}
whilst the two middle component fields satisfy the differential equations
\begin{subequations}
	\bea
	0&=&\phantom{-}\ri \m \nabla_{(\ad_1}{}^{\b}B_{\b\a(m)\ad_2\dots\ad_{n+1})}
	+
	\Big(\mathcal{Q}-\big[M^2-\frac{1}{2}(m-n+3)\mu\bar{\mu}\big]\Big)C_{\a(m)\ad(n+1)}
	~, \label{EOM1} \hspace{0.7cm}\\
	0&=& -\ri \mub \nabla_{(\a_1}{}^{\bd}C_{\a_2\dots\a_{m+1})\ad(n)\bd}
	+
	\Big(\mathcal{Q}-\big[M^2-\frac{1}{2}(n-m+3)\mu\bar{\mu}\big]\Big)B_{\a(m+1)\ad(n)}~. 
	~~~\label{EOM2}
	\eea
\end{subequations}
The latter imply
that only one of the two fields $B_{\a(m+1)\ad(n)}$ and $C_{\a(m)\ad(n+1)}$ is independent, and lead to
the higher derivative mass-shell equations
\begin{subequations}\label{MS2}
	\begin{align}
	0&=\Big(\mathcal{Q}-M^2_{+}\Big)\Big(\mathcal{Q}-M^2_{-}\Big)B_{\a(m+1)\ad(n)}~,\\
	0&=\Big(\mathcal{Q}-M^2_{+}\Big)\Big(\mathcal{Q}-M^2_{-}\Big)C_{\a(m)\ad(n+1)}~,
	\end{align}
\end{subequations}
where we have denoted
\begin{align}
M^2_{\pm}&= M^2-\mu\bar{\mu}\pm\frac{\mu\mub}{2}\sqrt{ 4\frac{M^2}{\mu\mub}-(m+n)(m+n+4)+1}~.
\end{align}

Using the relation \eqref{FASPhysicalMassEnergy}, one may confirm that the above component results 
are in agreement with the decomposition \eqref{E1},  
\begin{align}
\mathfrak{S}\big({E}_0,s\big) =
D\big(E_0+ \frac{1}{2}, s- \frac{1}{2}\big)  &\oplus D\big( E_0, s\big) \oplus D \big(E_0+1, s\big) \oplus  D\big(E_0+ \frac{1}{2}, s+ \frac{1}{2}\big)~, \label{Blah4}
\end{align}
as dictated by representation theory. 
Here 
$s=\frac{1}{2}(m+n+1)$ and $E_0\equiv E_0(M)$ is defined according to $\eqref{Energy}$. 


\subsubsection{Partially massless supermultiplets} \label{section 3.6}
In this section we restrict our attention to on-shell true partially massless supermultiplets with super-depth $2\leq t \leq \text{min}(m+1,n+1)$. The massless case with $t=1$ will be considered separately in the next section. Upon specifying the pseudo-mass to be given by \eqref{FASPMMass}, one can show that the mass-shell equations \eqref{MS1} and \eqref{MS2} become
\begin{subequations}
	\begin{align}
	0&=\Big(\mathcal{Q}-\tau_{(t-1,m,n)}\mu\bar{\mu}\Big)A_{\a(m)\ad(n)}~,\label{PMA2}\\
	0&=\Big(\mathcal{Q}-\tau_{(t-1,m+1,n)}\mu\bar{\mu}\Big)\Big(\mathcal{Q}-\tau_{(t,m+1,n)}\mu\bar{\mu}\Big)B_{\a(m+1)\ad(n)}~,\label{PMB2}\\
	0&=\Big(\mathcal{Q}-\tau_{(t-1,m,n+1)}\mu\bar{\mu}\Big)\Big(\mathcal{Q}-\tau_{(t,m,n+1)}\mu\bar{\mu}\Big)C_{\a(m)\ad(n+1)}~,\label{PMC2}\\
	0&=\Big(\mathcal{Q}-\tau_{(t,m+1,n+1)}\mu\bar{\mu}\Big)E_{\a(m+1)\ad(n+1)}~. \label{PMD2}
	\end{align}
\end{subequations}
Here $\tau_{(t,m,n)}$ are the non-supersymmetric partially massless values \eqref{FAPhysicalPseudoRel}, which are related to the supersymmetric ones \eqref{FASSuperPartiallyMasslessValues} through
\begin{align}
\lambda_{(t,m,n)}=\tau_{(t-1,m,n)}+\frac{1}{2}(m+n+4)~.
\end{align}

In accordance with the discussion of on-shell partially massless fields given in section \ref{FASPMSec}, the following remarks hold:
\begin{itemize}
	\item The pair of equations \eqref{PMA1} and \eqref{PMA2} admits a depth $t_{A}=t-1$ gauge symmetry
	\begin{align}
	\delta_{\zeta}A_{\a(m)\ad(n)}=\nabla_{(\a_1(\ad_1}\cdots\nabla_{\a_{t_A}\ad_{t_A}}\zeta_{\a_{t_A+1}\dots\a_{m})\ad_{t_A+1}\dots\ad_{n})}
	\end{align}
	when $2\leq t \leq \text{min}(m,n)$.
	\item The pair of equations \eqref{PMD1} and \eqref{PMD2} admits a depth $t_{E}=t$ gauge symmetry
	\begin{align}
	\delta_{\xi}E_{\a(m+1)\ad(n+1)}=\nabla_{(\a_1(\ad_1}\cdots\nabla_{\a_{t_E}\ad_{t_E}}\xi_{\a_{t_E+1}\dots\a_{m+1})\ad_{t_E+1}\dots\ad_{n+1})}
	\end{align}
	when $2\leq t \leq \text{min}(m+1,n+1)$.
	
	\item Equation \eqref{PMB1} and the first branch of \eqref{PMB2} admit a depth $t_B=t-1$ 
	\begin{align}
	\delta_{\rho}B_{\a(m+1)\ad(n)}=\nabla_{(\a_1(\ad_1}\cdots\nabla_{\a_{t_B}\ad_{t_B}}\rho_{\a_{t_B+1}\dots\a_{m+1})\ad_{t_B+1}\dots\ad_{n})}\label{GSB2}
	\end{align}
	gauge symmetry when $2\leq t\leq \text{min}(m+1,n+1)$. 
	
	\item Equation \eqref{PMB1} and the second branch of \eqref{PMB2} admit a depth $t'_B=t$ 
	\begin{align}
	\delta_{\rho}B_{\a(m+1)\ad(n)}=\nabla_{(\a_1(\ad_1}\cdots\nabla_{\a_{t'_B}\ad_{t'_B}}\rho_{\a_{t'_B+1}\dots\a_{m+1})\ad_{t'_B+1}\dots\ad_{n})}\label{GSB1}
	\end{align}
	gauge symmetry when $2\leq t \leq \text{min}(m+1,n)$.
	
	\item Comments similar to those above regarding $B_{\a(m+1)\ad(n)}$, apply to $C_{\a(m)\ad(n+1)}$.
	\item The above gauge symmetries hold only for gauge parameters which are transverse and which satisfy the same mass-shell equation as the parent gauge field. 
\end{itemize}

One can reverse the logic and ask the question: for what values of $M^2$ does there appear a gauge symmetry (of any depth) at the component level? The answer is precisely those values which appear in the superprojectors \eqref{FASSuperPartiallyMasslessValues} (though, as above, special care must be taken when deducing the upper and lower bounds on the range of possible depths). 

From the above component analysis one can see that the true partially massless representation $\mathfrak{P}\big(t,m,n \big)$ of $ \mf{osp} (1|4)$ decomposes into $\mathfrak{so}(3,2)$ subrepresentations as follows
\be
\mathfrak{P}\big(t,m,n \big)=P\big(t-1,m,n\big) \oplus P\big(t-1,m+1,n\big)  \oplus P\big(t,m+1,n\big) \oplus P\big(t,m+1,n+1\big)~.
\ee
For integer ($m=n+1=s$) and half-integer ($m=n=s$) superspin, this decomposition is in agreement with the results of \cite{G-SHR}.


\subsubsection{Massless supermultiplets} 

To study the component structure of an on-shell  massless supermultiplet of superspin $s>0$, 
it is advantageous to work with the gauge-invariant field strength $\mathfrak{W}_{\a(2s)}$ 
defined by \eqref{HSW1}, instead of the prepotential $\F_{\a(m) \ad(n)}$, 
where $2s = m+n+1$. The fundamental properties of $\mathfrak{W}_{\a(2s)}$,
\bea \label{FASPropoFFieldStrength}
\bar \cD_\bd \mathfrak{W}_{\a(2s)}=0~, \qquad
\mathcal{D}^{\b}\mathfrak{W}_{\b \a(2s-1)} =0
\quad \Leftrightarrow \quad \big( \cD^2 - 4(s+1) \bar \m \big) \mathfrak{W}_{\a(2s)} =0~,
\eea
imply the following equations:  
\begin{subequations} \label{FASFieldStrtengthCond}
	\bea
	\mathcal{D}^{\b \bd}\mathfrak{W}_{\b \a(2s-1)} &=&0~,\\
	\big(\mathbb{Q} -  
	\l_{(1,2s-1,0)} 
	\mu\mub\big)\mathfrak{W}_{\a(2s)} &=&0~.
	\eea
\end{subequations}

The chiral field strength $\mathfrak{W}_{\a(2s)}$ has two independent component fields
\begin{subequations} 
	\bea
	C_{\a(2s)} &:=& \mathfrak{W}_{\a(2s)} |~, \\
	C_{\a(2s+1)}&:=& \cD_{(\a_1} \mathfrak{W}_{\a_2 \dots \a_{2s+1})} |~.
	\eea
\end{subequations} 
Their properties follow from the equations \eqref{FASPropoFFieldStrength} and \eqref{FASFieldStrtengthCond} 
\begin{subequations}
	\bea
	\nabla^{\b \bd}C_{\b \a(2s-1)} &=&0 \quad \implies \quad 
	\big(\mathcal{Q} - 
	\t_{(1,2s,0)} 
	\m \bar \m\big) C_{\a(2s)} =0~, \\
	\nabla^{\b \bd}C_{\b \a(2s)} &=&0 \quad \implies \quad 
	\big( \mathcal{Q} - 
	\t_{(1,2s+1,0)} 
	\m \bar \m  \big) C_{\a(2s+1)} =0~.
	\eea 
\end{subequations}
These relations tell us that the component field strengths $C_{\a(2s)}$ and $C_{\a(2s+1)}$ furnish the massless representation 
$\mathfrak{S}\big(s+1,s\big)=D(s+1, s) \oplus  D\big(s+ \frac{3}{2}, s+ \frac{1}{2}\big) $.


\subsection{Superspin projection operators} \label{Superspin-projection operators}
Let us denote the space of tensor superfields of Lorentz type $(\frac{m}{2},\frac{n}{2})$ on AdS$^{4|4}$ by $\cV_{(m,n)}$. For any integers $m,n \geq 1$, the superspin projection operator $\bm{\P}^{\perp}_{(m,n)}$ is defined by its action on $\cV_{(m,n)}$ according to the rule
\bsubeq
\bea
\bm{\P}^{\perp}_{(m,n)}: \cV_{(m,n)} &\longrightarrow& \cV_{(m,n)} ~, \\
\F_{\a(m)\ad(n)} &\longmapsto& \bm{\P}^{\perp}_{(m,n)} \F_{\a(m)\ad(n)} =: \bm{\P}^{\perp}_{\a(m)\ad(n)}(\F)~.
\eea
\esubeq
For fixed $m$ and $n$, the operator $\bm{\P}^{\perp}_{(m,n)}$ is defined to satisfy the properties: 
\begin{enumerate}
	\item  \textbf{Idempotence:} The operator $\bm{\P}^{\perp}_{(m,n)}$ is idempotent
	\bsubeq \label{FASProjectorProperties}
	\be
	\bm{\P}^{\perp}_{(m,n)}\bm{\P}^{\perp}_{(m,n)} \F_{\a(m)\ad(n)} = \bm{\P}^{\perp}_{(m,n)} \F_{\a(m)\ad(n)}~.\\
	\ee
	\item \textbf{TLAL:} The operator $\bm{\P}^{\perp}_{(m,n)}$ maps $\F_{\a(m)\ad(n)}$ to a TLAL superfield
	\be
	\cD^\b  \bm{\P}^{\perp}_{\b\a(m-1)\ad(n)}(\F)=0~, \qquad \cDB^\bd \bm{\P}^{\perp}_{\a(m)\bd\ad(n-1)}(\F) = 0~. \label{FASProjectorTLAL}
	\ee
	\esubeq
\end{enumerate}
The superspin projector $\bm{\P}^{\perp}_{(m,n)}$ maps a superfield on $\cV_{(m,n)} $ which satisfies the mass-shell equation \eqref{FASOnShellMass}, but is otherwise unconstrained, to an on-shell superfield   \eqref{FASOnShellConditions}\footnote{ Off the mass-shell, $\bm{\P}^{\perp}_{(m,n)}$ maps a superfield to a conformal supercurrent in AdS$^{4|4}$ \cite{BHK}. }
\bsubeq
\begin{gather} \label{FASProjOnshell}
\cD^\b  \bm{\P}^{\perp}_{\b\a(m-1)\ad(n)}(\F)=0~, \qquad \cDB^\bd \bm{\P}^{\perp}_{\a(m)\bd\ad(n-1)}(\F) = 0~, \\
(\mb{Q} - M^2)\bm{\P}^{\perp}_{\a(m)\ad(n)}(\F) = 0~.
\end{gather}
\esubeq
In other words, the superspin projection operator $\bm{\P}^{\perp}_{(m,n)}$ selects out the physical component of an unconstrained superfield on the mass-shell  \eqref{FASOnShellMass}, which realises the irreducible representation $\mathfrak{S}\big(E_0(M),\frac{1}{2}(m+n+1)\big)$ of $\mathfrak{osp} (1|4)$.

The superspin projection operators  \eqref{FASProjectorProperties} take the following form
\bsubeq \label{FASSuperprojectors}
\begin{align}
\bm{\P}^{\perp}_{(m,n)}\Phi_{\a(m)\ad(n)} &:=\bigg [-8\prod_{t=1}^{n+1} \big ( \mathbb{Q}-\l_{(t,m,n)}\m \mub \big ) \bigg ]^{-1}\mathbb{P}_{\a(m)\ad(n)}(\Phi)~, \label{proj1}\\
\widehat{\bm{\P}}{}^{\perp}_{(m,n)}\Phi_{\a(m)\ad(n)}&:=\bigg [\hspace{0.47cm}8 \prod_{t=1}^{m+1} \big ( \mathbb{Q}-\l_{(t,m,n)}\m \mub \big ) \bigg ]^{-1} \widehat{\mathbb{P}}_{\a(m)\ad(n)}(\Phi)~,  \label{proj2}
\end{align}
\esubeq
where $\mathbb{P}_{\a(m)\ad(n)}(\Phi)$ and  $\widehat{\mathbb{P}}_{\a(m)\ad(n)}(\Phi)$ are given by
\begin{subequations}\label{strippedprojectors}
	\begin{align}
	\mathbb{P}_{\a(m)\ad(n)}(\Phi)&=\cD_{(\ad_1}{}^{\b_1}\dots \cD_{\ad_n)}{}^{\b_n}\cD^{\g}(\cDB^2-4\mu  )\cD_{(\b_1}{}^{\bd_1}\dots\cD_{\b_n}{}^{\bd_n}\cD_\g \Phi_{\a_1\dots\a_m)\bd(n)}~,\label{P1}\\
	\widehat{\mathbb{P}}_{\a(m)\ad(n)}(\Phi)&=\cD_{(\a_1}{}^{\bd_1}\dots \cD_{\a_m)}{}^{\bd_m}\cDB^{\gd} (\cD^2-4\mub  )\cD_{(\bd_1}{}^{\b_1}\dots\cD_{\bd_m}{}^{\b_m}\cDB_\gd \Phi_{\b(m)\ad_1\dots\ad_n)}~,\label{P2} 
	\end{align}
\end{subequations}
and $\l_{(t,m,n)}$ denotes the partially massless values \eqref{FASSuperPartiallyMasslessValues}. In the flat superspace limit, the superspin projection operators \eqref{FASSuperprojectors} reduce to those \eqref{FMSHighestSuperspinProjector} in $\mb{M}^{4|4}$.

Although not immediately apparent, it can be shown that the two types of superprojectors \eqref{proj1} and \eqref{proj2} coincide on $\cV_{(m,n)}$
\begin{align}
\bm{\P}^{\perp}_{\a(m)\ad(n)}(\Phi) = \widehat{\bm{\P}}{}^{\perp}_{\a(m)\ad(n)}(\Phi) ~. \label{coincidence}
\end{align}
In accordance with this equivalence, we will only consider the superspin projection operator $\bm{\P}^{\perp}_{(m,n)}$ on $\cV_{(m,n)}$.
Furthermore, it can be shown that $\bm{\P}^{\perp}_{(m,n)}$ acts like the identity operator on the space of TLAL superfields $\F^{\perp}_{\a(m)\ad(n)}$,
\be
\cD^{\b} \F^{\perp}_{\b\a(m-1)\ad(n)}=0~,  \quad  \cDB^{\bd} \F^{\perp}_{\a(m)\bd\ad(n-1)}=0 \quad \Rightarrow  \quad \bm{\P}^{\perp}_{(m,n)}\F^{\perp}_{\a(m)\ad(n)}  = \F^{\perp}_{\a(m)\ad(n)} ~. \label{FASSuperprojectorSurj}
\ee

The superspin projection operators \eqref{FASSuperprojectors} encode all the information necessary to deduce the partially massless supermultiplets in AdS$^{4|4}$. Specifically, the poles of these operators coincide with the partially massless values \eqref{FASSuperPartiallyMasslessValues}. This observation, with the known results in AdS$_4$ (see section \ref{FASpinProjectors}) leads one to conjecture that this relationship between partially massless (super)fields and  (super)spin projection operators is universal to AdS space, for any dimension, and any number of supersymmetries. In the subsequent chapter, we will probe the validity of this claim by studying the (super)spin projection operators in AdS$_3$. We would also like to point out that (non-unitary) massive modes, characterised by $\lambda_{(t,m,n)}$ with $\text{min}(m+1,n+1) < t \leq \text{max}(m+1,n+1)$, also appear in the poles of the superprojectors \eqref{FASSuperprojectors}.\footnote{For $m>n$,  the poles of the superprojector $\bm{\P}^{\perp}_{(m,n)}$ only contain partially massless modes $\lambda_{(t,m,n)}$ with $1 \leq t \leq \text{min}(m+1,n+1)$.  One arrives at an analogous result for $\widehat{\bm{\P}}^{\perp}_{(m,n)}$ when  $n>m$. }

In the case when $m>n=0$, the superprojectors analogous to \eqref{FASSuperprojectors}  take the form
\begin{subequations} \label{projectors-m}
	\begin{align}
	\bm{\P}^{\perp}_{(m)}\Phi_{\a(m)}\equiv \bm{\P}^{\perp}_{\a(m)}(\Phi)&:= \Big ( \mathbb{Q}-\l_{(1,m,0)}\m \mub \Big )^{-1}\mathbb{P}_{\a(m)}(\Phi)~, \label{3.33a} \\
	\widehat{\bm{\P}}{}^{\perp}_{(m)}\Phi_{\a(m)}\equiv \widehat{\bm{\P}}{}^{\perp}_{\a(m)}(\Phi)&:=\bigg [ \prod_{t=1}^{m+1} \big ( \mathbb{Q}-\l_{(t,m,0)}\m \mub \big ) \bigg ]^{-1} \widehat{\mathbb{P}}_{\a(m)}(\Phi)~,
	\end{align}
\end{subequations}
where $\mathbb{P}_{\a(m)}(\Phi)$ and $\widehat{\mathbb{P}}_{\a(m)}(\Phi)$ are given by
\begin{subequations}
	\begin{align}
	\mathbb{P}_{\a(m)}(\Phi)&=-\frac{1}{8}\cD^{\g}(\cDB^2-4\mu  )\cD_{(\g} \Phi_{\a_1\dots\a_m)}~,\\
	\widehat{\mathbb{P}}_{\a(m)}(\Phi)&=\frac{1}{8}\cD_{(\a_1}{}^{\bd_1}\dots \cD_{\a_m)}{}^{\bd_m}\cDB^{\gd} (\cD^2-4\mub  )\cD_{(\bd_1}{}^{\b_1}\dots\cD_{\bd_m}{}^{\b_m}\cDB_{\gd)} \Phi_{\b(m)}~.
	\end{align}
\end{subequations}
Note that we have introduced the notation  $\bm{\P}^{\perp}_{(m)} :=\bm{\P}{}^{\perp}_{(m,0)} $ and  $\widehat{\bm{\P}}{}^{\perp}_{(m)} := \widehat{\bm{\P}}{}^{\perp}_{(m,0)} $ in \eqref{projectors-m}.
The operators \eqref{projectors-m} square to themselves but, in contrast to \eqref{FASSuperprojectors}, they project onto the subspace of LTAL superfields,
\begin{subequations} \label{3.34}
	\begin{align}
	\mathcal{D}^{\b}\bm{\P}_{\b\a(m-1)}(\Phi)&=0~, \qquad ~~~~\phantom{..}\mathcal{D}^{\b}\widehat{\bm{\P}}{}^{\perp}_{\b\a(m-1)}(\Phi)=0~, \\
	\big(\bar{\mathcal{D}}^2-4\mu\big)\bm{\P}^{\perp}_{\a(m)}(\Phi)&=0~, \qquad \big(\bar{\mathcal{D}}^2-4\mu\big)\widehat{\bm{\P}}{}^{\perp}_{\a(m)}(\Phi)=0~.~~~~~~~
	\end{align}
\end{subequations}
Once again one may show that the two types of projectors coincide, $\bm{\P}^{\perp}_{\a(m)}(\Phi)=\widehat{\bm{\P}}{}^{\perp}_{\a(m)}(\Phi)$. In the case when $n>m=0$, the corresponding projectors $\overline{\bm{\P}}{}^{\perp}_{\ad(n)}(\bar \Phi)$ and $\widehat{\overline{\bm{\P}}}{}^{\perp}_{\ad(n)}(\bar \Phi)$ can be  obtained by complex conjugation and similar comments apply. This time however they project onto the subspace of simultaneously anti-linear and transverse linear superfields.

\subsubsection{Orthogonal complement of $\bm{\P}^{\perp}_{(m,n)}$}\label{FASOrthoSect}
Given $m,n > 0$, the orthogonal complement $\bm{\P}^{\parallel}_{(m,n)}$ of $\bm{\P}^{\perp}_{(m,n)}$ is given by
\be  \label{FASLongitudinalProjector}
\bm{\P}^{\parallel}_{(m,n)} = \mds{1} - \bm{\P}^{\perp}_{(m,n)} ~,
\ee
whereby construction, the operator $\bm{\P}^{\parallel}_{(m,n)} $ satisfies the following properties on $\mc{V}_{(m,n)}$ 
\be
\bm{\P}^{\parallel}_{(m,n)} \bm{\P}^{\parallel}_{(m,n)}  = \bm{\P}^{\parallel}_{(m,n)} ~, \qquad  \bm{\P}^{\parallel}_{(m,n)} \bm{\P}^{\perp}_{(m,n)} = \bm{\P}^{\perp}_{(m,n)}\bm{\P}^{\parallel}_{(m,n)} =0~.
\ee
It may be shown that $\bm{\P}^{\parallel}_{(m,n)} $ acting on $\Phi_{\a(m)\ad(n)}$ can be expressed as the following sum
\be
\bm{\P}^{\parallel}_{(m,n)}\Phi_{\a(m)\ad(n)} = \bar{\mathcal{D}}_{(\ad_1}\Psi_{\a(m)\ad_2\dots\ad_{n})} + \mathcal{D}_{(\a_1}\O_{\a_2\dots\a_m)\ad(n)}~,
\ee
for some unconstrained superfields $\Psi_{\a(m)\ad(n-1)}$ and $\O_{\a(m-1)\ad(n)}$. We see that $\bm{\P}^{\parallel}_{(m,n)}$ projects $\Phi_{\a(m)\ad(n)}$ onto the union of spaces of longitudinal linear \eqref{FASLongitudinalLinearCon} and longitudinal anti-linear \eqref{FASLongitduinalAntiLinear} superfields.

Let us introduce the superfield $\F^{\parallel}_{\a(m)\ad(n)} = \bar{\mathcal{D}}_{\ad}\Psi_{\a(m)\ad(n-1)} + \mathcal{D}_{\a}\O_{\a(m-1)\ad(n)} $,  where $\Psi_{\a(m)\ad(n-1)}$ and $\O_{\a(m-1)\ad(n)}$ are complex and unconstrained. We do not assume $\F^{\parallel}_{\a(m)\ad(n)}$ to be in the image of  $\bm{\P}^{\parallel}_{(m,n)}$. It can be shown that the superspin projection operators $\bm{\P}^{\perp}_{(m,n)}$ annihilate the superfield $\F^{\parallel}_{\a(m)\ad(n)}$ 
\be \label{FASProjKillLong}
\bm{\P}^{\perp}_{(m,n)}\F^{\parallel}_{\a(m)\ad(n)} = 0~.
\ee
It follows from \eqref{FASLongitudinalProjector}  and \eqref{FASProjKillLong} that $\bm{\P}^{\parallel}_{(m,n)}$ acts like the unit operator on $\F^{\parallel}_{\a(m)\ad(n)}$ 
\be
\bm{\P}^{\parallel}\F^{\parallel}_{\a(m)\ad(n)} = \F^{\parallel}_{\a(m)\ad(n)} ~.
\ee
These properties will be useful in the subsequent section.

If instead $m>0$ and $n=0$, the orthogonal complement may be written as the sum of a longitudinal anti-linear superfield and a chiral superfield,
\begin{align}
\bm{\P}^{\parallel}_{(m)}\Phi_{\a(m)}= \mathcal{D}_{\a}\O_{\a(m-1)}+\Lambda_{\a(m)}~, \qquad 
\bar{\mathcal{D}}_{\bd}\Lambda_{\a(m)}=0~,
\end{align}
where $\O_{\a(m-1)}$ is unconstrained.
and chiral   
$\Lambda_{\a(m)}$ superfields. 
On the other hand, if $n>0$ and $m=0$,  then the orthogonal complement splits into the sum of a longitudinal linear superfield and an anti-chiral superfield,
\begin{align}
\overline{\bm{\P}}^{\parallel}_{(n)}\bar{\Phi}_{\ad(n)}= \bar{\mathcal{D}}_{\ad}\bar{\Psi}_{\ad(n-1)}+\bar{\Lambda}_{\ad(n)}~, \qquad \mathcal{D}_{\b}\bar{\Lambda}_{\ad(n)}=0~,
\end{align}
where $\bar{\Lambda}_{\ad(n)}$ is unconstrained.

Using the fact that $\bm{\P}^{\perp}_{(m,n)}$ and $\bm{\P}^{\parallel}_{(m,n)}$ resolve the identity operator, one can decompose an arbitrary superfield $\Phi_{\a(m)\ad(n)}$ as follows
\begin{align}
\Phi_{\a(m)\ad(n)}=\F^{\perp}_{\a(m)\ad(n)}+\bar{\mathcal{D}}_{\ad}\z_{\a(m)\ad(n-1)}+\mathcal{D}_{\a}\x_{\a(m-1)\ad(n)}~, \label{decomp0}
\end{align}
where $\F^{\perp}_{\a(m)\ad(n)}$ is TLAL and is hence irreducible, whilst $\z_{\a(m)\ad(n-1)}$ and $\x_{\a(m-1)\ad(n)}$ are unconstrained and thus reducible. It is apparent from this decomposition that the superspin projection operator $\bm{\P}^{\perp}_{(m,n)}$ selects out the component $\F^{\perp}_{\a(m)\ad(n)}$ which describes the pure superspin state with maximal superspin $s=\hf (m+n+1)$, as a consequence of the identities \eqref{FASSuperprojectorSurj} and \eqref{FASProjKillLong}.

One can repeat the decomposition \eqref{decomp0} on the lower-rank unconstrained superfields in \eqref{decomp0}. After reiterating this procedure a finite number of times, one eventually arrives at a decomposition of $\Phi_{\a(m)\ad(n)}$ solely in terms of irreducible superfields.

In the $m>n$ case,  the result of this process is
\begin{align}
\Phi_{\a(m)\ad(n)}&=\sum_{t=0}^{n} (\mathcal{D}_{\a\ad})^t 
\F^{\perp}_{\a (m-t) \ad (n-t)} 
+\sum_{t=0}^{n-1}\big[\mathcal{D}_{\a},\bar{\mathcal{D}}_{\ad}\big]
(\mathcal{D}_{\a\ad})^t
\J_{\a (m-t-1) \ad (n-t-1) }\non\\
&+\sum_{t=0}^{n} (\mathcal{D}_{\a \ad} )^t \mathcal{D}_{\a}
\X_{\a (m-t-1 ) \ad(n-t) } 
+\sum_{t=0}^{n-1} (\mathcal{D}_{\a \ad} )^t  \bar{\mathcal{D}}_{\ad} 
\O_{\a (m-t) \ad (n-t-1) } \non\\
&+(\mathcal{D}_{\a \ad} )^n
\Big(\L_{\a (m-n)} + \mathcal{D}_{\a }\L_{\a(m-n-1) }\Big) ~,
\label{decomp1}
\end{align}
for some irreducible complex superfields $\F, \J, \X$ and $\O$ whose properties are summarised in table \ref{table 1}. The superfields
$\L_{\a (m-n)} $ and $\L_{\a(m-n-1) }$ in \eqref{decomp1} are chiral, 
\begin{align}
\bar{\mathcal{D}}_{\ad}\L_{\a(m-n)}=0~,\qquad \bar{\mathcal{D}}_{\ad}\L_{\a(m-n-1)}=0~.
\end{align}

\begin{table}[h]
	\begin{center}
		\begin{tabular}{|c|c|c|c|}
			\hline
			~  & $0\leq t \leq n-2$           & $t=n-1$ & $t=n$  \Tstrut\Bstrut\\ \hline
			$\F_{\a(m-t) \ad(n-t)}$       &  TLAL   & TLAL  & LTAL\Tstrut\\ 
			$\J_{\a(m-t-1) \ad(n-t-1)}$   &  TLAL   & TLAL  & --\\ 
			$\X_{\a(m-t-1) \ad(n-t)}$     &  TLAL   & TLAL  & LTAL \\ 
			$\O_{\a(m-t) \ad(n-t-1)}$  &  TLAL   & LTAL   & -- \\
			\hline
		\end{tabular}
	\end{center} 
	\vspace{-15pt} 
	\caption{Properties of the superfields appearing in \eqref{decomp1}.}
	\label{table 1}
\end{table}

If instead $m=n=s$, then we may further impose the reality condition 
\begin{align}
H_{\a(s)\ad(s)}:=\Phi_{\a(s)\ad(s)}=\bar{H}_{\a(s)\ad(s)}~. \label{reality}
\end{align}
In contrast to \eqref{decomp1}, $H_{\a(s)\ad(s)}$ now decomposes as
\begin{align}
H_{\a(s)\ad(s)}&=\sum_{t=0}^{s} (\mathcal{D}_{\a \ad} )^t
\F_{\a (s-t)\ad (s-t)} 
+\sum_{t=0}^{s-1}\big[\mathcal{D}_{\a},\bar{\mathcal{D}}_{\ad}\big]
(\mathcal{D}_{\a\ad} )^t
\J_{\a (s-t-1 )\ad (s-t-1)}\non\\
&+\sum_{t=0}^{s-1} (\mathcal{D}_{\a\ad})^t
\Big(\bar{\mathcal{D}}_{\ad}\X_{\a( s-t)\ad (s-t-1)}+\text{c.c.}\Big) 
+ (\mathcal{D}_{\a\ad})^s
\big(\L+\bar{\L}\big)~ ,\label{decomp2}
\end{align}
for some irreducible complex superfield $\X_{\a(s-t) \ad(s-t-1)}$ and 
irreducible real superfields
\begin{align} 
\F_{\a(s-t)\ad(s-t)}=\bar{\F}_{\a(s-t)\ad(s-t)}~, \qquad \J_{\a(s-t-1)\ad(s-t-1)}=\bar{\J}_{\a(s-t-1)\ad(s-t-1)}~,
\end{align}
whose properties are described in table \ref{table 2}. 
The scalar $\s$ in \eqref{decomp2} is chiral, 
$\bar{\mathcal{D}}_{\ad}\s=0$.
\begin{table}[h]
	\begin{center}
		\begin{tabular}{|c|c|c|c|}
			\hline
			~  & $0\leq t \leq s-2$     & $t=s-1$ & $t=s$  \Tstrut\Bstrut\\ \hline
			$\F_{\a(s-t) \ad(s-t)}$        &  TLAL   & TLAL  & LAL \Tstrut\\ 
			$\J_{\a(s-t-1) \ad(s-t-1)}$    &  TLAL   & LAL     & --\\ 
			$\X_{\a(s-t) \ad(s-t-1)}$      &  TLAL   & LTAL   & -- \\ 
			\hline
		\end{tabular}
	\end{center} 
	\vspace{-15pt} 
	\caption{Properties of the superfields appearing in \eqref{decomp2}.}
	\label{table 2}
\end{table}

The above decompositions were used, albeit without derivation,  
in \cite{BKS} for the covariant quantisation of 
the massless supersymmetric higher-spin models in AdS${}_4$ \cite{KS94}.


\subsection{Superconformal higher-spin theory}\label{FMSSCHSTheory}
For integers $m,n \geq 1$,\footnote{For the superconformal gauge multiplets with either $m=0$ or $n=0$ see \cite{Kuzenkokis2020}.} the SCHS action in AdS$^{4|4}$ is described in terms of the gauge prepotential  $H_{\a(m)\ad(n)}$, which is defined 
modulo gauge transformations \cite{KuzenkoManvelyanTheisen2017} (see also \cite{KuzenkoPonds2019})
\be
\delta_{\L,\O}H_{\a(m)\ad(n)}=\bar{\mathcal{D}}_{(\ad_1}\L_{\a(m)\ad_2\dots\ad_n)}+\mathcal{D}_{(\a_1}\O_{\a_2\dots\a_m)\ad(n)}~, \label{FMSGT}
\ee
where the gauge parameters $\L_{\a(m)\ad(n-1)}$ and $\O_{\a(m-1)\ad(n)}$ are unconstrained.

Associated with  $H_{\a(m)\ad(n)}$ and its complex conjugate  $\bar{H}_{\a(n)\ad(m)}$
are the the linearised higher-spin super-Weyl tensors \cite{KuzenkoManvelyanTheisen2017}
\begin{subequations} \label{FASHigherSpinWeylTensors}
	\begin{align}
	\mathfrak{W}_{\a(m+n+1)}(H)&= -\frac{1}{4}\big(\bar{\mathcal{D}}^2-4\mu\big)\mathcal{D}_{(\a_1}{}^{\bd_1}\cdots\mathcal{D}_{\a_n}{}^{\bd_n}\mathcal{D}_{\a_{n+1}}H_{\a_{n+2}\dots\a_{m+n+1})\bd(n)}~, \label{HSW1}\\
	\mathfrak{W}_{\a(m+n+1)}(\bar{H})&= -\frac{1}{4}\big(\bar{\mathcal{D}}^2-4\mu\big)\mathcal{D}_{(\a_1}{}^{\bd_1}\cdots\mathcal{D}_{\a_m}{}^{\bd_m}\mathcal{D}_{\a_{m+1}}\bar{H}_{\a_{m+2}\dots\a_{m+n+1})\bd(m)}~,\label{HSW2}
	\end{align}
\end{subequations}
which are  invariant under the gauge transformations \eqref{FMSGT}
\begin{align}
\delta_{\zeta,\xi}\mathfrak{W}_{\a(m+n+1)}(H)=0~,\qquad \delta_{\zeta,\xi}\mathfrak{W}_{\a(m+n+1)}(\bar{H})=0~.
\end{align}
For the cases $m=n=s$ and $m=n+1=s$, the SCHS gauge prepotentials $H_{\a(m)\ad(n)}$ first appeared in \cite{KS94} as dynamical variables in the massless superspin-$s$ and superspin-$(s+\hf)$ multiplets, respectively. Additionally, their corresponding gauge-invariant field strengths \eqref{FASHigherSpinWeylTensors} were also given in \cite{KS94}.

The gauge-invariant action $S_{\text{SCHS}}^{(m,n)}[H,\bar{H}]$, 
which describes the dynamics of $H_{\a(m)\ad(n)}$ and its conjugate  $\bar H_{\a(n)\ad(m)}$,
is typically written as a functional over the chiral subspace 
of the full superspace
\cite{KuzenkoManvelyanTheisen2017,KuzenkoPonds2019}.
The specific feature of AdS${}^{4|4}$ is the identity \cite{Siegel78}
\bea
\int\rd^4x\rd^2\q\, \cE \,\cL_{\rm c}
= \int\rd^4x\rd^2\q\rd^2\tb\, \frac{E}{\m} \,\cL_{\rm c}~, \qquad \bar \cD_\ad \cL_{\rm c} =0~,
\label{full-chiral}
\eea
which relates the integration over the chiral subspace to that over the full superspace;
here $\cE$ denotes the chiral integration measure  and $E^{-1} = {\rm Ber} (E_A{}^M)$. 
Keeping in mind  \eqref{full-chiral}, 
the gauge-invariant SCHS\footnote{For details concerning the superconformal properties of the SCHS action \eqref{FASSCHSchiral}, and its constituents, we direct the reader to the original work \cite{KuzenkoManvelyanTheisen2017} (see also \cite{KuzenkoPonds2019}). }  action is given by \cite{KuzenkoManvelyanTheisen2017}
\begin{align}
S_{\text{SCHS}}^{(m,n)}[H,\bar{H}]=\frac{1}{2}\text{i}^{m+n}\int 
\text{d}^{4|4}z \, \frac{E}{\m} \, 
\mathfrak W^{\a(m+n+1)}(H)\mathfrak{W}_{\a(m+n+1)}(\bar{H}) +{\rm c.c.}
~, \label{FASSCHSchiral}
\end{align}
where we have denoted $\rd^{4|4}z = \rd^4x\rd^2\q\rd^2\tb$. 

Upon integrating by parts, the action \eqref{FASSCHSchiral} may be written in the suggestive forms
\begin{subequations}
	\begin{align}
	S_{\text{SCHS}}^{(m,n)}[H,\bar{H}]&=\text{i}^{m+n}\int \text{d}^{4|4}z \, E \, \bar{H}^{\a(n)\ad(m)}\mathfrak{B}_{\a(n)\ad(m)}(H) +\text{c.c. } \label{SCHS1}\\
	&=\text{i}^{m+n}\int \text{d}^{4|4}z \, E \, \bar{H}^{\a(n)\ad(m)}\widehat{\mathfrak{B}}_{\a(n)\ad(m)}(H) +\rm{c.c.}\label{SCHS2}
	\end{align}
\end{subequations}
They are suggestive because, in addition to being gauge invariant, the linearised higher-spin super-Bach tensors
\cite{KPR}
\begin{subequations} \label{FASBach}
	\begin{align}
	\mathfrak{B}_{\a(n)\ad(m)}(H)&= \frac{1}{2}\mathcal{D}_{(\ad_1}{}^{\b_1}\cdots\mathcal{D}_{\ad_{m})}{}^{\b_m}\mathcal{D}^{\b_{m+1}}\mathfrak{W}_{\a(n)\b(m+1)}(H)~,\label{Bach1}\\
	\widehat{\mathfrak{B}}_{\a(n)\ad(m)}(H)&=\frac{1}{2}\mathcal{D}_{(\a_1}{}^{\bd_1}\cdots\mathcal{D}_{\a_{n})}{}^{\bd_n}\bar{\mathcal{D}}^{\bd_{n+1}}\overline{\mathfrak{W}}_{\ad(m)\bd(n+1)}(H)~, \label{Bach2}
	\end{align}
\end{subequations}
are simultaneously transverse linear and transverse anti-linear 
\begin{subequations}
	\begin{align}
	\mathcal{D}^{\b}\mathfrak{B}_{\b\a(n-1)\ad(m)}(H)&= 0~, \qquad \bar{\mathcal{D}}^{\bd}\mathfrak{B}_{\a(n)\bd\ad(m-1)}(H)=0~, \label{FASBachTensorWeylTensor}\\
	\mathcal{D}^{\b}\widehat{\mathfrak{B}}_{\b\a(n-1)\ad(m)}(H)&= 0~, \qquad \bar{\mathcal{D}}^{\bd}\widehat{\mathfrak{B}}_{\a(n)\bd\ad(m-1)}(H)=0~.
	\end{align}
\end{subequations}
It can be shown that the super-Bach tensors \eqref{FASBach} coincide
\be
\mathfrak{B}_{\a(n)\ad(m)}(H) = 	\widehat{\mathfrak{B}}_{\a(n)\ad(m)}(H)~.
\ee
Proving this equivalence is non-trivial and is closely related to the superprojector coincidence relation \eqref{coincidence}. As a by-product, it follows that the SCHS actions \eqref{SCHS1} and \eqref{SCHS2} are also equivalent.

The linearised higher-spin super-Bach tensors \eqref{FASBach} can be written in a manifestly TLAL form by recasting them in terms of the unique TLAL superprojectors \eqref{FASSuperprojectors} 
\bsubeq \label{FASBachTensorsProjectors}
\begin{align}
\mathfrak{B}_{\a(n)\ad(m)}(H) 
&=\prod_{t=1}^{n+1} \big ( \mathbb{Q}-\l_{(t,m,n)}\m \mub \big ) \big ( \cD_{\ad}{}^\b \big )^{m-n} \bm{\P}^{\perp}_{(m,n)} H_{\a(n)\b(m-n)\ad(n)}~,\\
\mathfrak{B}_{\a(n)\ad(m)}(H)&=\prod_{t=1}^{m+1} \big ( \mathbb{Q}-\l_{(t,m,n)}\m \mub \big )\big ( \cD_{\a}{}^\bd \big )^{n-m}  \bm{\P}^{\perp}_{(m,n)} H_{\a(m)\bd(n-m)\ad(m)}~,
\end{align}
\esubeq
for the cases $ m \geq n$ and $n \geq m$, respectively.
It follows from \eqref{FASBach} and \eqref{FASBachTensorsProjectors} that the higher-spin super-Weyl tensors \eqref{FASHigherSpinWeylTensors} can also be recast in terms of the superspin projection operator \eqref{proj1}.

We can immediately make use of these new representations for the higher-spin Bach tensors to recast the SCHS action \eqref{SCHS1} in terms of the superprojector \eqref{proj1}
\bsubeq \label{FASSCHSProjector}
\begin{align}
S_{\text{SCHS}}^{(m,n)}[H,\bar{H}]=\text{i}^{m+n}&\int\text{d}^{4|4}z \, E \, \bar{H}^{\a(n)\ad(m)}\prod_{t=1}^{n+1} \big ( \mathbb{Q}-\l_{(t,m,n)}\m \mub \big ) \non\\
&\times(\mathcal{D}_{\ad}{}^{\b})^{m-n} \bm{\P}^{\perp}_{(m,n)} H_{\a(n)\b(m-n)\ad(n)}+\text{c.c.}~, \hspace{1.37cm}  m \geq n~, \label{fact3} \\
S_{\text{SCHS}}^{(m,n)}[H,\bar{H}]=\text{i}^{m+n}&\int\text{d}^{4|4}z \, E \, \bar{H}^{\a(n)\ad(m)}\prod_{t=1}^{m+1} \big ( \mathbb{Q}-\l_{(t,m,n)}\m \mub \big ) \non\\
&\times(\mathcal{D}_{\a}{}^{\bd})^{n-m} \bm{\P}^{\perp}_{(m,n)}H_{\a(m)\bd(n-m)\ad(m)}+\text{c.c.}~, \hspace{1.2cm} n \geq m ~.\label{fact2} 
\end{align}
\esubeq
These actions reduce to their $\mb{M}^{4|4}$ counterpart \eqref{FMSSCHSProj} in the flat superspace limit. Moreover, they are automatically gauge-invariant because of the superprojector property \eqref{FASProjKillLong}.

One of the benefits of recasting the SCHS actions in terms of the superspin projection operators is that they factorise\footnote{The factorisation of the SCHS actions was obtained by Michael Ponds in \cite{BHKP}.} into products of minimal second order differential operators. Let us begin with the example $m=n=s$ and impose the reality condition \eqref{reality}. Now, one can employ the decomposition from section \ref{Superspin-projection operators} (which is valid for all  superfields off the mass-shell) to split the unconstrained prepotential $H_{\a(s)\ad(s)}$ in the action \eqref{SCHS1} into TLAL and longitudinal components,
\begin{align} \label{FASDecompSCHS}
H_{\a(s)\ad(s)}={H}^{\perp}_{\a(s)\ad(s)}+\bar{\mathcal{D}}_{\ad}\z_{\a(s)\ad(s-1)}-\mathcal{D}_{\a}\bar{\z}_{\a(s-1)\ad(s)}~.
\end{align}
Here  ${H}^{\perp}_{\a(s)\ad(s)}$ is TLAL, whilst $\z_{\a(s)\ad(s-1)}$ is complex and unconstrained.

Expressing  $H_{\a(s)\ad(s)}$ in the SCHS action \eqref{FASSCHSProjector} in the form \eqref{FASDecompSCHS}, it follows that the action reduces to the factorised form
\begin{align}
S_{\text{SCHS}}^{(s,s)}[H^{\perp}]=2(-1)^s\int\text{d}^{4|4}z \, E \,  {H}^{\perp \a(s)\ad(s)} \prod_{t=1}^{s+1} \big ( \mathbb{Q}-\l_{(t,s,s)}\m \mub \big ){H}^{\perp}_{\a(s)\ad(s)} ~,  \label{fact1}
\end{align}
by virtue of the projector properties \eqref{FASProjectorTLAL}, \eqref{FASSuperprojectorSurj} and \eqref{FASProjKillLong}.
This process is equivalent to fixing the gauge freedom \eqref{FMSGT} by imposing the gauge condition $H_{\a(s)\ad(s)}\equiv{ H}^{\perp}_{\a(s)\ad(s)} $, since the action \eqref{fact1} no longer possesses any gauge symmetry.

Following the steps employed in the previous case, one can obtain the following factorisation of the SCHS action \eqref{SCHS1} for $n >m$
\begin{align}
S_{\text{SCHS}}^{(m,n)}[H^{\perp},\bar{H}^{\perp}]=\text{i}^{m+n}&\int\text{d}^{4|4}z \, E \, \bar{H}^{\perp \a(n)\ad(m)}\prod_{t=1}^{m+1} \big ( \mathbb{Q}-\l_{(t,m,n)}\m \mub \big ) \non \\
&\times \big ( \mathcal{D}_{\a}{}^{\bd} \big )^{n-m}H^{\perp}_{\a(m)\bd(n-m)\ad(m)}+\text{c.c.} \label{FASSCHSMGN}
\end{align}
This SCHS action does not factorise completely into products of second-order operators due to the mismatch in the integers $m$ and $n$. However, one can instead show that the equation of motion obtained by varying the action \eqref{FASSCHSMGN} with respect to $\bar{H}^{\perp}_{\a(m)\ad(n)}$ can be factorised wholly in terms of products of second-order operators
\begin{align}
0=\prod_{t=1}^{n+1} \big ( \mathbb{Q}-\l_{(t,m,n)}\m \mub \big )H^{\perp}_{\a(m)\ad(n)}~. \label{eom1}
\end{align}
Note that one must act with the appropriate vector derivatives on the equation of motion in order to arrive at this result.
According to our definitions in section \ref{FASOnShellSupermultiplets}, it is apparent that extra non-unitary massive modes, corresponding to the values of $\lambda_{(t,m,n)}$ with $m+1< t \leq n+1$ enter the spectrum of the wave equation \eqref{eom1}.

Similarly, in the case $m \geq n$, the factorised action \eqref{fact3} takes the form 
\begin{align}
S_{\text{SCHS}}^{(m,n)}[H^{\perp},\bar{H}^{\perp}]=\text{i}^{m+n}&\int\text{d}^{4|4}z \, E \, \bar{H}^{\perp \a(n)\ad(m)}\prod_{t=1}^{n+1} \big ( \mathbb{Q}-\l_{(t,m,n)}\m \mub \big ) \non\\
&\times \big( \mc{D}_\ad{}^\b \big)^{m-n} H^{\perp}_{\a(n)\b(m-n)\ad(n)}+\text{c.c.}~,  \label{FASCHSMGN}
\end{align}
whilst the analogous equation of motion is
\begin{align}
0=\prod_{t=1}^{m+1} \big ( \mathbb{Q}-\l_{(t,m,n)}\m \mub \big )H^{\perp}_{\a(m)\ad(n)}~. \label{eom2}
\end{align}
Once again, non-unitary massive modes are present in the resulting wave equation.

\subsection{Off-shell models in AdS$^{4|4}$  }\label{Off-shellmodels}
In this section we study the off-shell models for the Wess-Zumino (superspin-$0$)  and massive vector (superspin-$\hf$) multiplets at the superfield and component level. We also present an off-shell model for the massive gravitino (superspin-$1$) multiplet.

\subsubsection{Wess-Zumino supermultiplet} \label{WZ section}

Various aspects of the Wess-Zumino model in AdS$_4$ were studied in the 1980s,  
see \cite{IS,BreitenF,BFDG,DF,BG1,BG2} and references therein. 
Here we will only discuss its group-theoretic aspects for completeness. 
In superspace, the Wess-Zumino model
is formulated in terms of a chiral scalar superfield $\Phi$ and its  conjugate $\bar{\Phi}$.
Without self-coupling, the model is described by  the action 
\bea
S_{\text{WZ}}[\Phi, \bar{\Phi}] = \int \rd^{4|4} z\, E\, \Big \{ \Phi \bar{\Phi} + \frac{\l}{2} \Phi^2 + \frac{\bar{\l}}{2}\bar{\Phi}^2 \Big \}~, \qquad \cDB_\ad \Phi = 0~.
\label{5.20}
\eea
Here $\l$ is a dimensionless complex parameter.\footnote{This parameter can be made real by applying the redefinition $\F \to \re^{\ri \g } \F$, where $ \g = \bar \g$ is constant.} 
The  equations of motion corresponding to  this model are
\begin{subequations} \label{EoMWZ}
	\bea
	- \frac{1}{4}\big(\cDB^2-4\m\big)\bar{\Phi} + \l \m \Phi &=&0~, \\
	- \frac{1}{4}\big(\cD^2-4\mub\big)\Phi + \bar{\l} \mub \bar{\Phi} &=&0~.
	\eea
\end{subequations}
Using \eqref{EoMWZ}, it can be shown that the superfield $\Phi$ satisfies the mass-shell equation
\be \label{MassShellWZ}
\big(\mathbb{Q} +\m \mub -|\l \m|^2\big)\Phi = 0~.
\ee

The on-shell chiral scalar  $\F$ contains 
two independent component fields, which are
\begin{subequations} \label{CompFields}
	\bea
	\varphi :&=& \Phi |~, \\
	\psi_\a :&=& \cD_\a \Phi|~.
	\eea
\end{subequations}
By making use of the superfield mass-shell equation \eqref{MassShellWZ}, it can be shown that the component fields \eqref{CompFields} satisfy the mass-shell equations
\begin{subequations} \label{MassShellWZComp}
	\bea
	\big(\mathcal{Q}-\r_{(1)}^2 \big)\big(\mathcal{Q}-\r_{(2)}^2 \big)\varphi &=& 0~,\label{MassShellWZComp-a}\\
	\big(\mathcal{Q}-\r_{(3)}^2\big)\psi_\a &=& 0~,
	\eea
\end{subequations}
where the pseudo-masses $\r_{(i)}^2 $ are 
\begin{subequations} \label{psuedomassWZ}
	\bea
	\r_{(1)}^2 &=&|\l \m |^2 - \m \mub \big(|\l| + 2\big) ~, \\
	\r_{(2)}^2 &=& |\l \m |^2 + \m \mub \big(|\l| - 2\big)~, \\
	\r_{(3)}^2 &=& | \l \m |^2 - \frac{3}{2} \m \mub~.
	\eea
\end{subequations}
In the massless limit $\l \rightarrow 0$, the mass-shell conditions \eqref{psuedomassWZ} take the form
\begin{subequations} 
	\bea
	\r_{(1)}^2 &=& \r_{(2)}^2 = - 2\mu \mub = \t_{(1,0,0)}\m \mub  ~, \\
	\r_{(3)}^2 &=&- \frac{3}{2} \m \mub = \t_{(1,1,0)}\m \mub~,
	\eea
\end{subequations}
which is consistent with the on-shell massless field conditions \eqref{2.17}. We wish to show that the model \eqref{5.20} describes two
Wess-Zumino representations 
\eqref{E2}. First, let us point out that the equation for $\varphi$, 
\eqref{MassShellWZComp-a},
is factorised 
into the product of two second-order differential equations
and, hence, describes two spin-$0$ modes with masses $\r_{(1)}$ and $\r_{(2)}$. This means that the Wess-Zumino model
describes three representations of $\mathfrak{so}(3,2)$.
To see that the structure of these representations is consistent with~\eqref{E2}
it is necessary compute the 
minimal energies $E^{(i)}_0$ associated with the pseudo-masses $\r_{(i)}$. Using the prescription advocated in section \ref{FAIrredFieldReps}, we find 
\begin{subequations}
	\bea
	\big (E_0^{(1)} \big )_\pm &=&\frac{3}{2}\pm \hf \big(2|\l|-1\big) ~, \\
	\big (E_0^{(2)} \big )_\pm &=& \frac{3}{2}\pm \hf \big(1+2|\l|\big)~, \\
	\big (E_0^{(3)} \big )_\pm &=& \frac{3}{2}\pm |\l|~.
	\eea
\end{subequations}
We see that there exist two branches of minimal energy solutions which furnish 
two Wess-Zumino representations $\mathfrak{S}\big(E_0,0\big)$, eq.  \eqref{E2}. 
We will call these branches the positive branch and the negative branch.  
These solutions are given in table \ref{table 4}. 
\begin{table}[h]
	\begin{center}
		\begin{tabular}{|c|c|c|}
			\hline
			~ Component fields  & $E_0$ (positive branch)   & $E_0$ (negative branch)   \Tstrut\Bstrut\\ \hline
			\phantom{\Big|} 		spin-$0$       &  $1+|\l|$    & $1 - |\l|$    \Tstrut\\ 
			\phantom{\Big|} 	spin-$0$  &  $2 +|\l|$  &  $2-|\l|$   \\ 
			\phantom{\Big|} 		spin-$\hf$     &  $\frac{3}{2} + |\l|$   & $\frac{3}{2} - |\l|$  
			\\ 
			\hline
		\end{tabular}
	\end{center} 
	\vspace{-15pt} 
	\caption{Two branches of solutions.}
	\label{table 4}
\end{table}
These branches furnish the Wess-Zumino representations $\mathfrak{S}\big(1+|\l|,0\big)$ and $\mathfrak{S}\big(1-|\l|,0\big)$, eq. \eqref{E2}. 
In order for these representations to be unitary, we require $E_0 > \hf$. 
Therefore, the positive branch describes unitary representations for all values of $\l$. 
The negative branch describes unitary representations for  $|\l| < \hf$. 
In the massless case $\l=0$ the two branches coincide and we have two massless scalars with energies $E_0=1$ and $E_0=2$ and a massless fermion 
with energy $E_0= \frac{3}{2}$.\footnote{As pointed out in section \ref{FASRepTheory}, the massless spin-$0$ representations of $\mathfrak{osp}(1|4)$ correspond to two 
	possible values of the minimal energy $E_0=1$ or $E_0=2$, 
	and the massless spin-$\frac{1}{2}$ representation corresponds to
	$E_0= \frac{3}{2}$.}  

There are three special values of $\l$. The choice $\l=0$ corresponds to the superconformal or massless model. For $\l=1$ the model has a dual formulation in terms of a tensor supermultiplet \cite{Siegel-tensor}. In this case the equations of motion 
\eqref{EoMWZ} can be recast in term of a real superfield $L=\F + \bar \F$ 
to take form 
\begin{subequations}\label{tensor}
	\bea
	\big(\bar \cD^2 - 4\m \big) L = \big( \cD^2 - 4 \bar \m \big) L = 
	0~, \label{tensor-a}
	\eea
	while the chirality of $\F$ gives 
	\bea
	\big(\bar \cD^2 - 4\m \big) \cD_\a L =0~. \label{tensor-b}
	\eea
\end{subequations}
The equations \eqref{tensor} imply that $L$ satisfies the mass-shell equation
\bea
\mathbb{Q}L=0~, \label{tensor2}
\eea
which together define an on-shell scalar superfield. 
Finally for $\l =1/2$ the on-shell chiral scalar $\F$ and its conjugate $\bar \F$ contain 
a sub-representation describing 
the super-singleton, as follows from table \ref{table 4}.


\subsubsection{Massive vector multiplet} \label{section5.5}

The off-shell massive vector multiplet in a supergravity background was formulated in
\cite{Siegel-tensor,VanProeyen}. In the superspace setting, it is naturally 
described in terms of a real scalar prepotential $V$.  
Its action functional  in AdS$^{4|4}$ is given by 
\bea
S [V]= \hf \int\rd^{4|4}z\, E \, V \Big\{ \frac{1}{8}
\cD^\a \big(\bar \cD^2-4\m\big) \cD_\a + M^2 \Big\} V~, \qquad \bar V = V~,
\label{vmaction}
\eea
with $M$ a non-zero real parameter. 
The corresponding equation of motion 
\bea
\Big( \frac{1}{8}
\cD^\a \big(\bar \cD^2-4\m\big) \cD_\a + M^2 \Big) V =0~,
\label{vmeom-0}
\eea
implies that $V$ is a linear superfield, 
\begin{subequations}\label{vmeom}
	\bea
	\big(\bar \cD^2 -4\m\big)  V = \big( \cD^2 -4\bar \m\big)  V=0~.
	\label{5.30} 
	\eea
	Then making use of \eqref{A.4a} gives 
	\bea 
	\big( \Box +2 \m \bar \m - M^2 \big) V = \big({\mathbb Q} - M^2 \big)V =0 ~.
	\label{5.31}
	\eea
\end{subequations}
The equations \eqref{5.30} and \eqref{5.31} define an on-shell supermultiplet.\footnote{It is instructive to compare the system of equations \eqref{tensor} with \eqref{vmeom-0} and its corollaries \eqref{vmeom}. The former describes a non-conformal tensor multiplet in AdS$_4$, which reduces to the free massless tensor multiplet in the flat-superspace limit.}
It is worth pointing out that the above model has a dual formulation, which is the massive tensor multiplet model \cite{Siegel-tensor}.

On the mass shell, the independent component fields of $V$ are:
\begin{subequations}
	\bea
	A &=& V|~, \\
	\j_\a &=& \cD_\a V|~,
	\\
	h_{\a\ad} &=& \frac{1}{2}\big[ \cD_\a , \bar \cD_\ad \big] V| ~, \qquad 
	\nabla^{\a\ad} h_{\a\ad} =0~.
	\eea
\end{subequations}
The corresponding equations of motion are:
\begin{subequations}
	\bea
	0&=&\big(\cQ+2\m \mub -M^2\big)A~, \label{5.34a} \\
	0&=&\big(\cQ+\frac{3}{2}\m \mub -M^2\big)\j_\a - \ri \mub \nabla_{\a}{}^{\bd}\bar{\j}_\bd~, \label{5.34b} \\
	0&=&\big(\cQ+\frac{3}{2}\m \mub -M^2\big)\bar{\j}_\ad + \ri \mu \nabla^{\b}{}_{\ad}\j_\b~,
	\label{5.34c}\\
	0&=&\big(\cQ-M^2\big)h_{\a\ad}~.
	\label{5.34d}
	\eea
\end{subequations}        
The equations \eqref{5.34b} and \eqref{5.34c} lead to the quartic equation
\bea
0&=&\big(\cQ+\frac{3}{2}\m \mub - M_{+}^2\big)
\big(\cQ+\frac{3}{2}\m \mub - M_{-}^2 \big)\j_\a~, \label{VMquartic}
\eea
where
\bea
M_{\pm}^2:=M^2+ \hf \m \mub \pm \hf \sqrt{\m\mub(4M^2+\m\mub)}
=\frac{1}{4} \Big (\sqrt{\m \mub} \pm \sqrt{4M^2+\m\mub}\Big )^2~.
\eea
The above results imply that the component fields furnish the massive superspin-$\hf$ representation of $\mathfrak{osp}(1,4)$
\bea
\mathfrak{S}\big({E}_0 ,\hf \big)=
D\big(E_0+ \frac{1}{2}, 0\big)  \oplus D\big( E_0, \hf \big) \oplus  
D \big( E_0+1, \hf \big) \oplus  D \big(E_0+ \hf, 1\big) ~, 
\eea
where
$E_0  =1  + \frac{1}{2}
\sqrt{ \frac{4M^2}{\mu\mub} +1} $.

The spinor equation \eqref{5.34b} and its conjugate \eqref{5.34c} are second-order partial differential equations.
First-order equations for the component spinor fields can be obtained 
if one makes use of a   
St\"uckelberg reformulation of the above model. It is obtained through the replacement 
\begin{align}
V\, \rightarrow \, V+\frac{1}{M}\big(\Phi+\bar{\Phi}\big)~,\qquad \bar{\mathcal{D}}_{\ad}\Phi=0~,
\end{align}
for some chiral superfield $\Phi$. This leads to the action
\bea
S [V,\Phi,\bar{\Phi}]= \hf \int\rd^{4|4}z\, E \, \Big\{  \frac{1}{8}V
\cD^\a \big(\bar \cD^2-4\m\big) \cD_\a V + \big(MV+\Phi+\bar{\Phi}\big)^2 \Big\} ~,
\eea
which is invariant under gauge transformations of the form
\begin{align}
\delta_{\Lambda}V=\Lambda+\bar{\Lambda}~,\qquad \delta_{\Lambda}\Phi=-M\Lambda~,\qquad \bar{\mathcal{D}}_{\ad}\Lambda=0~. \label{VMGF}
\end{align} 
The corresponding equations of motion are given by 
\begin{subequations}\label{VM}
	\begin{align}
	0&=\phantom{-}\frac{1}{8}\mathcal{D}^{\a}\big(\bar{\mathcal{D}}^2-4\mu\big)\mathcal{D}_{\a}V +M^2 V + M\big(\Phi+\bar{\Phi}\big)~, \label{VMa} \\
	0&= -\frac{1}{4}\big(\bar{\mathcal{D}}^2-4\mu\big)V+\frac{\mu}{M}\Phi-\frac{1}{4M}\big(\bar{\mathcal{D}}^2-4\mu\big)\bar{\Phi}~, \label{VMb}\\
	0&= -\frac{1}{4}\big(\mathcal{D}^2-4\mub\big)V+\frac{\mub}{M}\bar{\Phi}-\frac{1}{4M}\big(\mathcal{D}^2-4\mub\big)\Phi~. \label{VMc}
	\end{align}
\end{subequations}

Using the gauge freedom \eqref{VMGF}, one may impose the gauge condition $\Phi=0$ whereupon the original model \eqref{vmaction} is recovered. On the other hand, one can instead choose the following Wess-Zumino gauge
\begin{align}
V|=0~, \qquad \mathcal{D}_{\a}V|&=0~,\qquad \mathcal{D}^2V|=0~,\qquad \big(\Phi-\bar{\Phi}\big)|=0~, 
\end{align}
which exhausts the gauge freedom. The remaining non-zero component fields are
\begin{subequations}
	\begin{align}
	A&=\Phi|~,\\
	\psi_{\a}&=\mathcal{D}_{\a}\Phi |~, \label{5.43a}\\
	h_{\a\ad}&=\frac{1}{2}\big[\mathcal{D}_{\a},\bar{\mathcal{D}}_{\ad}\big]V|~, \qquad \nabla^{\a\ad}h_{\a\ad}=0~, \label{5.43c} \\
	\chi_{\a}&=-\frac{1}{4}\big(\bar{\mathcal{D}}^2-4\mu\big)\mathcal{D}_{\a}V|~.
	\end{align}
\end{subequations}
The spinor fields $\psi_{\a}$ and $\chi_{\a}$ and their conjugates 
prove to satisfy the following 
first-order differential equations
\begin{subequations} \label{FOspin}
	\bea
	\text{i}\nabla^{\a}{}_{\ad}\chi_{\a}+M\bar{\psi}_{\ad}=0~,  \quad && \quad 
	-\text{i}\nabla_{\a}{}^{\ad}\bar \chi_{\ad}+M{\psi}_{\a}=0~,
	\label{FOspin-a}\\ 
	\text{i}\nabla^{\a}{}_{\ad}\psi_{\a}+M\bar{\chi}_{\ad}+\mub\bar{\psi}_{\ad}=0~,
	\quad && \quad 
	-\text{i}\nabla_{\a}{}^{\ad}\bar{\psi}_{\ad}+M{\chi}_{\a}+\mu {\psi}_{\a}=0~.
	\label{FOspin-b}
	\eea
\end{subequations}
Making use of the equations \eqref{FOspin-b} allows us to express 
the fields $\c_\a$ and $\bar \c_\ad$ in terms of $\j_\a$ and $\bar \j_\ad$.
Then the latter fields prove 
to satisfy the equations \eqref{5.34b} and \eqref{5.34c}. 
As regards the bosonic fields \eqref{5.43a} and \eqref{5.43c}, they may be seen to obey the equations \eqref{5.34a} and \eqref{5.34d}.


\subsubsection{Massive gravitino supermultiplet}

As an illustration of our general discussion in section \ref{FASOnShellSupermultiplets}, here we present an off-shell  model for the massive gravitino (superspin-$1$) multiplet in AdS$_4$, 
\bea
\mathfrak{S}\big({E}_0,1\big):=
D\Big(E_0+ \frac{1}{2}, \hf \Big)  \oplus D( E_0, 1) \oplus  D (E_0+1, 1) \oplus  D\Big(E_0+ \frac{1}{2}, \frac{3}{2}\Big)~.
\label{gravitinorep}
\eea
The unitarity bound for the superspin-$1$ case is $E_0 \geq 2$, with the  $E_0 =2$ value corresponding to the massless gravitino multiplet. We point out that on-shell models 
(i.e. without auxiliary fields) for the 
the massive gravitino multiplet in AdS$_4$ have appeared in the literature \cite{AB}
(see also \cite{Zinoviev07}).

Two off-shell formulations for the massless gravitino in AdS$_4$ were introduced in \cite{KS94}. One of them  is described by the action\footnote{This model has several dual versions given in \cite{KS94,Butter:2011ym, BHK}. In the flat-superspace limit, the action \eqref{gravitinoaction} reduces to that derived in \cite{GS}.}
\bea \label{gravitinoaction}
&&S_{\text{massless}}[H,\J, \bar \J] = - \int \rd^{4|4}z\, E\,\Big 
\{ \frac{1}{16}H\cD^\a \big (\cDB^2-4\mu \big )\cD_\a H+\m \mub H^2
\non \\
&& \qquad
+\frac{1}{4}H\big (\cD^\a\cDB^\ad G_{\a\ad} - \cDB^\ad \cD^\a \bar{G}_{\a\ad} \big) 
+\bar{G}^{\a\ad}G_{\a\ad}+\frac{1}{4}\big (\bar{G}_{\a\ad}\bar{G}^{\a\ad}+ G^{\a\ad}G_{\a\ad} \big ) \Big \} ~,   \hspace{0.5cm}
\eea
where $H$ is a real scalar superfield, 
and $G_{\a\ad}$ is a longitudinal linear superfield constructed in terms of 
an unconstrained prepotential $\J_\a$, 
\bea
G_{\a\ad} = \cDB_\ad \J_\a~, \qquad \bar{G}_{\a\ad} = -\cD_\a \bar{\J}_\ad~.
\eea
We propose a massive extension of the action \eqref{gravitinoaction} by adding a mass term $S_{m}[H,\J, \bar \J] $
\bea
\label{MassiveAct}
S_{\text{massive}}[H,\J, \bar \J ]=S_{\text{massless}}[H,\J, \bar \J ] 
+ S_{m}[H,\J, \bar \J]~.
\eea
We propose the following mass term:
\bea
S_{m}[H,\J, \bar \J] &=& m \int \rd^{4|4}z\,E\,\Big \{ \J^\a \J_\a + \bar{\J}_\ad\bar{\J}^\ad  +\hf H\big (\cD^\a\J_\a + \cDB_\ad\bar{\J}^\ad \big ) \\
&&- \frac{1}{4}\big(m+2\m + 2\mub\big) H^2 \Big \}~, \non
\eea
where $m$ is a real parameter of unit mass dimension. 

The equations of motion corresponding to $S_{\text{massive}}[H,\J, \bar \J]$ \eqref{MassiveAct} are given by
\begin{subequations} \label{EoM}
	\bea
	0&=&\hf \cD^\a W_\a + \frac{1}{4}\big (\cD^\a \cDB^2\J_\a + \cDB_\ad \cD^2  \bar{\J}^\ad \big) + \frac{m}{2} \big ( \cD^\a \J_\a + \cDB_\ad \bar{\J}^\ad \big ) \label{EoM1}\\
	&&-\big( m \m + m \mub +2 \m\mub+\hf m^2\big)H ~, \non \\
	0 &=&-W_\a+ \big(\frac{m}{2}+\m\big)\cD_\a H + \cDB_\ad \cD_\a \bar{\J}^\ad -\hf \cDB^2 \J_\a -2m \J_\a ~, \label{EoM2}
	\eea
\end{subequations}
where we have introduced the field strength $W_\a = -\frac{1}{4}\big(\cDB^2-4\m\big)\cD_\a H$. Acting on \eqref{EoM1} and \eqref{EoM2} with $\cDB^2$ gives
\begin{subequations} \label{DBarSquareCond}
	\bea 
	\big(\cDB^2 - 4\m\big) \cD^\a \J_\a &=&  \big(m+2\mub\big) \big(\cDB^2 - 4\m\big) H~, \\
	\big(\cDB^2 -4\m\big)\J_\a &=& - W_\a~.
	\eea
\end{subequations}
Next, acting on \eqref{EoM2} with $\cD^\a$ yields
\bea \label{DCond}
\cDB_\ad\bar{\J}^\ad = \big(m+2\m\big)H-\frac{1}{\mub}\big(m+\m\big)\cD^\a\J_\a~.
\eea
Using relations \eqref{EoM1}, \eqref{DBarSquareCond} and \eqref{DCond}, 
one can show that 
\be \label{TransverseCondGrav}
\cD^\a \J_\a =0~.
\ee
Substituting \eqref{TransverseCondGrav}  into \eqref{DCond}, it immediately follows that 
\bea
H=0~,
\eea
on the mass shell. 
The above relations imply that the spinor superfield $\J_\a$ obeys 
the following constraints:
\begin{subequations}
	\bea
	\cD^\a \J_\a &=& 0~, \quad \big(\cDB^2 -4\mu\big)\J_\a = 0~, \\
	\ri \cD_{\a\ad} \bar{\J}^\ad + \big(m+\m\big) \J_\a &=&0 \quad \Longrightarrow \quad \big (\mathbb{Q}-(m+\mu)(m+\mub) - \m \mub  \big )\J_\a =0~. ~~~
	\label{MassShellCond}
	\eea
\end{subequations}
These are the on-shell conditions for a massive superspin-$1$ multiplet, as defined in section \ref{FASOnShellSupermultiplets}. The unitarity bound for the massive gravitino multiplet is 
\bea
|m +\m |^2 > \m \bar \m  \quad \implies \quad m +2\m \neq 0~.
\eea
These conditions were used in the above derivation. One may think of the 
first Dirac-type equation in \eqref{MassShellCond} as the  reality condition relating 
$\J_\a$ and $\bar \J_\ad$.

In conclusion, we give the explicit expression for 
the massive gravitino action \eqref{MassiveAct}
\bea
&&S_{\text{massive}}[H,\J, \bar \J ] =\int \rd^{4|4}z\, E\,\Big \{ -\frac{1}{16}H\cD^\a \big (\cDB^2-4\mu \big )\cD_\a H-\m \mub H^2
~ \non \\
&&\qquad  +\frac{1}{4}H\big (\cD^\a\cDB^2 \J_\a + \cDB_\ad \cD^2 \bar{\J}^\ad \big )
+\cD^\a \bar{\J}^\ad \cDB_{\ad}\J_\a
+\frac{1}{4}\big (\bar{\J}_{\ad}\cD^2\bar{\J}^{\ad}+ \J^{\a}\cDB^2 \J_\a \big )  ~ \non
\\
&&\qquad - \frac{m}{4}\big(m+2\m +2\mub\big) H^2 +\frac{m}{2} H\big (\cD^\a\J_\a + \cDB_\ad\bar{\J}^\ad \big )+ m \big(\J^\a \J_\a +  \bar{\J}_\ad \bar{\J}^\ad\big)
\Big \}~.~~~~~~
\eea
Note that in the flat superspace limit, the action reduces to the massive gravitino model given in \cite{BGKP}. 

In the massless limit, $m \rightarrow 0$, the second mass-shell condition \eqref{MassShellCond} becomes 
\be
\big(\mathbb{Q}-2\mu \mub \big)\J_\a = \big(\mathbb{Q}-\l_{(1,1,0)}\m \mub\big)\J_\a = 0~.
\ee
This agrees with the definition of on-shell massless supermultiplets 
given in section  \ref{FASOnShellSupermultiplets}.

On the mass shell, the independent component fields of $\J_\a$ are:
\begin{subequations}\label{GravComp}
	\bea
	\c_\a &=& \J_\a|~,\\
	A_{\a\ad}&=&\cDB_\ad \J_\a| ~, \qquad \nabla^{\a\ad} A_{\a\ad}=0~,\\
	B_{\a\b}&=& \cD_{(\a}\J_{\b)}|~,\\
	\varphi_{\a\b\ad}&=&\Big (\hf [\cD_{(\a},\cDB_{\ad} ] - \frac{\ri}{3}\mathcal{D}_{(\a\ad} \Big )\J_{\b)}|~, \qquad \nabla^{\a\ad}\varphi_{\a\b\ad}=0~.
	\eea
\end{subequations}
Here $A_{\a\ad}$ is a complex vector field.
The component equations of motion, which
follow from the first equation in \eqref{MassShellCond} are:
\begin{subequations}\label{6.16}
	\bea
	0&=&\ri \nabla^\b{}_{\ad}\c_\b +\big(m+\mub\big)\bar{\c}_\ad~, \\
	0&=& \ri \nabla^\b{}_{\ad} B_{\a\b}+\mub A_{\a\ad}-\big(m+\mub\big)\bar{A}_{\a\ad}~, \\
	0&=&\ri \nabla^\b{}_{(\ad}A_{\b\bd)}-\big(m+\mub\big)\bar{B}_{\ad\bd}~, \\
	0&=&\ri \nabla^\b{}_{(\ad}\varphi_{\a\b\bd)}+\big(m+\mub\big)\bar{\varphi}_{\a\ad\bd}~.
	\eea
\end{subequations}
These equations can be viewed as reality conditions expressing the component fields of 
$\bar \J_\ad$ in terms of those contained in $\J_\a$. 
Making use of the relations  \eqref{6.16} and their conjugates
leads to the following equations: 
\begin{subequations}
	\bea
	0&=&\big (\cQ+\frac{3}{2}\m\mub-\k^2 \big )\c_\a~,\\
	0&=&\big (\cQ-\k^2\big )B_{\a\b}-\ri \mub \nabla_{(\a}{}^\bd A_{\b)\bd}~, \label{EntG1}\\
	0&=&\big (\cQ+\m \mub -\k^2 \big ) A_{\a\ad}+\ri \m \nabla^\b{}_\ad B_{\a\b}~, \label{EntG2} \\
	0&=&\big (\cQ-\frac{3}{2}\m \mub -\k^2 \big )\varphi_{\a\b\ad}~,
	\eea
\end{subequations}
where $\k^2:=|m+\m |^2$.
It can be shown that the equations \eqref{EntG1} and \eqref{EntG2} imply the quartic equation
\begin{subequations}
	\bea
	0&=&\big (\cQ-\k^2-\sqrt{\m \mub}\k \big ) \big (\cQ-\k^2+\sqrt{\m \mub}\k \big )A_{\a\ad}~,\\
	0&=&\big (\cQ-\k^2-\sqrt{\m \mub}\k \big )\big (\cQ-\k^2+\sqrt{\m \mub}\k \big )B_{\a\b}~.
	\eea
\end{subequations}
The analysis above indicates that the component fields furnish the massive superspin-$1$ representation  \eqref{gravitinorep}, where $E_0=1+\sqrt{\frac{\k^2}{\m \mub}}$. Additionally, the mass-shell conditions are consistent with the results in section \ref{FASMassiveSupermultipletsComp}, for a superfield with the index structure $m-1=n=0$, upon the redefinition $\k^2=M^2-\mu \mub$.

\section{Four-dimensional $\cN=2$ anti-de Sitter superspace} \label{FAS2sec4dAdS}

Recently, free superconformal higher-spin gauge theories were formulated in AdS$^{4|8}$ by Kuzenko and Raptakis \cite{KR}. As discussed previously, the superspin projection operators are encoded within the SCHS action, since their presence ensures that the action describes a pure spin state with maximal superspin off the mass-shell. This section is devoted to extracting the superspin projection operators from the SCHS action of \cite{KR} and studying some of their properties.

\subsection{Facets of four-dimensional $\cN=2$ AdS superspace}\label{FAS2Superspace}
Four-dimensional $\cN=2$ AdS superspace was introduced in \cite{KLRTM} (see also \cite{KT-M08}) as the coset space 
\be
\text{AdS}^{4|8} = \frac{\mathsf{OSp}(2|4)}{\mathsf{SO}(3,1) \times \mathsf{SO}(2)}~. 
\ee

Let us denote by $z^{M} = (x^m, \q^\m_i,\tb^i_{\mud})$ the local coordinates which parametrise AdS$^{4|8}$, where $m=0,1,2,3$, $\m=1,2$, $\mud = \od , \td$ and $i = \1, \2$.\footnote{The Grassmann variables $\q^\m_i$ and $\tb^i_{\mud}$ are related via complex conjugation, $\overline{\q^\m_i}=\tb^{\mud i}$.} The geometry of AdS$^{4|8}$ is described in terms of the covariant derivatives of the form
\be
\cD_A = (\cD_a, \cD^i_\a , \cDB^\ad_i) = E_A + \O_A + \F_A ~, \qquad E_A = E_A{}^M \pa_M~,
\ee
where $E_A{}^M$ is the inverse superspace vielbein and 
\be
\O_A = \hf \O_A{}^{bc}M_{bc}  = \O_A{}^{\b \g}M_{\b \g} + \bar{\O}_A{}^{\bd \gd}\bar{M}_{\bd \gd}~, \qquad \F_A = \F_A{}^{jk}J_{jk}~, 
\ee
are the Lorentz and  $\mathsf{SU}(2)_{{R}}$ connections respectively. Note that the isospinor indices are raised and lowered by the antisymmetric matrix $\ve_{ij} = - \ve_{ji}$, which shares the same normalisation as the spinor metric tensor $\ve^{12}=\ve_{21} =1$ and $\ve_{ij}\ve^{jk}=\d_i{}^j$. The $\mathsf{SU}(2)_R$ generator $J_{ij}$ acts on isospinors in the following fashion
\be
J_{ij} \j_k = -\ve_{k(i}\j_{j)}~, 
\ee
while Weyl spinors are inert under the action of $J_{ij}$.

The covariant derivatives satisfy the following graded commutation relations
\bsubeq
\be
\{ \cD_\a^i , \cDB^\bd_j \} = -2 \ri \d_j{}^i\cD_\a{}^\bd~, \quad [\cD_{\a \ad}, \cD_{\b\bd}] = - 2\cS^2(\ve_{\a\b}\bar{M}_{\ad\bd} + \ve_{\ad\bd}M_{\a\b})~,
\ee
\be
\{\cD^i_\a , \cD^j_\b \} = 4\cS^{ij}M_{\a\b} + 2 \ve_{\a\b} \ve^{i j}\cS^{kl}J_{kl}~, \quad \{\cDB_i^\ad , \cDB^{\bd}_j \} = -4\cS_{ij}\bar{M}^{\ad\bd} - 2 \ve_{ij} \ve^{\ad \bd}\cS^{kl}J_{kl}~, \hspace{-0.1cm}
\ee
\be
[\cD_{\a\ad},\cD_\b^i] = - \ri \ve_{\a\b} \cS^{ij}\cDB_{\ad j}~, \quad [\cD_{\a\ad},\cDB^\bd_i] = -\ri \d_\ad{}^\bd\cS_{ij} \cD_\a^j~.
\ee
\esubeq
Here we have denoted $\cS^2: = \hf \cS^{ij}\cS_{ij}$, where $\cS_{ij} = \cS_{ji}$ is a covariantly constant iso-triplet which is   related to the scalar curvature $\mc{R}$ of AdS$^{4|8}$ by the rule $\mc{R}=-12\cS^2$. Note that we have chosen $\cS_{ij}$ to satisfy the reality condition $\overline{\cS^{ij}} = \cS_{ij}$, which ensures $\cS^2$ is real.\footnote{For details on the permissibility of the  reality condition $\overline{\cS^{ij}} = \cS_{ij}$, see \cite{KLRTM,KT-M08}.}

We will adopt the following condensed notation:
\be
\cD_{ij} = \cD^\b_{(i} \cD_{\b j)}~, \quad \cDB_{ij} = \cDB_{\bd{(i}} \cDB^\bd_{ j)}~, \quad \cD_{\a\b} = \cD^j_{(\a} \cD_{\b)j}~, \quad \cDB_{\ad\bd} = \cDB_{(\ad j} \cDB^j_{\bd)}~.
\ee

The quadratic Casimir operator $\mb{Q}$ of the $\cN=2$ AdS superalgebra $\mf{osp}(2|4)$ is realised on the space of tensor superfields as follows
\be \label{FAS2Casimir}
\mb{Q} : = \Box - \frac{1}{4}\cS^{ij} \big (\cD_{ij} + \cDB_{ij} \big ) -\cS^2 \big (M^2+\bar{M}^2 \big ) + \hf \cS^{ij}\cS^{kl}J_{ij}J_{kl}~,~~ ~[\mb{Q},\cD_A]=0~,
\ee
where $\Box =-\hf \cD^{\a\ad}\cD_{\a\ad}$ is the AdS$^{4|8}$ d'Alembertian.

Pertinent to our subsequent analyses are transverse (anti-)linear and (anti-)chiral superfields in AdS$^{4|8}$. Given integers $m,n \geq 1$, a complex superfield $\G_{\a(m)\ad(n)}$ of Lorentz type $(\frac{m}{2},\frac{n}{2})$ is said to be transverse linear if it satisfies the conditions
\be \label{FAS2TL}
\cDB^{\bd}_i \G_{\a(m)\bd\ad(n-1)} = 0 \quad \Longrightarrow  \quad \big ( \bar{\cD}_{ij} + 2(n+2)\cS_{ij} \big ) \G_{\a(m)\ad(n)} = 0~.
\ee
Alternatively, the superfield $\G_{\a(m)\ad(n)}$ is said to be transverse anti-linear if it obeys
\be \label{FAS2TAL}
\cD^{\b}_i \G_{\b\a(m-1)\ad(n)} = 0 \quad \Longrightarrow  \quad \big ( \cD_{ij} + 2(m+2)\cS_{ij} \big ) \G_{\a(m)\ad(n)} = 0~.
\ee

Given an integer $m \geq 0$, a tensor superfield $\F_{\a(m)}$ of Lorentz type $(\frac{m}{2},0)$ is said to be chiral if it obeys the constraint
\be
\cDB_{\bd}^i \F_{\a(m)} =0~.
\ee
Similarly,  a tensor superfield $\F_{\ad(n)}$ of Lorentz type $(0,\frac{n}{2})$ is  anti-chiral if it satisfies 
\be
\cD_{\b}^i \F_{\ad(n)} =0~.
\ee

Let us introduce the scalar operator $\cDB^4$ \cite{Muller}
\be \label{Chiraloperator}
\cDB^4:= \frac{1}{48}\big( \cDB^{ij} +4\cS^{ij} \big )\cDB_{ij}~,
\ee
which is chiral when restricted to the space of superfields $\F_{\a(m)}$ of Lorentz type $(\frac{m}{2},0)$, 
\be 
\cDB_\bd^i \cDB^4 \F_{\a(m)} = 0~.
\ee
Taking the complex conjugate of \eqref{Chiraloperator} will yield the anti-chiral analogue. 

\subsection{Superconformal higher-spin theory}\label{FAS2SCHS}
For integers $m,n \geq 1$,  the SCHS action in AdS$^{4|8}$ is described in terms of the gauge prepotential $H_{\a(m)\ad(n)}$, which is defined modulo gauge transformations of the form \cite{KR}
\be \label{FAS2SCHSGT2}
\d_{\z,\l}H_{\a(m)\ad(n)} = \cD^i_{\a}\z_{\a(m-1)\ad(n)i} + \cDB^i_{\ad}\l_{\a(m)\ad(n-1)i}~.
\ee
Here the gauge parameters $\z_{\a(m-1)\ad(n)i}$ and $\l_{\a(m)\ad(n-1)i}$ are complex and unconstrained. Associated with the superfield $H_{\a(m)\ad(n)}$ and its complex conjugate $\bar{H}_{\a(n)\ad(m)}$ are the linearised higher-spin super-Weyl tensors\cite{KR}
\bsubeq \label{Weyl}
\bea 
\mathfrak{W}_{\a(m+n+2)}{(H)} &=& \cDB^4 \cD_{(\a_1}{}^{\bd_1} \cdots \cD_{\a_n}{}^{\bd_n} \cD_{\a_{n+1} \a_{n+2}}H_{\a_{n+3} \cdots \a_{m+n+2})\bd(n)}~, \\
\mathfrak{W}_{\a(m+n+2)}{(\bar{H})} &=& \cDB^4 \cD_{(\a_1}{}^{\bd_1} \cdots \cD_{\a_m}{}^{\bd_m} \cD_{\a_{m+1} \a_{m+2}}\bar{H}_{\a_{m+3} \cdots \a_{m+n+2})\bd(m)}~.
\eea
The super-Weyl tensors \eqref{Weyl} are invariant under the gauge transformations \eqref{FAS2SCHSGT2}
\esubeq
\be
\d_{\z,\l} \mathfrak{W}_{\a(m+n+2)}{(H)} = 0~, \quad \d_{\z,\l} \mathfrak{W}_{\a(m+n+2)}{(\bar{H})} = 0~,
\ee
and are chiral 
\be
\cDB_\bd^i \mathfrak{W}_{\a(m+n+2)}{(H)}  = 0~, \qquad \cDB_\bd^i \mathfrak{W}_{\a(m+n+2)}{(\bar{H})} = 0~,
\ee
due to the presence of the chiral operator \eqref{Chiraloperator}.

The gauge-invariant SCHS action in AdS$^{4|8}$ is usually expressed as a functional over the chiral subspace of the full superspace \cite{Butter2011}
\be
\int \rd^4 x \rd^4 \q \rd^4 \tb \, E \, \cL = \int \rd^4 x \rd^4 \q \, \cE \,  \cDB^4 \cL~, \qquad E^{-1} = \text{Ber}(E_A{}^M)~,
\ee
where the Lagrangian $\cL$ is a real scalar superfield. Keeping this in mind, the gauge-invariant SCHS action \cite{KR} is given by\footnote{For details on the superconformal properties of the SCHS action \eqref{SCHSAction}, and its constituents, see \cite{KR}.}
\be \label{SCHSAction}
S_{\text{SCHS}}^{(m,n)}[H,\bar{H}] = \ri^{m+n} \int \rd^4 x \rd^4 \q  ~\cE~ \mathfrak{W}^{\a(m+n+2)}{(H)} \mathfrak{W}_{\a(m+n+2)}{(\bar{H})}  + \text{c.c.} 
\ee
Upon integrating by parts, one can rewrite the action \eqref{SCHSAction} in the alternative form
\bsubeq
\bea \label{FAS2SCHS2}
S_{\text{SCHS}}^{(m,n)}[H,\bar{H}] &=& \ri^{m+n} \int \rd^4 x \rd^4 \q \rd^4 \tb ~E~ \bar{H}^{\a(n)\ad(m)}\mathfrak{B}_{\a(n)\ad(m)}(H) + \text{c.c.} ~, \\
&=& \ri^{m+n} \int \rd^4 x \rd^4 \q \rd^4 \tb~E~ \bar{H}^{\a(n)\ad(m)}\widehat{\mathfrak{B}}_{\a(n)\ad(m)}(H) + \text{c.c.}~,
\eea
\esubeq
where the superfields $\mathfrak{B}_{\a(n)\ad(m)}(H)$  and $\widehat{\mathfrak{B}}_{\a(n)\ad(m)}(H) $ are known as the  linearised higher-spin Bach tensors
\bsubeq \label{BT}
\bea
\mathfrak{B}_{\a(n)\ad(m)}(H) &=& \phantom{-} \cD_{(\ad_1}{}^{\b_1} \cdots \cD_{\ad_m)}{}^{\b_m}\cD^{\b_{m+1}\b_{m+2}}\mathfrak{W}_{\a(n)\b(m+2)}(H)~, \\
\widehat{\mathfrak{B}}_{\a(n)\ad(m)}(H) &=& - \cD_{(\a_1}{}^{\bd_1} \cdots \cD_{\a_n)}{}^{\bd_n}\cDB^{\bd_{n+1}\bd_{n+2}}\overline{\mathfrak{W}}_{\ad(m)\bd(n+2)}(H)~.
\eea
\esubeq
In addition to being invariant under the gauge transformations \eqref{FAS2SCHSGT2}, the superfields \eqref{BT} are also both simultaneously transverse linear and transverse anti-linear 
\bsubeq \label{TLALB}
\bea
\cD^\b_i \mf{B}_{\b\a(n-1)\ad(m)}(H) &=& 0~, \qquad \cDB^\bd_i \mf{B}_{\a(n)\bd\ad(m-1)}(H) = 0~,
\\
\cD^\b_i \widehat{\mf{B}}_{\b\a(n-1)\ad(m)}(H) &=& 0~, \qquad \cDB^\bd_i \widehat{\mf{B}}_{\a(n)\bd\ad(m-1)}(H)= 0~.
\eea
\esubeq

\subsection{Superspin projection operators}\label{FAS2Superprojectors}
Our construction of the superspin projection operators in AdS$^{4|8}$ is based on the SCHS gauge theory.  
Recall, by definition, the (S)CHS action describes pure (super)spin states which carry maximal (super)spin. In accordance with this, the (S)CHS action will always encode the (super)spin projection operators since, by definition, they select out the component of an arbitrary (super)field which describes said pure (super)spin states with maximal (super)spin. 

Thus, our ansatz for the superspin projection operator is based on the general tensorial form of the higher-spin Bach tensors \eqref{BT}. However, if the operator is to be a projector, it must preserve the rank of the tensor field on which it acts. This can be achieved by appropriately removing or inserting vector derivatives in \eqref{BT} as a means to convert indices.

In accordance with the above considerations, for integers $m,n \geq 1$, let us define  two differential operators which act on an unconstrained superfield $\F_{\a(m)\ad(n)}$ via the rule
\bsubeq \label{TransverseOp}
\bea
\mb{P}_{\a(m)\ad(n)}(\F)&=& - \frac{1}{16}\cD_{(\ad_1}{}^{\b_1} \cdots \cD_{\ad_n )}{}^{\b_n}\cD^{\g(2)}\cDB^4\cD_{(\b_1}{}^{\bd_1} \cdots \cD_{\b_n}{}^{\bd_n} \cD_{\g(2)}\F_{\a_1 \ldots \a_m)\bd(n)}~, \\
\widehat{\mb{P}}_{\a(m)\ad(n)}(\F)&=& - \frac{1}{16}\cD_{(\a_1}{}^{\bd_1} \cdots \cD_{\a_m )}{}^{\bd_m}\cDB^{\gd(2)}\cD^4\cD_{(\bd_1}{}^{\b_1} \cdots \cD_{\bd_m}{}^{\b_m} \cDB_{\gd(2)}\F_{\b(m)\ad_1 \ldots \ad_n)}~.  \hspace{1.3cm}
\eea
\esubeq
Note that these superfields are TLAL, since they are built from the linearised higher-spin Bach tensors  \eqref{BT}. One can immediately introduce the operators $\bm{\P}^{\perp}_{(m,n)}$ and $\widehat{\bm{\P}}{}^{\perp}_{(m,n)}$
\bsubeq \label{Projectors}
\bea
\bm{\P}^{\perp}_{(m,n)}\F_{\a(m)\ad(n)} &\equiv& \bm{\P}^{\perp}_{\a(m)\ad(n)}(\F) = \Big [ \prod_{t=1}^{n+2}  \big (\mb{Q} - \l_{(t,m,n)}  \cS^2 \big ) \Big ]^{-1}  \mb{P}_{\a(m)\ad(n)}(\F)~, \\
\widehat{\bm{\P}}{}^{\perp}_{(m,n)}\F_{\a(m)\ad(n)} &\equiv& \widehat{\bm{\P}}{}^{\perp}_{\a(m)\ad(n)}(\F) = \Big [ \prod_{t=1}^{m+2}  \big (\mb{Q} - {\l}_{(t,m,n)}  \cS^2 \big ) \Big ]^{-1} \widehat{\mb{P}}_{\a(m)\ad(n)}(\F)~, \hspace{0.3cm}
\eea
\esubeq
where $\l_{(t,m,n)}  $ are  dimensionless constants defined by
\bsubeq
\bea
\l_{(t,m,n)}  &=&\hf  \Big ( (m+n-t+4)^2 + t(t-2) \Big )~.
\eea
\esubeq

The operators $\bm{\P}^{\perp}_{(m,n)}$ and $\widehat{\bm{\P}}{}^{\perp}_{(m,n)}$ map any superfield  $\F_{\a(m)\ad(n)}$ to a TLAL superfield
\bsubeq
\bea
\cDB^\bd_i \bm{\P}^{\perp}_{\a(m)\bd\ad(n-1)}(\F) &=& 0~, \qquad \cD^\b_i \bm{\P}^{\perp}_{\b\a(m-1)\ad(n)}(\F) = 0~, \\
\cDB^\bd_i \widehat{\bm{\P}}{}^{\perp}_{\a(m)\bd\ad(n-1)}(\F) &=& 0~,\qquad \cD^\b_i \widehat{\bm{\P}}{}^{\perp}_{\b\a(m-1)\ad(n)}(\F) = 0~.
\eea
\esubeq
Furthermore, the TLAL operators \eqref{Projectors} are idempotent
\bsubeq
\bea
\bm{\P}^{\perp}_{(m,n)}\bm{\P}^{\perp}_{(m,n)}\F_{\a(m)\ad(n)} &=& \bm{\P}^{\perp}_{(m,n)}\F_{\a(m)\ad(n)} ~,\\
\widehat{\bm{\P}}{}^{\perp}_{(m,n)}\widehat{\bm{\P}}{}^{\perp}_{(m,n)}\F_{\a(m)\ad(n)} &=& \widehat{\bm{\P}}{}^{\perp}_{(m,n)}\F_{\a(m)\ad(n)} ~.
\eea
\esubeq
Hence, the operators \eqref{Projectors} are TLAL superprojectors.\footnote{Recently, a new superprojector was constructed in \cite{HutchingsKuzenkoRaptakis2023} which maps an  isospinor superfield $\U_i$ in AdS$^{4|8}$ into a multiplet which has the properties of a conformal supercurrent.} In actuality, they can be identified as the superspin projection operators since they were extracted from the SCHS action \eqref{FAS2SCHS2}. Although not obvious, it can be shown that the projection operators $\bm{\P}^{\perp}_{\a(m)\ad(n)}(\F)$ and $\widehat{\bm{\P}}{}^{\perp}_{\a(m)\ad(n)}(\F)$ are actually equivalent on the space of superfields $\F_{\a(m)\ad(n)}$
\be
\bm{\P}^{\perp}_{\a(m)\ad(n)}(\F)  = \widehat{\bm{\P}}{}^{\perp}_{\a(m)\ad(n)}(\F)~.
\ee
Showing this result explicitly is non-trivial and computationally demanding.

In the case $m>n=0$, it can be shown that the operators analogous to \eqref{TransverseOp} are
\bsubeq
\bea
\mb{P}_{\a(m)}(\F)&:=& - \frac{1}{16}\cD^{\b(2)}\cDB^4\cD_{(\b_1 \b_2}\F_{\a_1 \cdots \a_m)}~, \\
\widehat{\mb{P}}_{\a(m)}(\F)&:=& - \frac{1}{16}\cD_{(\a_1}{}^{\bd_1} \cdots \cD_{\a_m )}{}^{\bd_m}\cDB^{\gd(2)}\cD^4\cD_{(\bd_1}{}^{\b_1} \cdots \cD_{\bd_m}{}^{\b_m} \cDB_{\gd_1 \gd_2)}\F_{\b(m)}~.
\eea
\esubeq
It follows that the corresponding projectors $\bm{\P}^{\perp}_{(m)}:= \bm{\P}^{\perp}_{(m,0)}$ and $\widehat{\bm{\P}}{}^{\perp}_{(m)}:= \widehat{\bm{\P}}{}^{\perp}_{(m,0)}$ are
\bsubeq
\bea
\bm{\P}^{\perp}_{(m)}\F_{\a(m)} &\equiv& \bm{\P}^{\perp}_{\a(m)}(\F) =    \Big [ \prod_{t=1}^{2}  \big (\mb{Q} - {\l}_{(t,m,0)}  \cS^2 \big ) \Big ]^{-1}  \mb{P}_{\a(m)}(\F)~, \\
\widehat{\bm{\P}}{}^{\perp}_{(m)}\F_{\a(m)} &\equiv& \widehat{\bm{\P}}{}^{\perp}_{\a(m)}(\F) = \Big [ \prod_{t=1}^{m+2}  \big (\mb{Q} - {\l}_{(t,m,0)}  \cS^2 \big ) \Big ]^{-1} \widehat{\mb{P}}_{\a(m)}(\F)~.
\eea
\esubeq
It can be shown that these operators are idempotent and project an arbitrary field $\F_{\a(m)}$ onto the space of simultaneously linear and transverse anti-linear superfields
\bsubeq
\bea
\big (\cDB_{ij} +4\cS_{ij} \big )\bm{\P}^{\perp}_{\a(m)}(\F) &=&0~, \qquad \cD^\b_i \bm{\P}^{\perp}_{\b\a(m-1)}(\F) =  0~,  \\
\big (\cDB_{ij} +4\cS_{ij} \big )\widehat{\bm{\P}}{}^{\perp}_{\a(m)}(\F) &=& 0~, \qquad \cD^\b_i \widehat{\bm{\P}}{}^{\perp}_{\b \a(m-1)}(\F) = 0~.
\eea
\esubeq

\subsection{On-shell superfields}\label{FAS2Onshell}
For integers $m, n \geq 1$, we say that a totally symmetric tensor superfield $\F_{\a(m)\ad(n)}$ on AdS$^{4|8}$ is on-shell if it satisfies the constraints\footnote{The on-shell constraints \eqref{FAS2OnshellConditions} are reminiscent of their AdS$^{4|4}$ cousins \eqref{FASOnShellConditions}.}
\bsubeq \label{FAS2OnshellConditions}
\begin{gather}
\cD^\b_i \F_{\b\a(m-1)\ad(n)}=0~, \qquad \cDB^\bd_i \F_{\a(m)\bd\ad(n-1)}=0~, \label{FAS2OnshellProp1} \\
( \mb{Q} - M^2) \F_{\a(m)\ad(n)} = 0~, \label{FAS2OnshellProp2}
\end{gather}
\esubeq
where $M$ is a parameter of mass dimension known as pseudo-mass. This definition was proposed based on the properties of the superspin projection operators \eqref{Projectors}. 

We conjecture that the space of on-shell superfields \eqref{FAS2OnshellConditions} furnish irreducible representations of $\mf{osp}(2|4)$. Irreducible representations of $\mf{osp}(2|4)$ are conveniently described by their decompositions into irreducible representations of $\mf{osp}(1|4)$  \cite{Nicolai:1984hb}, which were detailed in section \ref{FASRepTheory}. Thus in order to confirm the above claim, it is necessary to perform a superspace reduction on the on-shell superfields \eqref{FAS2OnshellConditions} to AdS$^{4|4}$. Once completed, one needs to demonstrate that the surviving superfields are on-shell superfields \eqref{FASScalarProjectors} in AdS$^{4|4}$. In particular, these on-shell superfields should furnish irreducible representations of $\mf{osp}(1|4)$, which are consistent with those appearing in the decomposition described in \cite{Nicolai:1984hb}. The superspace reduction of the on-shell superfields \eqref{FAS2OnshellConditions}, and their corresponding analysis, will be given in \cite{Hutchings2022}.

Recall the important observation that the poles of the (super)spin projection operators in AdS$_4$ coincide with partially massless values. In the spirit of this result, it is natural to expect that the $\cN=2$ partially massless supermultiplets are associated with the poles of the superspin projection operators \eqref{Projectors}. Investigating this idea further, let us define a partially massless supermultiplet\footnote{Note that the partially massless $\cN=2$ supermultiplets in AdS$_4$ have been recently studied in the component setting in \cite{Bittermann:2020xkl}.} $\F^{(t)}_{\a(m)\ad(n)}$ with super-depth $t$ as an on-shell superfield \eqref{FAS2OnshellConditions} carrying the pseudo-mass
\be
M^2 = \l_{(t,m,n)} \cS^2~, \qquad 1\leq t \leq \text{min}(m+2,n+2)~.
\ee

Analogous to the AdS$^{4|4}$ story \ref{FASOnShellSupermultiplets}, massless superfields should be identified as the partially massless superfield with the lowest super-depth. In accordance with this, we say that an on-shell superfield \eqref{FAS2OnshellConditions} is massless if it carries pseudo-mass 
\be \label{FAS2Massless}
M^2 = \l_{(1,m,n)} \cS^2 ~, \qquad \l_{(1,m,n)}  = \hf(m+n+2)(m+n+4)~.
\ee
The system of equations \eqref{FAS2OnshellConditions} with pseudo-mass \eqref{FAS2Massless} can be shown to be compatible with the gauge symmetry
\be  \label{FAS2MasslessGaugeSymmetry}
\d_{\z,\l}\F_{\a(m)\ad(n)} = \cD^i_{(\a_1}\z_{\a_2 \cdots \a_m)\ad(n)i} + \cDB^i_{(\ad_1}\l_{\a(m)\ad_2 \dots \ad_n)i}~,
\ee
given that the gauge parameters $\z_{\a(m-1)\ad(n)i}$ and $\l_{\a(m)\ad(n-1)i}$ are TLAL and obey the constraints
\bsubeq \label{FAS2GaugeCondtions}
\bea
\cD_{(\a_1}{}^\bd \z_{\a_2 \ldots \a_m)\bd \ad(n-1) i} &=& \ri(n+2) \cS_{ij} \l_{\a(m)\ad(n-1)}{}^j~, \\
\cD^\b{}_{(\ad_1} \l_{\b\a(m-1) \ad_2 \ldots \ad_n)i} &=& \ri (m+2)\cS_{ij} \z_{\a(m-1)\ad(n)}{}^j~.
\eea
\esubeq
The conditions \eqref{FAS2GaugeCondtions} imply that the gauge parameters satisfy
\be \label{FAS2MassGaugeparameter}
(\mb{Q} - \t_{(1,m,n)}\cS^2) \z_{\a(m-1)\ad(n)i} =0~,  \qquad 
(\mb{Q} - \t_{(1,m,n)}\cS^2) \l_{\a(m)\ad(n-1)i} =0~. 
\ee
Thus, it follows from \eqref{FAS2MassGaugeparameter} and the gauge variation of \eqref{FAS2OnshellProp2} that the gauge parameters are only non-zero if $M$ satisfies \eqref{FAS2Massless}. The gauge symmetry \eqref{FAS2MasslessGaugeSymmetry} coincides with the gauge symmetry \eqref{FAS2SCHSGT2} associated with the SCHS prepotential. This result supports our proposed definition for on-shell superfields since it is well-known that massless higher-superspin gauge theories are realised in terms of SCHS prepotentials (with additional compensators). The gauge symmetry  associated with partially massless superfields in  AdS$^{4|8}$ will be explored in \cite{Hutchings2022}.

Finally, in this framework, it is natural to define massive supermultiplets as an on-shell superfield \eqref{FAS2OnshellConditions} carrying pseudo-mass 
\be
M^2 > \l_{(1,m,n)} \cS^2~. 
\ee
In other words, the pseudo-mass carried by a massive superfield is greater than that carried by a massless superfield.

\section{Summary of results} \label{FASsecSummary}
The primary objective of this chapter was to construct the spin projection operators in four-dimensional anti-de Sitter (super)space and explore several of their applications. The AdS$_4$ spin projection operators were reviewed and expanded upon in section \ref{FAec4dAdS}, while their supersymmetric generalisations to $\cN=1$  and $\cN=2$ AdS superspace were studied in sections \ref{FASsec4dAdS} and \ref{FAS2sec4dAdS} respectively.

In section \ref{FASpinProjectors}, we derived a new realisation of the spin projection operators $\tilde{\P}^{\perp}_{(m,n)}$ in AdS$_4$. The novelty of this operator is that they can be written solely in terms of the Casimir operators of $\mathfrak{so}(3, 2)$, see e.q. \eqref{FAProjectorsCasimir}. This result appeared for the first time in this thesis. The structure of these projectors are interesting for two reasons. Firstly, they encode all information concerning partially massless fields in the poles. Moreover given $m \geq n$, the zeros of the projector $\tilde{\P}^{\perp}_{(m,n)}$ can be identified as the eigenvalues for the quartic Casimir operator \eqref{FAQuarticCasimirOn-ShellFields} when acting on the lower-rank transverse field $\f^{\perp}_{\a(m-k)\ad(n-k)}$ or $\f^{\perp}_{\a(n-k)\ad(m-k)}$, where $ 1 \leq k \leq n$. A similar result  holds in the analogous case $n \geq m$.

The form of the projector $\tilde{\P}^{\perp}_{(m,n)}$  is better suited for studying certain applications, which are not feasible, nor computationally tractable when using the known projectors \eqref{FAProjectionOperators} of \cite{KP20}. For example, they are useful in the computation of the bosonic and fermionic extractors, see e.q. \eqref{FAExtractors}. These operators are responsible for extracting the bosonic $\f^{\perp}_{\a(s-k)\ad(s-k)}$ and fermionic $\f^{\perp}_{\a(s-k)\ad(s-k-1)} $ fields from the decomposition \eqref{FADecomposition}.  In section \ref{FACHS}, a novel representation for the linearised higher-spin Weyl tensors was formulated in terms of the projectors $\tilde{\P}^{\perp}_{(m,n)}$, see e.q. \eqref{FASWeylTensors}. 

In section \ref{FASsec4dAdS} we constructed the superspin projection operators $\bm{\P}^{\perp}_{(m,n)}$ in AdS$^{4|4}$ for the first time \cite{BHKP}. They are given by eq. \eqref{FASSuperprojectors} for $m\geq n>0$, 
and by eq. \eqref{projectors-m} for $m>n=0$.  
The operator  $\bm{\P}^{\perp}_{(m,n)}$ maps a superfield $\F_{\a(m)\ad(n)}$ on the mass-shell \eqref{FASOnShellMass} to an on-shell superfield \eqref{FASProjOnshell},\footnote{The superprojector maps an arbitrary superfield to a superfield which realises the properties of a conserved conformal supercurrent.}  which were shown in section \ref{FASOnShellSupermultiplets} to realise irreducible representations of the ${\cal N}=1$ AdS${}_4$ superalgebra $\mathfrak{osp} (1|4)$.
In particular, we established a one-to-one correspondence between 
the on-shell partially massless supermultiplets with super-depth $t$ and the poles of $\bm{\P}^{\perp}_{(m,n)}$ which are determined by  $\lambda_{(t,m,n)}\mu\mub$ 
belonging to  the range \eqref{FASPMMass} \cite{BHKP}.

Analysing the poles of the $\bm{\P}^{\perp}_{(m,n)}$ leads to a systematic discussion on how to realise the unitary (both massive and massless) and the partially massless representations of the $\mathfrak{osp} (1|4)$ in terms of on-shell superfields. As an example,
we presented an off-shell model for the massive gravitino multiplet in AdS$^{4|4}$ \cite{BHKP}. We also demonstrated that the AdS$^{4|4}$ superspin projection operators can be used to decompose any arbitrary superfield $\F_{\a(m)\ad(n)}$ into its irreducible components. This decomposition is given by eq. \eqref{decomp1} for $m>n$ \cite{BHKP}. In addition, the superprojectors were essential in the factorisation of the superconformal higher-spin gauge actions in AdS$_4$. These factorised actions are given by  eqs. \eqref{FASSCHSMGN} and \eqref{FASCHSMGN}.\footnote{The factorisation of the SCHS action was computed by Michael Ponds in \cite{BHKP}.} 

In the final section \ref{FAS2sec4dAdS}, we derived the quadratic Casimir operator \eqref{FAS2Casimir} of the $\cN=2$ AdS superalgebra $\mf{osp}(2|4)$ in the superfield representation \cite{Hutchings2022}. This result was pertinent in the computation of the TLAL superspin projection operators \eqref{Projectors} in AdS$^{4|8}$ \cite{Hutchings2022}, which were extracted from the free SCHS action of \cite{KR}. 
Taking inspiration from the (super)spin projection operators in AdS$_4$ and AdS$^{4|4}$, the poles of these superprojectors were conjectured to be connected to  (partially-)massless supermultiplets. Under this assumption, we proposed a dictionary for on-shell superfields in section \ref{FAS2Onshell} \cite{Hutchings2022}. These superfields should, in principle, furnish irreducible representations of $\mf{osp}(2|4)$. To support this claim, we showed that an on-shell superfield carrying the lowest depth partially massless value \eqref{FAS2Massless} is compatible with the massless gauge symmetry \eqref{FAS2MasslessGaugeSymmetry} \cite{Hutchings2022}.

\begin{subappendices}

\section{AdS$^{4|4}$ superspace toolkit} \label{TASappendixA}
Below we collate a list of identities that were essential in the computation of many of the main results in section \ref{FASsec4dAdS}. Note that they can be readily derived from the algebra of covariant derivatives \eqref{FASDerivativeAlgebra}:
\begin{subequations} 
	\label{A.2}
	\bea 
	\cD_\a\cD_\b
	\!&=&\!\frac{1}{2}\ve_{\a\b}\cD^2-2{\bar \m}\,M_{\a\b}~,
	\quad\qquad \,\,\,
	{\bar \cD}_\ad{\bar \cD}_\bd
	=-\frac{1}{2}\ve_{\ad\bd}{\bar \cD}^2+2\m\,{\bar M}_{\ad\bd}~,  \label{A.2a}\\
	\cD_\a\cD^2
	\!&=&\!4 \bar \m \,\cD^\b M_{\a\b} + 4{\bar \m}\,\cD_\a~,
	\quad\qquad
	\cD^2\cD_\a
	=-4\bar \m \,\cD^\b M_{\a\b} - 2\bar \m \, \cD_\a~, \label{A.2b} \\
	{\bar \cD}_\ad{\bar \cD}^2
	\!&=&\!4 \m \,{\bar \cD}^\bd {\bar M}_{\ad\bd}+ 4\m\, \bar \cD_\ad~,
	\quad\qquad
	{\bar \cD}^2{\bar \cD}_\ad
	=-4 \m \,{\bar \cD}^\bd {\bar M}_{\ad\bd}-2\m\, \bar \cD_\ad~,  \label{A.2c}\\
	\left[\bar \cD^2, \cD_\a \right]
	\!&=&\!4\rm i \cD_{\a\bd} \bar \cD^\bd +4 \m\,\cD_\a = 
	4\rm i \bar \cD^\bd \cD_{\a\bd} -4 \m\,\cD_\a~,
	\label{A.2d} \\
	\left[\cD^2,{\bar \cD}_\ad\right]
	\!&=&\!-4\rm i \cD_{\b\ad}\cD^\b +4\bar \m\,{\bar \cD}_\ad = 
	-4\rm i \cD^\b \cD_{\b\ad} -4 \bar \m\,{\bar \cD}_\ad~,
	\label{A.2e}
	\eea
\end{subequations} 
where $\cD^2=\cD^\a\cD_\a$, and ${\bar \cD}^2={\bar \cD}_\ad{\bar \cD}^\ad$. Furthermore, the following identities also proved indispensable in the analyses of section \ref{FASsec4dAdS}:
\begin{subequations}\label{A.4}
	\begin{gather}
	\cD_\a{}^\bd \cD_\bd{}^\b = \d_\a{}^\b \Box -2\m\mub M_\a{}^\b~, \label{A.4a}\\
	\cD^\a{}_\ad \cD_\a{}^\bd = \d_\ad{}^\bd \Box - 2 \mu \mub \bar{M}_\ad{}^\bd~, \\
	\Box +2\m \bar \m = 
	- \frac 18 \mathcal{D}^{\a}\big(\bar{\mathcal{D}}^2-4\mu\big)\mathcal{D}_{\a} 
	+\frac{1}{16} \big\{\mathcal{D}^2 -4\bar \m ,\bar{\mathcal{D}}^2 -4 \m\big\} ~, \label{A.5a}\\
	\mathcal{D}^{\a}\big(\bar{\mathcal{D}}^2-4\mu\big)\mathcal{D}_{\a}
	=\bar{\mathcal{D}}_{\ad}\big(\mathcal{D}^2-4\mub\big)\bar{\mathcal{D}}^{\ad}~,\\
	\big[\mathcal{D}^2,\bar{\mathcal{D}}^2\big]=-4\text{i}\big[\mathcal{D}^\b,\bar{\mathcal{D}}^{\bd}\big]\mathcal{D}_{\b\bd}+8\mu\mathcal{D}^2-8\mub\bar{\mathcal{D}}^2  
	\end{gather}
\end{subequations}

The isometry group of AdS$^{4|4}$  is $\sOSp(1|4)$.
The  isometry transformations of AdS$^{4|4}$ are generated by the Killing supervector fields
$\x^A E_A$ which are defined to solve the Killing equation
\bea
\big[\X,\cD_{A} \big]=0~,\qquad
\X:=-\hf \x^{\b\bd } \cD_{\b\bd}+\x^\b \cD_\b+{\bar \x}_\bd {\bar \cD}^\bd 
+ \x^{\b\g}M_{\b\g} +\bar \x^{\bd \gd} \bar M_{\bd \gd} ~, 
\label{A.6}
\eea
for some Lorentz parameter  $\x^{\b\g}  = \x^{\g\b}$. 
Given a supersymmetric field theory in AdS$^{4|4}$ which is formulated in terms of the superfield dynamical variables $\mathfrak V$ (with suppressed indices), its action is invariant under 
the isometry transformations 
\bea
\d {\mathfrak V} = \X  \mathfrak V ~, 
\label{SUSY}
\eea
with $\x^B$ being  an arbitrary Killing supervector field.

The Killing equation \eqref{A.6} implies the following \cite{BuchbinderKuzenko1998}:
\begin{subequations} \label{A.7}
	\bea
	\cD_\a \x_{\b\bd} &=& 4\ri \ve_{\a\b} \bar \x_\bd~, \qquad \qquad 
	\bar \cD_\ad \x_{\b\bd} = -4\ri \ve_{\ad\bd}  \x_\b~, \\
	\cD_\a \x_\b &=&\x_{\a\b} ~,\qquad  \qquad \qquad \bar \cD_\ad \x_\b = -\frac{\ri}{2} \m \x_{\b \ad}~, \\
	\cD_\a \x_{\b\g} &=& -4\bar \m \ve_{\a (\b} \x_{\g)}~,  \qquad ~\bar \cD_\ad \x_{\b\g}=0~.
	\eea
\end{subequations}
It is seen that the parameters $\x_\b$, $\x_{\b\g}$ and their conjugates are expressed in terms of the vector parameter $\x_{\b \bd}$, and the latter obeys the equation 
\bea
\cD_{(\a}\x_{\b)\bd}=0 \quad \implies  \quad \cD_{(a} \x_{b)}=0~.
\label{A.8} 
\eea
It also follows from \eqref{A.7} that, for every element of the set of parameters $\U =\{ \x^B, \x^{\b\g}, \bar \x^{\bd \gd} \}$, its covariant derivative $\cD_A \U$ is a linear combination of the elements of $\U$. Therefore, all information about the superfield parameters
$\U$ is encoded in their bar-projections, $\U|$. 

Every Killing supervector superfield $\x^B$ on AdS$^{4|4}$, eq. \eqref{A.6}, can be uniquely decomposed as a sum of even and odd ones. The Killing supervector field $\xi^{B}$ is defined to be even if
\bea
\label{even}
v^{b} := \xi^{b}| \neq 0 ~, \quad \xi^\b| = 0~.
\eea
and odd if
\bea
\label{odd}
\xi^{b}|  = 0 ~, \quad \e^{\b} := \xi^{\b}| \neq 0~.
\eea
The fields $v^{b}(x) $ and $\e^{\b}(x) $ encode complete information about the parent conformal Killing vector superfield. It follows from \eqref{A.8} that $v^b$ is a Killing vector field on AdS$_4$, 
\bea
\nabla_{(a} v_{b)}=0~.
\eea
The relations \eqref{A.7} imply that $\e^\b$ is a Killing spinor field satisfying the equation
\bea
\nabla_{\a\ad} \e_\b = \frac{\ri}{4} \m \ve_{\a\b} \bar \e_\ad~.
\eea
Every Killing vector  $v^b$ on AdS$_4$ can be lifted to 
an even  Killing supervector field $\x^B$ on AdS$^{4|4}$ using the relations \eqref{A.7}.
A similar statement holds for Killing spinors.

Given a tensor superfield ${\mathfrak V}$ (with suppressed indices), its independent component fields are contained in the set of fields 
$\vf=\F|$, where  
$\F:=\big\{ {\mathfrak V}, \cD_{ \a} {\mathfrak V}, \bar \cD_{ \ad} {\mathfrak V}, \dots \big\}$.
In accordance with \eqref{SUSY}, the supersymmetry transformation of $\vf$ is 
\bea
\d_\e \vf =  \e^\b (\cD_\b \F)| + \bar \e_\bd (\bar \cD^\bd  \F )|~.
\label{A.14}
\eea

\section{Partially massless gauge symmetry} \label{TASappendixB}

This appendix is devoted to the derivation of partially massless gauge transformations in AdS$^{4|4}$. The results presented in this appendix were obtained by Michael Ponds in \cite{BHKP}.

Let us consider the real supermultiplet $\F_{\a\ad}=\bar{\F}_{\a\ad}$ which is on-shell \eqref{FASOnShellConditions}
\begin{subequations} \label{B.35}
	\begin{align}
	0&=\mathcal{D}^{\b}\F_{\b\ad}=\bar{\mathcal{D}}^{\bd}\F_{\a\bd}~, \label{B.35b} \\
	0&=\big(\mathbb{Q}-M^2\big)\F_{\a\ad}~.\label{B.35a}
	\end{align}
\end{subequations}
We would like to determine those values of $M$ for which the above system of equations
is compatible with a gauge symmetry. We will see that this occurs only at partially massless values. For the supermultiplet $\F_{\a\ad}$ there are only two such values, 
\begin{subequations} \label{B.355}
	\begin{align}
	M_{(1)}^2&\equiv \lambda_{(1,1,1)}\mu\mub=5\mu\mub~,\label{B.355a}\\
	M_{(2)}^2& \equiv \lambda_{(2,1,1)}\mu\mub=3\mu\mub~,\label{B.355b}
	\end{align}
\end{subequations}
corresponding to the super-depths $t=1$ (massless) and $t=2$ respectively.

We begin by postulating a gauge transformation of the form 
\begin{align}
\delta \F_{\a\ad}= \bar{\mathcal{D}}_{\ad}\Lambda_{\a}-\mathcal{D}_{\a}\bar{\Lambda}_{\ad}~,\label{B.36}
\end{align}
for an unconstrained gauge parameter  $\Lambda_{\a}$. From this we must deduce which constraints must be placed upon $\Lambda_{\a}$ in order for $\delta \F_{\a\ad}$ to be a solution to the equations \eqref{B.35}. 

Gauge invariance of \eqref{B.35a} requires $\Lambda_{\a}$ to also have pseudo-mass $M$,
\begin{align}
0&=\big(\mathbb{Q}- M^2\big)\Lambda_{\a}~. \label{B.37}
\end{align}
Next we decompose the gauge parameter into irreducible parts via the prescription given in section \ref{FASOrthoSect}. Performing this procedure, we find that \eqref{B.36} takes the form
\begin{align}
\delta \F_{\a\ad}=\bar{\mathcal{D}}_{\ad}\zeta_{\a}-\mathcal{D}_{\a}\bar{\z}_{\ad}+\mathcal{D}_{\a\ad}\xi + \big[\mathcal{D}_{\a},\bar{\mathcal{D}}_{\ad}\big]\z+\mathcal{D}_{\a\ad}\big(\s+\bar{\s}\big)~. \label{B.38}
\end{align}
In \eqref{B.38}, the gauge parameter $\z_{\a}$ is LTAL
\begin{align}
0=\big(\bar{\mathcal{D}}^2-4\mu\big)\z_{\a}~,\qquad 0=\mathcal{D}^{\a}\z_{\a}~~ \Leftrightarrow ~~ 0=\big(\mathcal{D}^2-6\mub\big)\z_{\a}~. \label{B.39}
\end{align}
The real parameters $\xi=\bar{\xi}$ and $\z=\bar{\zeta}$ are linear
\begin{align}
0=\big(\bar{\mathcal{D}}^2-4\mu\big)\z=\big(\mathcal{D}^2-4\mub\big)\z~,\qquad  0=\big(\bar{\mathcal{D}}^2-4\mu\big)\xi=\big(\mathcal{D}^2-4\mub\big)\xi~, \label{B.40}
\end{align}
whilst $\s$ is complex chiral, $\bar{\mathcal{D}}_{\ad}\s=0$.

Before solving the equation $0=\mathcal{D}^{\b}\delta \F_{\b\ad}$, we first analyse the higher-order equations $0=\mathcal{D}^{\b\bd}\delta \F_{\b\bd}$ and $0=\big[\mathcal{D}^{\b},\bar{\mathcal{D}}^{\bd}\big]\delta \F_{\b\bd}$. Respectively,  they yield the following equations\footnote{The (non-unitary) masses $M^2=2\mu\mub$ and $M^2=0$ should be considered separately. The result of this analysis is that there is no gauge-symmetry present.}
\begin{subequations}\label{B.41}
	\begin{align}
	0&=\big(M^2-\frac{1}{4}\mu\mathcal{D}^2-\frac{1}{4}\mub\bar{\mathcal{D}}^2\big)(\s+\bar{\s})+\big(M^2-2\mu\mub\big)\xi~,\label{B.41a}\\
	0&=\big(M^2+\frac{3}{4}\mu\mathcal{D}^2+\frac{3}{4}\mub\bar{\mathcal{D}}^2\big)(\s-\bar{\s})-6\text{i}\big(M^2-2\mu\mub\big)\zeta~.\label{B.41b}
	\end{align}
\end{subequations}
One may show that both equations in \eqref{B.41} are consistent with \eqref{B.40} only if 
\begin{align}
\zeta=0~,\qquad \xi=0~,
\end{align}
which holds for all values of $M^2$. In addition, we must also have 
\begin{align}
\sigma = \bar{\s}=0~,
\end{align}
unless $M^2=M^2_{(2)}$, in which case $\s$ must satisfy the equation 
\begin{align}
0=-\frac{1}{4}(\mathcal{D}^2-4\mub )\s + 2\mub \bar{\s}~. \label{B.43}
\end{align}
Finally, using the above information to solve the equation $0=\mathcal{D}^{\b}\delta \F_{\b\ad}$, one arrives at the reality conditions (valid for all $M^2$)
\begin{align}
\mathcal{D}_{\ad}{}^{\b}\z_{\b}=2\text{i}\mub \bar{\z}_{\ad}~,\qquad \mathcal{D}_{\a}{}^{\bd}\bar{\z}_{\bd}=-2\text{i}\mu \z_{\a}~, \label{B.44}
\end{align}
which imply the mass-shell condition
\begin{align}
0=\big(M^2-5\mu\mub\big)\z_{\a}~.
\end{align}
We see that unless $\F_{\a\ad}$ has pseudo-mass  $M^2=M^2_{(1)}$, then
\begin{align}
\zeta_{\a}=0~.
\end{align}

In conclusion, the on-shell conditions \eqref{B.35} are only compatible with a gauge symmetry when the pseudo-mass of $\F_{\a\ad}$ (and the corresponding gauge parameters) takes one of the partially-massless values \eqref{B.355a} or \eqref{B.355b}. For super-depth $t=1$, the relevant gauge symmetry is
\begin{align}
\delta_\z \F_{\a\ad}=\bar{\mathcal{D}}_{\ad}\zeta_{\a}-\mathcal{D}_{\a}\bar{\z}_{\ad}~,
\end{align}
where $\z_{\a}$ satisfies \eqref{B.39} and \eqref{B.44}. On the other hand, for super-depth $t=2$ it is
\begin{align}
\delta_\s \F_{\a\ad}= \mathcal{D}_{\a\ad}(\s+\bar{\s})~,
\end{align}
where $\s$ is chiral and satisfies \eqref{B.43}.

By conducting a similar analysis on the real supermultiplet $\F_{\a(2)\ad(2)}$, one arrives at the gauge transformations presented in table \ref{table 3}. In principle this procedure may be carried out for any on-shell superfield $\F_{\a(m)\ad(n)}$ with fixed $m$ and $n$, 
but in practice this procedure is computationally demanding. In section \ref{section 3.2} we use the above analysis to motivate an ansatz for the case arbitrary half-integer superspin. 

\end{subappendices}

\chapter{Three-dimensional anti-de Sitter (super)space} \label{TAChapter3d(super)space}
In the previous chapter we studied (super)spin projection operators in $4d$ AdS (super)space for the first time. In addition to the plethora of applications that the (super)projectors possess in flat (super)space, new features started to emerge in AdS$_4$. For example, they provided a laboratory to formulate the AdS$_4$ on-shell supermultiplets which were previously unknown. In particular, we were able to find superfield realisations for partially massless representations of the AdS superalgebra, and elucidate their defining features. These results were deduced from the information encoded within the poles of the superprojectors. In some sense this observation is not so surprising since in Minkowski (super)space, the poles of the (super)spin projection operators contain all information concerning fields which furnish massless UIRs (and hence possess a gauge symmetry). 

For this reason, we anticipate that this pattern will be a universal feature which persists in arbitrary dimensions (at least in maximally symmetric backgrounds), with any amount of supersymmetry. The purpose of this chapter is to investigate whether this claim holds true in the case $d=3$.
In particular, we will derive the (super)spin projection operators in AdS space, and $\cN=1$ AdS superspace. Additionally, we also review massless and massive higher-spin field theories on these (super)spaces, which will be pertinent for the analysis in the subsequent chapter.

This chapter is based on the publications \cite{HutchingsHutomoKuzenko,HutchingsKuzenkoPonds2021} and is organised as follows. Section \ref{TAsec3dAdS} is 
primarily focused on the construction of spin projection operators in AdS$_3$. In subsection \ref{IrrepsAdS3} we outline salient facts concerning the irreducible representations of three-dimensional anti-de Sitter space $\mf{so}(2,2)$. We then review the on-shell fields which realise these irreducible representations in subsection \ref{IrrepFieldsAdS3}. In subsection \ref{TAProjectors} we derive the spin projection operators in AdS$_3$ and explore several of their applications. In particular, we show that all information about (partially-)massless fields is encoded within the poles of the spin projection operators. These projectors were formulated solely in terms of the quadratic Casimir operators of $\mf{so}(2,2)$. This allows us to recast the AdS$_3$ higher-spin Cotton tensors and their corresponding conformal actions into a manifestly gauge-invariant and factorised form. This was done in section \ref{secCT}.

In subsection \ref{TAMasslessActions} we review the formulation of massless higher-spin field theories in AdS$_3$, which were first derived in \cite{KuzenkoPonds2018}. We then explore topologically massive higher-spin theories in section \ref{TAMassiveactions}. In particular, the higher-spin Cotton tensors of section \ref{secCT} are used to 
find new realisations for new topologically massive spin-$s$ gauge models, which are of order $2s$ in derivatives, where $s$ is a positive (half-)integer. In the case when $s$ is an integer, it is possible to construct NTM models of order $2s-1$. In $\mb{M}^3$ such models were recently proposed in \cite{DalmaziSantos2021}, here we extend them to AdS$_3$.

The main objective of section \ref{TASsec3dAdS} is to formulate the superspin projection operators in AdS$^{3|2}$. 
This section is organised in a similar fashion to that of section \ref{TAsec3dAdS}, with many of the results in AdS$_3$ being extended to the case of $\cN = 1$ AdS$_3$ supersymmetry. In subsection \ref{AppB}, we analyse the component structure of the massless superspin theories of \cite{HK19,KuzenkoPonds2018}, which will be reviewed in section \ref{TASMasslessHStheories}. We provide a summary of the novel results obtained in section \ref{TASsecSoR}. This chapter is accompanied by the single technical appendix \ref{TAAppendixA}. In appendix \ref{TAAppendixA} we review the generating function formalism, which is a convenient framework used to derive the non-supersymmetric results of section \ref{TAsec3dAdS}.

\section{Three-dimensional anti-de Sitter space} \label{TAsec3dAdS}
In this section we will compute the spin projection operators in AdS$_3$. These projectors are AdS$_3$ extensions of the spin projection operators in $\mb{M}^3$, which were reviewed in section \ref{SecThreeDimensionalMinkowskiSpace}. Before introducing the spin projection operators,  it is necessary to review the classification of the irreducible representations of $\mf{so}(2,2) $.


\subsection{Irreducible representations of the AdS$_3$ algebra}\label{IrrepsAdS3}
The AdS$_3$ isometry algebra $\mf{so}(2,2) = \mathfrak{sl}(2,\mathbb{R}) \oplus \mathfrak{sl}(2,\mathbb{R})$ is given by eq. \eqref{FAAlgebra} for the case $d=3$. The lowest-energy irreducible representations of $\mf{so}(2,2)$, which we denote by $D(E_0,s)$, are labelled by the lowest value $E_0$ of the energy $E$ and by the helicity $s$ (see, e.g. \cite{DKSS,BHRST} and references therein).\footnote{The energy $E$ is chosen to be dimensionless; it can be restored by rescaling $E\rightarrow l^{-1} E$.} Note that we define spin by $|s|$.

The algebra $\mf{so}(2,2)$ possesses the two quadratic Casimir operators $C_1$ and $C_2$ \cite{Evans,Fronsdal:1974ew}
\vspace{-\baselineskip}
\bsubeq \label{TACasimirOperatorsOperators}
\begin{align}
C_1 &= \hf J^{AB}J_{AB}~,  & [C_1,J_{AB}] &= 0~, \\
C_2 &= - \frac{1}{4} \ve_{A B C D} J^{AB} J^{CD} ,  & [C_2,J_{AB}] &= 0~.
\end{align}
\esubeq
The parameters $E_0$ and $s$ are determined by the eigenvalues of these Casimir operators, which for a lowest-energy irreducible representation of $\mf{so}(2,2)$ are \cite{DKSS}
\bsubeq \label{TACasimirOperators}
\bea
C_1 &=& E_0 (E_0 -2) +|s|^2~, \label{TAQuadraticCasimirLowestWeight} \\
C_2 &=&-2s(E_0 - 1)~. \label{TAQuadraticCasimirLowestWeight2}
\eea
\esubeq
We are only interested in those representations carrying integer or half-integer spin with $|s|\geq 1$ and, consequently, the allowed values of helicity are 
$s= \pm 1,\pm \frac{3}{2},\pm 2,\cdots$. In order for the irreducible representation $D(E_0, s)$ to be unitary, the inequality $E_0\geq |s|$, known as the unitarity bound, must be satisfied.


\subsection{Irreducible field representations}\label{IrrepFieldsAdS3}
The geometry of AdS$_3$ is described by the Lorentz covariant derivative,
\bea
\mf{D}_{a}
=e_a{}^{m}\pa_m+\frac{1}{2}\omega_{a}{}^{bc}M_{bc}
=e_a{}^{m}\pa_m+\frac{1}{2}\omega_{a}{}^{\b\g}M_{\b\g}~,
\eea
where $e_a{}^{m}$ is the inverse vielbein and $\omega_{a}{}^{bc}$ is the Lorentz connection. The covariant derivative $\mf{D}_{a}$ 
satisfies the commutation relation
\be \label{ADSAlg}
[ \mf{D}_a, \mf{D}_b ] = -4 |\m|^2 M_{ab} \quad \Longleftrightarrow \quad 
\ [ \mf{D}_{\a \b}, \mf{D}_{\g \d} ] = 4 |\m|^2 \big(\ve_{\g(\a}M_{\b)\d} + \ve_{\d(\a} M_{\b)\g}\big)~,
\ee
where the parameter $|\m|$ is related to the scalar curvature $\mc{R}$ and AdS$_3$ radius $l$ via $\mc{R}=-24|\m|^2$ and $l^{-1} = 2|\m|$, respectively.

In our subsequent analysis, we will make extensive use of the quadratic Casimir operators \eqref{TACasimirOperators} of $\mf{so}(2,2)$ in the field representation, for which we choose (see, e.g. \cite{BPSS})
\vspace{-\baselineskip}
\bsubeq \label{QC}
\begin{alignat}{2} 
\mathcal{F}:&=\mf{D}^{\a\b}M_{\a\b}~,  &[\mathcal{F},\mf{D}_{\a\b}]=0~, \label{QCF}\\
\mathcal{Q}:&= \Box -2|\m|^2M^{\a\b}M_{\a\b}~,  \qquad &[\mathcal{Q},\mf{D}_{\a\b}]=0~. \label{QCQ} 
\end{alignat}
\esubeq
Here $\Box:= - \hf \mf{D}^{\a\b}\mf{D}_{\a\b}$ is the d'Alembert operator 
in AdS$_3$. Note  that we denote the Casimir operators \eqref{TACasimirOperatorsOperators} by $\mc{Q} \equiv  l^{-2}C_1 $ and $\cF \equiv l^{-1}C_2$ in the field representation, where dimensions have been restored.

Let us denote by $\cV_{(n)}$ the space of totally symmetric rank-$n$ spinor fields $\phi_{\a(n)}:=\phi_{\a_1\dots\a_n}=\phi_{(\a_1\dots\a_n)}$ on AdS$_3$. It can be shown that the Casimir operators $\cF$ and $\cQ$ are related to each other on $\cV_{(n)}$ as follows
\bea \label{FSQ}
\mathcal{F}^2 \f_{\a(n)} = n^2 \big [ \mathcal{Q} - (n-2)(n+2)|\m|^2 \big ]\f_{\a(n)} + n(n-1)\mf{D}_{\a(2)}\mf{D}^{\b(2)} \f_{\b(2)\a(n-2)}~.
\eea
The structure $\mf{D}_{\a(2)}\mf{D}^{\b(2)} \f_{\b(2)\a(n-2)}$ in \eqref{FSQ}
is not defined for the cases $n=0$ and $n=1$. However, it is accompanied by a factor of $n(n-1)$ which vanishes for these particular cases.

For $n \geq 2$, a tensor field $\f_{\a(n)}$ on $\cV_{(n)}$ is said to be on-shell if it satisfies the conditions
\vspace{-\baselineskip}
\bsubeq \label{TAOnShellConditions}
\bea
\mf{D}^{\b(2)}\f_{\b(2)\a(n-2)}&=&0~, \label{OT} \\
(\cF - \s \r) \f_{\a(n)}&=& 0 ~, \label{TAFirstOrderMassConstraint}
\eea
\esubeq
where $\s:=\pm 1$ and $\r \geq 0$ is a real parameter of unit mass dimension. 
The first-order equation  \eqref{TAFirstOrderMassConstraint} is equivalent  to 
\be \label{TAFOMassConsA}
\mf{D}_{(\a_1}{}^{\b}\f_{\a_2 ... \a_n)\b} = \sigma\frac{\r}{n} \f_{\a(n)}~.
\ee
Any field $\phi_{\a(n)}$ satisfying both constraints \eqref{OT} and \eqref{TAFirstOrderMassConstraint}, is an eigenvector of the Casimir operator $\mc{Q}$
\begin{align}
\big(\mc{Q}-m^2\big)\phi_{\a(n)}=0~,\qquad m^2:= \Big (\frac{\r}{n} \Big )^2+(n-2)(n+2)|\m|^2~. \label{TASecondOrderMass}
\end{align}
Since the Casimir operators $\cF$ and $\cQ$ are multiples of the unit operator, \eqref{TAFirstOrderMassConstraint} and \eqref{TASecondOrderMass}, it follows that an on-shell field \eqref{TAOnShellConditions} furnishes the irreducible representation $D(E_0,s)\equiv D(E_0,\s |s|)$.  In accordance with this, we say that an on-shell field carries pseudo-mass $\r$, spin $|s| = \frac{n}{2}$ and helicity $s =  \frac{\s n}{2} $. The equations \eqref{OT} and \eqref{TAFOMassConsA} were first introduced in \cite{BHRST}. 

Comparing the eigenvalue \eqref{TASecondOrderMass} of $\mc{Q}$ with the eigenvalue \eqref{TAQuadraticCasimirLowestWeight} of  $l^{-2} C_1$  establishes a relationship between the pseudo-mass and the minimal energy,
\be \label{TAEnergyMassrelation}
\r^2 = 4 n^2 |\m|^2  \big [ E_0 \big ( E_0 -2 \big ) +1 \big ]~.
\ee
In terms of $\r$ and $n$, the unitarity bound $E_0\geq |s|$ reads 
\be \label{TAUnitaritybound}
\rho\geq n(n-2)|\m|~.
\ee
As a caveat, we see from \eqref{TAEnergyMassrelation} that there are two distinct values of $E_0$ leading to the same value of pseudo-mass $\r$
\begin{align} \label{TAEnergyMassRelation}
\big (E_0 \big )_{\pm} :=1 \pm  \frac{ \r}{2n|\m|}~.
\end{align}
Note that this relationship also follows from the comparison of the eigenvalue \eqref{TAFirstOrderMassConstraint} of $\cF$ with the eigenvalue \eqref{TAQuadraticCasimirLowestWeight2} of $l^{-1}C_2$.
Similar to the case in AdS$_4$ (see section \ref{FAIrredFieldReps}), a unique energy branch in \eqref{TAEnergyMassRelation} is chosen by requiring that the representation furnished by an on-shell field to be unitary. It follows from \eqref{TAEnergyMassRelation} that the solution $(E_0)_{-}$ violates the unitarity bound  $E_0\geq |s|$. Thus, when referring to a unitary representation with pseudo mass $\r$, we will always be referring to a representation with energy $(E_0)_+$ 
\begin{align} \label{TAEnergyMassRelationPositive}
\big (E_0 \big )_{+}=1 + \frac{\rho}{2n|\m|}~.
\end{align}
With this in mind, we will label the UIRs of $\mf{so}(2,2)$ using $\rho$ in place of $(E_0)_+$, and use the notation $ D(\rho,\s \frac{n}{2})$.

In place of \eqref{OT} and \eqref{TAFirstOrderMassConstraint}, one may instead consider tensor fields $\phi_{\a(n)}$ constrained by the equations \eqref{OT} and \eqref{TASecondOrderMass},
\bsubeq \label{OC2}
\bea
\mf{D}^{\b(2)}\f_{\b(2)\a(n-2)}&=&0~, \label{OT2} \\
\big(\mc{Q}-m^2\big)\phi_{\a(n)}&=&0~. \label{OMS2}
\eea
\esubeq
In this case, the equation \eqref{FSQ} becomes
\begin{align}
\big(\mc{F}-\rho\big)\big(\mc{F}+\rho\big)\phi_{\a(n)}=0~.
\end{align}
It follows that such a $\phi_{\a(n)}$ furnishes the reducible representation 
\be \label{red}
{ D}\Big (\rho,-\frac{n}{2} \Big )\oplus { D} \Big (\rho,\frac{n}{2} \Big )~.
\ee
In particular, representations with both signs of helicity $\pm\frac{n}{2}$ appear in this decomposition. 

Below we review the different types of on-shell fields in AdS$_3$, which are characterised by the pseudo-mass they carry.

\subsubsection{Massless fields}\label{TAmasslesssec}
For $n \geq 1$, we say that a field $\f_{\a(n)}$ on $\cV_{(n)}$ is massless with spin $|s|:=\frac{n}{2}$ and helicity\footnote{Unlike in $\mb{M}^3$, massless fields  in AdS$_3$ can be interpreted to carry helicity due to the presence of $|\m|$.} $\frac{\s n}{2}$ if it satisfies the on-shell conditions \eqref{TAOnShellConditions} with pseudo-mass
\bsubeq \label{TAMasslessOnShell}
\be 
\r = \r_{(1,n)}~, \qquad \r_{(1,n)} = n(n-2)|\m| ~. \label{TAMasslessOnShellPM}
\ee
It may be shown that the massless system of equations \eqref{TAOnShellConditions}, with pseudo-mass \eqref{TAMasslessOnShellPM}, is compatible with the gauge symmetry
\be
\d_{\z} \phi_{\a(n)}=\mf{D}_{\a(2)}\z_{\a(n-2)}~, \label{TAMasslessGaugeSymmetry}
\ee
given that the real gauge parameter $\z_{\a(n-2)}$ is also on-shell
\bea
\mf{D}^{\b(2)} \z_{\b(2)\a(n-4)} = 0~,  \label{TATrans}\\
\big ( \cF - \s \r_{(1,n)} \big ) \z_{\a(n-2)} =0~.
\eea
\esubeq
The minimal energy \eqref{TAEnergyMassRelationPositive} associated with a massless field is $E_0=|s|$. The on-shell conditions \eqref{OT} and \eqref{TAMasslessOnShell} can be identified as the on-shell conditions associated with the Fang-Fronsdal action \eqref{TAFangFronsdal} for the case $n=2s+1$.\footnote{The on-shell conditions \eqref{TAMasslessOnShell} are identical to those given by the Chern-Simons type action \eqref{CSA} for the case $n=2$.}

Analogous to the case in $\mb{M}^3$ (see section \ref{TMMasslessFieldsRepsSec}), the notion of higher-spin in AdS$_3$ is purely kinematical, and is well defined only in the massive case. For $n >3$, a field $\f_{\a(n)}$ on $\cV_{(n)}$ is said to be a massless higher-spin field if it does not propagate any physical degrees of freedom on-shell. Let us demonstrate that the massless on-shell field proposed above is consistent with this definition.

Counting the degrees of freedom, a field $\f_{\a(n)}$ on $\cV_{(n)}$ carries $n+1$ off-shell degrees of freedom. The transverse constraint \eqref{OT} removes $n-1$ degrees of freedom, leaving $2$ remaining. In order to remove these, we need to make use of the gauge symmetry. Specifically, the gauge parameter $\z_{\a(n-2)}$ is transverse, thus carrying $n-1-(n-3) =2$ gauge degrees of freedom. Hence, we can eliminate the final two physical degrees freedom by consuming the gauge symmetry, thus leaving zero propagating degrees of freedom on-shell.

For the massless field $\f_{\a(n)}$ \eqref{TAMasslessOnShell}, the second order equation \eqref{TASecondOrderMass} takes the form
\be \label{TASSecondOrderMassless}
\big ( \cQ -  \t_{(1,n)}|\m|^2 \big )\f_{\a(n)}=0~,\qquad \t_{(1,n)}= 2n(n-2) ~.
\ee
In place of the conditions \eqref{OC2}, one may say that a field $\f_{\a(n)}$ on $\cV_{(n)}$ is massless if it
satisfies the alternative constraints \eqref{TATrans} and \eqref{TASSecondOrderMassless}
\bsubeq \label{TAMasslessConditions2}
\bea
\mf{D}^{\b(2)}\f_{\b(2)\a(n-2)}&=&0~, \\
\big ( \cQ -  \t_{(1,n)}|\m|^2 \big )\f_{\a(n)}&=&0~.
\eea
It can be shown that the system of equations \eqref{TAMasslessConditions2} is compatible with the gauge symmetry
\be
\d_{\z} \phi_{\a(n)}=\mf{D}_{\a(2)}\z_{\a(n-2)}~, 
\ee
given that the real gauge parameter $\z_{\a(n-2)}$ is itself on-shell
\bea
\mf{D}^{\b(2)}\z_{\b(2)\a(n-4)}&=&0~, \\
\big ( \cQ -  \t_{(1,n)}|\m|^2 \big )\z_{\a(n-2)}&=&0~.
\eea
\esubeq 
By the same counting discussion given above, it follows that a field obeying the conditions \eqref{TAMasslessConditions2} does not propagate any physical degrees of freedom on-shell, and hence is massless. We will show in section \ref{TAFronsSec} that for $n=2s$, the equations \eqref{TAMasslessConditions2} appear in the on-shell analysis of the Fronsdal model \eqref{TAFronsdal} in AdS$_3$.\footnote{For $n=2$, the on-shell conditions \eqref{TAMasslessConditions2} are identical to those given by the Maxwell action in AdS$_3$.}

It is easier to show that massless on-shell fields do not propagate any physical degrees of freedom if one instead works with their corresponding gauge-invariant field strengths. These field strengths\footnote{The field strengths \eqref{TACott} are the linearised higher-spin Cotton tensors. See section \ref{secCT} for details on their properties. }  were recently derived in \cite{KP21}, where they were shown to take the explicit closed form 
\begin{subequations}\label{TACott}
	\begin{align}
	\mathfrak{C}_{\a(2s)}(\f)&=\frac{1}{2^{2s-1}}\sum_{j=0}^{s-1}2^{2j+1}\binom{s+j}{2j+1}\prod_{t=1}^{j}\Big(\cQ-\tau_{(s-t,2s)}|\m|^2\Big) \non\\
	&\phantom{\frac{1}{2^{2s-1}}\sum_{j=0}^{s-1}2^{2j+1}\binom{s+j}{2j+1}}\times
	\mf{D}_{\a(2)}^{s-j-1}\mf{D}_{\a}{}^{\b}\big(\mf{D}^{\b(2)}\big)^{s-j-1}\f_{\a(2j+1)\b(2s-2j-1)}~, \\
	\mathfrak{C}_{\a(2s+1)}(\f)&=\frac{1}{2^{2s}}\sum_{j=0}^{s}2^{2j}\binom{s+j}{2j}\frac{(2s+1)}{(2j+1)}\prod_{t=1}^{j}\Big(\cQ-\tau_{(s-t+1,2s+1)}|\m|^2\Big) ~~~~~~~~~~~~~~~~~~~~~~~~~~~\non\\
	&\phantom{\frac{1}{2^{2s}}\sum_{j=0}^{s}\binom{s+j}{2j}\frac{(2s+1)}{(2j+1)}}\times \mf{D}_{\a(2)}^{s-j}\big(\mf{D}^{\b(2)}\big)^{s-j}\f_{\a(2j+1)\b(2s-2j)}~.
	\end{align}
\end{subequations}
As will be treated in section \ref{secCT}, it follows from the gauge completeness property that the field $\f_{\a(n)}$ is pure gauge if the field strength $\mf{C}_{\a(n)}(\f)$ vanishes.
It is immediately apparent that for $n \geq 3$, the field strength $\mf{C}_{\a(n)}(\f)$ \eqref{TACott} is always vanishing when $\f_{\a(n)}$ satisfies either set of massless conditions \eqref{TAMasslessOnShell} or \eqref{TAMasslessConditions2}. Hence for $n \geq 3$, the on-shell field $\f_{\a(n)}$ carries no propagating degrees of freedom. 

Special care needs to be taken when considering $n=2$ case. Here, the field strength is given by
\be \label{TASRank2CT}
\mf{C}_{\a(2)}(\f) = \hf \cF \f_{\a(2)}~.
\ee
If the field $\f_{\a(2)}$ satisfies the first type of on-shell constraints \eqref{TAMasslessOnShell}, then it immediately follows that the field strength vanishes $\mf{C}_{\a(2)}(\f)=0$. In accordance with gauge completeness, the field $\f_{\a(n)}$ is pure gauge.

On the other hand,  if the field $\f_{\a(2)}$ satisfies the second type of on-shell constraints \eqref{TAMasslessConditions2}, then it is apparent that the field strength \eqref{TASRank2CT} is non-vanishing. Moreover, it can be shown that $\mf{C}_{\a(2)}(\f)$ obeys the constraints
\bsubeq\label{TACT2Prop}
\bea 
\mf{D}^{\b(2)} \mf{C}_{\b(2)}(\f) &=& 0~, \label{TACT2Prop1} \\
\cF \mf{C}_{\a(2)}(\f) &=& 0~. \label{TACT2Prop2}
\eea
\esubeq
Here, $\mf{C}_{\a(2)}(\f)$ is equivalent to the dual of the Maxwell field strength $G_a$. Condition \eqref{TACT2Prop1} is the off-shell Bianchi identity on $G_a$, i.e. $\mf{D}^aG_a =0$ whilst condition \eqref{TACT2Prop2} is the equation of motion for Maxwell's theory, i.e. $\mf{D}_{[a}G_{b]} = 0$.
The constraints \eqref{TACT2Prop} are analogous to those observed in $\mb{M}^3$ (cf. eq. \eqref{FMN2FieldStrength}), thus  $\mf{C}_{\a(2)}(\f)$ carries a single physical degree of freedom.

\subsubsection{Partially massless fields}
An on-shell field $\f_{\a(n)}^{(t)}$ \eqref{TAOnShellConditions} is said to be partially massless with depth-$t$ if it carries the fixed pseudo-mass 
\be
\r \equiv \r_{(t,n)}=  n(n-2t) |\m| ~, \qquad 2 \leq t \leq \lfloor n/2 \rfloor~.
\label{PMvals}
\ee
It may be shown that when the pseudo-mass assumes the value \eqref{PMvals}, then the corresponding representation ${D}(\rho,\s\frac{n}{2})$, with either sign for $\s$, shortens. At the field-theoretic level, this is manifested by the appearance of a depth-$t$ gauge symmetry
\begin{align}
\delta_{\z} \phi^{(t)}_{\a(n)}=\big(\mf{D}_{\a(2)}\big)^t\z_{\a(n-2t)}~, \label{PMGTs}
\end{align}
under which the system of equations \eqref{TAOnShellConditions}, with $\rho$ given by \eqref{PMvals} and $\s$ arbitrary, is invariant.\footnote{This is true when the gauge parameter satisfies conditions analogous to \eqref{TAOnShellConditions}, see \cite{KP21} for the details. } 
For the field $\phi^{(t)}_{\a(n)}$ the second order equation \eqref{TASecondOrderMass} takes the form
\be
\big ( \cQ - 
\t_{(t,n)}\mc{S}^2
\big )\f^{(t)}_{\a(n)}=0~,\qquad 
\t_{(t,n)}
= \big[2n(n-2t) +4(t-1)(t+1)\big]
~,\label{PMV}
\ee
where the parameters $\t_{(t,n)}$ are known as partially massless values. 
For $t>1$, the pseudo-mass  $\r_{(t,n)}$, eq. \eqref{PMvals} violates the unitarity bound \eqref{TAUnitaritybound} and hence the partially massless representations are non-unitary.


\subsubsection{Massive fields}
An on-shell field \eqref{TAOnShellConditions} is said to be massive if it carries the pseudo-mass which satisfies
\be \label{TAMassivePM}
\r > \r_{(1,n)} ~.
\ee
In other words, a massive field carries a pseudo-mass which is greater than that of a massless field. The pseudo-mass \eqref{TAMassivePM} does not violate the unitarity bound \eqref{TAUnitaritybound} and hence defines a unitary representation of $\mf{so}(2,2)$.

Note that there also exists another class of massive fields that realise non-unitary representations of $\mf{so}(2,2)$. These fields satisfy the on-shell conditions \eqref{TAOnShellConditions} and carry the pseudo-mass $\r < \r_{(1,n)}$, where $\r \neq \r_{(t,n)}$ for $1 \leq t \leq \lfloor n/2 \rfloor $. We will not be interested in such (non-unitary) massive fields in this chapter.


\subsection{Spin projection operators} \label{TAProjectors}
For $n \geq 2$,  the rank-$n$ spin projection operator $\P^{\perp}_{[n]}$ is defined by its action on $\cV_{(n)}$ according to the rule:
\bsubeq\label{Proj}
\bea 
\P^{\perp}_{[n]}: \cV_{(n)} &\longrightarrow& \cV_{(n)}~, \\
\f_{\a(n)} &\longmapsto&  \P^{\perp}_{[n]} \phi_{\a(n)}~  =:\P^{\perp}_{\a(n)}(\f)~.
\eea
\esubeq
For fixed $n$, the operator $\P^{\perp}_{[n]}$ is defined by the following properties:
\begin{enumerate}
	\item \textbf{Idempotence:} The operator $\P^{\perp}_{[n]}$ squares to itself, 
	\bsubeq \label{SPO}
	\be \label{ProjProp}
	\P^{\perp}_{[n]}\P^{\perp}_{[n]}=\P^{\perp}_{[n]}~.
	\ee
	\item \textbf{Transversality:} The operator $\P^{\perp}_{[n]}$ maps $\phi_{\a(n)}$ to a transverse field,
	\be \label{ProjTrans}
	\mf{D}^{\b(2)}\P^{\perp}_{\b(2)\a(n-2)}(\f) =0~.
	\ee
	\item \textbf{Surjectivity:} 
	Every transverse field $\f^{\perp}_{\a(n)}$ belongs to the image of $\P^{\perp}_{[n]}$, 
	\be \label{ProjU}
	\mf{D}^{\b(2)}\f^{\perp}_{\b(2)\a(n-2)}=0~\quad\implies\quad\P^{\perp}_{[n]} \f^{\perp}_{\a(n)} = \f^{\perp}_{\a(n)}~.
	\ee
	\esubeq
	In other words, $\P^{\perp}_{[n]}$ acts as the identity operator on the space of transverse fields.
\end{enumerate}

Let us consider a field $\f_{\a(n)}$ on $\cV_{(n)}$ which obeys the first-order differential constraint \eqref{TAFirstOrderMassConstraint}. It follows from the properties \eqref{SPO} that the spin projection operator $\P^{\perp}_{[n]}$ maps such a $\f_{\a(n)}$ to an on-shell field 
\bsubeq
\bea
\mf{D}^{\b(2)}\P^{\perp}_{\b(2)\a(n-2)}(\f) &=& 0~, \\
(\cF - \s \r) \P^{\perp}_{\a(n)}(\f)&=&0~.
\eea
\esubeq
Hence, $\P^{\perp}_{[n]}$ selects out the physical component of $\f_{\a(n)}$ which realises the irreducible representation $D(\rho,\s \frac{n}{2})$ of $\mf{so}(2,2)$.

Below we derive the spin projection operators $\P^{\perp}_{[n]}$.
For this purpose it is convenient to make use of the generating function formalism, which is described in appendix \ref{TAAppendixA}.
In this framework, the properties \eqref{ProjProp} and \eqref{ProjTrans} take the following form:
\be \label{PPGFF}
\P^{\perp}_{[n]}\P^{\perp}_{[n]}\f_{(n)} = \P^{\perp}_{[n]}\f_{(n)}~,\qquad \mf{D}_{(-2)} \P^{\perp}_{[n]} {\f}_{(n)}=0~. \qquad 
\ee
It is necessary to separately analyse the cases with $n$ even and $n$ odd. 

\subsubsection{Bosonic spin projection operator}\label{TASBosSPSec}

We will begin by studying the bosonic case, $n=2s$, for integer $s \geq 1$. Let us introduce the differential operator $\mathbb{T}_{[2s]}$
of order $2s$ in derivatives\footnote{When the upper bound in a product is less than the lower bound, we define the result to be unity. }
\be \label{BTO}
\mathbb{T}_{[2s]}=\sum_{j=0}^{s} 2^{2j}s\frac{(s+j-1)!}{(s-j)!}\prod_{t=1}^{j}\big (\mathcal{Q} - \t_{(s-t+1,2s)} |\m|^2 \big ) \mf{D}_{(2)}^{s-j}\mf{D}_{(-2)}^{s-j}~.
\ee
Here $\t_{(t,n)} $ denotes the partially massless values \eqref{PMV}. Given an arbitrary field $\f_{(2s)} \in \mc{V}_{(2s)}$, one may use the identity \eqref{ID15} to show that $\mathbb{T}_{[2s]}$ maps $\f_{(2s)} $ to a transverse field
\be \label{TATOpTrans}
\mf{D}_{(-2)} \mathbb{T}_{[2s]} {\f}_{(2s)}=0~.
\ee
However, it is not a projector on $\mc{V}_{(2s)}$ since it does not square to itself,
\be
\mathbb{T}_{[2s]} \mathbb{T}_{[2s]} {\f}_{(2s)} = 2^{2s-1}(2s)! \prod_{t=1}^{s} \big (\mathcal{Q} - \t_{(t,2s)} |\m|^2 \big )\mathbb{T}_{[2s]} {\f}_{(2s)} ~.\label{Tsquared}
\ee  
To prove this identity, we observe that only the $j=s$ term of the sum  in \eqref{BTO} survives when $\mathbb{T}_{[2s]}$ acts on a transverse field such as $\mathbb{T}_{[2s]} {\f}_{(2s)} $. 

To obtain a projector, we define the following dimensionless operator
\be \label{BP}
\widehat{\Pi}^{\perp}_{[2s]} := \Big [2^{2s-1}(2s)! \prod_{t=1}^{s} \big (\mathcal{Q} - \t_{(t,2s)} |\m|^2 \big )  \Big ]^{-1}\mathbb{T}_{[2s]} ~.
\ee
The operator $\widehat{\Pi}^{\perp}_{[2s]}$ is transverse and  idempotent on $\mc{V}_{(2s)}$, by virtue of the results \eqref{TATOpTrans} and \eqref{Tsquared}, respectively. In a fashion similar to the proof of \eqref{Tsquared}, it may also be shown that $\widehat{\Pi}^{\perp}_{[2s]}$ acts as the identity operator on the space of rank-$(2s)$ transverse fields.  Thus, $\widehat{\Pi}^{\perp}_{[2s]}$ satisfies the properties \eqref{SPO} and is hence the spin projection operator on $\mc{V}_{(2s)}$.
Making the indices explicit, the latter reads
\bea \label{BosSpin}
\widehat{\Pi}^{\perp}_{[2s]}\f_{\a(2s)}  &=&\Big [ \prod_{t=1}^{s} \big (\mathcal{Q} - \t_{(t,2s)} |\m|^2 \big ) \Big ]^{-1} \sum_{j=0}^{s} 2^{2j-2s}\frac{2s}{s+j}\binom{s+j}{2j}\non \\
&&\times \prod_{t=1}^{j}\big (\mathcal{Q} - \t_{(s-t+1,2s)} |\m|^2 \big ) \mf{D}_{\a(2)}^{s-j}\big (\mf{D}^{\b(2)}\big )^{s-j}\f_{\a(2j)\b(2s-2j)}~. 
\eea

It is possible to construct a spin projection operator solely in terms of the two quadratic Casimir operators \eqref{QC}. To this end, we introduce the operator
\begin{align} 
\P^{\perp}_{[2s]}= \frac{1}{2^{2s-1}(2s)!} \prod_{j=1}^{s}\frac{ \big ( \cF^2 -4(j-1)^2  (\cQ-4j(j-2)|\m|^2  ) \big )}{  (\mathcal{Q} - \t_{(j,2s)} |\m|^2  ) }   
~. \label{SimpBosProj}
\end{align}
Let us show that \eqref{SimpBosProj} satisfies the three defining properties \eqref{SPO} on $\mc{V}_{(2s)}$. Given an arbitrary transverse field $\f^{\perp}_{(2s)}$, $\mf{D}_{(-2)}\f^{\perp}_{(2s)}=0$, one may show that 
\bea
&&\prod_{j=1}^{s}\big ( \cF^2 - 4  (j-1)^2  ( \cQ - 4j(j-2)|\m|^2  ) \big )  \f^{\perp}_{(2s)} \non\\
&&= 2^{2s-1}(2s)!\prod_{j=1}^{s}\big ( \cQ -\t_{(j,2s)} |\m|^2 \big) \f^{\perp}_{(2s)} ~.\label{BosonicID1}
\eea
Note that the identity \eqref{FSQ} is useful in proving \eqref{BosonicID1}.
It follows that ${\P}^{\perp}_{[2s]}$ acts as the identity operator on the space of transverse fields,
\begin{align}
\mf{D}_{(-2)}\f^{\perp}_{(2s)}=0\quad \implies \quad \P^{\perp}_{[2s]}\f^{\perp}_{(2s)}=\f^{\perp}_{(2s)}~. \label{TTCPbos}
\end{align}
Next, the image of any unconstrained field $\f_{(2s)}$ under $\P^{\perp}_{[2s]}$ is transverse, which follows elegantly from \eqref{PTI}
\be
\mf{D}_{(-2)} \P^{\perp}_{[2s]} \f_{(2s)} = \P^{\perp}_{[2s]} \mf{D}_{(-2)} \f_{(2s)} \propto \mf{D}_{(2)}^{s}\mf{D}_{(-2)}^{s+1}\f_{(2s)}=0 ~. \label{TransBP}
\ee
Finally, using \eqref{TTCPbos} and \eqref{TransBP} one can 
show that $\P^{\perp}_{[2s]}$ squares to itself 
\begin{align}
\P^{\perp}_{[2s]}\P^{\perp}_{[2s]} \f_{(2s)} 
=\P^{\perp}_{[2s]} \f_{(2s)}~.  
\end{align}
Thus $\P^{\perp}_{[2s]}$ satisfies \eqref{ProjProp}, \eqref{ProjTrans} and \eqref{ProjU} and can also be identified as a spin projection operator. 
It follows from \eqref{BosSpin} that the poles of  ${\P}^{\perp}_{[2s]} $ correspond to the partially massless values $\t_{(j,2s)}$ defined by \eqref{PMV}.

Although it is not immediately apparent, the two projectors $\widehat{\P}^{\perp}_{[2s]}$ and $\P^{\perp}_{[2s]}$ actually coincide on $\mc{V}_{(2s)}$. The proof for this is analogous to the one used to show that the $\mb{M}^4$ spin projectors, \eqref{FMBFprojectors} and \eqref{FMBFProjectorCasimir}, are equivalent \eqref{FMBFEquivalence} (see section \ref{Spin-projection operators} for details).

So far our analysis of the spin projection operators $\widehat{\Pi}_{[2s]}^{\perp}$ and $\Pi_{[2s]}^{\perp}$ has been restricted to the linear space $\mc{V}_{(2s)}$. However, for fixed $s$,
the operator $\Pi_{[2s]}^{\perp}$ given by  eq. \eqref{SimpBosProj}  is also defined 
to act on the linear spaces $\mc{V}_{(2s')}$ with $s' < s$. 
In fact, making use of \eqref{FSQ} and \eqref{PTI}, it is possible to show that the following holds true 
\be \label{BosProjProp}
\P^{\perp}_{[2s]} \f_{(2s')}=0~, \qquad 1 \leq s' \leq s-1 ~.
\ee 
This important identity states that $\Pi^{\perp}_{[2s]}$ annihilates any lower-rank field $\f_{\a(2s')}\in \mc{V}_{(2s')}$. It should be mentioned that $\P^{\perp}_{[2s]} $ does not annihilate lower-rank fermionic fields
$\f_{\a(2s'+1)}$.
When acting on $\cV_{(2s')}$, the two operators $\widehat{\Pi}_{[2s]}^{\perp}$ and $\Pi_{[2s]}^{\perp}$ are no longer equal to each other, and in particular $\widehat{\Pi}_{[2s]}^{\perp}\phi_{(2s')}\neq 0 $. It is for this reason that we will continue to use different notation for the two operators.

\subsubsection{Fermionic spin projection operator}

We now turn our attention to the fermionic case, $n=2s+1$,  for integers $s\geq 1$. Let us introduce the differential operator $\mathbb{T}_{[2s+1]}$ of order $2s$ in derivatives
\be \label{FTO}
\mathbb{T}_{[2s+1]}=\sum_{j=0}^{s} 2^{2j}\frac{(s+j)!}{(s-j)!}\prod_{t=1}^{j}\big (\mathcal{Q} - \t_{(s-t+1,2s+1)} |\m|^2 \big ) \mf{D}_{(2)}^{s-j}\mf{D}_{(-2)}^{s-j}~.
\ee
Here $\t_{(t,n)}$ are the partially massless values \eqref{PMV}. The operator $\mathbb{T}_{[2s+1]}$ maps $\f_{(2s+1)}$ to a transverse field
\be
\mf{D}_{(-2)} \mathbb{T}_{[2s+1]} {\f}_{(2s+1)}=0~.
\ee
However, this operator does not square to itself on $\cV_{(2s+1)}$
\be \label{TFsquared}
\mathbb{T}_{[2s+1]}\mathbb{T}_{[2s+1]}\f_{(2s+1)} = 2^{2s}(2s)! \prod_{t=1}^{s} \big (\mathcal{Q} - \t_{(t,2s+1)} |\m|^2 \big ) \mathbb{T}_{[2s+1]}\f_{(2s+1)} ~.
\ee
As a result, one can immediately define the dimensionless operator
\be \label{FP}
\widehat{\P}^{\perp}_{[2s+1]}  := \Big [2^{2s}(2s)! \prod_{t=1}^{s} \big (\mathcal{Q} - \t_{(t,2s+1)} |\m|^2 \big ) \Big ]^{-1} ~\mathbb{T}_{[2s+1]} ~,
\ee
which is a transverse projector by construction. Following a derivation similar to that of \eqref{TFsquared}, it can be shown that  $\widehat{\P}^{\perp}_{[2s+1]}$ acts like the identity operator on the space of transverse fields. Hence, the operator $\widehat{\P}^{\perp}_{[2s+1]}$ satisfies properties \eqref{SPO}, and is thus a spin projection operator on $\cV_{(2s+1)}$. Converting \eqref{FP} to spinor notation yields
\bea \label{FermSpin}
\widehat{\P}^{\perp}_{[2s+1]} \f_{\a(2s+1)}&=&\Big [\prod_{t=1}^{s} \big (\mathcal{Q} - \t_{(t,2s+1)} |\m|^2 \big ) \Big ]^{-1} \sum_{j=0}^{s} 2^{2j-2s}\frac{2s+1}{2j+1}\binom{s+j}{2j}~ \non\\
&&\times \prod_{t=1}^{j}\big (\mathcal{Q} - \t_{(s-t+1,2s+1)} |\m|^2 \big )   \mf{D}_{\a(2)}^{s-j}\big (\mf{D}^{\b(2)}\big )^{s-j}\f_{\a(2j+1)\b(2s-2j)}~.
\eea

As in the bosonic case, one can construct a fermionic projector purely in terms of the quadratic Casimir operators \eqref{QC}. Let us introduce the operator 
\bea 
{\P}^{\perp}_{[2s+1]} 
= \frac{1}{2^{2s}(2s)!}  \prod_{j=1}^{s} \frac{\big ( \cF^2 -(2j-1)^2  (\cQ-(2j-3)(2j+1)|\m|^2  ) \big )}{ (\mathcal{Q} - \t_{(j,2s+1)} |\m|^2  ) }
~. \label{SimpFermProj}
\eea
We wish to show that \eqref{SimpFermProj} indeed satisfies the properties \eqref{SPO} on $\cV_{(2s+1)}$. Given an arbitrary transverse field $\f^{\perp}_{(2s+1)} $, 
one can make use of \eqref{FSQ} to derive the identity
\bea \label{RFP}
&&\prod_{j=1}^{s}\big ( \cF^2 -  (2j-1  )^2  ( \cQ - (2j-3)(2j+1)|\m|^2  )  \big ) \f^{\perp}_{(2s+1)} \\
&&= 2^{2s}(2s)!\prod_{j=1}^{s}\big ( \cQ -\t_{(j,2s+1)} |\m|^2 \big )  \f^{\perp}_{(2s+1)} ~. \non 
\eea
It follows that ${\P}^{\perp}_{[2s+1]}$ acts like the identity operator on the space of transverse fields
\be \label{FPI}
\mf{D}_{(-2)}\f^{\perp}_{(2s+1)} = 0 \quad \Longrightarrow \quad   {\P}^{\perp}_{[2s+1]}\f^{\perp}_{(2s+1)}  = \f^{\perp}_{(2s+1)} ~.
\ee
By making use of \eqref{PTI}, one can show that the operator $ {\P}^{\perp}_{[2s+1]}$ maps $\f_{(2s+1)}$ to a transverse field
\be
\mf{D}_{(-2)} {\P}^{\perp}_{[2s+1]} \f_{(2s+1)} = {\P}^{\perp}_{[2s+1]} \mf{D}_{(-2)} \f_{(2s+1)} \propto \mf{D}_{(2)}^{s}\mf{D}_{(-2)}^{s+1}\f_{(2s+1)}=0 ~.\label{TransFP}
\ee
Finally, using \eqref{FPI} in conjunction with \eqref{TransFP}, one can show that $ {\P}^{\perp}_{[2s+1]}$  is idempotent
\begin{subequations}
	\bea
	{\P}^{\perp}_{[2s+1]}{\P}^{\perp}_{[2s+1]} \f_{(2s+1)} ={\P}^{\perp}_{[2s+1]} \f_{(2s+1)} ~.
	\eea
\end{subequations}
Hence, $ {\P}^{\perp}_{[2s+1]}$ satisfies \eqref{SPO}, and can thus be classified as a spin projection operator on AdS$_3$. It follows from \eqref{SimpFermProj} that the poles of  $\P^{\perp}_{[2s+1]} $ correspond to the partially massless values $\t_{(j,2s+1)}$ defined by \eqref{PMV}.

It can be shown that $\widehat{\P}^{\perp}_{[2s+1]}$ and ${\P}^{\perp}_{[2s+1]}$ are equivalent on $\cV_{(2s+1)}$,
\be  \label{EFP}
\widehat{\Pi}^{\perp}_{[2s+1]}\f_{(2s+1)} ={\Pi}^{\perp}_{[2s+1]}\f_{(2s+1)}  ~.
\ee
Stepping away from $\cV_{(2s+1)}$, one can show that for fixed $s$, the projector $\Pi_{[2s+1]}^{\perp}$ annihilates any lower-rank field $\f_{(2s'+1)}\in \mc{V}_{(2s'+1)}$
\be \label{FermProjProp}
\P^{\perp}_{[2s+1]} \f_{(2s'+1)}=0~, \qquad 1 \leq s' \leq s-1 ~.
\ee 
The two operators $\widehat{\Pi}_{[2s+1]}^{\perp}$ and $\Pi_{[2s+1]}^{\perp}$ are not equivalent on $\mc{V}_{(2s'+1)}$. We remark that $\P^{\perp}_{[2s+1]} $ does not annihilate lower-rank bosonic fields $\f_{\a(2s'+2)}$.

An important property of the projectors \eqref{SimpBosProj} and  
\eqref{SimpFermProj} is that they are symmetric operators, that is 
\bea
\int\text{d}^3x\, e \, \j^{\a(n)} \P^{\perp}_{[n]} \f_{\a(n)} 
= \int\text{d}^3x\, e \, \f^{\a(n)} \P^{\perp}_{[n]} \j_{\a(n)} ~, 
\eea
for arbitrary well-behaved fields $\j_{\a(n)} $ and $\f_{\a(n)} $.

\subsubsection{Helicity projection operators}\label{TAHPA}

Given a tensor field $\f_{\a(n)}$ on AdS$_3$, the spin projection operator $\P^{\perp}_{[n]}$ with the defining properties \eqref{SPO}, selects the component $ \f^{\perp}_{\a(n)}$ of $\phi_{\a(n)}$ which is transverse. 
If, in addition, $\phi_{\a(n)}$ satisfies the second order mass-shell equation \eqref{TASecondOrderMass},
then $\P^{\perp}_{[n]}\phi_{\a(n)}$  furnishes the reducible representation ${\mathfrak D}(\rho,-\frac{n}{2})\oplus {\mathfrak D}(\rho,\frac{n}{2})$ of $\mf{so}(2,2)$ (cf. eq. \eqref{red}).

In order to isolate the component of $\phi_{\a(n)}$ describing an irreducible representation of $\mf{so}(2,2)$, it is necessary to split the spin projection operators $\Pi_{[n]}^{\perp}$ according to
\begin{align}
\Pi_{[n]}^{\perp}=\mathbb{P}^{(+)}_{[n]}+\mathbb{P}^{(-)}_{[n]}~.\label{splitting}
\end{align}    
Each of the helicity projectors $\mathbb{P}^{(\pm)}_{[n]}$ should satisfy the properties \eqref{ProjProp} and \eqref{ProjTrans}. In addition, they should also project out the component of $\phi_{\a(n)}$ carrying a single value of helicity. The last two requirements are equivalent to the equations
\begin{subequations}
	\begin{align}
	\mf{D}^{\b(2)}\f^{(\pm)}_{\b(2)\a(n-2)}&=0~,\\
	\big(\mc{F}\mp\rho\big)\f^{(\pm)}_{\a(n)}&=0~,\label{HPprop2}
	\end{align}
\end{subequations}
where we have denoted $\f^{(\pm)}_{\a(n)}:=\mathbb{P}^{(\pm)}_{[n]} \f_{\a(n)}$. It follows that $\f^{(\pm)}_{\a(n)}$ furnishes the irreducible representation ${\mathfrak D}(\rho,\pm \frac{n}{2})$ (cf. \eqref{TAOnShellConditions}).

It is not difficult to show that the following operator satisfies these requirements\footnote{In the flat-space limit, \eqref{TAhelicityproj} reduces to the helicity projectors \eqref{helicityproj} in $\mb{M}^3$.}
\be \label{TAhelicityproj}
\mathbb{P}^{(\pm)}_{[n]} :=\hf \bigg (\mathds{1} \pm \frac{\mathcal{F}}{n\sqrt{\mathcal{Q}-(n+2)(n-2)|\m|^2}} \bigg ) {\Pi}^{\perp}_{[n]}~.
\ee 
Here ${\Pi}^{\perp}_{[n]}$ are the spin projectors written in terms of the Casimir operators and are given by \eqref{SimpBosProj} and \eqref{SimpFermProj} in the bosonic and fermionic cases, respectively. 
Of course, on $\cV_{(n)}$, one could instead represent the latter in their alternate form, \eqref{BP} and \eqref{FP}.

Using the defining features of ${\Pi}^{\perp}_{[n]}$, it can be shown that the operators $\mathbb{P}^{(+)}_{[n]} $ and $\mathbb{P}^{(-)}_{[n]} $ are orthogonal projectors when restricted to $\mc{V}_{(n)}$:
\be \label{OHP}
\mathbb{P}^{(\pm)}_{[n]} \mathbb{P}^{(\pm)}_{[n]}  = \mathbb{P}^{(\pm)}_{[n]} ~, \qquad \mathbb{P}^{(\pm)}_{[n]} \mathbb{P}^{(\mp)}_{[n]}  = 0~.
\ee
It is also clear that \eqref{TAhelicityproj} projects onto the transverse subspace of $\mc{V}_{(n)}$-- it inherits this property from ${\Pi}^{\perp}_{[n]}$. Moreover, the field $\f^{(\pm)}_{\a(n)}$ satisfies the constraint
\be \label{helcityprojprop}
\Big(\mathcal{F} \mp n \sqrt{\mathcal{Q} - (n-2)(n+2)|\m|^2} \Big)\f^{(\pm)}_{\a(n)} =0 ~,
\ee
off the mass-shell.
If $\f^{(\pm)}_{\a(n)}$ is placed on the mass-shell, eq. \eqref{TASecondOrderMass}, then \eqref{helcityprojprop} reduces to \eqref{HPprop2}. Thus, given a field $\f_{\a(n)} \in \mc{V}_{(n)}$ satisfying \eqref{TASecondOrderMass}, it follows that the helicity projectors $\mathbb{P}^{(\pm)}_{[n]}$ \eqref{TAhelicityproj} single out the component $\f^{(\pm)}_{\a(n)}$ which realises the irreducible representation ${\mathfrak D}(\rho,\pm \frac{n}{2})$.

\subsubsection{Longitudinal projectors and lower-spin extractors} \label{secLP}
Let us introduce the operator $\Pi^{\parallel}_{[n]}$ which is the orthogonal complement of $\P^{\perp}_{[n]}$,
\be 
\Pi^{\parallel}_{[n]}  :=  \mathds{1} - \P^{\perp}_{[n]} ~.
\ee
By construction, the two operators $\P^{\perp}_{[n]}$ and $\Pi^{\parallel}_{[n]}$ resolve the identity, $\mathds{1} = \Pi^{\parallel}_{[n]} + \P^{\perp}_{[n]}$, and form an orthogonal set of projectors
\begin{subequations}
	\begin{align}
	\P^{\perp}_{[n]}\P^{\perp}_{[n]} &= \P^{\perp}_{[n]}~,\qquad \Pi^{\parallel}_{[n]}\Pi^{\parallel}_{[n]} = \Pi^{\parallel}_{[n]}~, \label{DualProps}\\
	\Pi^{\parallel}_{[n]} \P^{\perp}_{[n]} &=0~, \qquad  ~~~\phantom{.} \P^{\perp}_{[n]} \Pi^{\parallel}_{[n]}=0 ~.
	\end{align}
\end{subequations}
Moreover, it can be shown that $\Pi^{\parallel}_{[n]}$ projects a field $\f_{\a(n)}$ onto its longitudinal component. A rank-$n$ field $\f^{\parallel}_{\a(n)}$ in AdS$_3$ is said to be longitudinal if there exists a rank-$(n-2)$ field $\f_{\a(n-2)}$ such that $\f^{\parallel}_{\a(n)}$ may be expressed as $\f^{\parallel}_{\a(n)} = \mf{D}_{\a(2)}\f_{\a(n-2)}$. Such fields are also sometimes referred to as being pure gauge.
Therefore, we find that 
\begin{align}
\f^{\parallel}_{\a(n)}:=\Pi^{\parallel}_{[n]} \f_{\a(n)}=\mf{D}_{\a(2)}\f_{\a(n-2)}~,
\end{align}
for some unconstrained field $\phi_{\a(n-2)}$. 

Let $\f^{\parallel}_{\a(n)}$ be some longitudinal field, which we do not assume to be in the image of $\Pi^{\parallel}_{[n]}$. 
However, since $\P_{[n]}^{\perp}$ commutes with $\mf{D}_{\a(2)}$ and annihilates all lower-rank fields, eq. \eqref{BosProjProp}, it follows that it also annihilates any rank-$n$ longitudinal field\footnote{This also implies that $\widehat{\P}^{\perp}_{[n]}\psi_{\a(n)}=0$, since both $\widehat{\P}^{\perp}_{[n]}$ and $\P^{\perp}_{[n]}$ are equivalent on $\mc{V}_{(n)}$. }
\begin{align}\label{LongitudinalKiller}
\f^{\parallel}_{\a(n)}=\mf{D}_{\a(2)}\f_{\a(n-2)}\qquad \implies \qquad \P^{\perp}_{[n]}\f^{\parallel}_{\a(n)}=0~.
\end{align}
As a consequence, given two integers $m,n$ satisfying $2\leq m \leq n$, it immediately follows that $\Pi^{\parallel}_{[n]}$ acts as the identity operator on the space of rank-$m$ longitudinal fields $\f^{\parallel}_{\a(m)}$,
\begin{align} 
\f^{\parallel}_{\a(m)}=\mf{D}_{\a(2)}\f_{\a(m-2)} \qquad \implies \qquad \Pi^{\parallel}_{[m+2s]}\f^{\parallel}_{\a(m)}=\f^{\parallel}_{\a(m)}~,
\end{align}
with $s$ a non-negative integer.
These properties will be useful in section \ref{secCT}.

Using the fact that $\P^{\perp}_{[n]}$ and $\Pi^{\parallel}_{[n]}$ resolve the identity, one can decompose an arbitrary field $\f_{\a(n)}$ on $\mc{V}_{(n)}$ as follows
\be
\f_{\a(n)} = \f^{\perp}_{\a(n)} + \mf{D}_{\a(2)}\f_{\a(n-2)}~.
\label{2.51}
\ee
Here $\f^{\perp}_{\a(n)}$ is transverse and $\f_{\a(n-2)}$ is unconstrained. Repeating this process iteratively, we obtain the following decomposition
\bea\label{Decomp}
\f_{\a(n)} &=& \sum_{j=0}^{\lfloor n/2 \rfloor }  \big (\mf{D}_{\a(2)} \big )^j \f^{\perp}_{\a(n-2j)}~. 
\eea
Here each of the fields $\f^{\perp}_{\a(n-2j)}$ are transverse, except of course $\phi^{\perp}$ and $\f^{\perp}_{\a}$. We note that, using \eqref{splitting}, one may take the decomposition \eqref{Decomp} a step further and bisect each term into irreducible components which are transverse and have positive or negative helicity,
\begin{align} \label{TADecompIrred}
\f_{\a(n)} = \sum_{j=0}^{\lfloor n/2 \rfloor }  \big (\mf{D}_{\a(2)} \big )^j\Big( \f^{(+)}_{\a(n-2j)}+\f^{(-)}_{\a(n-2j)}\Big)~. 
\end{align}

One can make use of the spin projectors \eqref{SimpBosProj} and \eqref{SimpFermProj} to construct operators which extract the component $\phi_{\a(n-2j)}^{\perp}$ from the decomposition \eqref{Decomp}, where $1\leq j \leq \lfloor n/2 \rfloor$. In particular, we find that the spin $\frac{1}{2}(n-2j)$ component may be extracted via
\begin{align}
\phi_{\a(n)}\mapsto \phi^{\perp}_{\a(n-2j)}=\big(\mb{S}_{[n-2j]}^{\perp}\phi\big)_{\a(n-2j)}\equiv \mb{S}_{\a(n-2j)}^{\perp}(\phi)~,
\end{align}
where we have defined
\begin{align} \label{Extractors}
\mb{S}_{\a(n-2j)}^{\perp}(\phi)&=\frac{(-1)^j}{2^{2j}}\binom{n}{j} \prod_{k=1}^{j}\big (\cQ - \t_{(k,n-2j+2k)}|\m|^2 \big )^{-1} \P^{\perp}_{[n-2j]}\big(\mf{D}^{\b(2)}\big)^j\f_{\a(n-2j)\b(2j)}~.
\end{align}
From this expression, it is clear that $\mb{S}_{\a(n-2j)}^{\perp}(\phi)$ is transverse,
\be
0=\mf{D}^{\b(2)}\mb{S}_{\b(2)\a(n-2j-2)}^{\perp}(\phi)~.
\ee
Therefore it is appropriate to call $\mb{S}_{[n-2j]}^{\perp}$ the transverse spin $\frac{1}{2}(n-2j)$ extractor. It is not a projector, since it is dimensionful and reduces the rank of the field on which it acts.


\subsection{Linearised higher-spin Cotton tensors} \label{secCT}

Further applications of spin projection operators can be found in modern conformal higher-spin theories. In particular, we will show that the spin projectors can be used to obtain new realisations of the linearised higher-spin Cotton tensors, which were recently derived by Kuzenko and Ponds in \cite{KP21}.\footnote{The AdS$_3$ Cotton tensors $\mf{C}_{\a(n)}(h)$ were first computed in \cite{KuzenkoPonds2018} for $3 \leq n \leq 6$. They were later extended to a conformally flat spacetime for arbitrary spin in \cite{KuzenkoPonds2019}. However, degauging $\mf{C}_{\a(n)}(h)$ to AdS$_3$ via the method provided in \cite{KuzenkoPonds2019}  proves to be computationally demanding.}
For integer $n\geq 2$, the higher-spin bosonic and fermionic Cotton tensors $\mathfrak{C}_{\a(n)}(h)$ take the respective closed forms \cite{KP21}
\begin{subequations}\label{cotE}
	\begin{align}
	\mathfrak{C}_{\a(2s)}(h)&=\frac{1}{2^{2s-1}}\sum_{j=0}^{s-1}2^{2j+1}\binom{s+j}{2j+1}\prod_{t=1}^{j}\big(\cQ-\tau_{(s-t,2s)}|\m|^2\big) \non\\
	&\phantom{\frac{1}{2^{2s-1}}\sum_{j=0}^{s-1}2^{2j+1}\binom{s+j}{2j+1}}\times
	\mf{D}_{\a(2)}^{s-j-1}\mf{D}_{\a}{}^{\b}\big(\mf{D}^{\b(2)}\big)^{s-j-1}h_{\a(2j+1)\b(2s-2j-1)}~, \label{cotF}\\
	\mathfrak{C}_{\a(2s+1)}(h)&=\frac{1}{2^{2s}}\sum_{j=0}^{s}2^{2j}\binom{s+j}{2j}\frac{(2s+1)}{(2j+1)}\prod_{t=1}^{j}\big(\cQ-\tau_{(s-t+1,2s+1)}|\m|^2\big) ~~~~~~~~~~~~~~~~~~~~~~~~~~~\non\\
	&\phantom{\frac{1}{2^{2s}}\sum_{j=0}^{s}\binom{s+j}{2j}\frac{(2s+1)}{(2j+1)}}\times \mf{D}_{\a(2)}^{s-j}\big(\mf{D}^{\b(2)}\big)^{s-j}h_{\a(2j+1)\b(2s-2j)}~.\label{cotB}
	\end{align}
\end{subequations}
The Cotton tensors are primary\footnote{For details concerning the conformal properties of the higher-spin Cotton tensors and the conformal gauge field, see \cite{KP21}.}  descendents of  the conformal gauge field $h_{\a(n)}$, which is a real field defined modulo gauge transformations of the form
\be
\d_\z h_{\a(n)} = \mf{D}_{\a(2)}\z_{\a(n-2)}~,
\ee
for some real unconstrained gauge parameter $\z_{\a(n-2)}$. The higher-spin Cotton tensors \eqref{cotE} are characterised by the properties:
\begin{enumerate}
	\item $\mathfrak{C}_{\a(n)}(h)$ is transverse
	\bsubeq \label{CotProp}
	\be \label{CTP1}
	\mf{D}^{\b\g}\mathfrak{C}_{\b\g\a(n-2)}(h)=0~.
	\ee
	\item $\mathfrak{C}_{\a(n)}(h)$  is gauge-invariant
	\be \label{CTP2}
	\d_\z \mathfrak{C}_{\a(n)}( h) = 0~.
	\ee
	\esubeq
\end{enumerate}

The bosonic \eqref{BP} and fermionic \eqref{FP} spin projectors can be used to recast the higher-spin Cotton tensors \eqref{cotE} in the simple form:
\begin{subequations} \label{TACT}
	\bea
	\mathfrak{C}_{\a(2s)} (h) &=& \frac{1}{2s} \prod_{t=1}^{s-1}\big (\mathcal{Q} - \t_{(t,2s)} |\m|^2 \big ) \mathcal{F} \widehat{\Pi}^{\perp}_{[2s]}h_{\a(2s)}~, \label{CTB} \\
	\mathfrak{C}_{\a(2s+1)}(h) &=&\prod_{t=1}^{s}\big (\mathcal{Q} - \t_{(t,2s+1)} |\m|^2 \big )\widehat{\Pi}^{\perp}_{[2s+1]}h_{\a(2s+1)}~.\label{CTF}
	\eea 
\end{subequations}
The identity $\cF \mf{D}_{(-2)}^s \f_{\a(2s)} = 0$ proves useful in deriving \eqref{CTB}. In the flat space limit $|\m| \rightarrow 0$, the Cotton tensors \eqref{TACT} reduce to \eqref{CT}. Moreover, we can make use of the equivalent family of projectors $\P^{\perp}_{[n]}$ to recast $\mathfrak{C}_{\a(n)}(h)$ purely in terms of the quadratic Casimir operators \eqref{QC} of $\mf{so}(2,2)$. Explicitly, they read
\begin{subequations} \label{CT1}
	\bea
	\mathfrak{C}_{\a(2s)} (h) &=& \frac{\cF}{2^{2s-1}(2s-1)!}  \prod_{j=1}^{s-1} \Big ( \cF^2 -4j^2 \big (\cQ-4(j-1)(j+1)|\m|^2 \big ) \Big )   h_{\a(2s)}~,  \\
	\mathfrak{C}_{\a(2s+1)}(h) &=&\frac{1}{2^{2s}(2s)!}\prod_{j=0}^{s-1} \Big ( \cF^2 -(2j+1)^2 \big (\cQ-(2j-1)(2j+3)|\m|^2 \big ) \Big ) h_{\a(2s+1)}~. \hspace{1cm}
	\eea
\end{subequations}

There are many advantages to expressing the Cotton tensors in terms of spin projection operators. Firstly, in both \eqref{TACT} and \eqref{CT1}, the properties of transversality \eqref{CTP1} and gauge invariance \eqref{CTP2} are made manifest, as a consequence of the projector properties \eqref{ProjTrans}  and \eqref{LongitudinalKiller} respectively. Since the higher-spin Cotton tensor is gauge-invariant, we may impose the transverse gauge condition on $h_{\a(n)}$, 
\begin{align}
h_{\a(n)}\equiv h^{\perp}_{\a(n)}~, \qquad 0=\mf{D}^{\b(2)}h^{\perp}_{\b(2)\a(n-2)}~.
\end{align}
In this gauge, the Cotton tensors become manifestly factorised into products of second order differential operators involving all partial masses,
\bsubeq
\bea
\mf{C}_{\a(2s)}(h^{ \perp}) &=&  \frac{1}{2s} \prod_{t=1}^{s-1}\big (\mathcal{Q} - \t_{(t,2s)} |\m|^2 \big ) \mathcal{F} h^{ \perp}_{\a(2s)} ~, \\
\mathfrak{C}_{\a(2s+1)}(h^{ \perp}) &=&\prod_{t=1}^{s}\big (\mathcal{Q} - \t_{(t,2s+1)} |\m|^2 \big )h^{ \perp}_{\a(2s+1)}~,
\eea
\esubeq
on account of  \eqref{ProjU}.
This property was observed in \cite{KP21} without the use of projectors.
An interesting feature of the new realisation \eqref{CT1}, which was not observed in \cite{KP21}, is that the Cotton tensors are manifestly factorised in terms of second-order differential operators without having to enter the transverse gauge.  

By virtue of the above observations, it follows that the AdS$_3$ CHS action \cite{KuzenkoPonds2018, KuzenkoPonds2019}
\begin{align} 
S_{\text{CHS}}^{(n)}[h]=\frac{\text{i}^n}{2^{\lceil n/2 \rceil+1}}\int\text{d}^3x\, e \, h^{\a(n)}\mf{C}_{\a(n)}(h)
\label{CSA}
\end{align}
is manifestly gauge invariant and factorised when $\mf{C}_{\a(n)}(h)$ takes the form \eqref{CT1}.

Recently in \cite{KP21}, the gauge completeness property \cite{HHL,HHL2} (cf. \eqref{TMGaugeCompleteness}) was extended to conformally flat backgrounds. Specifically, in any conformally flat background, the Cotton tensor $\mf{C}_{\a(n)}(h)$ vanishes if and only if the gauge field $h_{\a(n)}$ is pure gauge
\be \label{TAgaugecompleteness}
\mf{C}_{\a(n)}(h) = 0 \qquad  \Longleftrightarrow \qquad  h_{\a(n)} = \mf{D}_{\a(2)} h_{\a(n-2)}~,
\ee
for some arbitrary field $h_{\a(n-2)}$. Moreover, the spin projection operators \eqref{TMTransverseProjector} were used in \cite{KP21} to develop an alternative proof for the gauge completeness property in $\mb{M}^3$. Following this philosophy, we will demonstrate that the spin projectors \eqref{TAProjectors} can be used to prove gauge completeness in AdS$_3$.

For convenience, we collect several results which will be necessary to prove gauge completeness in AdS$_3$. It follows from \eqref{2.51} that a general field $h_{\a(n)}$ can be decomposed in the following manner
\be \label{TADecomp}
h_{\a(n)} = \big ({\P}^{\perp}_{[n]} + \P^{\parallel}_{[n]} \big ) h_{\a(n)} = \tilde{h}^{\perp}_{\a(n)} +\mf{D}_{\a(2)} h_{\a(n-2)}~,
\ee
where $h_{\a(n-2)}$ is an arbitrary unconstrained field and $ \tilde{h}^{\perp}_{\a(n)} = \widehat{\Pi}^{\perp}_{[n]}h_{\a(n)}$.

Recall that the higher-spin Cotton tensors \eqref{TACT} are given by
\begin{subequations}  \label{TACT2}
	\bea
	\mathfrak{C}_{\a(2s)} (h) &=& \frac{1}{2s} \prod_{t=1}^{s-1}\big (\mathcal{Q} - \t_{(t,2s)} |\m|^2 \big ) \mathcal{F}  \tilde{h}^{\perp}_{\a(2s)}~, \\
	\mathfrak{C}_{\a(2s+1)}(h) &=&\prod_{t=1}^{s}\big (\mathcal{Q} - \t_{(t,2s+1)} |\m|^2 \big ) \tilde{h}^{\perp}_{\a(2s+1)}~,
	\eea 
\end{subequations}
which can be inverted to yield
\begin{subequations}  \label{TACTInversion}
	\bea
	\Big [ 2s \prod_{t=1}^{s}\big (\mathcal{Q} - \t_{(t,2s)} |\m|^2 \big ) \Big ]^{-1} \mathcal{F} 	\mathfrak{C}_{\a(2s)} (h) &=&  \tilde{h}^{\perp}_{\a(2s)}~, \\
	\Big [ \prod_{t=1}^{s}\big (\mathcal{Q} - \t_{(t,2s+1)} |\m|^2 \big ) \Big ]^{-1}	\mathfrak{C}_{\a(2s+1)}(h) &=& \tilde{h}^{\perp}_{\a(2s+1)}~.
	\eea 
\end{subequations}
We now have the ingredients needed to prove the gauge completeness \eqref{TAgaugecompleteness} in AdS$_3$. First, let us assume that the Cotton tensor vanishes, $\mf{C}_{\a(n)}(h) =0$. It immediately follows from \eqref{TACTInversion} that
\be
\tilde{h}^{\perp}_{\a(2s)} = 0~, \qquad \tilde{h}^{\perp}_{\a(2s+1)} = 0~.
\ee
Thus, it follows from \eqref{TADecomp} that $h_{\a(n)}$ is pure gauge, $h_{\a(n)} = \mf{D}_{\a(2)} h_{\a(n-2)}$.

Now let us assume that $h_{\a(n)}$ is pure gauge, $h_{\a(n)} = \mf{D}_{\a(2)} h_{\a(n-2)}$.  Demonstrating that the Cotton tensor \eqref{TACT2} vanishes is equivalent to showing that is gauge-invariant, which we already know to be true. This completes the proof for gauge completeness \eqref{TAgaugecompleteness} in AdS$_3$ and illustrates another useful application for the spin projection operators.


\subsection{(Fang-)Fronsdal-type actions}\label{TAMasslessActions}
We begin by reviewing the AdS$_3$ counterparts of the (Fang-)Fronsdal actions in AdS$_4$ \cite{Fronsdal1979Sing,Fronsdal1979}, mostly following the presentation in \cite{KuzenkoPonds2018}.

\subsubsection{Fang-Fronsdal action}
The Fang-Fronsdal action in AdS$_4$ \cite{Fronsdal1979} can be generalised to AdS$_3$ using two different gauge-invariant actions, 
$S_{(\frac{n}{2},+)}^{\text{FF}}$ and $S_{(\frac{n}{2},-)}^{\text{FF}}$. Given an integer $n\geq 4$, the actions $S_{(\frac{n}{2},\pm)}^{\text{FF}} $
is realised in terms of the three real fields $h_{\a(n)}$,  $y_{\a(n-2)}$ and $z_{\a(n-4)}$,
which are defined modulo gauge transformations of the form
\begin{subequations}\label{B.3}
	\bea
	\d_\z h_{\a(n)}&=&\mathfrak{D}_{\a(2)}\z_{\a(n-2)}~,\\
	\d_\z y_{\a(n-2)}&=&\frac{1}{n}\mathfrak{D}_{(\a_1}{}^\b \z_{\a_2 ... \a_{n-2})\b}\pm|\m|\z_{\a(n-2)}~, \label{TAFangFronsdalGaugeCondition2}\\
	\d_\z z_{\a(n-4)}&=&\mathfrak{D}^{\b\g} \z_{\b\g\a(n-4)}~. \label{TAFangFronsdalGaugeCondition3}
	\eea
\end{subequations}
Here, the gauge parameter $\z_{\a(n-2)}$ is real and unconstrained. Modulo an overall constant, the corresponding gauge-invariant  action is given by
\bea \label{TAFangFronsdal}
S_{(\frac{n}{2},\pm)}^{\text{FF}}[h,y,z]&=&\frac{\ri^n}{2^{\lfloor n/2 \rfloor +1}}\int \rd^3x\,e\,\bigg \{ h^{\b\a(n-1)}\mathfrak{D}_{\b}{}^{\g}h_{\g\a(n-1)}+2(n-2)y^{\a(n-2)}\mathfrak{D}^{\b\g}h_{\b\g\a(n-2)} ~ \non \\
&&+4(n-2)y^{\b\a(n-3)}\mathfrak{D}_\b{}^\g y_{\g\a(n-3)}+\frac{2n(n-3)}{n-1}z^{\a(n-4)}\mathfrak{D}^{\b\g}y_{\b\g\a(n-4)}~ \non \\
&&- \frac{(n-3)(n-4)}{(n-1)(n-2)}z^{\b\a(n-5)}\mathfrak{D}_\b{}^\g z_{\g\a(n-5)}
\pm |\m| \Big ( (n-2)  h^{\a(n)}h_{\a(n)}~ \non \\
&&- 4n(n-2) y^{\a(n-2)}y_{\a(n-2)}- \frac{n(n-3)}{n-1} z^{\a(n-4)}z_{\a(n-4)} \Big ) \bigg \} ~.
\eea
Only the action $S_{(\frac{n}{2},+)}^{\text{FF}}[h,y,z]$ was considered in \cite{KuzenkoPonds2018}.
Both actions $S_{(\frac{n}{2},+)}^{\text{FF}}[h,y,z]$ and $S_{(\frac{n}{2},-)}^{\text{FF}}[h,y,z]$ are  the $d=3$ counterparts of the Fang-Fronsdal action provided $n$ is odd, 
$n = 2s +1$, with $s\geq 2$ an integer. 

It must be noted that the action \eqref{TAFangFronsdal} can be expressed in a more compact way in terms of a reducible frame field,
\be
\bm h_{\b\g;\a{(n-2)}} :=(\g^b)_{\b\g} \bm h_{b;\a{(n-2)}}~,
\ee
which is defined modulo gauge transformations of the form
\bea
\d_\z \bm{h}_{\b\g;\a(n-2)}=\mathfrak{D}_{\b\g}\z_{\a(n-2)}\pm(n-2)|\m|\big ( \varepsilon_{\b(\a_1}\z_{\a_2 ... \a_{n-2})\g}+\varepsilon_{\g(\a_1}\z_{\a_2 ... \a_{n-2})\b} \big ) ~.
\label{frameB.6}
\eea
The corresponding gauge-invariant action takes the form
\bea \label{VT}
S_{(\frac{n}{2},\pm)}^{\text{FF}}[\bm{h}]&=&\frac{\ri^n}{2^{\lfloor n/2 \rfloor +1}}\int \rd^3x\,e\,\bigg \{ \bm{h}^{\b\g;\a(n-2)} \mathfrak{D}_\b{}^\d \bm{h}_{\g\d;\a(n-2)}\pm(n-2)|\m|\big ( \bm{h}^{\b\g;\a(n-2)}\bm{h}_{\b\g;\a(n-2)} ~\non \\
&&+2\bm{h}^{\b\g;\a(n-3)}{}_\b \bm{h}_{\g\d;}{}_{\a(n-3)}{}^\d \big ) \bigg \}~.
\eea
In the flat-space limit, the action \eqref{VT} reduces to the model considered
by Tyutin and Vasiliev 
in \cite{TyutinVasiliev1997} (see also \cite{KuzenkoPonds2018} for a review).
Ref. \cite{BSZ2} also made use of the action \eqref{VT} in the frame-like formulation, but only a particular choice of sign was considered. 
In order to relate the action \eqref{VT} to \eqref{TAFangFronsdal}, it is necessary to decompose the field $\bm{h}_{\b\g;\a(n-2)}$ into its irreducible components. 
The irreducible fields contained in $\bm{h}_{\b\g;\a(n-2)}$ can be defined as follows:
\begin{subequations} \label{Irred}
	\bea
	h_{\a(n)}&:=&\bm{h}_{(\a_1 \a_2;\a_3 ... \a_n)}~,\\
	y_{\a(n-2)}&:=&\frac{1}{n}\bm{h}^{\b}{}_{(\a_1 ; \a_2 ... \a_{n-2})\b}~, \\
	z_{\a(n-4)}&:=&\bm{h}^{\b\g;}{}_{\b\g\a(n-4)}~.
	\eea
\end{subequations}
The gauge transformation laws for these fields, which follow from \eqref{frameB.6},  are given by those in  \eqref{B.3}.   
Upon expressing the action \eqref{VT} in terms of the  fields \eqref{Irred}, it can be shown that the resulting action coincides with \eqref{TAFangFronsdal}.

Let us analyse the on-shell content of the action \eqref{TAFangFronsdal} for the case $n=2s+1$ in order to show that the theory describes a massless field with spin $s+\hf$. The equations of motion corresponding to \eqref{TAFangFronsdal} are
\bsubeq \label{TAFangFronsdalEoM}
\bea
0&=&\mf{D}_\a{}^\b h_{\b\a(2s)} - (2s-1) \mf{D}_{\a(2)} y_{\a(2s-1)} \pm (2s-1)|\m|h_{\a (2s+1)} ~, \label{TAFangFronsdalEoM1}\\
0&=& s(2s-1) \big ( \mf{D}^{\b\g} h_{\b\g\a(2s-1)}+ 4\mf{D}_\a{}^\b y_{\b \a(2s-2)} \mp 4(2s+1) |\m| y_{\a(2s-1)} \big ) \label{TAFangFronsdalEoM2}\non \\
&&-(s-1)(2s+1)\mf{D}_{\a(2)} z_{\a(2s-3)}~, \\
0&=&(2s-1)(2s+1) \big ( \mf{D}^{\b \g} y_{\b\g\a(2s-3)} \mp |\m| z_{\a(2s-3)} \big ) -(2s-3)\mf{D}_\a{}^\b z_{\b\a(2s-4)}~. \label{TAFangFronsdalEoM3}
\eea
\esubeq
It follows from \eqref{TAFangFronsdalGaugeCondition3} that we can impose the gauge condition 
\be \label{TAFangFronsdalGaugeCondition1}
z_{\a(2s-3)} = 0~.
\ee 
The residual gauge symmetry preserving \eqref{TAFangFronsdalGaugeCondition1}  is described by $\z_{\a(2s-1)}$ constrained by 
\be
\mf{D}^{\b \g} \z_{\b \g\a(2s-3)} = 0 \quad \Longrightarrow \quad \mf{D}_{(\a_1}{}^\b \z_{\a_2 \ldots \a_{2s-1})\b} = \mf{D}_{\a_1}{}^\b \z_{\a_2 \ldots \a_{2s-1}\b}~. \label{TAFangFronsdalGaugeParameterTransverse}
\ee
In the gauge \eqref{TAFangFronsdalGaugeCondition1}, the equation of motion \eqref{TAFangFronsdalEoM3} reduces to
\be \label{TAFangFronsdalEoM3Gauge1}
0=\mf{D}^{\b \g} y_{\b\g\a(2s-3)} \quad  \Longrightarrow  \quad \mf{D}_{(\a_1}{}^\b y_{\a_2 \ldots \a_{2s-1})\b} = \mf{D}_{\a_1}{}^\b y_{\a_2 \ldots \a_{2s-1}\b}~.
\ee
Due to \eqref{TAFangFronsdalEoM3Gauge1}, it follows that the gauge transformation \eqref{TAFangFronsdalGaugeCondition3} becomes 
\be \label{TAFangFronsdalGTGauge1}
\d_\z y_{\a(2s-1)}=\frac{1}{2s+1}\mathfrak{D}_{(\a_1}{}^\b \z_{\a_2 ... \a_{2s-1})\b}\pm|\m|\z_{\a(2s-1)}~,
\ee
Since $y_{\a(2s-1)}$ and $\z_{\a(2s-1)}$ are of the same functional form, it can be shown that the field $y_{\a(2s-1)}$ can be gauged away
\be \label{TAFangFronsdalGauge2}
y_{\a(2s-1)} = 0~.
\ee
The residual gauge freedom is described by $\z_{\a(2s-1)}$, which is further constrained by 
\be \label{TAFangFronsdalFirstOrderEquation}
\big ( \mc{F} \pm \r_{(1,2s+1)} \big )\z_{\a(2s-1)} =0~.
\ee
In the gauge \eqref{TAFangFronsdalGauge2}, the equations of motion \eqref{TAFangFronsdalEoM1} and \eqref{TAFangFronsdalEoM2} reduce to
\bsubeq \label{TAFangFronsdalEoMGauge2}
\bea
0&=& \mf{D}^{\b\g} h_{\b\g\a(2s-1)} ~, \\
0&=&\big ( \mc{F} \pm \r_{(1,2s+1)}  \big ) h_{\a(2s+1)}~.
\eea
\esubeq
Thus $h_{\a(2s+1)}$ is the only surviving field on-shell. In accordance with \eqref{TAMasslessOnShell}, it follows from the above analysis that on-shell, the Fang-Fronsdal action $S_{(s+\hf,\pm)}^{\text{FF}}[h,y,z]$ \eqref{TAFangFronsdal} describes a massless spin-$(s+\hf)$ field $h_{\a(2s+1)}$ with helicity $\pm (s+\hf)$.

\subsubsection{Fronsdal action}\label{TAFronsSec}

The Fronsdal action in AdS$_4$ \cite{Fronsdal1979Sing}  can be generalised to AdS$_3$ as follows. Given an integer $n\geq 4$, 
we introduce two real field variables $h_{\a(n)}$, $y_{\a(n-4)}$, which are defined modulo gauge transformations of the form
\begin{subequations} \label{fg00}
	\bea
	\d_\z h_{\a(n)}&=&\mathfrak{D}_{\a(2)}\z_{\a(n-2)}~ ,\\
	\d_\z y_{\a(n-4)}&=&\frac{n-2}{n-1}\mathfrak{D}^{\b\g}\z_{\b \g \a(n-4)}~,
	\eea
\end{subequations}
where the gauge parameter $\z_{\a(n-2)}$ is real and unconstrained. 
Modulo an overall constant, the gauge-invariant  action is given by \cite{KuzenkoPonds2018}
\bea \label{TAFronsdal}
S_{(\frac{n}{2})}^{\text{F}}[h,y]&=& \frac{\ri^n}{2^{\lfloor n/2 \rfloor +1}}\int \rd^3x\,e\,\bigg \{ h^{\a(n)} \cQ h_{\a(n)}-\frac{n}{4}\mathfrak{D}_{\b \g}h^{\b\g\a(n-2)}\mathfrak{D}^{\d\l}h_{\d\l\a(n-2)}~ \non\\
&&-\frac{n-3}{2}y^{\a(n-4)}\mathfrak{D}^{\b\g}\mathfrak{D}^{\d\l}h_{\b\g\d\l\a(n-4)}-2n(n-2)|\m|^2~   h^{\a(n)} h_{\a(n)} \non ~ \\
&&-\frac{n-3}{n} \bigg ( 2 y^{\a(n-4)} \cQ y_{\a(n-4)}-4(n^2+2)|\m|^2y^{\a(n-4)}y_{\a(n-4)} ~ \non \\
&&+\frac{(n-4)(n-5)}{4(n-2)}\mathfrak{D}_{\b\g}y^{\b\g\a(n-6)}\mathfrak{D}^{\d\l}y_{\d\l\a(n-6)} \bigg ) \bigg \}~, 
\eea
The model \eqref{TAFronsdal} is the $d=3$ counterpart of the Fronsdal action \cite{Fronsdal1979Sing}  provided $n$ is even, $n = 2s $, with $s\geq 2$. 

Performing a similar component analysis as was done for the Fang-Fronsdal action, it can be shown that the gauge symmetry \eqref{fg00} provides the freedom to impose the gauge conditions
\be \label{TASGCFF}
y_{\a(2s-4)} = 0~, \qquad \mf{D}^{\b\g} h_{\b\g\a(2s-2)} = 0~.
\ee
The residual gauge symmetry is described by $\z_{\a(2s-1)}$ constrained by
\be
\mf{D}^{\b\g}\z_{\b \g \a(2s-4)} = 0~, \qquad \big ( \mc{Q} -\t_{(1,2s)} |\m|^2 \big ) \z_{\a(2s-1)} = 0~.
\ee

In the gauge \eqref{TASGCFF}, the equation of motion obtained by varying the action \eqref{TAFronsdal} with respect to $h_{\a(2s)}$ reduces to
\be
0 =\big ( \mc{Q} -\t_{(1,2s)}|\m|^2 \big ) h_{\a(2s)} = \big ( \cF - \r_{(1,2s)} \big ) \big ( \cF + \r_{(1,2s)} \big ) h_{\a(2s)}~.
\ee
Thus, we are left with the remaining field $h_{\a(2s)}$ on-shell. In accordance with \eqref{TAMasslessConditions2}, it follows from the above analysis that the Fronsdal model $S_{(s)}^{\text{F}}$ describes two massless spin-$s$ modes, each with a differing sign of helicity $\pm s$.


\subsection{New topologically massive actions}\label{TAMassiveactions}
New topologically massive models were recently computed in AdS$_3$ by Kuzenko and Ponds \cite{KuzenkoPonds2018} for fields with arbitrary spin. These models are formulated solely in terms of the gauge prepotentials $h_{\a(n)}$ and the associated Cotton tensors $\mf{C}_{\a(n)}(h)$. In particular,
given an integer $n\geq 2$, the gauge-invariant NTM action for the field $h_{\a(n)}$ 
is  \cite{KuzenkoPonds2018}
\begin{align}
S_{\text{NTM}}^{(n)}[h]=\frac{\text{i}^n}{2^{\lceil n/2 \rceil+1}}\frac{1}{\rho}\int\text{d}^3x\, e \, h^{\a(n)}
\big(\mc{F}-\s\rho\big) \mf{C}_{\a(n)}(h)
~, \label{HSNTMG}
\end{align}
where $\rho$ is some positive mass parameter and $\s:=\pm 1$. Making use of  the new representation \eqref{CT1} for the super-Cotton tensor leads to a manifestly gauge invariant and factorised form for the action \eqref{HSNTMG}. 

The equation of motion obtained by varying \eqref{HSNTMG} with respect to the field $h_{\a(n)}$ is 
\begin{align}
0=\big(\mc{F}-\s \rho\big)\mf{C}_{\a(n)}(h)~. \label{TAEOM1}
\end{align}
By analysing \eqref{TAEOM1}, it can be shown that on-shell, the  action \eqref{HSNTMG} describes a propagating mode with pseudo-mass $\r$, spin $\frac{n}{2}$ and helicity $\frac{\s n}{2}$ given $\r \neq \r_{(t,2s)}$ (cf. \eqref{TAOnShellConditions}). For the case $\r = \r_{(t,2s)}$, the model describes only pure gauge degrees of freedom.

Next, we wish to compute the AdS$_3$ extension of the NTM model  \eqref{TANNTM}, which was recently derived in $\mb{M}^3$ \cite{DalmaziSantos2021}.  Given an integer $s\geq 1$,  the action  \eqref{TANNTM} may be readily extended to AdS$_3$ as follows
\bea \label{NNTM}
\widetilde{S}_{\text{NTM}}^{(2s)}[h] =\int \rd^3 x\, e\,  h^{\a(2s)} \big ( \cF - \s \r \big ) \mf{W}_{\a(2s)}(h)  ~,
\eea
where $\r$ is a positive mass parameter, $\s:=\pm 1$, and  $\mf{W}_{\a(2s)}(h)$ is the field strength,
\be
\mf{W}_{\a(2s)}(h):=\prod_{t=1}^{s-1}\big (\mathcal{Q} - \t_{(t,2s)} |\m|^2 \big )  {\Pi}^{\perp}_{[2s]}h_{\a(2s)}~.
\ee
Due to the properties of $\Pi^{\perp}_{[2s]}$, the action \eqref{NNTM} is manifestly gauge invariant and factorised.  The descendent $\mf{W}_{\a(2s)}(h)$ may be obtained from $\mf{C}_{\a(2s)}(h)$ by stripping off $\mc{F}$: 
\begin{align}
\mf{C}_{\a(2s)}(h)=\frac{1}{2s}\mc{F}\mf{W}_{\a(2s)}(h)~.
\end{align}
A similar construction does not appear to be possible in the fermionic case. 

The equation of motion obtained by varying \eqref{NNTM} with respect to the field $h_{\a(2s)}$ is
\bea \label{TA2EoM}
0 = (\cF - \s \r)\mf{W}_{\a(2s)}(h)~.
\eea
By analysing \eqref{TA2EoM}, it can be shown that on-shell, the model \eqref{NNTM} has the same particle content as the NTM model \eqref{HSNTMG}.

\section{Three-dimensional $\cN=1$ anti-de Sitter superspace}  \label{TASsec3dAdS}


The main objective of this section is to construct the superspin projection operators in $\cN=1$ AdS superspace AdS$^{3|2}$ and explore several of their applications. These operators can be considered supersymmetric generalisations of the AdS$_3$ spin projection operators, which were derived in section \ref{TAProjectors}. They can also be considered AdS$_3$ analogues of the superspin projectors in $\mb{M}^{3|2}$, which were computed in section \ref{TMSN1SUSY}. Before studying these operators, it is first necessary to outline the irreducible representations of  $\mathfrak{osp}(1|2;{\mathbb R} ) \oplus \mathfrak{sl}(2, {\mathbb R})$.

\subsection{Irreducible representations of the AdS$_3$ superalgebra}\label{TASIrreps}
The irreducible representations of the $\cN=1$ AdS$_3$ superalgebra  $\mathfrak{osp}(1|2;{\mathbb R} ) \oplus \mathfrak{sl}(2, {\mathbb R})$ are described by their decomposition into irreducible representations of $\mf{so}(2,2)$, which were reviewed in section \ref{IrrepsAdS3}. We denote the irreducible representations of $\mathfrak{osp}(1|2;{\mathbb R} ) \oplus \mathfrak{sl}(2, {\mathbb R})$  by $\mathfrak{S}(M, \s \frac{n}{2})$, where $M$ is the pseudo-mass, $s := \frac{n}{2}$ is the superspin and the parameter $\s:=\pm 1$ labels the sign of the superhelicity $\k = \frac{1}{2}(n+\hf)\s$. It can be shown that the representation $\mathfrak{S}(M, \s \frac{n}{2})$ decomposes into two irreducible representations of $\mathfrak{so}(2,2)$,
\be \label{decomp}
\mathfrak{S} \Big (M, \s \frac{n}{2} \Big ) = \mathfrak{D} \Big ( \r_A, \s_A \frac{n}{2} \Big ) \oplus \mathfrak{D} \Big ( \r_B, \s_B\frac{n+1}{2} \Big )~.
\ee
Here, the pseudo-masses $\r_A$ and $\r_B$ are given by
\bea
\r_A= \frac{n}{2n+1}\Big | \s M-(n+2)|\m| \Big |~, \qquad
\r_B= \frac{n+1}{2n+1} \Big | \s M + (n-1)|\m| \Big |~,
\eea
and the corresponding signs of the helicities $\s_A $ and $\s_B$ are  
\bea
\s_A &=& \frac{  \s M-(n+2)|\m| }{\big | \s M-(n+2)|\m| \big |}~, \qquad
\s_B = \frac{ \s M + (n-1)|\m|}{\big | \s M + (n-1)|\m| \big |}~.
\eea
Here $|\m| >0 $ is a real constant parameter which determines the curvature of AdS$^{3|2}$.\footnote{The parameter $|\m|$ was denoted $\mc{S}$ in \cite{KLT-M12}. However,  we prefer to make use of 
	the notation $|\m|$ since it is better suited in the $(1,1) \to (1,0)$ superspace reduction scheme, which will be considered later. } The representation $\mathfrak{S}(M, \s \frac{n}{2})$ is unitary if the parameter $M$ obeys the unitarity bound 
\be \label{TASUnitarityBound}
M\geq 2(n-1)(n+1)|\m|~.
\ee
This bound ensures that both representations appearing in the decomposition \eqref{decomp} are unitary.

\subsection{Geometry of AdS$^{3|2}$}\label{TASGeomAdS32}

Let us denote $z^\cM = (x^m, \q^\m)$ the local coordinates which parametrise  AdS$^{3|2}$.\footnote{We remind the reader that AdS$^{3|2}$ is the maximally supersymmetric solution of three-dimensional $\cN=1$ AdS supergravity \cite{GatesGrisaruRocekSiegel1983}.} The geometric structure of AdS$^{3|2}$ is described in terms of its covariant derivatives\cite{KLT-M12}\footnote{In the hope that no confusion arises, we use the same notation for the vector covariant derivative in AdS$_3$ and in AdS$^{3|2}$.
}
\be \label{SCD}
\nabla_A = (\nabla_a, \nabla_\a) = E_A{}^M {\partial}_M + \hf \O_A{}^{bc}M_{bc}~. 
\ee
Here $E_A{}^M$ is the inverse supervielbein and  $\O_A{}^{bc}$ the Lorentz connection. The covariant derivatives obey the following (anti-)commutation relations
\begin{subequations} \label{SA}
	\begin{gather}
	\{ \nabla_\a , \nabla_\b \} = 2\ri \nabla_{\a\b} - 4\ri|\m| M_{\a\b}~, \label{SA1} \\
	\ [ \nabla_{\a \b}, \nabla_\g ] = -2|\m| \ve_{\g(\a}\nabla_{\b)}~, \qquad \ [ \nabla_{\a \b}, \nabla_{\g \d} ] = 4 |\m|^2 \big(\ve_{\g(\a}M_{\b)\d} + \ve_{\d(\a} M_{\b)\g}\big)~, \label{SA2}
	\end{gather}
\end{subequations}
It should be remarked 
that the geometry of $\cN=1$ AdS superspace can also be described by 
the graded commutation relations which are obtained from \eqref{SA} 
by replacing $|\m| \to -|\m|$. 
The two choices, $|\m|$ and $-|\m|$,  correspond to the so-called $(1,0)$ and $(0,1)$  
AdS superspaces \cite{KLT-M12}, which are different realisations of AdS$^{3|2}$. The $(1,0)$ and $(0,1)$ AdS superspaces are naturally embedded 
in $(1,1)$ AdS superspace\footnote{ 
	As follows from \eqref{1.3}, in $(1,1)$ AdS superspace
	two subsets of covariant derivatives,  $( {\bm \na}_{a}, {\bm \na}_{\a}^{\1}) $ and 
	$( {\bm \na}_{a}, {\bm \na}_{\a}^{\2}) $, form closed algebras of the type \eqref{SA}, 
	with the curvature parameter being $|\m| $ and $ -|\m|$, respectively.}
and are related to each other by a parity transformation.

We list several identities which prove indispensable for calculations:
\begin{subequations}  \label{CDI}
	\bea 
	\nabla_\a \nabla_\b &=& \ri \nabla_{\a\b} - 2\ri|\m| M_{\a\b}+\frac{1}{2}\ve_{\a\b}\nabla^2~, \\
	\nabla^{\b} \nabla_\a \nabla_\b &=& 4\ri |\m|\nabla_\a~, \quad \{ \nabla^2, \nabla_\a  \} = 4\ri |\m|\nabla_\a~,\\
	\nabla^2 \nabla_\a &=& 2\ri |\m|\nabla_\a + 2\ri \nabla_{\a\b} \nabla^\b - 4\ri |\m| \nabla^\b M_{\a\b}~, \\
	\qquad \ [ \nabla_{\a} \nabla_{\b}, \nabla^2 ]
	&=& 0 \quad \Longrightarrow \quad \ [\nabla_{\a\b}, \nabla^2 ] = 0~, \\
	\ [ \nabla_\a ,  \Box] &=& 2|\m| \nabla_{\a\b} \nabla^\b + 3|\m|^2 \nabla_\a~.
	\eea
\end{subequations}
where we have denoted $\nabla^2 = \nabla^\a \nabla_\a$ and $\Box = \na^a \na_a = -\hf \na^{\a \b} \na_{\a \b}$. These relations can be derived from the algebra of covariant derivatives \eqref{SA}.

According to the general formalism of \cite{BuchbinderKuzenko1998}, the isometry transformations of AdS$^{3|2}$ are generated by the Killing supervector fields,
\bea
\x = \x^B\nabla_B = \x^b\nabla_b + \x^\b\nabla_\b~, 
\eea
which by definition solve the master equation \cite{KLT-M12}
\be \label{Kcondition1-1}
[\x + \frac{1}{2}\z^{bc}M_{bc}, \nabla_A] = 0~,
\ee
for some Lorentz  parameter $\z^{bc} = - \z^{cb}$. 
It can be shown that the Killing equation \eqref{Kcondition1-1} is equivalent to the set of relations \cite{KLT-M12} (see also \cite{KuzenkoPonds2018})
\begin{subequations}  \label{2.40d-1}
	\bea
	\nabla_{(\a}\x_{\b\g)} &=& 0~, \qquad \nabla_\b \x^{\b\a} =-6 \ri \x^\a~, \\
	\nabla_{(\a}\z_{\b\g)} &=& 0~, \qquad  \nabla_\b \z^{\b\a} = -12 \ri |\m| \x^\a ~, \\
	\nabla_\a \x_\b &=& \frac{1}{2} \z_{\a\b} +  |\m|   \x_{\a\b}~,
	\eea
\end{subequations}
which imply 
\bsubeq \label{2.40e}
\bea
\na_a \xi_b + \na_b \xi_a &=& 0~,\\
\na_{\a \b}\xi^{\b}+3 |\m| \xi_{\a} &=& 0~, \\
\big( \ri \na^2 
+12  |\m| \big) \xi_{\a} &=&0 ~.
\eea
\esubeq
Thus, $\xi^a$ is a Killing vector. 
Given a tensor superfield $U(x, \q)$ (with suppressed indices) on AdS$^{3|2}$, its infinitesimal isometry transformation law is given by 
\bea
\d_\x U= \big (  \x^a \nabla_a + \x^\a \nabla_\a + \hf \z^{ab}M_{ab} \big ) U ~. \label{211}
\eea

Crucial to our analysis are the two independent Casimir operators of $\mathfrak{osp}(1|2;{\mathbb R} ) \oplus \mathfrak{sl}(2, {\mathbb R})$. They take the following form in the superfield representation  \cite{KP21,KuzenkoPonds2018} 
\bsubeq \label{SCQ}
\begin{alignat}{2}
\mathbb{Q}:&=- \frac{1}{4}\nabla^2 \nabla^2 + \ri |\m| \nabla^2~, \qquad &[\mathbb{Q}, \nabla_A]= 0~, \label{QQC}\\
\mathbb{F}:&=- \frac{\ri}{2}\nabla^2 + 2\nabla^{\a\b}M_{\a\b}~,  &[\mathbb{F}, \nabla_A]= 0~. \label{FQC}
\end{alignat}
\esubeq
Making use of the identity
\bea
-\frac{1}{4}\nabla^2\nabla^2 &=&  \Box - 2\ri|\m| \nabla^2+2|\m| \nabla^{\a\b}M_{\a\b} -2|\m|^2M^{\a\b}M_{\a\b}~,
\eea
allows us to express $\mathbb{Q}$ in terms of the AdS$^{3|2}$ d'Alembertian $\Box$. Let us denote by $\mathds{V}_{(n)}$ the space of real symmetric tensor superfields $\F_{\a(n)}$ on AdS$^{3|2}$.
It can be shown that the Casimir operators $\mathbb{Q}$ and $\mathbb{F}$ are related to each other on $\mathds{V}_{(n)}$ via the relation
\bea \label{FSquared}
\mb{F}^2\F_{\a(n)} &=& \Big ( (2n +1)^2\mb{Q}  + (2n+1)(2n^2+2n-1)\ri |\m| \nabla^2 +4n^2(n+2)^2 |\m|^2\Big  ) \F_{\a(n)}  \non \\
&&+ 4 n(2n^2+n-2)\ri |\m| \nabla_{\a}\nabla^\b \F_{\b \a(n-1)} -4\ri n\nabla_{\a\b}\nabla^{\b} \nabla^{\g}\F_{\g\a(n-1)}~ \non \\
&&+4n(n-1)\nabla_{\a(2)}\nabla^{\b(2)}\F_{\b(2)\a(n-2)}~.
\eea


\subsection{Irreducible superfield representations} \label{TASirrepsAdS32}

We begin by reviewing aspects of on-shell superfields in AdS$^{3|2}$, as presented in \cite{KP21}.
Given an integer $n \geq 1$, a superfield $\F_{\a(n)}$ on $\mathds{V}_{(n)}$ is said to be on-shell if it satisfies the two constraints 
\bsubeq \label{SOC}
\bea
0 &=& \nabla^\b \F_{\b\a(n-1)}~, \label{STrans} \\
0 &=& \big ( \mathbb{F} - \s M \big ) \F_{\a(n)}~, \label{SFO}
\eea
\esubeq
where $\s:=\pm 1$ and $M\geq 0$ is a real parameter of unit mass dimension. Such a superfield furnishes the irreducible representation $\mathfrak{S}(M, \s \frac{n}{2})$ of $\mathfrak{osp}(1|2;{\mathbb R} ) \oplus \mathfrak{sl}(2, {\mathbb R})$. Thus, we say that an on-shell superfield carries pseudo-mass $M$, superspin $\frac{n}{2}$ and superhelicity $\hf(n+\hf)\s$.

It is necessary to study the component structure of an on-shell superfield \eqref{SOC} in order to demonstrate that it furnishes the irreducible representation $\mathfrak{S}(M, \s \frac{n}{2})$. An on-shell superfield describes two non-vanishing independent component fields,
\be
A_{\a(n)}: = \F_{\a(n)}|~, \qquad B_{\a(n+1)} = \ri^{n+1} \na_\a \F_{\a(n)}| ~. \label{TASCompStructOnsell}
\ee
It  follows from the superfield constraints \eqref{SOC} that the component fields are transverse,
\be
\na^{\b\g}A_{\b\g \a(n-2)} = 0~, \qquad \na^{\b\g}B_{\b\g \a(n-1)} = 0~.
\ee
Furthermore, the component fields can be shown to satisfy the first-order constraints,
\bsubeq
\bea
0 = \big (\cF -\s_A \r_A \big )A_{\a(n)}~, \qquad 0 = \big (\cF -\s_B \r_B \big )B_{\a(n+1)}~,
\eea
\esubeq
where the pseudo-masses $\r_A$ and $\r_B$ take the form
\bea
\r_A= \frac{n}{2n+1}\Big | \s M-(n+2)|\m| \Big |~, \qquad
\r_B= \frac{n+1}{2n+1} \Big | \s M + (n-1)|\m| \Big |~,
\eea
and the corresponding signs of the helicities $\s_A $ and $\s_B$ are  given by
\bea
\s_A &=& \frac{  \s M-(n+2)|\m| }{\big | \s M-(n+2)|\m| \big |}~, \qquad
\s_B = \frac{ \s M + (n-1)|\m|}{\big | \s M + (n-1)|\m| \big |}~.
\eea
In accordance with the definition \eqref{TAOnShellConditions} of an on-shell field in AdS$^3$, it follows from the above analysis  that the component fields $A_{\a(n)}$ and $B_{\a(n+1)}$ realise the irreducible representations $\mathfrak{D} \big ( \r_A, \s_A \frac{n}{2} \big )$ and $\mathfrak{D} \big ( \r_B, \s_B\frac{n+1}{2} \big )$ of $\mf{so}(2,2)$, respectively. Hence, the on-shell superfield \eqref{SOC} furnishes the irreducible representation $\mathfrak{S}(M, \s \frac{n}{2})$ of $\mathfrak{osp}(1|2;{\mathbb R} ) \oplus \mathfrak{sl}(2, {\mathbb R})$ (cf. eq. \eqref{decomp}). 

A superfield satisfying the first condition \eqref{STrans} is said to be transverse. Any transverse superfield may be shown to satisfy the following relation
\be \label{Transprop1}
-\frac{\ri}{2}\nabla^2\F_{\a(n)} = \nabla_{(\a_1}{}^\b\F_{\a_2 ... \a_n)\b} + (n+2)|\m| \F_{\a(n)}~.
\ee
From \eqref{Transprop1} it follows that an on-shell superfield \eqref{SOC} satisfies
\be \label{FOSM}
-\frac{\ri}{2}\nabla^2\F_{\a(n)} = \frac{1}{2n+1}\Big (\s M+2n(n+2)|\m| \Big ) \F_{\a(n)}~,
\ee
and hence  the second-order mass-shell equation
\bsubeq \label{SOMSOM}
\begin{align} \label{SOMS}
0 &= \big ( \mb{Q} -  \l^2\big ) \F_{\a(n)}~, \\
\l^2:=\frac{1}{(2n+1)^2} \big [\s M +2n&(n+2)|\m| \big ] \big [\s M+2(n-1)(n+1)|\m| \big ]~.
\end{align}
\esubeq
The equations \eqref{STrans} and \eqref{FOSM} were introduced in \cite{KuzenkoNovakTartaglino-Mazzucchelli2015}. On the other hand, one may instead consider a superfield $\F_{\a(n)} $ satisfying \eqref{STrans} and \eqref{SOMS}. In this case, using the identity \eqref{FSquared}, one can show that \eqref{SOMS} becomes
\bea
0=\Big ( \mb{F} - \s_{(-)}| M_{(-)}|\Big ) \Big ( \mb{F} - \s_{(+)}|M_{(+)}| \Big )\F_{\a(n)}~,
\eea
where we have defined $\s_{(\pm)}=\text{sgn}(M_{(\pm)}) $ and
\bea
M_{(\pm)} := -(2n^2+2n-1)|\m| \pm (2n+1)\sqrt{\l^2+|\m|^2} ~.
\eea
It follows that such a field furnishes the reducible representation 
\be \label{redS}
\mf{S} \Big (|M_{(-)}|, \s_{(-)} \frac{n}{2} \Big ) \oplus \mf{S} \Big (|M_{(+)}|, \s_{(+)}\frac{n}{2} \Big )~.
\ee

\subsubsection{Massless superfields}
In this section we will introduce the notion of massless on-shell supermultiplets in AdS$^{3|2}$. For $n \geq 3$, an  on-shell superfield $\F_{\a(n)}$ \eqref{SOC} is said to be a massless higher-spin superfield if it does not propagate any physical degrees of freedom. In AdS$^{3|2}$, there are two distinct classes of massless higher-spin superfields which are distinguished by the sign of their superhelicity.

Given $\s =-1$, an on-shell superfield  \eqref{SOC} is said to be massless if it carries the pseudo-mass
\be \label{TASMasslessPMass1}
M = M_{(0,n)}^{(-)}=2(n-1)(n+1)|\m|~.
\ee
It may be shown that the system of equations \eqref{SOC} with pseudo-mass \eqref{TASMasslessPMass1} is compatible with the gauge symmetry
\be \label{TASMasslessGS1}
\d_\L \F_{\a(n)} = \ri^n \nabla_\a \L_{\a(n-1)}~,
\ee
given that the real gauge parameter $\L_{\a(n-1)}$ is also on-shell 
\bsubeq
\bea
0 &=& \nabla^\b \L_{\b\a(n-2)}~, \\
0 &=& \big ( \mb{F} + M^{(-)}_{(0,n)} \big ) \L_{\a(n-1)}~.
\eea
\esubeq

Given $\s =+1$, we say that an on-shell supermultiplet \eqref{SOC} is massless if it carries pseudo-mass
\be
M=M^{(+)}_{(1,n)} = 2n(n-1)|\m|~.
\ee
It can be shown that such a superfield is compatible with the gauge symmetry 
\be
\d_\L \F_{\a(n)} =  \nabla_{\a(2)} \L_{\a(n-2)}~,
\ee
given that the real gauge parameter $\L_{\a(n-2)}$ is also on-shell 
\bsubeq
\bea
0 &=& \nabla^\b \L_{\b\a(n-2)}~, \\
0 &=& \big ( \mb{F} - M^{(+)}_{(1,n)} \big ) \L_{\a(n-2)}~.
\eea
\esubeq

Analysing the component structure of these massless higher-spin superfields, it can be shown that, upon imposing the appropriate Wess-Zumino gauges, the surviving component fields satisfy the massless on-shell conditions in AdS$_3$ (see section \ref{TAmasslesssec}). Thus, massless higher-spin superfields do not propagate physical degrees of freedom in AdS$^{3|2}$. Note that \eqref{TASMasslessGS1} is the gauge symmetry commonly associated with massless (half-)integer superspin actions (see section \ref{TASMasslessHStheories}).

\subsubsection{Partially massless superfields}

In AdS$^{3|2}$ there exist two distinct types of on-shell partially massless superfields \cite{KP21}, which are distinguished by the sign $\s$ of their superhelicity. More specifically, they are described by an on-shell superfield \eqref{SOC} whose pseudo-mass and parameter $\s$ assume the special combinations \cite{KP21}\footnote{In \cite{KP21}, partially massless superfields carrying negative $\s=-1$ and positive $\s=1$ sign of superhelicity were referred to as type A and type B partially massless superfields respectively.  }
\bsubeq \label{PsudeoM}
\begin{align}
\s=+1~,\quad M &\equiv M^{(+)}_{(t,n)} = 2 \big [ n(n-2t+1)-(t-1) \big ] |\m|~,  &1&\leq t \leq \lfloor n/2 \rfloor ~, \\
\s=-1~,\quad M &\equiv M^{(-)}_{(t,n)} = 2\big [n(n-2t)-(t+1) \big ] |\m| ~,  &0&\leq t \leq \lceil n/2 \rceil -1~. \label{typeAPM}
\end{align}
\esubeq
The integer $t$ is called the (super)depth and the corresponding supermultiplets are denoted by $\Phi^{(t,+)}_{\a(n)}$ and $\Phi^{(t,-)}_{\a(n)}$ respectively. For $t>1$, the pseudo-mass $M^{(+)}_{(t,n)}$ violates the unitarity bound \eqref{TASUnitarityBound}. Further more, the pseudo-mass $M^{(-)}_{(t,n)}$ also violates the unitarity bound for $t >0$. Thus, partially massless representations are non-unitary.

It was shown in \cite{KP21} that the gauge symmetry associated with positive and negative superhelicity partially massless superfields of depth-$t$ is
\begin{subequations} \label{SPMG}
	\begin{align}
	\delta_{\L}\F^{(t,+)}_{\a(n)}&=
	\phantom{\rm{i}^n} \big(\mc{D}_{\a(2)}\big)^t \L_{\a(n-2t)}~, &&1 \leq t \leq \lfloor n/2 \rfloor~, \label{SPMG2} \\
	\delta_{\L}\F^{(t,-)}_{\a(n)}&=
	\text{i}^n \big(\mc{D}_{\a(2)}\big)^t\mc{D}_{\a}\L_{\a(n-2t-1)}~, &&0 \leq t \leq \lceil n/2 \rceil - 1~.  \label{SPMG1}
	\end{align}
\end{subequations}
In particular, the system of equations \eqref{SOC} and \eqref{PsudeoM} is only invariant under these transformations for an on-shell real gauge parameter which carries the same pseudo-mass as its parent field. Note that strictly massless superfields with superhelicity sign $\s = +1$ and $\s = -1$  correspond to partially massless superfields $\Phi^{(t,+)}_{\a(n)}$ and $\Phi^{(t,-)}_{\a(n)}$ with minimal super-depth $t=1$ and $t=0$, respectively.

Partially massless superfields satisfy the second order equations \eqref{SOMSOM} 
\be
0 = \big ( \mb{Q} - \l_{(t,n)}^{(+)} |\m|^2 \big ) \F_{\a(n)}^{(t,+)}~, \quad 0= \big ( \mb{Q} - \l_{(t,n)}^{(-)} |\m|^2 \big ) \F_{\a(n)}^{(t,-)}~,
\ee
where the dimensionless constants
\be
\l_{(t,n)}^{(+)}=4(n-t)(n-t+1)~, \qquad \l_{(t,n)}^{(-)} = 4t(t+1)~,
\ee
correspond to the partially massless values.

\subsubsection{Massive superfields}
An on-shell superfield \eqref{SOC} is said to be massive if it carries pseudo-mass satisfying
\be \label{TASMassiveMass}
M > M^{(-)}_{(0,n)}~.
\ee
This pseudo-mass does not violate the unitarity bound \eqref{TASUnitarityBound}, hence the corresponding representation is unitary.  We can also introduce the notion of a (non-unitary) massive supermultiplet, which carries the defining pseudo-mass $M < M^{(-)}_{(0,n)}$ such that $M \neq $ $M^{(-)}_{(t,n)} $ for  $1 \leq t \leq \lceil n/2 \rceil -1$ and $M \neq $ $M^{(+)}_{(t,n)} $  for $1\leq t \leq \lfloor n/2 \rfloor$. We will not be interested in (non-unitary) massive superfields in this chapter.

\subsection{Superspin projection operators} \label{TASSuperprojectorsAdS32}

In this section we study the supersymmetric generalisations of 
the AdS$_3$ spin projection operators and their corresponding applications. 
For any integer $n\geq 1$, the
rank-$n$ superspin projection operator $\bm\P^{\perp}_{[n]}$ is defined by its action on $\mathds{V}_{(n)}$ 
according to the rule:
\bsubeq
\bea
\bm \Pi^{\perp}_{[n]}: \mathds{V}_{(n)} &\longrightarrow& \mathds{V}_{(n)}~, \\
\F_{\a(n)}  &\longmapsto&  \bm \Pi^{\perp}_{[n]} \F_{\a(n)}~  =:\bm{\P}^{\perp}_{\a(n)}(\F)~.
\eea
\esubeq
For fixed $n$, the superspin projection operator $\bm\P^{\perp}_{[n]}$ is defined by the following properties:
\begin{enumerate} 
	\item \textbf{Idempotence:} The operator $\bm \Pi^{\perp}_{[n]}$ squares to itself, 
	\bsubeq  \label{SProp}
	\be 
	\bm \Pi^{\perp}_{[n]}\bm \Pi^{\perp}_{[n]}=\bm \Pi^{\perp}_{[n]}~.
	\ee
	\item \textbf{Transversality:} The operator $\bm \Pi^{\perp}_{[n]}$ maps $\F_{\a(n)}$ to a transverse superfield,
	\be 
	\nabla^{\b}\bm{\P}^{\perp}_{\b\a(n-1)}(\F) =0~.
	\ee
	\item \textbf{Surjectivity:} Every transverse superfield  $\F^{\perp}_{\a(n)}$  belongs to the image of $\bm \Pi^{\perp}_{[n]}$,
	\be \label{SIT}
	\mc{D}^{\b}\F^{\perp}_{\b\a(n-1)}=0~\quad\implies\quad\bm \Pi^{\perp}_{[n]} \F^{\perp}_{\a(n)} = \F^{\perp}_{\a(n)}~.
	\ee
	\esubeq
\end{enumerate}

Let us consider a superfield $\F_{\a(n)}$ on $\mathds{V}_{(n)}$ which satisfies the first-order differential constraint \eqref{SFO}. It follows that the superspin projection operator $\bm \Pi^{\perp}_{[n]}$ maps $\F_{\a(n)}$ to an on-shell superfield \eqref{SOC},
\bsubeq
\bea
\nabla^{\b}\bm{\P}^{\perp}_{\b\a(n-1)}(\F) &=& 0~, \\
\big ( \mathbb{F} - \s M \big ) \bm{\P}^{\perp}_{\a(n)}(\F)&=&0~.
\eea
\esubeq
In other words, $\bm \Pi^{\perp}_{[n]}$ extracts the physical component from $\F_{\a(n)}$ which realises the irreducible representation $\mathfrak{S}(M, \s \frac{n}{2})$ of $\mathfrak{osp}(1|2;{\mathbb R} ) \oplus \mathfrak{sl}(2, {\mathbb R})$.

To obtain a superprojector, we introduce the operator $\D^{\a}{}_\b$ \cite{KuzenkoPonds2018}
\be \label{TransOp}
\D^{\a}{}_\b:=-\frac{\ri}{2}\nabla^\a \nabla_\b - 2|\m| \d^{\a}{}_{\b}~, \quad \nabla^\b \D^{\a}{}_\b =  \D^{\a}{}_\b \nabla_\a = 0~,
\ee
and its corresponding extensions \cite{KP21} 
\be \label{ExTransOp}
\D^{\a}_{[j]}{}_\b := -\frac{\ri}{2}\nabla^\a \nabla_\b - 2j|\m| \d^{\a}{}_{\b}~. 
\ee
Note that for the case $j=1$, \eqref{ExTransOp} coincides with \eqref{TransOp}. It can be shown that the operator \eqref{ExTransOp} has the following properties
\bsubeq \label{Ident}
\bea
[\D^{\a_1}_{[j]}{}_{\b_1},\D^{\a_2}_{[k]}{}_{\b_2}] &=& \varepsilon_{\b_1 \b_2}|\m| \big ( \nabla^{\a(2)}-|\m| M^{\a(2)} \big ) - \varepsilon^{\a_1 \a_2}|\m| \big ( \nabla_{\b(2)}-|\m| M_{\b(2)} \big ) ~,  \hspace{1cm} \label{Ident1}\\
\varepsilon^{\b_1 \b_2}\D^{\a_1}_{[j]}{}_{\b_1}\D^{\a_2}_{[j+1]}{}_{\b_2}&=&-j  \varepsilon^{\a_1 \a_2}|\m| \big (\ri \nabla^2 + 4(j+1)|\m|^2 \big )~, \label{Ident2} \\
\varepsilon_{\a_1 \a_2}\D^{\a_1}_{[j+1]}{}_{\b_1}\D^{\a_2}_{[j]}{}_{\b_2}&=& j  \varepsilon_{\b_1 \b_2}|\m| \big (\ri \nabla^2 + 4(j+1)|\m|^2 \big )~, \label{Ident3}\\
\D^{\b}_{[j]}{}_{\a}\D^{\g}_{[k]}{}_{\b}&=& - \frac{\ri}{2}\nabla^2 \D^{\g}_{[1]}{}_{\a} +  (j+k-1) \ri  |\m|  \nabla^\g \nabla_\a + 4jk|\m|^2 \d_\a{}^\g ~,\label{Ident4}  \\
\ [ \D^{\a}_{[j]}{}_\b, \nabla^2 ] &=& 0~,
\eea
\esubeq
for arbitrary integers $j$ and $k$. 

Let us define the operator $\mathbb{T}_{[n]}$, which acts on $\mathds{V}_{(n)}$ by the rule
\be \label{TSO}
\mathbb{T}_{[n]} \F_{\a(n)} \equiv \mathbb{T}_{\a(n)}(\F) =\D^{\b_1}_{[1]}{}_{(\a_1}\D^{\b_2}_{[2]}{}_{\a_2} \cdots \D^{\b_n}_{[n]}{}_{\a_n)}\F_{\b(n)}~. 
\ee
This operator maps $\F_{\a(n)}$ to a transverse superfield
\be \label{STransprop}
\nabla^\b \mathbb{T}_{\b\a(n-1)}( \F)=0~.
\ee
To see this, one needs to open the symmetrisation in \eqref{TSO}
\bea
\nabla^\g \mathbb{T}_{\g\a(n-1)}( \F) &=&  \nabla^\g\D^{\b_1}_{[1]}{}_{(\g}\D^{\b_2}_{[2]}{}_{\a_1} \cdots \D^{\b_n}_{[n]}{}_{\a_{n-1})}\F_{\b(n)}~ \non \\
&\propto&\nabla^\g \big (\D^{\b_1}_{[1]}{}_{\g}\D^{\b_2}_{[2]}{}_{\a_1} \cdots \D^{\b_n}_{[n]}{}_{\a_{n-1}}+ (n!-1)~\text{permutations} \big ) \F_{\b(n)}~.  \hspace{1cm} 
\eea
By making use of \eqref{Ident2}, it can be shown that the remaining $(n!-1)$ terms can be expressed in the same form as the first. Then transversality follows immediately as a consequence of property \eqref{TransOp}. However, $\mathbb{T}_{[n]}$ does not square to itself on $\mathds{V}_{(n)}$
\bea 
\mathbb{T}_{[n]} \mathbb{T}_{[n]} \F_{\a(n)} = \frac{1}{(2n+1)^n} \prod_{t=0}^{\lceil n/2 \rceil - 1}\big (\mathbb{F}+M^{(-)}_{(t,n)} \big )\prod_{t=1}^{\lfloor n/2 \rfloor }\big (\mathbb{F}-M^{(+)}_{(t,n)} \big )\mathbb{T}_{[n]} \F_{\a(n)}  ~,
\eea
where $M^{(\pm)}_{(t,n)}$ denotes the pseudo-masses associated with a partially massless superfield \eqref{PsudeoM}. We can immediately introduce the dimensionless operator  
\bea \label{superprojector}
\bm \P^{\perp}_{[n]}\F_{\a(n)} := (2n+1)^n \bigg [\prod_{t=0}^{\lceil n/2 \rceil - 1}\big (\mathbb{F}+M^{(-)}_{(t,n)} \big )\prod_{t=1}^{\lfloor n/2 \rfloor }\big (\mathbb{F}-M^{(+)}_{(t,n)} \big ) \bigg ]^{-1} \mathbb{T}_{[n]}\F_{\b(n)}~,
\eea
which is idempotent and transverse by construction. In addition, it can be shown that the operator $\bm \P^{\perp}_{[n]}$ acts as the identity on the space of transverse superfields \eqref{SIT}. Hence, $\bm \P^{\perp}_{[n]}$ satisfies properties \eqref{SProp} and can be identified as a rank-$n$ superprojector on AdS$^{3|2}$. So far, we have been unable to obtain an expression for ${\bm \P}^\perp_{[n]}$ which is purely in terms of the Casimir operators $\mb{F}$ and $\mb{Q}$.

An alternative form of the superprojector $\bm \P^{\perp}_{[n]}$ can be derived, which instead makes contact with the Casimir operator $\mb{Q}$. Let us introduce the dimensionless operator 
\bea \label{altsupproj}
\widehat{\bm \P}{}^\perp_{[n]}\F_{\a(n)} &=& \bigg [\prod_{t=0}^{n-1}\big (\mathbb{Q}+\ri t |\m| \nabla^2 \big ) \bigg ]^{-1} \widehat{\D}^{\b_1}_{[1]}{}_{(\a_1}\widehat{\D}^{\b_2}_{[2]}{}_{\a_2} . . . \widehat{\D}^{\b_n}_{[n]}{}_{\a_n)}\F_{\b(n)}~,
\eea
where we denote  $\widehat{\D}^{\b}_{[j]}{}_{\a}$ as
\be \label{altop}
\widehat{\D}^{\b}_{[j]}{}_{\a}:= - \frac{\ri}{2}\nabla^2 {\D}^{\b}_{[j]}{}_{\a}~.
\ee
Making use of the properties of $\bm \P^{\perp}_{[n]}$ and the identity
\be \label{SCasF}
- \frac{\ri}{2}\nabla^2 \F^{\perp}_{\a(n)} = \frac{1}{2n+1} \big (\mathbb{F}  + 2n(n+2) |\m| \big )\F^{\perp}_{\a(n)}~,
\ee
where $\F^{\perp}_{\a(n)} $ is an arbitrary transverse superfield, it can be shown that $\widehat{\bm \P}{}^\perp_{[n]}\F_{\a(n)}$ satisfies properties \eqref{SProp} and thus is also a superprojector on AdS$^{3|2}$. In the flat-superspace limit $|\m| \rightarrow 0$, the superspin projection operator $\widehat{\bm \P}{}^\perp_{[n]}$ coincides with the superprojector \eqref{TMSN1ProjectorE} in $\mb{M}^{3|2}$. It can be shown that ${\bm \P}^\perp_{[n]}$ and $\widehat{\bm \P}{}^\perp_{[n]}$ are equivalent on $\mathds{V}_{(n)}$. 

We recall that applying the AdS$_3$ spin projection operator  $\P^{\perp}_{[n]}$ to a field $\f_{\a(n)}$ on the mass-shell \eqref{OMS2} yields the projection $\P^{\perp}_{[n]}\f_{\a(n)}$ which  furnishes the reducible representation \eqref{red}. A single irreducible representation can be singled out from the decomposition \eqref{red} via the application of the helicity projectors \eqref{TAhelicityproj}. The significance of the condition \eqref{OMS2} is that it allows one to resolve the poles in both types of projectors.

In the supersymmetric case, the equation analogous to \eqref{OMS2} which $\F_{\a(n)}$ should satisfy is \eqref{SOMS}. Applying $\bm \P^{\perp}_{[n]}$ to such a $\F_{\a(n)}$, one obtains the reducible representation \eqref{redS}. However, it appears that the imposition of \eqref{SOMS} does not allow one to resolve the poles of the superprojector in either of the forms \eqref{superprojector} or \eqref{altsupproj}. Therefore, rather then imposing \eqref{SOMS}, one must start with a superfield $\F_{\a(n)}$ obeying the first-order constraint \eqref{SFO}, which does allow for resolution of the poles. In this case, after application of $\bm \P^{\perp}_{[n]}$, the superfield $\F_{\a(n)}$ already corresponds to an irreducible representation with fixed superhelicity, relinquishing the need for superhelicity projectors. Thus, it suffices to provide only the superspin projection operators  $\bm \P^{\perp}_{[n]}$.

\subsubsection{Longitudinal projectors}\label{TASLongProj22}
For $n\geq 1$, let us define the orthogonal complement of $\bm \P^{\perp}_{[n]}$  on $\mds{V}_{(n)}$ by the rule
\be \label{SLongProj}
\bm \P^\parallel_{[n]}\F_{\a(n)} =  \big (\mathds{1}-\bm \P^{\perp}_{[n]} \big )\F_{\a(n)} ~.
\ee
By construction, the operators $\bm \P^{\perp}_{[n]}$ and $\bm \P^\parallel_{[n]}$ resolve the identity, $\mathds{1} = \bm \Pi^{\parallel}_{[n]} + \bm \Pi^{\perp}_{[n]}$, and are orthogonal projectors 
\be \label{SOrthoProjProp}
\bm \Pi^{\perp}_{[n]}\bm \Pi^{\perp}_{[n]} = \bm \Pi^{\perp}_{[n]}~, \qquad \bm \Pi^{\parallel}_{[n]}\bm \Pi^{\parallel}_{[n]} = \bm \Pi^{\parallel}_{[n]}~, \qquad \bm \Pi^{\parallel}_{[n]} \bm \Pi^{\perp}_{[n]} = \bm \Pi^{\perp}_{[n]} \bm \Pi^{\parallel}_{[n]}  = 0~. 
\ee
It can be shown that $\bm \P^\parallel_{[n]}$ extracts the longitudinal component of a superfield $\F_{\a(n)}$. A rank-$n$ superfield  $\F^{\parallel}_{\a(n)}$ on $\mds{V}_{(n)}$ is said to be longitudinal if there exists a rank-$(n-1)$ superfield $\F_{\a(n-1)}$ such that $\F^{\parallel}_{\a(n)}$ can be expressed as $\F^{\parallel}_{\a(n)}= \ri^n\nabla_\a \F_{\a(n-1)}$. Thus, we find
\be
\bm \P^\parallel_{[n]}\F_{\a(n)} = \ri^n \nabla_{\a}\F_{\a(n-1)}~,
\ee
for some unconstrained real superfield $\F_{\a(n-1)}$. In order to see this, it proves beneficial to make use of the superprojector $\widehat{\bm \P}{}^{\perp}_{[n]}$, and express the operator $\widehat{\D}^{\b}_{[j]}{}_{\a}$ in the form
\be 
\widehat{\D}^{\b}_{[j]}{}_{\a}: = -\frac{1}{4}   \nabla_\a \nabla^\b \nabla^2 + \big (\mathbb{Q}+\ri (j-1)|\m|\nabla^2\big )\d_\a{}^\b ~. 
\ee

Let us consider the action of $\bm \P^{\perp}_{[n]}$ on the longitudinal superfield $\F^{\parallel}_{\a(n)}$. Opening the symmetrisation present in $\bm \P^{\perp}_{[n]}$ gives
\bea
\bm \P^{\perp}_{[n]}\F^{\parallel}_{\a(n)}&=& \ri^n\D^{\b_1}_{[1]}{}_{(\a_1}\D^{\b_2}_{[2]}{}_{\a_2} . . . \D^{\b_n}_{[n]}{}_{\a_n)}\nabla_{(\b_1} \F_{\b_2 ... \b_n)}~ \\
&=&\frac{\ri^n}{n!}\D^{\b_1}_{[n]}{}_{(\a_1}\D^{\b_2}_{[n-1]}{}_{\a_2} . . . \D^{\b_n}_{[1]}{}_{\a_n)}\big (\nabla_{\b_n} \F_{\b_1 ... \b_{n-1}} + (n!-1)~\text{permutations} \big )~. \non \hspace{1cm}
\eea 
Note that we have made use of the identity \eqref{Ident1} to rearrange the operators $\D^{\b}_{[j]}{}_{\a}$.
Making use of the relation \eqref{Ident3} allows us to express the other $(n!-1)$ permutations in the same form as the first. Then due to the property \eqref{TransOp}, it follows that 
\be \label{SPKL}
\F^{\parallel}_{\a(n)}=\ri^n \nabla_{\a}\F_{\a(n-1)}\qquad \implies \qquad  \bm \P^{\perp}_{[n]}\F^{\parallel}_{\a(n)} = 0~.
\ee
Thus, the superprojector $\bm \P^{\perp}_{[n]}$ annihilates any longitudinal superfield. 
Consequently,  $\bm \P^\parallel_{[n]}$ acts as the identity operator on the space of rank-$n$ longitudinal superfields 
\be \label{LongIden}
\F^{\parallel}_{\a(n)}=\ri^n \nabla_{\a}\F_{\a(n-1)}\qquad \implies \qquad \bm \P^{\parallel}_{[n]}\F^{\parallel}_{\a(n)}= \F^{\parallel}_{\a(n)}~.
\ee

Using the fact that the $\bm \Pi^{\parallel}_{[n]}$ and $\bm \Pi^{\perp}_{[n]}$ resolve the identity, it follows that one can decompose any superfield $\F_{\a(n)}$ in the following manner
\be
\F_{\a(n)} = \F^{\perp}_{\a(n)} +\ri^n\nabla_{\a}\F_{\a(n-1)}~.
\ee
Here, $\F^{\perp}_{\a(n)}$ is transverse and $\F_{\a(n-1)}$ is unconstrained. 
Repeating this prescription iteratively yields the decomposition
\bsubeq \label{TAS22Decomp}
\bea
\F_{\a(n)}&=& \sum_{j=0}^{\lfloor n/2 \rfloor} \big (  \nabla_{\a(2)} \big )^j \F^{\perp}_{\a(n-2j)} + \ri^{n}\sum_{j=0}^{\lceil n/2 \rceil -1} \big ( \nabla_{\a(2)} \big )^j \nabla_\a \F^{\perp}_{\a(n-2j-1)}~.
\eea
\esubeq
Here, the real superfields $\F^{\perp}_{\a(n-2j)}$ and $\F^{\perp}_{\a(n-2j-1)}$  are transverse, except for $\F^{\perp}$.

\subsection{Linearised higher-spin super-Cotton tensors}\label{TASSCHSSec}
In this section we make use of the rank-$n$ superspin projection operators $\bm \P^{\perp}_{[n]}$ to construct the linearised higher-spin super-Cotton tensors in AdS$^{3|2}$. These super-Cotton tensors were recently derived in \cite{KP21}, where they were shown to take the explicit form
\be \label{SCT}
\mathfrak{W}_{\a(n)}(H) =\D^{\b_1}_{[1]}{}_{(\a_1}\D^{\b_2}_{[2]}{}_{\a_2} \cdots \D^{\b_n}_{[n]}{}_{\a_n)}H_{\b(n)}~,
\ee
which is a real primary descendent of the SCHS superfield $H_{\a(n)}$. The latter is defined modulo gauge transformations of the form
\be \label{SCHSGT}
\d_{\L}H_{\a(n)} = \ri^n \nabla_\a \L_{\a(n-1)}~,
\ee
where the gauge parameter $\L_{\a(n-1)}$ is a real unconstrained superfield.
The super-Cotton tensor \eqref{SCT} satisfies the defining properties: 
\begin{enumerate}
	\bsubeq \label{SCTP}
	\item $\mathfrak{W}_{\a(n)}(H)$ is transverse,
	\be
	\nabla^\b\mathfrak{W}_{\b\a(n-1)}(H) = 0~.
	\ee
	\item $\mathfrak{W}_{\a(n)}(H)$ is invariant under the gauge transformations \eqref{SCHSGT},
	\be
	\d_{\L}	\mathfrak{W}_{\a(n)}(H)= 0~.
	\ee
	\esubeq	
\end{enumerate}

The superprojectors \eqref{superprojector} can be used to recast the
super-Cotton tensors \eqref{SCT} in the simple form
\be \label{SCTPr}
\mathfrak{W}_{\a(n)}(H) = \frac{1}{(2n+1)^n}\prod_{t=0}^{\lceil n/2 \rceil - 1}\big (\mathbb{F}+M^{(-)}_{(t,n)} \big )\prod_{t=1}^{\lfloor n/2 \rfloor }\big (\mathbb{F}-M^{(+)}_{(t,n)} \big )  \bm \P^{\perp}_{[n]}H_{\a(n)} ~,
\ee
where $M^{(\pm)}_{(t,n)}$ denotes the partially massless pseudo-masses \eqref{PsudeoM}. In the flat-superspace limit,  the super-Cotton tensor \eqref{SCTPr} reduces to its counterpart \eqref{3.99} in $\mb{M}^{3|2}$. Expressing $\mathfrak{W}_{\a(n)}(H) $ in the form \eqref{SCTPr} is beneficial for the following reasons: (i) transversality of $\mathfrak{W}_{\a(n)}(H)$ is made manifest on account of property \eqref{STransprop}; (ii)  gauge invariance is also manifest as a consequence of \eqref{SPKL}; and (iii) in the transverse gauge
\be \label{TransG}
H_{\a(n)} \equiv H^{\perp}_{\a(n)}~, \qquad \nabla^\b H^{\perp}_{\b\a(n-1)} = 0~,
\ee
it follows from \eqref{SIT} that $\mathfrak{W}_{\a(n)}(H)$ factorises as follows
\be \label{KinF}
\mathfrak{W}_{\a(n)}(H^{\perp}) = \frac{1}{(2n+1)^n}\prod_{t=0}^{\lceil n/2 \rceil - 1}\big (\mathbb{F}+M^{(-)}_{(t,n)} \big )\prod_{t=1}^{\lfloor n/2 \rfloor }\big (\mathbb{F}-M^{(+)}_{(t,n)} \big )  H^{\perp}_{\a(n)} ~.
\ee

From the above observations, it follows that the action \cite{KuzenkoPonds2018,KuzenkoPonds2019} for the superconformal higher-spin prepotential $H_{\a(n)}$
\begin{align}
\mb{S}^{(n)}_{\rm SCHS}[H] = 
- \frac{\ri^n}{2^{\left \lfloor{n/2}\right \rfloor +1}} \int \rd^3 x \rd^2 \theta  \, E \, H^{\a(n)} \mf{W}_{\a(n)}(H) ~, 
\label{SCS}
\end{align}
is manifestly gauge-invariant. In the transverse gauge \eqref{TransG}, the kinetic operator in \eqref{SCS} factorises into wave operators associated with partially massless superfields of all depths, in accordance with \eqref{KinF}.

\subsection{Massless higher-spin theories} \label{TASMasslessHStheories}
According to \cite{HK19}, for each superspin value $\hat{s}\geq 1$, where $\hat{s}$ is either integer $(\hat{s}=s)$ or half-integer $(\hat{s}=s+\frac{1}{2})$, there exist two off-shell formulations for a massless $\cN=1$ superspin-$\hat{s}$ multiplet in 
AdS$_3$. These two formulations are referred to as being either longitudinal or transverse.
In this section, we review the explicit formulation of these respective theories, which were derived in \cite{HK19,KuzenkoPonds2018}.

An off-shell massless superspin-$\frac{n}{2}$ gauge theory in AdS${}_3$ can be realised in terms of two superfields: a superconformal gauge prepotential (see section \ref{TASSCHSSec}); and a compensating multiplet. The longitudinal and transverse formulations are distinguished by the compensating multiplet and its corresponding gauge transformations. Given an integer $n \geq 4$, a massless higher-spin theory is said to have integer superspin or half-integer superspin if the associated superconformal prepotential $H_{\a(n)}$ has an even $n=2s$  or odd $n=2s+1$ number of indices, respectively. 

To study the dynamics of $\cN = 1$ supersymmetric field theories
in AdS$_3$, a manifestly supersymmetric action principle is required. 
Such an action is associated with a real scalar Lagrangian $L$ and has the form
\bea
S = \int \rd^3x\,\rd^2\q E\, L~, \qquad E^{-1} = {\rm Ber}(E_A{}^{M})~.
\label{212}
\eea
In what follows, we make use of the notation $\rd^{3|2}z := \rd^3x \rd^2 \theta$.  Making use of the $\cN=1$  AdS transformation law $\d_\xi L = \xi L$ and the identities \eqref{2.40d-1} and \eqref{2.40e}, it can be shown that the action
\eqref{212}  is invariant under the isometry group of AdS$^{3|2}$.


\subsubsection{Longitudinal formulation for the massless superspin-$(s+\hf)$ multiplet} 
The longitudinal formulation for the massless superspin-$(s+\hf)$ multiplet is realised in terms of the real unconstrained variables
\be \label{MLHI}
\cV^\parallel_{(s+\frac{1}{2})} = \big \{ H_{\a(2s+1)} , L_{\a(2s-2)} \big \}~,
\ee
which are defined modulo gauge transformations of the form
\begin{subequations} \label{LFHIGT}
	\bea
	\d_\z H_{\a(2s+1)} &=& \ri \nabla_{\a} \z_{\a(2s)}~, \\
	\d_\z L_{\a(2s-2)} &=& - \frac{s}{2(2s+1)}\nabla^{\b\g}\z_{\b\g\a(2s-2)}~,
	\eea
\end{subequations}
where the real gauge parameter $\z_{\a(2s)}$ is unconstrained. 
The unique gauge-invariant action 
formulated 
in terms of the superfields \eqref{MLHI} takes the following form
\bea \label{LongHalfIntAct}
&&{S^{\parallel}_{(s+\hf)}[H_{\a(2s+1)} ,L_{\a(2s-2)}] = \Big(-\hf \Big)^{s} 	\int \rd^{3|2}z \, E\, \bigg\{-\frac{\ri}{2} H^{\a(2s+1)} {\mathbb{Q}} H_{\a(2s+1)} }
\non \\
&& -\frac{\ri}{8} \nabla_{\b} H^{\b \a(2s)} \nabla^2 \nabla^{\g}H_{\g \a(2s)} +\frac{\ri s}{4}{\nabla}_{\b \g}H^{\b \g \a(2s-1)} {\nabla}^{ \d \l}H_{ \d \l \a(2s-1)}  \non \\
&&+ (2s-1) L^{\a(2s-2)}\nabla^{\b \g} \nabla^{\d} H_{\b \g \d \a(2s-2)}
\non \\
&&  + 2 (2s-1)\Big( L^{\a(2s-2)} (\ri \nabla^2 - 4 |\m|) L_{\a(2s-2)}
- \frac{\ri}{s}(s-1) \nabla_{\b} L^{\b \a(2s-3)} \nabla^{\g}L_{\g \a(2s-3)}\Big)
\non \\
&&  + |\m| \Big(s \,\na_{\b}H^{\b \a(2s)} \na^{\g} H_{\g \a(2s)}+ \hf (2s+1)H^{\a(2s+1)}( \nabla^2-4 \ri |\m|)H_{\a(2s+1)} \Big)
\bigg\}~,
\eea
where $\mathbb{Q}$ is the quadratic Casimir operator 
\eqref{QQC}.
The action \eqref{LongHalfIntAct} was first derived
in \cite{KuzenkoPonds2018}. In the flat-superspace limit $|\m|\rightarrow 0$,  the action \eqref{LongHalfIntAct} coincides with the model derived in \eqref{TMSSecondOrderMassless} upon the field redefinition $L_{\a(2s-2)} = - \hf Y_{\a(2s-2)}$.


\subsubsection{Transverse formulation for the massless superspin-$(s+\hf)$ multiplet} 
The transverse formulation for the massless superspin-$(s+\hf)$ multiplet  is constructed in terms of the real superfields
\be \label{THI}
\cV^\perp_{(s+\frac{1}{2})} = \big \{ H_{\a(2s+1)}, \U_{\b;\a(2s-2)} \big \}~, 
\ee
where $\U_{\b;\a(2s-2)}$ is a reducible superfield pertaining to the representation $\bm{ 2 \otimes (2s-1)}$ of $\mathsf{SL}(2, \mathbb{R})$. The superfield $\U_{\b;\a(2s-2)}$ can be decomposed into irreducible components by the following rule
\be
\U_{\b;\a(2s-2)} = \U_{(\b;\a_1 ... \a_{2s-2})} + \frac{1}{2s-1} \sum_{k=1}^{2s-2}\varepsilon_{\b\a_k}\U^{\g;}{}_{\g\a_1 ... \hat{\a}_k ... \a_{2s-2}}~.
\ee
where the hatted index of $\U^{\g;}{}_{\g\a_1 ... \hat{\a}_k ... \a_{2s-2}}$ is omitted.
The superfields \eqref{THI}  are defined modulo gauge transformations of the form
\begin{subequations}  \label{GTTHI}
	\bea
	\d_\z H_{\a(2s+1)} &=& \ri \nabla_{\a}\z_{\a(2s)}~, \\
	\d_{\z,\eta} \U_{\b ; \a(2s-2)} &=& \frac{\ri}{2s+1}\Big (\nabla^\g \z_{\g\b\a(2s-2)} + (2s+1)\nabla_\b \eta_{\a(2s-2)} \Big )~,
	\eea	
\end{subequations}	
where the gauge parameters $\z_{\a(2s)}$ and $\eta_{\a(2s-2)}$ are real unconstrained. The unique action which is invariant under the gauge transformations \eqref{GTTHI} takes the form
\bea \label{TransHIAct}
&&S^{\perp}_{(s+\hf)}[{H}_{\a(2s+1)} ,\U_{\b ; \a(2s-2)}  ]
= \Big(-\hf \Big)^{s} 
\int \rd^{3|2}z \, E \,\bigg\{-\frac{\ri}{2} H^{\a(2s+1)} {\mathbb{Q}} H_{\a(2s+1)}
\non \\
&&   -\frac{\ri}{8} \nabla_{\b} H^{\b \a(2s)} \nabla^2 \nabla^{\g}H_{\g \a(2s)}+\frac{\ri}{8}{\nabla}_{\b \g}H^{\b \g \a(2s-1)} {\nabla}^{ \d \l}H_{ \d \l \a(2s-1)}
\non \\
&&   -\frac{\ri}{4}(2s-1) \O^{\b; \,\a(2s-2)} \nabla^{\g \d}H_{\g \d \b \a(2s-2)}  + \ri s (2s-1) |\m| \,\U^{\b ;\, \a(2s-2)} \O_{\b ; \, \a(2s-2)}
\non \\
&&   -\frac{\ri}{8}(2s-1)\Big(\O^{\b ;\, \a(2s-2)} \O_{\b ;\, \a(2s-2)}
-2(s-1)\O_{\b;}\,^{\b \a(2s-3)} \O^{\g ;}\,_{\g \a(2s-3)}  \Big) 
\non \\
&&   
+ |\m| \Big( H^{\a(2s+1)} \big( \nabla^2 - 4\ri |\m|\big) H_{\a(2s+1)} 
+ \hf \ \na_{\b}H^{\b \a(2s)} \na^{\g}H_{\g \a(2s)}
\Big) \bigg\}~,
\eea
where $\O_{\b ; \a(2s-2)}$ corresponds to the real $\cN=1$ field strength
\bea
\O_{\b;\, \a(2s-2)}
= -\ri \big( \nabla^{\g} \nabla_{\b} - 4\ri |\m| \delta_{\b}\,^{\g}\big){\U}_{\g; \,\a(2s-2)} ~,
\qquad \nabla^\b  \O_{\b;\, \a(2s-2)}=0~.
\eea
The above theory was recently introduced in \cite{HK19}.



\subsubsection{Longitudinal formulation for the massless superspin-$s$ multiplet} 
The longitudinal formulation for the massless superspin-$s$ multiplet is realised in terms of the real unconstrained variables
\be
\cV^\parallel_{(s)} = \big \{ H_{\a(2s)}, V_{\a(2s-2)} \big  \}~,
\ee
which are defined modulo gauge transformations of the form
\begin{subequations} \label{LongGTI}
	\bea 
	\d_\z H_{\a(2s)} &=& \nabla_{\a}\z_{\a(2s-1)}~, \\
	\d_\z V_{\a(2s-2)} &=& \frac{1}{2s}\nabla^\b \z_{\b\a(2s-2)}~,
	\eea
\end{subequations}
where the gauge parameter $\z_{\a(2s-1)}$ is real unconstrained. The unique action which is invariant under the gauge transformations \eqref{LongGTI} takes the following form
\bea\label{action-t3}
\lefteqn{S^{\parallel}_{(s)}[H_{\a(2s)} ,V_{\a(2s-2)} ]
	= \Big(-\hf \Big)^{s} 
	\int 
	\rd^{3|2}z
	\, E \, \bigg\{
	\frac{1}{2} H^{\a(2s)} \big(\ri \na^2  +4 |\m|\big)H_{\a(2s)} }
\non \\
&&  
- \frac{\ri}{2}\na_{\b}H^{\b \a(2s-1)} \na^{\g}H_{\g \a(2s-1)}
-(2s-1) V^{\a(2s-2)} \nabla^{\b \g} H_{\b \g \a(2s-2)}
\\
&&  +(2s-1)\Big(\hf V^{\a(2s-2)} \big( \ri \nabla^2 +8s|\m|\big)V_{\a(2s-2)}
+ \ri (s-1) \nabla_{\b}V^{\b \a(2s-3)} \nabla^{\g}V_{\g \a(2s-3)} \Big) \bigg\}
~.
\non
\eea
Up to an overall normalisation factor, the action \eqref{action-t3} coincides with the off-shell $\cN=1$  action for the massless superspin-$s$ multiplet in \cite{KuzenkoPonds2018}. In the flat-superspace limit, the action \eqref{action-t3} reduces to \eqref{TASFirstOrderAction} in $\mb{M}^{3|2}$.


\subsubsection{Transverse formulation for the massless superspin-$s$ multiplet} 
The transverse formulation for the massless superspin-$s$ multiplet is constructed in terms of the real unconstrained superfields
\be
\cV^{\perp}_{(s)} = \big \{ H_{\a(2s)}, \J_{\b ; \a(2s-2)} \big \}~,
\ee
which are defined modulo gauge transformations of the form
\begin{subequations} \label{TIGT}
	\bea
	\d_\z H_{\a(2s)} &=& \nabla_{\a} \z_{\a(2s-1)}~, \\
	\label{psigt}
	\d_{\z , \eta} \J_{\b ; \a(2s-2)} &=& - \z_{\b\a(2s-2)} + \ri \nabla_\b \eta_{\a(2s-2)}~,
	\eea
\end{subequations}
where the gauge parameters $\z_{\a(2s-1)}$ and $\eta_{\a(2s-2)}$ are real unconstrained. The unique gauge-invariant action is given by 
\bea \label{TIAction}
&&S^{\perp}_{(s)}[H_{\a(2s)} , \J_{\b ; \a(2s-2)} ]
= \Big(-\hf \Big)^{s} 
\int \rd^{3|2}z \,
E\, \bigg\{\frac{1}{2} H^{\a(2s)} (\ri \nabla^2 +8 s |\m|) H_{\a(2s)}
\non \\
&&  - \ri s \nabla_{\b} H^{\b \a(2s-1)} \nabla^{\g}H_{\g \a(2s-1)} 
-(2s-1) \cW^{\b ;\,\a(2s-2)} \nabla^{\g} H_{\g \b \a(2s-2)}
\non \\
&&   -\frac{\ri}{2} (2s-1)\Big(\cW^{\b ;\, \a(2s-2)} \cW_{\b ;\, \a(2s-2)}+\frac{s-1}{s} \cW_{\b;}\,^{\b \a(2s-3)} \cW^{\g ;}\,_{\g \a(2s-3)} \Big) 
\non\\
&&  
-2 \ri (2s-1) |\m| \Psi^{\b ;\, \a(2s-2)} \cW_{\b ; \, \a(2s-2)}
\bigg\}~,
\eea
where $\cW_{\b; \, \a(2s-2)}$ denotes the real  $\cN=1$ field strength
\bea
\cW_{\b;\, \a(2s-2)} =
-\ri \big( \nabla^{\g} \nabla_{\b} - 4\ri |\m| \delta_\b{}^\g \big){\Psi}_{\g; \,\a(2s-2)} ~,
\qquad \nabla^\b  \cW_{\b;\, \a(2s-2)}=0~.
\eea

For $s > 1$, the $\z$-gauge freedom \eqref{psigt}
can be used to impose the gauge condition
\bea
\J_{(\a_1;\, \a_2 \dots \a_{2s-1})} =0 \quad \Longleftrightarrow \quad 
\J_{\b ;\, \a(2s-2)} =  \sum_{k=1}^{2s-2}\ve_{\b \a_k} \vf_{\a_1 \dots \hat{\a}_k \dots \a_{2s-2}}~,
\label{3.355}
\eea
for some superfield $\varphi_{\a(2s-3)}$. 
The residual gauge freedom is characterised by 
\bea
\z_{\a(2s-1)} = \ri \nabla_{\a}  \eta_{\a(2s-2)}~,
\eea
which means that $\eta_{\a(2s-2)}$ is the only independent gauge parameter. As a consequence, the model can be reformulated in terms of the following gauge superfields
\bea
\big\{H_{\a(2s)},~\vf_{\a(2s-3)}\big\}~,
\eea
which are defined modulo gauge transformations of the form
\bea
\d_{\eta} H_{\a(2s)} &=& -\nabla_{\a(2)} \eta_{\a(2s-2)}~,\\
\d_{\eta} \vf_{\a(2s-3)}&=& \ri \nabla^{\b} \eta_{\b \a(2s-3)}~.
\eea
The corresponding gauge-invariant action is given by
\bea \label{GaugedTransIntAction}
&&S^{\perp}_{(s)}[H_{\a(2s)}, \varphi_{\a(2s-3)}] =  \Big ( - \frac{1}{2} \Big )^s   \int \rd^{3|2}z~E~ \bigg \{ \frac{1}{2}H^{\a(2s)}(\ri \nabla^2 +8s|\m|)H_{\a(2s)}~\non \\
&&-\ri s \nabla_\b H^{\b\a(2s-1)}\nabla^\g H_{\g\a(2s-1)} -2(s-1) \varphi^{\a(2s-3)}\nabla^\b\nabla^{\g\d}H_{\b\g\d\a(2s-3)} ~ \non \\
&&+ \frac{1}{s}(s-1) \Big ( 8 \ri (2s-1)|\m|^2\varphi^{\a(2s-3)}\varphi_{\a(2s-3)}+2(2s-3)|\m|\varphi^{\a(2s-3)}\nabla^2\varphi_{\a(2s-3)} \non \\
&&-2\ri \varphi^{\a(2s-3)}\mathbb{Q}\varphi_{\a(2s-3)} + \frac{\ri}{2(2s-1)}(2s-3)\nabla_\b \varphi^{\b\a(2s-4)}\nabla^2\nabla^\g \varphi_{\g\a(2s-4)} ~ \non \\
&&+ \frac{4}{2s-1}(2s-3)(3s-2)|\m|\nabla_\b \varphi^{\b \a(2s-4)}\nabla^\g \varphi_{\g \a(2s-4)}~ \non \\
&&-\frac{\ri}{2s-1}(2s-3)(s-2)\nabla_{\b\g}\varphi^{\b\g\a(2s-5)}\nabla^{\d \l}\varphi_{\d\l\a(2s-5)} \Big ) \bigg \}~.
\eea
The above theory was recently proposed in \cite{HK19}.
In the flat-superspace limit, the action \eqref{GaugedTransIntAction} coincides with the $\cN=1$ model given in \cite{HutomoKuzenkoOgburn2018}.

\subsection{Component structure of $\cN=1$ supersymmetric higher-spin actions in AdS$_3$} \label{AppB}

In this section, we study the component structure of the longitudinal and transverse formulations for the (half-)integer superspin multiplets, which were reviewed in section \ref{TASMasslessHStheories}. Any 
supersymmetric action in $\rm AdS^{3|2}$ can be reduced to components by the rule  
\bea
S = \frac{1}{4} \int \rd^3x \, e \, (\ri \nabla^2 + 8|\m|) L\,\big|_{\,\q = 0}~.
\eea
In what follows, we will denote the torsion-free covariant derivative on  AdS$_3$ by
${\mathfrak D}_a$. It is related to the vector covariant derivative $\nabla_a$ in \eqref{SCD}
by the simple rule ${\mathfrak D}_{a} := \nabla_{a}|_{\q = 0}$, provided an appropriate Wess-Zumino gauge is chosen.

\subsubsection{Transverse formulation for the superspin-$(s+\hf)$ multiplet}
\label{apB3}

The transverse formulation for the massless superspin-$(s+\hf)$ multiplet is described by the action \eqref{TransHIAct}, which is invariant under the gauge transformations \eqref{GTTHI}. The gauge freedom \eqref{GTTHI} can be used to impose the following Wess-Zumino gauge
\be \label{CGT}
H_{\a(2s+1)}|=0~, \qquad \nabla^{\b}H_{\b\a(2s)}|=0~, \qquad \U_{\b;\a(2s-2)}|=0~, \qquad \nabla^\b \U_{\b;\a(2s-2)}|=0~.
\ee
The residual gauge freedom preserving the gauge conditions \eqref{CGT} is given by
\begin{subequations}
	\bea
	\nabla_{(\a_1}\z_{\a_2...\a_{2s+1})}|&=&0~,\\
	\nabla^2\z_{\a(2s)}|&=&-\frac{2\ri s}{s+1}\big (\nabla_{(\a_1}{}^\b \z_{\a_2 ... \a_{2s})\b}+2(s+1)|\m|\z_{\a(2s)} \big )|~, \\
	\nabla_{\b}\eta_{\a(2s-2)}|&=&\nabla_{(\b}\eta_{\a_1 \ldots \a_{2s-2})}|=-\frac{1}{2s+1}\nabla^\g \z_{\b\g\a(2s-2)}|~, \\
	\nabla^2 \eta_{\a(2s-2)}|&=&-\frac{\ri}{2s+1}\nabla^{\b\g}\z_{\b\g\a(2s-2)}|~.
	\eea
\end{subequations}
This implies that there are three real independent gauge parameters at the component level, which we define as
\bea
\x_{\a(2s)}:=\z_{\a(2s)}|~, \quad \l_{\a(2s-1)}:=-\frac{\ri s}{2s+1}\nabla^\b \z_{\b\a(2s-1)}|~, \quad  \r_{\a(2s-2)}:= -\eta_{\a(2s-2)}|~.
\eea
The next task is to identify the remaining independent component fields of $H_{\a(2s+1)}$ and $\U_{\b;\a(2s-2)}$  in the Wess-Zumino gauge \eqref{CGT}. 

Let us first consider the fermionic sector. We begin by decomposing the superfield $\U_{\b;\a(2s-2)}$ into irreducible components
\be \label{DI}
\U_{\b;\a(2s-2)}:= Y_{\b\a(2s-2)}+\sum_{k=1}^{2s-2}\varepsilon_{\b\a_k}Z_{\a_1 ... \hat{\a}_k ... \a_{2s-2}}~,
\ee
where we have introduced the irreducible superfields
\begin{subequations}
	\bea
	Y_{\a(2s-1)}&:=& \U_{(\a_1;\a_2 \dots \a_{2s-1})}~,\\
	Z_{\a(2s-3)}&:=& \frac{1}{2s-1}\U^{\b;}{}_{\b\a(2s-3)}~.
	\eea
\end{subequations}
We find that the remaining independent fermionic fields are given by
\begin{subequations}
	\bea
	h_{\a(2s+1)}&:=&\frac{\ri}{4}\nabla^2H_{\a(2s+1)}|~, \\
	y_{\a(2s-1)}&:=& \frac{\ri}{8}\nabla^2 Y_{\a(2s-1)}|~, \\
	z_{\a(2s-3)}&:=& \frac{\ri s}{2}(2s-1) \nabla^2 Z_{\a(2s-3)}|~,
	\eea
\end{subequations}
Their gauge transformations are given by
\begin{subequations}
	\bea
	\d_\l h_{\a(2s+1)}&=& \mathfrak{D}_{\a(2)}\l_{\a(2s-1)}~, \\
	\d_\l y_{\a(2s-1)}&=&\frac{1}{2s+1}\mathfrak{D}_{(\a_1}{}^\b \l_{\a_2 ... \a_{2s-1})\b} + |\m|\l_{\a(2s-1)}~, \\
	\d_\l z_{\a(2s-3)}&=& \mathfrak{D}^{\b\g}\l_{\b\g\a(2s-3)}~.
	\eea
\end{subequations}
Upon component reduction of the action \eqref{TransHIAct}, we find that the fermionic sector coincides with the Fang-Fronsdal-type action $S_{(s+\hf,+)}^{\text{FF}}[h,y,z]$, which is given by eq. \eqref{TAFangFronsdal}.

We now turn to the bosonic sector. To start with, we point out that the fourth  condition in \eqref{CGT} is equivalent to
\be
\nabla_\b \U_{\g;\a(2s-2)} |=\nabla_\g \U_{\b;\a(2s-2)} |~.
\ee
We then choose to define the remaining independent bosonic fields as follows
\begin{subequations}
	\bea
	h_{\a(2s+2)}&:=&-\nabla_{\a}H_{\a(2s+1)}|~,\\
	\f_{\b\g;\a(2s-2)}&:=&\nabla_\b \U_{\g;\a(2s-2)}|=\nabla_{(\b} \U_{\g);\a(2s-2)}|~.
	\eea
\end{subequations}
Performing the component reduction on the action \eqref{TransHIAct}, we arrive at the bosonic action
\bea \label{BosComp}
S_{\text{bos}}[h,\f] &=& \Big (-\frac{1}{2}\Big )^s \int \rd^3 x\,e\, \bigg \{-\frac{1}{4}h^{\a(2s+2)}\mathcal{Q} h_{\a(2s+2)}+\frac{3}{16}\mathfrak{D}_{\b\g}h^{\b\g\a(2s)}
\mathfrak{D}^{\d\r}h_{\d\r\a(2s)}~ \non \\
&&+\frac{1}{4}(2s-1)F^{(\b\g;\a(2s-2))}\mathfrak{D}^{\d\l}h_{\b\g\d\l\a(2s-2)} -\frac{1}{4}(2s-1)F^{\b\g;\a(2s-2)}F_{\b\g;\a(2s-2)}~ \non \\
&&+\hf(s-1)(2s-1)F^{\b\g ; \a(2s-3)}{}_\b F_{\d\g ; \a(2s-3)}{}^\d \non\\
&&-\hf (2s-1)|\m| \Big( h^{\a(2s+2)}\mathfrak{D}_{(\a_1}{}^\b h_{\a_2 \dots \a_{2s+2})\b} - (s-1) \f^{\b \g; \, \a(2s-2)} \mathfrak{D}^{\d \l} h_{\d \l \b \g \a(2s-2)} \non\\
&& -2 F^{\b \g;\, \a(2s-2)} \f_{\b \g; \, \a(2s-2)} + 8(s-1) F^{\b \g;\, \a(2s-3)}\,_{\b} \f_{\d \g;\, \a(2s-3)}{}^{\d} \non\\
&&-2 (s-1)(2s-3) F_{\b \g;}{}^{\b \g \a(2s-4)} \f^{\d \l;}{}_{\d \l \a(2s-4)} \Big) \non\\
&&+|\m|^2 (s-1) (2s-1)\Big( \frac{3s}{(s-1) (2s-1)}\, h^{\a(2s+2)} h_{\a(2s+2)} \non\\
&&+ (s+1) \f^{\b \g; \, \a(2s-2)} \f_{\b \g; \, \a(2s-2)} + 2(2s^2-s+3) \f^{\b \g; \, \a(2s-3)}\,_{\b} \f_{\g \d;\, \a(2s-3)}{}^{\d} \non\\
&&+(s-2)(2s-3) \f_{\b \g;}{}^{\b \g \a(2s-4)} \f^{\d \l;}{}_{\d \l \a(2s-4)} \Big)
\bigg \}~,
\eea
where we have introduced the field strength $F_{\b\g ; \a(2s-2)}$, 
\be
F_{\b\g ; \a(2s-2)} := \mathfrak{D}_{(\b}{}^\d \f_{\g)\d;\a(2s-2)}~.
\ee
The bosonic action \eqref{BosComp} is invariant under the gauge transformations
\begin{subequations} \label{rt0}
	\bea
	\d_\x h_{\a(2s+2)} &=& \mathfrak{D}_{\a(2)}\x_{\a(2s)}~, \\
	\d_{\x, \r} \f_{\b\g;\a(2s-2)} &=& \mathfrak{D}_{\b\g}\r_{\a(2s-2)} + 2(s-1)|\m| \big( \varepsilon_{\b  (\a_1}\r_{\a_2 ... \a_{2s-2})\g} + \varepsilon_{\g  (\a_1}\r_{\a_2 ... \a_{2s-2})\b} \big) ~   \non\\
	&&-\frac{1}{(2s+1)(s+1)}\Big ( (s+2)\mathfrak{D}_{(\b}{}^\d \x_{\g)\d\a(2s-2)}+(s-1)\mathfrak{D}_{(\a_1}{}^\d \x_{\a_2 ... \a_{2s-2})\b\g\d} \non \\
	&&+2(s+1)(2s+1)|\m|\x_{\b\g\a(2s-2)}\Big ) ~. 
	\label{rt}
	\eea
\end{subequations}
From \eqref{rt}, it can be shown that $F_{\b\g;\a(2s-2)}$ transforms in the following manner
\bea \label{rt2}
\d_{\x, \r}  F_{\b\g;\a(2s-2)} &=&-2(s-1)|\m| \Big (\varepsilon_{(\a_1|(\b}\mathfrak{D}_{\g)|}{}^\d\r_{\a_2 ... \a_{2s-2})\d} + \mathfrak{D}_{(\b|(\a_1}\r_{\a_2 ... \a_{2s-2})|\g)} ~\\
&&-4|\m|\varepsilon_{(\b|(\a_1}\r_{\a_2 ... \a_{2s-2})|\g)} \Big ) \non \\
&&-\frac{1}{(2s+1)(s+1)}\Big ( \hf (s+2)\mathfrak{D}_{\b\g}\mathfrak{D}^{\d\r}\x_{\d\r \a(2s-2)}+(2s+1)\cQ\x_{\g\b \a(2s-2)} \non ~\\
&&+(s-1)\mathfrak{D}_{(\b|(\a_1}\mathfrak{D}^{\d\r}\x_{\a_2 ... \a_{2s-2})|\g)\d\r} +2(s+1)(2s+1)|\m|\mathfrak{D}_{(\b}{}^\d \x_{\g)\a(2s-2)\d} ~ \non \\
&&-4(s-1)(s+1)(2s+1)|\m|^2\x_{\b\g\a(2s-2)} \Big ) ~ \non~.
\eea
Inspecting the gauge transformation \eqref{rt2}, one sees that the field strength $F_{\b\g; \a(2s-2)}$ is invariant under $\r$-gauge transformations in the flat-superspace limit.

The complete component action of our transverse half-integer model takes the form
\bea
S^{\perp}_{(s+\hf)}[{H} ,\U ] = S_{\rm bos} [h,\f] 
+ S_{(s+\hf,+)}^{\text{FF}}[h,y,z]~.
\label{b27}
\eea
By applying the duality transformation described in 
subsection \ref{apB2}, it can be shown that the flat-space counterpart of \eqref{BosComp} is 
the dual formulation \eqref{b03} of 
the flat-space counterpart of Fronsdal action $S^{\rm F}_{(s+1)}$  \eqref{TAFronsdal} provided we fix the coefficients $A$ and $B$ as
\bea
A= -\hf (s-1)~, \qquad B = \frac{3}{4s} (s-1)~.
\eea
That is, the dual action given in \eqref{b03} coincides with the bosonic action \eqref{BosComp} in Minkowski space
\be
S_{\rm bos} [h, \f] \Big|_{ |\m|=0}
=S_{\rm A}^{(s+1)} [h , \f]~.
\label{eqb43}
\ee
However, it must be noted that this duality does not hold in the presence of a non-vanishing AdS curvature, since \eqref{BosComp} cannot be written solely in terms of the field strength $F_{\b \g; \, \a(2s-2)}$.


\subsubsection{Longitudinal formulation for the superspin-$(s+\hf)$ multiplet}
\label{apB4}

The longitudinal formulation for the massless superspin-$(s+\hf)$ multiplet is described by the action \eqref{LongHalfIntAct}, which  is invariant under the gauge transformations \eqref{LFHIGT}.
The component structure of this model was studied in\cite{KuzenkoPonds2018} 
only in the flat-superspace limit. Here we extend these results to AdS$_3$.

The gauge freedom \eqref{GTTHI} can be used to impose the following Wess-Zumino gauge
\be \label{gc}
H_{\a(2s+1)}| = 0~, \qquad \nabla^\b H_{\b\a(2s)}|=0~.
\ee
The residual gauge freedom which preserves the conditions \eqref{gc} is described by 
\begin{subequations}
	\bea
	0 &=& \nabla_{(\a_1}\z_{\a_2 ... \a_{2s+1})}|~,\\
	\nabla^2 \z_{\a(2s)}| &=&- \frac{2 \ri s}{s+1} \big ( \nabla_{(\a_1}{}^\b \z_{\a_2 ... \a_{2s})\b} +2(s+1)|\m|\z_{\a(2s)}\big )|~.
	\eea
\end{subequations}
These conditions imply that there are only two independent gauge parameters, which we define as follows:
\be
\x_{\a(2s)} := \z_{\a(2s)}|~, \qquad \l_{\a(2s-1)}:=-\frac{\ri s}{2s+1}\nabla^\b \z_{\b\a(2s-1)}|~.
\ee
Upon imposing the gauge \eqref{gc}, we are left with the following  component fields:
\begin{subequations} 
	\bea
	h_{\a(2s+1)}&:=&\frac{\ri}{4}\nabla^2 H_{\a(2s+1)}|~, \\
	h_{\a(2s+2)}&:=&-\nabla_{\a}H_{\a(2s+1)}|~, \\
	y_{\a(2s-2)}&:=&-4L_{\a(2s-2)}|~ \\
	y_{\a(2s-1)} &:=& (2s+1)\ri \nabla_{\a}L_{\a(2s-2)}|~, \\
	z_{\a(2s-3)}&:=&2\ri \nabla^\b L_{\b \a(2s-3)}|~, \\
	F_{\a(2s-2)}&:=&\frac{\ri}{4}\nabla^2 L_{\a(2s-2)}|~.
	\eea
\end{subequations} 

Performing a component reduction on the higher-spin model \eqref{LongHalfIntAct}, 
we find that the theory decouples into a bosonic and a fermionic sector, respectively. The bosonic action is given by
\bea 
&&S_{\text{bos}}[h,y,F] = \Big ( -\frac{1}{2} \Big )^s \int \rd^3 x\,e\, \bigg \{ - \frac{1}{4}h^{\a(2s+2)} \mathcal{Q} h_{\a(2s+2)} +\frac{1}{8}(s+1)\mathfrak{D}_{\b\g}h^{\b\g\a(2s)} \mathfrak{D}^{\d\l}h_{\d\l\a(2s)} ~ \non \\
&&+\frac{1}{8}(2s-1)y^{\a(2s-2)}\mathfrak{D}^{\b\g}\mathfrak{D}^{\d\l}h_{\b\g\d\l\a(2s-2)}+\frac{1}{16s}(2s-1)(s+1)y^{\a(2s-2)}\mathcal{Q} y_{\a(2s-2)}~ \non \\
&&+ \frac{1}{s}(s-1)(2s-1)F^{\b\a(2s-3)}\mathfrak{D}_\b{}^\g y_{\g\a(2s-3)}+\frac{4}{s}(s+1)(2s-1)F^{\a(2s-2)}F_{\a(2s-2)}~ \non \\
&& +\frac{1}{4}(s-1)(2s-1)|\m|y^{\b\a(2s-3)}\mathfrak{D}_\b{}^\g y_{\g\a(2s-3)}+2(s-1)(2s-1)|\m|F^{\a(2s-2)}y_{\a(2s-2)} ~ \non \\
&&+2s(s+1)|\m|^2h^{\a(2s+2)} h_{\a(2s+2)}-\hf (2s-1)(s+1)|\m|^2 y^{\a(2s-2)}y_{\a(2s-2)}  \bigg \}~. \label{BA00}
\eea
The bosonic field $F_{\a(2s-2)}$ is auxiliary, so upon elimination via its equation of motion
\be 
F_{\a(2s-2)}=-\frac{1}{8(s+1)}(s-1)\mathfrak{D}_{(\a_1}{}^\b y_{\a_2 ... \a_{2s-2})\b} - \frac{s}{4}|\m|y_{\a(2s-2)} ~,
\ee
we find that the resulting action coincides with the massless spin-$(s+1)$ action, 
$S_{(s+1)}^{\text{F}}$, which is obtained from \eqref{TAFronsdal} by setting $n = 2s+2$.

It can be shown that the fermionic action emerging from the reduction procedure coincides with the Fang-Fronsdal type action $S^{\text{FF}}_{(s+\hf,-)}$ \eqref{TAFangFronsdal}
Therefore, our component actions take the form
\bea
S^{\parallel}_{(s+\hf)}[H,L ] = S_{(s+1)}^{\text{F}}[h, y] 
+ S^{\text{FF}}_{(s+\hf,-)}[h, y, z]~.
\label{b40}
\eea

Comparing \eqref{b40} with \eqref{b27}, it is seen that the fermionic sector resulting from the reduction of the longitudinal model is given by $S^{\text{FF}}_{(s+\hf,+)}[ h, y , z]$. 
In flat-space, the two bosonic actions, $S_{(s+1)}^{\text{F}}[h, y]$ and $S_{\rm bos} [h,\f]$, are related to each other by a duality transformation as described in section \ref{apB2}. The fermionic actions are now given by $S^{\text{FF}}_{(s+\hf)}[h,y,z]$, thus they coincide identically.


\subsubsection{Transverse formulation for the superspin-$s$ multiplet}

The transverse formulation for the massless superspin-$s$ multiplet is described by the action \eqref{TIAction} and is invariant under the gauge transformations \eqref{TIGT}. The gauge freedom \eqref{TIGT} can be used to impose the following Wess-Zumino gauge conditions:
\be \label{TICGT}
H_{\a(2s)}|=0~, \qquad \nabla^{\b}H_{\b\a(2s-1)}|=0~, \qquad \J_{\b;\a(2s-2)}|=0~, \qquad \nabla^\b \J_{\b;\a(2s-2)}|=0~.
\ee
The residual gauge symmetry preserving \eqref{TICGT} is given by
\begin{subequations}
	\bea
	\nabla_{(\a_1}\z_{\a_2...\a_{2s})}|&=&0~,\\
	\nabla^2\z_{\a(2s-1)}|&=&-\frac{2\ri(2s-1)}{2s+1}\big (\nabla_{(\a_1}{}^\b \z_{\a_2 ... \a_{2s-1})\b}+(2s+1)|\m|\z_{\a(2s-1)} \big )|~, \\
	\nabla_{\b}\eta_{\a(2s-2)}|&=&\nabla_{(\b}\eta_{\a_1 \ldots \a_{2s-2})}|=-\ri  \z_{\b\a(2s-2)}|~, \\
	\nabla^2 \eta_{\a(2s-2)}|&=&-\ri \nabla^{\b}\z_{\b\a(2s-2)}|~.
	\eea
\end{subequations}
Thus, there are three real independent gauge parameters, which we define as
\begin{subequations}
	\begin{gather}
	\x_{\a(2s-1)}:=\z_{\a(2s-1)}|~, \qquad \r_{\a(2s-2)}:=-\eta_{\a(2s-2)}|~, \\ \l_{\a(2s-2)}:=-\frac{2s-1}{4s}\nabla^\b \z_{\b\a(2s-2)}|~.
	\end{gather}
\end{subequations}
We now wish to find the remaining independent component fields of $H_{\a(2s)}$ and $\J_{\b;\a(2s-2)}$  in the Wess-Zumino gauge \eqref{TICGT}. The last Wess-Zumino gauge condition in \eqref{TICGT} yields the relation
\be
\nabla_\b \J_{\g;\a(2s-2)}=\nabla_\g \J_{\b;\a(2s-2)}~.
\ee
Thus, we choose to describe the bosonic sector in terms of the independent fields
\begin{subequations}
	\bea
	X_{\a(2s)}&:=&\frac{\ri}{4}\nabla^2 H_{\a(2s)}|~,\\
	\f_{\b\g;\a(2s-2)}&:=&\nabla_\b \J_{\g;\a(2s-2)}|=\nabla_{(\b} \J_{\g);\a(2s-2)}|~.
	\eea
\end{subequations}
Performing a component reduction on the action \eqref{TransHIAct}, we obtain the bosonic action
\bea \label{TIB}
&&S_{\text{bos}}[\f,X]= -2(2s-1) \Big (-\frac{1}{2}\Big )^s  \int \rd^3 x~e~ \bigg \{ \frac{s-1}{2s-1}X^{\a(2s)}X_{\a(2s)}+F^{\b\g;\a(2s-2)}X_{\b\g\a(2s-2)} \non ~ \\
&&+\hf F^{\b\g;\a(2s-2)}F_{\b\g;\a(2s-2)}+\frac{1}{2s}(s-1)F^{\b \g; \a(2s-3)}{}_\b F_{\g \d ; \a(2s-3)}{}^\d ~ \non \\
&&+2(s-1)|\m|\f^{\b\g;\a(2s-2)}X_{\b\g\a(2s-2)}+(2s-1)|\m|F^{\b\g;\a(2s-2)}\f_{\b\g;\a(2s-2)}~ \non \\
&&+2(s-1)|\m|F^{\b \g; \a(2s-3)}{}_\b \f_{\g \d; \a(2s-3)}{}^\d+\frac{2}{s}(s-1)^2|\m|F^{\b \g; \a(2s-3)}{}_{\b}\f_{\d (\g;\a_1 \ldots \a_{2s-3})}{}^\d ~ \non \\
&&+2s(s-1)|\m|^2\f^{\b\g;\a(2s-2)}\f_{\b\g;\a(2s-2)}+2s(s-1)|\m|^2\f^{\b \g; \a(2s-3)}{}_\b \f_{\g \d; \a(2s-3)}{}^\d ~\non \\
&&+\frac{2}{s}(s-1)^2(s-2)|\m|^2 \f^{\b \g; \a(2s-3)}{}_\b \f_{\d (\g;\a_1 \ldots \a_{2s-3})}{}^\d  \bigg \}~,
\eea
where we have introduced the field strength $F_{\b\g ; \a(2s-2)}$, 
\be
F_{\b\g ; \a(2s-2)} := \mathfrak{D}_{(\b}{}^\d \f_{\g)\d;\a(2s-2)}~.
\ee
It can be shown that the action \eqref{TIB} is invariant under the gauge transformations\begin{subequations}
	\bea
	\d_\l X_{\a(2s)} &=& \mathfrak{D}_{\a(2)}\l_{\a(2s-2)}~, \\
	\d_{\l, \r} \f_{\b\g;\a(2s-2)} &=& \mathfrak{D}_{\b\g}\r_{\a(2s-2)} +4 (s-1)|\m|\varepsilon_{(\b | (\a_1}\r_{\a_2 ... \a_{2s-2})|\g)} ~  \label{rgt1} \\
	&&+\frac{4}{2s-1}(s-1) \varepsilon_{(\b | (\a_1}\l_{\a_2 ... \a_{2s-2})|\g)} ~. \non
	\eea 
\end{subequations}
The field $X_{\a(2s)}$ is auxiliary, so upon elimination via its equation of motion
\be
X_{\a(2s)}=-\frac{2s-1}{2(s-1)}\big ( F_{\a(2s)}+2(s-1)|\m|\f_{(\a_1 \a_2;\, \a_3 \dots \a_{2s})} \big )~,
\ee
we obtain
\bea \label{BWA}
&&S_{\text{bos}}[\f]= \frac{2s-1}{s(2s-2)} \Big (-\frac{1}{2}\Big )^s  \int \rd^3 x~e~ \bigg \{ s \, F^{\b\g;\a(2s-2)}F_{\b\g;\a(2s-2)}  \non ~ \\
&&+ 2s(s-1)F^{\b \g; \a(2s-3)}{}_\b F_{\g \d ; \a(2s-3)}{}^\d +(2s-3)(s-1)F_{\b \g;}\,^{\b \g \a(2s-4)} F^{\d \l;}\,_{\d \l \a(2s-4)}  ~ \non \\ 
&&-4 (s-1)^2|\m|^2 \Big( s\, \f^{\b\g;\a(2s-2)}\f_{\b\g;\a(2s-2)} + 2 \,\f^{\b \g; \a(2s-3)}{}_\b \f_{\d \g; \a(2s-3)}{}^\d \non\\
&&- (2s-3)\f_{\b \g;}\,^{\b \g \a(2s-4)} \f^{\d \l;}\,_{\d \l \a(2s-4)} \Big) \bigg\}~.
\eea
We can express the field $\f_{\b\g;\a(2s-2)}$ in terms of its irreducible components
\begin{subequations}
	\bea
	h_{\a(2s)}&:=&\f_{(\a_1 \a_2;\a_3 ... \a_{2s})}~, \label{CF1} \\
	y_{\a(2s-2)}&:=&\f^{\b}{}_{(\a_1 ; \a_2 ... \a_{2s-2})\b}~, \\
	y_{\a(2s-4)}&:=&\frac{2(s-1)}{2s-1}\f^{\b\g;}{}_{\b\g\a(2s-4)}~. \label{CF2}
	\eea
\end{subequations}
The gauge freedom \eqref{rgt1} can be used to gauge away $y_{\a(2s-2)}$. It follows that the remaining fields, \eqref{CF1} and \eqref{CF2}, have the corresponding gauge transformations\begin{subequations}
	\bea
	\d_\r h_{\a(2s)} &=& \mathfrak{D}_{\a(2)}\r_{\a(2s-2)}~, \\
	\d_\r y_{\a(2s-4)}&=& \frac{2(s-1)}{2s-1}\mathfrak{D}^{\b\g}\r_{\b\g\a(2s-4)}~.
	\eea
\end{subequations}
Recasting \eqref{BWA} in terms of the irreducible fields \eqref{CF1} and \eqref{CF2}, the resulting action coincides with, up to an overall factor, with the massless spin-$s$ action \eqref{TAFronsdal}.

The study of the fermionic sector requires the decomposition of the reducible superfield $\J_{\b;\a(2s-2)}$ into irreducible parts. This procedure is completely analogous to that of the prepotential $\U_{\b;\a(2s-2)}$ \eqref{DI}. 
In this case, we find that the remaining independent fermionic fields are given by
\begin{subequations}
	\bea
	h_{\a(2s+1)}&:=&- \ri \nabla_{\a}H_{\a(2s)}|~, \\
	y_{\a(2s-1)}&:=& \frac{\ri}{4}\nabla^2 Y_{\a(2s-1)}|~, \\
	z_{\a(2s-3)}&:=& \frac{\ri }{2}(2s-1) \nabla^2 Z_{\a(2s-3)}|~.
	\eea
\end{subequations}
Upon the component reduction of the action \eqref{TIAction}, it can be shown that the fermionic action obtained coincides with the Fang-Fronsdal-type action, $S_{(s+\hf,-)}^{\text{FF}}$ \eqref{TAFangFronsdal}.
As a result, we find that the action corresponding to the transverse formulation of the massless superspin-$s$ multiplet in AdS$_3$ decomposes into
\bea \label{b65}
S^{\perp}_{(s)}[H, \J] = \frac{2s-1}{s-1}S_{(s)}^{\text{F}}[h , y] + \, S_{(s+\hf,-)}^{\text{FF}}[h,y,z]~. 
\eea

\subsubsection{Longitudinal formulation for the superspin-$s$ multiplet}
The longitudinal formulation for the massless superspin-$s$ multiplet is described by the action \eqref{action-t3} and is invariant under gauge transformations \eqref{LongGTI}. This formulation corresponds to the massless first-order model, whose component reduction in the flat-superspace limit has been studied in \cite{KuzenkoPonds2018}. 

The gauge freedom \eqref{LongGTI} can be used to impose the following Wess-Zumino gauge
\be  \label{GCLI}
H_{\a(2s)}| = 0~, \qquad \nabla^\b H_{\b\a(2s-1)}|=0~, \qquad V_{\a(2s-2)}| =0~.
\ee
The residual gauge symmetry which preserves the gauge transformations \eqref{GCLI} is
\begin{subequations} \label{RGLI}
	\bea
	0 &=& \nabla_{(\a_1}\z_{\a_2 ... \a_{2s})}|~,\\
	0 &=& \nabla^\b \z_{\b\a(2s-2)}|~,\\
	\nabla^2 \z_{\a(2s-1)}| &=&- \frac{2 \ri }{2s+1}(2s-1) \big ( \nabla_{(\a_1}{}^\b \z_{\a_2 ... \a_{2s-1})\b} +(2s+1)|\m|\z_{\a(2s-1)}\big )|~.
	\eea
\end{subequations}
The relations \eqref{RGLI} indicate that there is only one independent gauge parameter, which we choose to define as
\be
\x_{\a(2s-1)} := \z_{\a(2s-1)}|~.
\ee
Thus, we are left with the remaining independent component fields in the gauge \eqref{gc}:
\begin{subequations} 
	\bea
	h_{\a(2s+1)}&:=&- \ri \nabla_{\a}H_{\a(2s)}|~, \\
	h_{\a(2s)}&:=&\frac{\ri}{4}\nabla^2 H_{\a(2s)}|~, \\
	y_{\a(2s-2)}&:=&\frac{\ri}{4}\nabla^2 V_{\a(2s-2)}|~ \\
	y_{\a(2s-1)} &:=& \frac{\ri}{2} \nabla_{\a}V_{\a(2s-2)}|~, \\
	z_{\a(2s-3)}&:=&-2\ri s \nabla^\b V_{\b \a(2s-3)}|~. 
	\eea
\end{subequations}
It is easy to see that the gauge transformation laws for the bosonic fields are
\bea
\d h_{\a(2s)} = 0~, \qquad \d y_{\a(2s-2)} = 0~.
\eea
Upon reduction of the higher-spin model \eqref{LongHalfIntAct}, it can be shown that these bosonic fields appear without derivatives and hence, they are pure auxiliary fields. Indeed, we find that 
\bea
S^{\parallel}_{(s)}[H ,V ] &=& \Big ( -\frac{1}{2} \Big )^s \int \rd^3 x \, e\, \Big \{ h^{\a(2s)} h_{\a(2s)} + 2s(2s-1)y^{\a(2s-2)}y_{\a(2s-2)} \Big \}  \non \\
&&+ S^{\text{FF}}_{(s+\hf,+)}\,[h,y,z]~.
\eea
Unlike the reduction of the transverse superspin-$s$ multiplet described by \eqref{b65}, here the fermionic action comes with a positive sign in the $|\m|$-dependent terms,  $S^{\text{FF}}_{(s+\hf,+)}$.

\subsection{Topologically massive theories} \label{TASTMSec}
Topologically massive higher-spin theories were constructed in AdS$^{3|2}$ in  \cite{KuzenkoPonds2018,HK19}. 
In accordance with \cite{HK19,KuzenkoPonds2018}, there exist four off-shell formulations for topologically massive higher-spin ${\cN}=1$ multiplets in AdS${}_3$. For a positive integer $s$, there are two off-shell gauge-invariant models for a topologically 
massive superspin-$(s+\hf)$ multiplet in AdS${}_3$:
\begin{subequations} \label{8.5}
	\bea
	S^{\parallel}_{(s+\hf)}[H_{\a(2s+1)},L_{\a(2s-2)}|m] &=& \kappa
	S^{(s+\hf)}_{\text{SCHS}}[H_{\a(2s+1)}] \label{N2Massive1}  \\ 
	&&+m^{2s-1}S^{\parallel}_{(s+\hf)}[H_{\a(2s+1)},L_{\a(2s-2)}]~, \non
	\\
	S^{\perp}_{(s+\hf)}[H_{\a(2s+1)},\U_{\b;\a(2s-2)}|m] &=& \kappa S^{(s+\hf)}_{\text{SCHS}}[H_{\a(2s+1)}] \label{TASN2Mass}\\
	&&+m^{2s-1}S^{\perp}_{(s+\hf)}[H_{\a(2s+1)},\U_{\b;\a(2s-2)}]~. \non
	\eea
\end{subequations}
Here $\kappa$ is a dimensionless parameter and $m$ is a real massive parameter.
For a topologically massive superspin-$s$ multiplet in AdS${}_3$, we have  the following models:
\begin{subequations} \label{8.4}
	\bea
	S^{\parallel}_{(s)}[H_{\a(2s)} ,V_{\a(2s-2)} |m]
	&=& {S}_{\rm{SCHS}}^{(s)} [ H_{\a(2s)}] 
	+m^{2s-1}S^{\parallel}_{(s)}[H_{\a(2s)} ,V_{\a(2s-2)} ]~, \label{8.4a} \\
	S^{\perp}_{(s)}[H_{\a(2s)} ,{\Psi}_{\b; \,\a(2s-2)} |m]
	&=& {S}_{\rm{SCHS}}^{(s)} [ H_{\a(2s)}] 
	+m^{2s-1}S^{\perp}_{(s)}[H_{\a(2s)} ,{\Psi}_{\b; \,\a(2s-2)}]~. \label{8.4b}
	\eea
\end{subequations}
Note that the massive actions \eqref{N2Massive1} and \eqref{8.4a} were derived by Kuzenko and Ponds in \cite{KuzenkoPonds2018}, while the remaining models, \eqref{TASN2Mass} and \eqref{8.4b}, were derived by Hutomo and Kuzenko in \cite{HK19}.
It is an instructive exercise to show that the actions \eqref{8.5} and \eqref{8.4}  describe massive dynamics on-shell, in accordance with \eqref{SOC}. 

It is worth noting that the new topologically massive action \eqref{HSNTMG} in AdS$_3$ was recently extended to AdS$^{3|2}$ in \cite{KuzenkoPonds2018,KP21}. For $n \geq 1$, the gauge-invariant NTM action is given by \cite{KuzenkoPonds2018,KP21}
\be \label{TASNTMAction}
S^{(n)}_{\text{NTM}}[H_{\a(n)}] = - \frac{\ri^n}{2^{\lfloor \frac{n}{2} \rfloor +1}}\frac{1}{M} \int \rd^{3|2}z~E~\mf{W}^{\a(n)}(H) \big (\mb{F} - \s M \big )H_{\a(n)}~,
\ee
where $\s = \pm 1$, $H_{\a(n)}$ is the SCHS gauge superfield and $\mf{W}_{\a(n)}(H)$ is the linearised super-Cotton tensor \eqref{SCT}. 

The equation of motion obtained by varying \eqref{TASNTMAction} with respect to $H_{\a(n)}$ is
\be \label{TASNTMEoM}
0 = \big (\mb{F} - \s M \big )\mf{W}_{\a(n)}(H)~.
\ee
In accordance with \eqref{TASNTMEoM} and the fact that $\mf{W}_{\a(n)}(H)$ is transverse \eqref{SCTP}, it follows from \eqref{SOC} that the field strength $\mf{W}_{\a(n)}(H)$ describes a massive superfield with pseudo-mass $M$, superspin $\frac{n}{2}$ and superhelicity $\hf(n+\hf)\s$.

\section{Summary of results} \label{TASsecSoR}
The main goal of this chapter was to compute the spin projection operators in three-dimensional anti-de Sitter (super)space and study their corresponding applications. In section \ref{TAProjectors}, we computed the AdS$_3$ spin projection operators $\widehat{\P}^{\perp}_{(n)}$, for $n \geq 2$. These operators were computed for the first time in \cite{HutchingsKuzenkoPonds2021} and are given by eqs. \eqref{BosSpin} and \eqref{FermSpin} for the bosonic $(n=2s)$ and fermionic $(n=2s+1)$ cases, respectively. We also derived an equivalent\footnote{Recall that the spin projection operators $\widehat{\P}^{\perp}_{(n)}$ and ${\P}^{\perp}_{(n)}$ are only equivalent on the space $\cV_{(n)}$.} spin projection operator ${\P}^{\perp}_{(n)}$ which is written solely in terms of the Casimir operators  of $\mf{so}(2,2)$ \cite{HutchingsKuzenkoPonds2021}. These bosonic $(n=2s)$ and fermionic $(n=2s+1)$ projectors are given by eqs. \eqref{SimpFermProj} and \eqref{SimpBosProj}, respectively. 

The spin projection operator ${\P}^{\perp}_{(n)}$ maps a field $\f_{\a(n)}$ satisfying the first-order equation \eqref{TAFirstOrderMassConstraint}, but otherwise unconstrained, to an on-shell superfield \eqref{TAOnShellConditions} which realises the irreducible representation $D(E_0,\s |s|)$ of $\mf{so}(2,2)$. If instead the field $\f_{\a(n)}$  obeys the Klein-Gordon equation \eqref{OMS2}, then the projected field furnishes the reducible representation ${\mathfrak D}(\rho,-\frac{n}{2})\oplus {\mathfrak D}(\rho,\frac{n}{2})$ of $\mf{so}(2,2)$. In order to isolate the component with fixed helicity, one needs to make use of the helicity projectors $\mathbb{P}^{(\pm)}_{[n]}$ \cite{HutchingsKuzenkoPonds2021}. They are given by eq. \eqref{TAhelicityproj}.

The spin projectors had many fruitful applications. Similar to the case in AdS$_4$, we demonstrated that these projectors are connected to partially massless fields in AdS$_3$. In particular, we established a correspondence between partially massless on-shell  fields $\f^{(t)}_{\a(n)}$ with depth $t$ and the  poles of ${\P}^{\perp}_{(n)}$, which are determined by $\t_{(t,n)}|\m|^2$, with $2 \leq t \leq \lfloor n/2 \rfloor$. In section \ref{secLP}, the spin projection operators ${\P}^{\perp}_{(n)}$ were used to decompose any field $\f_{\a(n)}$ into its irreducible components \eqref{Decomp} \cite{HutchingsKuzenkoPonds2021}. They were also used construct the lower-spin extractors, which single out the spin $\frac{1}{2}(n-2j)$ component from the decomposition \eqref{Decomp} \cite{HutchingsKuzenkoPonds2021}. In section \ref{secCT} we made use of ${\P}^{\perp}_{(n)}$  to obtain new representations for the linearised higher-spin super Cotton tensors \eqref{TACT} and their corresponding conformal actions in AdS$_3$ \cite{HutchingsKuzenkoPonds2021}. The significance of these new realisations is that the properties of gauge invariance, transversality and factorisation are all made manifest. 

In section \ref{TAMasslessActions} we reviewed free massless higher-spin field theories in AdS$_3$. The corresponding actions were shown not to propagate any physical degrees of freedom on-shell, both at the level of superfields and component fields.  We also studied new topologically massive theories in section \ref{TAMassiveactions}. In particular, new realisations for the NTM massive models \eqref{HSNTMG} were found which immediately appear in a factorised form \cite{HutchingsKuzenkoPonds2021}. We also computed the AdS$_3$ extension of the new NTM model for bosonic fields, which was recently derived in $\mb{M}^3$ \cite{DalmaziSantos2021}. This is given by eq. \eqref{NNTM} \cite{HutchingsKuzenkoPonds2021}.

In section \ref{TASsec3dAdS}, we generalised many of the results of the previous section to AdS$^{3|2}$. In section \ref{TASSuperprojectorsAdS32} we derived the superspin projection operators $\bm \P^{\perp}_{[n]}$ in AdS$^{3|2}$ for the first time \cite{HutchingsKuzenkoPonds2021}. These are given by eq. \eqref{superprojector}. These operators map a superfield $\F_{\a(n)}$ satisfying the first-order equation \eqref{SFO} to an on-shell superfield \eqref{SOC}, which realises the irreducible representation $\mathfrak{S}(M, \s \frac{n}{2})$ of $\mathfrak{osp}(1|2;{\mathbb R} ) \oplus \mathfrak{sl}(2, {\mathbb R})$. We demonstrated that the poles of $\bm \P^{\perp}_{[n]}$  are intimately related to partially massless superfields, supporting our claim that this is a universal feature of anti-de Sitter space.

In section \ref{TASLongProj22} we made use of the superspin projection operators to decompose an arbitrary superfield into irreducible components \cite{HutchingsKuzenkoPonds2021}. This decomposition is given by eq. \eqref{TAS22Decomp}. In section \ref{TASSCHSSec}, the superprojectors were also used to recast the linearised super-Cotton tensors, and their corresponding SCHS actions, which were recently derived in \cite{KP21}. These new representations are given by eq. \eqref{SCTPr}. Recasting the higher-spin super-Cotton tensors in terms of $\bm \P^{\perp}_{[n]}$ makes the properties of gauge-invariance and transversality manifest.

In section \ref{TASMasslessHStheories} we reviewed the structure of the four series of off-shell massless higher-spin $\cN = 1$ supersymmetric models in AdS$_3$. These models were derived in \cite{HK19,KuzenkoPonds2018}. In section \ref{AppB} we studied the corresponding component structure of these $\cN = 1$ actions for the first time \cite{HutchingsHutomoKuzenko}. At the component level, it was shown that these actions decoupled into a bosonic and a fermionic sector. The bosonic sector coincided with either the Fronsdal theory or a pure auxiliary sector, while the fermionic part always coincided with a Fang-Fronsdal action.  In section \ref{TASTMSec} we reviewed the topologically massive higher-spin theories of \cite{KuzenkoPonds2018,HK19}, which are built from the SCHS actions of section \ref{TASSCHSSec} and the massless theories of section \ref{TASMasslessHStheories}.


\begin{subappendices}

	\section{Generating function formalism} \label{TAAppendixA}

	We employ the generating function formalism which was developed in \cite{KP21}. Within this framework, a one-to-one correspondence between a homogenous polynomial $\f_{(n)}(\U)$ of degree $n$ and a rank-$n$ spinor field $\f_{\a(n)}$ is established via the rule
	\be
	\f_{(n)}(\U):=\U^{\a_1} \cdots\U^{\a_n}\f_{\a(n)}~.
	\ee
	Here, we have introduced the commuting real auxiliary variables $\U^{\a}$, which are inert under the action of the Lorentz generators $M_{\a\b}$. 
	
	Making use of the auxiliary fields $\U^{\a}$, and their corresponding partial derivatives, $\partial_\b := \frac{\partial}{\partial \U^\b}$, we can realise the AdS$_3$ derivatives as index-free operators on the the space of homogenous polynomials of degree $n$. We introduce the differential operators which increase and decrease the degree of homogeneity by $2$, $0$ and $-2$ respectively:
	\be \label{Op}
	\nabla_{(2)} : = \U^\a \U^\b \nabla_{\a\b}~, \quad \nabla_{(0)}:= \U^\a \nabla_\a{}^\b\partial_\b, \quad \nabla_{(-2)}:= \nabla^{\a\b}\pa_\a \pa_\b.
	\ee
	Note that the action $\nabla_{(0)}$ is equivalent to that of the Casimir operator $\cF$.
	Making use of the algebra \eqref{ADSAlg}, one can derive the important identities
	\begin{subequations}
		\bea
		\big[\nabla_{(2)}, \nabla^{\phantom{.}t}_{(-2)}\big]\f_{(n)}&=&4t(n-t+2)\big(\cQ-\tau_{(t,n+2)}\cS^2\big)\nabla_{(-2)}^{t-1}\f_{(n)}~,\label{ID14}\\
		\big[\nabla_{(-2)}, \nabla^{\phantom{.}t}_{(2)}\big]\f_{(n)}&=&-4t(n+t)\big(\cQ-\tau_{(t,n+2t)}\cS^2\big)\nabla_{(2)}^{t-1}\f_{(n)}~,\label{ID15}\\
		\nabla^t_{(2)}\nabla^t_{(-2)}\f_{(n)} &=& \prod_{j=0}^{t-1}\Big (\cF^2 - \big(n-2j \big )^2 \big (\cQ-(n-2j-2) (n-2j+2)\cS^2 \big )  \Big ) \f_{(n)}~, \label{PTI} \hspace{1.35cm}
		\eea
	\end{subequations}
	via induction on $t$. Here $\cQ$ and $\cF$ are the quadratic Casimir operators \eqref{QC} and $\t_{(t,n)}$ are the partially massless values \eqref{PMV}.

\end{subappendices}

\chapter{Higher-spin gauge models with $(1,1)$ supersymmetry
	in AdS$_3$}\label{TAS211AdS}
There exist several incarnations of $\cN$-extended AdS supersymmetry
\cite{AT}, which are known as $(p,q)$ AdS supersymmetry types,
where the integers $p \geq q\geq 0$ are such that $\cN=p+q$. In principle, field theories possessing $(p,q)$ AdS supersymmetry may
be realised in $(p,q)$ AdS superspace
${\rm AdS}^{(3|p,q)} $ \cite{KLT-M12}, 
which can be understood as a maximally supersymmetric solution of $(p,q)$ AdS supergravity \cite{AT}.\footnote{In order 
	to realise field theories with the  $(p,q)$ AdS supersymmetry in ${\rm AdS}^{(3|p,q)} $
	for $p+q\leq 4$, one can employ
	the off-shell supergravity methods developed in \cite{KLT-M11,KT-M11,KLRST-M}.}

Specifically, within the supergravity framework of \cite{KLT-M12,KLT-M11},
the superspace ${\rm AdS}^{(3|p,q)} $
originates as  a maximally symmetric supergeometry possessing covariantly constant torsion and curvature generated by a real symmetric torsion $S^{IJ}= S^{JI}$. Here the indices $I, J$  of the structure group $\sSO(\cN)$ take values from 1 to $\cN$.  
The torsion $S^{IJ}$ is nonsingular with signature determined by the parameters $p$ and $q= \cN-p$.
Since the isometry group of ${\rm AdS}^{(3|p,q)} $ is 
${\sOSp} (p|2; {\mathbb R} ) \times  {\sOSp} (q|2; {\mathbb R} )$ and $S^{IJ}$ is 
invariant under its compact subgroup  $ \sSO(p) \times {\sSO}(q)$, the global realisation 
of $(p,q)$ AdS superspace is\footnote{As detailed in \cite{KLT-M12}, the superspaces \eqref{1.2} are conformally flat. The superconformal flatness of AdS  superspaces in diverse dimensions, including the $d=3$ case, was studied in  \cite{BILS}. }
\bea
{\rm AdS}^{(3|p,q)} = \frac{ {\sOSp} (p|2; {\mathbb R} ) \times  {\sOSp} (q|2; {\mathbb R} ) } 
{ {\sSL}( 2, {\mathbb R}) \times {\sSO}(p) \times {\sSO}(q)}~.
\label{1.2}
\eea

Consider a $(p,q)$ supersymmetric field theory formulated  in ${\rm AdS}^{(3|p,q)} $,
with $ p+q\geq 3$ and $p\geq q$.
As argued in \cite{BKT-M}, it can always be recast
as a supersymmetric theory realised in  $(2,0)$ AdS superspace, with $(p+q -2)$ 
supersymmetries being hidden. In the case $ p+q\geq 3$ and $p\geq q>0$, 
every supersymmetric field theory  in ${\rm AdS}^{(3|p,q)} $, can be reformulated as 
a theory in $(1,1)$ AdS superspace.\footnote{Such reformulations were developed in \cite{BKT-M} 
	for general $(p,q)$ supersymmetric nonlinear $\s$-models in AdS${}_3$, with $p+q \leq4$.}
In accordance with this discussion, the comparison of higher-spin field theories living in arbitrary $(p,q)$ superspaces for allowed values of $p$ and $q$, can be boiled down to the problem of comparing supersymmetric field theories possessing $(1,1)$ and $(2,0)$ AdS supersymmetry. 

The main motivation of this chapter is to investigate the differences between massless higher-spin supersymmetric field theories realised on ${\rm AdS}^{(3|2,0)} $ and 
${\rm AdS}^{(3|1,1)} $. Specifically, the off-shell structure of the 
half-integer superspin
multiplets with $(2,0)$ AdS supersymmetry \cite{HK18} drastically differs to their $(1,1)$ counterparts \cite{HutomoKuzenkoOgburn2018}.\footnote{It is pertinent to mention here that the $(2,0)$ and $(1,1)$ AdS supersymmetries support different supercurrent multiplets \cite{KT-M11}.}     Direct comparison of these theories is difficult in a manifestly supersymmetric setting
since they are each formulated in different superspaces,  ${\rm AdS}^{(3|2,0)} $ and 
${\rm AdS}^{(3|1,1)} $, respectively.  However, both families of higher-spin theories can be reformulated in the same $(1,0)$ AdS superspace, upon where the precise difference between the $(2,0)$ and $(1,1)$ higher-spin supermultiplets can be elucidated. 

The off-shell massless higher-spin multiplets in ${\rm AdS}^{(3|2,0)} $  \cite{HK18} were reduced to AdS$^{3|2}$ in \cite{HK19}.  This chapter is devoted to applying the $(1,1) \to (1,0)$ AdS reduction prescription to the off-shell higher-spin multiplets in AdS$^{(3|1,1)}$ (both massless and massive), which were computed in \cite{HutomoKuzenkoOgburn2018} by Hutomo, Kuzenko and Ogburn.\footnote{The massless higher-spin theories proposed in \cite{HutomoKuzenkoOgburn2018,KuzenkoOgburn2016} have natural counterparts in four dimensions \cite{KPS,KS93,KS94}, see also \cite{BuchbinderKuzenko1998} for a review.} We will demonstrate that every massless higher-spin theory possessing $(1,1)$ or $(2,0)$ AdS supersymmetry decomposes into a sum of two off-shell $(1,0)$ supermultiplets
which belong to the four series of inequivalent higher-spin gauge models constructed in \cite{HK19}.

This chapter is based on the publication \cite{HutchingsHutomoKuzenko} and is organised as follows. In section \ref{TAS2Reduction}, a formalism to reduce every field theory in AdS$^{(3|1,1)}$ to AdS$^{3|2}$ is developed. As an aside, we also initiate a program to formulate the superspin projection operators in AdS$^{(3|1,1)}$.  The formalism developed in section \ref{TAS2Reduction} is then applied throughout sections \ref{Section5}$-$\ref{Section8} to carry out the $(1,1) \to (1,0)$ AdS superspace reduction of all known massless higher-spin models with $(1,1)$ AdS supersymmetry. These are detailed in sections \ref{Section5}$-$\ref{Section6} for the half-integer  superspin-$(s+\hf)$ case, and in sections \ref{Section7}$-$\ref{Section8} for the integer superspin-$s$ case. In addition to summarising our results in section \ref{TAS2Discussion}, we also detail key differences between the $(2,0)$ and $(1,1)$ higher-spin (massless and massive) supermultiplets. In appendix \ref{TAS2appendixA}, we present some identities which prove indispensable when performing the $(1,1) \to (1,0)$ AdS superspace reduction in sections \ref{Section5}$-$\ref{Section8}.

\section{ (1,1) $\rightarrow$ (1,0) AdS superspace reduction} \label{TAS2Reduction}
In this section we review key aspects of $(1,1)$ AdS superspace 
\cite{KT-M11, HutomoKuzenkoOgburn2018, KLT-M12} 
which are crucial in the development of a consistent reduction procedure for field theories from AdS$^{(3|1,1)}$ to AdS$^{3|2}$. In addition, we also comment on the form of the superspin projection operators in AdS$^{(3|1,1)}$. Specifically, we provide closed form expressions for the superprojectors which act on the space of rank-$1$ and rank-$2$ tensor superfields in ${\rm AdS}^{(3|1,1)}$.

\subsection{(1,1) AdS superspace: Complex basis} \label{ss21}
We begin by summarising salient facts concerning $(1,1)$ AdS superspace \cite{KT-M11, HutomoKuzenkoOgburn2018} and superfield representations of the associated isometry group ${{\sOSp}(1|2; {\mathbb{R}})} \times {{\sOSp}(1|2; {\mathbb{R}})}$, following the presentation in \cite{HutomoKuzenkoOgburn2018}. The geometry of $(1,1)$ AdS
superspace can be described using either a real or complex basis for the spinor covariant derivatives. In this section we consider the formulation in terms of the complex basis.

Three-dimensional $(1,1)$ AdS superspace is parametrised by the local complex superspace coordinates $z^{\cM} = (x^m, \theta^{\m}, \bar \theta_{\mu})$, where $m = 0,1,2$ and $\m = 1,2$.\footnote{The fermionic coordinates $(\theta^{\m}, \bar \theta_{\mu})$ are related to each other via complex conjugation $\overline{\q^\m} = \tb^\m$. } The geometry of ${\rm AdS}^{(3|1,1)}$ is described in terms of the  covariant derivatives 
\bea \label{TAS2ComplexDerivatives}
{{\bm{\cD}}}_{\cA}= \big ({\bm{\cD}}_{a}, {\bm{\cD}}_{\a}, {{\bm{\cDB}}}{}^{\a} \big )
=E_{\cA} + \O_{{\cA}}~, 
\eea
where  $E_{\cA}$ and $\O_{\cA}$ denote the inverse supervielbein and Lorentz connection,
\bea
E_{\cA} = E_{\cA}{}^{\cM}\frac{\pa}{\pa z^{\cM}}~, \qquad \O_{\cA} = \hf \O_{\cA}{}^{bc}M_{bc} = \hf \O_{\cA}{}^{\b \g}M_{\b \g}~.
\eea

The covariant derivatives ${{\bm{\cD}}}_{\cA}$ satisfy the following algebra
\begin{subequations}  \label{1.1}
	\bea
	&& \qquad \{ {\bm{\cD}}_\a ,  {\bm{\cDB}}_\b \} = -2\rm i {\bm{\cD}}_{\a \b} ~, \\
	&& \qquad \{{\bm{\cD}}_\a, {\bm{\cD}}_\b \} = -4\bar \m\, M_{\a \b}~, \qquad \hspace{0.82cm}
	\{ { {\bm{\cDB}}}_\a, { {\bm{\cDB}}}_\b \} = 4\m\,M_{\a \b}~, \\
	&& \qquad [ {\bm{\cD}}_{ \a \b }, {\bm{\cD}}_\g ] = -2 \rm i \bar \m\,\ve_{\g (\a}  {\bm{\cDB}}_{\b)}~,  \qquad 
	\,\,[{\bm{\cD}}_{ \a \b }, {  {\bm{\cDB}}}_{\g} ] = 2 \rm i \m\,\ve_{\g (\a} {\bm{\cD}}_{\b)}~,   \\
	&&\quad \,\,\,\,[ {\bm{\cD}}_{\a \b} , {\bm{\cD}}_{ \g \d } ] = 4 \bar \m \m \big(\ve_{\g (\a} M_{\b) \d}+ \ve_{\d (\a} M_{\b) \g}\big)~,  
	\eea
\end{subequations} 
where $\m\neq 0$ is a complex parameter which determines the curvature of 
${\rm AdS}^{(3|1,1)} $. The phase of $\m = |\m | \re^{\ri \vf} $ can be given any fixed value by a re-definition ${\bm{\cD}}_\a \to \re^{\ri \r} {\bm{\cD}}_\a$ and $ {\bm{\cDB}}_\a \to \re^{-\ri \r}  {\bm{\cDB}}_\a$, with $\r$ constant. In appendix \ref{TAS2appendixA}, we collate a list of identities involving the covariant derivatives $\bm{\cD}_{\cA}$ which will be helpful in the subsequent analysis.

Let us compile a dictionary of important constrained superfields in AdS$^{(3|1,1)}$ which will appear frequently in the subsequent analysis. Given an integer $n \geq 1$, a superfield $\G_{\a(n)}$ on $\bm{\mc{V}}_{(n)}$ is called transverse linear if it satisfies the following constraint
\begin{subequations} \label{TAS2Transverse}
	\be
	{\bm{\cDB}}{}^\b \G_{\b\a(n-1)} = 0 \qquad \Longrightarrow \qquad ({\bm{\cDB}}{}^2 -2(n+2)\m )\G_{\a(n)} = 0~. \label{TAS2TransL}
	\ee
	On the other hand, a superfield ${\G}_{\a(n)}$ on $\bm{\mc{V}}_{(n)}$ is said to transverse anti-linear if it obeys
	\be
	{\bm{\cD}}^\b {\G}_{\b\a(n-1)} = 0 \qquad  \Longrightarrow \qquad ({\bm{\cD}}^2 -2(n+2)\mub ){\G}_{\a(n)} = 0~. \label{TAS2TransA}
	\ee
\end{subequations}
In the $n=0$ case, the constraints \eqref{TAS2Transverse} are clearly not defined. However their corollaries \eqref{TAS2TransL} and \eqref{TAS2TransA} are perfectly consistent, 
\bsubeq \label{TAS2LinearAntiLinear}
\bea
\big(  {\bm{\cDB}}{}^2- 4\m\big){\G} = 0~, \label{TAS2Linear} \\
\big( {\bm{\cD}}^2- 4\mub \big){\G} = 0~, \label{TAS2AntiLinear}
\eea
\esubeq 
and defines a covariantly linear \eqref{TAS2Linear} and anti-linear \eqref{TAS2AntiLinear} scalar superfield, respectively.

Given $n \geq 1$, a superfield ${G}_{\a(n)}$ on $\bm{\mc{V}}_{(n)}$ is said to be longitudinal linear if it obeys the following first-order constraint
\begin{subequations}
	\be
	{\bm{\cDB}}_{(\a_1} {G}_{\a_2 \dots \a_{n+1} )} = 0  \qquad \Longrightarrow \qquad \big( {\bm{\cDB}}{}^2+2n\m \big){G}_{\a(n)} = 0~. \label{TAS2LongLin}
	\ee
	Similarly, a superfield ${G}_{\a(n)}$ on $\bm{\mc{V}}_{(n)}$ is said to be longitudinal anti-linear if it satisfies the condition
	\be
	{\bm{\cD}}_{(\a_1} {G}_{\a_2 \dots \a_{n+1} )} = 0  \qquad \Longrightarrow \qquad \big( {\bm{\cD}}^2+2n\mub \big){G}_{\a(n)} = 0~. \label{TAS2LongAntLin}
	\ee
\end{subequations}
In the scalar case, $n=0$, the conditions \eqref{TAS2LongLin} and \eqref{TAS2LongAntLin} reduce to the chiral and anti-chiral constraints, respectively
\bsubeq
\bea
{\bm{\cDB}}_\a {G}&=&0 ~, \label{TAS2Chiral} \\
{\bm{\cD}}_\a {G}&=&0~. \label{TAS2AntiChiral}
\eea
\esubeq

The transverse \eqref{TAS2TransL} and longitudinal \eqref{TAS2LongLin} linear constraints may be solved in terms of the complex unconstrained
prepotentials ${\L}_{\a(n+1)}$ and ${\O}_{\a(n-1)}$ as follows\footnote{Taking the complex conjugate of  \eqref{TAS2SolveLinCons} yields the solution to the transverse \eqref{TAS2AntiLinear} and longitudinal \eqref{TAS2LongAntLin} anti-linear constraints in terms of complex unconstrained prepotentials.}
\begin{subequations} 	\label{TAS2SolveLinCons}
	\bea
	{\G}_{\a(n)}&=&  {\bm{\cDB}}{}^\b
	{\L}_{\b \a(n)} ~,
	\label{TAS2SolveTransCons} \\
	{G}_{\a(n)}&=&  {\bm{\cDB}}_{\a}
	{\O}_{ \a(n-1) } ~.
	\label{TAS2SolveLongCons}
	\eea
\end{subequations}
The superfields ${\L}_{\a(n+1)}$ and ${\O}_{\a(n-1)}$ are defined modulo
gauge transformations of the form\begin{subequations} 
	\bea
	\d_\x {\L}_{\a(n+1)} &=&   {\bm{\cDB}}{}^\b
	{\x}_{\b \a(n+1)} ~, \\
	\d_\z {\O}_{\a(n-1)} 
	&=&   {\bm{\cDB}}_{\a }
	{\z}_{\a(n-2)} ~, 
	\eea
\end{subequations}
where the gauge parameters ${\x}_{\a(n+2)}$ and ${\z}_{\a(n-2)}$
are complex unconstrained.

Let us consider a superfield $\F_{\a(n)}$ on $\bm{\mc{V}}_{(n)}$ which is simultaneously transverse linear and transverse anti-linear \eqref{TAS2Transverse}. It can be shown that the $\boldsymbol{\D}$ operator \eqref{TAS2DeltaOperator} preserves the TLAL nature of $\F_{\a(n)}$
\be
{\bm{\cD}}^{\b I} \F_{\b\a(n-1)} = 0 \qquad  \Longrightarrow \qquad  {\bm{\cD}}^{\b I}\boldsymbol{\D} \F_{\b\a(n-1)}  = 0~.
\ee
Here, we have introduced the notation ${\bm{\cD}}^{\b I}$ which represents both ${\bm{\cD}}^\b$ and ${\bm{\cDB}}{}^\b$.

Given a positive integer $n$, an arbitrary complex tensor superfield  ${V}_{\a(n)} $  can be uniquely represented as a sum of transverse linear and longitudinal linear multiplets \cite{HutomoKuzenkoOgburn2018}
\bea \label{TASFieldDecomposition}
{V}_{ \a(n)} = &-& 
\frac{1}{2 \mu (n+2)} {\bm{\cDB}}^\g {\bm{\cDB}}_{(\g} {V}_{ \a_1 \dots  \a_n)} 
- \frac{1}{2 \mu (n+1)} {\bm{\cDB}}_{(\a_1} {\bm{\cDB}}{}^{|\g|} {V}_{ \a_2 \dots \a_{n} ) \g} 
~ ,~~~
\eea

There exists projection operators $\mc{P}^{\perp}_{(n)}$ and $\mc{P}^{\parallel}_{(n)}$ which map $\bm{\mc{V}}_{(n)}$ to the the space of transverse linear \eqref{TAS2TransL} and longitudinal linear \eqref{TAS2LongLin} superfields respectively \cite{HutomoKuzenkoOgburn2018}. These operators have the form\footnote{The projectors \eqref{TAS2TLLLProjetors} are the three-dimensional cousins of the AdS$^{4|4}$ projectors \eqref{TLLL} of \cite{IS}. }
\begin{subequations} \label{TAS2TLLLProjetors}
	\bea
	\mc{P}^{\perp}_{(n)}&=& \frac{1}{4 (n+1)\m} ( {\bm{\cDB}}{}^2+2n\m) ~,\\
	\mc{P}^{\parallel}_{(n)}&=&- \frac{1}{4 (n+1)\m} ( {\bm{\cDB}}{}^2-2(n+2)\m ) ~,
	\eea
\end{subequations} 
and satisfy the projector properties 
\bea \label{TAS2TLLLProjetorsProps}
\mc{P}^{\perp}_{(n)}\mc{P}^{\perp}_{(n)} =\mc{P}^{\perp}_{(n)} ~, \quad 
\mc{P}^{\parallel}_{(n)}\mc{P}^{\parallel}_{(n)}=\mc{P}^{\parallel}_{(n)}~,
\quad \mc{P}^{\perp}_{(n)} \mc{P}^{\parallel}_{(n)}=\mc{P}^{\parallel}_{(n)}\mc{P}^{\perp}_{(n)}=0~,  \quad	\mc{P}^{\perp}_{(n)} + \mc{P}^{\parallel}_{(n)} = \mds{1}~. \hspace{0.2cm}
\eea
Note that the projectors \eqref{TAS2TLLLProjetors} are not superspin projection operators, as the corresponding independent superfields obtained via a $(1,1) \to (1,0)$ superfield reduction are unconstrained, and thus cannot furnish irreducible representations of the $\cN=1$ AdS$_3$ superalgebra. The properties \eqref{TAS2TLLLProjetorsProps} imply that any superfield $V_{\a(n)}$ on $\bm{\mc{V}}_{(n)}$ can be uniquely represented as the sum of transverse linear and longitudinal linear superfields,
\be
V_{\a(n)} = \G_{\a(n)} + G_{\a(n)}~,
\ee
which is consistent with \eqref{TASFieldDecomposition}.

The isometry transformations of ${\rm AdS}^{(3|1,1)}$ are generated by real supervector fields ${\l}^{\cA} E_{\cA} $ which solve 
the Killing equation 
\bea
\Big{[}{\L}+\hf l^{ab}M_{ab},{\bm{\cD}}_{\cC}\Big{]}=0~,
\eea
where 
\bea \label{KillingSVC}
\L= \l^{\cA} {\bm{\cD}}_{\cA} =\l^a{\bm{\cD}}_a+\l^\a{\bm{\cD}}_\a+\bar \l_\a {\bm{\cDB}}{}^\a~,
\qquad\bar{\l}^a=\l^a~,
\eea 
and $l^{ab}$ is some local Lorentz parameter.
As shown in \cite{KT-M11},
this equation implies that the parameters $\l^\a$ and $l^{ab}$ 
can be uniquely determined in terms of the vector  $\l_{\a \b}$,
\bea
\l_\a =\frac{\ri}{6}  {\bm{\cDB}}{}^\b \l_{\a\b}~,\qquad
l_{\a\b} =2{\bm{\cD}}_{(\a}\l_{\b)}~,
\label{Killing-1}
\eea
where the vector parameter obeys the equation
\bea
{\bm{\cD}}_{(\a}\l_{\b\g)}=0 \quad \Longleftrightarrow \quad 
{\bm{\cDB}}_{(\a}\l_{\b\g)}=0~.
\eea
A specific feature of ${\rm AdS}^{(3|1,1)} $ is that any two of the three parameters 
$\{ \l_{\ab}, \l_\a, l_{\a\b}\}$ can be expressed in terms of the third parameter. In particular,
\bea
\l_{\a\b} =\frac{\ri}{\m}  {\bm{\cDB}}_{(\a}\l_{\b)}~,\qquad 
\l_\a =
\frac{1}{12 \bar \m} {\bm{\cD}}^\b   l_{\a\b}
~.
\label{2.7}
\eea
From \eqref{Killing-1} and \eqref{2.7} we deduce
\bea
{\bm{\cDB}}_\a\l^\a=
{\bm{\cD}}^\a \l_\a=0~.
\label{1,1-SK_1}
\eea
The solution to these equations is given in \cite{KT-M11}.

The Casimir operators of the ${\rm AdS}^{(3|1,1)} $ isometry algebra ${\mf{osp}(1|2; {\mathbb{R}})} \oplus {\mf{osp}(1|2; {\mathbb{R}})}$ take the following form in the superfield representation\footnote{The Casimir operator \eqref{TAS2Casimiroperator1} is reminiscent of its counterpart \eqref{Cas1} in AdS$^{4|4}$.}
\bsubeq \label{TAS2Casimiroperators}
\begin{align} 
\mathbb{Q}&:=\Box + \hf (\m {\bm{\cD}}^2+ \mub {\bm{\cDB}}{}^2) - 2|\m|^2 M^{\b \g}M_{\b \g} ~,  && [\mathbb{Q}, {\bm{\cD}}_{\cA}] =0~, \label{TAS2Casimiroperator1} \\
\mathbb{F}&:= 2{\bm{\cD}}^{\b\g}M_{\b\g }  + 2 \boldsymbol{\D} ~,  &&[\mathbb{F}, {\bm{\cD}}_{\cA}] =0~,   \label{TAS2Casimiroperator2}
\end{align}
\esubeq
where $\Box := -\hf {\bm{\cD}}^{\a\b} {\bm{\cD}}_{\a\b}$ and $\boldsymbol{\D}$ is the real scalar operator which we define as 
\be \label{TAS2DeltaOperator}
\boldsymbol{\boldsymbol{\D}} := \frac{\ri}{2} {\bm{\cD}}^\b {\bm{\cDB}}_\b = \frac{\ri}{2} {\bm{\cDB}}{}^\b {\bm{\cD}}_\b ~.
\ee
The operator $\boldsymbol{\D}$ can be shown to satisfy the property
\be
[ \boldsymbol{\D}, {\bm{\cD}}^\g {\bm{\cDB}}_\b + {\bm{\cDB}}{}^\g {\bm{\cD}}_\b ] = 0~.
\ee
Note that the overall normalisation factor of the  Casimir operator \eqref{TAS2Casimiroperator2} was chosen so that it coincides with its $\mb{M}^{3|4}$ counterpart  \eqref{TMSSuperhelicityCasimir} in the flat-superspace limit $\m \to 0$. Note that  \eqref{TMSSuperhelicityCasimir} needs to be converted to the complex basis.

Let us denote by ${\bm{\mc{V}}}_{(n)}$ the space of symmetric complex rank-$n$ tensor superfields $\F_{\a(n)}$ on AdS$^{(3|1,1)}$. The Casimir operators \eqref{TAS2Casimiroperators} are related to each other on $\bm{\mc{V}}_{(n)}$ as follows
\bea
\mathbb{F}^2 \F_{\a(n)} =&& 4(2n+1) \boldsymbol{\D}^2 \F_{\a(n)} + 4n^2 \mathbb{Q}\F_{\a(n)}+ 4 \ri n \boldsymbol{\D} \big ( \bm{\cD}_\a \bm{\cDB}{}^\b + \bm{\cDB}_\a \bm{\cD}^\b \big ) \F_{\b\a(n-1)} \non  \\
&&+4n(n-1){\bm{\cD}}_{\a(2)} {\bm{\cD}}^{\b(2)} \F_{\b(2)\a(n-2)}- 2 n^2 \big (\m {\bm{\cD}}^2 + \mub {\bm{\cDB}}{}^2 \big ) \F_{\a(n)}  \non \\
&&- 4n^2(n-2)(n+2)|\m|^2 \F_{\a(n)}~. \label{TAS2RelatingCasimirs}
\eea

In general, the operator $\boldsymbol{\D}$ is not invertible, as seen by the following relation
\be \label{TAS2DeltaInvertibleCondition}
\mathbb{Q} = \frac{1}{16}\{ {\bm{\cD}}^2,{\bm{\cDB}}{}^2\}  + \boldsymbol{\D}^2 - \frac{1}{4} \big ( \m {\bm{\cD}}^2 + \mub {\bm{\cDB}}{}^2 \big ) - |\m|^2M^{\b\g}M_{\b\g}~.
\ee
Let us introduce the operators
\be \label{ScalProj}
\cP_{(0)} := \frac{1}{\mathbb{Q}} \boldsymbol{\D}^2~, \quad 
\cP_{(+)}:=  \frac{1}{16 \mathbb{Q}} \big ( {\bm{\cDB}}{}^2 -4 \m \big ){\bm{\cD}}^2~, \quad 
\cP_{(-)}:=  \frac{1}{16 \mathbb{Q}} \big ( {\bm{\cD}}^2 -4 \mub \big ){\bm{\cDB}}{}^2~,
\ee
which can be extracted from the identity \eqref{TAS2DeltaInvertibleCondition}.
The operators \eqref{ScalProj} can be shown to be orthogonal projectors 
\be
\sum_{i }^{}\cP_{(i)} = \mathds{1}~, \qquad \cP_{(i)} \cP_{(j)} = \d_{ij} \cP_{(i)}~,
\ee
for $i = \{ 0,+,- \}$.
When restricted to the space of scalar superfields, the projectors $\cP_{(+)}$ and $\cP_{(-)}$ map to the subspace of chiral \eqref{TAS2Chiral} and anti-chiral \eqref{TAS2AntiChiral} superfields, respectively, while $\cP_{(0)}$ projects onto the space of linear and anti-linear superfields \eqref{TAS2LinearAntiLinear}. 

\subsubsection{On-shell superfields}
Given a positive integer $n$, a superfield $\F_{\a(n)}$ on $\bm{\mc{V}}_{(n)}$  is said to be on-shell if it obeys the conditions \cite{KuzenkoNovakTartaglino-Mazzucchelli2015}
\bsubeq \label{TAS1OnshellConditions}
\bea
{\bm{\cD}}^{\b I} \F_{\b \a(n-1)} &=&0~, \label{TAS1OnShellCondition1}\\ 
\bm{\D} \F_{\a(n)} &=& \s \bm{M} \F_{\a(n)}~. \label{TAS2OnShellCondition2}
\eea
\esubeq
Here, the real parameter $\bm{M}$ is the pseudo-mass which carries mass dimension one and $\s:= \pm 1$. In the flat-superspace limit, the on-shell conditions \eqref{TAS1OnshellConditions} reduce to those \eqref{TMSMassiveOnshell} in $\mb{M}^{3|4}$ when converted to the complex basis.
It follows from the constraints \eqref{TAS1OnshellConditions} that an on-shell superfield also satisfies the condition
\be \label{TAS2OnshellDelta}
\big ( \mb{F}  - 2(n+1) \s \bm{M} \big ) \F_{\a(n)} = 0 ~.
\ee
Using the identity \eqref{TAS2RelatingCasimirs} in conjunction with \eqref{TAS2OnshellDelta}, it can be shown that an on-shell superfield \eqref{TAS1OnshellConditions} is an eigenvector of the Casimir operator \eqref{TAS2Casimiroperator1},
\be \label{TAS2SecondOrderCas}
\big ( \mb{Q} - \bm{\r}^2 \big ) \F_{\a(n)} = 0~, \qquad \bm{\r}^2:= \bm{M}^2 +n(n+2)|\m|^2~.
\ee
It follows from \eqref{TAS2OnShellCondition2} and \eqref{TAS2SecondOrderCas} that the Casimir operators \eqref{TAS2Casimiroperators} are proportional to the unit operator. Hence an on-shell superfield \eqref{TAS1OnshellConditions} realises an irreducible representation of ${\mf{osp}(1|2; {\mathbb{R}})} \oplus {\mf{osp}(1|2; {\mathbb{R}})}$, which we denote by $\bm{\mf{S}}(\bm{M},\s \frac{n}{2})$. We say that an on-shell superfield \eqref{TAS1OnshellConditions} carries pseudo-mass $\bm{M}$, spin $|s|:=\frac{n}{2}$ and superhelicity $s = \hf(n +1)\s$. 

Recall in anti-de Sitter (super)space in three and four dimensions, we classified the on-shell superfields into two classes: i) partially massless superfields which carry certain pseudo-masses that give rise to a higher depth gauge symmetry;\footnote{Recall that strictly massless (super)fields correspond to the partially massless (super)fields with the maximal gauge symmetry i.e. in general  the lowest depth gauge transformation.} ii) massive superfields which possess a pseudo-mass which is not compatible with a gauge symmetry. Currently, such a dictionary for on-shell supermultiplets is not available in AdS$^{(3|1,1)}$. In particular, the pseudo masses characteristic of partially massless superfields are unknown.

In order to complete the picture of on-shell superfields in AdS$^{(3|1,1)}$, it is necessary to compute the partially massless values. As heavily emphasised throughout this thesis, the information concerning partially massless pseudo-masses is encoded within the poles of the (super)spin projection operators. To conclude this subsection, we will initiate the program to compute the superspin projection operators in AdS$^{(3|1,1)}$.

\subsubsection{Superspin projection operators}
For any integer $n \geq 1$, we define the rank-$n$ superspin projection operator $\bm{\P}_{[n]}^{\perp}$ by its action on $\bm{\mc{V}}_{(n)}$ by the rule
\bsubeq \label{TAS2SpinProjectors}
\bea
\bm{\P}^{\perp}_{[n]} : \bm{\mc{V}}_{(n)} &\longrightarrow& \bm{\mc{V}}_{(n)}, \\
\F_{\a(n)} &\longmapsto& \bm{\P}^{\perp}_{[n]} \F_{\a(n)} = : \bm{\P}^{\perp}_{\a(n)}(\F)~.
\eea
The operator $\bm{\P}^{\perp}_{[n]}$ satisfies the defining properties: 
\begin{enumerate} \label{TAS1ProjProp}
	\item \textbf{Idempotence:} The operator $\bm{\P}^{\perp}_{[n]}$ squares to itself,
	\be
	\bm{\P}^{\perp}_{[n]}\bm{\P}^{\perp}_{[n]} = \bm{\P}^{\perp}_{[n]}~.
	\ee
	\item \textbf{TLAL:} The  operator $\bm{\P}^{\perp}_{[n]}$ maps $\F_{\a(n)}$ to a TLAL superfield,
	\be
	{\bm{\cD}}^{\b I} \bm{\P}^{\perp}_{\b\a(n-1)}(\F) = 0~.
	\ee
	\item \textbf{Surjectivity:} Every transverse superfield $\F^{\perp}_{\a(n)}$ belongs to the image of $\bm{\P}^{\perp}_{[n]}$,
	\be
	{\bm{\cD}}^{\b I}\F^{\perp}_{\b\a(n-1)} = 0 \qquad \Longrightarrow \qquad  \bm{\P}^{\perp}_{[n]} \F^{\perp}_{\a(n)} = \F^{\perp}_{\a(n)}~.
	\ee
\end{enumerate}
\esubeq

Let us consider a superfield $\F_{\a(n)}$ which satisfies the first-order differential constraint \eqref{TAS2OnShellCondition2}. By virtue of the properties \eqref{TAS1ProjProp}, it follows that the superspin projection operator $\bm{\P}^{\perp}_{[n]}$ maps such a superfield to the space of on-shell superfields \eqref{TAS1OnshellConditions}
\bsubeq 
\bea
{\bm{\cD}}^{\b I} \bm{\P}^{\perp}_{\b \a(n-1)} (\F)&=&0~, \\ 
\bm{\D} \bm{\P}^{\perp}_{\a(n)} (\F) &=& \s \bm{M} \bm{\P}^{\perp}_{\a(n)} (\F)~. 
\eea
\esubeq

In order to quench our apparent addiction of deriving $\cN$-extended superprojectors in diverse backgrounds and dimensions,  we will initiate the program of formulating the superprojectors \eqref{TAS2SpinProjectors} in AdS$^{(3|1,1)}$. This will be done by applying a minimal uplift prescription to the $\cN=2$ Minkowski superprojectors \eqref{TMSN2ProjectorComplex}, which involves promoting the $\mb{M}^{3|4}$ covariant derivatives in  \eqref{TMSN2ProjectorComplex} to covariant derivatives in AdS$^{(3|1,1)}$. This will ensure that the superprojectors have the correct form in the flat-superspace limit.
As of yet, we have only been able to compute the superprojectors $\bm{\P}^{\perp}_{[n]}$ for the cases $n=1$ and $n=2$. They take the explicit form
\bsubeq \label{TAS11Projetors}
\bea 
\bm{\P}^{\perp}_{[1]}\F_\a &=& \frac{1}{2 \big (\mathbb{Q} - 4|\m|^2 \big )}\Big ( \boldsymbol{\D} \big ( {\bm{\cD}}_\a{}^\b + \boldsymbol{\D} \d_\a{}^\b \big ) - \frac{1}{4} \big (\mub {\bm{\cDB}}{}^2 + \m {\bm{\cD}}^2 - 4 |\m|^2 \big )\d_\a{}^\b \Big ) \F_{\b}~, \hspace{0.8cm} \label{TAS11Projetors1}\\
\bm{\P}^{\perp}_{[2]}\F_{\a(2)}&=& \frac{1}{4 \big (\mathbb{Q} - 8|\m|^2 \big )\big (\mathbb{Q} - 12|\m|^2 \big )}\boldsymbol{\D} \Big ( {\bm{\cD}}_{(\a_1}{}^{\b_1} + \boldsymbol{\D} \d_{(\a_1}{}^{\b_1} \Big ) \non \\
&&\times \Big ( \boldsymbol{\D} \big ( {\bm{\cD}}_{\a_2)}{}^{\b_2} + \boldsymbol{\D} \d_{\a_2)}{}^{\b_2} \big ) - \frac{1}{2} \big (\mub {\bm{\cDB}}{}^2 + \m {\bm{\cD}}^2  \big )\d_{\a_2)}{}^{\b_2} \Big ) \F_{\b(2)}~. \label{TAS11Projetors2}
\eea
\esubeq
In the flat superspace limit $\mu \rightarrow 0$, the superspin projection operators \eqref{TAS11Projetors1} and \eqref{TAS11Projetors2} coincide with the superprojectors \eqref{TMSN2ProjectorComplex} in $\mb{M}^{3|4}$, for the cases $n=1$ and $n=2$, respectively.
Computing the higher-rank generalisations of the superspin projection operators \eqref{TAS11Projetors} in AdS$^{(3|1,1)}$  remains an open problem.


\subsection{(1,1) AdS superspace: Real basis}

It proves beneficial to realise the $(1,1)$ AdS covariant derivatives in a real basis when performing the reduction procedure to $\cN=1$ AdS superspace. In accordance with \cite{KLT-M12}, the algebra of covariant derivatives \eqref{1.1} can be converted to the real basis by (i) making the convenient choice  $\m =-\ri |\m|$;  and (ii) 
replacing the complex operators ${\bm{\cD}}_\a ,  {\bm{\cDB}}_\a$ with ${\bm \na}^I_{\a} = ({\bm \na}^{\1}_{\a} , {\bm \na}^{\2}_{\a} )$ defined by
\bea \label{realrep}
&& \qquad {\bm{\cD}}_\a = \frac{1}{\sqrt{2}}({\bm \na}^{\1}_\a - \ri{\bm \na}^{\2}_\alpha) ~, 
\qquad {{\bm{\cDB}}}_\a = - \frac{1}{\sqrt{2}}({\bm \na}^{\1}_\a + \ri{\bm \na}^{\2}_\alpha) ~.
\eea
Furthermore, we introduce the real coordinates $z^{\cM}= (x^m, \theta^{\mu}_I)$ which are used to parametrise ${\rm AdS}^{(3|1,1)} $. Choosing to define ${\bm \na_a} = {\bm{\cD}}_a$, it can be shown that the algebra of $(1,1)$ AdS covariant derivatives assumes the following form in the real basis \eqref{realrep}
\vspace{-\baselineskip}
\begin{subequations}  \label{1.3}
	\begin{alignat}{3}
	\qquad \{{\bm \na}^{\1}_\a	, {\bm \na}^{\2}_\b \} &= 0 ~,& \\
	\qquad \{ {\bm \na}^{\1}_\a , {\bm \na}^{\1}_\b \} &= 2 \ri {\bm \na}_{\a \b} - 4 \ri |\m| M_{\a \b} ~,&
	\qquad \{ {\bm \na}^{\2}_\a , {\bm \na}^{\2}_\b \} &= 2 \ri {\bm \na}_{\a \b} + 4 \ri |\m| M_{\a \b} ~,\\
	\qquad [ {\bm \na}_{a} , {\bm \na}^{\1}_\a ] &= |\m| (\g_a)_\a{}^\b{\bm \na}^{\1}_\b ~,&
	\qquad [ {\bm \na}_{a} , {\bm \na}^{\2}_\a] &= - |\m| (\g_a)_\a{}^\b {\bm \na}^{\2}_\b ~, \\
	\qquad [{\bm \na}_{a}, {\bm \na}_{b}] &= -4|\m|^2 M_{ab}~.&
	\end{alignat}
\end{subequations} 
We collect an array of identities involving the $(1,1)$ AdS covariant derivatives in section \ref{TAS2appendixA}, which will be useful in deriving many of the results in this chapter.

It follows from \eqref{1.3} that 
the operators ${\bm \na}_{a}$ and ${\bm \na}_{\a}^{\1}$ possess the property:
\begin{enumerate}
	\item These operators form a closed algebra given by 
	\begin{subequations} \label{1.4}
		\be
		\{ {\bm \na}_{\a}^{\1} , {\bm \na}^{\1}_\b \} = 2 \ri {\bm \na}_{\a\b} - 4 \ri |\m| M_{\a \b} ~,
		\ee
		\be
		\quad [ {\bm \na}_{a} , {\bm \na}_{\b}^{\1} ] = |\m| (\g_a)_\b{}^\g {\bm \na}^{\1}_\g ~,
		\quad [{\bm \na}_a, {\bm \na}_b] = -4|\m|^2 M_{ab}~.
		\ee
	\end{subequations}
	Relations \eqref{1.4} coincide with the algebra \eqref{SA} of covariant derivatives of AdS$^{3|2}$.
\end{enumerate}
This property implies that AdS$^{3|2}$ is naturally realised as a surface embedded in $(1,1)$ AdS superspace. One can make an appropriate choice in the real Grassmann variables 
$\q^\m_I = (\q^{\m}_{\1},\q^\m_{\2})$ such that AdS$^{3|2}$ can be identified as the surface defined by $\q^\m_{\2}=0$ in ${\rm AdS}^{(3|1,1)} $. These properties enable the consistent reduction of any field theory in ${\rm AdS}^{(3|1,1)} $ to AdS$^{3|2}$.

We now wish to recast the fundamental properties of the Killing supervector fields of $(1,1)$ AdS superspace \eqref{KillingSVC} in the real representation \eqref{realrep}. The isometries of $(1,1)$ AdS superspace are generated by the $(1,1)$ AdS Killing supervector fields,
\be \label{11KillingVector}
\L : = \l^\cA\bm \nabla_\cA = \l^a \bm \nabla_a + \l^\a_I \bm \nabla^I_\a~, \qquad I= \1, \2~,
\ee
which are defined to satisfy the Killing equation
\be \label{Kconstraint}
[\L + \frac{1}{2}l^{ab}M_{ab}, \bm \nabla_\cA] = 0~,
\ee
for some real Lorentz parameter $l^{ab} = - l^{ba}$. It can be shown that equation \eqref{Kconstraint} is equivalent to the set of equations
\begin{subequations} \label{KillingEqnSD}
	\bea 
	\bm \nabla^{\1}_\a \l^{\1}_\b &=& \frac{1}{2}l_{\a\b}+|\m|\l_{\a\b}~, \qquad \bm \nabla^{\1}_\a \l^{\2}_\b = 0 ~,\\
	\bm \nabla^{\2}_\a \l^{\2}_\b &=& \frac{1}{2}l_{\a\b}-|\m|\l_{\a\b} ~,  \qquad \bm \nabla^{\2}_\a \l^{\1}_\b = 0 ~,\\
	\bm \nabla^{\1}_\a   l_{\b\g} &=& 8 \ri |\m| \varepsilon_{\a(\b}\l^{\1}{}_{\g)}~, \qquad ~\bm \nabla^{\2}_\a   l_{\b\g} = -8 \ri |\m| \varepsilon_{\a(\b}\l^{\2}{}_{\g)}~, \\
	\bm \nabla^I_\a \l_a &=& 2\ri (\g_a)_{\a\b} \l^{I\b}~,
	\eea
\end{subequations}
and
\begin{subequations} \label{KillingEqnVD}
	\bea \label{2.26a}
	\bm \nabla_a \l_b &=&l_{ab}~, \\
	\bm \nabla_a \l^{\1}_\a &=& |\m|(\g_a)_\a{}^\b \l ^{\1}_\b~, \qquad \bm \nabla_a \l^{\2}_\a = - |\m|(\g_a)_\a{}^\b \l ^{\2}_\b~, \label{SKeqn}\\
	\bm \nabla_a l_{bc} &=& 8 |\m|^2 \eta_{a[b} \l_{c]}~.
	\eea
\end{subequations}
Equations \eqref{KillingEqnSD} and \eqref{KillingEqnVD} can be recast in the equivalent form
\begin{subequations} \label{K-equiv}
	\bea
	\bm \nabla^I_{(\a} \l_{\b\g)} &=& 0~, \hspace{3cm} \bm \nabla^I_{(\a} l_{\b\g)} = 0~, \\
	\bm \nabla^{\1}_{(\a} \l^{\1}_{\b)} &=& \frac{1}{2}l_{\a\b} + |\m|\l_{\a\b}~, \qquad \bm \nabla^{\2}_{(\a} \l^{\2}_{\b)} = \frac{1}{2}l_{\a\b} - |\m|\l_{\a\b}      ~,\\
	\bm \nabla^{\1}_{(\a} \l^{\2}_{\b)} &=& \bm \nabla^{\2}_{(\a} \l^{\1}_{\b)} = 0, \qquad ~~ \bm \nabla^{\a(I} \l^{J)}_\a = 0~, \\
	\l^{\1 \a} &=& \frac{\ri}{6} \bm \nabla^{\1}_\b \l^{\a\b} = \frac{\ri}{12|\m|} \bm \nabla^{\1}_\b l^{\a\b}~, \label{2.27d} \\
	\l^{\2 \a} &=& \frac{\ri}{6} \bm \nabla^{\2}_\b \l^{\a\b} = - \frac{\ri}{12|\m|} \bm \nabla^{\2}_\b l^{\a\b}~. \label{2.27e}
	\eea
\end{subequations}
It follows from \eqref{2.26a} that the parameter $\l_a$ is a Killing vector field
\be
\bm \nabla_a \l_b + \bm \nabla_b \l_a = 0~,
\ee
and relations \eqref{SKeqn} are Killing spinor equations.

\subsection{Reduction from $(1,1) \to (1,0)$ AdS superspace}
Given a tensor superfield $U(x,\q_I)$ on $(1,1)$ AdS superspace, where indices have been suppressed, its bar-projection to $\cN=1$ AdS superspace is defined by the rule
\be
U|:= U(x,\q_I)|_{\q_{\2}=0}
\ee
in a special coordinate system which will be detailed below. Given the $(1,1)$ covariant derivative in the the real representation \eqref{realrep}
\be
\bm \nabla_\cA = (\bm \nabla_a, \bm \nabla^I_\a) = E_{\cA}{}^{\cM}\frac{\pa}{\pa z^{\cM}} + \hf \O_{\cA}{}^{bc}M_{bc} ~,
\ee
we define its $\cN=1$ projection by the rule
\be
\bm \nabla_\cA| = E_{\cA}{}^{\cM}|\frac{\pa}{\pa z^{\cM}} + \hf \O_{\cA}{}^{bc}|M_{bc} ~.
\ee
We use the freedom to perform general coordinate and local Lorentz transformations to impose the gauge
\be \label{reductiongauge}
\bm \nabla_a | = \nabla_a~, \qquad \bm \nabla^{\1}_\a| = \nabla_\a ~,
\ee
where $\nabla_A = (\nabla_a, \nabla_\a)$ is the set of covariant derivatives for AdS$^{3|2}$, see eq.\,\eqref{SCD}. 
In the chosen coordinate system, the operator $\bm \nabla^{\1}_\a$ does not involve any partial derivatives with respect to $\q_{\2}$. Thus for any positive integer $k$, it follows that $(\bm \nabla^{\1}_{\a_1}\ldots\bm \nabla^{\1}_{\a_k} U)| = \nabla_{\a_1}\ldots \nabla_{\a_k} U|$. 

We now consider the $\cN=1$  projection of the $(1,1)$ AdS Killing supervector \eqref{11KillingVector}
\be
\L | = \x^a \nabla_a + \x^\a \nabla_\a + \e^\a \bm \nabla^{\2}_\a |~,
\ee
where we have introduced the $\cN=1$ superfields
\be
\x^a := \l^a |~, \qquad \x^\a := \l^\a_{\1}|~, \qquad \e^\a : = \l^\a_{\2}|~.
\ee
Additionally, we define the $\cN=1$ projection of the Lorentz parameter $l^{ab}$ to be
\be
\z^{ab} := l^{ab}|~.
\ee
It is important to note that the superfields $(\x^a,\x^\a, \z^{ab})$ parametrise the infinitesimal isometries of AdS$^{3|2}$. Such transformations are generated by
the Killing supervector fields, $\x = \x^a\nabla_a + \x^\a \nabla_\a$, satisfying the ${\cN}=1$ Killing equation, eq.~\eqref{Kcondition1-1}. Indeed, the relations \eqref{2.40d-1} and \eqref{2.40e} automatically follow from the $(1,1)$ AdS Killing equations \eqref{K-equiv}, upon projection.
The parameter $\e_\a$, which generates the second supersymmetry transformation, has the property
\bea
\nabla_a\e_\a &=& -|\m|(\g_a)_\a{}^\b \e_\b~. \label{2.40d}
\eea

Given the transformation law of a tensor superfield $U(x,\q_I)$ on $(1,1)$ AdS superspace
\be
\d_\L U= \big ( \L + \frac{1}{2}l^{ab}M_{ab} \big ) U~,
\ee
we find its projection to $\cN=1$ AdS superspace to be
\be \label{PTL}
\d_\L U| = \d_\x U| + \d_\e U| ~,
\ee
where 
\begin{subequations} \label{PTL2}
	\bea \label{PTL2A}
	\d_\x U| &=& \big (  \x^a \nabla_a + \x^\a \nabla_\a + \hf \z^{ab}M_{ab} \big ) U| ~, \\
	\d_\e U| &=&  \e^\a \big (\bm \nabla^{\2}_\a U \big ) |~. \label{PTL2B}
	\eea
\end{subequations}
The first transformation \eqref{PTL2A} coincides with the infinitesimal transformation generated by a Killing supervector in AdS$^{3|2}$, eq.~\eqref{211}. Thus, $U|$ can be identified as a tensor superfield on $\cN=1$ AdS superspace. The other transformation \eqref{PTL2B} corresponds to the second supersymmetry transformation, which is generated by $\e^\a$.


\subsection{The (1,1) AdS supersymmetric actions in AdS$^{3|2}$} \label{Subsect44}

Every supersymmetric field theory in $(1,1)$ AdS superspace can be reduced to $\cN=1$ AdS superspace. In the following section, we explore the necessary mathematical framework which will be employed to develop such a reduction procedure. 
As presented in \cite{KLT-M11,KT-M11,KLRST-M,BKT-M}, manifestly supersymmetric actions in $(1,1)$ AdS superspace can be constructed by either:
\begin{enumerate}
	\item Integrating a real scalar Lagrangian $\cL$ over the full $(1,1)$ AdS superspace,
	\bea \label{real11}
	&&\int \rd^3 x \, \rd^2 \q\, \rd^2 \bar \q \, \bm E\,\cL = \frac{1}{16}\int \rd^3 x\,e\, ({\bm{\cD}}^2- 16 \bar \m)( {\bm{\cD}}^2 - 4 \mu) \cL \Big |_{\q=0} \non\\
	&&= \int \rd^3 x\,e\,\Big ( \frac{1}{16}{\bm{\cD}}^\a ({{\bm{\cDB}}}^2 - 6 \m){\bm{\cD}}_\a - \frac{\m}{4}{\bm{\cD}}^2 - \frac{\bar{ \m}}{4}{ {\bm{\cDB}}}^2+4\m \bar{\m} \Big )  \cL \Big |_{\q=0}~.
	\eea
	\item Integrating a covariantly chiral Lagrangian $\cL_c$ over the chiral subspace,
	\be \label{CI}
	\int \rd^3 x\,\rd^2 \q \,\cE\,\cL_c  = - \frac{1}{4}\int \rd^3 x\,e\,({\bm{\cD}}^2 - 16\bar{ \m})\cL_c\Big |_{\q=0}~, \qquad { {\bm{\cDB}}}_{\a} \cL_c = 0~.
	\ee	
	
	The two supersymmetric invariants are related by the rule
	\bea
	\int \rd^3 x \, \rd^2 \q\, \rd^2 \bar \q \, \bm E\,\cL = \int \rd^3 x\,\rd^2 \q \,\cE\,\cL_c~, \qquad \cL_c := -\frac{1}{4} ( {\bm{\cDB}}{}^2 - 4 \mu) \cL~. 
	\eea
\end{enumerate}
In $(1,1)$ AdS superspace, every chiral action can always be recast as an integral over the full superspace 
\be \label{CFRelation}
\int \rd^3 x\,\rd^2 \q \,\cE\,\cL_c = \frac{1}{\mu} \int \rd^3 x \, \rd^2 \q\, \rd^2 \bar \q \, \bm E \,\cL_c~.
\ee
We will use the notation $\rd^{3|4}z :=\rd^3x\, \rd^2 \q \rd^2 \bar \q$ to denote the full superspace measure.

Instead of reducing the above supersymmetric actions to components, we wish to obtain a prescription which allows for their reduction to $\cN=1$ AdS superspace. The supersymmetric action in $\rm AdS^{3|2}$ is described by a real scalar Lagrangian $L$
\be \label{1.8}
S = \int \rd^{3|2}z \, E \, L = \frac{1}{4} \int \rd^3x \, e \, (\ri \nabla^2 + 8|\m|) L\,\big|_{\,\q = 0}~.
\ee
The action \eqref{real11} reduces to $\rm AdS^{3|2}$ as follows
\be \label{N1rule}
{\mathbb S} = \int \rd^{3|4}z \, \bm E \, \cL = -\frac{\ri}{4}\int \rd^{3|2} z \, E \, \Big ( (\bm \nabla^{\2})^2+8\ri|\m| \Big ) \cL \big|~.
\ee
By making use of the Killing equations \eqref{K-equiv} and \eqref{2.40d}, it can be shown that the action \eqref{N1rule} is invariant under $(1,1)$ AdS isometry transformations given by equation \eqref{PTL}.


\section{Massless half-integer superspin: Transverse formulation} \label{Section5}
In $(1,1)$ AdS superspace, there exist two off-shell formulations for the massless multiplet of half-integer superspin-$(s+\frac{1}{2})$, with $s \geq 2$ \cite{HutomoKuzenkoOgburn2018}. These two theories, which are called transverse and longitudinal, prove to be dual to each other. In the following section, we develop the $(1,1) \rightarrow (1,0)$ reduction procedure for the transverse formulation.

\subsection{Transverse formulation}
According to \cite{HutomoKuzenkoOgburn2018}, the transverse formulation for the massless superspin-$(s+\frac{1}{2})$ multiplet is described by the dynamical variables
\be
\cV^\perp_{(s+\hf )} = \big\{ \mathfrak{H}_{\a(2s)}, \G_{\a(2s-2)}, \bar{\G}_{\a(2s-2)} \big\} ~, \label{2.1}
\ee
where the real superfield $\mathfrak{H}_{\a(2s)}$ is unconstrained, while the complex superfield $\G_{\a(2s-2)}$ is transverse linear \eqref{TAS2TransL}. 
The superfields $\mathfrak{H}_{\a(2s)}$ and $ \G_{\a(2s-2)}$ are defined modulo gauge transformations of the form 
\begin{subequations} \label{tr-gauge-half}
	\bea 
	\d_\l \mathfrak{H}_{\a(2s)}&=& 
	{ {\bm{\cDB}}}_{(\a_1} \l_{\a_2 \dots \a_{2s})}-{{\bm{\cD}}}_{(\a_1}\bar {\l}_{\a_2 \dots \a_{2s})}
	\equiv
	g_{\a(2s)}+ \bar { g}_{\a(2s)} ~, \label{H-gauge} \\ 
	\d_\l \G_{\a(2s-2)}&=&
	-\frac{1}{4}{{\bm{\cDB}}}^{\b} 
	\big( {{\bm{\cD}}}^2  +2(2s-1) \bar \m \big)\bar{\l}_{\b\a(2s-2)}
	\equiv \frac{s}{2s+1}{{\bm{\cDB}}}^{\b}{\bm{\cD}}^{\g}\bar{g}_{\b \g \a{(2s-2)}}
	~,
	\label{gamma-gauge}
	\eea
\end{subequations}
where the gauge parameter $\l_{\a(2s-1)}$ is complex unconstrained. By construction, the complex gauge parameter $g_{\a(2s)} : = { {\bm{\cDB}}}_{(\a_1} \l_{\a_2 \dots \a_{2s})}$ is longitudinal linear \eqref{TAS2LongLin}. Note that the gauge transformation \eqref{H-gauge} indicates that  $\mathfrak{H}_{\a(2s)}$ corresponds to the $\cN=1$ superconformal gauge prepotential \cite{KuzenkoOgburn2016}. 

Up to normalisation, there exists a unique quadratic action which is invariant under the gauge transformations \eqref{tr-gauge-half}. This action takes the explicit form
\bea \label{2.5}
&&\mathbb{S}^{\perp}_{(s+\hf)} [\mathfrak{H}_{\a(2s)},\G_{\a(2s-2)}, \bar{\G}_{\a(2s-2)}]
= \Big(-\hf \Big)^s \int \rd^{3|4}z \, \bm E \,
\bigg\{ 2s(2s-1)\bar{\m}\m \mathfrak{H}^{\a(2s)} \mathfrak{H}_{\a(2s)} ~ \non \\
&&+\frac{1}{8} \mathfrak{H}^{\a(2s)}  {\bm{\cD}}^\b \big ({ {\bm{\cDB}}}{}^2- 6\mu \big ){\bm{\cD}}_\b \mathfrak{H}_{\a(2s)} + \mathfrak{H}^{\a(2s)}\big({\bm{\cD}}_{(\a_1} { {\bm{\cDB}}}_{\a_2} {\G}_{\a_3 \dots \a_{2s})} - { {\bm{\cDB}}}_ {(\a_1} {\bm{\cD}}_{\a_2} \bar {\G}_{\a_3 \dots \a_{2s})} \big)
~ \non \\
&&+ \frac{2s-1}{s} \bar {\G}^{\a(2s-2)} {\G}_{\a(2s-2)} + \frac{2s+1}{2s} 
\big({\G}^{\a(2s-2)}{\G}_{\a(2s-2)} + \bar {\G}^{\a(2s-2)} \bar {\G}_{\a(2s-2)}\big) \bigg\}~.
\eea
The above construction does not take into consideration the case where $s=1$, since the transverse linear constraint \eqref{TAS2TransL} is ill-defined for $n=0$. However, corollary \eqref{TAS2Linear} is consistent for $n=0$ and defines a covariantly linear scalar superfield. The superfield $\G$ and its complex conjugate $\bar{\G}$ can be interpreted as compensators, which have the corresponding gauge transformations
\begin{subequations}
	\bea
	\d_\l \mathfrak{H}_{\a\b} &=& {{\bm{\cDB}}}_{(\a}\l_{\b)} - {\bm{\cD}}_{(\a} \bar{\l}_{\b)}~, \\
	\d_\l \G &=& - \frac{1}{4} {{\bm{\cDB}}}^\b({\bm{\cD}}^2+2\bar{\m})\bar{\l}_\b~, \label{variationgamma}
	\eea
\end{subequations}
as a result of \eqref{tr-gauge-half}. It is simple to show that transverse linear constraint \eqref{TAS2TransL} satisfied by $\G$ is consistent with the variation \eqref{variationgamma}. Choosing $s=1$ in \eqref{2.5} yields the linearised action for non-minimal $(1,1)$ AdS supergravity \cite{KT-M11}.

\subsection{Reduction of gauge prepotentials to $\text{AdS}^{3|2}$} \label{TASReductionGaugePrep}
We wish to reduce the gauge prepotentials \eqref{2.1} to $\cN=1$ AdS superspace. We start by reducing the superconformal gauge multiplet $\mathfrak{H}_{\a(2s)}$. Converting the longitudinal linear constraint  of $g_{\a(2s)}$ \eqref{TAS2LongLin} to the real representation \eqref{realrep} yields  
\be \label{2.6}
\bm \nabla^{\2}{}_{(\a_1} g_{\a_2\ldots\a_{2s+1})} = \ri \bm \nabla^{\1}{}_{(\a_1} g_{\a_2\ldots\a_{2s+1})}~.
\ee
Taking a Taylor expansion of $g_{\a(2s)}(\q^I)$ about $\q^{\2}$, and using \eqref{2.6}, we find the independent $\q^{\2}$~-components of $g_{\a(2s)}$ to be
\be
g_{\a(2s)} |~, \qquad \bm \nabla^{\underline{2}\beta}g_{\b\a(2s-1)}|~.
\ee
The gauge transformation \eqref{H-gauge} allows us to impose the gauge conditions
\be
\mathfrak{H}_{\a(2s)}|=0~, \qquad \bm \nabla^{\2\beta}\mathfrak{H}_{\b\a(2s-1)}|=0~. \label{2.8}
\ee
In this Wess-Zumino (WZ) gauge, we stay with the unconstrained real $\cN=1$ superfields\begin{subequations} 
	\label{2.9}
	\bea 
	H_{\a(2s+1)} :&=&\ri \bm \nabla^{\2}{}_{(\a_1}\mathfrak{H}_{\a_2\ldots\a_{2s+1})}|~,  \\
	H_{\a(2s)} :&=&\frac{\ri}{4}(\bm \nabla^{\2})^2\mathfrak{H}_{\a(2s)}|~.
	\eea
\end{subequations}
The residual gauge freedom which preserves the gauge conditions \eqref{2.8} is described by the real $\cN = 1$ unconstrained superfields
\begin{subequations}\label{2.10}
	\begin{align}
	g_{\a(2s)}|&= - \frac{\ri}{2}\z_{\a(2s)}~, & \bar \z_{\a(2s)} &= \z_{\a(2s)}~, \\
	\bm \nabla^{\2\b}g_{\b\a(2s-1)}|&= \frac{2s+1}{2s}\z_{\a(2s-1)}~, & \bar \z_{\a(2s-1)} &=  \z_{\a(2s-1)}~.
	\end{align}
\end{subequations}
From \eqref{2.10}, we can readily determine the gauge transformations of the superfields  \eqref{2.9}
\vspace{-\baselineskip}
\begin{subequations} \label{2.11}
	\bea 
	\d_\z H_{\a(2s+1)} &=& \ri \nabla_{(\a_1} \z_{\a_2 \ldots \a_{2s+1})}~, \\
	\d_\z H_{\a(2s)} &=& \nabla_{(\a_1}\z_{\a_2 \ldots \a_{2s})}~.
	\eea
\end{subequations}

Next, we wish to reduce $\G_{\a(2s-2)}$ to $\cN =1$ AdS superspace. The superfield $\G_{\a(2s-2)}$ obeys the transverse linear constraint \eqref{TAS2TransL}, which takes the following form in the real representation \eqref{realrep}
\be \label{2.12}
\bm \nabla^{\2\b}\G_{\b\a(2s-3)} = \ri \bm \nabla^{\1\b} \G_{\b\a(2s-3)}~.
\ee
It follows that $\G_{\a(2s-2)}(\q^I)$ has two independent $\q^{\2}$ -components
\be \label{2.13}
\G_{\a(2s-2)}| \ , \qquad  \bm \nabla^{\2}{}_{(\a_1}\G_{\a_2 \ldots \a_{2s-1})}|~.
\ee
Utilising the gauge transformation of $\G_{\a(2s-2)}$ \eqref{gamma-gauge} and the real representation \eqref{realrep}, we find
\begin{subequations} \label{2.14}
	\bea 
	\d_\l \G_{\a(2s-2)} &=& \frac{\ri s}{2s+1}\Big (\bm \nabla^{\1\b} \bm \nabla^{\2\g} - \bm \nabla^{\b\g}\Big ) \bar {g}_{\b\g\a(2s-2)} ~, \\ 
	\bm \nabla^{\2}{}_{(\a_1}\d_\l \G_{\a_2\ldots\a_{2s-1})} &=& \frac{s}{2s+1}\Big ( \bm \nabla^{\1 \b} \bm \nabla_{(\a_1}{}^\g \bar {g}_{\a_2 \ldots \a_{2s-1})\b\g}-(4s+1)|\m| \bm \nabla^{\1 \b} \bar {g}_{\b\a(2s-1)}~\\ \non
	&+&  \bm {\nabla}_{\b}{}^\g \bm {\nabla}^{\1\b}\bar {g}_{\g\a(2s-1)} - \ri \bm {\nabla}_\b{}^\g \bm \nabla^{\2\b} \bar{g}_{\g\a(2s-1)}-\frac{1}{2}( \bm \nabla^{\1})^2 \bm \nabla^{\2\b} \bar {g}_{\b\a(2s-1)}~ \non \\
	&-& (2s+1)\ri|\m|\bm \nabla^{\2\b}\bar {g}_{\b\a(2s-1)} -\ri \bm \nabla^{\b\g} \bm \nabla^{\2}{}_{(\a_1} \bar {g}_{\a_2 \ldots \a_{2s-1})\b\g}\Big )~. \non
	\eea
\end{subequations}
From \eqref{2.14}, we can immediately read off the gauge transformations of the complex $\cN =1$ superfields \eqref{2.13}
\begin{subequations} \label{GTHITrans}
	\bea \label{2.15a}
	\d_\l \G_{\a(2s-2)}| &=& - \frac{\ri}{2}\nabla^\b\z_{\b\a(2s-2)} + \frac{s}{2(2s+1)}\nabla^{\b\g}\z_{\b\g\a(2s-2)} ~, \\ 
	\bm \nabla^{\2}{}_{(\a_1}\d_\l \G_{\a_2\ldots\a_{2s-1})}| &=& \frac{s}{2(2s+1)}\Big ( \ri \nabla^\b\nabla_{(\a_1}{}^\g \z_{\a_2\ldots\a_{2s-1})\b\g} + \ri \nabla_\b{}^\g \nabla^\b \z_{\g\a(2s-1)}~\\ \non
	&-& (4s+1)\ri |\m|\nabla^\b\z_{\b\a(2s-1)} + \frac{2s+1}{s}\big ((2s+1)\ri|\m|\z_{\a(2s-1)}  ~\\
	&+&\ri \nabla_{(\a_1}{}^\b \z_{\a_2\ldots\a_{2s-1})\b} + \frac{1}{2}\nabla^2\zeta_{\a(2s-1)} \big) \Big ) ~. \label{2.15b} \non
	\eea
\end{subequations}

Let us express the ${\cN}=1$ superfields \eqref{2.13} in terms of their real and imaginary parts,
\vspace{-\baselineskip}
\begin{subequations}
	\bea
	\G_{\a(2s-2)}| &=&-L_{\a(2s-2)} -  \ri s V_{\a(2s-2)} ~,\\
	\bm \nabla^{\2}{}_{(\a_1} \G_{\a_2\ldots\a_{2s-1})}| &=&\frac{1}{2} \big (\F_{\a(2s-1)} + \ri \O_{\a(2s-1)} \big )~.
	\eea
\end{subequations}

It then follows from the gauge transformations \eqref{2.11} and \eqref{GTHITrans} that we are in fact dealing with two different gauge theories. The first model is formulated in terms of the real unconstrained gauge superfields
\be \label{2.17}
\cV^\parallel_{(s+\frac{1}{2})}= \{ H_{\a(2s+1)}, L_{\a(2s-2)}, \F_{\a(2s-1)} \}~,
\ee
which are defined modulo gauge transformations of the form
\begin{subequations}\label{2.18}
	\bea \label{2.18a}
	\d_\z H_{\a(2s+1)} &=& \ri \nabla_{(\a_1}\z_{\a_2 \ldots \a_{2s+1})}~, \\ \label{2.18b}
	\d_\z L_{\a(2s-2)} &=&- \frac{s}{2(2s+1)}\nabla^{\b\g}\z_{\b\g\a(2s-2)}~,\\
	\d_\z \F_{\a(2s-1)} &=& \frac{\ri s}{2s+1}\Big (\nabla^\b \nabla_{(\a_1}{}^\g \z_{\a_2\ldots\a_{2s-1})\b\g} - (4s+1)|\m|\nabla^\b\z_{\b\a(2s-1)} ~ \\
	&-& \nabla^{\b\g}\nabla_{\b} \z_{\g\a(2s-1)}\Big )~, \non
	\eea
\end{subequations}
where the gauge parameter $\z_{\a(2s)}$ is real unconstrained. The other gauge theory is constructed in terms of the superfields
\be \label{2.19}
\cV^\parallel_{(s)} = \{ H_{\a(2s)}, V_{\a(2s-2)}, \O_{\a(2s-1)} \}~,
\ee
which possess the gauge freedom
\begin{subequations} \label{2.20}
	\bea \label{2.20a}
	\d_\z H_{\a(2s)} &=&  \nabla_{(\a_1}\z_{\a_2 \ldots \a_{2s})} ~, \\ \label{2.20b}
	\d_\z V_{\a(2s-2)} &=&  \frac{1}{2s}\nabla^\b \z_{\b\a(2s-2)}~, \\
	\d_\z \O_{\a(2s-1)} &=& \nabla_{(\a_1}{}^\b \z_{\a_2 \ldots \a_{2s-1})\b} - \frac{\ri}{2}\nabla^2\z_{\a(2s-1)} + (2s+1)|\m|\z_{\a(2s-1)} ~,
	\eea
\end{subequations}
where the parameter $\z_{\a(2s-1)}$ is real unconstrained. 

Applying the superspace reduction procedure to the action \eqref{2.5} yields two decoupled $\cN =1$ supersymmetric theories, each formulated in terms of the gauge fields \eqref{2.17} and \eqref{2.19} respectively
\bea \label{ReductionTransHI}
\mathbb{S}^{\perp}_{(s+\hf)} [\mathfrak{H}_{\a(2s)},\G_{\a(2s-2)}, \bar{\G}_{\a(2s-2)}] &=& S^\parallel_{(s+\frac{1}{2})}[H_{\a(2s+1)},L_{\a(2s-2)},  \F_{\a(2s-1)}] \non ~\\ 
&&+ S^\parallel_{(s)}[H_{\a(2s)},V_{\a(2s-2)}, \O_{\a(2s-1)}]~.
\eea
Explicit expressions for these decoupled $\cN=1$ supersymmetric actions are given in the subsequent section.


\subsection{Massless higher-spin $\cN =1$ supermultiplets}
The first of the decoupled $\cN=1$ supersymmetric actions, which is realised in terms of the superfields \eqref{2.17}, takes the following form
\bea \label{2.22}
&&S^\parallel_{(s+\frac{1}{2})}[H_{\a(2s+1)},L_{\a(2s-2)},\F_{\a(2s-1)}]= \Big (-\frac{1}{2} \Big )^s \Big ( - \frac{\ri}{8}\Big  )\int \rd^{3|2}z~ E ~\bigg \{ 2H^{\a(2s+1)}\mathbb{Q} H_{\a(2s+1)} \non ~\\
&&- 8(s-2)(2s+1)|\m|^2H^{\a(2s+1)}H_{\a(2s+1)} +  (2s+1)\ri|\m|H^{\a(2s+1)}\nabla^2H_{\a(2s+1)}~ \non \\
&&-4(2s+1)|\m|H^{\b\a(2s)}\nabla_\b{}^\g H_{\g\a(2s)}-\ri H^{\b\a(2s)}\nabla^2\nabla_\b{}^\g H_{\g\a(2s)}  - 4 H^{\b\g\a(2s-1)}\nabla_{\b\g}\F_{\a(2s-1)}~ \non \\
&&- 8\ri H^{\b\g\d\a(2s-2)}\nabla_{\b\g} \nabla_{\d}L_{\a(2s-2)} - 16\ri L^{\a(2s-2)}\nabla^\b \F_{\b\a(2s-2)} + \frac{2}{s}\F^{\a(2s-1)}\F_{\a(2s-1)} ~ \non \\
&&+ \frac{8\ri}{s(2s-1)} \Big ( 2(s-1)(4s-1)L^{\b\a(2s-3)}\nabla_\b{}^\g L_{\g\a(2s-3)}  ~ \non \\ && - (3s-1)\ri L^{\a(2s-2)}\nabla^2L_{\a(2s-2)}  -4s(4s^2-3s+1)|\m|L^{\a(2s-2)}L_{\a(2s-2)} 
\Big ) \bigg \}~.
\eea
Upon inspection, it is apparent that the superfield $\F_{\a(2s-1)}$ is auxiliary. So by making use of its equation of motion, 
\be
\F_{\a(2s-1)} = -s \nabla^{\b\g}H_{\b\g\a(2s-1)}-4\ri s \nabla_{(\a_1}L_{\a_2 \ldots \a_{2s-1})}~,
\ee
we can eliminate the auxiliary field $\F_{\a(2s-1)}$ from the action \eqref{2.22}. It can be shown that the resulting action coincides with the  action for the massless superspin-$(s+\hf)$ multiplet, which is given by  \eqref{LongHalfIntAct}.

The other $\cN =1$ action is formulated in terms of the gauge superfields \eqref{2.19}  
\bea \label{2.23}
&&S^\parallel_{(s)}[H_{\a(2s)},V_{\a(2s-2)}, \O_{\a(2s-1)}] = \Big (-\frac{1}{2} \Big )^s \int \rd^{3|2}z~ E ~\bigg\{ \frac{\ri}{2} H^{\a(2s)}\nabla^2H_{\a(2s)} \non \\
&&+2 |\m| H^{\a(2s)}H_{\a(2s)}  + 2s H^{\b\g\a(2s-2)}\nabla_{\b\g}V_{\a(2s-2)} - H^{\b\a(2s-1)}\nabla_\b \O_{\a(2s-1)} \non ~ \\
&&+ \frac{\ri}{2}\O^{\a(2s-1)}\O_{\a(2s-1)} + \frac{1}{2s-1} \Big (4s(s-1)^2V^{\b\a(2s-3)}\nabla_\b{}^\g V_{\g\a(2s-3)}~ \non \\
&&+\ri s (2s^2-2s+1)V^{\a(2s-2)}\nabla^2 V_{\a(2s-2)} +4s(2s^3-2s+1)|\m|V^{\a(2s-2)}V_{\a(2s-2)}  \non \\
&&+(2s-1)V^{\a(2s-2)}\nabla^\b\O_{\b\a(2s-2)} \Big ) \bigg \}~.
\eea
The superfield $\O_{\a(2s-1)}$ is auxiliary, so upon elimination via its equation of motion
\be
\O_{\a(2s-1)} = - \ri \nabla^\b H_{\b\a(2s-1)} - \ri \nabla_{(\a_1}V_{\a_2 \ldots \a_{2s-1})}~,
\ee
we find that the resulting action coincides with the action for the massless superspin-$s$ multiplet \eqref{action-t3}.


\subsection{Second supersymmetry transformations} \label{Subsect5.4}
As discussed in section \ref{Subsect44}, by construction, the $\cN=1$ actions \eqref{2.22} and \eqref{2.23} are invariant under the second supersymmetry transformations \eqref{PTL2B}. For convenience, we recall the form of these transformations
\be \label{SST}
\d_\e  U| = \e^\a (\bm \nabla^{\2}_\a U)|~,
\ee
where $U(x,\q^I)$ is a superfield in AdS$^{(3|1,1)}$, with indices suppressed. The second supersymmetry transformation acts on the remaining $\cN=1$ superfields \eqref{2.17} and \eqref{2.19} in the following fashion
\begin{subequations}
	\bea
	\d_\e \mathfrak{H}_{\a(2s)} | &=& -\ri \e^\b H_{\b\a(2s)} ~, \label{SST1} \\
	\d_\e \bm \nabla^{\2\b} \mathfrak{H}_{\b\a(2s-1)}| &=& 2 \ri \e^\b H_{\b\a(2s-1)}~, \label{SST2} \\
	\d_{\e} H_{\a(2s+1)} &=& -2\e_{(\a_1}H_{\a_2\ldots\a_{2s+1})} ~, \\
	\d_{\e} H_{\a(2s)} &=& - \frac{\ri}{2} \Big ( (2s+1)|\m|\e^\b H_{\b\a(2s)} - \e^\b \nabla_\b{}^\g H_{\g\a(2s)} \Big ) ~,  \\
	\d_{\e}L_{\a(2s-2)}&=& - \frac{\ri}{2}\e^\b\O_{\b\a(2s-2)} + \frac{2s}{2s-1}(s-1) \e_{(\a_1} \nabla^\b V_{\a_2 \ldots \a_{2s-2})\b}~,  ~ \\
	\d_\e V_{\a(2s-2)}&=& \frac{\ri}{2s}\e^\b \F_{\b\a(2s-2)} - \frac{2}{s(2s-1)}(s-1)\e_{(\a_1} \nabla^\b L_{\a_2 \ldots \a_{2s-2})\b}~, \\
	\d_\e \F_{\a(2s-1)} &=& \frac{1}{2s-1} \Big ( 2s(2s-1) \e^\b \nabla_{\b(\a_1}V_{\a_2 \ldots \a_{2s-1})} + \ri s \e_{(\a_1}\nabla^2 V_{\a_2 \ldots \a_{2s-1})} ~ \\
	&-& 4s(s-1)\e_{(\a_1}\nabla_{\a_2}{}^\b V_{\a_3 \ldots \a_{2s-1})\b} + (2s-1)\e_{(\a_1}\nabla^\b \O_{\a_2 \ldots \a_{2s-1})\b} ~ \non \\
	&+&4s(4s^2-3s+1)|\m|\e_{(\a_1}V_{\a_2 \ldots \a_{2s-1})} \Big ) ~, ~ \non\\
	\d_\e \O_{\a(2s-1)} &=& - \frac{1}{2s-1} \Big (2(2s-1) \e^\b \nabla_{\b(\a_1}L_{\a_2 \ldots \a_{2s-1})} + \ri  \e_{(\a_1}\nabla^2 L_{\a_2 \ldots \a_{2s-1})} ~ \\
	&-& 4(s-1)\e_{(\a_1}\nabla_{\a_2}{}^\b L_{\a_3 \ldots \a_{2s-1})\b} + (2s-1)\e_{(\a_1}\nabla^\b \F_{\a_2 \ldots \a_{2s-1})\b} \non ~ \\
	&+&4(4s^2-3s+1)|\m|\e_{(\a_1}L_{\a_2 \ldots \a_{2s-1})}  \Big ) ~.~ \non
	\eea
\end{subequations}

It is apparent from \eqref{SST1} and \eqref{SST2}  that the second supersymmetry transformation \eqref{SST} breaks the WZ gauge conditions \eqref{2.8}, which we recall for convenience
\be \label{GC}
\mathfrak{H}_{\a(2s)}|=0~, \qquad \bm \nabla^{\2\beta}\mathfrak{H}_{\b\a(2s-1)}|=0~.
\ee
In order to resolve this, it is necessary to supplement the variation \eqref{SST} with the $\e$-dependent gauge transformations:
\bsubeq \label{edepgt}
\bea 
\d_{g(\e)} \mathfrak{H}_{\a(2s)} &=& g_{\a(2s)}(\e) + \bar{g}_{\a(2s)}(\e)~, \\
\d_{g(\e)} \G_{\a(2s-2)} &=& \frac{s}{2s+1}  {\bm{\cDB}}^{\b} {\bm{\cD}}^{\g} \bar g_{\b \g \a(2s-2)} (\e)~.
\eea
\esubeq
The modified second supersymmetry transformations now take the following form
\bsubeq \label{MSST}
\bea 
\hat{\d}_\e \mathfrak{H}_{\a(2s)} &=&  \d_\e  \mathfrak{H}_{\a(2s)}  + \d_{g(\e)} \mathfrak{H}_{\a(2s)} ~, \\
\hat{\d}_\e \G_{\a(2s-2)} &=&  \d_\e  \G_{\a(2s-2)}  + \d_{g(\e)} \G_{\a(2s-2)} ~.
\eea
\esubeq
The sole purpose of introducing the $\e$-dependent gauge transformations \eqref{edepgt} is to restore the original Wess-Zumino gauge \eqref{GC}. Fixing the form of the ${\cN}=1$ components of $g_{\a(2s)} (\e)$ as 
\begin{subequations} \label{FRGP}
	\bea 
	g_{\a(2s)}(\e)| &=& \frac{\ri}{2}\e^\b H_{\b\a(2s)}~,\\
	\bm \nabla^{\2\b}g_{\b\a(2s-1)}(\e)| &=& - \ri \e^\b H_{\b\a(2s-1)}~,
	\eea
\end{subequations}
ensures that \eqref{MSST} takes us back to the original gauge \eqref{GC}.
The modified transformations \eqref{MSST} act on the ${\cN}=1$ superfields \eqref{2.17} and \eqref{2.19}  in the following manner
\vspace{-\baselineskip}
\begin{subequations} \label{MSSTU}
	\bea
	\hat{\d}_{\e} H_{\a(2s+1)} &=&  -2\e_{(\a_1}H_{\a_2\ldots\a_{2s+1})} ~, \\
	\hat{\d}_{\e} H_{\a(2s)} &=&- \frac{\ri}{4(2s+1)} \Big ( \ri \e^\b \nabla^2 H_{\b\a(2s)} 
	+4s\e_{(\a_1}\nabla^{\b\g}H_{\a_2 \ldots \a_{2s})\b\g} \\
	&-&2\e^\b \nabla_\b{}^\g H_{\g\a(2s)} +2(2s+1)(4s+1)|\m|\e^\b H_{\b\a(2s)}   \Big )~, \non \\
	\hat{\d}_{\e}L_{\a(2s-2)}&=& \frac{s}{2s+1}\e^\b \nabla^\g H_{\b\g\a(2s-2)}  + \frac{2s}{2s-1}(s-1) \e_{(\a_1} \nabla^\b V_{\a_2 \ldots \a_{2s-2})\b} \\
	&-&  \frac{\ri}{2}\e^\b\O_{\b\a(2s-2)} ~, \non \\
	\hat{\d}_\e V_{\a(2s-2)}&=& \frac{\ri}{2(2s+1)}\e^\b \nabla^{\g\d}H_{\b\g\d\a(2s-2)}  - \frac{2}{s(2s-1)}(s-1)\e_{(\a_1} \nabla^\b L_{\a_2 \ldots \a_{2s-2})\b} ~ \\
	&+&  \frac{\ri}{2s}\e^\b \F_{\b\a(2s-2)}~, \non \\
	\hat{\d}_\e \F_{\a(2s-1)} &=& - \frac{s}{2s+1} \e^\b \Big ( 2\nabla_{(\a_1}{}^\g H_{\a_2 \ldots \a_{2s-1})\b\g} - \ri \nabla^2 H_{\b\a(2s-1)}  + 4(s+1)|\m|  H_{\b\a(2s-1)} \Big ) \non \\ 
	&+&\frac{1}{2s-1} \Big ( 2s(2s-1) \e^\b \nabla_{\b(\a_1}V_{\a_2 \ldots \a_{2s-1})} + \ri s \e_{(\a_1}\nabla^2 V_{\a_2 \ldots \a_{2s-1})} ~ \\
	&+&4s(4s^2-3s+1)|\m|\e_{(\a_1}V_{\a_2 \ldots \a_{2s-1})} + (2s-1)\e_{(\a_1}\nabla^\b \O_{\a_2 \ldots \a_{2s-1})\b} ~ \non \\
	&-&4s(s-1)\e_{(\a_1}\nabla_{\a_2}{}^\b V_{\a_3 \ldots \a_{2s-1})\b} \Big ) ~ ,\non\\
	\hat{\d}_\e \O_{\a(2s-1)} &=& - \frac{1}{2s-1} \Big (2(2s-1) \e^\b \nabla_{\b(\a_1}L_{\a_2 \ldots \a_{2s-1})} + \ri  \e_{(\a_1}\nabla^2 L_{\a_2 \ldots \a_{2s-1})} ~ \\
	&+&4(4s^2-3s+1)|\m|\e_{(\a_1}L_{\a_2 \ldots \a_{2s-1})} + (2s-1)\e_{(\a_1}\nabla^\b \F_{\a_2 \ldots \a_{2s-1})\b} \non ~ \\
	&-&4(s-1)\e_{(\a_1}\nabla_{\a_2}{}^\b L_{\a_3 \ldots \a_{2s-1})\b} \Big )  + \frac{s}{2s+1}\e^\b\Big ( 2(2s-1)|\m| \nabla^\g H_{\b\g\a(2s-1)} \non ~ \\
	&-&  \nabla_{(\a_1}{}^\g \nabla^\d H_{\a_2 \ldots \a_{2s-1})\b\g\d} + \nabla^{\g \d} \nabla_{\g} H_{\b\d\a(2s-1)} \Big )~. \non
	\eea
\end{subequations}	
\section{Massless half-integer superspin: Longitudinal formulation} \label{Section6}
In this section, we describe the superspace reduction of the longitudinal formulation for the massless multiplet of half-integer superspin, following the prescription advocated in section \ref{Section5}.

\subsection{Longitudinal formulation}
For $s \geq 2$, the longitudinal formulation 
for the massless superspin-$(s+\hf)$ multiplet  
is described in terms of the variables 
\bea \label{3.1}
\cV^{\|}_{(s+\hf)} = \big\{ \mathfrak{H}_{\a(2s)}, G_{\a(2s-2)}, \bar{G}_{\a(2s-2)} \big\} ~,
\eea
Here, the real superfield $\mathfrak{H}_{\a(2s)}$ is identical to that of \eqref{2.1},
and the complex superfield ${G}_{\a(2s-2)}$ is longitudinal linear \eqref{TAS2LongLin}. The superfields \eqref{3.1} are defined modulo gauge transformations of the form
\begin{subequations} \label{3.4}
	\bea \label{3.4a}
	\d_\l \mathfrak{H}_{\a(2s)}&=& 
	{ {\bm{\cDB}}}_{(\a_1} \l_{\a_2 \dots \a_{2s})}-{{\bm{\cD}}}_{(\a_1}\bar {\l}_{\a_2 \dots \a_{2s})} \equiv g_{\a(2s)} + \bar {g}_{\a(2s)} ~,\\ \label{3.4b}
	\d_\l {G}_{\a(2s-2)}&=& 
	-\frac{1}{4}\big( {{\bm{\cDB}}}{}^{2} -4s\m\big){\bm{\cD}}^{\b}\l_{\b \a(2s-2)}
	+\ri (s-1){{\bm{\cDB}}}_{(\a_{1}} {\bm{\cD}}^{|\b\g|} \l_{\a_2 \dots \a_{2s-2})\b\g} \non ~\\
	&\equiv& \Big(\frac{s}{2s+1}{\bm{\cD}}^\b  {\bm{\cDB}}{}^\g + \ri s {\bm{\cD}}^{\b\g}\Big )g_{\b\g\a(2s-2)}~,
	\eea
\end{subequations}
where $\l_{\a(2s-1)}$ is complex unconstrained and $g_{\a(2s)}$ is  longitudinal linear, as in \eqref{H-gauge}.

Modulo an overall normalisation factor, the longitudinal formulation is uniquely described by the gauge-invariant action \cite{HutomoKuzenkoOgburn2018}
\bea \label{3.3}
&&\mathbb{S}^{\|}_{(s+\hf)}[\mathfrak{H}_{\a(2s)}, G_{\a(2s-2)}, \bar{G}_{\a(2s-2)}]= \Big(-\hf \Big)^{s}\int 
\rd^{3|4}z\, \bm E~ \bigg\{\frac{1}{8}\mathfrak{H}^{\a(2s)}
{\bm{\cD}}^{\b}({{\bm{\cDB}}}{}^{2}-6\mu){\bm{\cD}}_{\b}
\mathfrak{H}_{\a(2s)} \non \\
&&+ 2s(s-1)\mu\bar{\mu} \mathfrak{H}^{\a(2s)}\mathfrak{H}_{\a(2s)}
-\frac{1}{16} \big ([{\bm{\cD}}_{\b},{{\bm{\cDB}}}_{\g}]\mathfrak{H}^{\b \g \a(2s-2)} \big )
[{\bm{\cD}}^{\d},{{\bm{\cDB}}}{}^{\l}]\mathfrak{H}_{\d \l \a(2s-2)}
\non \\
&&+ \frac{s}{2} {\bm{\cD}}_{\b \g}\mathfrak{H}^{\b \g \a(2s-2)}
{\bm{\cD}}^{\d \l}\mathfrak{H}_{\d \l \a(2s-2)}
\non \\
&&+ \frac{2s-1}{2s} \Big ( \ri
{\bm{\cD}}_{\b \g}\mathfrak{H}^{\b \g \a(2s-2)} 
\left( G_{\a(2s-2)}-\bar{G}_{\a(2s-2)} \right)
+\frac{1}{s}
\bar{G}^{\a(2s-2)} G_{\a(2s-2)} \Big )
\non \\
&&-\frac{2s+1}{4s^{2}}\big ( G^{\a(2s-2)}G_{\a(2s-2)}+\bar{G}^{\a(2s-2)} \bar {G}_{\a(2s-2)} \big )
\bigg\}~.
\label{long-action-half}
\eea
In the case where $s=1$, the compensator $G$ is covariantly chiral \eqref{TAS2Chiral}. If we introduce the field redefinition $G=3\s$ in action \eqref{3.3}, along with choosing $s=1$, then the corresponding model coincides with the linearised action for minimal $(1,1)$ AdS supergravity \cite{KT-M11}. In the case $s=1$,  the gauge transformations \eqref{3.4} become
\bea
\d_\l \mathfrak{H}_{\a\b} &=& {{\bm{\cDB}}}_{(\a}\l_{\b)} - {\bm{\cD}}_{(\a} \bar{\l}_{\b)}~, \\
\d_\l G &=& - \frac{1}{4}({{\bm{\cDB}}}{}^2-4\bar{\m}){\bm{\cD}}^\b \l_\b~. \label{Gvariation}
\eea
It is an easy exercise to show that $\d_\l G$ is covariantly chiral \eqref{TAS2Chiral}. 


\subsection{Reduction of  gauge prepotentials to AdS$^{3|2}$} 

The reduction of the superconformal gauge prepotential $\mathfrak{H}_{\a(2s)}$ was addressed in section \ref{Section5}. We need only reduce the compensator $G_{\a(2s-2)}$ to AdS$^{3|2}$. We start by converting the longitudinal linear constraint of $G_{\a(2s-2)}$  to the real basis \eqref{realrep}
\be \label{3.5}
\bm \nabla^{\2}{}_{(\a_1}G_{\a_2\ldots\a_{2s-1})}=\ri \bm \nabla^{\1}{}_{(\a_1}G_{\a_2\ldots\a_{2s-1})}~.
\ee
Taking a Taylor expansion of $G_{\a(2s-2)}(\q^I)$ about $\q^{\2}$, and using \eqref{3.5}, we find the independent $\q^{\2}$  -components of $G_{\a(2s-2)}$ to be
\be \label{3.6}
G_{\a(2s-2)}|~, \qquad   \bm \nabla^{\2\b}G_{\b\a(2s-3)}|~.
\ee
Utilising the gauge transformations \eqref{3.4b} and the real representation \eqref{realrep}, we find
\begin{subequations}
	\bea
	\d_\l G_{\a(2s-2)} &=& \frac{\ri s}{2s+1}\big (2s\bm \nabla^{\b\g} - \bm \nabla^{\1 \b}\bm\nabla^{\2\g}\big )g_{\b\g\a(2s-2)}~,\\
	\bm \nabla^{\2\b}\d_\l G_{\b\a(2s-3)} &=& \frac{s}{2s+1}\big (2\ri s \bm \nabla^{\b\g}\bm \nabla^{\2\d}- \bm \nabla^{\b\g} \bm \nabla^{\1 \d} \big ) g_{\b\g\d\a(2s-3)}~.
	\eea
\end{subequations}
We can use the residual gauge freedom \eqref{2.10} to compute the gauge transformations of the complex $\cN=1$ superfields \eqref{3.6}
\begin{subequations} \label{3.8}
	\bea
	\d_\z G_{\a(2s-2)}| &=& \frac{s^2}{2s+1} \nabla^{\b\g}\z_{\b\g\a(2s-2)} - \frac{\ri}{2}\nabla^\b\z_{\b\a(2s-2)}~ \label{3.8a} ,\\
	\bm \nabla^{\2\b}\d_\z G_{\b\a(2s-3)}| &=& \ri s \nabla^{\b\g}\z_{\b\g\a(2s-3)}  + \frac{\ri s}{2(2s+1)}\nabla^{\b\g} \nabla^\d \z_{\b\g \d \a(2s-3)}~. \label{3.8b}
	\eea
\end{subequations}
It is useful to separate the complex superfields \eqref{3.6} into its real and imaginary parts\begin{subequations}
	\bea
	G_{\a(2s-2)}| &=&-2sL_{\a(2s-2)} - \ri s V_{\a(2s-2)} ~,\\
	\bm \nabla^{\2\b} G_{\b\a(2s-3)}|  &=&\frac{1}{2}\big ( \F_{\a(2s-3)} + \ri \O_{\a(2s-3)} \big )~.
	\eea
\end{subequations}

It then becomes apparent from the gauge transformations \eqref{2.11} and \eqref{3.8} that we are in fact dealing with two different gauge theories. The first gauge theory is formulated in terms of the real dynamical superfields
\be \label{3.10}
\cV^\parallel_{(s+\frac{1}{2})} = \big \{ H_{\a(2s+1)}, L_{\a(2s-2)}, \F_{\a(2s-3)} \big \}~,
\ee
which are defined modulo gauge transformations of the form
\begin{subequations} \label{3.11}
	\bea
	\d_\z H_{\a(2s+1)} &=& \ri \nabla_{(\a_1}\z_{\a_2 \ldots\a_{2s+1})} ~, \label{3.11a} \\
	\d_\z L_{\a(2s-2)} &=& -\frac{s}{2(2s+1)} \nabla^{\b\g}\z_{\b\g\a(2s-2)}~,  \label{3.11b}\\
	\d_\z \F_{\a(2s-3)} &=& \frac{\ri s}{2s+1}\nabla^{\b\g}\nabla^\d\z_{\b\g\d\a(2s-3)}~.
	\eea
\end{subequations}
The other gauge model is described in terms of the real superfields
\be \label{3.12}
\cV^\parallel_{(s)} = \big \{H_{\a(2s)}, V_{\a(2s-2)}, \O_{\a(2s-3)} \big \}~,
\ee
which possess the following gauge freedom
\begin{subequations} \label{3.13}
	\bea
	\d_\z H_{\a(2s)} &=& \nabla_{(\a_1}\z_{\a_2 \ldots\a_{2s})} ~, \label{3.13a}\\
	\d_\z V_{\a(2s-2)} &=& \frac{1}{2s}\nabla^\b\z_{\b\a(2s-2)}~, \label{3.13b}\\
	\d_\z \O_{\a(2s-3)} &=& 2s\nabla^{\b\g}\z_{\b\g\a(2s-3)}~.
	\eea
\end{subequations}
After carrying out the reduction to $\cN=1$ AdS superspace, the action \eqref{3.3} decouples into two $\cN =1$ theories, which are formulated in terms of the superfields \eqref{3.10} and \eqref{3.12}, respectively
\bea \label{RATHI}
\mathbb{S}^{\parallel}_{(s+\hf)} [\mathfrak{H}_{\a(2s)},G_{\a(2s-2)}, \bar{G}_{\a(2s-2)}] =&& S^\parallel_{(s+\frac{1}{2})} [H_{\a(2s+1)},L_{\a(2s-2)},  \F_{\a(2s-3)}] \non ~\\ 
&&+ S^\parallel_{(s)}[H_{\a(2s)},V_{\a(2s-2)}, \O_{\a(2s-3)}]~.
\eea
In the following subsection, we provide the explicit forms of these decoupled $\cN=1$ actions.


\subsection{Massless higher-spin $\cN=1$ supermultiplets}
The first $\cN = 1$ gauge theory, which is  realised in terms of the superfield variables \eqref{3.10}, is described by the action 
\bea \label{3.15}
&&S^\parallel_{(s+\frac{1}{2})}[H_{\a(2s+1)}, L_{\a(2s-2)}, \F_{\a(2s-3)}] = \Big(-\hf \Big)^{s} \Big ( - \frac{\ri}{8} \Big ) \int 
\rd^{3|2}z \, E ~\bigg \{2H^{\a(2s+1)} \mathbb{Q} H_{\a(2s+1)}~ \non \\
&&+(2s+1)\ri|\m|H^{\a(2s+1)}\nabla^2 H_{\a(2s+1)}   -4(2s+1)|\m|H^{\b\a(2s)}\nabla_\b{}^\g H_{\g\a(2s)}-\ri H^{\b\a(2s)}\nabla^2\nabla_\b{}^\g H_{\g\a(2s)}~ \non \\
&&-8(s-2)(2s+1)|\m|^2H^{\a(2s+1)}H_{\a(2s+1)} - 2s\nabla_{\b\g}H^{\b\g\a(2s-1)}\nabla^{\d\l}H_{\d\l\a(2s-1)}  ~ \non \\
&&-\frac{4}{2s-1} \Big ( 16\ri |\m| (2s^3-2s+1)L^{\a(2s-2)}L_{\a(2s-2)} + 4(2s^2-1)L^{\a(2s-2)}\nabla^2L_{\a(2s-2)}~ \non \\
&&+16\ri  (s-1)^2 L^{\b\a(2s-3)}\nabla_\b{}^\g L_{\g\a(2s-3)}  + 2\ri (2s-1)^2\nabla_{\b\g}H^{\b\g\d\a(2s-2)}\nabla_\d L_{\a(2s-2)} ~ \non \\
&&+\frac{4\ri}{s} (s-1)L^{\b\a(2s-3)}\nabla_\b \F_{\a(2s-3)} + \frac{1}{s}(s-1)\F^{\a(2s-3)}\F_{\a(2s-3)} \Big ) \bigg \}~.
\eea
The action \eqref{3.15} is invariant under the gauge transformations \eqref{3.11}. We can eliminate the auxiliary field $\F_{\a(2s-3)} $ from \eqref{3.15} by its equation of motion
\be
\F_{\a(2s-3)} = - 2 \ri \nabla^\b L_{\b\a(2s-3)}~.
\ee
The resulting action, up to an overall factor,  coincides with the massless superspin-$(s+\hf)$ multiplet \eqref{LongHalfIntAct}.

The other $\cN=1$ model, which is constructed in terms of the superfields \eqref{3.12}, is described by the action
\bea \label{3.16}
&&S^\parallel_{(s)}[H_{\a(2s)},V_{\a(2s-2)}, \O_{\a(2s-3)}] = - \Big (-\frac{1}{2} \Big )^s \int \rd^{3|2}z~ E ~\bigg\{ (s-1)|\m|H^{\a(2s)}H_{\a(2s)} ~\non \\
&&-\frac{\ri}{4} H^{\a(2s)}\nabla^2H_{\a(2s)} + \frac{1}{2}H^{\b\a(2s-1)}\nabla_\b{}^\g H_{\g\a(2s-1)}-(2s-1)H^{\a(2s)}\nabla_{(\a_1 \a_2}V_{\a_3 \ldots \a_{2s})} \non ~ \\
&&-\frac{1}{4(2s-1)}\Big (2\ri s V^{\a(2s-2)}\nabla^2 V_{\a(2s-2)} -4 (s-1)(4s-1)V^{\b\a(2s-3)}\nabla_\b{}^\g V_{\g\a(2s-3)} ~ \non \\
&&+8s(4s^2-3s+1)|\m|V^{\a(2s-2)}V_{\a(2s-2)} +8(s-1)V^{\b\a(2s-3)}\nabla_\b \O_{\a(2s-3)}~ \non \\
&&-\frac{\ri}{s^2}(s-1)\O^{\a(2s-3)}\O_{\a(2s-3)} \Big ) \bigg \}~,
\eea
which is invariant under the gauge transformations \eqref{3.13}. It is evident that the superfield $\O_{\a(2s-3)}$ is auxiliary, so upon elimination via its equation of motion
\be
\O_{\a(2s-3)} = - 4 \ri s^2 \nabla^\b V_{\b\a(2s-3)}~,
\ee
the action \eqref{3.16} reduces to the massless superspin-$s$ multiplet \eqref{action-t3}.

\subsection{Second supersymmetry transformations}
Let us note that the reduced actions \eqref{RATHI}
\bea
\mathbb{S}^{\parallel}_{(s+\hf)} [\mathfrak{H}_{\a(2s)},G_{\a(2s-2)}, \bar{G}_{\a(2s-2)}] =&& S^\parallel_{(s+\frac{1}{2})} [H_{\a(2s+1)},L_{\a(2s-2)},  \F_{\a(2s-3)}] \non ~\\ 
&&+ S^\parallel_{(s)}[H_{\a(2s)},V_{\a(2s-2)}, \O_{\a(2s-3)}]~,
\eea
are invariant under the second supersymmetry transformations \eqref{PTL2B}. More explicitly, the second supersymmetry transformation acts on the dynamical superfields \eqref{3.10} and \eqref{3.12} by the following rule
\begin{subequations}
	\bea	
	\d_\e \mathfrak{H}_{\a(2s)} | &=& -\ri \e^\b H_{\b\a(2s)} ~, \label{SSTHIL1}\\
	\d_\e \bm \nabla^{\2\b} \mathfrak{H}_{\b\a(2s-1)}| &=& 2 \ri \e^\b H_{\b\a(2s-1)}~,  \label{SSTHIL2}\\
	\d_{\e} H_{\a(2s+1)} &=& -2\e_{(\a_1}H_{\a_2\ldots\a_{2s+1})} ~, \\
	\d_{\e} H_{\a(2s)} &=&  \frac{\ri}{2} \big (  \e^\b \nabla_\b{}^\g H_{\g\a(2s)}  - (2s+1)|\m|\e^\b H_{\b\a(2s)} \big ) ~,  \\
	\d_\e L_{\a(2s-2)} &=& -\hf \e^\b\nabla_{(\b}V_{\a_1 \ldots \a_{2s-2})} + \frac{\ri (s-1)}{2s(2s-1)}\e_{(\a_1}\O_{\a_2\ldots\a_{2s-2})}~, \\
	\d_\e V_{\a(2s-2)} &=& 2 \e^\b \nabla_{(\b}L_{\a_1 \ldots \a_{2s-2})} - \frac{\ri (s-1)}{s(2s-1)}\e_{(\a_1}\F_{\a_2\ldots\a_{2s-2})} ~, \\
	\d_\e \F_{\a(2s-3)} &=& \frac{1}{2s-1}\e^\b \Big ( 2s(2s-3) \nabla_{(\a_1}{}^\g V_{\a_2 \ldots \a_{2s-3})\b\g} - \ri s \nabla^2 V_{\b\a(2s-3)} \non ~ \\
	&+&4s^2 \nabla_\b{}^\g V_{\g\a(2s-3)} - 4s(4s^2-5s+2)|\m| V_{\b\a(2s-3)}    ~ \non \\
	&-& 2(s-1)\nabla_{(\b}\O_{\a_1 \ldots \a_{2s-3})} \Big )~,  \\
	\d_\e \O_{\a(2s-3)} &=& \frac{1}{2s-1} \e^\b \Big ( 2\ri s \nabla^2 L_{\b\a(2s-3)} - 4s(2s-3) \nabla_{(\a_1}{}^\g L_{\a_2\ldots\a_{2s-3})\b\g}  \non ~ \\
	&-&8s^2  \nabla_{\b}{}^\g L_{\g\a(2s-3)} + 8s(4s^2-5s+2)|\m| L_{\b\a(2s-3)}~  \non \\
	&+&2(s-1)\nabla_{(\b}\F_{\a_1 \ldots \a_{2s-3})} \Big )~.
	\eea
\end{subequations}
It is evident from the variations \eqref{SSTHIL1} and \eqref{SSTHIL2} that the second supersymmetry transformations are not compatible with the WZ gauge conditions \eqref{GC}.

Using a similar approach as in subsection \ref{Subsect5.4}, the WZ gauge conditions \eqref{GC} in the longitudinal formulation can be restored, provided we accompany the second supersymmetry transformations with the following $\e$-dependent gauge transformations:
\bsubeq \label{edepgt-HIL}
\bea 
\d_{g(\e)} \mathfrak{H}_{\a(2s)} &=& g_{\a(2s)}(\e) + \bar{g}_{\a(2s)}(\e)~, \\
\d_{g(\e)} G_{\a(2s-2)} &=& \Big ( \frac{s}{2s+1} {\bm{\cD}}^{\b}  {\bm{\cDB}}^{\g} + \ri s {\bm{\cD}}^{\b \g} \Big) g_{\b \g \a(2s-2)} (\e)~, 
\eea
\esubeq
where
\begin{subequations} 
	\bea 
	g_{\a(2s)}(\e)| &=& \frac{\ri}{2}\e^\b H_{\b\a(2s)}~,\\
	\bm \nabla^{\2\b}g_{\b\a(2s-1)}(\e)| &=& - \ri \e^\b H_{\b\a(2s-1)}~.
	\eea
\end{subequations}
The modified second supersymmetry transformations now read
\bsubeq \label{MSST-HIL}
\bea 
\hat{\d}_\e \mathfrak{H}_{\a(2s)} &=&  \d_\e  \mathfrak{H}_{\a(2s)}  + \d_{g(\e)} \mathfrak{H}_{\a(2s)} ~, \\
\hat{\d}_\e G_{\a(2s-2)} &=&  \d_\e  G_{\a(2s-2)}  + \d_{g(\e)} G_{\a(2s-2)} ~.
\eea
\esubeq
The transformations \eqref{MSST-HIL} act on the $\cN=1$ superfields \eqref{3.10} and \eqref{3.12} as follows
\begin{subequations}
	\bea	
	\hat{\d}_{\e} H_{\a(2s+1)} &=&  -2\e_{(\a_1}H_{\a_2\ldots\a_{2s+1})} ~, \\
	\hat{\d}_{\e} H_{\a(2s)} &=&- \frac{\ri}{4(2s+1)} \Big ( \ri \e^\b \nabla^2 H_{\b\a(2s)} 
	+4s\e_{(\a_1}\nabla^{\b\g}H_{\a_2 \ldots \a_{2s})\b\g} \\
	&-& 2\e^\b \nabla_\b{}^\g H_{\g\a(2s)} + 2(2s+1)(4s+1)|\m|\e^\b H_{\b\a(2s)} \Big )~, \non \\
	\hat{\d}_\e L_{\a(2s-2)} &=& -\hf \e^\b\nabla_{(\b}V_{\a_1 \ldots \a_{2s-2})} + \frac{\ri (s-1)}{2s(2s-1)}\e_{(\a_1}\O_{\a_2\ldots\a_{2s-2})}\\ 
	&-& \frac{1}{2(2s+1)}\e^\b\nabla^\g H_{\b\g\a(2s-2)}~, \non \\
	\hat{\d}_\e V_{\a(2s-2)} &=& 2 \e^\b \nabla_{(\b}L_{\a_1 \ldots \a_{2s-2})} - \frac{\ri (s-1)}{s(2s-1)}\e_{(\a_1}\F_{\a_2\ldots\a_{2s-2})} ~ \\
	&-& \frac{\ri s}{2s+1} \e^\b \nabla^{\g\d} H_{\b\g\d \a(2s-2)}~, \non \\
	\hat{\d}_\e \F_{\a(2s-3)} &=& \frac{1}{2s-1}\e^\b \Big ( 2s(2s-3) \nabla_{(\a_1}{}^\g V_{\a_2 \ldots \a_{2s-3})\b\g} - \ri s \nabla^2 V_{\b\a(2s-3)}  ~ \\
	&+&4s^2 \nabla_\b{}^\g V_{\g\a(2s-3)} - 4s(4s^2-5s+2)|\m| V_{\b\a(2s-3)}  ~ \non \\
	&-&2(s-1)\nabla_{(\b}\O_{\a_1 \ldots \a_{2s-3})} \Big ) + \frac{4s^2}{2s+1} \e^\b \nabla^{\g\d} H_{\b\g\d\a(2s-3)}~, \non \\
	\hat{\d}_\e \O_{\a(2s-3)} &=& \frac{1}{2s-1} \e^\b \Big ( 2\ri s \nabla^2 L_{\b\a(2s-3)} - 4s(2s-3) \nabla_{(\a_1}{}^\g L_{\a_2\ldots\a_{2s-3})\b\g}   ~ \\
	&-&8s^2  \nabla_{\b}{}^\g L_{\g\a(2s-3)} + 8s(4s^2-5s+2)|\m| L_{\b\a(2s-3)}~  \non \\
	&+&2(s-1)\nabla_{(\b}\F_{\a_1 \ldots \a_{2s-3})} \Big ) + \frac{s}{2s+1}\e^\b \nabla^{\g\d} \nabla^\l H_{\b\g\d\l\a(2s-3)}~. \non
	\eea
\end{subequations}

\section{Massless integer superspin: Longitudinal formulation} \label{Section7}
In \cite{HutomoKuzenkoOgburn2018}, two dually equivalent off-shell formulations, called transverse and longitudinal, for the massless multiplets of integer superspin were developed in $(1,1)$ AdS superspace. In the following section, we reduce the longitudinal model to AdS$^{3|2}$.

\subsection{Longitudinal formulation}
Given an integer $s\geq1$, the longitudinal formulation 
for the massless superspin-$s$ multiplet 
is realised in terms of the following dynamical variables
\bea
\cV^{\|}_{(s)} = \big \{U_{\a(2s-2)}, G_{\a(2s)}, \bar{G}_{\a(2s)} \big \} ~.
\label{4.1}
\eea
Here, the real superfield $U_{\a(2s-2)}$ is unconstrained, while the complex superfield $G_{\a(2s)}$ is longitudinal linear \eqref{TAS2LongLin}. The  superfields $U_{\a(2s-2)}$ and $G_{\a(2s)}$ are defined modulo gauge transformations of the form 
\begin{subequations}\label{4.3}
	\bea \label{4.3a}
	\d_L U_{\a(2s-2)}
	&=&{\bm{\cD}}^{\b}L_{\b \a(2s-2)}-{{\bm{\cDB}}}{}^{\b}\bar{L}_{\b \a(2s-2)} 
	\equiv \bar {\g}_{\a(2s-2)}+{\g}_{\a(2s-2)}
	~,  \\
	\d_L G_{\a(2s)} 
	&=&-\frac{1}{2}{{\bm{\cDB}}}_{(\a_{1}}\big({\bm{\cD}}^{2}
	-2(2s+1) \bar \m \big)
	L_{\a_2 \dots \a_{2s})}
	\equiv {{\bm{\cDB}}}_{(\a_{1}} {\bm{\cD}}_{\a_{2}}\bar{\g}_{\a_3 \dots \a_{2s})}
	~.
	\label{4.3b}
	\eea
\end{subequations} 
Here the gauge parameter $L_{\a(2s-1)}$ is complex and unconstrained, while 
${\g}^{\a(2s-2)}:= { {\bm{\cDB}}}_{\b}\bar{L}^{\b \a(2s-2)}$ is transverse linear \eqref{TAS2TransL}.

Modulo an overall  normalisation factor, the longitudinal formulation for the massless superspin-$s$ multiplet is described by the action
\bea \label{4.4}
&&\mathbb{S}_{(s)}^{\|}[U_{\a(2s-2)},G_{\a(2s)}, \bar{G}_{\a(2s)}]
= \Big(-\hf\Big)^{s}\int 
\rd^{3|4}z\, \bm E \,
\bigg\{\frac{1}{8}U^{\a(2s-2)}{\bm{\cD}}^{\g}({ {\bm{\cDB}}}{}^{2}-6\mu){\bm{\cD}}_{\g}U_{\a(2s-2)}
\non \\
&&+\frac{s}{2s+1}U^{\a(2s-2)}\Big({\bm{\cD}}^{\b} { {\bm{\cDB}}}{}^{\g}
G_{\b \g \a(2s-2) }
-{ {\bm{\cDB}}}{}^{\b}{{\bm{\cD}}}^{\g}\bar{G}_{\b \g \a(2s-2)} \Big) +\frac{s}{2s-1}  \bar{G}^{\a(2s)} G_{\a(2s)} \non \\
&&+ \frac{s}{2(2s+1)}\Big(G^{\a(2s)}G_{\a(2s)}+\bar{G}^{\a(2s)}\bar{G}_{\a(2s)}\Big) +2s(s+1)\mu\bar{\mu} U^{\a(2s-2)}U_{\a(2s-2)}
\bigg\}~,
\label{long-action-int}
\eea
which is invariant under the gauge transformations \eqref{4.3}.

\subsection{Reduction of gauge prepotentials to AdS$^{3|2}$}
We wish to reduce the gauge superfields \eqref{4.1} to AdS$^{3|2}$. We start by reducing the superfield $U_{\a(2s-2)}$. Converting the transverse linear constraint \eqref{TAS2TransL} of $\g_{\a(2s-2)}$  to the real basis \eqref{realrep} gives
\be \label{TLReal}
\bm \nabla^{\2\b}\g_{\b\a(2s-3)}=\ri \bm \nabla^{\1\b} \g_{\b\a(2s-3)}~.
\ee
Taking a Taylor expansion of $\g_{\a(2s-2)}(\q^I)$ about $\q^{\2}$, then using \eqref{TLReal}, we find the independent $\q^{\2}$~-components of $\g_{\a(2s-2)}$ to be
\be
\g_{\a(2s-2)}|, \qquad \qquad \bm \nabla^{\2}{}_{(\a_1}\g_{\a_2\ldots\a_{2s-1})}|~.
\ee

The gauge transformation \eqref{4.3a} allows us to impose the gauge condition
\be \label{4.8}
U_{\a(2s-2)}| = 0~.
\ee
It must be noted that the gauge condition \eqref{4.8} is less restrictive than those proposed in appendix B of \cite{HutomoKuzenkoOgburn2018}, where the analogous reduction procedure was carried out in $\mb{M}^{3|2}$. This was done in order to ensure that one of the decoupled $\cN=1$ actions coincides with \eqref{TIAction}. It can be shown that if one chooses to impose the same gauge conditions as detailed in \cite{HutomoKuzenkoOgburn2018}, i.e
\be
U_{\a(2s-2)}| = 0~,  \qquad \bm \nabla^{\2}{}_{(\a_1}U_{\a_2 \ldots \a_{2s-1})}| = 0~,
\ee
then one will obtain the $\cN=1$ action \eqref{GaugedTransIntAction} after reduction, rather than \eqref{TIAction}.

The residual gauge symmetry which preserves the gauge condition \eqref{4.8} is described by the real unconstrained $\cN =1$ superfield
\be
\g_{\a(2s-2)}| = - \frac{\ri}{2}\z_{\a(2s-2)}~, \qquad \bar \z_{\a(2s-2)} = \z_{\a(2s-2)}~. \label{4.10}
\ee 

We now wish to reduce the superfield $\bar{G}_{\a(2s)}$ to AdS$^{3|2}$. Converting the longitudinal linear constraint of $\bar{G}_{\a(2s)}$ \eqref{TAS2LongLin} to the real basis yields
\be \label{4.12}
\bm \nabla^{\2}{}_{(\a_1}\bar{G}_{\a_2\ldots\a_{2s+1})} = - \ri \bm \nabla^{\1}{}_{(\a_1}\bar{G}_{\a_1\ldots\a_{2s+1})}~.
\ee
Performing the superspace reduction to AdS$^{3|2}$, it follows from constraint \eqref{4.12} that $\bar{G}_{\a(2s)}$ has two independent $\q^{\2}$ -components:
\be \label{4.13}
\bar{G}_{\a(2s)}|~, \qquad \bm \nabla^{\2\b}\bar{G}_{\b\a(2s-1)}|~.
\ee
Making use of the gauge transformations \eqref{4.3b} and the real representation \eqref{realrep}, we find
\begin{subequations} \label{LIGvar}
	\bea
	\d_L \bar{G}_{\a(2s)} &=& \ri \big (\bm \nabla^{\1}{}_{(\a_1}\bm \nabla^{\2}{}_{\a_2} + \bm \nabla_{(\a_1\a_2}\big )\g_{\a_3\ldots{\a_{2s})}}~,\\
	\bm \nabla^{\2\b}\d_L \bar{G}_{\b\a(2s-1)} &=& \frac{\ri}{2s} \big ( 2s\bm \nabla_{(\a_1}{}^\b \bm \nabla^{\2}{}_{\a_2}\g_{\a_3\ldots\a_{2s-1})\b} + (4s^2-1)|\m| \bm \nabla^{\2}{}_{(\a_1}\g_{\a_2 \ldots \a_{2s-1})}~  \\ 
	&+& 2\ri(s+1) \bm \nabla_{\b(\a_1} \bm \nabla^{\1\b} \g_{\a_2 \ldots \a_{2s-1})} + 6\ri s(2s+1)|\m| \bm \nabla^{\1}{}_{(\a_1} \g_{\a_2 \ldots \a_{2s-1})} \non ~ \\
	&+&\frac{\ri}{2}(2s+1)(\bm \nabla^{\1})^2\bm \nabla^{\2}{}_{(\a_1}\g_{\a_2 \ldots \a_{2s-1})} - 2\ri (s-1)\bm \nabla_{(\a_1}{}^\b\bm \nabla^{\1}{}_{\a_2}\g_{\a_3 \ldots \a_{2s-1})\b} \big ) ~. \non
	\eea
\end{subequations}
Using the residual gauge freedom \eqref{4.10}, we can directly compute the gauge transformations of the $\cN =1$ superfields \eqref{4.13} from \eqref{LIGvar}
\begin{subequations} \label{4.15}
	\bea
	\d_L \bar {G}_{\a(2s)}| &=& \frac{1}{2} \Big (\nabla_{(\a_1\a_2}\z_{\a_3 \ldots \a_{2s})} + 2 \ri \nabla_{(\a_1}\big (\t_{\a_2\ldots\a_{2s})} + \ri \tilde{\t}_{\a_2 \dots \a_{2s})} \big ) \Big )~, \\
	\bm \nabla^{\2\b} \d_L \bar {G}_{\b\a(2s-1)}| &=& \frac{\ri}{2s} \Big ( 2s(2s+1)|\m|\nabla_{(\a_1}\z_{\a_2\ldots\a_{2s-1})}  + (s+1)\nabla_{\b(\a_1}\nabla^\b \z_{\a_2 \ldots \a_{2s-1})} ~  \\
	&+& (4s^2-1)|\m| \big ( \t_{\a(2s-1)} + \ri  \tilde{\t}_{\a(2s-1)} \big ) + \frac{\ri}{2}(2s+1)\nabla^2\big ( \t_{\a(2s-1)} + \ri  \tilde{\t}_{\a(2s-1)} \big ) ~ \non \\ 
	&+& (2s-1) \nabla_{(\a_1}{}^\b \big ( \t_{\a_2 \ldots \a_{2s-1})\b} + \ri \tilde{\t}_{\a_2 \ldots \a_{2s-1})\b} \big ) - (s-1)\nabla_{(\a_1}{}^\b \nabla_{\a_2} \z_{\a_3 \ldots \a_{2s-1})\b} \Big ) ~ , \non 
	\eea
\end{subequations}
where we have defined
\be \label{RIGcomp}
\bm \nabla^{\2}{}_{(\a_1} \g _{\a_2 \ldots \a_{2s-1})} | = \t_{\a{(2s-1)}} + \ri \tilde { \t}_{\a{(2s-1)}}~.
\ee

It proves useful to decompose the complex $\cN = 1$ superfields \eqref{4.13} into real and imaginary parts as follows
\begin{subequations}
	\bea
	\bar {G}_{\a(2s)}| &=& -\frac{1}{2} \big ( H_{\a(2s)} - 2\ri \tilde{H}_{\a(2s)} \big  )~,\\
	\bm \nabla^{\2\b}\bar{ G}_{\b\a(2s-1)}|  &=& \F_{\a(2s-1)} -2 \ri \O_{\a(2s-1)}  ~.
	\eea
\end{subequations}
The gauge transformations of these four real $\cN=1$ superfields immediately follow from gauge transformations \eqref{4.15}
\begin{subequations}  \label{LongGaugeTrans}
	\bea
	\d H_{\a(2s)} &=& \nabla_{(\a_1} \r_{\a_2 \ldots \a_{2s})}~ , \\
	\d \tilde{H}_{\a(2s)} &=&  \nabla_{(\a_1} \t_{\a_2 \ldots \a_{2s})}~, \\
	\d \F_{\a(2s-1)}&=& - \frac{1}{8s} \Big ( 2(2s-1)\nabla_{(\a_1}{}^\b \r_{\a_2 \ldots \a_{2s-1})\b} + 2(4s^2-1)|\m| \r_{\a(2s-1)} \\
	&&+ (2s+1)\ri \nabla^2 \r_{\a(2s-1)} - 8\ri(4s^2-1)|\m|\nabla_{(\a_1}\z_{\a_2\ldots\a_{2s-1})} \Big ) \non~, \\
	\d \O_{\a(2s-1)} &=& - \frac{1}{8s} \Big ( 2(2s-1) \nabla_{(\a_1}{}^\b \t_{\a_2 \ldots \a_{2s-1})\b} +2 (4s^2-1)|\m|\t_{\a(2s-1)} ~ \\
	&&+ \ri(2s+1)\nabla^2 \t_{\a(2s-1)} \Big )~, \non 
	\eea
\end{subequations}
where we have introduced the gauge parameter redefinition
\be
\r_{\a(2s-1)} := 2 \tilde{\t}_{\a(2s-1)} + \ri \nabla_{(\a_1}\z_{\a_2 \ldots \a_{2s-1})}~.
\ee

We are now left with the remaining unconstrained real $\cN = 1$ superfields
\begin{subequations} \label{TAS24.9}
	\bea
	\cU_{\b;\a(2s-2)} &:=& \ri \bm \nabla^{\2}_\b U_{\a(2s-2)}|~,\\
	V_{\a(2s-2)} &:=& -\frac{\ri}{8s}(\bm \nabla^{\2})^2U_{\a(2s-2)}|~.
	\eea
\end{subequations}

Utilising the gauge transformations \eqref{4.3a}, in conjunction with the residual gauge symmetry \eqref{4.10} and definition \eqref{RIGcomp}, we find that the superfields \eqref{TAS24.9} possess the following gauge freedom
\begin{subequations}\label{4.11}
	\bea \label{ULong}
	\d \cU_{\b;\a(2s-2)}  &=& -\r_{\b\a(2s-2)} + \ri \nabla_\b \z_{\a(2s-2)}~,\\
	\d V_{\a(2s-2)} &=& \frac{1}{2s}\nabla^\b\t_{\b\a(2s-2)}~.
	\eea
\end{subequations}

It then becomes apparent from gauge transformations \eqref{LongGaugeTrans} and \eqref{4.11} that we are dealing with two different gauge theories. The first of these gauge theories is formulated in terms of superfields
\be \label{4.17}
\cV^\perp_{(s)} = \big \{H_{\a(2s)}, \cU_{\b;\a(2s-2)} , \F_{\a(2s-1)} \big \}~,
\ee
which are defined modulo gauge transformations of the form
\begin{subequations} \label{4.18}
	\bea\label{5.18a}
	\d H_{\a(2s)} &=& \nabla_{(\a_1} \r_{\a_2 \ldots \a_{2s})}  ~,\\
	\d \cU_{\b;\a(2s-2)}  &=& -\r_{\b\a(2s-2)} + \ri \nabla_\b \z_{\a(2s-2)}~,\label{5.18b}	\\
	\d \F_{\a(2s-1)}&=& -\frac{1}{8s} \Big ( 2(2s-1)\nabla_{(\a_1}{}^\b \r_{\a_2 \ldots \a_{2s-1})\b} + (2s+1)\ri \nabla^2 \r_{\a(2s-1)}  \\
	&&+ 2(2s-1)(2s+1)|\m| \big (  \r_{\a(2s-1)} - 4 \nabla_{(\a_1}\z_{\a_2\ldots\a_{2s-1})} \big ) \Big ) ~. \non
	\eea
\end{subequations}
The other gauge model is described in terms of the superfields
\be \label{4.19}
\cV^\parallel_{(s)} = \big \{\tilde{H}_{\a(2s)}, V_{\a(2s-2)}, \O_{\a(2s-1)} \big \}~,
\ee
which possess the following gauge freedom
\begin{subequations} \label{4.20}
	\bea 
	\d \tilde{H}_{\a(2s)} &=& \nabla_{(\a_1}\t_{\a_2\ldots\a_{2s})}\label{4.20b} ~,\\
	\d V_{\a(2s-2)} &=& \frac{1}{2s}\nabla^\b\t_{\b\a(2s-2)}~, \label{4.20a}\\
	\d \O_{\a(2s-1)} &=& - \frac{1}{8s} \Big ( 2(2s-1) \nabla_{(\a_1}{}^\b \t_{\a_2 \ldots \a_{2s-1})\b} + \ri(2s+1)\nabla^2 \t_{\a(2s-1)} ~ \\
	&&+ 2(2s-1)(2s+1) |\m|\t_{\a(2s-1)} \Big )~. \non
	\eea
\end{subequations}
Applying the reduction prescription to the action \eqref{4.4}, we obtain two decoupled $\cN =1$ supersymmetric actions, which are formulated in terms of superfields \eqref{4.17} and \eqref{4.19}, respectively
\bea
\mathbb{S}^{\parallel}_{(s)} [U_{\a(2s-2)},G_{\a(2s)}, \bar{G}_{\a(2s)}] &=& S^\perp_{(s)}[H_{\a(2s)}, \cU_{\b;\a(2s-2)} ,  \F_{\a(2s-1)}] \non ~\\ 
&+& S^\parallel_{(s)}[\tilde{H}_{\a(2s)}, V_{\a(2s-2)}, \O_{\a(2s-1)}]~. \label{723}
\eea
The explicit form of the $\cN =1$ decoupled actions are provided in the following subsection.

Let us point out that there also appeared a new off-shell formulation for the massless integer superspin multiplet in $(1,1)$ AdS superspace in \cite{HutomoKuzenkoOgburn2018}. This formulation proves to be a generalised version of the longitudinal action \eqref{long-action-int}, as the gauge-invariant action involves not only $U_{\a(2s-2)}, \J_{\a(2s-1)}$ and $\bar \J_{\a(2s-1)}$, but also new compensating superfields. Furthermore, the prepotential $\J_{\a(2s-1)}$ (associated to the longitudinal linear field strength $G_{\a(2s)}$) enjoys a larger gauge symmetry, which is that of the superconformal complex superspin-$s$ multiplet \cite{HutomoKuzenkoOgburn2018}. Upon reduction to ${\cN}=1$ AdS superspace, we found that this new formulation decoupled into two ${\cN}=1$ supersymmetric higher-spin models which coincide with the right-hand side of eq.~\eqref{723}. 


\subsection{Massless higher-spin $\cN=1$ supermultiplets} \label{TAS2Massless}
The first $\cN=1$ supersymmetric action, which is described in terms of the superfields \eqref{4.17}, takes the form
\bea 
&&S^\perp_{(s)}[H_{\a(2s)}, \cU_{\b;\a(2s-2)} ,  \F_{\a(2s-1)}] = \bigg ( -\frac{1}{2} \bigg )^s \frac{1}{4}  \int \rd^{3|2}z\,E\, \bigg \{-\ri \,\cU^{\b; \, \a(2s-2)} \Box \cU_{\b; \, \a(2s-2)} \non\\
&&-\hf \cU^{\b; \, \a(2s-2)} \na^2 \na_{\b}{}^{\g}\cU_{\g; \, \a(2s-2)} + \hf  (2s-3)|\m| \cU^{\b; \, \a(2s-2)} \na^2 \cU_{\b; \, \a(2s-2)}  \non\\
&&+ 2  (s-1)|\m| \cU_{\b;}{}^{\b \a(2s-3)} \na^2 \cU^{\g;}{}_{\g \a(2s-3)} + \ri  (4s^2-1) |\m|^2\cU^{\b; \, \a(2s-2)}\cU_{\b; \, \a(2s-2)}  \non\\
&&- 8 \ri  (s-1)|\m|^2 \cU_{\b;}{}^{\b \a(2s-3)} \cU^{\g;}{}_{\g \a(2s-3)} \non \\
&&- \frac{2s}{2s+1} \cU^{\b; \, \a(2s-2)} \Big( -2 \na^{\g \d} \na_{\g}H_{\b \d  \a(2s-2)} + \na^{\g \d} \na_{\b} H_{\g \d \a(2s-2)} + \na^2 \F_{\b \a(2s-2)} \non\\
&&+ 2 \ri \, \na_{\b}{}^{\g} \F_{\g \a(2s-2)} -4  s |\m| \na^{\g}H_{\g \b \a(2s-2)} + 2\ri (2s+1) |\m|  \F_{\b \a(2s-2)} \Big)\non\\
&&-\frac{s}{(2s+1)^2 (2s-1)}\Big( -\ri(3s+1) H^{\a(2s)} \na^2 H_{\a(2s)} \non\\
&&+ 2s(4s+1) H^{\b \a(2s-1)} \na_{\b}{}^{\g} H_{\g \a(2s-1)}+ 16s^2 H^{\a(2s)} \na_{(\a_1}\F_{\a_2 \dots \a_{2s})} \non\\
&&- 8\ri s \F^{\a(2s-1)} \F_{\a(2s-1)}-4s (4s^2+3s+1)|\m| H^{\a(2s)}H_{\a(2s)}\Big )
\bigg\}~.
\eea
It is apparent that the superfield $\F_{\a(2s-1)}$ is auxiliary, thus, upon elimination via its equation of motion,
\bea
\F_{\a(2s-1)} &=& -\frac{\ri}{8s}\Big (  (4s^2-1)\nabla^2 \cU_{\a(2s-1)} - 2\ri(4s^2-1) \nabla_{\b(\a_1} \cU^{\b;}{}_{\a_2 \ldots\a_{2s-1})}  ~ \\
&&+ 2\ri (2s+1)(4s^2-1)|\m| \cU_{\a(2s-1)} + 8s^2\nabla^\b H_{\b\a(2s-1)} \Big ) ~,\non
\eea
we arrive at the resulting action
\bea \label{5.24}
&&S^\perp_{(s)}[H_{\a(2s)}, \cU_{\b;\a(2s-2)}]= \bigg ( -\frac{1}{2} \bigg )^s \frac{1}{4}  \int \rd^{3|2}z\,E\, \bigg \{-\frac{\ri s(s-1)}{2s-1}H^{\a(2s)} \na^2 H_{\a(2s)} \non\\
&&- \frac{2s^2}{2s-1}H^{\b \a(2s-1) } \na_{\b}{}^{\g} H_{\g \a(2s-1)} -  \frac{4s^2(s-1)}{2s-1}|\m|H^{\a(2s)}H_{\a(2s)} \non\\
&&-2s \,\cU^{\d\a(2s-2)} \na^{\b \g} \Big( \na_{(\d} H_{\a_1 \dots \a_{2s-2} ) \b \g} - 2 \na_{\b} H_{\g \d \a(2s-2)}\Big)\non\\
&&+ \frac{4s(s-1)}{2s-1} \cU_{\l;}{}^{\l \a(2s-3)} \na^{\b \g} \na^{\d} H_{\b \g \d \a(2s-3)} - 8s^2|\m| \, \cU^{\b;\, \a(2s-2)} \na^{\g} H_{\b \g \a(2s-2)}\non\\
&&+ \cU^{\a(2s-1)} \bigg( -2\ri s \Box \cU_{\a(2s-1)} -s \na^2 \na^{\b}{}_{(\a_1} \cU_{\a_2 \dots \a_{2s-1})\b}\non\\
&&-\ri(s-1) \na_{(\a_1 \a_2}\na^{\b \g}\cU_{\a_3 \dots \a_{2s-1}) \b \g} + 2\ri \frac{(s-1)(2s-3)}{2s-1}  \na_{(\a_1 \a_2} \na_{\a_3}{}^{\b} \cU^{\g;}{}_{\a_4 \dots \a_{2s-1}) \b \g} \non\\
&&+ \frac{(s-1)(2s+1)}{2s-1} \na^2 \na_{(\a_1 \a_2} \cU^{\b;}{}_{\a_3 \dots \a_{2s-1}) \b} -s(2s+1)|\m|  \na^2 \cU_{\a(2s-1)} \non\\
&&- 4 \ri\,  s(2s-1) |\m|\na^{\b}{}_{(\a_1} \cU_{\a_2 \dots \a_{2s-1}) \b} + 2\ri  (s-1)(2s+1) |\m|\na_{(\a_1 \a_2} \cU^{\b;}{}_{\a_3 \dots \a_{2s-1}) \b} \non\\
&&- 2 \ri  s(2s-1)(2s+1)|\m|^2 \cU_{\a(2s-1)}\bigg) \non\\
&&+ \cU_{\l;}{}^{\l \a(2s-3)} \bigg(2 \ri \frac{(s-1)(6s-5)}{(2s-1)^2} \Box \cU^{\b;}{}_{\b \a(2s-3)} \non\\
&&- \frac{(s-1)(2s-3)}{(2s-1)^2} \Big(2 \ri (s-2) \na^{\b \g} \na_{(\a_1 \a_2} \cU^{\d;}{}_{\a_3 \dots \a_{2s-3}) \b \g \d} + \na^2 \na^{\b}{}_{(\a_1} \cU^{\g;}{}_{\a_2 \dots \a_{2s-3}) \b \g}\Big) \non\\
&& + \frac{s-1}{2s-1}|\m| \Big ( (2s+1)  \na^2 \cU^{\b;}{}_{\b \a(2s-3)}  - 2 \ri (4s^2 +16s-17)|\m| \, \cU^{\b;}{}_{\b \a(2s-3)} \Big ) \bigg) \bigg\}~. 
\eea
Note that we have introduced the notation $\cU_{\a(2s-1)} = \cU_{(\a_1;\a_2 \ldots \a_{2s-1})}$. It is a tedious but straightforward exercise to verify that the above ${\cN}=1$ action coincides with the action for the massless superspin-$s$ multiplet \eqref{TIAction}. To prove this, one  needs to introduce the ${\cN}=1$ field strength $\cW_{\b;\, \a(2s-2)}$ associated with the prepotential $\cU_{\b; \a(2s-2)}$,
\bea
\cW_{\b;\, \a(2s-2)} = -\ri \big ( \na^{\g} \na_{\b} - 4\ri |\m| \d_{\b}{}^{\g} \big )\, \cU_{\g; \, \a(2s-2)}~, \qquad \nabla^\b  \cW_{\b;\, \a(2s-2)}=0~.
\eea
The action \eqref{5.24} can then be written in the compact form
\bea
&&S^\perp_{(s)}[H_{\a(2s)}, \cU_{\b;\a(2s-2)}]
= \Big(-\hf \Big)^{s} \bigg (\frac{2s}{2s-1} \bigg )
\int \rd^{3|2}z\,
E\, \bigg\{\frac{1}{2} H^{\a(2s)} (\ri \nabla^2 +8 s |\m|) H_{\a(2s)}
\non \\
&& - \ri s \nabla_{\b} H^{\b \a(2s-1)} \nabla^{\g}H_{\g \a(2s-1)} 
-(2s-1) \cW^{\b ;\,\a(2s-2)} \nabla^{\g} H_{\b \g \a(2s-2)}
\non \\
&& -\frac{\ri}{2} (2s-1)\Big(\cW^{\b ;\, \a(2s-2)} \cW_{\b ;\, \a(2s-2)}+\frac{s-1}{s} \cW_{\b;}\,^{\b \a(2s-3)} \cW^{\g ;}\,_{\g \a(2s-3)} \Big) 
\non\\
&&
-2 \ri (2s-1) |\m| \, \cU^{\b ;\, \a(2s-2)} \cW_{\b ; \, \a(2s-2)}
\bigg\}~,
\eea
which is exactly the action describing the massless superspin-$s$ multiplet \eqref{TIAction}, modulo an overall normalisation factor. 

The other decoupled $\cN=1$ action, which is formulated in terms of the dynamical variables \eqref{4.19}, assumes the form
\bea \label{long_int_aux_action_11}
&&S^\parallel_{(s)}[\tilde{H}_{\a(2s)}, V_{\a(2s-2)}, \O_{\a(2s-1)}] = \bigg ( -\frac{1}{2} \bigg )^s \int \rd^{3|2}z\,E\, \bigg \{2s^2V^{\a(2s-2)} \big ( \ri \nabla^2 + 4|\m| \big ) V_{\a(2s-2)}~ \non\\
&&-\frac{4s^2}{2s+1}V^{\a(2s-2)} \big( {\nabla}^{\b \g}\tilde{H}_{\b \g \a(2s-2)} + 2{\nabla}^{\b} \O_{\b \a(2s-2)} \big )\non\\
&&+\frac{s}{(2s+1)^2 (2s-1)} 
\Big((2s+1)^2 \tilde{H}^{\a(2s)} (\ri \na^2 + 4 |\m|) \tilde{H}_{\a(2s)} - 8s \tilde{H}^{\a(2s)} \na_{(\a_1}\O_{\a_2 \dots \a_{2s})} \non \\
&&- 4\ri s(s+1) \na_{\b} \tilde{H}^{\b \a(2s-1)}\na^{\g}\tilde{H}_{\g \a(2s-1)} +16 \ri s^2 \O^{\a(2s-1)}\O_{\a(2s-1)}\Big) \bigg \}~. 
\eea
It is clear from the action \eqref{long_int_aux_action_11} that the superfield $\O_{\a(2s-1)}$ is auxiliary. Integrating it out by using its equation of motion,
\be
\O_{\a(2s-1)} = \frac{\ri}{4s} \Big ( (4s^2-1) \nabla_{(\a_1}V_{\a_2 \ldots\a_{2s-1})} - \nabla^\b \tilde{H}_{\b\a(2s-1)} \Big )~,
\ee
we find that the resulting action, modulo a normalisation factor, coincides with the action for the massless superspin-$s$ multiplet \eqref{action-t3}.


\subsection{Second supersymmetry transformations}
Let us compute the second supersymmetry transformations \eqref{SST} for the superfields \eqref{4.17} and \eqref{4.19} 
\begin{subequations}
	\bea
	\d_\e U_{\a(2s-2)}| &=& -\ri \e^\b \cU_{\b;\a(2s-2)}~, \label{SSTLI1}\\
	\d_\e \cU_{\b;\a(2s-2)} &=& 4s \e_\b V_{\a(2s-2)} ~, \\
	\d_\e V_{\a(2s-2)} &=& - \frac{\ri}{4s} \Big (   \e^\b \nabla_\b{}^\g \cU_{\g;\a(2s-2)}  - (2s-1)|\m|\e^\b \cU_{\b;\a(2s-2)} ~ \\
	&-& 4(s-1)|\m|\e_{(\a_1}\cU^{\b;}{}_{\a_2\ldots\a_{2s-2})\b} \Big ) ~, \non \\
	\d_\e H_{\a(2s)} &=& -2 \e^\b \nabla_{(\b}\tilde{H}_{\a_1 \ldots \a_{2s})} - \frac{8\ri s}{2s+1}\e_{(\a_1} \O_{\a_2 \ldots \a_{2s})}~,  \\
	\d_\e \tilde{H}_{\a(2s)} &=& \hf \e^\b \nabla_{(\b}H_{\a_1 \ldots \a_{2s})}+\frac{2\ri s}{2s+1}\e_{(\a_1} \F_{\a_2 \ldots \a_{2s})} ~, \\ 
	\d_\e \F_{\a(2s-1)} &=& \frac{1}{2(2s+1)} \e^\b \Big (   \ri \nabla^2 \tilde{H}_{\b\a(2s-1)} - 2(4s+3) \nabla_\b{}^\g \tilde{H}_{\g\a(2s-1)}~\\
	&-& 2(2s-1) \nabla_{(\a_1}{}^\g \tilde{H}_{\a_2\ldots\a_{2s-1})\b\g}  + 2 (4s^2+3s+1)|\m|  \tilde{H}_{\b\a(2s-1)}~ \non \\ 
	&+&8s \nabla_{(\b}\O_{\a_1 \ldots \a_{2s-1})}  \Big ) ~, \non \\ 
	\d_\e \O_{\a(2s-1)} &=& -\frac{1}{8(2s+1)}\e^\b \Big ( \ri \nabla^2 H_{\b\a(2s-1)} - 2 (4s+3) \nabla_\b{}^\g H_{\g\a(2s-1)}  ~ \\
	&-& 2(2s-1) \nabla_{(\a_1}{}^\g H_{\a_2\ldots\a_{2s-1})\b\g} + 4(4s^2+3s+1)|\m| H_{\b\a(2s-1)}   ~ \non \\
	&+&8s \nabla_{(\b} \F_{\a_1 \ldots \a_{2s-1})} \Big ) ~. \non
	\eea
\end{subequations}

It is clear from the variation \eqref{SSTLI1} that the second supersymmetry transformation \eqref{SST} breaks the WZ gauge \eqref{4.8}. To restore this gauge condition, it is necessary to supplement the second supersymmetry transformation with the $\e$-dependent gauge transformations:
\bsubeq
\bea
\d_{\g (\e)} U_{\a(2s-2)} &=& \g_{\a(2s-2)}(\e) + \bar{\g}_{\a(2s-2)}(\e) ~,\\
\d_{\g (\e)} G_{\a(2s)} 
&=& {{\bm{\cDB}}}_{(\a_{1}} {\bm{\cD}}_{\a_{2}}\bar{\g}_{\a_3 \dots \a_{2s})} (\e) \non\\
&=& \ri \Big (\bm \nabla^{\1}{}_{(\a_1}\bm \nabla^{\2}{}_{\a_2} + \bm \nabla_{(\a_1\a_2}\Big)\g_{\a_3\ldots{\a_{2s})}} (\e)~,
\eea
\esubeq
where
\be
\g_{\a(2s-2)}(\e)| = \frac{\ri}{2}\e^\b \cU_{\b;\a(2s-2)}~.
\ee
The modified second supersymmetry transformations take the form
\bsubeq \label{MSSTLI}
\bea 
\hat{\d}_\e U_{\a(2s-2)} &=& \d_\e U_{\a(2s-2)} + \d_{\g (\e)} U_{\a(2s-2)}~,\\
\hat{\d}_\e G_{\a(2s)} &=& \d_\e G_{\a(2s)} + \d_{\g (\e)} G_{\a(2s)}~.
\eea
\esubeq
They act on the ${\cN}=1$ gauge superfields \eqref{4.17} and \eqref{4.19} in the following manner
\begin{subequations}
	\bea
	\hat{\d}_\e \cU_{\b;\a(2s-2)} &=& 4s \e_\b V_{\a(2s-2)} ~, \\
	\hat{\d}_\e V_{\a(2s-2)} &=& - \frac{\ri}{8s} \Big (2 \e^\b \nabla_\b{}^\g \cU_{\g;\a(2s-2)}   + \ri \e^\b \nabla^2 \cU_{\b;\a(2s-2)} \\
	&-&8(s-1)|\m|\e_{(\a_1}\cU^{\b;}{}_{\a_2 \ldots \a_{2s-2})\b} + 2(2s+1)|\m|\e^\b \cU_{\b;\a(2s-2)} \Big )~, \non \\
	\hat{\d}_\e H_{\a(2s)} &=& -2 \e^\b \nabla_{(\b}\tilde{H}_{\a_1 \ldots \a_{2s})} - \frac{8\ri s}{2s+1}\e_{(\a_1} \O_{\a_2 \ldots \a_{2s})} ~, \non \\
	\hat{\d}_\e \tilde{H}_{\a(2s)} &=& \hf \e^\b \nabla_{(\b}H_{\a_1 \ldots \a_{2s})}+\frac{2\ri s}{2s+1}\e_{(\a_1} \F_{\a_2 \ldots \a_{2s})} - \frac{\ri}{2}\e_\b \nabla_{(\a_1 \a_2}\cU^{\b;}{}_{\a_3\ldots\a_{2s})}
	\\ 
	&+& \ri |\m| \e_{(\a_1}\cU_{\a_2 \ldots \a_{2s})} ~ , \non \\
	\hat{\d}_\e \F_{\a(2s-1)} &=& \frac{1}{2(2s+1)} \e^\b \Big (  \ri \nabla^2 \tilde{H}_{\b\a(2s-1)} -2(2s-1) \nabla_{(\a_1}{}^\g \tilde{H}_{\a_2\ldots\a_{2s-1})\b\g}  ~\\
	&-& 2(4s+3) \nabla_\b{}^\g \tilde{H}_{\g\a(2s-1)}+ 2 (4s^2+3s+1)|\m|  \tilde{H}_{\b\a(2s-1)} ~ \non \\ 
	&+&8s \nabla_{(\b}\O_{\a_1 \ldots \a_{2s-1})}  \Big ) ~, \non \\ 
	\hat{\d}_\e \O_{\a(2s-1)} &=& -\frac{1}{8(2s+1)}\e^\b \Big ( \ri \nabla^2 H_{\b\a(2s-1)} - 2(2s-1) \nabla_{(\a_1}{}^\g H_{\a_2\ldots\a_{2s-1})\b\g} ~ \\
	&-&2 (4s+3) \nabla_\b{}^\g H_{\g\a(2s-1)}   +4(4s^2+3s+1)|\m| H_{\b\a(2s-1)} ~ \non \\
	&+&8s \nabla_{(\b} \F_{\a_1 \ldots \a_{2s-1})} \Big ) ~ \non \\
	&+& \frac{1}{4s}\Big ( (s+1)\e_\b \nabla_{\g(\a_1}\nabla^\g\cU^{\b;}{}_{\a_2\ldots\a_{2s-1})}  - (s-1) \e_\b \nabla_{(\a_1}{}^\g \nabla_{\a_2}\cU^{\b;}{}_{\a_3 \ldots\a_{2s-1})\g} \non ~ \\
	&+&2(s-1)|\m| \e_{(\a_1} \nabla_{\a_2} \cU^{\b;}{}_{\a_3 \ldots\a_{2s-1})\b} + (2s+1) |\m| \e^\b \nabla_\b \cU_{\a(2s-1)} \non \\
	&+&(6s^2+3s+2)|\m|\e_\b \nabla_{(\a_1}\cU^{\b;}{}_{\a_2\ldots\a_{2s-1})} \Big )~. \non
	\eea
\end{subequations}

\section{Massless integer superspin: Transverse formulation} \label{Section8}
We now apply the same reduction procedure to the off-shell transverse formulation for the massless multiplet of integer superspin \cite{HutomoKuzenkoOgburn2018}.
\subsection{Transverse formulation}
For any integer $s\geq 1$, the transverse formulation for the massless superspin-$s$ multiplet is realised in terms of the following superfields
\be
\cV^\perp_{(s)} = \big \{U_{\a(2s-2)}, \G_{\a(2s)},\bar{\G}_{\a(2s)} \big  \}~.
\ee
The real superfield $U_{\a(2s-2)}$ is the same as in \eqref{4.1} and the complex superfield $\G_{\a(2s)}$ is transverse linear \eqref{TAS2TransL}. Modulo an overall factor, the transverse formulation for the massless superspin-$s$ multiplet is described by the action\footnote{The transverse formulation for the massless superspin-$s$ multiplet was derived in \cite{HutomoKuzenkoOgburn2018}  by performing a superfield duality transformation to the longitudinal action \eqref{long-action-int}.}
\bea \label {5.3}
&&\mathbb{S}^\perp_{(s)}[U_{\a(2s-2)}, \G_{\a(2s)}, \bar{\G}_{\a(2s)} ] = \Big (-\frac{1}{2} \Big )^s \int \text{d}^{3|4}z~\bm E~\bigg \{ 2s(s+1)\m\bar{\m}U^{\a(2s-2)}U_{\a(2s-2)} \non \\
&&+\frac{1}{8} U^{\a(2s-2)}{\bm{\cD}}^\b \big ({{\bm{\cDB}}}{}^2-6\m \big ){\bm{\cD}}_\b U_{\a(2s-2)} + \ri{\bm{\cD}}^{(\a_{1}\a_{2}} U^{\a_3 \ldots \a_{2s})} \big (\G_{\a(2s)}-\bar{\G}_{\a(2s)} \big ) ~\non \\
&&-\frac{2s-1}{16(2s+1)}\Big ( 8s{\bm{\cD}}^{(\a_1\a_2}U^{\a_3 \ldots\a_{2s})}{\bm{\cD}}_{(\a_1\a_2}U_{\a_3\ldots\a_{2s})} ~ \non \\
&&+ [{\bm{\cD}}^{(\a_1},{{\bm{\cDB}}}{}^{\a_2}]U^{\a_3 \ldots\a_{2s})}[{\bm{\cD}}_{(\a_1},\bar{{\bm{\cD}}}_{\a_2}]U_{\a_3 \ldots\a_{2s})}\Big )~~ \non \\
&&- \frac{2}{2s-1}\bar{\G}^{\a(2s)}\G_{\a(2s)} +\frac{1}{2s+1}\big (\G^{\a(2s)}\G_{\a(2s)} + \bar{\G}^{\a(2s)}\bar{\G}_{\a(2s)}\big )\bigg \}~,
\eea
which is invariant under the gauge transformations
\begin{subequations}
	\bea \label{7.3a}
	\d_L U_{\a(2s-2)} &&= {\bm{\cD}}^\b L_{\b\a(2s-2)} - {{\bm{\cDB}}}{}^\b \bar{L}_{\b\a(2s-2)} \equiv \bar{\g}_{\a(2s-2)} +\g_{\a(2s-2)}~, \\
	\d_L \G_{\a(2s)} &&= -\frac{1}{4}({{\bm{\cDB}}}{}^2 + 4s\m){\bm{\cD}}_{(\a_1}\bar{L}_{\a_2 \ldots \a_{2s})} + \frac{\ri}{2}(2s+1){{\bm{\cDB}}}{}^\b {\bm{\cD}}_{(\b\a_1}\bar{L}_{\a_2 \ldots \a_{2s})} \non \\ 
	&&\equiv \frac{1}{2}{\bm{\cD}}_{(\a_1}{{\bm{\cDB}}}_{\a_2}\g_{\a_3 \ldots\a_{2s})} - \frac{\ri}{2}(2s-1){\bm{\cD}}_{(\a_1\a_2}\g_{\a_3 \ldots \a_{2s})}~, \label{5.4b}
	\eea
\end{subequations}
where $L_{\a(2s-1)}$ is complex unconstrained and $\g_{\a(2s-2)}$ is transverse linear, as in \eqref{4.3a}.

\subsubsection{Reduction of gauge prepotentials to AdS$^{3|2}$}
We begin by reducing the real superfield $U_{\a(2s-2)}$. Following the prescription employed in section \ref{Section7}, we have the freedom to impose the gauge condition
\be \label{GcU}
U_{\a(2s-2)}| = 0~.
\ee
The residual gauge symmetry which preserves this gauge is described by the real unconstrained $\cN=1$ superfield
\be
\g_{\a(2s-2)} | = -\frac{\ri}{2}\z_{\a(2s-2)}~, \qquad \z_{\a(2s-2)} = \bar{\z}_{\a(2s-2)}~.
\ee
As a result of choosing the gauge condition \eqref{GcU}, we are left with the remaining independent $\cN=1$ superfields with respect to $U_{\a(2s-2)}$
\begin{subequations} \label{TVar}
	\bea 
	\cU_{\b;\a(2s-2)} : &=& \ri \bm \nabla^{\2}_\b U_{\a(2s-2)}| ~, \\
	V_{\a(2s-2)} :&=& - \frac{\ri}{8s}(\bm \nabla^{\2})^2 U_{\a(2s-2)}|~.
	\eea
\end{subequations}
Using \eqref{7.3a} enables the computation of the corresponding gauge transformations
\begin{subequations}
	\bea \label{OldUTrans}
	\d \cU_{\b;\a(2s-2)} &=& -2s \r_{\b\a(2s-2)} + \ri \nabla_\b \z_{\a(2s-2)} + \ri (2s-1)\nabla_{(\b}\z_{\a(2s-2))}~, \\
	\d V_{\a(2s-2)} &=& \frac{1}{2s}\nabla^\b \t_{\b\a(2s-2)}~, \label{Vtrans}
	\eea 
\end{subequations}
where we have introduced the $\cN=1$ gauge parameters $\t_{\a(2s-1)}$ and $ \tilde{\t}_{\a(2s-1)} $
\bea
\bm \nabla^{\2}{}_{(\a_1}\g_{\a_2 \ldots \a_{2s-1})}| = \t_{\a(2s-1)}
+ \ri s \tilde{\t}_{\a(2s-1)} ~, 
\eea
and the gauge parameter redefinition
\be
\r_{\a(2s-1)} := \tilde{\t}_{\a(2s-1)} + \ri \nabla_{(\a_1}\z_{\a_2 \ldots \a_{2s-1})}~.
\ee

It is immediately obvious that the gauge transformation of $\cU_{\b;\a(2s-2)}$ does not coincide with its analogue from the longitudinal formulation \eqref{ULong}. Addressing this, we introduce the following ${\cN}=1$ superfield redefinition
\be 
\tilde{\cU}_{\b;\a(2s-2)} := \cU_{\b;\a(2s-2)} - \frac{2s-1}{2s}\cU_{\b\a(2s-2)}~.
\ee
By using the gauge transformation \eqref{OldUTrans}, it can be shown that $\tilde{\cU}_{\b;\a(2s-2)}$ possesses the desired gauge freedom
\be\label{NewUGT}
\d \tilde{\cU}_{\b;\a(2s-2)}  = - \r_{\b\a(2s-2)} + \ri \nabla_\b \z_{\a(2s-2)}~.
\ee

Next, we wish to go about reducing $\G_{\a(2s)}$ to $\cN=1$ AdS superspace. We begin by converting the transverse linear constraint $\G_{\a(2s)}$\eqref{TAS2TransL} to the real representation \eqref{realrep}
\be \label{5.5}
\bm \nabla^{\2\b}\G_{\b\a(2s-1)} = \ri \bm \nabla^{\1\b} \G_{\b\a(2s-1)}~.
\ee
Taking a Taylor expansion of $\G_{\a(2s)}(\q^I)$ about $\q^{\2}$, and using the constraint \eqref{5.5}, we find that the independent $\q^{\2}$~-components of $\G_{\a(2s)}$ are
\be \label{5.6}
\G_{\a(2s)}|~, \qquad \bm \nabla^{\2}{}_{(\a_1}\G_{\a_2 \ldots \a_{2s+1})}|~.
\ee
Using the gauge transformations \eqref{5.4b} and the real representation \eqref{realrep}, we find\begin{subequations} \label{gaugetransgamma}
	\bea 
	\d \G_{\a(2s)} &=& -\frac{\ri}{2} \bm \nabla^{\1}{}_{(\a_1}\bm \nabla^{\2}{}_{\a_2}\g_{\a_3 \ldots \a_{2s})} - \ri s \bm \nabla_{(\a_1 \a_2}\g_{\a_3 \ldots \a_{2s})}~, \\
	\bm  \nabla^{\2}{}_{(\a_1} \d \G_{\a_2 \ldots \a_{2s+1})} &=& - \frac{1}{2}\bm \nabla_{(\a_1\a_2}\bm \nabla^{\1}{}_{\a_3}\g_{\a_4 \ldots\a_{2s+1})} - \ri s\bm \nabla_{(\a_1 \a_2}\bm \nabla^{\2}{}_{\a_3} \g_{\a_4 \ldots \a_{2s+1})}~.
	\eea
\end{subequations}
Making use of \eqref{gaugetransgamma}  and the gauge freedom \eqref{4.10} allows us to directly compute the gauge transformations of the $\cN=1$ complex superfields \eqref{5.6}
\begin{subequations} \label{gtgamma}
	\bea
	\d \G_{\a(2s)}| &=& - \frac{\ri}{2} \nabla_{(\a_1}\t_{\a_2 \ldots\a_{2s})} + \frac{s}{2} \nabla_{(\a_1} \tilde{\t}_{\a_2 \ldots \a_{2s})}- \frac{s}{2} \nabla_{(\a_1 \a_2}\z_{\a_3 \ldots \a_{2s})} ~, \\
	\bm \nabla^{\2}{}_{(\a_1}\d  \G_{\a_2 \ldots \a_{2s+1})}| &=& \frac{\ri}{4} \nabla_{(\a_1 \a_2} \nabla_{\a_3}\z_{\a_4 \ldots \a_{2s+1})} - \ri s \nabla_{(\a_1 \a_2}\t_{\a_3 \ldots \a_{2s+1})} \\
	&+& s^2 \nabla_{(\a_1 \a_2}\tilde \t_{\a_3 \ldots \a_{2s+1})}~. \non 
	\eea
\end{subequations}
Let us split the complex $\cN=1$ superfields \eqref{5.6} into real and imaginary parts\begin{subequations} \label{5.8}
	\bea
	\G_{\a(2s)}| = \frac{1}{2} \big ( sH_{\a(2s)} - \ri \tilde{H}_{\a(2s)} \big )~, \\
	\bm \nabla^{\2}{}_{(\a_1}\G_{\a_2 \ldots \a_{2s+1})}| = \F_{\a(2s+1)} - \ri s \O_{\a(2s+1)}~.
	\eea
\end{subequations}
The gauge transformations of these four real $\cN=1$ superfields can be read off from the gauge transformations \eqref{gaugetransgamma}
\begin{subequations} \label{ReducedTransGC}
	\bea 
	\d H_{\a(2s)} &=& \nabla_{(\a_1}\r_{\a_2 \ldots \a_{2s})}~, \\
	\d \tilde{H}_{\a(2s)} &=& \nabla_{(\a_1}\t_{\a_2 \ldots \a_{2s})}~, \\
	\d \F_{\a(2s+1)} &=& s^2 \nabla_{(\a_1\a_2}\r_{\a_3 \ldots \a_{2s+1})} -\frac{\ri}{4}(4s^2-1)\nabla_{(\a_1\a_2}\nabla_{\a_3}\z_{\a_4\ldots\a_{2s+1})}~,\\
	\d \O_{\a(2s+1)} &=& \nabla_{(\a_1\a_2}\t_{\a_3\ldots\a_{2s+1})}~.
	\eea
\end{subequations}

It then follows from gauge transformations \eqref{Vtrans}, \eqref{NewUGT} and \eqref{ReducedTransGC} that we are indeed dealing with two different gauge theories. The first of these models is formulated in terms of the superfield variables
\be \label{5.10}
\cV^\perp_{(s)} = \big \{  H_{\a(2s)}, \tilde{\cU}_{\b;\a(2s-2)} , \F_{\a(2s+1)} \big \}~,
\ee
which are defined modulo gauge transformations
\begin{subequations} \label{7.21}
	\bea
	\d_\r H_{\a(2s)} &=&  \nabla_{(\a_1}\r_{\a_2\ldots\a_{2s})}~, \label{5.11b} \\
	\d_{\r , \z} \tilde{\cU}_{\b;\a(2s-2)}  &=& - \r_{\b\a(2s-2)} + \ri \nabla_\b \z_{\a(2s-2)}~, \label{5.11a} \\
	\d_{\r , \z} \F_{\a(2s+1)} &=& s^2\nabla_{(\a_1\a_2}\r_{\a_3 \ldots \a_{2s+1})} - \frac{\ri}{4}(4s^2-1)\nabla_{(\a_1\a_2}\nabla_{\a_3}\z_{\a_4 \ldots\a_{2s+1})}~.
	\eea
\end{subequations}
The other gauge theory is described in terms of the superfields
\be \label{5.12}
\cV^\parallel_{(s)} = \big \{ \tilde H_{\a(2s)}, V_{\a(2s-2)}, \O_{\a(2s+1)} \big \}~,
\ee
which possess the following gauge freedom
\begin{subequations} \label{7.22}
	\bea
	\d_\t \tilde{H}_{\a(2s)} &=& \nabla_{(\a_1}\t_{\a_2 \ldots \a_{2s})}~, \label{5.13a} \\
	\d_\t V_{\a(2s-2)} &=& \frac{1}{2s}\nabla^\b \t_{\b\a(2s-2)}~, \label{5.13b} \\
	\d_\t \O_{\a(2s+1)} &=& \nabla_{(\a_1\a_2}\t_{\a_3 \ldots \a_{2s+1})}~.
	\eea
	From the perspective of $\cN=1$ AdS superspace, the reduction of the action \eqref{5.3} gives rise to two decoupled supersymmetric actions, which are formulated in terms of superfields \eqref{5.10} and \eqref{5.12}, respectively
\end{subequations}
\bea \label{t-int}
\mathbb{S}^\perp_{(s)}[U_{\a(2s-2)}, \G_{\a(2s)},\bar{\G}_{\a(2s)} ] &=& S^\perp_{(s)}[ H_{\a(2s)}, \tilde{\cU}_{\b;\a(2s-2)} , \F_{\a(2s+1)} ] ~ \non \\
&&+ S^\parallel_{(s)}[\tilde{H}_{\a(2s)}, V_{\a(2s-2)}, \O_{\a(2s+1)}]~.
\eea
We provide the explicit realisations of these two $\cN=1$ decoupled actions in the following subsection.


\subsection{Massless higher-spin $\cN=1$ supermultiplets}
The first of the decoupled $\cN=1$ supersymmetric actions, which is formulated in terms of the superfields \eqref{5.10}, takes the form 
\bea \label{5.15}
&&S^\perp_{(s)}[ H_{\a(2s)}, \tilde{\cU}_{\b;\a(2s-2)} , \F_{\a(2s+1)} ] = \bigg ( - \frac{1}{2} \bigg )^s \frac{1}{4}  \int \rd^{3|2}z~E~ \bigg \{ \tilde{\cU}^{\a(2s-1)} \Big ( -2\ri s (4s-1) \Box \tilde{\cU}_{\a(2s-1)}  ~ \non \\ 
&&-s \na^2 \na^{\b}{}_{(\a_1} \tilde{\cU}_{\a_2 \dots \a_{2s-1})\b} +4\ri s (s-1)^2 \na_{(\a_1 \a_2}\na^{\b \g}\tilde{\cU}_{\a_3 \dots \a_{2s-1}) \b \g} -s(2s+1)|\m|\nabla^2\tilde{\cU}_{\a(2s-1)} ~ \non\\
&&+  \frac{2\ri (s-1)(2s-3)}{2s-1}  \na_{(\a_1 \a_2} \na_{\a_3}{}^{\b} \tilde{\cU}^{\g;}{}_{\a_4 \dots \a_{2s-1}) \b \g} + \frac{(s-1)(2s+1)}{2s-1} \na^2 \na_{(\a_1 \a_2} \tilde{\cU}^{\b;}{}_{\a_3 \dots \a_{2s-1}) \b} ~  \non \\
&&- 4 \ri  s(2s-1) |\m|\na_{(\a_1}{}^{\b} \, \tilde{\cU}_{\a_2 \dots \a_{2s-1}) \b} + 2\ri  (s-1)(2s+1) |\m|\na_{(\a_1 \a_2} \tilde{\cU}^{\b;}{}_{\a_3 \dots \a_{2s-1}) \b} \non \\
&&+ 2\ri s(2s-1)(2s+1)(4s-3)|\m|^2 \, \tilde{\cU}_{\a(2s-1)}\Big ) ~ \non \\
&&+ \tilde{\cU}_{\b;}{}^{\b \a(2s-3)} \bigg(2 \ri \frac{(s-1)(6s-5)}{(2s-1)^2} \Box \tilde{\cU}^{\g;}{}_{\g \a(2s-3)}+ \frac{(s-1)(2s+1)}{(2s-1)}|\m| \na^2 \tilde{\cU}^{\g;}{}_{\g \a(2s-3)} ~ \non \\
&&- \frac{(s-1)(2s-3)}{(2s-1)^2} \Big(2 \ri (s-2) \na^{\g \d} \na_{(\a_1 \a_2} \tilde{\cU}^{\l;}{}_{\a_3 \dots \a_{2s-3}) \g \d \l} + \na^2 \na^{\g}{}_{(\a_1} \tilde{\cU}^{\d;}{}_{\a_2 \dots \a_{2s-3}) \g\d}\Big) \non\\
&&- \frac{2\ri(s-1)}{2s-1} (4s^2 +16s-17)|\m|^2 \, \tilde{\cU}^{\g;}{}_{\g \a(2s-3)} \bigg) \non \\
&&-\frac{8s}{(2s+1)^2(2s-1)}  \Big ( s(2s^3-2s-1)|\m|H^{\a(2s)}H_{\a(2s)} +s^2(s+1)H^{\b\a(2s-1)}\nabla_\b{}^\g H_{\g\a(2s-1)}\non \\
&&+ \frac{\ri s}{4}(2s^2-1)H^{\a(2s)}\nabla^2H_{\a(2s)}  + (2s+1)H^{\a(2s)}\nabla^\b \F_{\b\a (2s)} + 2\ri  (2s+1)\F^{\a(2s+1)}\F_{\a(2s+1)} \Big ) \non \\
&&+\frac{8s}{2s+1}\tilde{\cU}^{\a(2s-1)} \Big ( s\nabla^{\b\g}\nabla_\b H_{\g\a(2s-1)}-s\nabla^{\b\g}\nabla_{(\a_1}H_{\a_2 \ldots \a_{2s-1})\b\g}+s(2s+1)|\m|\nabla^\b H_{\b\a(2s-1)} \non \\
&&+ \ri (2s+1) \nabla^{\b\g}\F_{\b\g\a(2s-1)} \Big ) + \frac{s(s-1)}{2s-1}\tilde{\cU}_{\b;}{}^{\b\a(2s-3)}\nabla^{\b\g}\nabla^\d H_{\b\g\d\a(2s-3)}  \bigg \} ~, 
\eea
which is invariant under the gauge transformations \eqref{7.21}. It is apparent that the superfield $\F_{\a(2s+1)}$ is auxiliary, so upon elimination via its equation of motion,
\be
\F_{\a(2s+1)} = - \frac{1}{4}\Big ( (4s^2-1)\nabla_{(\a_1\a_2}\tilde{\cU}_{\a_3\ldots\a_{2s+1})} + \ri \nabla_{(\a_1}H_{\a_2 \ldots \a_{2s+1})} \Big ) ~,
\ee
one arrives at the action which coincides with \eqref{5.24}, up to an overall normalisation factor. Recall that in section \ref{TAS2Massless}, the action \eqref{5.24} was shown to coincide with the action for the massless superspin-$s$ multiplet \eqref{TIAction}.

The second $\cN=1$ action in \eqref{t-int}, which is formulated in terms of the superfields \eqref{5.12}, takes the form
\bea \label{5.16}
&&S^\parallel_{(s)}[\tilde{H}_{\a(2s)}, V_{\a(2s-2)}, \O_{\a(2s+1)}] = \Big ( - \frac{1}{2} \Big )^s \int \rd^{3|2}z~ E ~ \bigg \{ -2sV^{\a(2s-2)}\nabla^{\b\g}\tilde{H}_{\b\g\a(2s-2)} \hspace{1.3cm}\\
&&+2s(s-1)V^{\b \a(2s-3)}\nabla_\b{}^\g V_{\g\a(2s-3)} + \ri s^2 V^{\a(2s-2)}\nabla^2 V_{\a(2s-2)} + 4s^2(s+1)|\m|V^{\a(2s-2)}V_{\a(2s-2)}  \non \\
&&-\frac{1}{2(2s+1)^2(2s-1)} \Big ( 4\ri s^2(2s+1)\O^{\a(2s+1)}\O_{\a(2s+1)} +2s(4s+1)\tilde{H}^{\b \a(2s-1) }\nabla_\b{}^\g \tilde{H}_{\g\a(2s-1)} \non ~\\
&&- \ri s \tilde{H}^{\a(2s)}\nabla^2 \tilde{H}_{\a(2s)}+ 8s^2(2s+1)\tilde{H}^{\a(2s)}\nabla^\b\O_{\b\a(2s)} -4s(4s^2+3s+1)|\m|\tilde{H}^{\a(2s)}\tilde{H}_{\a(2s)} \Big ) \bigg \}~. \non
\eea
The superfield $\O_{\a(2s+1)}$ is auxiliary, so upon integrating it out via its equation of motion, 
\be
\O_{\a(2s+1)} = -\ri \nabla_{(\a_1}\tilde{H}_{\a_2\ldots\a_{2s+1})}~,
\ee
one finds that up to an overall factor, the resulting action coincides with the action for the massless superspin-$s$ multiplet \eqref{action-t3}.


\subsection{Second supersymmetry transformations}
We begin by computing the second supersymmetry transformations for the real $\cN=1$ fields \eqref{5.10} and \eqref{5.12}
\begin{subequations}
	\bea
	\d_\e U_{\a(2s-2)} |&=&-2\ri s \e^\b \tilde{\cU}_{\b\a(2s-2)} + \frac{2\ri(s-1)}{2s-1}\e_{(\a_1}\tilde{\cU}^{\b;}{}_{\a_2\ldots\a_{2s-2})\b} ~, \label{TLBG}\\
	\d_\e  \tilde{\cU}_{\b;\a(2s-2)} &=& 2(2s-1)\e_\b V_{\a(2s-2)} -4(s-1)\e_{(\a_1}V_{\a_2\ldots\a_{2s-2})\b}~, \\
	\d_\e V_{\a(2s-2)} &=& \frac{\ri}{2s} \Big ( \frac{s-1}{2s-1}\e^\b \nabla_{\b(\a_1}\tilde{\cU}^{\g ;}{}_{\a_2 \ldots \a_{2s-2})\g}  -s\e^\b \nabla_\b{}^\g \tilde{\cU}_{\g\a(2s-2)}   ~ \\
	&+&  s(2s-1)|\m|\e^\b \tilde{\cU}_{\b\a(2s-2)}+(s-1)|\m|\e_{(\a_1} \tilde{\cU}^{\b ;}{}_{\a_2 \ldots \a_{2s-2})\b} \Big ) ~, \non \\
	\d_\e H_{\a(2s)} &=& -2\ri \e^\b \O_{\b\a(2s)} - \frac{2 }{2s+1}\e_{(\a_1}\nabla^\b \tilde{H}_{\a_2 \ldots\a_{2s})\b}   ~, \\
	\d_\e \tilde{H}_{\a(2s)} &=& 2 \ri \e^\b\F_{\b\a(2s)} + \frac{2 s^2}{2s+1}\e_{(\a_1}\nabla^\b H_{\a_2 \ldots \a_{2s})\b} ~ , ~\\
	\d_\e \F_{\a(2s+1)} &=& \hf \e^\b \nabla_{\b(\a_1} \tilde{H}_{\a_2 \ldots \a_{2s+1})} -s \e_{(\a_1}\nabla^\b \O_{\a_2 \ldots \a_{2s+1})\b} ~\\
	&+&\frac{1}{4(2s+1)} \Big ( \ri \e_{(\a_1}\nabla^2 \tilde{H}_{\a_2 \ldots \a_{2s+1})} -4s \e_{(\a_1} \nabla_{\a_2}{}^\b \tilde{H}_{\a_3 \ldots \a_{2s+1})\b} \non ~\\
	&+& 4(4s^2 + 5s +2)|\m| \e_{(\a_1}\tilde{H}_{\a_2 \ldots \a_{2s+1})} \Big )~, \non \\
	\d_\e \O_{\a(2s+1)} &=& - \hf \e^\b \nabla_{\b(\a_1}H_{\a_2 \ldots \a_{2s+1})} + \frac{1}{s}\e_{(\a_1}\nabla^\b \F_{\a_2 \ldots \a_{2s+1})\b} ~ \\
	&-& \frac{1}{4(2s+1)} \Big ( \ri \e_{(\a_1} \nabla^2 H_{\a_2 \ldots \a_{2s+1})}
	-4s\e_{(\a_1}\nabla_{\a_2}{}^\b H_{\a_3 \ldots\a_{2s+1})\b} ~ \non  \\
	&+&4(4s^2+5s+2)|\m|\e_{(\a_1}H_{\a_2 \ldots \a_{2s+1})}\Big )~. \non 
	\eea
\end{subequations}

It is clear from the variation \eqref{TLBG} that the second supersymmetry transformation \eqref{SST} breaks the WZ gauge condition \eqref{GcU}. Thus, we need to supplement the second supersymmetry transformation with the $\e$-dependent gauge transformations:
\bsubeq
\bea
\d_{\g (\e)} U_{\a(2s-2)} &=& \g_{\a(2s-2)}(\e) + \bar{\g}_{\a(2s-2)}(\e) ~,\\
\d_{\g (\e)} \G_{\a(2s)} 
&=& \hf {{\bm{\cD}}}_{(\a_{1}}  {\bm{\cDB}}_{\a_{2}}{\g}_{\a_3 \dots \a_{2s})} (\e)-\frac{\ri}{2} (2s-1) {\bm{\cD}}_{(\a_1 \a_2} \g_{\a_3 \dots \a_{2s})} (\e) ~,
\eea
\esubeq
where
\be
\g_{\a(2s-2)}(\e)| = \ri s \e^\b \tilde{\cU}_{\b\a(2s-2)} - \frac{\ri(s-1)}{2s-1}\e_{(\a_1}\tilde{\cU}^{\b;}{}_{\a_2\ldots\a_{2s-2})\b}~.
\ee
The modified second supersymmetry transformations have the form
\bsubeq 
\bea 
\hat{\d}_\e U_{\a(2s-2)} &=& \d_\e U_{\a(2s-2)} + \d_{\g (\e)} U_{\a(2s-2)}~,\\
\hat{\d}_\e \G_{\a(2s)} &=& \d_\e \G_{\a(2s)} + \d_{\g (\e)} \G_{\a(2s)}~.
\eea
\esubeq
They act on the ${\cN}=1$ gauge superfields \eqref{5.10} and \eqref{5.12} as follows
\begin{subequations}
	\bea
	\hat{\d}_\e  \tilde{\cU}_{\b;\a(2s-2)} &=&  2(2s-1)\e_\b V_{\a(2s-2)} -4(s-1)\e_{(\a_1}V_{\a_2\ldots\a_{2s-2})\b}~, \\
	\hat{\d}_\e V_{\a(2s-2)} &=& -\frac{\ri}{4} \Big ( 2 \e^\b \nabla_\b{}^\g \tilde{\cU}_{\g\a(2s-2)} +\ri\e^\b \nabla^2 \tilde{\cU}_{\b\a(2s-2)} ~ \\
	&+&  2(2s+1)|\m| \e^\b \tilde{\cU}_{\b\a(2s-2)}- \frac{(s-1)}{s(2s-1)} \big (  2 \e^\b \nabla_{\b (\a_1}\tilde{\cU}^{\g;}{}_{\a_2\ldots\a_{2s-2})\g}  \non ~ \\
	&+&\ri \e_{(\a_1}\nabla^2\tilde{\cU}^{\b;}{}_{\a_2\ldots\a_{2s-2})\b} + 2(6s-1)|\m| \e_{(\a_1}\tilde{\cU}^{\b;}{}_{\a_2 \ldots \a_{2s-2})\b}\big ) \Big )~, \non \\
	\hat{\d}_\e H_{\a(2s)} &=&-2\ri \e^\b \O_{\b\a(2s)} - \frac{2 }{2s+1}\e_{(\a_1}\nabla^\b \tilde{H}_{\a_2 \ldots\a_{2s})\b}  ~, \\
	\hat{\d}_\e \tilde{H}_{\a(2s)} &=&  2 \ri \e^\b\F_{\b\a(2s)} + \frac{2 s^2}{2s+1}\e_{(\a_1}\nabla^\b H_{\a_2 \ldots \a_{2s})\b}  + 2\ri s^2 \e^\b \nabla_{(\a_1 \a_2} \tilde{\cU}_{\a_3 \ldots \a_{2s})\b}~ \\
	&-&  \frac{2\ri s}{2s-1}(s-1)\e_{(\a_1}\nabla_{\a_2 \a_3} \tilde{\cU}^{\b;}{}_{\a_4 \ldots \a_{2s})\b}+4\ri s^2 |\m| \e_{(\a_1}\tilde{\cU}_{\a_2 \ldots \a_{2s})}~,  \non \\
	\hat{\d}_\e \F_{\a(2s+1)} &=& \hf \e^\b \nabla_{\b(\a_1} \tilde{H}_{\a_2 \ldots \a_{2s+1})} -s \e_{(\a_1}\nabla^\b \O_{\a_2 \ldots \a_{2s+1})\b} ~\\
	&+&\frac{1}{4(2s+1)} \Big ( \ri \e_{(\a_1}\nabla^2 \tilde{H}_{\a_2 \ldots \a_{2s+1})} -4s \e_{(\a_1} \nabla_{\a_2}{}^\b \tilde{H}_{\a_3 \ldots \a_{2s+1})\b} \non ~\\
	&+& 4(4s^2 + 5s +2)|\m| \e_{(\a_1}\tilde{H}_{\a_2 \ldots \a_{2s+1})} \Big )~, \non \\
	\hat{\d}_\e \O_{\a(2s+1)} &=& - \hf \e^\b \nabla_{\b(\a_1}H_{\a_2 \ldots \a_{2s+1})} + \frac{1}{s}\e_{(\a_1}\nabla^\b \F_{\a_2 \ldots \a_{2s+1})\b} ~ \\
	&-& \frac{1}{4(2s+1)} \Big ( \ri \e_{(\a_1} \nabla^2 H_{\a_2 \ldots \a_{2s+1})}
	-4s\e_{(\a_1}\nabla_{\a_2}{}^\b H_{\a_3 \ldots\a_{2s+1})\b} ~  \non \\
	&+&4(4s^2+5s+2)|\m|\e_{(\a_1}H_{\a_2 \ldots \a_{2s+1})}\Big ) -\hf \e^\b \nabla_{(\a_1 \a_2}\nabla_{\a_3}\tilde{\cU}_{\a_4 \ldots \a_{2s+1})\b} \non \\
	&+&  \frac{1}{2s(2s-1)}(s-1)\e_{(\a_1}\nabla_{\a_2 \a_3}\nabla_{\a_4} \tilde{\cU}^{\b;}{}_{\a_5 \ldots \a_{2s+1})\b} - |\m|\e_{(\a_1}\nabla_{\a_2}\tilde{\cU}_{\a_3 \ldots \a_{2s+1})} ~. \non
	\eea
\end{subequations}

\section{Topologically massive theories}\label{TAS2TopMassSec}
In this section we study the $(1,1) \to (1,0)$  AdS reduction of the off-shell topologically massive higher-spin supersymmetric theories, which were first given in \cite{HutomoKuzenkoOgburn2018}. In particular, we will only be interested in employing the reduction prescription to the massive longitudinal and transverse half-integer theories.

The $(1,1)$ massless higher-spin theories presented in sections \ref{Section5} - \ref{Section8} do not propagate any physical degrees of freedom.\footnote{This follows from the fact that these theories describe two massless higher-spin actions in AdS$^{3|2}$, which were shown explicitly in section \ref{TASMasslessHStheories} to propagate no physical degrees of freedom on-shell.}  However, they can be deformed in order to generate off-shell topologically massive higher-spin supersymmetric theories. The gauge-invariant actions for such massive multiplets are obtained
by adding a superconformal and massless higher-spin action together, following the philosophy of topologically massive theories \cite{Siegel,JS,DJT1,DJT2}.

Given a positive integer $n$, the superconformal higher-spin action in AdS$^{(3|1,1)}$ \cite{KuzenkoOgburn2016,HutomoKuzenkoOgburn2018,KuzenkoPonds2018,BHHK} is given by 
\be \label{SCSAct}
\mathbb{S}^{(n)}_{\text{SCHS}}[\mathfrak{H}_{\a(n)}] = -\frac{\ri^n}{2^{\lfloor n/2 \rfloor+1}} \int \rd^{3|4}z\, \bm E \,\mathfrak{H}^{\a(n)}\mathfrak{W}_{\a(n)}(\mathfrak{H})~,
\ee
where $\mathfrak{H}_{\a(n)}$ is the superconformal gauge multiplet, 
and $\mathfrak{W}_{\a(n)}(\mathfrak{H})$
is the higher-spin super-Cotton tensor in AdS$^{(3|1,1)}$.
The action \eqref{SCSAct} is invariant under the gauge transformations
\be \label{N2GT}
\d_\l \mathfrak{H}_{\a(n)}= {\bm{\cDB}}_{(\a_1}\l_{\a_2\ldots\a_n)}-(-1)^n {\bm{\cD}}_{(\a_1}\bar{\l}_{\a_2\ldots\a_n)}~,
\ee
where the gauge parameter $\l_{\a(n-1)}$ is complex unconstrained. To check that the action \eqref{SCSAct} is gauge invariant, one needs to make use of the properties of the super-Cotton tensor $\mathfrak{W}_{\a(n)}(\mathfrak{H})$. In particular, $\mathfrak{W}_{\a(n)}(\mathfrak{H})$ is gauge invariant $\d_\l \mathfrak{W}_{\a(n)}(\mathfrak{H}) = 0$, and is simultaneously transverse linear \eqref{TAS2TransL} and transverse anti-linear \eqref{TAS2TransA}.\footnote{Recall that the closed form expression for
	$\mathfrak{W}_{\a(n)}(\mathfrak{H})$ in $\mb{M}^{3|4}$ is given by \eqref{4.188}. It was generalised to an arbitrary conformally flat superspace in \cite{KuzenkoPonds2019} by using the formalism of $\cN=2$ conformal superspace
	\cite{BKNT-M1}. However, it has yet to be degauged to AdS$^{(3|1,1)}$.}

In $(1,1)$ AdS superspace, let us consider the two dually equivalent off-shell formulations for a massive superspin-$(s+\hf)$ multiplet \cite{HutomoKuzenkoOgburn2018}\footnote{
	As pointed out in \cite{HutomoKuzenkoOgburn2018}, it is expected that the topologically massive actions \eqref{11MassiveTrans} and \eqref{11MassiveLong} describe the on-shell massive supermultiplets in AdS$^{(3|1,1)}$ \cite{KuzenkoNovakTartaglino-Mazzucchelli2015}.
	Recall that a massive  on-shell supermultiplet is realised in terms of a superfield on $\bm{\mc{V}}_{(n)}$ which satisfies the on-shell conditions \eqref{TAS1OnshellConditions}.}
\begin{subequations} \label{11Massive}
	\bea
	\mathbb{S}^{\perp}_{(s+\hf)}[\mathfrak{H}_{\a(2s)},\G_{\a(2s-2)},\bar{\G}_{\a(2s-2)}|m] &=& \k \mathbb{S}^{(2s)}_{\text{SCHS}}[\mathfrak{H}_{\a(2s)}] ~ \label{11MassiveTrans} \\ 
	&&+m^{2s-1}\mathbb{S}^{\perp}_{(s+\hf)}[\mathfrak{H}_{\a(2s)},\G_{\a(2s-2)},\bar{\G}_{\a(2s-2)}]~, \non \\
	\mathbb{S}^{\parallel}_{(s+\hf)}[\mathfrak{H}_{\a(2s)},G_{\a(2s-2)},\bar{G}_{\a(2s-2)}|m]&=&\k \mathbb{S}^{(2s)}_{\text{SCHS}}[\mathfrak{H}_{\a(2s)}]~  \label{11MassiveLong} \\
	&&+m^{2s-1}\mathbb{S}^{\parallel}_{(s+\hf)}[\mathfrak{H}_{\a(2s)},G_{\a(2s-2)},\bar{G}_{\a(2s-2)}]~, \non
	\eea
\end{subequations}
where $\k$ is dimensionless and $m$ has dimension of mass.
The superconformal action $\mathbb{S}^{(2s)}_{\text{SCHS}}[\mathfrak{H}_{\a(2s)}]$ is given by \eqref{SCSAct}, for the case $n=2s$. The transverse $\mathbb{S}^{\perp}_{(s+\hf)}[\mathfrak{H}_{\a(2s)},\G_{\a(2s-2)},\bar{\G}_{\a(2s-2)}]$ and longitudinal $\mathbb{S}^{\parallel}_{(s+\hf)}[\mathfrak{H}_{\a(2s)},G_{\a(2s-2)},\bar{G}_{\a(2s-2)}]$ actions are given by \eqref{2.5} and \eqref{3.3}, respectively.

We now turn to describing the $(1,1) \to (1,0)$ AdS reduction of the massive models \eqref{11Massive}. The reduction of the massless transverse \eqref{2.5} and longitudinal \eqref{3.3} formulations was achieved in sections \ref{Section5} and \ref{Section6}, respectively. Thus, we need only carry out the reduction for the superconformal action.

The superconformal gauge multiplet $\mathfrak{H}_{\a(2s)}$ was reduced in subsection \ref{TASReductionGaugePrep}. Upon reduction, the superfield $\mathfrak{H}_{\a(2s)}$ is equivalent to two unconstrained real $\cN=1$ superfields \eqref{2.9}, defined modulo gauge transformations of the form \eqref{2.11}. Since $\mathfrak{W}_{\a(2s)}$ is a real transverse linear superfield, it follows that $\mathfrak{W}_{\a(2s)}$ is equivalent to two real $\cN=1$ superfields, defined as:
\be
W_{\a(2s)}(H) = -\mathfrak{W}_{\a(2s)}(\mathfrak{H})|~, \qquad W_{\a(2s+1)}(H) = \frac{\ri}{2}\nabla_{(\a_1}\mathfrak{W}_{\a_2 \ldots \a_{2s+1})}(\mathfrak{H})|~.
\ee
Here each of the ${\cN}=1$ superfields $W_{\a(2s)}(H)$ and $W_{\a(2s+1)}(H)$ are transverse linear and gauge invariant. 

Applying the reduction procedure to the superconformal action \eqref{SCSAct} yields two decoupled $\cN=1$ supersymmetric theories
\be \label{SCSRed}
\mathbb{S}^{(2s)}_{\text{SCHS}}[\mathfrak{H}_{\a(2s)}] = - \frac{(-1)^s}{2^{s+1}}
\int \rd^{3|2}z\,E\, \big \{ H^{\a(2s)}W_{\a(2s)} 
+ \ri H^{\a(2s+1)}W_{\a(2s+1)} \big \}~.
\ee
These two gauge-invariant theories can be identified as the $\cN=1$ superconformal actions in AdS$_3$ \eqref{TASNTMAction}, for the cases $n=2s$ and $n=2s+1$, respectively. Recalling the reduction results \eqref{92a}, it is apparent that both \eqref{11MassiveTrans} and \eqref{11MassiveLong} decouple into two off-shell topologically massive $\cN=1$ theories. The first $S^{\parallel}_{(s+\hf)}[H_{\a(2s+1)} ,L_{\a(2s-2)} |m]$ coincides with \eqref{N2Massive1}, while the second model $S^{\parallel}_{(s)}[H_{\a(2s)} ,V_{\a(2s-2)} |m]$ can be identified with \eqref{8.4a}. 

One can also construct a new topologically massive model which leads directly to the equations
\eqref{TAS1OnshellConditions}. The corresponding action is given by
\bea
{\mathbb{S}}^{(n)}_{\rm{NTM}}
[ {\mathfrak H}_{\a(n)}] 
= - \frac{\ri^n}{2^{\left \lfloor{n/2}\right \rfloor +1}} \frac{\k}{M} 
\int \rd^{3|4}z\, \bm E\,{\mathfrak W}^{\a(n) }( {\mathfrak H}) 
\big ( \mb{F} - \s M  \big ) {\mathfrak H}_{\a(n)}~.
\label{5.13}
\eea
In the flat-superspace limit, the action reduces to \eqref{TMSNewTopoMassive} in the case $\cN=2$.
It is easy to see that the equation of motion with respect to $\mf{H}_{\a(n)}$ is
\be \label{TAS2NTMAction}
0 = \big ( \bm{\D} - \s M  \big ) {\mathfrak W}_{\a(n)}~.
\ee
Since the field strength ${\mathfrak W}_{\a(n)}(\mf{H})$ is TLAL and satisfies the constraint \eqref{TAS2NTMAction}, it follows from \eqref{TAS1OnshellConditions} that on-shell, the action \eqref{5.13} describes a massive superfield with mass $M$ and superhelicity $\hf (n+1) \s$.

\section{Discussion and summary of results} \label{TAS2Discussion}
We are now in the position to compare massless and massive theories with $(1,1)$ AdS supersymmetry to their $(2,0)$ counterparts. Before doing this, it is worth summarising the main results of this chapter.

In section \ref{TAS2Reduction}, we initiated the program to derive the superspin projection operators in AdS$^{(3|1,1)}$. These superprojectors map an arbitrary superfield in AdS$^{(3|1,1)}$ satisfying the constraint \eqref{TAS2OnShellCondition2} to an on-shell superfield \eqref{TAS1OnshellConditions}. Making use of the minimal uplift method, we computed the explicit form of the rank-$1$ and rank-$2$ superprojectors for the first time. They are given by eqs. \eqref{TAS11Projetors1} and \eqref{TAS11Projetors2}, respectively. Important to the construction of these superspin projection operators were the Casimir operators of the ${\rm AdS}^{(3|1,1)} $ isometry algebra.  These Casimir operators  \eqref{TAS2Casimiroperators} were derived in the superfield representation for the first time in this thesis.

Sections \ref{Section5} and \ref{Section6} detail the $(1,1) \rightarrow (1,0)$ AdS superspace reduction of the transverse and longitudinal formulations for the massless half-integer superspin multiplets, respectively \cite{HutchingsHutomoKuzenko}. When reduced to $\cN=1$ AdS superspace, we found that the actual difference between the transverse and longitudinal models lies in the structure of auxiliary superfields. Let us refer to eq.~\eqref{ReductionTransHI} which shows that the transverse formulation gives rise to two decoupled $\cN=1$ supersymmetric theories \cite{HutchingsHutomoKuzenko}:
\bea 
\mathbb{S}^{\perp}_{(s+\hf)} [\mathfrak{H}_{\a(2s)},\G_{\a(2s-2)}, \bar{\G}_{\a(2s-2)}] &=& S^\parallel_{(s+\frac{1}{2})}[H_{\a(2s+1)},L_{\a(2s-2)},  \F_{\a(2s-1)}] \non ~\\ 
&&+ S^\parallel_{(s)}[H_{\a(2s)},V_{\a(2s-2)}, \O_{\a(2s-1)}]~.
\eea
In the longitudinal case, from eq.~\eqref{RATHI} we have that \cite{HutchingsHutomoKuzenko}
\bea 
\mathbb{S}^{\parallel}_{(s+\hf)} [\mathfrak{H}_{\a(2s)},G_{\a(2s-2)}, \bar{G}_{\a(2s-2)}] =&& S^\parallel_{(s+\frac{1}{2})} [H_{\a(2s+1)},L_{\a(2s-2)},  \F_{\a(2s-3)}] \non ~\\ 
&&+ S^\parallel_{(s)}[H_{\a(2s)},V_{\a(2s-2)}, \O_{\a(2s-3)}]~.
\eea
We further showed that the superfields $\F_{\a(2s-1)}, \O_{\a(2s-1)}, \F_{\a(2s-3)}, \O_{\a(2s-3)}$ are all auxiliary. Once they are eliminated, each formulation then leads to the same ${\cN}=1$ supersymmetric higher-spin actions \cite{HutchingsHutomoKuzenko}:
\bsubeq
\begin{equation}
\begin{tikzpicture}[level distance=40mm, baseline=(current  bounding  box.center)]
\tikzstyle{level 2}=[sibling distance=10mm, ->,black]
\node (a) at (0,0) {Transverse:~~~~~~~~} [parent anchor=east]
[child anchor=west,grow=east]
child {node (c) at (0,0) {$\mathbb{S}^{\perp}_{(s+\hf)}[\mathfrak{H}_{\a(2s)}, \G_{\a(2s-2)}, \bar \G_{\a(2s-2)}]$}  edge from parent[draw=none]
	[child anchor=west,grow=east]
	child [level distance=61mm] {node {$S^{\parallel}_{(s)}[H_{\a(2s)}, V_{\a(2s-2)}]~,$}}
	child [level distance=64.5mm] {node {$S^{\parallel}_{(s+\hf)}[H_{\a(2s+1)}, L_{\a(2s-2)}]$}}
}; \label{92a}
\end{tikzpicture}
\end{equation}
\begin{equation}
\begin{tikzpicture}[level distance=43mm, baseline=(current  bounding  box.center)]
\tikzstyle{level 2}=[sibling distance=10mm, ->,black]
\node (b) at (0,0) {Longitudinal: } [parent anchor=east]
[child anchor=west,grow=east]
child {node {$\mathbb{S}^{\parallel}_{(s+\hf)}[\mathfrak{H}_{\a(2s)}, G_{\a(2s-2)}, \bar G_{\a(2s-2)}]$} edge from parent[draw=none]
	child [level distance=61mm] {node {$S^{\parallel}_{(s)}[H_{\a(2s)}, V_{\a(2s-2)}]~.$}}
	child [level distance=64.5mm] {node {$S^{\parallel}_{(s+\hf)}[H_{\a(2s+1)}, L_{\a(2s-2)}]$}}
};
\end{tikzpicture}
\end{equation}
\esubeq

Sections \ref{Section7} and \ref{Section8} concern the $(1,1) \rightarrow (1,0)$ AdS superspace reduction of the longitudinal and transverse formulations for the massless \textit{integer} superspin multiplets, respectively. In $(1,1)$ AdS superspace, they are dual to each other. As in the half-integer case, we demonstrated that upon reduction to $\cN=1$ AdS superspace, these formulations differ by the structure of the auxiliary superfields. 
As shown in eq.~\eqref{723}, the longitudinal model is equivalent to \cite{HutchingsHutomoKuzenko}
\bea
\mathbb{S}^{\parallel}_{(s)} [U_{\a(2s-2)},G_{\a(2s)}, \bar{G}_{\a(2s)}] &=& S^\perp_{(s)}[H_{\a(2s)}, \cU_{\b;\a(2s-2)} ,  \F_{\a(2s-1)}] \non ~\\ 
&&+ S^\parallel_{(s)}[\tilde{H}_{\a(2s)}, V_{\a(2s-2)}, \O_{\a(2s-1)}]~.
\eea
On the other hand, eq.~\eqref{t-int} shows that the transverse model yields \cite{HutchingsHutomoKuzenko}
\bea
\mathbb{S}^\perp_{(s)}[U_{\a(2s-2)}, \G_{\a(2s)},\bar{\G}_{\a(2s)} ] &=& S^\perp_{(s)}[ H_{\a(2s)}, \tilde{\cU}_{\b;\a(2s-2)} , \F_{\a(2s+1)} ] ~ \non \\
&&+ S^\parallel_{(s)}[\tilde{H}_{\a(2s)}, V_{\a(2s-2)}, \O_{\a(2s+1)}]~.
\eea
The superfields $\F_{\a(2s-1)}, \O_{\a(2s-1)}, \F_{\a(2s+1)}, \O_{\a(2s+1)}$ are all auxiliary and thus can be integrated out from their corresponding actions. We showed that the resulting ${\cN}=1$ actions are the same in both formulations \cite{HutchingsHutomoKuzenko}:
\bsubeq \label{TLRed}
\begin{equation}
\begin{tikzpicture}[level distance=33.5mm, baseline=(current  bounding  box.center)]
\tikzstyle{level 2}=[sibling distance=10mm, ->,black]
\node (b) at (0,0) {Transverse:~~~~~~~~~~~~~~~~~~} [parent anchor=east]
[child anchor=west,grow=east]
child {node {$\mathbb{S}^{\perp}_{(s)}[U_{\a(2s-2)}, \G_{\a(2s)}, \bar \G_{\a(2s)}]$} edge from parent[draw=none]
	child [level distance=57.6mm] {node {$S^{\parallel}_{(s)}[H_{\a(2s)}, V_{\a(2s-2)}]~,$}}
	child [level distance=59mm] {node {$S^{\perp}_{(s)}[H_{\a(2s)}, \J_{\b;\,\a(2s-2)}]$}}
};
\end{tikzpicture}
\end{equation}
\begin{equation}
\begin{tikzpicture}[level distance=44mm, baseline=(current  bounding  box.center)]
\tikzstyle{level 2}=[sibling distance=10mm, ->,black]
\node (a) at (0,0) {Longitudinal: } [parent anchor=east]
[child anchor=west,grow=east]
child {node (c) at (0,0) {$\mathbb{S}^{\parallel}_{(s)}[U_{\a(2s-2)}, G_{\a(2s)}, \bar G_{\a(2s)}]$}  edge from parent[draw=none]
	[child anchor=west,grow=east]
	child [level distance=57.6mm] {node {$S^{\parallel}_{(s)}[H_{\a(2s)}, V_{\a(2s-2)}]~.$}}
	child [level distance=59mm] {node {$S^{\perp}_{(s)}[H_{\a(2s)}, \J_{\b;\,\a(2s-2)}]$}}
};
\end{tikzpicture}
\end{equation}
\esubeq

We are now in a position to compare the above $(1,1) \rightarrow (1,0)$ AdS reduction results with the ${\cN}=1$ supersymmetric higher-spin gauge theories obtained via $(2,0) \to (1,0)$ AdS reduction \cite{HK19}. There exist two off-shell formulations for massless half-integer superspin multiplets with $(2,0)$ AdS supersymmetry, which are not dual to each other.\footnote{Off-shell constructions for the massless integer superspin $(2,0)$ multiplets remains an open problem.} They are known as type II and type III series \cite{HK18}. Upon reduction to ${\cN}=1$ superspace, the type II series yields
\begin{equation} \label{type2red}
\begin{tikzpicture}[level distance=35mm, baseline=(current  bounding  box.center)]
\tikzstyle{level 2}=[sibling distance=10mm, ->,black]
\node (a) at (0,0) {Type \RomanNumeralCaps{2}:~~~~~~~} [parent anchor=east]
[child anchor=west,grow=east]
child{node (c) at (0,0) {$\mathbb{S}^{\text{\RomanNumeralCaps{2}}}_{(s+\hf)}[\mathfrak{H}_{\a(2s)}, \mathfrak{L}_{\a(2s-2)}]$}  edge from parent[draw=none]
	[child anchor=west,grow=east]
	child [level distance=56.8mm] {node {$S^{\perp}_{(s)}[H_{\a(2s)}, \J_{\b ;\a(2s-2)}]~.$}}
	child [level distance=59.3mm] {node {$S^{\parallel}_{(s+\hf)}[H_{\a(2s+1)}, L_{\a(2s-2)}]$}}
};
\end{tikzpicture}
\end{equation}
On the other hand, the type III series leads to
\begin{equation} \label{type3red}
\begin{tikzpicture}[level distance=40mm, baseline=(current  bounding  box.center)]
\tikzstyle{level 2}=[sibling distance=10mm, ->,black]
\node (b) at (0,0) {~~~~Type \RomanNumeralCaps{3}:} [parent anchor=east]
[child anchor=west,grow=east]
child {node {$\mathbb{S}^{\text{\RomanNumeralCaps{3}}}_{(s+\hf)}[\mathfrak{H}_{\a(2s)}, \mathfrak{V}_{\a(2s-2)}]$} edge from parent[draw=none]
	child [level distance=55mm] {node {$S^{\parallel}_{(s)}[H_{\a(2s)}, V_{\a(2s-2)}]~.$}
	}
	child [level distance=61mm] {
		node {$S^{\perp}_{(s+\hf)}[H_{\a(2s+1)}, \U_{\b ;\a(2s-2)}]$}
}};
\end{tikzpicture}
\end{equation}
We point out that unlike the situation for the $(1,1)$ AdS multiplets, the $\cN=1$ models obtained from the $(2,0) \to (1,0)$ AdS reduction do not involve any auxiliary superfields.
Additionally, reductions of the $(1,1)$ AdS multiplets only produce three of the four series of off-shell $\cN=1$ multiplets, with the exception of the transverse half-integer model, $S^{\perp}_{(s+\hf)}[ H_{\a(2s+1)}, \U_{\b;\, \a(2s-2)}]$. The latter can only be derived via $(2,0) \to (1,0)$ AdS reduction of the type III series. This formulation corresponds to the linearised supergravity model (for $s=1$), which does not admit a non-linear extension \cite{KT-M11}. 

In section \ref{TAS2TopMassSec} we studied the $(1,1) \to (1,0)$ AdS reduction of the two dually equivalent off-shell formulations for a massive superspin-$(s+\hf)$ multiplet, which are described by the actions \eqref{11Massive} \cite{HutomoKuzenkoOgburn2018}. It was shown in \cite{HutchingsHutomoKuzenko} that these two gauge-invariant models decouple into two off-shell topologically massive $\cN=1$ theories. The first $S^{\parallel}_{(s+\hf)}[H_{\a(2s+1)} ,L_{\a(2s-2)} |m]$ coincides with \eqref{N2Massive1}, while the second model $S^{\parallel}_{(s)}[H_{\a(2s)} ,V_{\a(2s-2)} |m]$ can be identified with \eqref{8.4a}. 
Thus, the $(1,1) \to (1,0)$ AdS reduction of the massive actions \eqref{11Massive} is only able to generate two out of the four
gauge-invariant formulations for topologically massive $\cN = 1$ supermultiplets.

In $(2,0)$ AdS superspace, the following off-shell gauge-invariant massive models were proposed in \cite{HK18}:
\begin{subequations} \label{20mass}
	\bea
	\mathbb{S}_{(s+\hf)}^{\rm II}[{\mathfrak H}_{\a(2s)} ,{\mathfrak L}_{\a(2s-2)}|m]&=& 
	\k {\mathbb{S}}^{(2s)}_{\rm SCHS} [ {\mathfrak H}_{\a(2s)}] 
	+ m^{2s-1} \mathbb{S}^{\rm II}_{(s+\hf)} [{\mathfrak H}_{\a(2s)} ,{\mathfrak L}_{\a(2s-2)} ]~, 
	\label{5.7a} \\
	\mathbb{S}_{(s+\hf)}^{\rm III}[{\mathfrak H}_{\a(2s)} ,{\mathfrak V}_{\a(2s-2)}|m]&=& 
	\k {\mathbb{S}}^{(2s)}_{\rm SCHS} [ {\mathfrak H}_{\a(2s)}] 
	+ m^{2s-1} \mathbb{S}^{\rm III}_{(s+\hf)} [{\mathfrak H}_{\a(2s)} ,{\mathfrak V}_{\a(2s-2)} ]~. \hspace{0.5cm}
	\label{5.7b}
	\eea
\end{subequations} 
Here ${\mathbb{S}}^{(2s)}_{\rm{SCHS}}
[ {\mathfrak H}_{\a(2s)}] $ is the superconformal higher-spin action \cite{HutomoKuzenkoOgburn2018},
\bea
{\mathbb{S}}^{(2s)}_{\rm{SCHS}}
[ {\mathfrak H}_{\a(2s)}] 
= - \frac{(-1)^s}{2^{s+1}}
\int 
\rd^{3|4}z 
\, \bm E\,
{\mathfrak H}^{\a(2s)} 
{\mathfrak W}_{\a(2s) }( {\mathfrak H})~,
\label{2.34}
\eea
where
${\mathfrak W}_{\a(2s) }( {\mathfrak H}) = \bar {\mathfrak W}_{\a(2s) }( {\mathfrak H}) $ 
is the higher-spin super-Cotton tensor of AdS$^{(3|2,0)}$.

Recalling the reduction result \eqref{type2red}, it is clear that \eqref{5.7a} decouples into two off-shell massive ${\cN}=1$ models: $S^{\parallel}_{(s+\hf)}[H_{\a(2s+1)} ,L_{\a(2s-2)} |m]$ and $S^{\perp}_{(s)}[H_{\a(2s)} ,{\Psi}_{\b; \,\a(2s-2)} |m]$.
On the other hand, it follows directly from \eqref{type3red} that the reduction of \eqref{5.7b} leads to $S^{\perp}_{(s+\hf)}[H_{\a(2s+1)} ,\U_{\b;\, \a(2s-2)}| m]$ and $S^{\parallel}_{(s)}[H_{\a(2s)} ,V_{\a(2s-2)} |m]$.
Thus, the $(2,0) \to (1,0)$ AdS reduction of massless half-integer actions \eqref{20mass} allows us to recover all four off-shell gauge-invariant formulations for topologically massive ${\cN}=1$ supermultiplets, which are described by \eqref{8.5} and \eqref{8.4}.

\begin{subappendices}
	
	\section{$(1,1)$ AdS superspace toolkit} \label{TAS2appendixA}
In this appendix, we provide a summary of essential identities for (1,1) AdS covariant derivatives in both the complex ${{\bm{\cD}}}_{\cA}=({\bm{\cD}}_{a}, {\bm{\cD}}_{\a}, {{\bm{\cDB}}}{}^{\a})$ and real $\bm \nabla_A = (\bm \nabla_a, \bm \nabla^{\1}_\a, \bm \nabla^{\2}_\a)$ basis.

Making use of the algebra in the complex basis \eqref{1.1}, one can derive the following useful identities: 
\begin{subequations} 
	\label{CovIdent}
	\bea 
	{\bm{\cD}}_\a{\bm{\cD}}_\b
	\!&=&\!\frac{1}{2}\ve_{\a\b}{\bm{\cD}}^2-2{\bar \m}\,M_{\a\b}~,
	\quad\qquad \,\,\,
	{ {\bm{\cDB}}}_\a{ {\bm{\cDB}}}_\b
	=-\frac{1}{2}\ve_{\a \b}{ {\bm{\cDB}}}{}^2+2\m\,{ M}_{\a \b}~,  \label{1.4a}\\
	{\bm{\cD}}_\a{\bm{\cD}}^2
	\!&=&\!4 \bar \m \,{\bm{\cD}}^\b M_{\a\b} + 4{\bar \m}\,{\bm{\cD}}_\a~,
	\quad\qquad
	{\bm{\cD}}^2{\bm{\cD}}_\a
	=-4\bar \m \,{\bm{\cD}}^\b M_{\a\b} - 2\bar \m \, {\bm{\cD}}_\a~, \label{1.4b} \\
	{ {\bm{\cDB}}}_\a{ {\bm{\cDB}}}{}^2
	\!&=&\!4 \m \,{ {\bm{\cDB}}}{}^\b { M}_{\a \b}+ 4\m\,  {\bm{\cDB}}_\a~,
	\quad\qquad
	{ {\bm{\cDB}}}^2{ {\bm{\cDB}}}_\a
	=-4 \m \,{ {\bm{\cDB}}}{}^\b {M}_{\a \b}-2\m\,  {\bm{\cDB}}_\a~,  \label{1.4c}\\
	\ [ {\bm{\cDB}}{}^2, {\bm{\cD}}_\a ]
	\!&=&\!4\ri {\bm{\cD}}_{\a\b}  {\bm{\cDB}}{}^\b +6 \m\,{\bm{\cD}}_\a = 
	4\rm i  {\bm{\cDB}}{}^\b {\bm{\cD}}_{\b\a} -6 \m\,{\bm{\cD}}_\a~,
	\label{1.4d} \\
	\ [ {\bm{\cD}}^2,{ {\bm{\cDB}}}_\a ]
	\!&=&\!-4\ri {\bm{\cD}}_{\b\a}{\bm{\cD}}^\b +6\bar \m\,{ {\bm{\cDB}}}_\a = 
	-4\rm i {\bm{\cD}}^\b {\bm{\cD}}_{\b\a} - 6 \bar \m\,{ {\bm{\cDB}}}_\a~
	\label{1.4e}
	\eea
\end{subequations} 
Here, we have made use of the shorthand notation ${\bm{\cD}}^2={\bm{\cD}}^\a{\bm{\cD}}_\a$ and ${ {\bm{\cDB}}}^2={ {\bm{\cDB}}}_\a { {\bm{\cDB}}}^\a$. 
These relations imply the identity 
\bea
{\bm{\cD}}^\a ( {\bm{\cDB}}^2- 6 \mu) {\bm{\cD}}_\a = {\bm{\cDB}}_\a ({\bm{\cD}}^2 - 6 \mub) {\bm{\cDB}}{}^\a ~.
\label{1.44}
\eea

Utilising the algebra in the real basis \eqref{1.3} allows for the computation of the important relations:
\begin{subequations}  
	\bea \label{A.9a}
	\bm \nabla^{{\1}}_\a \bm \nabla^{\1}_\b &=& \ri \bm \nabla_{\a\b} - 2\ri|\m|M_{\a\b}+\frac{1}{2}\ve_{\a\b}\bm (\bm \nabla^{\1})^2~,\\
	\bm \nabla^{{\1}\b} \bm \nabla^{\1}_\a \bm \nabla^{\1}_\b &=& 4\ri |\m|\bm \nabla^{\1}_\a~, \qquad \ [ \bm \nabla^{\1}_\a,\bm \nabla_{\b\g}] = 2|\m|\ve_{\a(\b}\bm \nabla^{\1}{}_{\g)}~,\\
	\{ \bm (\bm \nabla^{\1})^2, \bm \nabla^{\1}_\a  \} &=& 4\ri |\m|\bm \nabla^{\1}_\a \qquad	\ [\bm \nabla^{\1}_\a ,  \Box] = 2|\m|\bm \nabla_{\a\b}\bm \nabla^{{\1}\b} + 3|\m|^2\bm \nabla^{\1}_\a~,  \\
	\bm (\bm \nabla^{\1})^2 \bm \nabla^{\1}_\a &=& 2\ri |\m|\bm \nabla^{\1}_\a + 2\ri \bm \nabla_{\a\b}\bm \nabla^{{\1}\b} - 4\ri |\m|\bm \nabla^{{\1}\b} M_{\a\b}~, \label{A.9d}\\
	\label{A.9e}
	-\frac{1}{4}\bm (\bm \nabla^{\1})^2 \bm (\bm \nabla^{\1})^2 &=&  \Box - 2\ri|\m|\bm (\bm \nabla^{\1})^2+2|\m|\bm \nabla^{\a\b}M_{\a\b} -2|\m|^2M^{\a\b}M_{\a\b}~,\\
	\bm \nabla^{\2}_\a\bm \nabla^{\2}_\b &=&  \ri \bm \nabla_{\a\b} + 2\ri|\m|M_{\a\b}+\frac{1}{2}\ve_{\a\b}(\bm \nabla^{\2})^2~,\\
	\bm \nabla^{{\2}\b}\bm \nabla^{\2}_\a \bm \nabla^{\2}_\b &=& -4\ri|\m|\bm \nabla^{\2}_\a,~  \qquad \ [ \bm \nabla^{\2}_\a,\bm \nabla_{\b\g}] = -2|\m|\ve_{\a(\b}\bm \nabla^{\2}{}_{\g)}~, \\
	\{ (\bm \nabla^{\2})^2 , \bm \nabla^{\2}_\a \} &=& -4\ri|\m|\bm \nabla^{\2}_\a ~, \qquad \ [\bm \nabla^{\2}_\a ,  \Box] = - 2|\m|\bm \nabla_{\a\b}\bm \nabla^{{\2}\b} + 3|\m|^2\bm \nabla^{\2}_\a~, \\
	(\bm \nabla^{\2})^2\bm \nabla^{\2}_\a&=& -2\ri|\m|\bm \nabla^{\2}_\a + 2\ri \bm \nabla_{\a\b} \bm \nabla^{{\2}\b}+4\ri |\m|\bm \nabla^{{\2}\b}M_{\a\b}~, \\
	-\frac{1}{4}(\bm \nabla^{\2})^2 (\bm \nabla^{\2})^2 &=&  \Box + 2\ri|\m|(\bm \nabla^{\2})^2- 2|\m|\bm \nabla^{\a\b}M_{\a\b} -2|\m|^2M^{\a\b}M_{\a\b}~,	
	\eea
\end{subequations} 
Here, we have employed the shorthand notation  $(\bm \nabla^{\1})^2 = \bm{\nabla}^{{\1}\a} \bm{\nabla}^{\1}_\a$ and $(\bm \nabla^{\2})^2 = \bm{\nabla}^{{\2}\a} \bm{\nabla}^{\2}_\a$.
	
\end{subappendices}

\chapter{Conclusion and outlook} \label{TheEnd}


The main objective of this thesis was to present a systematic discussion of (super)spin projection operators and their corresponding applications in various (super)space backgrounds in three and four dimensions. This chapter summarises the general themes underpinning this thesis, and is accompanied by some concluding remarks on possible future research directions. For detailed statements concerning the original results obtained, we direct the reader to the last section of each of the research chapters \ref{ChapThreeDimensionalExtendedMinkowskiSuperspace}-\ref{TAS211AdS}.

Given a maximally (super)symmetric (super)space background, an unconstrained (super)field satisfying some Klein-Gordon type equation realises a collection of irreducible representations of the isometry (super)algebra. The purpose of the (super)spin projection operator is to extract the component of such a (super)field corresponding to the irreducible representation with maximal (super)spin. In other words, the (super)spin projection operator projects out the physical component of an arbitrary (super)field that can be identified as an elementary (super)particle. In the absence of the Klein-Gordon type equation, the (super)projection operators extract out the pure maximal (super)spin component from any unconstrained (super)field. The derivation of the explicit form of the (super)spin projection operators was completed for the first time in this thesis for the (super)space backgrounds: $\mb{M}^{3|2 \cN}$ \cite{BHHK}; AdS$_3$ \cite{HutchingsKuzenkoPonds2021}; AdS$^{3|2}$ \cite{HutchingsKuzenkoPonds2021}; AdS$_4$;\footnote{We found new realisations of the projectors of \cite{KP20} in terms of the Casimir operators of $\mf{iso}(3,1)$.} AdS$^{4|4}$ \cite{BHKP}; AdS$^{4|8}$ \cite{Hutchings2022}.

The computation of the (super)spin projection operators lead to an array of interesting applications enjoyed by all the operators, independent of the (super)space background on which they act. The first immediate application is obtained directly from the structural form of the (super)projectors. The (super)spin projection operators are differential operators which are non-local, as a direct consequence of idempotency. It was found that the poles of the (super)spin projection operators provide non-trivial information concerning the families of on-shell (super)fields which are compatible with gauge symmetry.

The poles of the higher-rank (super)spin projectors in three- and four-dimensional Minkowski (super)space correspond to a (super)field satisfying the massless Klein-Gordon equation, which we know to be the defining field equation of a massless gauge (super)field with maximal gauge symmetry. The story becomes considerably more interesting in the context of anti-de Sitter (super)space.
It was shown in \cite{KP20,BHKP} that the poles of the (super)spin projection operators encode all necessary information concerning partially massless (super)fields. One of the defining features of these exotic (super)fields is that they are compatible with higher-depth gauge symmetries. Thus, the poles of the (super)spin projection operators provide a novel program to generate all necessary information to construct a complete dictionary of on-shell (super)fields in any maximally (super)symmetric background. This program was essential in the formulation of partially massless supermultiplets in $\cN=1$ AdS superspace, which was achieved in \cite{BHKP}.

A motif of this thesis was the relationship between (super)conformal higher-spin theory and (super)spin projection operators. This connection is to be expected, as by definition, a linearised (S)CHS theory describes a pure (super)spin state off the mass-shell. Consequently, the (super)projectors are ubiquitous in the formulation of (S)CHS theories, as they are the operators which extract the pure (super)spin component from any arbitrary (super)field. In accordance with this correspondence, it follows that: the (super)spin projection operators can be used to derive free (S)CHS theories or; the (super)spin projection operators can be extracted from (S)CHS theories. This connection was studied extensively in this thesis.

In particular, it was shown in this thesis that all (S)CHS theories in three- and four-dimensional Minkowski and anti-de Sitter backgrounds can be constructed in terms of (super)spin projection operators \cite{BHHK,BHKP,HutchingsKuzenkoPonds2021,Hutchings2022}. The benefits of recasting the (S)CHS in terms of (super)projectors is that they naturally appear in a manifestly gauge-invariant and factorised form. Moreover, the superspin projection operators were vital in the construction of linearised SCHS actions in $3d$ $\cN$-extended Minkowski superspace,  for $ 3 \leq \cN \leq 6$,  which appeared for the first time in \cite{BHHK}.  The superspin projection operators were also extracted directly from the SCHS action in $\cN=2$ AdS superspace \cite{Hutchings2022},  which was recently computed in \cite{KR}.

As demonstrated above, the (super)spin projection operators have found many useful applications within the landscape of high-energy physics. In accordance with this, a natural research direction to consider would be their generalisations to other maximally symmetric (super)space backgrounds which were not considered in this thesis. The immediate goal would be to construct the superspin projection operators in $\cN$-extended anti-de Sitter superspace in three and four dimensions. The next step in this program would be to complete the construction of the 4$d$ $\cN=2$ and 3$d$ $\cN=(1,1)$ AdS superprojectors, which was initiated in this thesis. We conjecture that the poles of these operators in $\cN$-extended AdS superspace will be intimately related to partially massless supermultiplets, as was consistently observed in the $\cN=0$ and $\cN=1$ cases in three and four dimensions \cite{BHKP,HutchingsKuzenkoPonds2021,KP20}. It would also be interesting to explore the formulation of these operators in the following (super)spaces: $d$-dimensional (A)dS space\footnote{The AdS$_d$ spin projection operators have been recently constructed in  \cite{HutchingsPonds2023}.} and their corresponding $\cN$-extended generalisations and; other conformally flat backgrounds, such as the Robertson-Walker spacetime.

It would also be interesting to study further applications of the many (super)spin projection operators which we already have at our disposal \cite{Fronsdal1958,BehrendsFronsdal1957,IsaevPodoinitsyn2017,IsaevPodoinitsyn2018,AuriliaUmezawa1969,AuriliaUmezawa1967,SiegelGates1981,GatesGrisaruRocekSiegel1983,Segal2003,RittenbergSokatchev1981,Sokatchev1981,BHKP,HutchingsKuzenkoPonds2021,KP20,Hutchings2022}. The potential of these operators have yet to be fully realised, with many known research avenues (and hopefully many unknown) still requiring further investigation. The two applications which immediately come to mind are; the systematic construction of massive and massless actions via (super)spin projection operators and; a detailed study of the role that (super)spin projection operators play in determining the propagators of higher-spin particles, in the spirit of the works \cite{Singh1981,IsaevPodoinitsyn2018}.

In the case of the former, it was demonstrated by Fronsdal \cite{Fronsdal1958} (and later extended by Chang \cite{Chang1967zzc}) that the projection operators of $\mb{M}^4$ \cite{Fronsdal1958,BehrendsFronsdal1957} can be used to construct free actions describing the propagation of massive particles. It would be interesting to generalise the prescription of Fronsdal and Chang to $\cN=1$ Minkowski superspace in order to construct the superspace actions for free massive superfields of arbitrary superspin. A superspace formulation for massive  multiplets of half-integer superspin in $\mb{M}^{4|4}$ was recently computed by Koutrolikos \cite{Koutrolikos2020tel} in 2020, however, the integer superspin counterpart still remains unknown.

Moreover, the superspin projection operators of \cite{SiegelGates1981,GatesGrisaruRocekSiegel1983} were instrumental in the complete classification of all off-shell formulations of the massless gravitino \cite{GS} and graviton (linearised supergravity) \cite{GKP} supermultiplets. In these respective works, the authors constructed the most general quadratic action for the unconstrained spinor superfield $\J_\a$ and its complex conjugate $\bar{\J}_\ad$, in the case of the gravitino supermultiplet, and the real vector superfield $H_{\a \ad}$ for the graviton supermultiplet. This general action was rewritten explicitly in terms of the $\cN=1$ superprojectors \cite{SiegelGates1981,GatesGrisaruRocekSiegel1983}, where it was observed that the omission of certain projection operators leads to different gauge-invariant models. In particular, the Lagrangian formulations obtained from this prescription describe the same physical superfield content, but differ in their auxiliary superfield structure. In the spirit of these works \cite{GS,GKP}, it would be interesting to extend their prescription to complete the classification of all superfield formulations for massless multiplets of arbitrary spin in $\mb{M}^{4|4}$.

A portion of this thesis was devoted to the study of massless higher-spin supermultiplets possessing $\cN=2$ AdS supersymmetry. Supersymmetric field theories with $\cN=2$ AdS supersymmetry can be realised on two non-equivalent superspaces, namely $(1,1)$ and $(2,0)$ AdS superspace. Off-shell gauge formulations for the massless multiplet of (half-)integer superspin have already been constructed on $(1,1)$ AdS superspace in \cite{HutomoKuzenkoOgburn2018}, while only the half-integer superspin cases have been constructed on $(2,0)$ AdS superspace \cite{HK18}. The off-shell structure of
the half-integer superspin multiplets with (2,0) AdS supersymmetry drastically differs from that of their (1,1) counterparts. Direct comparison of these theories is difficult in a manifestly supersymmetric setting, since they are formulated in different superspaces, AdS$^{(3|2,0)}$ and AdS$^{(3|1,1)}$, respectively. However, both families of higher-spin theories can be reformulated in the same $\cN=1$ AdS superspace, and then the precise difference between the $(2,0)$ and $(1,1)$ higher-spin supermultiplets can be elucidated. The $(2, 0) \rightarrow (1, 0)$ reduction of the $(2,0)$ supersymmetric higher-spin theories of \cite{HK18}  was completed  in\cite{HK19},  where they were shown to decompose into a sum of two off-shell supermultiplets in AdS$^{3|2}$ which belong
to four series of inequivalent higher-spin gauge models.

In this thesis, we reduced all known $(1,1)$ theories  (both massless and massive) of \cite{HutomoKuzenkoOgburn2018} to AdS$^{3|2}$ \cite{BHHK}. It was seen that the reduction of the massless $(1,1)$ AdS multiplets of half-integer superspin could only reproduce two out of the four off-shell $\cN = 1$ multiplets which appeared in the reduction of their $(2,0)$ counterparts. Additionally, we reduced all four off-shell $\cN = 1$ multiplets to AdS$_3$. Here, it was shown that the $(2,0)$ and $(1,1)$ massless supermultiplets of half-integer superspin do not coincide at the component level. As a future exercise, it would be interesting to derive the massless multiplet of integer superspin in AdS$^{(3|2,0)}$, and perform the $(2, 0) \rightarrow (1, 0)$ reduction to see if a similar result to the half-integer case holds.

To conclude, in addition to the original results obtained we hope that this thesis will serve as a useful resource on the fundamental aspects and applications of (super)spin projection operators in various (super)space backgrounds. We also hope that we achieved our goal of illustrating the importance, and thus necessity, of deriving and studying these (super)projectors. We look forward to seeing whether these operators may be of some potential use in other areas of theoretical physics which are unbeknownst to us.


	\end{document}